\documentclass[preprint,12pt]{elsarticle}

\usepackage{graphicx}
\usepackage{amssymb}

\usepackage{lineno}

\usepackage[T1]{fontenc}

\usepackage{xurl}

\usepackage{amsmath}

\usepackage{enumitem}

\usepackage[dvipsnames]{xcolor}

\usepackage{placeins}

\makeatletter
\DeclareRobustCommand{\iscircle}{\(\mathord{\mathpalette\is@circle\relax}\)}
\newcommand\is@circle[2]{%
  \begingroup
  \sbox\z@{\raisebox{\depth}{$\m@th#1\bigcirc$}}%
  \sbox\tw@{$#1\square$}%
  \resizebox{!}{\ht\tw@}{\usebox{\z@}}%
  \endgroup
}
\makeatother

\newcommand{\dup}[1]{\mathrm{d}#1}
\newcommand{\Dup}[1]{\mathrm{D}#1}
\newcommand{\dd}[2]{\frac{\dup{#1}}{\dup{#2}}}
\newcommand{\DD}[2]{\frac{\Dup{#1}}{\Dup{#2}}}
\newcommand{\pp}[2]{\frac{\partial#1}{\partial#2}}
\newcommand{\lra}[1]{\langle#1\rangle}

\journal{Physics Reports}

\begin{document}

\begin{frontmatter}


\title{Social physics}



\author[aff1,aff2]{Marko Jusup}
\address[aff1]{Tokyo Tech World Hub Research Initiative (WRHI), Institute of Innovative Research,\\Tokyo Institute of Technology, Tokyo 152-8550, Japan}
\address[aff2]{Department of Physics, University of Rijeka, HR-51000 Rijeka, Croatia}

\author[aff1]{Petter Holme}

\author[aff3,aff4]{Kiyoshi Kanazawa}
\address[aff3]{Faculty of Engineering, Information, and Systems, The University of Tsukuba,\\Tennodai, Tsukuba 305-8573, Japan}
\address[aff4]{Japan Science and Technology Agency, PRESTO, Kawaguchi, Saitama 332-0012, Japan}

\author[aff1]{Misako Takayasu}

\author[aff1,aff5]{Ivan Romi\'{c}}
\address[aff5]{Statistics and Mathematics College, Yunnan University of Finance and Economics, Kunming 650221, China}

\author[aff6]{Zhen Wang}
\address[aff6]{School of Artificial Intelligence, Optics, and Electronics (iOPEN),\\Northwestern Polytechnical University, Xi'an 710072, China}

\author[aff7]{Sun\v{c}ana Ge\v{c}ek}
\address[aff7]{Division for Marine and Environmental Research, Ru{\dj}er Bo\v{s}kovi\'{c} Institute,\\HR-10002 Zagreb, Croatia}

\author[aff8]{Tomislav Lipi\'{c}}
\address[aff8]{Division of Electronics, Ru{\dj}er Bo\v{s}kovi\'{c} Institute, HR-10002 Zagreb, Croatia}

\author[aff9,aff10,aff11a,aff11b]{Boris Podobnik}
\address[aff9]{Faculty of Civil Engineering, University of Rijeka, 51000 Rijeka, Croatia}
\address[aff10]{Faculty of Information Studies in Novo Mesto, 8000 Novo Mesto, Slovenia}
\address[aff11a]{Zagreb School of Economics and Management, 10000 Zagreb, Croatia}
\address[aff11b]{Luxembourg School of Business, 1450 Luxembourg}

\author[aff12]{Lin Wang}
\address[aff12]{Department of Genetics, University of Cambridge,\\Cambridge CB2 3EH, United Kingdom}

\author[aff13]{Wei Luo}
\address[aff13]{Department of Geography, Faculty of Arts \& Social Sciences National,\\University of Singapore, Singapore 117570, Singapore}

\author[aff7]{Tin Klanj\v{s}\v{c}ek}

\author[aff14a,aff14b]{Jingfang Fan}
\address[aff14a]{School of Systems Science, Beijing Normal University, Beijing 100875, China}
\address[aff14b]{Potsdam Institute for Climate Impact Research (PIK), 14412 Potsdam, Germany}

\author[aff15,aff16,aff17]{Stefano Boccaletti}
\address[aff15]{CNR---Institute for Complex Systems, 50019 Florence, Italy}
\address[aff16]{Universidad Rey Juan Carlos, 28933 M\'ostoles, Madrid, Spain}
\address[aff17]{Moscow Institute of Physics and Technology, National Research University,\\141701 Moscow Region, Russia}

\author[aff18,aff19,aff20]{Matja{\v z} Perc\corref{cor1}}
\address[aff18]{Faculty of Natural Sciences and Mathematics, University of Maribor,\\2000 Maribor, Slovenia}
\address[aff19]{Complexity Science Hub Vienna, 1080 Vienna, Austria}
\address[aff20]{Department of Medical Research, China Medical University Hospital,\\China Medical University, Taichung 404332, Taiwan{\newpage}}
\cortext[cor1]{matjaz.perc@gmail.com}

\begin{abstract}
Recent decades have seen a rise in the use of physics methods to study different societal phenomena. This development has been due to physicists venturing outside of their traditional domains of interest, but also due to scientists from other disciplines taking from physics the methods that have proven so successful throughout the 19th and the 20th century. Here we dub this field `social physics' and pay our respect to intellectual mavericks who nurtured it to maturity. We do so by reviewing the current state of the art. Starting with a set of topics that are at the heart of modern human societies, we review research dedicated to urban development and traffic, the functioning of financial markets, cooperation as the basis for our evolutionary success, the structure of social networks, and the integration of intelligent machines into these networks. We then shift our attention to a set of topics that explore potential threats to society. These include criminal behaviour, large-scale migrations, epidemics, environmental challenges, and climate change. We end the coverage of each topic with promising directions for future research. Based on this, we conclude that the future for social physics is bright. Physicists studying societal phenomena are no longer a curiosity, but rather a force to be reckoned with. Notwithstanding, it remains of the utmost importance that we continue to foster constructive dialogue and mutual respect at the interfaces of different scientific disciplines.
\end{abstract}


\end{frontmatter}

\tableofcontents


\section{Prologue: The physical roots of multidisciplinarity}
\label{S:1&2}

The present text is perhaps best described as a journey through what has become an extremely active and diverse research field known under the umbrella term \textit{social physics}. Before an interested reader embarks on this journey with us, it is only fair to inform them of our motivation and rationale. Doing so, at the very least, requires (i) defining what social physics means (to us), (ii) outlining the case for its importance, and (iii) elaborating the underlying line of thought that connects the topics covered henceforward.

The methods of probability and statistics first flourished among social scientists who sought quantitative regularities revealing the inner workings of society~\cite{ball2002physical}. This inspired the founders of statistical physics in the 19th century to move away from Newtonian determinism and embrace a probabilistic description of ideal gases. Today, however, physicists are completing the circle by applying physical methods (oftentimes those of statistical physics) to quantify social phenomena~\cite{ball2002physical}. For our purposes here, we decide to adopt a broad, operational definition of social physics. Specifically, social physics is a collection of active research topics aiming to resolve societal problems to which scientists with formal training in physics have contributed and continue to contribute substantially. Although the precision and rigour of such a definition may be questioned, we are in a good company when relying on what physicists actually do to define (an aspect of) physics~\cite{sinatra2015century}. We also believe that being inclusive and practical about what constitutes social physics makes us appreciate more the broad position of physics in modern science.

The twentieth century has often been called ``a century of physics''~\cite{bromley2002century}, and for good reasons too. The scientific method as practised by physicists achieved enormous success on all scales of reality. On the small end of things, quantum electrodynamics as the theory of the interaction between light and matter has been tested to within ten parts per billion (i.e., 10$^{-8}$)~\cite{gabrielse2006new} by examining if the dimensionless magnetic moment of the electron relates to the fine structure constant as predicted. On the large end of things, general relativity as the prevailing theory of gravitation has been tested, among others, by putting satellites in space, measuring the geodetic effect with an error of 0.2\,\% and the frame dragging effect caused by Earth's rotation with an error of about 19\,\%~\cite{everitt2011gravity} relative to predictions. The latter error, incidentally, amounts to 37\,mas which is best put into perspective by the words of investigator Francis Everitt that 1\,mas ``is the width of a human hair seen at the distance of 10 miles.'' What is important for us is that these enormous successes of physics have caught the attention of scientists from other disciplines, and have led to attempts to generate similar successes using physics-like quantitative methods. This is explicitly admitted by some disciplines; in ecology, for example, metabolic theory~\cite{nisbet2000molecules, brown2004toward, kooijman2010dynamic, jusup2017physics} and mechanistic niche modelling~\cite{kearney2009mechanistic, kearney2020nichemapr} draw heavily from thermodynamics. More generally, though, the adoption of physics-like quantitative methods is best reflected in the proliferation of physical and mathematical modelling in disciplines as diverse as epidemiology~\cite{grassly2008mathematical, sun2016pattern}, virology~\cite{wodarz2002mathematical, beauchemin2011review}, neuroscience~\cite{kaplan2011explanatory, bassett2018nature}, medicine~\cite{michor2004dynamics, eftimie2016mathematical},
psychology~\cite{abelson1967mathematical, rodgers2010epistemology}, sociology~\cite{sorensen1978mathematical, carrington2005models}, and countless others. But if the ultimate goal is to replicate the success of physics, whom better to call for help than physicists themselves? All this explains to a decent degree why physics is at the roots of the modern shift to multidisciplinarity, and why physicists publish more multidisciplinary physics articles than articles in physics journals~\cite{sinatra2015century}.

Although the success of physics over the past hundred year or so is impossible to dispute, signs of a progress slowdown~\cite{schweber1993physics, lykken2014supersymmetry} and dissatisfaction with fundamentals~\cite{smolin2007trouble, hossenfelder2018screams} have been brewing, consequently setting in motion multiple searches for `new physics'~\cite{alves2012simplified, safronova2018search}. Until such physics is found, however, many a young physicist may seek to employ their strong quantitative skills elsewhere. A well-known example is attempts by physicists to, both through research in academia~\cite{farmer2000physicists} and practice on Wall Street, enrich the world of finance.

The described state of affairs places physics squarely at the roots of multidisciplinarity. It is then hardly surprising that much of the multidisciplinary work conducted by physicists aims at resolving societal problems. We call that work social physics, believing that otherwise we would be diminishing the rightful role of physics in today's multidisciplinary movement. Provided the reader is willing to agree with us or, at least, give us the benefit of the doubt, it becomes glaringly obvious that the scope of social physics is enormous. How did we go about narrowing down a relevant set of topics for the present text?

Our focus was twofold, asking what enables or constitutes the modern way of living and what perturbs or threatens it. The majority of human population now lives in cities~\cite{un2018world} because of more healthcare, education, and employment opportunities. Despite their advantages, cities suffer from many problems, traffic being among the more acute ones~\cite{caves2005encyclopedia}. We therefore started by overviewing the contributions of physicists to research in \textit{urban dynamics} and \textit{traffic flows}. The prosperity of cities is in many ways tied to markets, and over the past two decades financial markets in particular had an enormous impact on urban life, innovation, and planning~\cite{cohen2011cities}. This warrants a better understanding of financial markets, which is the aim of the chapter on \textit{econophysics}. Life in cities, and civilised life in general, is based on widespread cooperativeness. The evolution of \textit{cooperation} accordingly deserves a chapter of its own, even more so given that this is a research domain in which physicists have been especially active~\cite{perc2017statistical}. Human population is furthermore organised in social networks, whose structure is entwined not only with the evolutionary dynamics of cooperation, but also many other dynamical processes of societal relevance~\cite{harush2017dynamic}. Probing network structure and their separation into \textit{communities} could therefore not be overlooked. An important realisation in this context is that computers increasingly take part in shaping social networks, especially so with the advent of human-like artificial intelligence. The present state of affairs and current technological trends, in fact, necessitate a candid discussion about \textit{human-machine networks}.

Among phenomena that perturb or threaten the modern way of living, crime is a conspicuous one, with impacts at large, societal~\cite{vandijk2007mafia} and small, community~\cite{taylor1995impact} scales. Interestingly, as the chapter on \textit{criminology} will demonstrate, both the evolutionary dynamics of cooperation and network-structure analyses prove useful in gaining insights into crime fighting and criminal organisations. In contrast to crime, \textit{migrations} per se come with positive effects, such as helping to alleviate labour-force deficits and age-structure imbalances in ageing populations~\cite{bijak2008replacement}, but there are many caveats. Developing countries, for example, were supposed to receive demographic dividends form their favourable workforce-to-dependants ratios, but substantial value has been lost to brain drain, that is, emigration of highly educated young adults~\cite{williamson2013demographic}. Much more consequential is when population displacements are triggered by environmental shifts or geopolitical instabilities; on the one hand, people losing homes is a humanitarian crisis, while on the other hand, countries quickly absorbing sizeable immigration fuels nationalism and xenophobia~\cite{postelnicescu2016europe, holmes2016representing}. Movements of people across large distances, especially at a fast pace of today, make humankind vulnerable to contagions~\cite{fineberg2014pandemic}. The chapter on \textit{contagion phenomena} covers disease transmissions on global and local scales, and overviews the budding subfield of digital epidemiology. Before closing off this review, we shift the focus to natural surroundings that support human life in the first place. The chapter on \textit{environment} puts emphasis on the proliferation of chemicals whose effects, particularly synergistic ones, remain partly understood at best~\cite{eggen2004challenges}. Even more critical is \textit{global climate change}~\cite{kirilenko2007climate, louis2008climate, wheeler2013climate}. Somewhat surprising in this context is the use of network science to unravel the intricacies of Earth's climate system. This only goes to show how versatile physics and its methods are, which we hope will inspire physicists to further nurture multidisciplinarity, as well as scientists from other disciplines to maintain a dialogue with physicists when resorting to quantitative approaches and tools.

\section{Urban dynamics}
\label{S:UD}

Cities are archetypal examples of complex systems~\cite{batty2008size}. They are, to some extent, self-organised, in other aspects planned. They need hierarchical and interdependent distribution systems. They exist across an extraordinary range of scales. They interact in complex networks, they consist of complex networks, and they are built by people interacting through complex networks. Many aspects of urban science do not fit the methodologies of physicists. This section will discuss the current state of urban science~\cite{batty2013new}, in particular the topics that interest physicists~\cite{barthelemy2019statistical}.

\subsection{The definition of a city}

Assume that we know the locations of all humans, at all times, and all buildings and infrastructures, then how can we decide what a city is? This is far from an easy question, and one soon realises that any simple solution will not overlap perfectly with the existing conventions. Most of the research that we present here use an administrative definition of a city.  We will, however, look at attempts to define cities from the population distribution.

Even if one starts from population density data, one can usually be completely independent of pre-defined administrative borders. For example, Ref.~\cite{arcaute2015constructing} uses the population of the most fine-grained subdivision of the United Kingdom (wards) for this purpose. The actual population count comes from the home address reported in a census.

\begin{figure}[!t]
\makebox[\textwidth][c]{\includegraphics[scale=1.0]{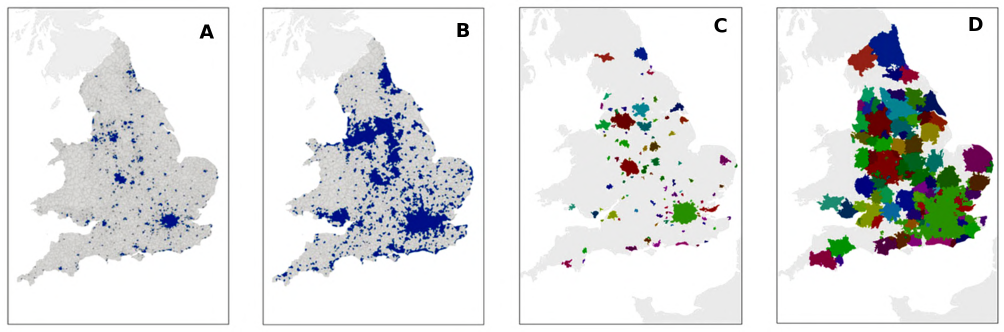}}
\caption{Two methods of defining cities in England. \textbf{A,\,B,} Results of the density thresholding method with thresholds $\rho_0=24$\,ha$^{-1}$ and $\rho_0=2$\,ha$^{-1}$, respectively. \textbf{C,\,D,} Results of the commuting-based method with minimum populations of 50,000 and commuting flow thresholds 40\,\% and 5\,\%, respectively.\newline
Source: Reprinted figure from Ref.~\cite{arcaute2015constructing} under the Creative Commons Attribution 4.0 International (CC BY 4.0).}
\label{fig:defining_cities}
\end{figure}

One idea for the definition of a city, akin to percolation theory~\cite{rozenfeld2008laws}, is to identify regions with a population density $\rho$ above a particular threshold $\rho_0$. Ref.~\cite{arcaute2015constructing} defines a city by requiring that $\rho_k>\rho_0$, for all wards $k$ at the boundary. If wards belonging to the city surround a ward $k$, we consider it a part of the city even if $\rho_k<\rho_0$. See Fig.~\ref{fig:defining_cities} for the results of this algorithm when applied to the population density of England.

Another approach to defining cities is to group smaller divisions with larger ones if at least a fraction of the population of the smaller commutes to the larger, and there is not a larger division that attracts even more commuters. In order for this approach to give sensible results, one needs to start from some seed regions more populous than a given threshold. From such a seed region, the algorithm of Ref.~\cite{arcaute2015constructing} recursively adds wards to a region that at least a certain fraction of people commute to. If there is more than one region that draws a fraction of commuted above the threshold, then the ward is added to the region attracting most commuters. When such a recursive procedure has converged, one is left with a city pattern such as in Fig.~\ref{fig:defining_cities}C,\,D.

\subsection{Size of cities}

George Kingsley Zipf noted in his 1949 \textit{The Principle of Least Effort}~\cite{zipf1949human} that city sizes $m$ follow a power-law distribution
\begin{linenomath}
\begin{equation}
    P(m) = \text{const.} \times m^{-\nu}.
\end{equation}
\end{linenomath}
Zipf found the exponent $\nu$ to be one and assumed that it was a universal value, but more recent studies argue that different regions of the world have different $\nu$~\cite{zanette1997role, soo2005zipf, malevergne2011testing}.

There is a vast number of mechanisms proposed for Zipf's law of city sizes, starting from Zipf himself~\cite{zipf1949human}. We will mention a few from the physics literature. First, Zanette and Manrubia suggested a multiplication-diffusion mechanism model operating on a square grid~\cite{zanette1997role}. From an initially even distribution of population density, the population $m$ at a random site $i$ is updated as
\begin{linenomath}
\begin{equation}\label{eq:zanette_manrubia_reaction}
    m_i(t)=\begin{cases}
    (1-q)p^{-1} & \text{with probability $p$}\\
    q^{-1}(p-1) & \text{otherwise}
    \end{cases}
\end{equation}
\end{linenomath}
Then a fraction $\alpha$ of the population is redistributed to the surrounding cells. This model produces emergent power-laws in agreement with Zipf, for a broad range of parameter values.

Ref.~\cite{marsili1998interacting} proposes a model in which the arrival $w_a$ and departure rates $w_d$ from a city of size $m$ depend on $m$ according to
\begin{linenomath}
\begin{subequations}
\begin{align}
    w_a &= \frac{m^2}{m_0}+m\\
    w_d &= e^{1/m*}\left[\frac{m^2}{m_0}+a\right]
\end{align}
\end{subequations}
\end{linenomath}
where $m^*$, $m_0$, and $a$ are parameters. These rules are repeatedly applied in combination with a growth of the number of cities (by occasionally adding cities of population one). This gives, for some parameter range, an emergent city-size distribution of
\begin{linenomath}
\begin{equation}
    \nu=\frac{1}{1+(a-1)m_0}.
\end{equation}
\end{linenomath}

In both of the above models, there is an element of `rich-gets-richer' (often called `Gibrat principle'~\cite{simon1955class}, sometimes the `Matthew effect'~\cite{merton1968matthew}, or `cumulative advantage'~\cite{cumulative_advantage}) that larger cities manage to attract more people and thus grow faster than smaller cities. Thus, many authors cite Herbert Simon's model for emergent power-law distribution~\cite{simon1955class} as an explanation for city size distributions. This mechanism has been revived and adapted for city growth in the relatively recent economics literature~\cite{gabaix1999zipf}. Indeed, out of all mechanisms generating power-laws~\cite{newman2005power}, the models specifically trying to explain city growth, that we are aware of, all seem to incorporate a rich-gets-richer mechanism. There is also some direct empirical evidence for a rich-gets-richer growth of cities~\cite{decker2007global}.

\begin{figure}[!t]
\centering\includegraphics[scale=1.0]{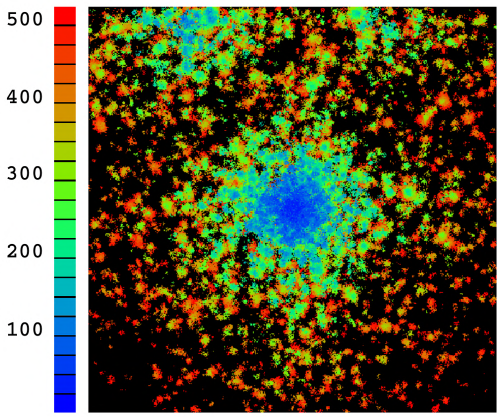}
\caption{Growth of a modelled city. Shown is an artificial city plan generated by the Manrubia-Zanette-Sol\'e model~\cite{manrubia1999transient}. The colour bar encodes a timeline for the development of the city. The model reproduces many features of real cities, but there are also striking differences compared to reality; in the model, for example, the oldest regions are completely embedded in newer ones.\newline
Source: Reprinted figure from Ref.~\cite{manrubia1999transient}.}
\label{fig:city_growth}
\end{figure}

\subsection{City growth}

Another issue about cities that has interested physicists is the spatial growth of cities. Indeed, the Zanette-Manrubia model has also been proposed as a model for the spatial growth of cities~\cite{manrubia1999transient}. This is maybe not so surprising because, in the spirit of self-similarity, the population distribution within a city could be similar to that of a region containing many cities.
Fig.~\ref{fig:city_growth} shows one example of a result of this model. Although the model manages to reproduce many features of real city growth, one immediately spot discrepancies when comparing the model output to real data. The most striking difference is that the oldest regions of the Zanette-Manrubia are completely embedded in newer built environments. However, in real data, they could border non-built land-use types.

In addition to reaction-diffusion type models of city growth, some models are somewhat similar to diffusion-limited aggregation (DLA)~\cite{witten1981diffusion}. For example, Ref.~\cite{andersson2002urban} follows a Markov random field framework, but adds many rules from the urban planning literature or the authors' observations. This model could be coupled with geographic data or similar to improve its predictive power. Another influential paper motivated by DLA, or rather its weaknesses, to model city growth, is Ref.~\cite{makse1995modelling}. In this model, the nodes are successively added to the cluster (representing a city) with a logarithmically decaying probability of the distance to occupied areas.

Finally, we note that predicting city growth patterns does not only interest physicists. For example, see Ref.~\cite{seto2012global} for a recent model by geographers of the co-evolution of land-use and population density.

\begin{figure}[!t]
\centering\includegraphics[scale=1.0]{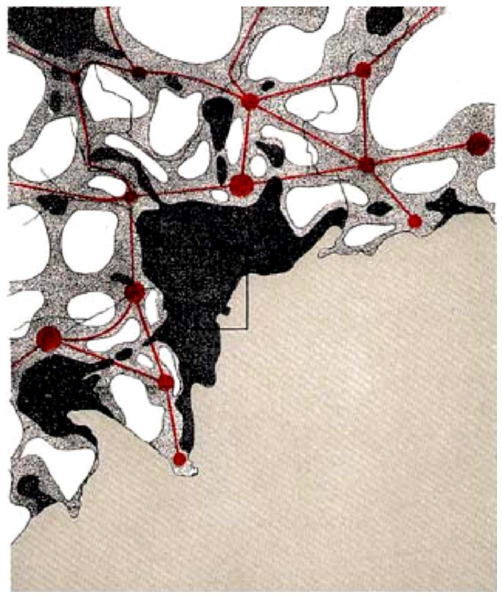}
\caption{A figure from a 1963 essay by the Greek-American architect C.\,A. Doxiadis to argue that for a city to grow without losing its vital functions, its centre needs to be supported by a network of subcenters---``to create a new network of lines of transportation and communication which do not lead towards the centre of existing cities but towards completely new nodal points''~\cite{doxiadis1963ekistics}.\newline
Source: Courtesy of Constantinos A. Doxiadis Archives.\newline
\textcopyright~Constantinos and Emma Doxiadis Foundation.}
\label{fig:ekistics_network}
\end{figure}

\subsection{Networks within and of cities}

As already alluded to, it is not straightforward to demarcate cities from their surrounding. Therefore many of the principles that relate different cities also apply to the organisation of cities themselves. Geographers had invented simple models to explain existing patterns and determine the optimal spatial networks long before physicists turned to this problem (for example, see Fig.~\ref{fig:ekistics_network}). We recommend reading Ref.~\cite{abler1971spatial}, which is an almost 50 years old textbook but will feel very modern for anyone working on spatial, temporal, or higher-order network structures, or the modelling of complex socioeconomic systems.

One influential early model for the network of cities was Walter Christaller's 1933 `central place theory'~\cite{christaller1966central}. It assumes an underlying featureless landscape of uniformly distributed resources. In such a scenario, larger cities would primarily organise in a hexagonal lattice. Secondary, smaller cities would fill the gaps around the central places, and so on. The economist August L\"osch derived a more flexible and more economics-favoured location theory than Christaller's in his 1940 \textit{The Economics of Location}~\cite{losch1954economics}. L\"osch also concludes that in a structureless world, human settlements would be organised into a hexagonal pattern. Furthermore, cities would have a fat-tailed size distribution~\cite{parr1973structure}, although deriving Zipf's law was not an explicit goal of L\"osch.

\begin{figure}[!t]
\centering\includegraphics[scale=1.0]{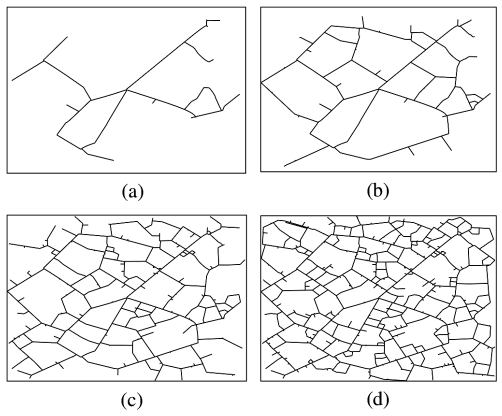}
\caption{Street patterns generated by an algorithm. Panels correspond to time progression in the model: (a) $t=100$; (b) $t=500$; (c) $t=2000$; and (d) $t=4000$. Initially, constructed road form a tree-like structure, but loops start appearing as the density of roads increases.\newline
Source: Reprinted figure from Ref.~\cite{barthelemy2008modeling}.}
\label{fig:bf_street_patterns}
\end{figure}

Physicists have spent more effort trying to understanding the evolution of the networks within cities~\cite{barthelemy2016structure,bettencourt:book} than networks of cities. One example is Barth{\'e}lemy and Flammini's model of the growth of street patterns~\cite{barthelemy2008modeling}. This model works by adding `centres' that are then connected according to the following rule. Say that $A$ and $B$ are neighbouring centres, and $M$ is a tip of a nearby road under construction. The road will grow from the tip $M$ in the direction of the vector
\begin{linenomath}
\begin{equation}
    \frac{\overrightarrow{MA}}{|MA|}+\frac{\overrightarrow{MB}}{|MB|},
\end{equation}
\end{linenomath}
such that the cumulative distance from the centres $A$ and $B$ to the road network is minimised (Fig.~\ref{fig:bf_street_patterns}). When the road $M$ reaches a point on the line between $A$ and $B$, a straight road segment is added from $A$ to $B$. For further details about this model, see Ref.~\cite{barthelemy2008modeling}.

Other models of spatial networks typically also operate by successively adding points and connecting these to the existing network~\cite{barthelemy2018morphogenesis}. A more general model of spatial growth that could work as a model for road networks is Gastner and Newman's model in Ref.~\cite{gastner2006shape}. This algorithm associates a cost to all the links that is proportional to
\begin{linenomath}
\begin{equation}
\label{eq:gastner_newman1}
c_{ij}=\lambda \sqrt{N}d_{ij}+(1-\lambda),
\end{equation}
\end{linenomath}
where $d_{ij}$ is the distance between points $i$ and $j$, and $\lambda$ is a parameter governing the relative cost of the distance of the link to its existence. Then the algorithm seeks a set of links $E$ that minimises
\begin{linenomath}
\begin{equation}
\label{eq:gastner_newman2}
\sum_{(i,j)\in E} d_{ij} \text{~given~} \sum_{(i,j)\in E} c_{ij} < C
\end{equation}
\end{linenomath}
for a parameter $C$ representing the total budget of the project. If $\lambda$ is large, the networks become more like urban infrastructures, otherwise the networks are rather like airline maps. The Gastner-Newman model is similar to Fabrikant-Koutsoupias-Papadimitriou model~\cite{fabrikant2002heuristically} of Internet evolution in the sense that it balances the cost of physical links and the presence of the link in the network.

\begin{figure}[!t]
\centering\includegraphics[scale=1.0]{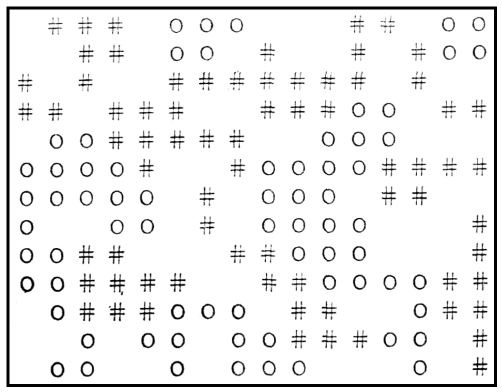}
\caption{Snapshot of the converged state of Schelling's segregation model. Symbols `\iscircle' or `\#' correspond to individuals of two different ethnicities occupying cells in an $L\times{}L$ square grid. Convergence is achieved by repeatedly following a simple rule such that individuals who have more than a fraction $b$ of neighbours belonging to the other ethnicity move to a another, randomly selected empty cell.\newline
Source: Reprinted figure from Ref.~\cite{schelling1971dynamic}.}
\label{fig:schelling}
\end{figure}

\subsection{Segregation}

Studies of segregation by physicists is essentially equal to studies of the model by Schelling~\cite{schelling1971dynamic}. Just one glance at Schelling's paper should be enough for the reader to understand why this particular model is popular among physicists (Fig.~\ref{fig:schelling}). The background of the model was the racial tensions in the 1960s USA in general and residential segregation in particular. Schelling used the model to argue that even if people are mostly tolerant of living close to others of another ethnicity, spatial constraints accentuate segregation. The model works as follows:
\begin{enumerate}
\item Consider an $L\times L$ square grid in which every cell can be empty or occupied by a resident of one of two ethnicities (`\iscircle' or `\#' in Fig.~\ref{fig:schelling}).
\item Initially, distribute $\lfloor{fL^2/2}\rfloor$ \iscircle{s} and equally many \#s randomly on the grid, where $0<f<1$ quantifies occupancy.
\item Update the configuration by picking an occupied square and, if this is surrounded by more than a fraction $b$ of the opposite ethnicity, move it to a random unoccupied square. Schelling considered the eight nearest neighbours.
\item Repeat the previous step until all occupied squares are below the threshold. If $f$ is sufficiently small, the procedure will converge. Otherwise, the problem is ill-defined.
\end{enumerate}
Essentially, the final level of segregation, measured by the average fraction of neighbours of the same ethnicity, will be far from the threshold, and this increases non-linearly with both the threshold $b$ and occupancy $f$.

There are many papers in the physics literature dealing with this model, typically without interpreting the results in terms of residential segregation. For example, Ref.~\cite{vinkovic2006physical} reinterprets Schelling's model as a model of crystal growth, while Ref.~\cite{dallasta2008statistical} studies scaling properties of the interfaces of the emergent clusters. See also Ref.~\cite{stauffer2013biased} for an amusing account by Dietrich Stauffer.

\subsection{Scaling theory of cities}

A small city is not a small version of a large city. As mentioned, there has been a considerable hype around \textit{self similarity} and \textit{scale-free} patterns~\cite{bak1996how} that, to some extent, has fuelled the development of models we have discussed. However, beyond the power-law size distribution, the way a city operates depends on its size. How things depend on size has, for long, been a common theme in biology and ecology. Note the difference to finite-size scaling in statistical physics where the goal is to extrapolate the results to the infinity limit (to study critical phenomena).

Scaling theory has recently come to the attention of physicists~\cite{west2017scale}. This interest comes from the physical theories of allometric scaling~\cite{west1997general}. One of the most influential papers is Ref.~\cite{bettencourt2007growth} that found that different sectors of the economy depend differently on the size of cities. For example, sectors that need people to collaborate---like research and development---scale superlinearly with city size. In contrast, facilities that need to exist relatively close in space---like gas stations---scale sublinearly. Ref.~\cite{um2009scaling} reestablishes these results in a framework more suitable for physics style modelling by using population density rather than city size as a basis for the scaling analysis. Ref.~\cite{bettencourt2013origins} proposes a model that relates many scaling exponents and finds the regions of parameter space where a city can exist.

\subsection{Human mobility}

Mobility studies mostly concern how many people move between two locations at a specific time. One can break down this topic in many ways. One can divide the people according to age, sex, or socioeconomic indices. One can separate different times of day, different months, or long-term trends. One can consider moving of residence, work, or the individual themself.

The origin of human mobility studies is Ernst Ravenstein's 1885 \textit{The laws of human migration}~\cite{ravenstein1885laws}. Ravenstein noticed, among other things, that the distance people move (their home) follows a skewed distribution---most people move only a short distance.

A more quantitative mobility law is the \textit{gravity law} stating that the number of people $T_{ij}$ travelling between two locations $i$ and $j$ is
\begin{linenomath}
\begin{equation}
    T_{ij} = c \frac{P_iP_j}{d_{ij}^\delta},
\end{equation}
\end{linenomath}
where $c$ and $\delta$ are constants, $d_{ij}$ is the distance between $i$ and $j$, and $P_i$ the population at location $i$. This relation was first studied by Zipf~\cite{zipf1946hypothesis}, who only considered the exponent $\delta = 1$. The phrase `gravity law' was coined later in the transportation literature, so it does not appear like Newtonian mechanics played any deeper role in this development than providing a namesake. Subsequent studies have tried to measure and explain the exponent~\cite{chen2018scaling} and otherwise improve the gravity law by adding information about the locations~\cite{lee2015relating}.

The gravity law was recently improved by the \textit{radiation model} of human travel stating that
\begin{linenomath}
\begin{equation}
    T_{ij} = T_i\frac{P_iP_j}{(P_i + s_{ij})(P_i + P_j + s_{ij})},
\end{equation}
\end{linenomath}
where $T_i=\sum_j T_{ij}$ and $s_{ij}$ is the number of people in the circle centred on $i$ and $j$ at the perimeter. The radiation model's main advantage is that it builds on some simple mechanistic assumptions, whereas the gravity model is merely a statistical relation. Indeed, the radiation model's assumptions are rather reminiscent of Stouffer's theory of intervening opportunities~\cite{stouffer1940intervening}, stating:
\begin{quote}
The number of persons going a given distance is directly proportional to the number of opportunities at that distance and inversely proportional to the number of intervening opportunities.
\end{quote}
Stouffer had moving to change work in mind, and the opportunities in question were job opportunities.

\subsection{Trajectory analysis}

With the advances in position tracking technology over the last couple of decades, researchers have gotten access to large datasets of people's trajectories. Most commonly, researchers have used datasets from cellphones where people's locations are identified by the location of the cellphone tower their phone is connected to. From such studies, often comprising hundreds of thousands of individuals, the overarching discovery is just how predictable people are~\cite{gonzalez2008understanding, park2010eigenmode, song2010limits}. In most situations of our daily lives, given a sequence of locations visited, one could guess the next location by a probability of around 90\,\%~\cite{song2010limits}. This phenomenon is also observed in disasters, where peoples' routines could be forced to change completely. Still people have been observed to settle into new, highly predictive movement patterns~\cite{lu2012predictability}.

Another type of trajectory analysis is based on the shape of vehicular travel routes. The most fundamental quantity is the actual travel distance divided by the Euclidean distance between origin and destination. The average value of this quantity, often measured as a function of the Euclidean distance, has many names in the literature, here we follow Ref.~\cite{louf2013emergence} and call it \textit{detour index}. For very short travel distances, the detour index could be above two (the travel distance is over twice the straight distance). As the distance increases, the detour index converges after 20-30\,km to a value of around 1.3. Many generative models of city maps can reproduce this observation~\cite{minjin2018imbalance}.

Another type of study based on the car-travel routes is focusing on how the city shapes the trajectories. Ref.~\cite{lee2017morphology}, for example, investigates whether the fastest travel routes by cars between points at equal distance from the city centre, tend to move first in, then out, or vice versa. This tendency could be quantified by the \textit{inness}---the area enclosed by the trajectory and the shortest path from origin to destination on the same side the city centre, minus the corresponding area on the other side. Cities dominated by highway ring roads tend to have negative inness because people travel out to the ring road, follow it, then travel in towards the city centre to reach the goal. Ref.~\cite{lee2017morphology} measures inness for close to a hundred cities worldwide and relates it to socioeconomic indicators.

\begin{figure}[!t]
\centering\includegraphics[scale=1.0]{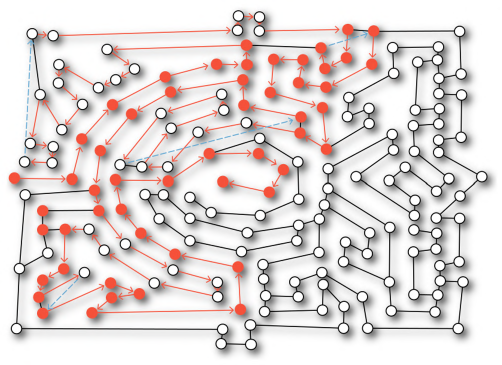}
\caption{Greedy navigators, agents following a stylised navigation strategy, getting lost in the Leeds Castle maze network. The graph is a `visibility graph' constructed by choosing as few nodes as possible so that the entire study area is visible from at least one node~\cite{deberg1997visibility}. The edges connect nodes that are visible from one another. The red arrows indicate the actual route travelled by a greedy navigator. The red solid circles denote the shortest path discovered by the greedy navigator. \newline
Source: Reprinted figure from Ref.~\cite{lee2012exploring}.}
\label{fig:maze}
\end{figure}

\subsection{Navigability}

Physicists have not only been interested in the structure of actual trajectories in urban car travel, but also how to find the destination when one does not have full information. The way people navigating their surroundings is an active area in cognitive science~\cite{wolbers2010determines}, and it is accepted that some cities, or buildings, are much easier to get lost in than others.

Any attempt to quantify the navigability of a city or building must rest on a model of how people exploit contextual information. Ref.~\cite{lee2012exploring}, for example, uses a framework called \textit{greedy navigators} in which the individuals have a notion of the direction to their destinations. At every intersection, a greedy navigator chooses the street, not previously travelled, that points most directly towards the target. In Fig.~\ref{fig:maze}, we show a proof of concept of how greedy navigators fail to find a short path in a garden maze (designed to be hard to navigate). Using greedy navigators one can obtain a navigability index similar to the detour factor---the average distance found by the greedy navigators divided by the actual average shortest distance.

\subsection{Future outlook}

So far, the physics of urban systems has not been driven by a paramount goal. Instead, it has been building on a collection of observations from data that physicists can, and do, try to explain. However, urban science~\cite{batty2013new}, in a multidisciplinary sense, has some general directions. From an engineering perspective, one would like to make cities sustainable and energy-efficient; whereas seen from social science---because of the ongoing urbanisation of our planet---one would like to foresee the problems and tap into the opportunities of ever-larger metropolises, perhaps via the spatio-socio-semantic analysis framework~\cite{luo2019cities}.

\FloatBarrier

\section{Traffic flows}
\label{S:TF}

Understanding and predicting traffic flow is important for social engineering in general and urban planning in particular~\cite{ni2015traffic, kessels2019traffic}. The study of traffic flow in the engineering sciences is thus old, dating back to the 1930s~\cite{greenshields1935study}. Outside of the engineering sciences, however, scientists have only recently discovered vehicular traffic as an interesting complex system that perhaps could be described with a few simple laws, and is thus worthy of study with scientific methods~\cite{chowdhury2000statistical, helbing2001traffic, nagatani2002physics}.

For physicists, traffic-flow models became a topic in the 1990s. In the recent decade, this topic has cooled down somewhat, but nonetheless remains an active field of research in physics and elsewhere, for instance, machine learning~\cite{mihaita2019motorway}. It is probably fair to say that the main motivation for physicists has never been to provide practical advice for urban planners. Rather, the attraction was that vehicular traffic exhibits several types of self-organised, collective phenomena also seen in statistical physics. It is no coincidence that the founding era of the physics studies of traffic flow was in the 1990s. This was a time when there was a prevailing idea that many phenomena in nature and society were connected by underlying ubiquitous organisational principles such as self-organised criticality~\cite{bak1996how}, manifested by many quantities that follow power laws. Even if this view has lately fallen out of fashion~\cite{newman2005power}, the idea that vehicular traffic is an archetypal complex systems---self-organised, decentralised, and with emergent behaviours that connect short and long spatial and temporal scales---still prevails.

Another reason for studying traffic flows is that the models themselves are interesting. They are among the simplest possible models of emergent phenomena in non-equilibrium systems. Furthermore, they bridge several different modelling frameworks (although none of them originally from physics)~\cite{fvwk2015genealogy}---from continuous models of traffic density~\cite{lighthill1955kinematic}, via discrete particle models (called `car-following theories' in this context)~\cite{reuschel1950fahrzeugbewegungen}, to cellular automata~\cite{nagel1992cellular}.

\subsection{Observed phenomena}

As mentioned, the primary focus of physicists interested in modelling traffic flow has not been to make accurate forecasting, but rather to qualitatively explain emergent phenomena. So what phenomena can be studied by models? In this section, we will go through some of these observations. Unless stated otherwise, we will discuss phenomena at continuous sections of highways.

\begin{figure}[!t]
\centering
\centering\includegraphics[scale=1.0]{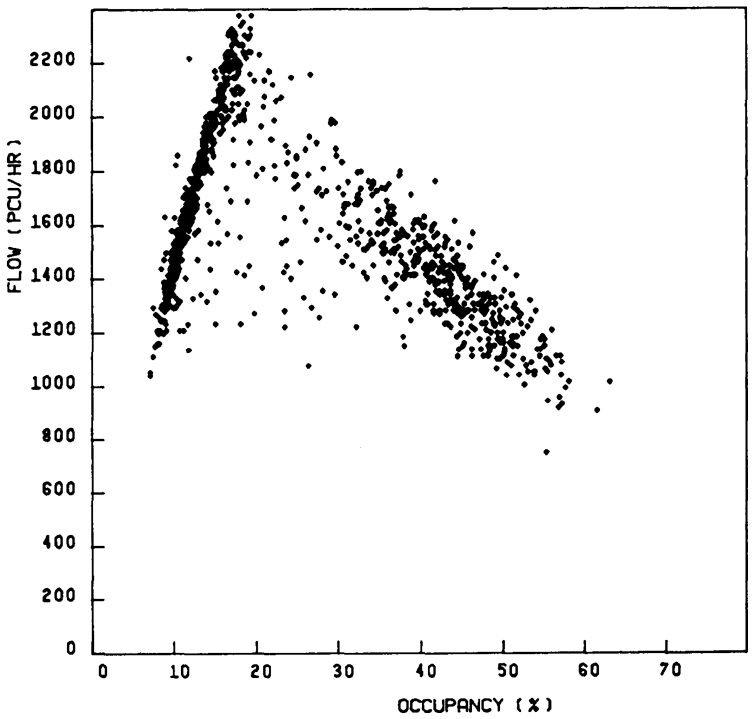}
\caption{Observed flow-occupancy relation in highway traffic in Ontario, Canada, 1985. Occupancy is measured (in percentage) as the fraction of time a sensor is blocked by a car. Flow is the number of cars passing per hour. Every data point is an average over 5 minutes.\newline
Source: Reprinted figure from Ref.~\cite{hall1986empirical}.}
\label{fig:flow_occupancy}
\end{figure}

\paragraph*{Occupancy-flow relations} If the traffic is light, a higher density of cars means that the flow increases---cars move at about the same speed so twice as many cars means twice as large flow. As the traffic gets denser, however, the average speed decreases. Eventually the flow starts decreasing as well. It has been known for over half a century that these quantities do not have a smooth relationship. Since Ref.~\cite{hall1986empirical} it is rather thought to be tent-shaped (Fig.~\ref{fig:flow_occupancy}), or inverse $\lambda$-shaped. This suggests the existence of two dynamic phases---a \textit{free-flow} and a \textit{congested} state~\cite{chowdhury2000statistical}, with an intermediate maximum flow~\cite{helbing2001traffic}.

Sometimes the congested state is divided into \textit{synchronised flow} in which cars are following each other at a relatively constant speed, and \textit{stop-and-go} motion (the name explains the concept) at even higher densities~\cite{kerner2017breakdown}. Some authors go further into dividing the synchronised flow depending on whether the speed and separation of the vehicles is stationary or not~\cite{kerner1996experimental}.

\begin{figure}[!t]
\centering
\centering\includegraphics[scale=1.0]{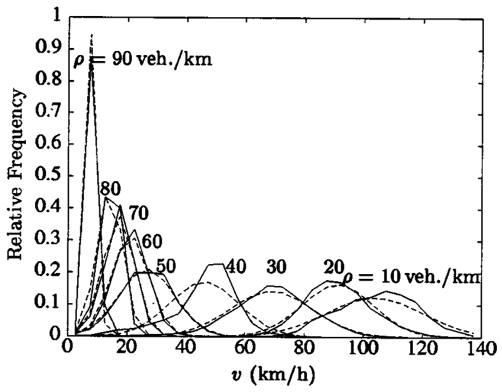}
\caption{The distribution of speeds at different densities. The dashed lines are the best fitting Gaussian distributions.\newline
Source: Reprinted figure from Ref.~\cite{helbing2001traffic}.}
\label{fig:distributions}
\end{figure}

\begin{figure}[!t]
\centering
\centering\includegraphics[scale=1.0]{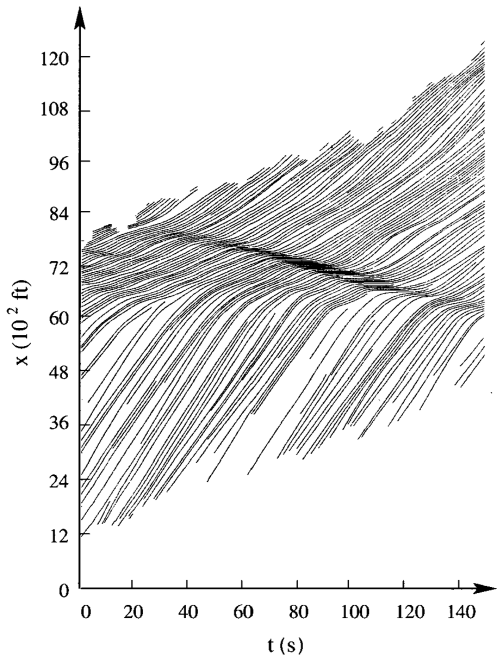}
\caption{Phantom traffic jam in highway traffic. Each line represents one vehicle in a particular lane (trajectories suddenly appearing mean a vehicle made a lane change). Original data from Ref.~\cite{treiterer1974hysteresis}.\newline
Source: Reprinted figure from Ref.~\cite{helbing2001traffic}.}
\label{fig:jam}
\end{figure}

\paragraph*{Phantom traffic jams} The distribution of speeds is fairly well described by a Gaussian distribution for all almost all traffic densities (Fig.~\ref{fig:distributions}). There could be some anomaly for intermediate speeds (40\,km/h in Fig.~\ref{fig:distributions}) which could ring a bell for physicists familiar with critical phenomena~\cite{tadaki2013phase}. Note however that there is nothing scale-free about the speed distribution at this point (scale-free, or power-law, distributions are usually the hallmark of phase transitions~\cite{newman2005power}). Still, there is one supposedly self-organised, emergent, phenomenon believed to explain this anomaly---\textit{phantom traffic jams}. These are jams that happen seemingly without an external trigger. Fig.~\ref{fig:jam} shows the classical figure illustrating phantom jams with data from aerial photography from Australia in 1967~\cite{treiterer1974hysteresis} and reproduced in almost every review paper or book on the subject~\cite{chowdhury2000statistical, helbing2001traffic, leutzbach1988intro}, and also original research papers~\cite{sugiyama2008traffic}. Apart from the existence of phantom traffic jams this figure also shows that the jam moves in a direction opposite to the traffic, a finding that has been established by other measurements~\cite{kesting2008agents}.

\paragraph*{Hysteresis} Hysteresis in traffic flow refers to the observation that the average speed-up as traffic gets lighter does not follow the same curve as when traffic gets denser. In the former situation, the average speeds are lower. This was first studied rigorously in Ref.~\cite{treiterer1974hysteresis} (although mentioned in earlier works).

\begin{figure}[!t]
\centering
\centering\includegraphics[scale=1.0]{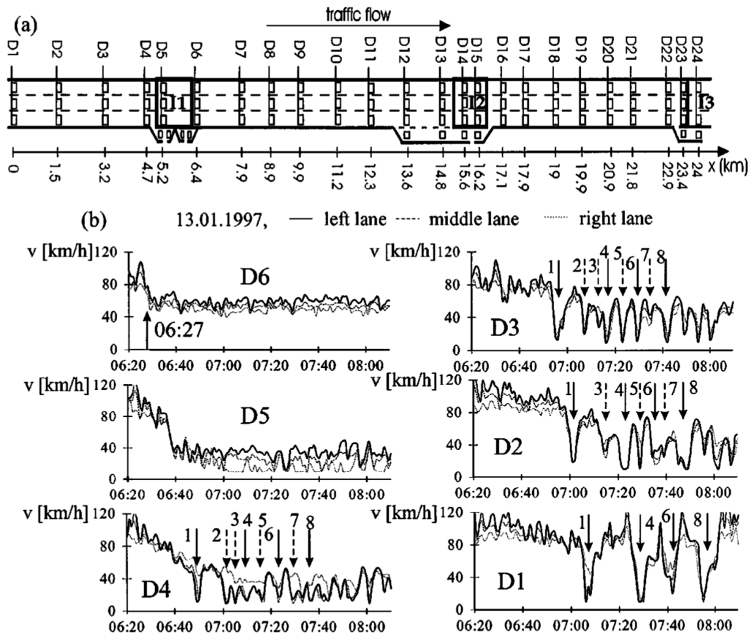}
\caption{The pinch effect. Panel (a) shows the setup of sensors along a highway. Panel (b) shows the readings (average speeds) of some of these sensors. The pinch effect is manifested in narrow dips (jams) that are created around D5 and aggregate to `wide jams' down the road.\newline
Source: Reprinted figure from Ref.~\cite{kerner1998experimental}.}
\label{fig:pinch}
\end{figure}

\paragraph*{Pinch effect} There is some evidence that stop-and-go type congestion waves are triggered at special locations along roads where small jams are formed, that later merge to form larger jams (Fig.~\ref{fig:pinch}). This is called the \textit{pinch effect} and the larger jams are called \textit{wide moving jams} (although, from a driver's perspective they are rather long than wide).

\subsection{Traffic-flow models}

Next, we describe several types of traffic-flow models that have been used to explain the observed phenomena. Some of these models have true physics origins, whereas others became popular among physicists, although their origins lie elsewhere (e.g., computer science and mathematics).

\paragraph*{Fluid-dynamical models} \textit{Macroscopic}, or \textit{fluid-dynamic}, models of traffic only use traffic density $\rho$, flow $Q$, and average velocity $v$ as variables describing the system. These are related by definition as
\begin{linenomath}
\begin{equation}\label{eq:flow_density_velocity}
    Q(x,t)=\rho(x,t) v(x,t) ,
\end{equation}
\end{linenomath}
where $x$ is the location along the road and $t$ is the time. Assuming continuity (that no cars are generated, or disappearing, along the road) we get the following equation describing mass conservation
\begin{linenomath}
\begin{equation}\label{eq:mass_conservation}
    \frac{\partial \rho}{\partial t} + \frac{\partial Q}{\partial x} = 0 .
\end{equation}
\end{linenomath}

Eq.~(\ref{eq:mass_conservation}) should be a part of all fluid-dynamical traffic flow theories, but we need one more equation to make it a full theory. The oldest approach is to assume $Q$ is only a function of $\rho$ and the functional relationship is to be inferred by data
\begin{linenomath}
\begin{equation}\label{eq:lighthill_whitham}
    Q(x,t)=Q\left[\rho(x,t)\right] ,
\end{equation}
\end{linenomath}
leading to
\begin{linenomath}
\begin{equation}\label{eq:lighthill_whitham2}
\frac{\partial \rho}{\partial t} + C(\rho)\frac{\partial \rho}{\partial x} = 0
\end{equation}
\end{linenomath}
where $C(\rho)$ comes from data. This \textit{Lighthill-Whitham theory}~\cite{lighthill1955kinematic} describes kinematic waves that travel in the opposite direction of the traffic flow (according to observations). In the solution of the Lighthill-Whitham equations, shock-waves of infinite density build up. These should be interpreted as jams, and are a challenge numerically, but not conceptually. There are several more sophisticated theories following the footsteps of Lighthill and Whitham. All of them replace Eq.~(\ref{eq:lighthill_whitham}) by a more elaborate equation.

\paragraph*{Kinetic models} \textit{Kinetic models} of traffic flow are inspired by the kinetic theory of gasses. They use a distribution of car speeds as their main variable describing the state of the system. The original kinetic theory of car traffic was proposed by Prigogine and co-workers in the 1970s~\cite{prigogine1960boltzmann}. It was quite similar to the original model from physics and to little surprises it shows many discrepancies with car traffic (for example, all cars would drive with the same average speed). This model was later heavily modified by Paveri-Fontana~\cite{paveri1975boltzmann}. Like above, Refs.~\cite{chowdhury2000statistical, helbing2001traffic} give a summary of these theories.

\paragraph*{Car-following theories} We have mentioned traffic flow theories inspired by fluid dynamic and kinetic gas theory, maybe to little surprise, there are also theories inspired by Newtonian mechanics. Such, \textit{car-following theories} are based on equations for the individual drivers and their response to the behaviour of the preceding car. The simplest car-following equation, due to Reuschel~\cite{reuschel1950fahrzeugbewegungen}, is
\begin{linenomath}
\begin{equation}\label{eq:car_following}
    \tau\ddot{x}_n(t)=\dot{x}_{n+1}(t)-\dot{x}_n(t) .
\end{equation}
\end{linenomath}
This equation is derived from assumptions that a driver wants to drive as fast as the preceding car, but not get closer than a safety distance. Later, improved, car-following theories have assumed each car has an internal desired speed, that it follows unless it needs to avoid a collision. These more sophisticated theories can explain the mirrored-$\lambda$ shape of the flow-density curves and hysteresis effects.

\paragraph*{Coupled-map lattice models} The models we have seen so far have all been continuous in both time and space. So called \textit{coupled-map lattice models} share many assumptions of car-following theories, but use a discrete time. In general, such models have the form
\begin{linenomath}
\begin{subequations}
\begin{align}
    v_n(t+1) & = \text{Map}_n(v_n(t),v_{n,{\rm des}},\Delta x_n)\\
    x_n(t+1) & = v_n(t) + x_n(t),
\end{align}
\end{subequations}
\end{linenomath}
where $\text{Map}_n(\cdot)$ is a dynamical map that takes into account the speed and position of the $n$th vehicle, $v_n$ and $x_n$, the desired speed of the $n$th vehicle, $v_{n,{\rm des}}$, and the headway to the $n+1$th vehicle, $\Delta x_n$. This versatile framework can accommodate different personalities of drivers, and different classes of vehicles, etc. Several of the empirical characteristics of traffic flow (such as flow-density relations) can be reproduced by coupled-map lattice models. Popular models of this kind includes those of Yukawa and Kikuchi~\cite{yukawa1995coupled} and Krauss, Wagner, and Gawron~\cite{krauss1996continuous}.

\paragraph*{Cellular automata models} One further abstraction from coupled-map lattice models is to discretise space as well as time. This leads to so called cellular automata models. These are the most well-studied type of models in the physics literature, which might seem surprising since it is a type of models derived from computer science and mathematics rather than physics (whereas the kinetic and car-following models above have a much stronger physics flavour). The explanation is probably that physicists became interested in traffic models as a part of a general hype around complex systems, where cellular automata models of artificial life are among the most iconic theories.

\begin{figure}
\centering\includegraphics[scale=1.0]{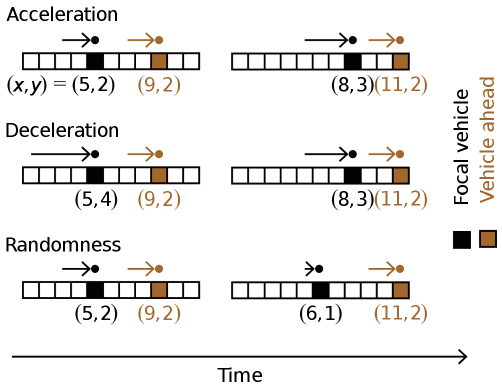}
\caption{The three rules for changing the speed in the Nagel-Schreckenberg model. The rules are illustrated for the focal vehicle, while the vehicle ahead keeps its speed. The arrow illustrates the speed $v$; the circle illustrates the position $x$.}
\label{fig:nasch}
\end{figure}

The most well-studied cellular automata model of traffic flow is the Nagel-Schreckenberg model~\cite{nagel1992cellular}. In this model, the road is represented by a one-dimensional discrete lattice. There are $N$ vehicles on this road. Each cell of the road is occupied by maximally one vehicle. All vehicles are updated in parallel according to the following rules (to be followed in order):
\begin{enumerate}
\item\textit{Acceleration.} If $v_n<v_{\rm max}$, then the speed of vehicle $n$ is increased by one unit, otherwise the speed is unchanged:
\begin{linenomath}
\begin{equation}
    v_n(t+1)=\min(v_n(t)+1,v_{\rm max}) .
\end{equation}
\end{linenomath}
\item\textit{Deceleration.} If $x_{n+1}\leq x_n+v_n$---that is, the car ahead is so close that vehicle $n$ would reach its position (or further) the next time step---then the $n$th vehicle brakes:
\begin{linenomath}
\begin{equation}
    v_n(t+1)=\min(v_n(t),x_{n+1}-x_n-1) .
\end{equation}
\end{linenomath}
\item\textit{Randomisation.} By chance, that is, with probability $p$, the speed of some cars is decreased:
\begin{linenomath}
\begin{equation}
   v_n(t+1)=\max(v_n(t)-1,0).
\end{equation}
\end{linenomath}
\item\textit{Vehicle movement.} Each vehicle moves forward according to its new speed:
\begin{linenomath}
\begin{equation}
   x_n(t+1) = x_n(t) + v_n(t+1) .
\end{equation}
\end{linenomath}
\end{enumerate}
See Fig.~\ref{fig:nasch} for an illustration of the Nagel-Schreckenberg rules.

The Nagel-Schreckenberg model can, despite its simplicity, reproduce many features of real traffic, such as the flow-density curves and phantom traffic jams. With this model as a starting point the research has branched out in many directions. Some of the research has striven to include more realism~\cite{horiguchi1998numerical, helbing1998coherent}, while other~\cite{takayasu1993noise} has shown that it takes only a small modification to turn it to a model of self-organized criticality (the Nagel-Schreckenberg model itself does not have the necessary meta-stable state). Yet others studied further simplified models as discussed below, although these simplified models are incapable of reproducing the above-mentioned full statistical characteristics of traffic.

\paragraph*{Connections to non-linear statistical mechanics} If one gives up on trying to reproduce all statistical features of highway traffic, then one can further simplify models like the Nagel-Schreckenberg cellular automaton. This will typically reduce the models to standard models of non-equilibrium statistical mechanics, like the \textit{totally asymmetric simple exclusion process} (TASEP)~\cite{macdonald1968kinetics} or the Burgers' equation~\cite{burgers1948mathematical} (in particular, its noisy version~\cite{forster1976long}). These more stylised studies are often focused on finding dynamical critical exponents that relate the size of a system to its dynamics and the critical point separating the free flow and congested phases (see Ref.~\cite{biham1992self} for a typical example).

\subsection{Pedestrian traffic flows}

Pedestrian traffic is a related, but far from an equivalent type of socio-physical system compared to vehicular flow. It could also be thought of as a self-organised granular flow of semi-intelligent particles. The main research questions concern the formation of trails in open landscapes~\cite{helbing1997modelling}, the formation of lanes in dense pedestrian traffic~\cite{feliciani2016empirical}, and escape panic behaviour~\cite{helbing2000simulating}.

Most models of pedestrian flows take their inspiration in physics and model the individuals as particles driven by forces~\cite{henderson1971statistics}. Several things are different from real gasses---there is, for example, no conservation of momentum. Typically one have to assume that people repel each other by two forces; one is social---the desire not to be too close to another person---and one is physical---crowded conditions such that people actually have to be in physical contact~\cite{helbing2001traffic}. To accurately model escape panic, one has to break down the physical forces into tangential and radial components.

\begin{figure}[!t]
\centering\includegraphics[scale=1.0]{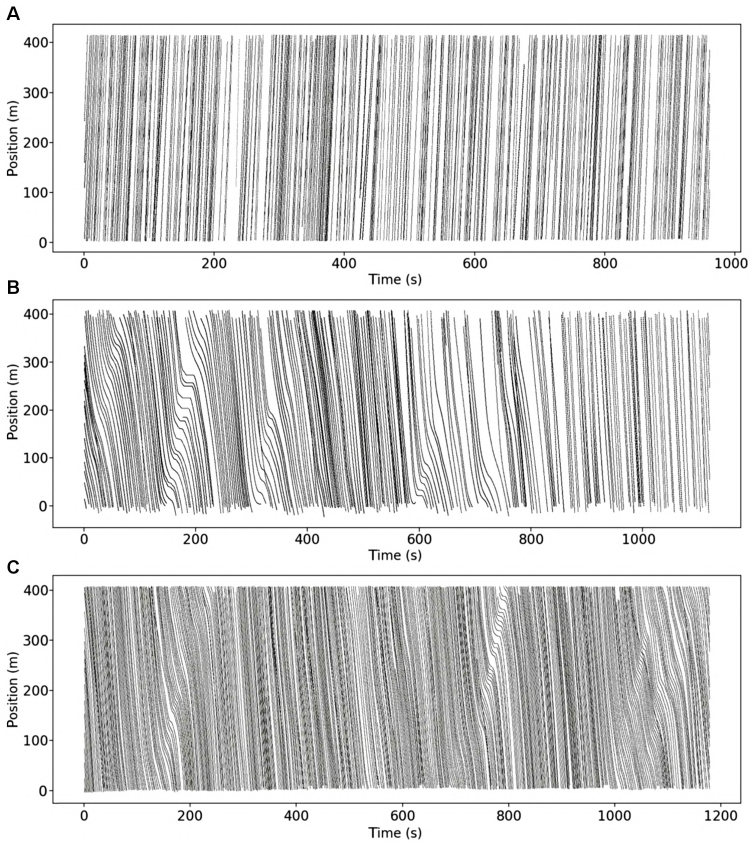}
\caption{Trajectories, from the highD dataset~\cite{krajewski2018highd} plotted in the style of Fig.~\ref{fig:jam}. \textbf{A,} Free-flow state (recording 60, lane 8). \textbf{B,\,C,} More congested traffic with stop-and-go waves (recordings 26 and 25, lanes 2 and 4, respectively).}
\label{fig:highd}
\end{figure}

\subsection{Future outlook}

From a physics point of view, the field of traffic flow modelling seems to have somewhat cooled down at the time of writing. Elsewhere, it is a topic of emergent interest. In particular, the recent boom in research on autonomous vehicles has renewed the interest in applying machine learning to these topics~\cite{mihaita2019motorway}. The current interest in self-driving cars produces an enormous amount of data. Most of it arguably useless for this type of research, but probably eventually enough to discover new statistical laws of vehicular traffic. Even if data does not come as a side product from the automotive industry, it is nowadays easier and cheaper to collect. There have been projects to this end that rely on drones~\cite{krajewski2018highd} or tower-mounted cameras~\cite{halkias2006ngsim}. In Fig.~\ref{fig:highd}, we plot individual trajectories of some of the recordings from one of these datasets~\cite{krajewski2018highd}.

When we---by new observations from new datasets---have created new qualitative statistical laws to replace the current qualitative observations, then the question will once again be to find minimal models recreating the observations. In particular, with high-quality data on the onset of the congested phase, we could measure how often phantom traffic jams actually occur and whether the current mechanistic models of these can explain the observation, or if we have to go back to the drawing board.

\FloatBarrier

\section{Econophysics}
\label{S:EconoPhys}

Unlike traditional economics, which is built upon a rational-choice model, econophysics borrows the particle model from statistical physics to explain the behaviour of an agent. Such a model assumes that the agent’s tastes and preferences are not fixed, but instead depend on the interactions with other agents~\cite{andersen2013financial}. In other words, econophysics puts a greater emphasis on the social environment of the agent~\cite{ormerod2016ten}. Some other physics models and concepts commonly applied to economics include the kinetic theory of gases, chaos theory, percolations, and self-organised criticality.

Empirical work in econophysics is mostly focused on the analysis of firm growth and competition, industry entry and exit rates, money flows, financial markets, and international trade~\cite{dragulescu2000statistical, aoyama2010econophysics, braha2011corporate, luo2014geovisual, harmon2015anticipating, abergel2017econophysics, wachs2019network}, that is, on areas in which huge datasets are available, and the application of statistical-physics tools and methods proves useful. The areas of economics with scarce data availability, such as macroeconomics in which datasets are short and noisy, has not attracted much attention among econophysicists. However, with the increasing acceptance of networks in the mainstream economics, econophysics may still play an important role in the future development of macroeconomics~\cite{ormerod2016ten}.

\subsection{The advent of econophysics}

In 1991, Mantegna published a paper in a physics journal, Physica A~\cite{mantegna1991levy}, in which a time series of daily financial market prices was analysed and price changes were shown to follow a power-law distribution. At that time, power-law distributions and scaling relations were attractive key topics to statistical physicists after the big booms of \textit{fractals} in the early 1980s~\cite{mandelbrot1983fractal, vicsek1992fractal, takayasu1990fractals} and \textit{self-organised criticality} in the late 1980s~\cite{bak1987self, bak1996how}. Research targets of interest to physicists were extended widely to general complexity in nature, thus crossing the traditional boundaries between research fields. Market price changes were a part of this extension and got accepted as one of physically interesting phenomena that exhibit power-law behaviour.

The next pioneering interdisciplinary paper appeared in the same journal in 1992 by H.~Takayasu et al.~\cite{takayasu1992statistical}. A simple artificial model of the market was proposed comprising mathematically defined dealers in the form of dynamical particles in a one-dimensional space of prices. The model's non-linear dynamics caused chaotic time evolution resulting in almost random price movements (see the next subsection for details).

In 1995, Mantegna and Stanley analysed the time series of stock-market prices recorded at a one-minute interval, and found that the price-change distributions at different time scales, upon re-scaling, conform to a function with symmetric power-law tails~\cite{mantegna1995scaling}. Such a data-analysis method was familiar from the study of critical phenomena involving phase transitions. Because their paper was published in a high-impact journal, the new physics approach to financial markets attracted wide attention. In the same year, Stanley coined the term \textit{econophysics} to represent an interdisciplinary research field that focuses on economic phenomena from a physics point of view. He introduced this term at a conference on statistical physics held in Kolkata, India.

In 1997, the first international meeting with the title `econophysics' was held in Budapest, Hungary. Most of the gathered researchers were specialists in statistical physics, although there was notable participation from fields as diverse as high-energy experiments. In this same year, the journal Physical Review Letters also opened the door to econophysics, and a theoretical paper explaining the generating mechanism for power-law distributions in financial markets was published~\cite{takayasu1997stable}. Prior to this acceptance, the journal was rejecting econophysics manuscripts on the basis of the topic being out of scope. The acceptance thus marked the promotion of econophysics to a status of a fully fledged field in applied physics, such as biophysics or geophysics. The number of econophysics researchers subsequently increased, as did the variety of research topics that began to cover more than just financial markets.

In 1999, the first monograph on econophysics was published~\cite{mantegna1999introduction}, and textbooks on financial markets from the physics viewpoint followed~\cite{sornette2003stock, bouchaud2003theory}. Multiple workshops and conferences on econophysics were held annually thereafter, with many economists and finance researchers joining to discuss practical problems~\cite{takayasu2013empirical, takayasu2012application, takayasu2010econophysics}.

The rest of this chapter focuses on the development of econophysics research stemming from the afore-mentioned simple physical model of financial markets~\cite{takayasu1992statistical}. We first describe the historical background of the dealer model, and show how more advanced dealer models have arisen. Then we introduce an empirically derived time-series model called the PUCK model, and proceed to demonstrate the relation between multiple dealer models. We also outline a recent analysis of comprehensive market data that includes all microscopic orders appearing on the Foreign Exchange market at a millisecond interval. The mechanism of financial Brownian motion is compared with the physical phenomenon of colloidal Brownian motion, and the most advanced dealer model, which is reconstructed directly from the data, is solved by applying the classical kinetic theory. We close off the chapter with a discussion of a novel emerging perspective, that of an ecosystem of strategic dealers.

\subsection{Agent-based modelling: The dealer model}
\label{S:EconoPhys:Dealer}

Mandelbrot's inspiration for introducing the concept of fractals, that is, the scale invariance of complicated shapes in nature, originated during an examination of historical cotton-price charts at various time scales. Market prices have thus become the very first example of fractals. In part through their interactions with Mandelbrot, H.~Takayasu and Hamada---a physicist and an economist---joined forces to create a model of financial markets that would explain why market prices fluctuate in a scale-invariant manner~\cite{takayasu1992statistical, hirabayashi1993behavior, yamada2007characterization}. A prevailing view in economics at the time was that if all market dealers were rational and possessed enough information, then the market price would be uniquely determined and stable. Yet, this view could not be further from the real-world price fluctuations, which prompted H.~Takayasu and Hamada to construct their model borrowing ideas from physics. The model thus had to incorporate the essential underlying mechanisms and processes in a way that is as simple as possible, but non-trivial.

Let us envision an artificial financial market comprising $N$ dealers who buy and sell financial instruments such as stocks. Every dealer is assumed to try to buy at a low price and sell at a high price, hoping to earn the price difference. The $i$th dealer's trading action at time $t$ is described by introducing two threshold prices, the buying price, $b_i(t)$, and the selling price, $s_i(t)$. The dealer hopes to buy at the former price or lower, and to sell at the latter price or higher. The difference, $L_i(t)=s_i(t)-b_i(t)$>0, called the spread characterises greediness of the dealer, which is set to a constant value $L$ in the simplest case. All dealers' buying and selling prices are gathered to make the market's order-book. A deal occurs if the condition $s_i(t)\leq{}b_j(t)$ is fulfilled for a pair of dealers $i$ and $j$, in which case the $i$th dealer sells to the $j$th dealer, or equivalently, the $j$th dealer buys from the $i$th dealer. An interesting point is that no deals occur if all dealers' buying prices are within the distance $L$ from the minimum buying price. Only when the distance between the farthest pair of dealers equals or exceeds $L$ can a deal (between this particular pair) take place. Deals thus represent a strong, non-linear, attractive interaction that makes a group of dealers compact in the price space.

The model is further simplified by assuming that dealers can possess at most one stock at a time. If a dealer possesses a stock, they are a seller quoting only the selling price. If the dealer does not possess a stock, then they are a buyer quoting only the buying price. A seller hopes to sell at a high price, but if there is no buyer whose buying price is equal or higher, then the seller should compromise by lowering the price until a trade becomes possible. This situation is described by a differential equation
\begin{linenomath}
\begin{equation}
\dd{b_i(t)}{t}=\sigma_i(t)c_i,
\label{eq:51}
\end{equation}
\end{linenomath}
where $\sigma_i(t)=1$ ($\sigma_i(t)=-1$) signifies the buyer (seller) state of the $i$th dealer, while $c_i>0$ quantifies the dealer's (initially random) hastiness. A deal between the seller $i$ and the buyer $j$ occurs when $s_i(t)=b_i(t)+L\leq{}b_j(t)$, at which moment the state functions $\sigma_i(t)$ and $\sigma_j(t)$ change their signs. The resulting market price, $P(t)$, which takes the value of the latest deal, evolves deterministically in time. The model is initialised such that all dealers start with the same buying price, and some dealers start as buyers and others as sellers.

With $N=2$ dealers, one seller and one buyer, the time evolution of the market price is almost trivial. The two dealers periodically alternate their state and the resulting market price oscillates regularly. With $N\geq{}3$ dealers, the time evolution of the market price becomes highly non-linear. There is, in fact, an underlying chaotic effect that magnifies small initial differences. We thus learn from this simple model that even fully deterministic dealer behaviour can cause noisy dynamics. The model is, nonetheless, insufficient to explain the fractal properties of market prices.

A minimal modification of the described model accounts for an effect called trend following. A dealer who follows the trend expects that price movements keep moving in the same direction as in the immediate past. This can be mathematically formulated using a moving average of length $T$
\begin{linenomath}
\begin{equation}
\lra{\Delta{}P(t)}=\int_0^T{w(u)\dd{}{t}P(t-u)\dup{u}},
\label{eq:52}
\end{equation}
\end{linenomath}
where $w(u)$ is a weight function that satisfies $\int_0^T{w(u)\dup{u}}=1$. Eq.~(\ref{eq:51}) is then appended with a trend-following term
\begin{linenomath}
\begin{equation}
\dd{b_i(t)}{t}=\sigma_i(t)c_i+d_i\lra{\Delta{}P(t)}.
\label{eq:53}
\end{equation}
\end{linenomath}
Coefficients $d_i$ quantify the extent of trend following. They are usually positive for dealers who are trend followers, but can also be negative for dealers called contrarians. The simplest possible variant of the model is obtained by assuming that all dealers are trend followers with the same coefficient $d>0$. Despite being simplistic, this assumption has a drastic effect, yielding deterministic price dynamics that resemble scale-invariant fluctuations of stochastic random walks~\cite{takayasu1992statistical}. The model thus identifies two mechanisms likely to be responsible for some of the key characteristics of realistic market-price time series. Namely, market prices fluctuate almost randomly due to non-linear, chaos-inducing interactions between dealers, while scale-invariance emerges from the dealer tendency to follow trends.

The dealer model with trend following as defined by Eq.~(\ref{eq:53}) can be studied analytically to some extent. The dynamics of the centre of dealer mass follows a Langevin equation~\cite{hirabayashi1993behavior}, which is an equation that is well-known in the context of colloidal Brownian motion. The distribution of market-price changes obeys a power law with an exponent that depends on the value of the parameter $d$~\cite{sato1998dynamic}, in line with known empirical facts~\cite{mantegna1991levy, mantegna1999introduction} and a previous theoretical analysis~\cite{takayasu1997stable}. The dealer model can also be used as a basis for deriving the autoregressive conditional heteroskedasticity (ARCH) model of Engle~\cite{engle1982autoregressive}, thus offering a mechanistic explanation for the origins of volatility clustering observed in financial markets~\cite{sato2002derivation}. Finally, by tuning parameters values, the model reproduces the phenomenon of abnormal diffusion, as well as the statistical properties of dealing time intervals~\cite{yamada2007characterization}.

A stochastic variant of the dealer model was introduced in 2009~\cite{yamada2009solvable}. In the model, the function $\sigma_i(t)$ is random, such that---at each moment of time $t$---either $\sigma_i(t)=1$ or $\sigma_i(t)=-1$ with the probability of 0.5. This modification improves upon already favourable properties of the dealer model even in the case of $N=2$, which is easily solvable using both analytical and numerical methods. The stochastic dealer model can, for example, generate bubble-like behaviours that cause the market price to grow exponentially if the trend-following coefficient, $d$, is above a certain value.

Further generalisation of the stochastic dealer model has enabled capturing the characteristics of an intervention by the Bank of Japan in the foreign exchange market between the US dollar and the Japanese yen~\cite{matsunaga2012construction}. Aside from ordinary dealers responsible for usual market-price fluctuations, the model also includes a special dealer that takes the role of the Bank of Japan. The special dealer can cause large market-price changes that, according to empirical analyses of market data, are accompanied by risk-averse responses such as bid-ask spread widening, loss cutting, and profit booking. Ordinary dealers in the model, with some adjustment, can mimic risk-averse responses and thus generate said empirical phenomena in simulations. The increased realism makes it possible to assimilate financial time-series data into the model. This opens the door to the planning of intervention strategies, as well as predicting subsequent market responses.

More recent extensions of the stochastic dealer model help clarify the cross-currency correlations between the US dollar, the Japanese yen, and the Euro~\cite{ciacci2020microscopic}. A new type of dealers, who pursue what is known as triangular arbitrage, is introduced into the model. Such dealers earn profit by quick circular exchange transactions from, for example, the US dollar to the Japanese yen to the Euro and back to the US dollar. Interestingly, triangular arbitrage in currency markets was first reported in 2002 in an econophysics study~\cite{aiba2002triangular}. New evidence shows that such arbitrage still exists in the present financial markets in which automated trading systems dominate~\cite{ito2020execution}. The stochastic dealer model has clarified that it is, in fact, a small number of triangular-arbitrage dealers who boost the cross-currency correlations in a manner consistent with empirical observations.

\subsection{Time-series modelling: The PUCK model}

Trend following introduced in the dealer model is based on the idea that dealers make their decision referring to the latest market data using the moving average. Applying a similar idea to the time series of deal intervals, known for their temporal-clustering behaviour~\cite{takayasu2002transaction}, it was found that the occurrence of deals in markets is modelled well by a Poisson process with a time-dependent mean value. This value is given by the moving average of the latest deal intervals over a time period $T$, where the best estimate of $T=150$\,s was obtained from the dollar-yen exchange-market data at the time. The finding clarifies the mechanism underlying the temporal clustering of deal intervals. When random fluctuations cause a few short intervals to repeat, the moving average value becomes smaller, making shorter intervals more likely to appear in the Poisson process. A dense period with short deal intervals ensues. Converse is true when random fluctuations cause a few long intervals to repeat. The described time-dependent Poisson process is a self-modulation process whose fluctuations have a $1/f$ power spectrum bordering between stationary and non-stationary processes~\cite{takayasu2003self}.

The idea of using the moving average was also extended to the time-series analysis of market prices~\cite{takayasu2006temporal, takayasu2006potential}. For a given time series of real market prices with a fixed sampling interval, $\left\{p(t)\right\}$, the following time-evolution model is defined
\begin{linenomath}
\begin{equation}
p(t+1)-p(t)=-\frac{b(t)}{M-1}[p(t)-p_M(t)]+g(t),
\label{eq:PUCKvulgaris}
\end{equation}
\end{linenomath}
where $M$ is the length of the moving average, $p_M(t)=\frac{1}{M}{\sum_{k=0}^{M-1}p(t-k)}$, and $g(t)$ is independent noise with a zero mean. The coefficient $b(t)$ denotes a slowly changing parameter estimated from the time-series data. The case of $b(t)=0$ corresponds to the ordinary random walk. In the case of $b(t)>0$, the future market price, $p(t+1)$, is likely to be attracted to the latest moving average price, $p_M(t)$, signifying stable market-price movements. In the case of $b(t)<0$, the future market price is likely to be repelled from the moving-average price, signifying unstable market-price movements.

The described model can be generalised with a time-dependent market-potential function, $\Phi_M(t)$, as follows
\begin{linenomath}
\begin{equation}
p(t+1)-p(t)=-\left.\frac{\partial}{\partial p}\Phi_M(t)\right|_{p=\frac{p(t)-p_M(t)}{M-1}}+g(t).
\label{eq:marketpotential}
\end{equation}
\end{linenomath}
This is the Potential of Unbalanced Complex Kinetics (PUCK) model. Eq.~(\ref{eq:PUCKvulgaris}) is a special case of the PUCK model with the quadratic potential $\Phi_M(t)=\frac{1}{2}b(t)p^2$. The PUCK model with the quadratic potential function has been shown to apply to various financial markets, successfully reproducing most of empirical stylised facts such as the power-law distribution of price changes, abnormal diffusion over short time scales, as well as volatility clustering~\cite{takayasu2010theoretical}.

\begin{figure}[!t]
\makebox[\textwidth][c]{\includegraphics[scale=1.0]{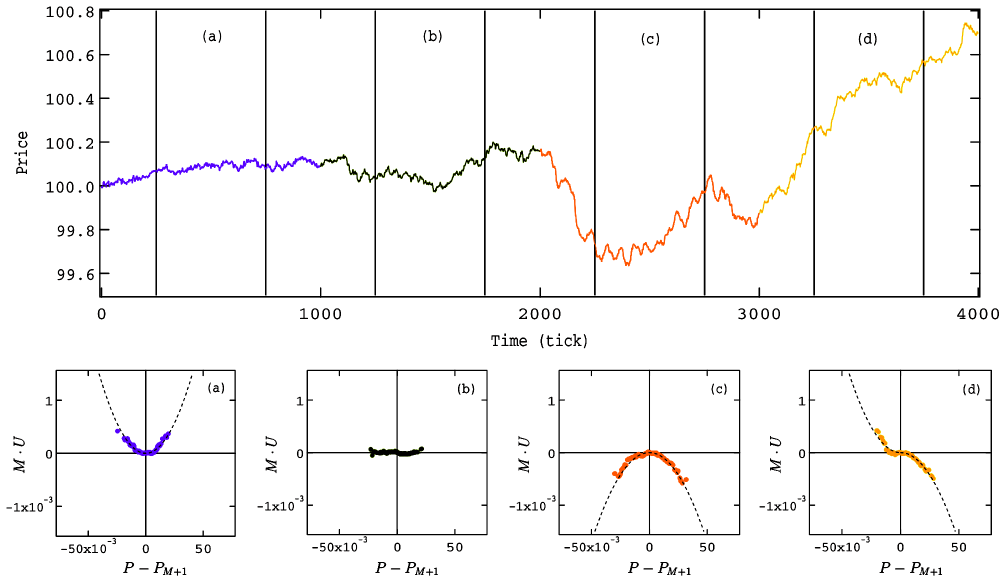}}
\caption{An example of artificial market-price fluctuations produced by the stochastic dealer model with the corresponding estimates of the potential functions. The parameter $d=d_1=d_2$ from Eq.~(\ref{eq:53}) varies such that $d=-1$ during the first 1000 time steps, $d=0$ during the second 1000 time steps, $d=1$ during the third 1000 time steps, and $d=-1$ if $\lra{\Delta P}_M<0$ whereas $d=1$ if $\lra{\Delta P}_M\geq0$ during the last 1000 time steps, where $\lra{\Delta P}_M$ is the moving average from Eq.~(\ref{eq:52}) of length $M=10$ ticks with uniform weights.\newline
Source: Reprinted figure from Ref.~\cite{yamada2009solvable}.}
\label{fig:potentialfunction}
\end{figure}

The market-potential function is estimable for any market-price time series, including those artificially produced by the dealer model described in Section~\ref{S:EconoPhys:Dealer}. A stable quadratic potential is obtained for contrarian dealers ($d<0$), an unstable quadratic potential is obtained for trend followers ($d>0$), and an asymmetric higher-order potential appears when trend following is asymmetric (Fig.~\ref{fig:potentialfunction}). Moreover, the value of the market-potential coefficient $b(t)$ can be theoretically derived from the dealer model, thus demonstrating that the origin of the market-potential function in Eq.~(\ref{eq:marketpotential}) comes from the trend-following behaviour of dealers.

A merit of the PUCK model is its wide applicability; the model describes market-price time series in various circumstances. This ranges from nearly random walks under normal market conditions, an exponential divergence in the case of bubbles or crashes, and even a double exponential divergence or a finite-time singularity in the case of hyper inflation~\cite{takayasu2009continuum, watanabe2009observation}. The threshold between the normal and the abnormal random walk is the value $b(t)=-2$ of the quadratic-potential coefficient; namely, when $b(t)<-2$, Eq.~(\ref{eq:PUCKvulgaris}) becomes linearly unstable causing price fluctuations to grow or decline exponentially as is observed in bubbles or crashes, respectively. If a cubic potential function is detected, it generally corresponds to asymmetric price movements~\cite{watanabe2009random}.

In a short time-scale limit, Eq.~(\ref{eq:PUCKvulgaris}) reduces to the Langevin equation for Brownian motion containing a mass term and a viscosity term~\cite{takayasu2009continuum}. Interestingly, the mass term is proportional to $-b(t)$, showing that trend following works as inertia. Also, the viscosity term becomes negative when $b(t)<-2$, suggesting that bubbles, crashes, and inflation should be regarded as negative viscosity phenomena, that is, as being under the influence of an accelerating instead of a decelerating force.

\subsection{Order-book modelling: Financial Brownian motion}

Another approach that physicists brought to financial markets is data analysis and modelling of an order book. Such a book lists buy orders (bids) and sell orders (asks) gathered in a market. Ref.~\cite{maslov2000simple} introduced a theoretical model in which bids and asks are injected randomly onto a price axis. A deal occurs when a new bid price is equal to or higher than the lowest ask price, or vice versa, when a new ask price is equal to or lower than the highest bid price. The deal signifies that the corresponding pair of orders annihilate forming the latest market price. Otherwise, injected orders accumulate on the price axis of the order book. The mechanism was pointed out to be similar to a one-dimensional catalytic chemical reaction in which the reaction front moves randomly. Ref.~\cite{bouchaud2002statistical} described the details of a stock-market order book, documenting empirical statistical laws about the injection of bids and asks, as well as cancel orders. A simple mathematical model called Zero Intelligence was proposed to capture the empirical findings, with the name of the model stemming from the fact that no intelligent dealer strategy was needed.

\begin{figure}[!t]
\centering
\includegraphics[scale=1.0]{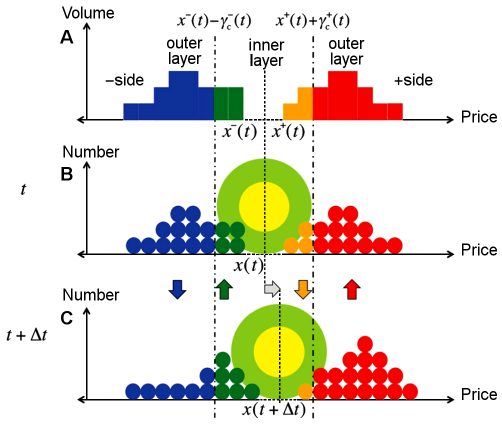}
\caption{Schematic representation of the Financial Brownian Particle model. \textbf{A,} An order-book configuration of buy orders (blue in the outer layer and green in the inner layer) and sell orders (red in the outer layer and orange in the inner layer) on the price axis. \textbf{B,} Corresponding configuration of outer-layer particles (blue and red disks), inner-layer particles (green and orange disks), the colloidal Brownian particle's interaction range (green ring), and the core (yellow circle). \textbf{C,} After time $\Delta t$ the configuration of surrounding particles changes.\newline
Source: Reprinted figure from Ref.~\cite{yura2014financial} under the Creative Commons Attribution 3.0 Unported (CC BY 3.0).}
\label{fig:colloidalparticle}
\end{figure}

Ref.~\cite{yura2014financial} introduced a novel data analysis of order books, focusing on an analogy between colloidal random walks and financial market-price movements. Accumulated bid and ask orders are regarded as water molecules in this analysis, while an imaginary colloidal particle is assumed to exist in the gap between bids and asks centred right in the middle between the highest bid and the lowest ask (Fig.~\ref{fig:colloidalparticle}). This colloidal-particle picture is intuitive in the following sense. As the particle gets displaced, say, to the right (i.e., towards higher prices), the opened up space in its wake gets quickly filled with water molecules (i.e., bids) from further back where the number of molecules decreases (Fig.~\ref{fig:colloidalparticle}B,\,C). In front of the colloidal particle, by contrast, water molecules get pushed forward, decreasing their number next to the particle, but increasing the number further away (Fig.~\ref{fig:colloidalparticle}B,\,C). This intuitive picture is fully consistent with the dealer model and trend following by which pairs of buy and sell orders move together with the market price.

\begin{figure}[!t]
\centering
\includegraphics[scale=1.0]{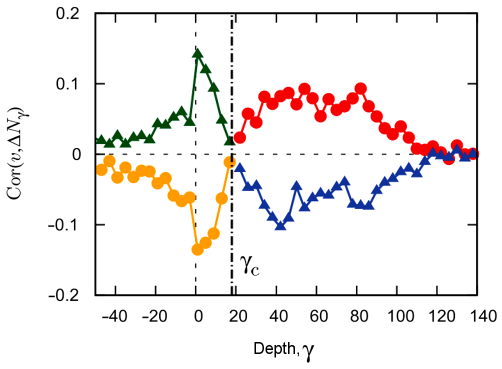}
\caption{Cross-correlation function between the velocity of market-price movements and the rate of change of orders as a function of depth. Buy orders are indicated with green and blue triangles, whereas sell orders are indicated with orange and red circles. At $\gamma_\mathrm{c}=18$ the nature of the cross-correlation changes, thus revealing the distinction between inner and outer layers. A positive inner-layer (outer-layer) cross-correlation means that orders in this layer act as a driving force (drag resistance) for market movements.\newline
Source: Reprinted figure from Ref.~\cite{yura2014financial} under the Creative Commons Attribution 3.0 Unported (CC BY 3.0).}
\label{fig:colloidalcorrelation}
\end{figure}

A more conventional picture treats all buy orders (and separately all sell orders) on an equal footing, but this is incorrect. As the colloidal-particle picture shows, buy and sell orders should be categorised into an inner and an outer layer of opposite behaviour. If bids (or asks) increase in the inner layer they decrease in the outer layer and vice versa. Inner-layer orders furthermore play a role of a driving force behind market-price movements, as indicated by a high positive cross-correlation between the velocity of price movements and the rate of change of orders in the inner layer (Fig.~\ref{fig:colloidalcorrelation}). Interestingly, outer-layer orders exhibit a negative cross-correlation between the velocity of price movements and the rate of change of orders, and thus can be considered as drag resistance for market-price movements (Fig.~\ref{fig:colloidalcorrelation}). All this implies a fluctuation-dissipation relation for the colloidal particle, which is modelled by the Langevin equation. The value of the drag coefficient normalised by the colloidal-particle mass can then be estimated from market-price data~\cite{yura2014financial}.

An often overlooked aspect of modelling financial markets using continuous-price models, such as the Langevin equation, is whether the continuous-price assumption can be justified. To this end, Ref.~\cite{yura2015financial} uses the described analogy between a market-order book and a molecular fluid to estimate the financial Knudsen number. Because the Knudsen number is generally defined as a ratio of the mean free path to a representative length scale of the system, in the case of financial markets, the former is given by the average distance of price movements in one direction, while the latter is the diameter of the colloidal particle in terms of the inner-layer width for both buy and sell sides (Fig.~\ref{fig:colloidalparticle}). The continuous-price assumption is valid if the Knudsen number is smaller than 0.01, whereas a discrete-time description is needed if the Knudsen number is larger than 0.1 (with transitional regimes in between). The estimated value of the Knudsen number for dollar-yen and dollar-Euro markets fluctuates around 0.05 most of the time, becoming larger than 0.1 in the times of market turmoil. This result indicates that the continuous-price assumption is questionable for modelling financial markets, especially when large market-price fluctuations take place.

\subsection{A kinetic approach to financial market microstructure}

We have reviewed the dealer model as a financial microscopic model describing decision-making process on the level of individual agents. This model has the advantage that (i) it can capture the strategic decision-making process of individual dealers (such as trend following), and (ii) it reduces to the PUCK model as its macroscopic dynamics for $N=2$ and thus can replicate the empirical facts seen in the price time series. The dealer microscopic model, however, has a disadvantage that it could not be directly validated from data, because it requires the truly microscopic data to track all traders' decision-making dynamics. Indeed, the trend-following mechanism is theoretically assumed in the model as a strategy of individual traders, which could not be directly confirmed. Also, this model requires calibration of the buy-sell spread distribution. This situation is in contrast to other mesoscopic models, such as purely-random order-book models~\cite{maslov2000simple, bouchaud2002statistical, daniels2003quantitative, smith2003statistical} (see Refs.~\cite{slanina2013essentials, bouchaud2018trades} for reviews), which require fewer calibration parameters although they cannot capture the strategic decision-making process of individual traders.

Recently, the situation with respect to data availability has drastically changed; truly microscopic data has become available due to the big-data revolution,	which has enabled confirming various theoretical assumptions of the dealer model, such as the trend-following mechanism and the buy-sell spread distributions. In the following, we review several results of the microscopic empirical analyses as performed in Refs.~\cite{kanazawa2018derivation, kanazawa2018kinetic, sueshige2018ecology} that summarise the trading strategies employed by real high-frequency traders. In particular, we focus on the trend-following strategies implemented by market makers that directly validate the dealer model with the microscopic data.

\paragraph*{Microscopic data: trading logs of individual traders} Here we describe the microscopic data used in the analyses in Refs.~\cite{kanazawa2018derivation, kanazawa2018kinetic, sueshige2018ecology}. The trading-log data originates from the Electronic Broking Services market, one of the biggest foreign exchange markets in the world managed by the CME Group. This data includes the decision-making process of traders, such as order submissions, cancellations, and executions, with anonymised trader identifiers and anonymised bank codes. Our focus, in particular, is on the exchange market between the US dollar (USD) and the Japanese yen (JPY) from 18:00 GMT on 5 June to 22:00 GMT on 10 June 2016, with the minimum volume unit being \$1M\,USD, the minimum price precision (called tick size) \yen0.005\,JPY, and the minimum time precision 1\,ms. For brevity, \yen0.001\,JPY is used as a price unit called tenth pip or simply tpip.

Our attention is centred around high-frequency traders (HFTs), typically machines that frequently submit and cancel their orders according to some strategic algorithm. HFTs are defined according to the total number of limit-order submissions. Specifically, a trader who submits more than 2,500 orders weekly qualifies as an HFT. This definition is similar to the one from a previous study~\cite{schmidt2012ecology} of the Electronic Broking Service market. There are many potential alternative definitions that could be considered, but ours offers clarity and the ease of implementation. With this definition in mind, we identified 134 HFTs during the week under consideration. The total number of traders was 1,015.

\begin{figure}[!t]
\centering
\makebox[\textwidth][c]{\includegraphics[scale=1.0]{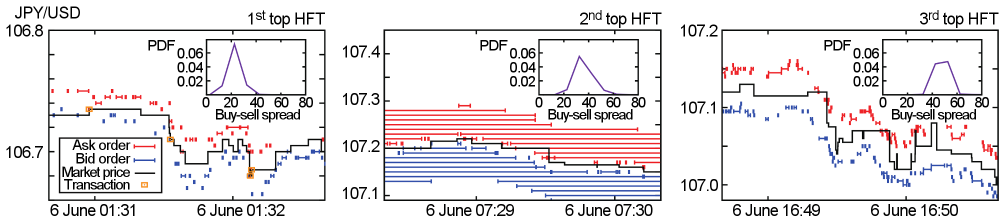}}
\caption{Sample trajectories of the three top HFTs. Plotted are the limit order prices for both bid (blue) and ask (red) sides. The insets illustrate the probability density functions of the buy-sell spread $\hat{L}_i$ for each HFT individually.\newline
Source: Reprinted figure from Ref.~\cite{kanazawa2018derivation} under the Creative Commons Attribution 4.0 International (CC BY 4.0).}
\label{fig:EcoPhysFinKZ_sample_trajectories_HFTs}
\end{figure}

\paragraph*{HFTs as liquidity providers} By plotting three sample trajectories of HFTs in terms of their limit orders (Fig.~\ref{fig:EcoPhysFinKZ_sample_trajectories_HFTs}), we observe that the HFTs typically maintain two-sided (buy-low and sell-high) quotes. Generally, two-sided quotes attempt to profit from the bid-ask spread, but are also subject to liquidity rebates that may exceed the trading fees, allowing HFTs to trade with zero marginal cost~\cite{poirer2012high}. In our case, the HFT behaviour is indeed interpreted as liquidity provision (i.e., market making) in response to the request by the platform managers. HFTs have an incentive to play the role of liquidity providers according to the rule book of the Electronic Broking Services market~\cite{unknown2016ebs}.

Here, we denote the best bid and ask prices of the $i$th HFT by $\hat{b}_i$ and $\hat{a}_i$, respectively, where the index $i$ is allocated according to the number of submissions during the week. Any variable with a hat (e.g., $\hat{A}$) implies a stochastic variable, to distinguish from a real number (such as $A$). The difference between the best bid and ask prices $\hat{L}_i= \hat{b}_i-\hat{a}_i$ is called the buy-sell (sometimes also bid-ask) spread of the $i$th HFT. The buy-sell spread $\hat{L}_i$ can be directly measured in our dataset at the level of individual HFTs (see the insets in Fig.~\ref{fig:EcoPhysFinKZ_sample_trajectories_HFTs}). In addition, we can define the mid-price of the $i$th HFT as $\hat{z}_i=\frac{1}{2}(\hat{b}_i+\hat{a}_i)$. The mid-price $\hat{z}_i$ can be interpreted as the appropriate price in the eyes of the $i$th HFT at the time, while $\hat{L}_i$ can be interpreted as a profit estimate for a round-trip trade, or alternatively, a risk evaluation against adverse selection (i.e., a possibility that the HFT misses some information).

\paragraph*{Trend-following behaviour} As discussed in Section~\ref{S:EconoPhys:Dealer}, the dealer model was originally constructed with the assumption of trend-following behaviour of traders in mind. We have validated this theoretical assumption by direct observation, that is, by identifying a statistical correlation in microscopic data that can be interpreted as trend following at the level of individual HFTs. The analytical framework in this context can be described as follows.

\begin{figure}[!t]
\centering
\makebox[\textwidth][c]{\includegraphics[scale=1.0]{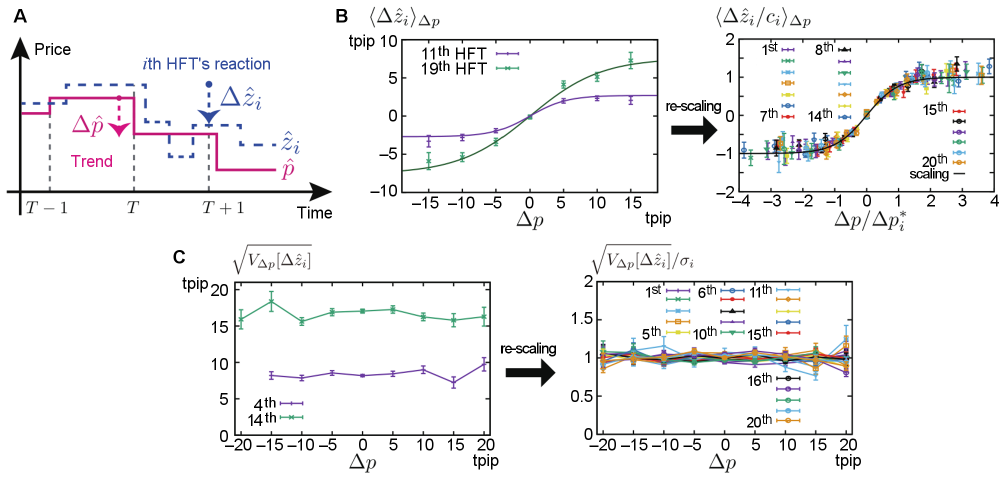}}
\caption{Trend-following analysis. \textbf{A,} Tick time $T$ is an integer time measure incremented by 1 at every transaction. $\Delta \hat{z}_i[T]$ is the future mid-price change of the $i$th HFT and $\Delta \hat{p}[T-1]$ is the historical market-price change. \textbf{B,} Statistical correlation suggests trend-following behaviour on average at the level of individual HFTs (left panel emphasises the 11th and the 19th HFT). Upon re-scaling, the same master curve Eq.~(\ref{eq:trend-following_1tick}) is seen to be valid for at least the top 20 HFTs (right panel). \textbf{C,} Variance $V_{\Delta p}[\Delta \hat{z}_i]$ conditional on the historical price change before and after scaling (left and right panel, respectively), implying that the variance is irrelevant to the historical price change.\newline
Source: Reprinted figure from Ref.~\cite{kanazawa2018derivation} under the Creative Commons Attribution 4.0 International (CC BY 4.0).}
\label{fig:EcoPhysFinKZ_Trend-following}
\end{figure}

First, let us introduce the tick time $T$ as an integer time incremented by 1 when a transaction is executed (see Fig.~\ref{fig:EcoPhysFinKZ_Trend-following}A). We note that the tick time can be mapped onto the physical time as $t=\hat{t}[T]$, where the square bracket stresses that the argument of the stochastic variable is based on the tick time. The mid-price of the $i$th HFT at the tick time $T$ is represented by $\hat{z}_i[T]$ and the market transaction price is represented by $\hat{p}[T]$.

Then, let us study the correlation between the one-tick future change of mid-price for the $i$th HFT, $\Delta \hat{z}_i[T]=\hat{z}_i[T+1]-\hat{z}_i[T]$, and the one-tick historical market-price change, $\Delta \hat{p}[T-1]=\hat{p}[T]-\hat{p}[T-1]$ (see Fig.~\ref{fig:EcoPhysFinKZ_Trend-following}A). The average of $\Delta \hat{z}_i[T]$ conditional on $\Delta \hat{p}[T-1]$, denoted $\lra{\Delta \hat{z}_i}_{\Delta p}$, for two sample HFTs (Fig.~\ref{fig:EcoPhysFinKZ_Trend-following}B, left panel) shows that the correlation is linear for $\Delta p\to 0$, but saturates for $\Delta p\to \infty$. This suggests, on average, a hyperbolic-tangent scaling relation
\begin{linenomath}
\begin{equation}
\lra{\Delta \hat{z}_i[T]}_{\Delta p} \approx c_i\tanh{\frac{\Delta p}{\Delta p^*_i}}
\label{eq:trend-following_1tick}
\end{equation}
\end{linenomath}
with the parameters $c_i$ and $\Delta p^*_i$ unique to the $i$th HFT. Here, the bracket $\lra{\hat{A}}_{\Delta p}=\lra{\hat{A}}_{\Delta \hat{p}[T-1]=\Delta p}$ implies the ensemble average of $\hat{A}$ conditional on the previous price change begin fixed to $\Delta \hat{p}[T-1]=\Delta p$ and on the HFT being active, that is, $\Delta \hat{z}_i[T] \neq 0$. Indeed, by re-scaling the horizontal axis to $\Delta p/\Delta p^*_i$ and the vertical axis to $\lra{\Delta \hat{z}_i/c_i}_{\Delta p}$, we observe a clear master curve among the top 20 HFTs (Fig.~\ref{fig:EcoPhysFinKZ_Trend-following}B, right panel), suggesting the universal validity of the formula~\eqref{eq:trend-following_1tick} for the top HFTs in the studied market.

There is also evidence of another scaling relation that holds for the variance of an HFT's one-tick future change of mid-price conditional on the historical market-price change being $\Delta p$ (Fig.~\ref{fig:EcoPhysFinKZ_Trend-following}C). We specifically have
\begin{linenomath}
\begin{equation}
V_{\Delta p}[\Delta \hat{z}_i[T]]= \lra{(\Delta \hat{z}_i[T]-\lra{\hat{z}_i[T]}_{\Delta p})^2}_{\Delta p} \approx \sigma^2_i,
\label{eq:trend-following_variance_1tick}
\end{equation}
\end{linenomath}
where the quantity $\sigma^2_i$ is a constant unique to the $i$th HFT. This relation suggests that the variance, unlike the mean, is independent of the historical market-price change $\Delta p$; HFTs follow the trend, but how much they adjust their mid-price in doing so is solely their intrinsic property.

The scaling relations described here are statistical laws that reveal strategic trading behaviour beyond the previously mentioned zero-intelligence models. Such behaviour holds for HFTs as market makers, but does it differ from what low-frequency traders (LFTs) do?

\begin{figure}[!t]
\centering
\includegraphics[scale=1.0]{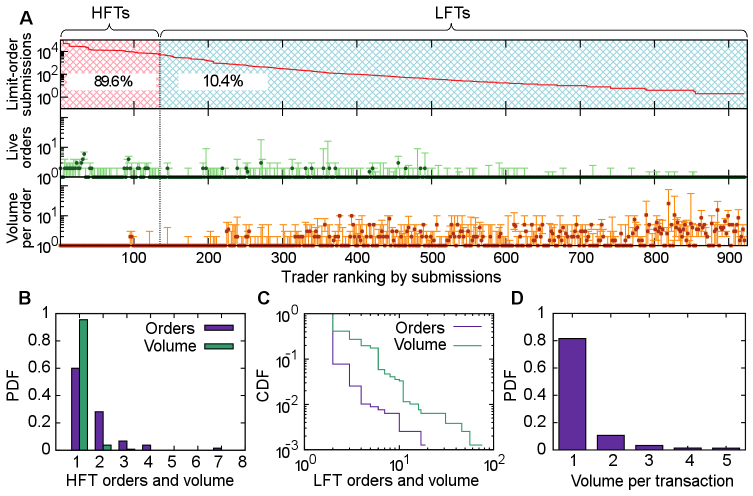}
\caption{How HFTs and LFTs differ. \textbf{A,} Plots are an overview of limit-order submissions (top), live orders (middle), and volume assigned per order (bottom), according to trader ranking by submissions. Traders submitting more than 2,500 limit orders weekly are defined as HFTs; there have been 135 such traders in the available dataset, responsible for almost 90\,\% of total submissions. \textbf{B,} Probability density functions reveal that HFTs maintain only a few live orders at a time, and that the volume per order is overwhelmingly one unit. \textbf{C,} Cumulative distribution functions reveal that LFTs use a wider range of volumes per order, suggesting a more diverse set of strategies than the one used by HFTs. \textbf{D,} Probability density function of the volume per transaction for orders that get filled shows that 81.5\,\% of transactions are one-to-one, while the volume is less than five units for 98.2\,\% of transactions.\newline
Source: Reprinted figure from Ref.~\cite{kanazawa2018kinetic} under the Creative Commons Attribution 4.0 International (CC BY 4.0).}
\label{fig:EcoPhysFinKZ_HFTvsLFT}
\end{figure}

Indeed, there are noticeable differences between HFTs and LFTs. The former keep a few live orders, typically less than 10, while allocating one unit of volume per order (see Fig.~\ref{fig:EcoPhysFinKZ_HFTvsLFT}A,\,B). This volume per order is in contrast to LFTs who allocate enough that the corresponding distribution follows a power law (Fig.~\ref{fig:EcoPhysFinKZ_HFTvsLFT}A,\,C). Overall, statistics illustrate that HFTs vary less in terms of trading strategies than LFTs. Classifying traders into these two distinct groups is therefore justified.

\begin{figure}[!t]
\centering
\makebox[\textwidth][c]{\includegraphics[scale=1.0]{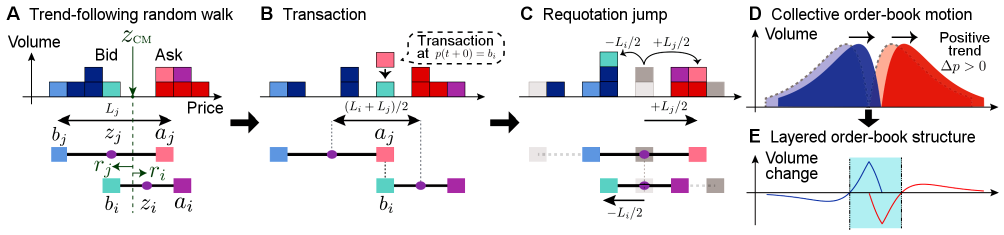}}
\caption{
Schematic of the microscopic HFT dynamics and the corresponding order-book dynamics. \textbf{A,} Trend-following random walks take place in the absence of transactions. \textbf{B,} Transactions occur at the instance of price matching $\hat{a}_j=\hat{b}_i$. \textbf{C,} After a transaction between a pair of HFTs takes place, they resubmit their prices far from the transaction price. \textbf{D,} Trend-following random walks induce the collective order-book motion. \textbf{E,} Collective order-book motion consistently causes the layered order-book structure.\newline
Source: Reprinted figure from Ref.~\cite{kanazawa2018kinetic} under the Creative Commons Attribution 4.0 International (CC BY 4.0).}
\label{fig:EcoPhysFinKZ_DealerModel}
\end{figure}

\paragraph*{Modelling based on microscopic evidence} Focusing again on HFTs, we here construct a mathematical model that reflects the microscopic empirical evidence described heretofore. Let $N\gg1$ denote the number of HFTs. The model's assumptions are:
\begin{enumerate}
\item \textit{Order and volume.} Every HFT submits a single order at a time with a single unit of volume (Fig.~\ref{fig:EcoPhysFinKZ_HFTvsLFT}A,\,B).
\item \textit{Liquidity provision.} Every HFT keeps both bid and ask orders to play the role of a liquidity provider (Fig.~\ref{fig:EcoPhysFinKZ_sample_trajectories_HFTs}).
\item \textit{Frequent price updates.} HFTs frequently update their prices by successive order submissions and cancellations. This implies that the price trajectory is approximately continuous except at the times of transactions. The continuous Markov stochastic processes for price trajectories are modelled as an It\^{o} process (i.e., a Gaussian stochastic process)~\cite{gardiner1985handbook}.
\item \textit{Trend following.} HFTs exhibit the trend-following behaviour in accordance to the empirical laws in Eq.~(\ref{eq:trend-following_1tick}) and Eq.~(\ref{eq:trend-following_variance_1tick}) (see also Fig.~\ref{fig:EcoPhysFinKZ_Trend-following}). For simplicity, we assume the uniformity of model parameters in these equations such that $c_i=c$, $\Delta p^*_i=\Delta p^*$, and $\sigma_i^2=\sigma^2$ for all $i$.
\item \textit{Spread.} The buy-sell spread of the $i$th HFT is defined by $\hat{L}_i=\hat{a}_i-\hat{b}_i$. Because the probability density functions of buy-sell spreads have a single peak (insets in Fig.~\ref{fig:EcoPhysFinKZ_sample_trajectories_HFTs}A--C), the spread is assumed unique to the HFT and constant, that is, $\hat{L}_i = L_i$.
This assumption implies that the mid-price $\hat{z}_i=\frac{1}{2}(\hat{b}_i+\hat{a}_i)$ is sufficient to characterise the $i$th HFT. The empirical probability density function of the buy-sell spread $\rho(L)$ is measurable from the available dataset using the relationship $\rho(L)=\frac{1}{N}\sum_{i=1}^N \delta(L-L_i)$, and thus describes the order-book distribution.
\end{enumerate}

\begin{subequations}
\label{eq:dealermodel_revisted_summary}

Based on the listed assumptions, we model the microscopic HFT dynamics in the absence of transactions as the trend-following random walks (Fig.~\ref{fig:EcoPhysFinKZ_DealerModel}A)
\begin{linenomath}
\begin{equation}
\frac{d\hat{z}_i}{dt} = c\tanh\frac{\Delta \hat{p}}{\Delta p^*_i} + \sigma \hat{\eta}_i^{\mathrm{R}},
\label{eq:trend_following_random_walks_kineticEcoPhys}
\end{equation}
\end{linenomath}
where the white noise $\hat{\eta}_i^{\mathrm{R}}$ is independent of the white noise $\hat{\eta}_j^{\mathrm{R}}$ for $j \neq i$. This is the minimal It\^o process~\cite{gardiner1985handbook} satisfying the empirical relations we have described so far.

At the instant of price matching (Fig.~\ref{fig:EcoPhysFinKZ_DealerModel}B)
\begin{linenomath}
\begin{equation}
\hat{a}_j = \hat{b}_i, \>\>\> i\neq j,
\label{eq:contanct_DealerModel}
\end{equation}
\end{linenomath}
the pair of HFTs $i$ and $j$ resubmit their prices far from the transaction price (Fig.~\ref{fig:EcoPhysFinKZ_DealerModel}C)
\begin{linenomath}
\begin{equation}
\hat{a}_j^{\mathrm{pst}} = \hat{a}_j + \frac{L_j}{2}, \>\>\> \hat{b}_i^{\mathrm{pst}} = \hat{b}_i - \frac{L_i}{2},
\label{eq:resubmission_DealerModel}
\end{equation}
\end{linenomath}
where $\hat{a}_j^{\mathrm{pst}}$ and $\hat{b}_i^{\mathrm{pst}}$ are the updated prices after the transaction. These transaction conditions can be rewritten as
\begin{linenomath}
\begin{equation}
\left|\hat{z}_i-\hat{z}_j\right|=\frac{L_i+L_j}{2} \>\>\> \Longrightarrow \>\>\> \hat{z}_i^{\mathrm{pst}} = \hat{z}_i - \frac{L_i}{2}\mathrm{sgn}(\hat{z}_i-\hat{z}_j),
\label{eq:transaction_rule_z}
\end{equation}
\end{linenomath}
where the sign function $\mathrm{sgn}(x)$ is defined by $\mathrm{sgn}(0)=0$, $\mathrm{sgn}(x)=1$ for $x>0$, and $\mathrm{sgn}(x)=-1$ for $x<0$. The market price $\hat{p}$ and the trend signal $\Delta \hat{p}$ at a transaction instant are updated with the post-transaction values
\begin{linenomath}
\begin{equation}
\hat{p}^{\mathrm{pst}} = \hat{z}_i - \frac{L_i}{2}\mathrm{sgn}(\hat{z}_i-\hat{z}_j), \>\>\> \Delta \hat{p}^{\mathrm{pst}} = \hat{p}^{\mathrm{pst}} - \hat{p}.
\label{eq:transaction_rule_p}
\end{equation}
\end{linenomath}
We note that the transaction condition in Eq.~(\ref{eq:transaction_rule_z}) and the resubmission rule in Eq.~(\ref{eq:transaction_rule_p}), respectively, bear mathematical resemblance to the contact condition and the momentum-exchange rule in conventional kinetic theory. This analogy will be revisited again later to formulate a statistical-physics description of financial markets. We also note that one-to-one transactions are the basic interaction mode between HFTs (Fig.~\ref{fig:EcoPhysFinKZ_HFTvsLFT}D), which is consistent with binary collisions.

In statistical physics, an appropriate separation of spatio-temporal scales is often used to formulate successful micro-macro theories; an example is the enslaving principle in Haken's synergistics~\cite{haken2013synergetics}. Here we introduce the centre of mass as a key macroscopic variable
\begin{linenomath}
\begin{equation}
\hat{z}_\mathrm{CM}=\frac{1}{N}\sum_{i=1}^N \hat{z}_i,
\end{equation}
\end{linenomath}
which is expected to play the role of a slow variable in the `thermodynamic limit' when $N\to \infty$. Indeed, the diffusion of the centre of mass turns out to be slow (i.e., proportional to $N^{-1}$ in the absence of trend following~\cite{kanazawa2018kinetic}), confirming the appropriateness of this particular variable selection.
\end{subequations}

A complete set of the system variables is given by $\hat{\Gamma}= (\hat{z}_1,\dots,\hat{z}_N; \hat{z}_\mathrm{CM}, \hat{p}, \Delta \hat{p})$, with $\hat{Z} = (\hat{z}_\mathrm{CM}, \hat{p}, \Delta \hat{p})$ being the subset of macroscopic variables. $\hat{\Gamma}$ can be regarded as a phase point in the space $\mathcal{S}$ such that $\hat{\Gamma} \in \mathcal{S} = \prod_{i=1}^{N+3}(-\infty,\infty)$. We have thus defined a Markovian stochastic processes whose dynamics is characterised by the set of Eqs.~(\ref{eq:dealermodel_revisted_summary}).

The order-book dynamics associated with the described microscopic model (Fig.~\ref{fig:EcoPhysFinKZ_DealerModel}A--C) relates the trend-following behaviour to the layered order-book structure found in Ref.~\cite{yura2014financial}. Specifically, trend following induces the collective motion of orders (Fig.~\ref{fig:EcoPhysFinKZ_DealerModel}D), which in turn accounts for the order book's layered structure (Fig.~\ref{fig:EcoPhysFinKZ_DealerModel}E).

\subsection{Solving the microscopic model via kinetic theory}

Having presented a model of the decision-making process among HFTs based on empirical microscopic evidence, we proceed to solve this microscopic model in order to understand its macroscopic behaviour. Mathematically, we are dealing with a high-dimensional stochastic dynamical system exhibiting a structure similar to the Hamiltonian dynamics. We therefore rely on the methods of statistical physics in general, and kinetic theory in particular, to crack the problem.

\begin{figure}[!t]
\centering
\makebox[\textwidth][c]{\includegraphics[scale=1.0]{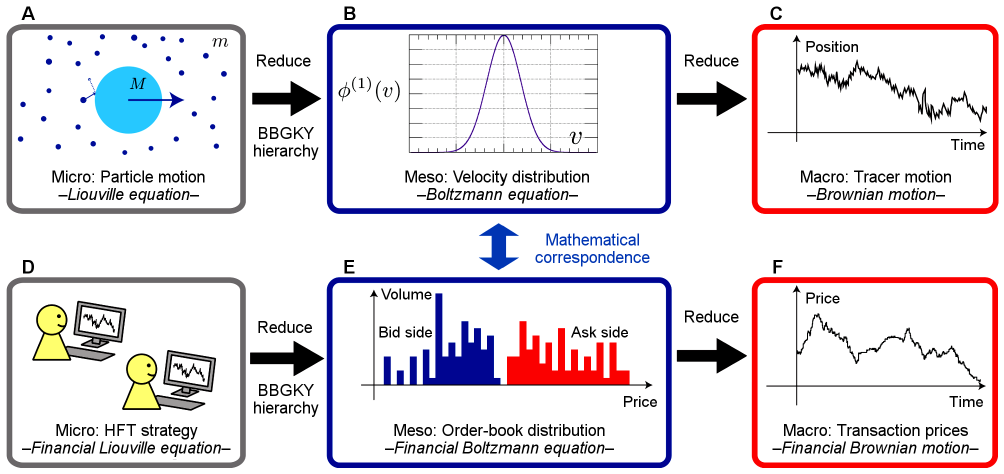}}
\caption{Schematic representation of the correspondence between kinetic theories for physical and financial Brownian motion. \textbf{A,} The starting point of physical kinetic theory is the Liouville equation for microscopic particle dynamics. \textbf{B,} The Liouville equation is reduced via the BBGKY hierarchy to the Boltzmann equation, which can be interpreted as the mesoscopic description of the dynamical system. \textbf{C,} The Brownian motion, described by the Langevin equation, is derived after further coarse-graining to yield the system's macroscopic dynamics. \textbf{D,} The microscopic dynamics of financial markets is driven by the decision-making process of individual HFTs. This is captured by the financial Liouville equation~(\ref{eq:financial_Liouville_eq}), which is mathematically equivalent to the trend-following random-walks model in Eqs.~(\ref{eq:dealermodel_revisted_summary}). \textbf{E,} By reducing the microscopic market dynamics via the BBGKY hierarchy, we obtain the order-book dynamics in terms of the financial Boltzmann equation~(\ref{eq:Boltzmann_fin}). The financial Boltzmann equation can be seen as the mesoscopic description of financial markets. \textbf{F,} Further coarse-graining reveals the market-price diffusion in the form of the financial Langevin equation~(\ref{eq:financial_Langevin_fin}).\newline
Source: Reprinted figure from Ref.~\cite{kanazawa2018kinetic} under the Creative Commons Attribution 4.0 International (CC BY 4.0).}
\label{fig:EcoPhysFinKZ_Hierarchy}
\end{figure}

\paragraph*{Financial Liouville equation} Here, we start by briefly reviewing the conventional kinetic theory whose goal is to reduce the original high-dimensional dynamical system composed of $N$ particles to a few-dimensional dynamical system. The microscopic dynamics (Fig.~\ref{fig:EcoPhysFinKZ_Hierarchy}A) is generally characterised by Newton's equation of motion in $6N$ dimensions, which is mathematically equivalent to the Liouville equation in analytical mechanics~\cite{landau1976course, evans2008statistical}. The Liouville equation reduces to the Boltzmann equation via the Bogoliubov-Born-Green-Kirkwood-Yvon (BBGKY) hierarchy~\cite{hansen1990theory} by applying the mean-field approximation called molecular chaos, thus offering a mesoscopic description of the system (Fig.~\ref{fig:EcoPhysFinKZ_Hierarchy}B). This reduction is powerful in the sense that the original $6N$-dimensional dynamics is approximated by a $6$-dimensional dynamics. Further reduction of the dynamics yields the Brownian motion of a tracer particle~\cite{vankampen1992stochastic, spohn1980kinetic} as the macroscopic description of the system (Fig.~\ref{fig:EcoPhysFinKZ_Hierarchy}C).

We have retraced the steps of the conventional kinetic theory with financial markets and the HFT behaviour in mind. From Eqs.~(\ref{eq:dealermodel_revisted_summary}), we derive the financial Liouville equation (equivalently, the Chapman-Kolmogorov equation~\cite{gardiner1985handbook} or the master equation~\cite{vankampen1992stochastic}) as the time-evolution equation for the $N$-body probability density function $P_t(\Gamma)$
\begin{linenomath}
\begin{equation}
\frac{\partial P_t(\Gamma)}{\partial t} = \mathcal{L} P_t(\Gamma)
\label{eq:financial_Liouville_eq}
\end{equation}
\end{linenomath}
with an appropriate linear operator $\mathcal{L}$ called the Liouville operator. We note that this equation is derived as an identity without any approximation
and is equivalent to the original stochastic model described by Eqs.~(\ref{eq:dealermodel_revisted_summary}). This is an exact starting point for our statistical-physics theory.

\paragraph*{Financial BBGKY hieararchy} While the financial Liouville equation~(\ref{eq:financial_Liouville_eq}) is exact, it cannot be solved analytically because it describes genuine many-body dynamics of a complex system. Here, we reduce this dynamical equation following the ideas behind the BBGKY hierarchy, which was historically invented for the purpose of a systematic derivation of the Boltzmann equation from microscopic particle dynamics. The derivation is based on reducing the dimension of the original high-dimensional dynamics by integrating out all variables except for a few dominating ones.

Because the derivation following the BBGKY hierarchy is long and technical~\cite{kanazawa2018kinetic}, we only present the final results. Let us first define prices relative to the centre of mass, $\hat{r}_i= \hat{z}_i - \hat{z}_\mathrm{CM}$, and the corresponding one-body and two-body distributions $\phi_L(r)$ and $\phi_{LL}(r,r')$. Here, $\phi_L(r)$ is the probability density function of the relative mid-price $r$ for an HFT with the spread $L$, while $\phi_{LL'}(r,r')$ is the joint probability density function of a pair of HFTs with the spreads $L$ and $L'$.

By integrating all but one variable out of Eq.~(\ref{eq:financial_Liouville_eq}), we obtain the lowest-order BBGKY hierarchical equation
\begin{linenomath}
\begin{subequations}
\label{eq:BBGKY_fin}
\begin{align}
\frac{\partial \phi_L(r)}{\partial t} &\approx \frac{\sigma^2}{2}\frac{\partial \phi_L(r)}{\partial r^2} + N\sum_{s=\pm 1}\int_0^\infty \dup{L'}\rho(L')[J_{LL'}^s(r+sL/2)-J_{LL'}^s(r)] \label{eq:BBGKY_fin_1}\\
J^s_{LL'}(r) &= \frac{\sigma^2}{2}|\tilde{\partial}_{rr'}|\phi_{LL'}(r,r')\left|_{r-r'=s(L+L')/2}\right.,
\end{align}
\end{subequations}
\end{linenomath}
where $|\tilde{\partial}_{rr'}|f = |\partial f/\partial r| + |\partial f/\partial r'|$.
The second term of the right-hand side of Eq.~(\ref{eq:BBGKY_fin_1}) represents the effect of transactions ($s=+1$ for bids and $s=-1$ for asks) and corresponds to the collision integral in physical kinetic theory. The three-body `collision' integral is dropped here because it is becomes irrelevant for a large $N$.

\paragraph*{Financial Boltzmann equation} The BBGKY hierarchical equation~(\ref{eq:BBGKY_fin}) is a formalism that needs a closure, that is, further approximation is necessary to find the solution. We apply the `molecular chaos' assumption, which is a standard mean-field approximation in kinetic theory
\begin{linenomath}
\begin{equation}
\phi_{LL'}(r,r') \approx \phi_L(r)\phi_{L'}(r'),
\label{eq:molecular_chaos_fin_Boltzmann}
\end{equation}
\end{linenomath}
to obtain the financial Boltzmann equation
\begin{linenomath}
\begin{subequations}
\label{eq:Boltzmann_fin}
\begin{align}
\frac{\partial \phi_L(r)}{\partial t} &\approx \frac{\sigma^2}{2}\frac{\partial \phi_L(r)}{\partial r^2} + N\sum_{s=\pm 1}\int_0^\infty \dup{L'}\rho(L')[\tilde{J}^{s}_{LL'}(r+sL/2)-J^{s}_{LL'}(r)] \label{eq:Boltzmann_fin_1}\\
\tilde{J}^s_{LL'}(r) &= \frac{\sigma^2}{2}|\tilde{\partial}_{rr'}|\phi_{L}(r)\phi_{L'}(r')\left|_{r-r'=s(L+L')/2}\right.,
\end{align}
\end{subequations}
\end{linenomath}
which is closed in terms of $\phi_{L}(r)$.
This equation can be analytically solved under an appropriate boundary condition for large $N$. Indeed, the leading-order steady solution is given by the tent function
\begin{linenomath}
\begin{equation}
\lim_{N\to \infty}\phi_L(r) = \frac{4}{L^2}\max\left\{\frac{L}{2}-|r|,0\right\}.
\end{equation}
\end{linenomath}
Notably, even the next-to-leading order solution is accessible, which is necessary for detailed mean-field analyses.

The average order-book profile follows from the financial Boltzmann equation via a formula
\begin{linenomath}
\begin{equation}
f^{\mathrm{A}}(r) = \frac{1}{N}\left< \sum_{i=1}^N \delta (r-\hat{r}_i)\right> \approx \int_0^\infty \dup{L}\rho(L)\phi_L(r-L/2),
\label{eq:order_book_profile_general}
\end{equation}
\end{linenomath}
where the summation is approximated by the integral over all spreads $L$.
Remarkably, the book profile is analytically derived for any spread distribution $\rho(L)$, suggesting that the microscopic model of the HFT behaviour is an analytically well-tractable model.

\paragraph*{Financial Langevin equation} By additional coarse-graining, we obtain the financial Langevin equation
\begin{linenomath}
\begin{equation}
\Delta \hat{p}[T+1]\approx c\hat{\tau}[T]\tanh \frac{\Delta \hat{p}[T]}{\Delta p^*} + \hat{\zeta}[T]
\label{eq:financial_Langevin_fin}
\end{equation}
\end{linenomath}
for the macroscopic dynamics of the studied system. Here, $\hat{\tau}[T] = \hat{t}[T+1]-\hat{t}[T]$ is the time interval between the ticks $T$ and $T+1$, whereas $\hat{\zeta}[T]$ is a random noise term. The mean-field approximation permits obtaining all the statistics for $\hat{\tau}$ and $\hat{\zeta}$ analytically. The time interval $\hat{\tau}$ exhibits the exponential distribution with the mean time interval $\tau^* \approx L^*_{\rho}/(2N\sigma^2)$ and $L^*_{\rho}=1/\int_0^\infty L^{-2}\rho(L)\dup{L}$, implying the Poissonian statistics asymptotically.

The dynamical characteristics of the financial Langevin equation~(\ref{eq:financial_Langevin_fin}) depend on two dimensionless parameters, $\tilde{c} = cL^*_{\rho}/(\sigma^2\sqrt{2N})$ and $\Delta \tilde{p}^* = \Delta p^*/(c\tau^*)$. Focusing on a regime in which $\tilde{c}\gtrapprox 1$ and $\Delta \tilde{p}^* \lessapprox 1$, called the marginal-to-strong trend following~\cite{kanazawa2018kinetic}, the statistics of the price change $\Delta \hat{p}$ asymptotically obeys the exponential law
\begin{linenomath}
\begin{equation}
P(\geq |\Delta p|; \kappa) \sim e^{-|\Delta p|/\kappa} \>\>\> \mbox{for large }|\Delta p|,
\label{eq:price_change_dist.}
\end{equation}
\end{linenomath}
where $P(\geq |\Delta p|; \kappa)$ is a complementary cumulative distribution function with the decay length $\kappa$.

\paragraph*{Numerical confirmation} The validity of the described analytical predictions, such as the order-book profile in Eq.~(\ref{eq:order_book_profile_general}) and the exponential price-change distribution in Eq.~(\ref{eq:price_change_dist.}), can be directly checked by means of numerical simulations~\cite{kanazawa2018kinetic}. The results are encouraging; especially the order-book profile formula in Eq.~(\ref{eq:order_book_profile_general}) agrees with the numerical results very precisely.
		
Such a precise agreement might be somewhat counter-intuitive because mean-field approximations are generally expected to be valid only for high-dimensional spaces while the price space is one-dimensional. This counter-intuitive result can be understood from the viewpoint of the `collision rule': low-dimensional spaces are special for physical systems in the sense that geometry restricts movements of particles after collisions. The mean-field approximation then fails because the same pair of particles collides many times and the two-body correlation persists. In the case of our microscopic model, the `collision rule' pushes the limit-order prices of a pair of HFTs far from the market price, making successive `collisions' between the same HFT pair highly unlikely for a large $N$. The two-body correlation thus quickly decays, which to a large degree validates the assumption of the molecular chaos in our kinetic formulation. In fact, the kinetic formulation might work better for some social systems than for physical systems because the continuity of paths is often unnecessary in social dynamics but strictly required in physical dynamics.

\subsection{Consistency between theory and data}

We have analysed the model for trend-following random walks in Eqs.~\eqref{eq:dealermodel_revisted_summary} within the kinetic framework. It is time to check the model's consistency against mesoscopic and macroscopic data.

\begin{figure}[!t]
\centering
\makebox[\textwidth][c]{\includegraphics[scale=1.0]{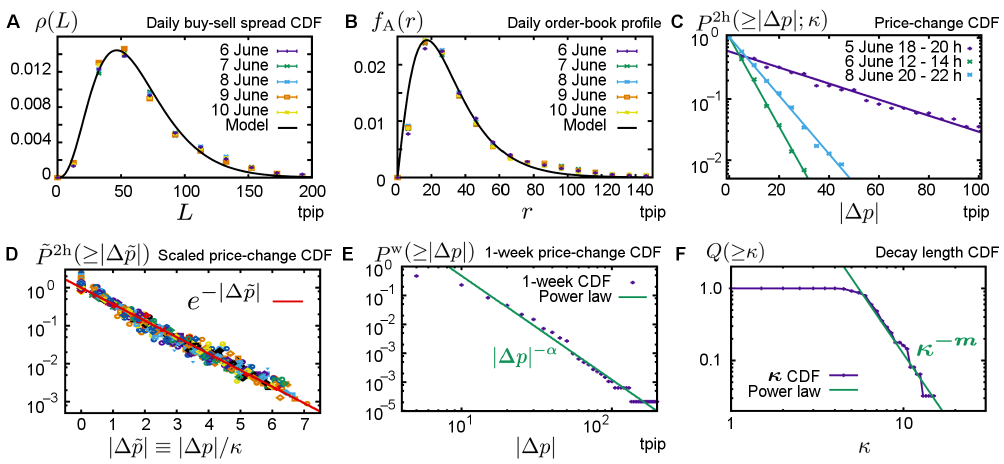}}
\caption{Theory is consistent with data. \textbf{A,} Daily buy-sell spread distribution for individual HFTs is closely approximated by the $\gamma$-distribution in Eq.~(\ref{eq:buy-sell_spread_HFT_finEcoPhys}). \textbf{B,} Daily average order-book profile composed of the best HFT prices agrees with the theoretical curve in Eq.~(\ref{eq:average_order_book_finEcoPhys}). \textbf{C,} Two-hourly segmented price-change cumulative distribution functions for three different time periods follow the exponential law in Eq.~(\ref{eq:price_change_dist.}) with the decay length $\kappa$ being time-dependent. \textbf{D,} After scaling, the two-hourly segmented price-change cumulative distribution functions collapse onto a single exponential master curve in Eq.~(\ref{eq:price_change_dist.}). \textbf{E,} Weekly segmented price-change cumulative distribution function exhibits fat tails with a power-law exponent $\alpha$. \textbf{F,} Decay length cumulative distribution function exhibits a fat tail with a power-law exponent $m$ such that $Q(\geq \kappa)\sim \kappa^{-m}$.\newline
Source: Reprinted figure from Ref.~\cite{kanazawa2018kinetic} under the Creative Commons Attribution 4.0 International (CC BY 4.0).}
\label{fig:EcoPhysFinKZ_PriceDiff}
\end{figure}

First, we have measured the buy-sell spreads of each individual HFT and estimated the corresponding daily distribution (Fig.~\ref{fig:EcoPhysFinKZ_PriceDiff}A). The buy-sell spread distribution is well approximated by the $\gamma$ distribution
\begin{linenomath}
\begin{equation}
\rho(L) \approx \frac{L^3}{6L^{*4}}e^{-L/L^*}
\label{eq:buy-sell_spread_HFT_finEcoPhys}
\end{equation}
\end{linenomath}
with $L^*\approx 15.5 \pm 0.2$\,tpip. This buy-sell distribution together with the general order-book formula in Eq.~(\ref{eq:order_book_profile_general}) implies that the average order-book profile is given by
\begin{linenomath}
\begin{equation}
f_{\mathrm{A}}(r) \approx \frac{4e^{-\frac{3r}{2L^*}}}{3L^*} \left[ \left(2+\frac{r}{L^*}\right)\sinh\frac{r}{2L^*} -\frac{re^{-\frac{r}{2L^*}}}{2L^*}  \right].
\label{eq:average_order_book_finEcoPhys}
\end{equation}
\end{linenomath}
Of note is that our model addresses the dynamics of the best bid and ask prices of individual HFTs. The average order-book profile after normalisation (Fig.~\ref{fig:EcoPhysFinKZ_PriceDiff}B) shows an excellent agreement with the theoretical curve from Eq.~(\ref{eq:average_order_book_finEcoPhys}) without any additional fitting of model parameters.

Turning to the macroscopic perspective in terms of the time series of price changes, we have seen that the financial Langevin equation~(\ref{eq:financial_Langevin_fin}) predicts the exponential law in Eq.~(\ref{eq:price_change_dist.}). This prediction turns out to be consistent with the available dataset (Fig.~\ref{fig:EcoPhysFinKZ_PriceDiff}C,\,D). The price-change distribution at a time scale of one tick indeed follows said exponential law with the decay length $\kappa$ depending on the chosen time period (Fig.~\ref{fig:EcoPhysFinKZ_PriceDiff}C). Re-scaling price changes by $\kappa$ eliminates the time-period dependence, causing all data to collapse onto a single master curve (Fig.~\ref{fig:EcoPhysFinKZ_PriceDiff}D).

Although price changes obey the exponential law at short time scales, a power law emerges as time scales get longer, which is in line with previous findings~\cite{mantegna1995scaling, lux1996stable, longin1996asymptotic, guillaume1997bird, plerou1999scaling}. The price-change distribution for the studied week has fat tails, fitted with a power-law whose exponent is $\alpha=3.6\pm 0.13$ (Fig.~\ref{fig:EcoPhysFinKZ_PriceDiff}E). The power law emerges here as a superposition of 2-hourly exponential distributions $P^{\mathrm{2h}}(\geq |\Delta p|; \kappa)$ such that
\begin{linenomath}
\begin{equation}
P^{\mathrm{w}}(\geq |\Delta p|) = \int_0^\infty \dup{\kappa} Q(\kappa)P^{\mathrm{2h}}(\geq |\Delta p|; \kappa ) \propto |\Delta p|^{-m},
\end{equation}
\end{linenomath}
where $P^{\mathrm{w}}(\geq |\Delta p|)$ is the weekly price-change cumulative distribution function, and $Q(\kappa)\propto \kappa^{-(m+1)}$ is the decay-length probability density function with the exponent being estimated at $m=3.5\pm 0.13$ (Fig.~\ref{fig:EcoPhysFinKZ_PriceDiff}F). The fact that $\alpha\approx m$ additionally confirms the consistency of the results.

\subsection{Future outlook: towards market ecology}

We have examined in detail the trend-following behaviour of HFTs at the time scale of one tick. This examination had yielded a microscopic model specified by Eqs.~(\ref{eq:dealermodel_revisted_summary}) and solved using the kinetic theory of statistical physics. Ref.~\cite{sueshige2019classification} extends the analysis, offering direct evidence that the (exponential) moving-average technique---a common tool in the repertoire of `technical analysts' or `chartists'---is applied in practice by HFTs. To achieve this feat, a regression relation inspired by Eq.~(\ref{eq:trend_following_random_walks_kineticEcoPhys}) is set up as follows
\begin{subequations}
\label{eq:NL_regression_trend-following}
\begin{linenomath}
\begin{equation}
\Delta \hat{z}_i [T] = c_i\tanh \left( \Delta \hat{p}^{(j_i)}_{\mathrm{Trend}}[T] + \alpha_i\right),
\end{equation}
\end{linenomath}
where the trend signal includes time delays
\begin{linenomath}
\begin{equation}
\Delta \hat{p}^{(j_i)}_{\mathrm{Trend}}[T] = \sum_{k=1}^{K_i}w_i[k]\Delta \hat{p}^{(j_i)}[T-k] + \sigma_i \hat{\epsilon}_i[T].
\label{eq:NL_regression_trend-parameters}
\end{equation}
\end{linenomath}
Here, $c_i$, $\alpha_i$, $w_i[k]$, $\sigma_i$, and $K_i$ are regression coefficients, $\hat{\epsilon}_i[T]$ is the white noise, and $\{ \Delta \hat{p}^{(j_i)}[T-k]\}_k$ is a coarse-grained price time series defined as
\begin{linenomath}
\begin{equation}
\Delta \hat{p}^{(j_i)}[T-k]= \hat{p}[T-j_i(k-1)] - \hat{p}[T-j_i k],
\end{equation}
\end{linenomath}
\end{subequations}
where $k$, $K_i$, and $j_i$ are respectively called the time lag, the maximum time lag of the $i$th HFT, and the coarse-graining parameter. The regression coefficients are estimated using iterative non-linear multiple-regression methods.

The weights $\{w_i[k]\}_{k=1,\dots,K_i}$ determine the nature of the moving average applied by HFTs. The results show that an exponential scaling law is satisfied
\begin{linenomath}
\begin{equation}
w_i[k] \approx d_i\exp\left(-\frac{k}{\tau_i}\right),
\label{eq:expon_w_i_scaling}
\end{equation}
\end{linenomath}
where parameters $d_i$ and $\tau_i$ characterise the $i$th HFT. Among the examined HFTs, 85\,\% were shown to conform to this exponential moving average. Further subdivision of HFTs was possible according to their typical trend-following time scale; (i) short time-scale HFTs follow the trend for about 4 ticks (30\,s), (ii) intermediate time-scale HFTs follow the trend for about 20 ticks (3\,min), and (iii) long time-scale HFTs follow the trend for about 40 ticks (6\,min). The remaining 15\,\% of HFTs use other trading strategies.

Identifying market strategies as described here is a first step in the direction of understanding market ecology~\cite{farmer2002market, farmer2013ecological}, that is, interactions among various trading strategies (e.g., how strategies contribute to market liquidity or price formation). If such interactions are precisely understood, designing market simulators for regulatory purposes becomes entirely plausible. Regulators could then plan interventions to enhance market liquidity and stability. The scarcity of microscopic data has so far precluded in-depth analyses of market ecology (see Refs.~\cite{odean1998investors, grinblatt2000investment, lee2004order, delachapelle2010turnover, toth2012does, tumminello2012identification, lillo2015news, musciotto2018long} for notable attempts so far), but we firmly believe that the situation is changing and that market ecology is within our grasp.

\FloatBarrier

\section{Cooperation}
\label{S:Coop}

Conspecific organisms mutually interact to a lesser or greater degree. Some species are highly individualistic, associating only for courtship and mating. Others are extremely social with overlapping adult generations, reproductive division of labour, cooperative care of young, and sometimes even a biological caste system. Sociality, or living in groups, implies a coexistence of two opposing forces: conflict over local resources and cooperation with neighbours~\cite{frank2007life}. These two opposing forces form the basis of evolutionary game theory whose aim is to understand the evolution and pervasiveness of cooperation in biological and social systems. More specifically, the goal is to answer how natural selection can favour costly, cooperative behaviours that benefit others.

Ever since the mathematical framework of game theory was applied to evolution~\cite{smith1982evolution}, the research on cooperation attracted the attention of fields as varied as biology, psychology, economics, physics, and others~\cite{sachs2004evolution, cooper2016other, henrich2021origins, perc2017statistical}. Such a variety brought together a plethora of unique perspectives on the problem of cooperation, giving rise to a reasonably good understanding of the origins and stability of cooperativeness. It is safe to say that we are at a point at which the evolution of cooperation is much less of a puzzle than it used to be~\cite{akcay2020deconstructing}.

Hereafter, we take a look at some of the central tenets of evolutionary game theory, and then review the main interests and contributions of physics to this field. A particular focus is on networks that define the topology of interactions among humans, as well as on a gap between theoretical models and empirical facts. We conclude with ideas for reconciling the gap between theory and experiments while heeding the initiatives from relevant behavioural disciplines (e.g., psychology and behavioural economics) for better research practices.

\subsection{Social dilemmas}

Social dilemmas are situations in which the process of selection favours defection over cooperation while reducing population welfare compared to when everybody cooperates~\cite{dawes1980social}. Many real-world social dilemmas fit the format of dyadic games in which a pair of players simultaneously choose between cooperation $C$ or defection $D$. Depending on the choices made, the outcome is defined with the following four payoff-matrix elements
\begin{linenomath}
\begin{equation}
\bordermatrix{
\text{ } & C & D \cr
       C & R & S \cr
       D & T & P
}.
\label{eq:payoffmatrix}
\end{equation}
\end{linenomath}
This payoff matrix signifies that mutual cooperation generates reward $R$, whereas mutual defection generates punishment $P$ for both players. Additionally, if one player defects and other cooperates, the former receives temptation $T$ and the latter the sucker's payoff $S$. Payoff ordering determines the nature of the dilemma. For example, $T>R>P>S$ indicates the prisoner's dilemma, that is, the archetypal dilemma for studying the emergence of cooperation between selfish individuals~\cite{kuhn2019prisoner}. Two other common dilemmas are stag hunt and snowdrift (also known as hawk-dove or chicken), obtained by setting $R>T$ and $S>P$, respectively.

The above-mentioned social dilemmas were considered static prior to the work of John Maynard Smith, who introduced the notion of repetitions (i.e., iterations) and thus laid the foundations of evolutionary game theory~\cite{smith1982evolution, weibull1997evolutionary, tanimoto2015fundamentals}. Traditionally set in populations in which all players have equal probability to interact with one another (i.e., well-mixed populations), the frequency $x_i$ ($0\leq x_i\leq1$) of players resorting to strategy $i$ is traced with the differential equation~\cite{taylor1978evolutionary}
\begin{linenomath}
\begin{equation}
\dd{x_{i}}{t} =  x_{i} \left[\varphi_i(\mathbf{x}) - \varPhi(\mathbf{x})\right]
\label{eq:replicator}
\end{equation}
\end{linenomath}
known as the replicator equation, where $\mathbf{x}$ is the vector of frequencies for all strategies satisfying $\sum_i x_i=1$, $\varphi_i(\cdot)$ is the per-capita payoff attained by resorting to the $i$th strategy, and $\varPhi(\cdot)$ is the average per-capita payoff. Because dilemmas differ in payoff ordering, they also reach different stationary points under Eq.~(\ref{eq:replicator}). If a stationary point is stable, it is called an evolutionarily stable state or a Nash equilibrium. If a Nash equilibrium is monomorphic, that is, $x_i=1$ and $x_j=0$ for all $i\neq j$, then the $i$th strategy is called an evolutionarily stable strategy (ESS). In prisoner's dilemma, defection is ESS. In a snowdrift dilemma, however, both $C$ and $D$ strategies coexist in an evolutionarily stable state~\cite{doebeli2005models} because cooperating with a defector is still better than mutually defecting.

\begin{figure}[!t]
\centering\includegraphics[scale=1.0]{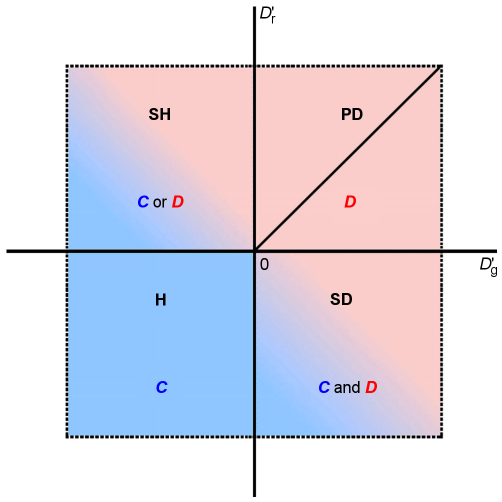}
\caption{Dilemma-strength parameters reflect the nature of social dilemmas that fit the format of dyadic games. Specifically, $D'_\mathrm{g}, D'_\mathrm{r}>0$ indicate the prisoner's dilemma; $D'_\mathrm{g}<0, D'_\mathrm{r}>0$ indicate the stag-hunt dilemma; $D'_\mathrm{g}>0, D'_\mathrm{r}<0$ indicate the snowdrift dilemma; and $D'_\mathrm{g}, D'_\mathrm{r}<0$ indicate no dilemma (in which case the dyadic game is called harmony).}
\label{fig:ds_coop}
\end{figure}

Social dilemmas that fit the format of dyadic games can be re-scaled in terms of the dilemma-strength parameters, one of which ($D'_\mathrm{g}$) measures how lucrative defection is in the presence of a cooperator, whereas the other ($D'_\mathrm{r}$) measures how hazardous cooperation is in the presence of a defector~\cite{wang2015universal, ito2018scaling, arefin2020social}. Precisely, the two parameters are defined by
\begin{linenomath}
\begin{subequations}
\begin{align}
D'_\mathrm{g}&=\frac{T-R}{R-P}, \label{eq:dsa}\\
D'_\mathrm{r}&=\frac{P-S}{R-P}. \label{eq:dsb}
\end{align}
\label{eq:ds}
\end{subequations}
\end{linenomath}
In Eq.~(\ref{eq:dsa}), the positive value of $D'_\mathrm{g}$ increases as $T$ increases relative to $R$, which facilitates defection by making the temptation payoff for defecting against a cooperator much larger than the reward payoff for mutual cooperation. In Eq.~(\ref{eq:dsb}), the positive value of $D'_\mathrm{r}$ increases as $S$ decreases relative to $P$, which again facilitates defection, but this time by making the sucker's payoff for cooperating with a defector much more negative than the punishment payoff for mutual defection. The normalisation factor in both equations, $R-P$, works in the opposite direction (i.e., facilitates cooperation) through more generous reward for cooperators and more stringent punishment for defectors. When $D'_\mathrm{g}, D'_\mathrm{r}>0$, then the payoff ordering of the prisoner's dilemma holds; if instead $D'_\mathrm{g}<0$ ($D'_\mathrm{r}<0$), then the payoff ordering of the stag-hunt (snowdrift) dilemma holds (Fig.~\ref{fig:ds_coop}).

The importance of the dilemma-strength parameters lies in the fact that they reduce the dimensionality of dyadic games from four (payoffs $R$, $S$, $T$, and $P$) to two. This is achieved by affine-transforming the payoff matrix in Eq.~(\ref{eq:payoffmatrix}) into
\begin{linenomath}
\begin{equation}
\bordermatrix{
\text{ } & C & D \cr
       C & 1 & -D'_\mathrm{r} \cr
       D & 1+D'_\mathrm{g} & 0
}.
\label{eq:payoffmatrixscaled}
\end{equation}
\end{linenomath}
Such a transform leaves the process of selection as specified by Eq.~(\ref{eq:replicator}) unchanged~\cite{weibull1997evolutionary}. Of note is that infinitely many four-payoff matrices can be mapped into a single two-parameter matrix. Consequently, all dyadic games that have the same dilemma strength, even if their payoff matrices wildly differ, are equivalent in terms of evolutionary outcomes.

The prisoner's dilemma is ubiquitous because by default it leads to defection. Cooperation can prevail only through various extensions of dyadic games based on this dilemma. Such extensions are then said to be cooperation-promoting mechanisms. Kin and group selection, as well as direct, indirect, and network reciprocity are seen as general mechanisms that act as the promoters of cooperation~\cite{nowak2006five}. Interestingly, numerical results show that even when dyadic games incorporate these cooperation-promoting mechanisms, evolutionary outcomes are predetermined by the dilemma-strength parameters~\cite{wang2015universal}. Why would that be? One way to understand why the quantities $D'_\mathrm{g}$ and $D'_\mathrm{r}$ work for extended dyadic games is to recognise that such games can all be transformed and reinterpreted as standard (i.e., non-extended) dyadic games but with suitably adjusted payoff matrices~\cite{taylor2007transforming}. In the case of direct reciprocity, for example, the payoff matrix in Eq.~(\ref{eq:payoffmatrix}) is transformed into
\begin{linenomath}
\begin{equation}
\bordermatrix{
\text{ } & C & D \cr
       C & \frac{1}{q}R & S+\frac{1-q}{q}P \cr
       D & T+\frac{1-q}{q}P & \frac{1}{q}P
},
\label{eq:payoffmatrixdirect}
\end{equation}
\end{linenomath}
where $q$ is the probability of terminating play with a given individual (and $1-q$ is the probability of continuing play with this individual). Using the same re-scaling as in Eq.~(\ref{eq:payoffmatrixscaled}), we get
\begin{linenomath}
\begin{equation}
\bordermatrix{
\text{ } & C & D \cr
       C & \frac{1}{q} & -D'_\mathrm{r} \cr
       D & 1+D'_\mathrm{g} & 0
},
\label{eq:payoffmatrixdirectscaled}
\end{equation}
\end{linenomath}
which shows that for a fixed $q$, the evolutionary outcome is determined by the dilemma-strength parameters. If the probability $q$ is small enough that \smash{$D'_\mathrm{g}<\frac{1-q}{q}$}, then the original prisoner's dilemma turns due to direct reciprocity to a stag-hunt dilemma, and cooperation is an ESS.

A more parsimonious explanation for the success of the dilemma-strength parameters is to recognise that all five cooperation-promoting mechanisms (i.e., kin selection, group selection, direct reciprocity, indirect reciprocity, and network reciprocity) have one crucial feature in common; one way or another, these mechanisms enable positive assortment by which cooperative acts occur more often between cooperators than expected based on population averages~\cite{eshel1982assortment, newton2018evolutionary, kay2020evolution}. Once the role of positive assortment is recognised, the affine transformation of the payoff matrix as specified by Eq.~(\ref{eq:payoffmatrixscaled}) must not interfere with such assortment, that is, the two must be compatible. This indeed is the case~\cite{tudge2015tale}. In other words, all five cooperation-promoting mechanisms are manifestations of positive assortment which itself is preserved by the affine transformation that parametrises dyadic games in terms of the dilemma-strength parameters.

A limitation of the dilemma-strength parameters is that they can be defined solely for social dilemmas that fit the format of dyadic games. However, a generalisation to more complex social dilemmas is possible in terms of an efficiency deficit defined as the fitness difference between the socially optimal steady state that maximises individual fitness and the current evolutionarily stable state~\cite{arefin2020social}. This definition implies that society evolving to a suboptimal equilibrium incurs an opportunity cost. If the opportunity cost is small (large), it can (cannot) be tolerated, and a societal change for the better is more difficult (easier) to accomplish.

\subsection{Cooperation in networks with pairwise interactions}

A population of players can be structured using graphs or networks such that vertices (i.e., nodes) represent players and edges (i.e., links) indicate pairwise interactions. In this picture, the usual well-mixed populations of evolutionary game theory are represented by complete networks in which all nodes are linked to one another. Structured populations, however, are specified to spatially constrain interactions, that is, prescribe who interacts with whom. In a square lattice, for example, players interact with their four or eight nearest physical neighbours (often called von Neumann and Moore neighbourhoods, respectively). Payoffs accumulated from such interactions are then used to update the lattice through either reproduction or imitation and learning, depending on whether biological or social evolution is of interest.

When the focus is on biological evolution, `death-birth' updating is commonly applied, meaning that at each time step a random player is chosen to die, followed by the offspring of neighbours competing for the empty site in proportion to their fitness~\cite{ohtsuki2006simple, ohtsuki2006evolutionary, ohtsuki2006replicator}. An alternative is `birth-death' updating by which a player is selected for reproduction proportional to fitness, followed by offspring replacing a randomly chosen neighbour. For social-evolution scenarios, `imitation' updating is used, meaning that at each time step a random player is chosen to decide whether to keep their current strategy or imitate one of the neighbours depending on the difference in fitness between the neighbour and the player.

Structured populations shot to fame through the work of Nowak and May~\cite{nowak1992evolutionary} who observed that repeated games in a square lattice generate spatial chaos. An even more influential finding was that cooperators could expand by forming clusters~\cite{nowak1993spatial} that enable reaping the benefits of cooperation despite exploitation by defectors at cluster boundaries. This finding spurred further studies on whether cooperators survive, or even thrive, in different types of network structures. Beside lattices, random and scale-free networks were featured prominently~\cite{gomez2007dynamical, poncela2007robustness, gomez2008natural, devlin2009evolution, poncela2009cooperative}. The cited studies, inspired in large part by the result in Ref.~\cite{santos2005scale}, established that more heterogeneous networks provide the best conditions for the evolution of cooperation by securing that large cooperative clusters remain little exposed to defection at cluster boundaries. The promise of scale-free networks strongly boosting cooperation succeeded in attracting much attention in the field, but has subsequently been proven to lack robustness to theoretical model assumptions~\cite{szolnoki2008towards, yamauchi2010controls}.

The proliferation of research on evolutionary games in structured populations begat a search for general rules that explain the evolution of cooperation in various networks. Ref.~\cite{ohtsuki2006simple} describes one such rule for weak selection. The term `weak selection' refers to the idea that many different factors affect a player's overall fitness, with the game under consideration being just another factor among the many. For this reason, a player is characterised by a baseline fitness that is large relative to payoffs earned throughout the game. Let the game be a variant of the prisoner's dilemma in which $R=b-c$, $S=-c$, $T=b$, and $P=0$ (with $b>c>0$). It turns out that, under weak selection, cooperation is favoured in pairwise networks if the benefit of altruistic acts, $b$, divided by the cost, $c$, exceeds the average number of neighbours, $k$, or $b/c>k$. This simple rule closely resembles Hamilton's rule, according to which kin selection favours cooperation if $b/c>1/r$, where $r$ represents the coefficient of genetic relatedness between individuals. The similarity of the two rules can be intuitively understood by considering that the average node degree is an inverse measure of social relatedness between players. A player's fate is loosely bound to that of the neighbours if there are many neighbours, whereas the opposite is true if there are a few neighbours. Similarity notwithstanding, Ref.~\cite{nowak2010evolutionary} argues that network reciprocity under the condition $b/c>k$ is fundamentally different from kin selection under the condition $b/c>1/r$. Importantly, the network-reciprocity rule is rather robust to theoretical model assumptions~\cite{yamauchi2010controls}.

Ref.~\cite{allen2017evolutionary} expanded the aforementioned research to obtain a general condition for the evolution of cooperation in any network under weak selection. Writing the condition using our notation, we have $\sigma R + S > T + \sigma P$, where \smash{$\sigma=\frac{-t_1+t_2+t_3}{t_1+t_2-t_3}$} is a structural coefficient that quantifies a network's propensity to support cooperation. The quantities $t_1$, $t_2$, and $t_3$ respectively denote the expected times at which the first, second, and third neighbours of an initial cooperator become cooperators. To revert back to the simple rule $b/c>k$, in addition to setting the payoffs to $R=b-c$, $S=-c$, $T=b$, and $P=0$, it is necessary that the network is large enough in order for all nodes (whose average degree is $k$) to be sparsely connected.

The research on the evolution of cooperation in structured populations described so far has maintained a rigid assumption of node-based selection. This means that any given node is either a cooperator or a defector and acts as such towards its whole neighbourhood. In reality, it is crucial for many simple and complex organisms alike to differentiate between cooperative and defecting neighbours. By refocusing selection on links instead of nodes, a series of recent works has enabled examining situations in which the same node can cooperate with some neighbours and defect against others. The results show that with link-based selection, the frequency of cooperation can be high for a wide range of game setups~\cite{su2016interactive}. A novel dynamic state has been observed between $b/c\approx\lra{k}$ and $b/c\gg\lra{k}$ in which cooperation and defection dynamically interchange with one another as a dominant strategy~\cite{sendina2020diverse}. In mixed populations with both node- and link-based selection, cooperation either increases monotonically as the link-based selection becomes more prevalent~\cite{su2018evolution}, or there is a clear separation of roles by which node-based selection spawns cooperative clusters, while link-based selection protects these cluster from defectors~\cite{jia2020evolutionary}.

\begin{figure}[!t]
\centering
\includegraphics[scale=1.0]{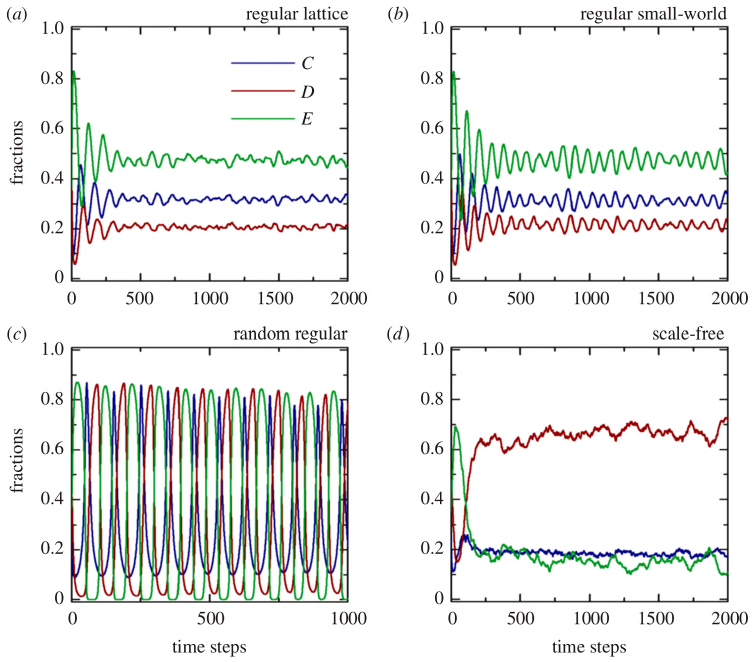}
\caption{Evolutionary dynamics of cooperators, defectors, and exiters depending on network topology. Panel (a) shows that abundances of cooperators, defectors, and exiters oscillate locally in regular lattices. Panel (b) further shows that local oscillations are somewhat amplified by regular small-world networks. Panel (c) reveals that global oscillations occur in random regular networks. In scale-free networks, shown in panel (d), the presence of hub nodes turns oscillations into random fluctuations.\newline
Source: Reprinted figure from Ref.~\cite{shen2021exit}.}
\label{fig:cyclicdynamics}
\end{figure}

When evolutionary games in structured populations are enriched with a third strategy on top of usual cooperation and defection, a commonly observed phenomenon is that of cyclic dominance~\cite{szolnoki2014cyclic}. The term `cyclic dominance' refers to an intransitive relationship between objects A, B, and C by which A in some aspect dominates B, B dominates C, but C dominates A. The third strategy leading to cyclic dominance can be as simple as that of loners~\cite{szabo2002evolutionary}, who always stay out of the game, thus settling for a small payoff no matter whom they were supposed to interact with. Similar results are seen with exiters~\cite{shen2021exit}, who pay a small cost to find out if they are supposed to interact with cooperators or defectors. In the former case, exiters stay in the game and cooperate. In the latter case, they exit the game to receive a small payoff before getting exploited through defection. Yet another example of a strategy leading to cyclic dominance is that of hedgers~\cite{guo2020novel}, who also pay a small cost to find out if they are supposed to interact with cooperators or defectors, but then cooperate in the former case and defect in the latter. Cyclic dominance is of interest from a dynamical point of view because evolutionary dynamics can greatly differ depending on the topology of the underlying network (Fig.~\ref{fig:cyclicdynamics}).

To conclude, network structure may be a powerful cooperation-promoting mechanism in the prisoner's dilemma, but this is not the case in the snowdrift dilemma. Structure, surprisingly, decreases the frequency of cooperators relative to well-mixed populations~\cite{hauert2004spatial, doebeli2005models}. Instead of large, compact clusters common in spatial prisoner's dilemma games, clusters in spatial snowdrift dilemma games are small and filament-like~\cite{doebeli2005models} due to the fact that two interacting snowdrift players should adopt the opposite strategies to one another.

\subsection{Cooperation in multilayer networks}

Many complex systems can be seen as a network of networks. Organisms, for example, comprise gene regulatory networks on the sub-cellular scale, neuronal networks on the cellular scale, and vascular networks on the scale of cellular collectives (i.e., organs)~\cite{gosak2018network}. Ecosystems comprise trophic networks and host-parasite networks in habitat patches accessible to individuals~\cite{pilosof2017multilayer}. Infrastructure comprises many interdependent networks such as power, communications, transportation, etc. Interdependence in particular implies that processes occurring in one network may affect what happens in other networks to the point that small and seemingly irrelevant changes have unexpected and catastrophic consequences~\cite{buldyrev2010catastrophic}. This possibility has sparked substantial interest in the robustness of networks of networks in general, as well as in many specific contexts~\cite{gao2011robustness, pocock2012robustness, dong2013robustness, evans2013robustness}. Interestingly, a wide variety social interactions fit a network of networks representation. For example, people mutually interact and transfer information both within and between online social networks. It is therefore natural to study the evolution of cooperation in interdependent networks~\cite{wang2015evolutionary}.

A rigorous way of representing networks of networks is via the multilayer-network formalism~\cite{dedomenico2013mathematical, kivela2014multilayer}. Keeping the discussion semi-formal, we can define a multilayer network as a quadruplet $M=(V_M, E_M, V, L)$, where $V$ is a set of physical nodes, $L=L_1\times\ldots\times L_d$ is a set of layers comprising $d$ elementary-layer sets $L_1$ to $L_d$, $V_M \subseteq V\times L$ is a set of state nodes encoding whether a physical node $v\in V$ is found in layer $l\in L$, and $E_M\subseteq V_M\times V_M$ is a set of intralayer and interlayer links. To exemplify, an elementary set could be an online social network (e.g., $L_1=\{\mathrm{Facebook}, \mathrm{Twitter}\}$), while the other elementary set could be a region (e.g., $L_2=\{\mathrm{US}, \mathrm{EU}\}$). Then the set of layers is $L=L_1\times L_2=\{(\mathrm{FB}, \mathrm{US}), (\mathrm{FB}, \mathrm{EU}), (\mathrm{TWTR}, \mathrm{US}), (\mathrm{TWTR}, \mathrm{EU})\}$. To indicate that the person $v_1$ accesses Facebook and Twitter from the US, we would write $(v_1,(\mathrm{FB},\mathrm{US}))\in V_M$ and similarly $(v_1, (\mathrm{TWTR},\mathrm{US}))$, which could be shortened to $(v_1,(1,1))$ and $(v_1,(2,1))$. The person $v_1$ is also an interlink between the two social networks $\{(v_1,(1,1)), (v_1,(2,1))\}\in E_M$. To indicate that the person $v_2$ is a Twitter user who travels between the US and the EU, we would write $(v_2,(2,1))$ and $(v_2,(2,2))$, automatically forming an interlink $\{(v_2,(2,1)), (v_2,(2,2))\}$. Finally, if the person $v_1$ follows the person $v_2$ on Twitter, this can be represented as the intralink $\{(v_1,(2,1)), (v_2,(2,1))\}$ and the interlink $\{(v_1,(2,1)), (v_2,(2,2))\}$. Another intuitive example of a multilayer network could have a collection of cities as physical nodes; the first elementary layer could be transport-mode availability (e.g., high-speed railway station or airport), the second could be city size (e.g., less than 1 million inhabitants, between 1 and 5 million inhabitants, or more than 5 million inhabitants), and the third elementary layer could be a country. Instances of intralayer links would then be domestic railways or flights connecting the cities of the same size. Instances of interlayer links would be any international railways or flights, domestic railways or flights between the cities of different size, but also cities that posses both a high-speed railway station and an airport where transferring between the two transport modes is possible.

From the perspective of evolutionary game theory, the focus is on interdependence through the coupled player utilities or the flow of information between players, although other proposals have been made too~\cite{wang2015evolutionary}. Typically, a single player occupies only one of the available layers (but see Ref.~\cite{gomez2012evolution}), while gameplay and strategy transfers take place only between players residing in the same layer. If the latter were not the case, and we dealt with truly interconnected networks (as opposed to just interdependent ones), then the social-dilemma game would essentially unravel in a single-layer network (albeit with two communities), meaning that the same cooperation-promoting mechanism would operate everywhere. Ref.~\cite{jiang2013spreading} presents a game setup designed along these lines, that is, the prisoner's dilemma is played within and between two communities. In Ref.~\cite{gomez2012evolutionary}, the prisoner's dilemma is played within communities and the snowdrift dilemma between them.

\begin{figure}[!t]
\centering
\includegraphics[scale=1.0]{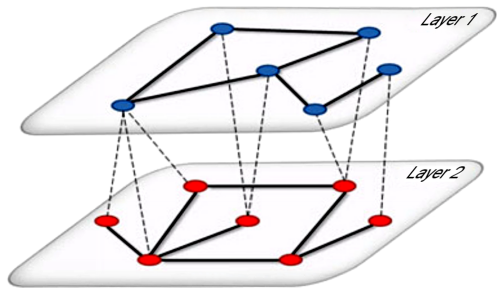}
\caption{Two-layer network for a social-dilemma game. Nodes in different layers represent different players, which is denoted by blue and red colours. Although players reside in different layers, there are interdependencies between them, which is denoted by the dashed interlinks. The topology of one layer, as indicated by solid intralinks, may differ from that of another layer, meaning that social connectivity is layer-specific.\newline
Source: Reprinted figure from Ref.~\cite{wang2015evolutionary}.}
\label{fig:twolayer}
\end{figure}

The simplest and most common social-dilemma games in multilayer networks are those unfolding in two-layer networks (Fig.~\ref{fig:twolayer}). Ref.~\cite{wang2012evolution} exemplifies such a game with coupled player utilities. Specifically, let $x$ denote a player from layer 1 whose payoff is $P_x$, and similarly, let $x'$ denote an interdependent player from layer 2 whose payoff is $P'_x$. Then the utility determining the course of evolutionary dynamics for both players is given by
\begin{linenomath}
\begin{equation}
U_x = U'_x = \alpha P_x + (1-\alpha) P'_x,
\end{equation}
\end{linenomath}
where $0\leq\alpha<\frac{1}{2}$. Furthermore, instead of playing the prisoner's dilemma with their first neighbours individually, the players $x$ and $x'$ participate in public goods games with all their first neighbours simultaneously. In a lattice, the public goods game can be centred around the player $x$ in layer 1 ($x'$ in layer 2), but also around each of the first neighbours, meaning that the players $x$ and $x'$ participate in five public good games to collect their payoffs, $P_x$ and $P'_x$, respectively. The gameplay rules are such that a cooperator contributes 1 unit to a pool, while a defector contributes nothing. The total contribution to the pool is multiplied by a return factor $r>1$, and divided equally between players participating in the same public goods game. The results show that cooperation is strongly promoted in layer 1, but not in layer 2, due to a dampening effect that coupled utilities have on the exploitation of cooperators by defectors in the former layer.

Aiming to resolve the social dilemma in both layers, Ref.~\cite{shen2018coevolutionary} redefined the coupled utilities
\begin{linenomath}
\begin{align}
U_x &= P_x + \alpha_x P'_x,\\
U'_x &= P'_x + \alpha'_x P_x,
\end{align}
\end{linenomath}
where $0\leq\alpha_x, \alpha'_x\leq1$ are directed interdependencies between the layers. These interdependencies are adaptive such that if the player $x$ ($x'$) earns a payoff $P_x$ ($P'_x$) greater than some threshold $E$, that is, $P_x\geq E$ ($P'_x\geq E$), then the interdependency $\alpha_x$ ($\alpha'_x$) is strengthened by an amount $\delta>0$; if conversely the payoff falls short of the threshold, the interdependency is reduced by the same amount.

\begin{figure}[!t]
\centering\includegraphics[scale=1.0]{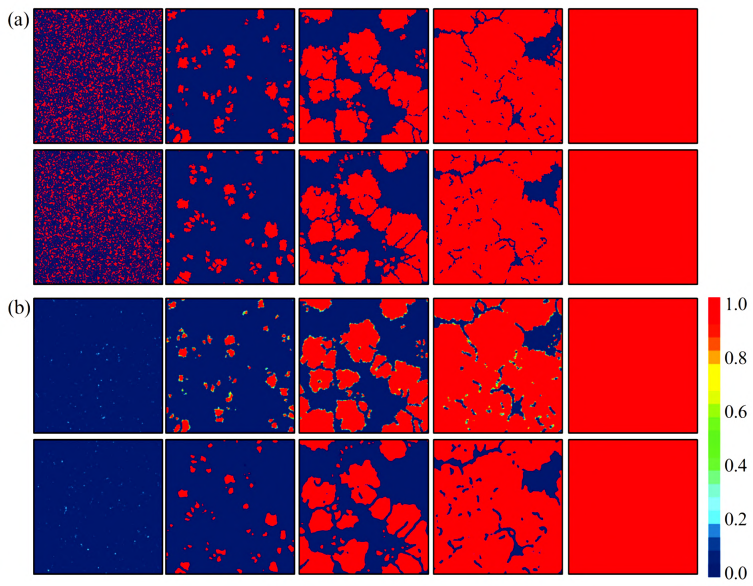}
\caption{Snapshots of co-evolutionary dynamics in a two-layer lattice. Interlayer links occupied by cooperator-cooperator pairs acts as seeds for the growth of cooperative clusters. Panels (a) show cooperation frequency in the two lattice layers, whereas panels (b) show directional interdependency strength between the two lattice layers. From left to right, the panels display Monte Carlo Steps 0, 100, 300, 500, and 9999, respectively.\newline
Source: Reprinted figure from Ref.~\cite{shen2018coevolutionary}.}
\label{fig:snapshots}
\end{figure}

It turns out that under the described setup, a large threshold value $E$ effectively keeps the two layers disconnected, meaning that there cannot be any synergy between them. Somewhat surprisingly, if the threshold value $E$ is too small, suboptimal synergy is achieved. This is because small $E$ allows defectors (alongside cooperators) to develop strong interdependencies, and then, if a defector in one layer gets to exploit some cooperators, then this defector can sustain their counterpart defector in the other layer. Only for the intermediate values of the threshold $E$ is full synergy between the layers achieved (Fig.~\ref{fig:snapshots}). In this case, predominantly cooperators create strong interdependencies which allows two interdependent cooperators to support one another. Such cooperators seed clusters of cooperation that later on take over the whole population in both layers (Fig.~\ref{fig:snapshots}), thus successfully resolving the social dilemma at hand.

\subsection{Cooperation in temporal networks}

Complex networks have heretofore been used as if social interactions were static in time. Although this is a reasonable approximation in many circumstances, the ephemeral nature of human contacts eventually needs to be accounted for. After all, when two persons engage in an activity, this often happens in short bursts followed by periods of relative lull~\cite{barabasi2005origin, vazquez2005exact, starnini2013modeling}.

Temporal networks have emerged as a convenient tool for representing time-varying social interactions~\cite{holme2012temporal, holme2015modern, masuda2016guide}. In such networks, any pair of momentarily disconnected nodes may get connected by the next time instant and vice versa. Interest in temporal networks stems from their ability to affect network-science fundamentals, among others, general dynamical processes~\cite{masuda2013temporal, scholtes2014causality, masuda2017random}, epidemiological dynamics~\cite{valdano2015analytical, rocha2011simulated}, and network controllability~\cite{li2017fundamental}.

In the study of the evolution of cooperation in networked populations, network temporality has found a natural place in co-evolutionary models~\cite{perc2010coevolutionary}. The term `co-evolution' implies that beside the usual cooperative trait, one or more other traits evolve in parallel. This could be, for instance, a homophilic trait such that cooperative individuals tend to connect with other cooperative individuals or, at least, shun connections with defecting individuals. Be it homophily or some other psycho-social mechanism, a consequence is that network topology changes over time~\cite{zimmermann2004coevolution, eguiluz2005cooperation, fu2009partner, du2011partner, melamed2020homophily}. Psycho-social mechanisms, however, imply an active screening for suitable contacts, whereas network temporality in the real world is oftentimes more serendipitous. This begs the question of how temporal contact networks, exogenous of psycho-social mechanisms, affect the evolution of cooperation.

\begin{figure}[!t]
\centering
\includegraphics[scale=1.0]{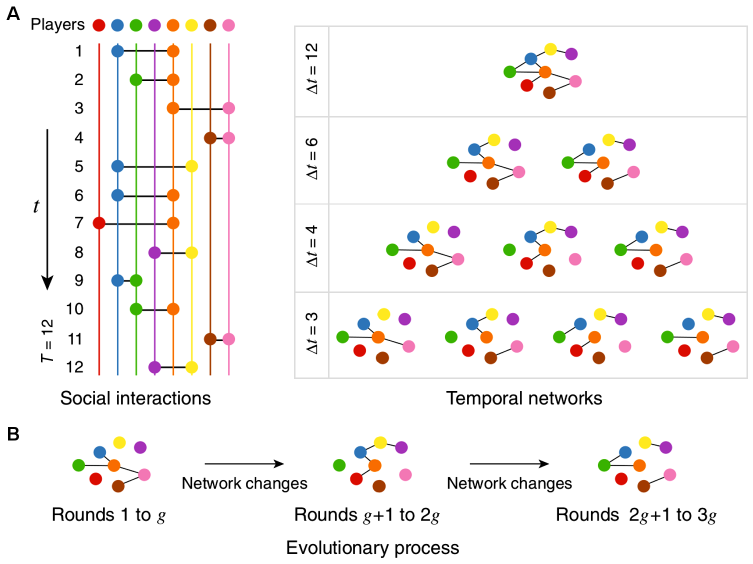}
\caption{Evolutionary games in temporal networks. \textbf{A,} Social interactions change from one time instant to another (left). This is represented using temporal networks after aggregating all social interactions over the time period of length $\Delta t$ (right). The longer the $\Delta t$ is, the coarser the picture of social interactions. When $\Delta t$ is large, we get only the fully aggregated (i.e., static) interaction network. \textbf{B,} The aggregation-length parameter $\Delta t$ captures one facet of network temporality. Another facet is captured by a parameter $g$, quantifying how fast evolutionary dynamics compared to network-structural dynamics. The parameter $g$ is defined as the number of evolutionary-game rounds between any two consecutive changes in the network topology.\newline
Source: Reprinted figure from Ref.~\cite{li2020evolution} under the Creative Commons Attribution 4.0 International (CC BY 4.0).}
\label{fig:temporal}
\end{figure}

Early work in the described context has indicated that temporal networks may be favouring selfish behaviour~\cite{cardillo2014evolutionary}. Using a temporal network of $N=100$ contacts recorded every five minutes over a period of six months (resulting in 41,291 snapshot graphs), it was shown that when snapshots are aggregated over a short time period of $\Delta t=1$\,h, then much lower cooperation frequencies ensue compared to longer aggregation periods of $\Delta t>1$\,wk. Here, aggregation means taking all snapshot graphs over the period $\Delta t$, and if two nodes $i$ and $j$ interacted in any of the snapshots, then these nodes are assumed to be momentarily connected; otherwise, the nodes are disconnected (Fig.~\ref{fig:temporal}A). The time period $\Delta t$ is a facet of network temporality, where small (large) $\Delta t$ values mark frequent (infrequent) changes in topology. As $\Delta t\rightarrow\infty$, the network becomes fully aggregated (i.e., static). Interestingly, randomising the time ordering of snapshots improves cooperativeness. Because such randomisation effectively removes the aforementioned burstiness of human interactions, it would seem that the bursts of activity, in particular, disfavour the evolution of cooperation.

\begin{figure}[!t]
\centering
\includegraphics[scale=1.0]{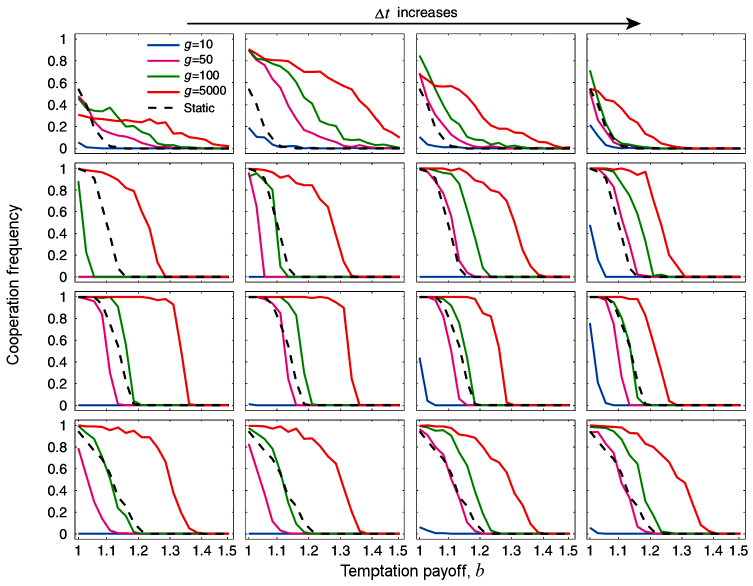}
\caption{Temporal networks can promote the evolution of cooperation beyond their static counterparts. Panels show the cooperation frequency as a function of the temptation payoff for several values of the parameter $g$. Each row of panels corresponds to one real-world temporal network: 1$^\mathrm{st}$ row is a contact network of attendees at a scientific conference, 2$^\mathrm{nd}$ and 3$^\mathrm{rd}$ rows are contact networks of students at a high school in Marseilles, France, in 2012 and 2013, and 4$^\mathrm{th}$ row is a contact network of workers in an office building in France. Irrespective of the network or the value of the aggregation parameter $\Delta t$, there is always some $g$ value for which the cooperation frequency in the temporal network is larger than in the corresponding fully-aggregated (i.e., static) network.\newline
Source: Reprinted figure from Ref.~\cite{li2020evolution} under the Creative Commons Attribution 4.0 International (CC BY 4.0).}
\label{fig:temporal_coop}
\end{figure}

In contrast to the outlined early work, the current state of the art~\cite{li2020evolution} paints a more nuanced picture of cooperativeness in temporal networks. This is achieved by considering an additional facet of network temporality; how fast evolutionary dynamics is relative to network-structural dynamics. A parameter $g$ quantifying the second facet of temporality is defined as the number of evolutionary-game rounds that take place during the time period $\Delta t$ (Fig.~\ref{fig:temporal}B). The increasing values of the parameter $g$ improve cooperativeness in temporal networks even beyond what is possible in static networks (Fig.~\ref{fig:temporal_coop}). The improvement occurs despite the unfavourable effects of burstiness.

The numerical results in Ref.~\cite{li2020evolution} point to a threshold for the outbreak of defection that reaches a maximum for intermediate values of the aggregation parameter $\Delta t$. This result can be understood via activity-driven modelling~\cite{perra2012activity}, which shows that defectors successfully spread if the following condition is satisfied
\begin{linenomath}
\begin{equation}
\frac{\lambda k}{\mu}\geq D^*,
\label{eq:definvasioncond}
\end{equation}
\end{linenomath}
where $\lambda$ ($\mu$) is the average probability of a cooperator (defector) turning into a defector (cooperator) in the next round. The quantity $k=2l\lra{a}$ is the average degree given in terms of the average number of links $l$ that an active temporal-network node randomly creates in the current time step, as well as in terms of the average activity $\lra{a}$. To clarify, in activity-driven models, each node is assigned a probability $a_i$ of being active in a particular time step; active nodes create $l$ random links to other (active or passive) nodes, while also being able to receive additional links from other active nodes. The quantity on the right-hand side of the invasibility condition in Eq.~(\ref{eq:definvasioncond}) is
\begin{linenomath}
\begin{equation}
D^*=\frac{1}{1+\sqrt{1+\frac{\mathrm{Var}(a)}{\lra{a}^2}}}.
\end{equation}
\end{linenomath}
This relationship shows that the threshold $D^*$ should decrease as the aggregation period $\Delta t$ gets very short because then temporal contact networks are sparse (see Fig.~\ref{fig:temporal}A for $\Delta t=3$), which makes the average activity $\lra{a}$ also small. A sparse contact network removes the benefits of network reciprocity. Nodes end up playing pairwise repeated prisoner's dilemma games in which a likely Nash equilibrium is defection (depending on game payoffs and repetitions). Furthermore, the threshold $D^*$ should decrease as the aggregation period $\Delta t$ gets very long because then temporal contact networks are highly heterogeneous (Fig.~\ref{fig:temporal}A for $\Delta t=12$), which makes the variance $\mathrm{Var}(a)$ large. A heterogeneous contact network undergoes big topological transformations from one time step to another. Any such transformation can destabilise cooperative clusters, thus making cooperators more vulnerable to defectors.

In summary, temporal networks may promote cooperation beyond what is possible in the corresponding fully aggregated (i.e., static) networks, but only if the evolutionary-game dynamics is fast relative to the network-structural dynamics, that is, the parameter $g$ is large. An additional important result is that temporal networks are most resistant to defection for the intermediate values of the aggregation period $\Delta t$. These positive effects may, however, be nullified by the burstiness of human interactions in instances when the two facets of temporality, $g$ and $\Delta t$, are unfavourable.

\subsection{Cooperation in networks with higher-order interactions}

As illustrated thus far, studying the evolution of cooperation in graphs or networks has a long tradition. Intriguingly though, cooperation in groups that are themselves embedded in a higher-order network structure has remained an open question until recently~\cite{alvarez2021evolutionary}. This is not to say that the question has been entirely ignored; special cases have been examined and have, in fact, left their mark on the field~\cite{santos2008social, szolnoki2009topology}. These early works would select a focal node and follow this node's pairwise links to determine which other nodes form a group. Thus determined group members would then participate in the same public goods game. A problem here is that determining group members based on pairwise links is rather unsatisfactory. Pairwise links are, after all, supposed to denote interactions between node pairs, not node groups, leading to many natural questions. Can every node be focal or, if not, how do we select focal nodes? Should a group extend only to the focal node's first neighbours or should it include second, third, or even more distant neighbours? Does the higher-order network of groups that emerges by clumping together focal nodes and their neighbours have the most general structure possible or are there substantial limitations to what is achievable?

Among the most straightforward generalisations of `classical' networks to higher-order ones is to allow more than two nodes to be connected via the same link, in which case the term hyperlink becomes customary~\cite{battiston2020networks}. The resulting higher-order network is often referred to as a hypergraph. Formally, a hypergraph $H$ is a pair of sets $H=(N,L)$, where $N=\{n_1,\ldots,n_n\}$ is a set of nodes and $L=\{l_1,\ldots,l_l | l_j\subseteq N\}$ is a set of hyperlinks. Because hyperlinks themselves are subsets of the set $N$, a hyperlink's number of elements (i.e., its cardinality) is a well-defined concept. The cardinality, $|l_j|=g$, is usually called the order of the hyperlink and is used in generalising the idea of the node degree. Specifically, if $k_i^g$ denotes the number of order-$g$ hyperlinks containing the $i$th node, then this node's hyperdegree is \smash{$k_i=\sum_{g=g_\mathrm{min}}^{g_\mathrm{max}} k_i^g$}, whereas the average hyperdegree is \smash{$\lra{k}=\frac{1}{n}\sum_{n_i \in N} k_i$}. These definitions permit introducing two types of heterogeneity into hypergraphs. First, hyperdegrees can differ from one node to another, while all links have the same order $g$. Second, and more generally, links of multiple orders can intermix in such a way that $k^g / k$ is the probability that a randomly chosen hyperlink is of the order $g$, where $g_\mathrm{min}\leq g\leq g_\mathrm{max}$. This means that a node of hyperdegree $k$ on average belongs to $k^g$ order-$g$ hyperlinks. For a hypergraph to be uniform, the node hyperdegree must be the same across all nodes and the hyperlink order must be the same across all hyperlinks.

Hypergraphs are an ideal setting to study cooperativeness in groups of individuals in a general situation when each individual potentially belongs to more than one group. An individual is represented by a hypergraph node, whereas a group comprises all nodes connected via a single hyperlink. Because every group has two or more individuals, it is natural to consider cooperation in the public goods game, that is, a social-dilemma game that is the multiplayer generalisation of the prisoner's dilemma~\cite{hauert2003prisoner, arefin2020social}. The usual game-rules apply. Cooperators pay the cost $c$ that is pooled, multiplied by a return factor, and then split equally among all game participants, even if they are defectors and refuse to contribute to the pool. If we set $c=1$ and denote with $r$ the return factor divided by the number of game participants, then the per-capita payoff of a cooperator is $\pi_\mathrm{C}=\nu_\mathrm{C}r-1$, whereas that of a defector is $\pi_\mathrm{D}=\nu_\mathrm{C}r$, where $\nu_\mathrm{C}$ denotes the number of cooperators. It is clear that $\pi_\mathrm{C}<\pi_\mathrm{D}$ for $\nu_\mathrm{C}>0$, but if nobody cooperates, nobody gets any return either. Therein lies the dilemma. Strategy selection proceeds such that a focal node $n_i$ is chosen randomly, as is a hyperlink $l_j$ to which this node belongs. Then all members of the hyperlink $l_j$ (i.e., $n_{i'}\in l_j$) play one public goods game in each of the hyperlinks they belong to. This provides the average payoff that nodes $n_{i'}$ earn per game played. The focal node finally adopts the strategy of the best performing neighbour with the probability \smash{$\frac{1}{\Delta}\left(\max\pi_{i'}-\pi_i\right)$}, where $\Delta$ is a normalising quantity equal to the absolute maximal payoff difference over all the possible strategies.

\begin{figure}[!t]
\centering\includegraphics[scale=1.0]{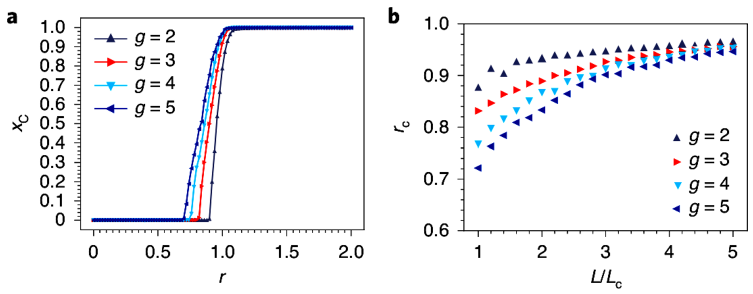}
\caption{Cooperation in uniform hypergraphs. Panel \textbf{a} shows that unlike in well-mixed populations, non-zero cooperation is possible for return-factor values $r_\mathrm{c}\leq r<1$. However, full cooperation evolves only for $r>1$. In these simulations, the number of hyperlinks equals the critical number that guarantees hypergraph connectedness, $l=l_\mathrm{c}$. Panel \textbf{b} reveals that as the number of hyperlinks $l$ increases above the critical value $l_\mathrm{c}$, the critical return factor $r_\mathrm{c}$ necessary for non-zero cooperativeness tends to unity. A quick rate of convergence indicates that even relatively sparse hypergraphs mimic well-mixed populations.\newline
Source: Reprinted figure from Ref.~\cite{alvarez2021evolutionary}.}
\label{fig:unihypercoop}
\end{figure}

It is a widely known result that in a well-mixed population, for cooperation to evolve in a public goods game, the condition $r>1$ must be satisfied (e.g., see Refs.~\cite{arefin2020social, alvarez2021evolutionary}). For $r<1$, defection prevails. Do hypergraphs help to promote cooperation in the sense of relaxing the condition $r>1$? Although the answer to this question is technically positive, the cooperation-promoting effect of hypergraphs is limited. Starting with uniform hypergraphs, for a given order $g$, there is a critical number of hyperlinks $l_\mathrm{c}$ that is needed to guarantee the existence of a giant connected component. When the actual number of hyperlinks $l$ is of the order of $l_\mathrm{c}$, non-zero cooperation does appear for the critical return factor $r_\mathrm{c}<1$, but full cooperation is possible only for $r>1$ (Fig.~\ref{fig:unihypercoop}A). In fact, the critical return factor tends rather quickly to unity, $r_\mathrm{c}\rightarrow 1$, as the actual-to-critical hyperlink ratio, $l/l_\mathrm{c}$, increases (Fig.~\ref{fig:unihypercoop}B), indicating that even relatively sparse hypergraphs mimic the well-mixed population when it comes to the evolutionary dynamics of cooperation. It is important to note that, for a constant $l/l_\mathrm{c}$, the increasing value of $g$ makes hypergraphs sparser, which fully explains the dependence of the results in Fig.~\ref{fig:unihypercoop} on the hyperlink order. It is furthermore of interest that introducing moderate hyperdegree heterogeneity has no impact on the results~\cite{alvarez2021evolutionary}. Only in scale-free hypergraphs new patterns emerge, but unlike in classical networks, establishing cooperation becomes more difficult because large-hyperdegree nodes connect hypergraph parts that would otherwise be disconnected, thus effectively counteracting sparsity~\cite{alvarez2021evolutionary}.

The other type of hypergraph heterogeneity, that is, order heterogeneity, is not so much of interest in the context of promoting cooperation as much as there are insights to be gained about the performance of collaborative groups~\cite{wu2019large}. Ref.~\cite{alvarez2021evolutionary} assumes that the return factor depends on hyperlink order, $r=r(g)$. If this is the case, the usual difference between the expected per-capita payoff of cooperators and defectors, $\overline{\pi}_\mathrm{C}-\overline{\pi}_\mathrm{D}=r-1$, changes to \begin{linenomath}
\begin{equation}
\overline{\pi}_\mathrm{C}-\overline{\pi}_\mathrm{D}=\sum\limits_{g=g_\mathrm{min}}^{g_\mathrm{max}}\frac{k^g}{k}\left[r(g)-1\right],
\end{equation}
\end{linenomath}
where for each hyperlink order $g$, we take into account the corresponding probability (given by the fractional multiplier) and the corresponding outcome (given by the square-bracket multiplicand). It is natural to assume that the return factor consists of two parts, a benefit part $\alpha g^{(\beta-1)}$ that increases with the hyperlink order due to synergies of working with collaborators, and a cost part $\exp\left[-\gamma(g-1)\right]$ that decreases with the hyperlink order due to difficulties of coordinating large groups. We thus have
\begin{linenomath}
\begin{equation}
r(g)=\alpha g^{(\beta-1)} e^{-\gamma(g-1)},
\label{eq:returnbenefitcost}
\end{equation}
\end{linenomath}
where the parameters $\alpha$, $\beta$, and $\gamma$ need to be estimated. Such estimation is doable if the return factor is extracted from data using the following two-step procedure:
\begin{enumerate}
\item Set \smash{$r(g)\propto\frac{k^g}{k}$} based on the intuition that a node hyperdegree distribution should align with potential returns, that is, the most probable hyperlink orders are also the ones that return the most.
\item Set the expected per-capita payoff for cooperating and defecting to be the same, that is, $\overline{\pi}_\mathrm{C}=\overline{\pi}_\mathrm{D}$. This guarantees that cooperators and defector coexist, as is often the case in real-world collaborations.
\end{enumerate}

\begin{figure}[!t]
\centering
\makebox[\textwidth][c]{\includegraphics[scale=1.0]{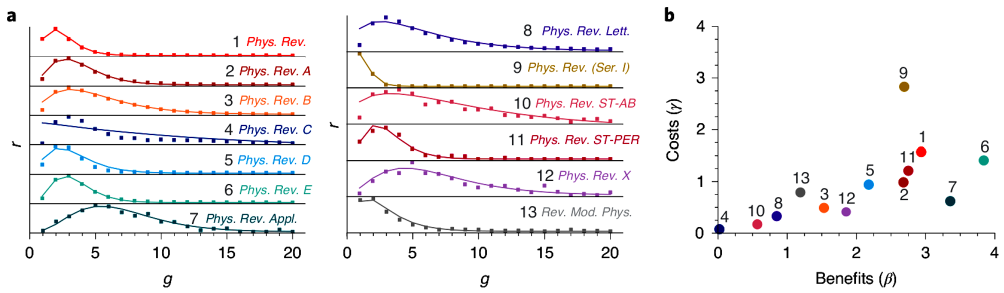}}
\caption{Optimal size of collaborative groups in physics. Panel \textbf{a} shows the relationships between the return factor and the hyperlink order as implied by 13 hypergraphs constructed for the same number of the American Physical Society (APS) journals. Nodes are individual authors and hyperlinks are articles. As the hyperlink order (and thus collaborative group size) increases, at first the return factor increases as well, signifying benefits from distributing research work. Thereafter, the return factor decreases, signifying costs of coordinating large collaborations. Panel \textbf{b} shows that cost and benefit parameters follow approximately linear correlation, which ultimately keeps the optimal size of collaborative groups between two and five.\newline
Source: Reprinted figure from Ref.~\cite{alvarez2021evolutionary}.}
\label{fig:collabsinphys}
\end{figure}

The two-step procedure is illustrated in Ref.~\cite{alvarez2021evolutionary} on 577,886 papers published by the American Physical Society (APS) from 1904 to 2015. Specifically, a hypergraph is constructed for each of 13 society journals such that nodes represent scientists publishing in a chosen APS journal, whereas hyperlinks represent articles. The hypergraph then provides the probabilities $k^g/k$, while the condition $\overline{\pi}_\mathrm{C}=\overline{\pi}_\mathrm{D}$ is used to fix the proportionality constant between $r(g)$ and $k^g/k$. Thus determined values of $r(g)$ are finally used to fit the parameters of Eq.~(\ref{eq:returnbenefitcost}). Several valuable lessons about collaboration in physics follow. First, collaborations between two to three scientists offer the best cost-benefit performance in most journals (Fig.~\ref{fig:collabsinphys}A). Collaborating is therefore beneficial, but the costs of coordinating larger groups become substantial rather quickly. Exceptions are \textit{Physical Review Series I}, which includes publications up to the year 1913 when publishing alone was still very much feasible, as well as \textit{Physical Review Applied} and \textit{Physical Review X}, which show that applied and interdisciplinary research profit from assembling larger collaborative groups. The benefit parameter $\beta$ and the cost parameter $\gamma$ approximately exhibit a linear correlation (Fig.~\ref{fig:collabsinphys}B). When benefits increase with group size, costs do too, keeping the optimal group size within two to five scientists across all modern APS journals.

\subsection{Empirical facts about human cooperation}

Studies on the evolution of cooperation, especially those based on modelling, have proliferated over the past two decades. To exemplify, Ref.~\cite{roca2009evolutionary} and Ref.~\cite{wang2015universal} are two reviews of the field with similar scope published six years apart in the same journal, but the former cites `only' 155 items, whereas the latter cites as many as 314 items. All this effort notwithstanding, the final word as to why cooperation evolves (among organisms in general and humans in particular) is yet to be uttered. Why is that?

Answering the posed question may not be straightforward, but some contributing factor can be singled out with confidence. Even after choosing a social dilemma of interest, modelling studies need to make a series of assumptions about the modelled population and evolutionary dynamics in this population. Typical choices include finite or infinite~\cite{nowak2004emergence, hindersin2019computation} and unstructured or structured~\cite{lieberman2005evolutionary, santos2006evolutionary, maciejewski2014evolutionary} populations with fitness-based or pairwise-comparison~\cite{wu2013interpretations, doebeli2017point} dynamics. What is crucial here is that differing choices in a contextually similar situation can lead to widely different results~\cite{wu2013extrapolating, wu2015fitness}. Models are, moreover, often proposed using cursory rationalisations that struggle to withstand the scrutiny of empirical tests~\cite{sanchez2015theory, sanchez2018physics}, which still somehow fails to discourage recycling similar rationalisations after the fact. Cases in point are peer punishment and network reciprocity. Peer punishment clearly promotes cooperation in theoretical models because getting punished is detrimental to payoff, but experimental results have been much less straightforward; despite an early confirmation~\cite{dreber2008winners}, later research generated mixed~\cite{wu2009costly} and negative~\cite{li2018punishment} results, likely due to intimidatory and retaliatory uses of punishment. Similarly, network reciprocity has shown promising cooperation-promoting effects in theoretical models~\cite{ohtsuki2006replicator}, particularly those built upon node-degree heterogeneous networks~\cite{santos2005scale, santos2008social}, but experimental confirmations of such a promise had failed to materialise at first~\cite{gracia2012heterogeneous, grujic2014comparative} and remained constrained afterwards~\cite{rand2014static}.

Theoretical and experimental advances in physics go hand in hand, and studying the evolution of cooperation should follow the same path. Why then have empirical tests so far failed to steer theoretical-model development? Perhaps this is in part because experiments involving human volunteers are far more ambiguous than experiments involving, say, elementary particles or gravitational waves. The above-mentioned examples of peer punishment and network reciprocity already show that behavioural experiments can go both ways. Ambiguity in empirical results is particularly unwelcome when central tenets of a theory are put to the test. One such tenet when considering the evolution of human behaviour is the imitation (i.e., update) rule~\cite{roca2009evolutionary}. Among early and popular imitation rules is the Fermi rule~\cite{szabo1998evolutionary, traulsen2006stochastic} given by
\begin{linenomath}
\begin{equation}
p_{i\leftarrow j}=\frac{1}{1+\exp\left(-\frac{\Pi_j-\Pi_i}{K}\right)},
\end{equation}
\end{linenomath}
where $p_{i\leftarrow j}$ denotes the probability that the $i$th individual imitates the $j$th individual, with $\Pi_i$ and $\Pi_j$ being the respective payoffs of these individuals, and $K$ denoting the irrationality of selection in the sense that when $K\rightarrow0$, then
\begin{linenomath}
\begin{equation}
p_{i\leftarrow j}=
\begin{cases}
0, & \mathrm{for}~\Pi_i>\Pi_j\\
\frac{1}{2}, & \mathrm{for}~\Pi_i=\Pi_j\\
1, & \mathrm{for}~\Pi_i<\Pi_j
\end{cases},
\end{equation}
\end{linenomath}
whereas when $K\rightarrow\infty$, then $p_{i\leftarrow j}=\frac{1}{2}$. Empirical evidence from economic game theory seems to provide strong support for imitation guided by payoff differences to the extent that volunteers consciously perceive themselves as imitators~\cite{apesteguia2007imitation}. This being said, the nature and complexity of economic games precludes an immediate interpretation of actions by others as sympathetic or antagonistic to oneself (as is the case with cooperation or defection), which ultimately leaves little to go on apart from payoffs. In a comparative analysis of three spatial prisoner's dilemma experiments in which actions taken by others could readily be interpreted as sympathetic or antagonistic, payoff differences seem to have had no decision-making value~\cite{grujic2014comparative}. Interestingly, one of the experiments analysed in Ref.~\cite{grujic2014comparative} was used in a prior publication~\cite{traulsen2010human} to argue in favour of the Fermi rule. The latest work on the subject~\cite{grujic2020people} is once again in favour of imitation because individuals confronted with more successful others imitate the behaviour of those others in accordance with the experienced payoff difference.

Just as it was the case with peer punishment and network reciprocity, empirical evidence favouring imitation driven by payoff differences comes with a degree of ambiguity. A plausible reason why this is so is that payoffs simply do not tell the whole story. For example, experiments reveal the existence of behavioural phenotypes~\cite{poncela2016humans}, meaning that volunteer behaviours are not idiosyncratic, but rather exhibit recognisable characteristics. This enables classifying volunteers into a relatively small number of distinct groups or phenotypes. The cooperative phenotype, in particular, has been shown to possess remarkable robustness with respect to the form of cooperativeness and the passage of time~\cite{peysakhovich2014humans}. Different people are, therefore, likely to be predisposed to cooperate to a different degree in the same social-dilemma situation. Volunteers may also directly respond to the actions of others instead of being concerned with payoffs. An antagonistic (respectively, sympathetic) action may provoke an antagonistic (resp., sympathetic) response. This is related to the theoretical concepts of the tit-for-tat strategy~\cite{axelrod1981evolution} and conditional cooperation~\cite{bendor1997evolutionary}, and indeed seems to regularly occur in behavioural experiments~\cite{keser2000conditional, fischbacher2001people, grujic2010social, shi2020freedom}. Somewhat similarly, volunteers have been shown to cooperate more (respectively, less) after cooperating (resp., defecting) previously even if the contextual situation is the same~\cite{grujic2010social, gutierrez2014transition, horita2017reinforcement}.

Additional reasons why payoffs may not tell the whole story in the context of cooperative behaviours is that human decision making is prone to some `peculiarities'. The field of behavioural economics made it a point to tease out such peculiarities, that is, document behaviours that deviate from the predictions of economic models based on a rational drive to maximise the expected utility~\cite{tversky1972elimination, kahneman1979prospect, schoemaker1982expected}. Experiments focusing on cooperativeness also pinpoint peculiarities that are similar in spirit or related to those emphasised by behavioural economists. Ref.~\cite{wang2017onymity}, for example, shows that acquaintances cooperate significantly more than strangers in a social-dilemma game in which there is absolutely no incentive to do so. Volunteers were, in fact, incentivised to score as much as possible for themselves following gameplay rules designed to be inconsequential to any sort of real-world interactions that may occur afterwards. The experiment was thus akin to point-gathering competitive gameplay that is so common among friends worldwide without ruining their friendships. That mere identifiability spurs altruistic behaviour has been observed elsewhere~\cite{hoffman1996social}, but it is hard to incorporate into theoretical models when the corresponding incentives are entirely absent.

In another experiment, aimed at discerning how rewards promote cooperation, it was found that an unexpected and convoluted mechanism is at play~\cite{wang2018exploiting}. Specifically, volunteers almost ignore the opportunity to reward one another, and yet cooperativeness doubles compared to the control in which there is no reward. To make matters even more perplexing, improved cooperation frequencies are observed right from the start, before any rewarding could ever happen. This peculiar behaviour can ultimately be traced to a known cognitive bias called the decoy effect~\cite{pettibone2000examining}. Specific to multiple-choice situations, `decoy' is a choice that shares some defining characteristics with another `target' choice, but is inferior in one defining characteristic. Such inferiority makes the target look disproportionately more attractive, and thus preferred over all other choices. Because rewarding as envisioned in Ref.~\cite{wang2018exploiting} was just an extra-demanding form of cooperating compared to the usual cooperative option, the decoy effect made cooperation appear much more attractive than defection, ultimately causing a surge in the cooperation frequency. On a more general note, cognitive biases may provoke not only behaviours for which there are no apparent incentives, but also behaviours that outright go against incentives implied by a particular social dilemma. This is particularly hard to model outside of broader evolutionary contexts in which cognitive biases may make more sense~\cite{group2014evolution}.

We have already mentioned in passing that from an empirical perspective, static networks offer a limited scope for promoting cooperation~\cite{rand2014static}. Consequently, the focus has shifted to dynamic networks in which volunteers could initiate links with cooperative others or sever links with defecting others~\cite{rand2011dynamic}. Just as cooperation fails in many static networks, the same happens in networks that are shuffled randomly every round, or in networks that are rewired infrequently (albeit freely) by volunteers. Only when network links can be updated frequently and freely, a high degree of cooperation is maintained because cooperators shun links with defectors, while preferentially linking with other cooperators. This form of shunning defectors is often interpreted as a sort of punishment that, to make things even better, comes with two positive side effects. First, there is no obvious cost to severing links with defectors, which avoids what is known as the second-order social dilemma by which cooperators who refuse punishing are better off than cooperators who do punish~\cite{gardner2004cooperation, hauert2007via, melis2010human}. Second, once the link is severed, there are no opportunities (at least not immediate ones) for retaliation, which avoids diminishing the willingness to punish defectors due to the threat of being retaliated against~\cite{nikiforakis2008punishment, nikiforakis2011altruistic, balafoutas2014third}. On a more fundamental level, however, dynamic networks can be seen as a means to foster positive assortment~\cite{fehl2011coevolution}. The importance of this perspective is that positive assortment has often been emphasised as a common denominator for many major cooperation-promoting mechanisms~\cite{eshel1982assortment, pepper2002mechanism, fletcher2006unifying, fletcher2009simple, momeni2013spatial}. If positive assortment is indeed key to resolving social dilemmas, then enriching particularly hard social dilemmas with novel degrees of freedom that facilitate assortment should universally promote cooperation~\cite{shi2020freedom}. Thinking in terms of degrees of freedom is, of course, dear to physicists.


To summarise, empirical evidence shows that:
\begin{itemize}
\item In situations in which rationality dictates a clear course of action (e.g., behave to avoid punishment because it is bad for payoff), psychological factors may prompt another course of action (e.g., use punishment to intimidate or retaliate).
\item The identification of distinct and stable behavioural phenotypes suggests that different individuals are predisposed to respond to the same social-dilemma situation with a different degree of cooperativeness.
\item One's performance relative to those of others (in terms of payoffs) may be more influential in guiding decisions when the complexity of contextual situations conceals whether an action by another is sympathetic or antagonistic to oneself. When this is not the case, however, performance may be secondary to a direct response to the action by another.
\item Intuitions may override incentives. For example, acquaintances are mutually more cooperative than strangers in exactly the same social-dilemma situation.
\item Human decision making is fraught with peculiarities (i.e., cognitive biases) that make sense only in broader evolutionary contexts. These peculiarities may also (partly) override incentives. For example, a mere presence of reward promotes cooperativeness although no one rewards anyone.
\item Introducing new degrees of freedom into hard social dilemmas may facilitate positive assortment and, by extension, promote cooperation. For example, static networks struggle to improve and maintain cooperativeness, but dynamic networks with frequent and free link updating work like a charm.
\end{itemize}

\subsection{Future outlook}

Having reviewed a wide variety of theoretical models pertaining to the evolution of human cooperation, and then summarising a number of empirical facts on the subject, we emphasised the need to reconnect theory and experiments. Doing so, however, faces challenges that have been first raised in relevant behavioural disciplines. Psychology and behavioural economics alike suffer from a deep replication crisis~\cite{klein2014investigating, open2015estimating, camerer2018evaluating, klein2018many, serra2021nonreplicable}. This state of affairs has triggered calls for an overhaul of the scientific process that had led to so many irreproducible results in the first place~\cite{nosek2015promoting, munafo2017manifesto, shrout2018psychology}. Many of the proposed measures are methodological; for example, the criterion for statistical significance should be more stringent~\cite{benjamin2018redefine}, much larger samples are required to ensure high statistical power~\cite{brandt2014replication, maxwell2015psychology}, and preregistration should become a norm to curb statistical manipulations and avoid some cognitive biases that interfere with sound research practices~\cite{nosek2018preregistration, nosek2019preregistration, soderberg2021initial}. All these measures are, if not necessary, then at least  absolutely welcome, but they do complicate the logistics of conducting experimental studies (e.g., recruiting thousands of volunteers), and increase time and effort from conceptualising to publishing a study (e.g., preparing preregistration and calculating power). If physicists are to scrutinise their theoretical models by means of behavioural experiments, the same high standards will apply as for psychologists, behavioural economists, and the like. Just keeping up with the standards will likely require widening multidisciplinary collaborations.

On top of methodological problems contributing to the replication crisis in behavioural disciplines, Ref.~\cite{muthukrishna2019problem} makes a compelling argument that a lack of a strong theoretical backbone plagues the field even more. When such a backbone exists, it paints a bigger picture, guiding researchers to formulate useful expectations. Ref.~\cite{muthukrishna2019problem} itself refers to an example from physics; early analyses of the data on neutrinos coming from CERN (near Geneva, Switzerland) to Gran Sasso National Laboratory (in the province of L'Aquila, Italy) suggested that these elementary particles move faster than light~\cite{brumfiel2011particles}. Because this would go against the special theory of relativity, the news of faster-than-light neutrinos was received with a healthy dose of scepticism. Later analyses indeed confirmed that neutrinos obey the limitation set by the speed of light~\cite{adam2012measurement}. It is almost a certainty that without firm theoretical expectations, the finding would be put to much less scrutiny, perhaps taking years to set the record straight.

Even more important is that, without firm theoretical expectations, an empirical behavioural scientist faces an almost infinite set of possible hypotheses that could be put to an experimental test. Experiments are then bound to be designed based on intuitions and guesswork about what may or may not be widely considered interesting (or worse yet, `hot') among peers. A major consequence is that, even with impeccable experimental methodology, only bits and pieces of disconnected knowledge can be acquired. The research on the evolution of human cooperation is thankfully nowhere near such a dire state, but there are some elements reminiscent of what Ref.~\cite{muthukrishna2019problem} refers to. Namely, evolutionary game theory specifies a blueprint for model construction that can always be extended with yet another cooperation-enhancing tweak. It is true that nuanced social-dilemma scenarios lead to fascinating dynamics that is of substantial interest in itself~\cite{guo2020novel, shen2021exit}, but do humans really behave as the models predict? Should we even empirically test countless modelled scenarios if they hardly get us any closer to general principles such as the aforementioned positive assortment? The answer would probably be negative even without the rising costs of behavioural experiments in terms of logistics, time, and effort. We therefore foresee the need for two types of models. One is models aimed at analysing incentives for \textit{cooperative behaviour} in specific situations of societal interest such as fighting corruption~\cite{lee2015games, lee2019social} or encouraging vaccination~\cite{arefin2019interplay, kabir2019behavioral}. The other is models aimed at explaining the evolution of \textit{cooperative trait} in humankind based on as general principles as possible such as robust paradigms for reciprocal altruism~\cite{axelrod1981evolution, nowak1993strategy} or robust social norms~\cite{ohtsuki2006leading, ohtsuki2007global}. Overall, this should lead to a narrower space of experimental hypotheses, theoretical models of cooperation that are backed by substantial empirical evidence, and ultimately more definitive answers as to when (circumstances) and why (forces) humans cooperate.

\FloatBarrier

\section{Networks and communities}
\label{S:Net}

Networks are a pillar of social physics. They permeate all aspects of the field, and more. Applications include---but are not limited to---online, physical, and even animal social networks~\cite{holme2003network, newman2003social, holme2004structure, rocha2010information, pinter2014dynamics, jalili2017link}, finance~\cite{battiston2012debtrank}, retail~\cite{rocha2010network}, supply chains~\cite{kito2014structure, inoue2019firm}, transport infrastructure~\cite{holme2003congestion, delaurentis2008network, hu2009empirical, xie2009modeling, derrible2011applications, du2016physics}, power grids~\cite{rosas2007topological, pagani2013power, kim2015community}, climate and Earth systems~\cite{donges2009complex, fan2021statistical}, medical and clinical investigations~\cite{barabasi2007network, hofmann2018complex}, nutrition~\cite{barabasi2020unmapped, herrera2020contribution}, and sports~\cite{buldu2018using, buldu2019defining}. Studies focusing on dynamics in networks are also ubiquitous, ranging from general dynamical patterns~\cite{harush2017dynamic} to random walks~\cite{noh2004random, perkins2014scaling} to synchronisation~\cite{wu2005synchronization, scardovi2009synchronization, delgenio2016synchronization}, epidemiological dynamics~\cite{bansal2007individual, danon2011networks, brockmann2013hidden}, evolutionary dynamics of cooperation (see Section~\ref{S:Coop}), social-balance dynamics~\cite{hummon2003some, antal2005dynamics, antal2006social}, innovation dynamics~\cite{iacopini2018network}, and many others. Network science applied to social systems has, in fact, grown to the point at which its branches are sufficiently broad to be a topic of massive standalone reviews~\cite{boccaletti2006complex, fortunato2010community, fortunato2016community, barthelemy2011spatial, holme2012temporal, boccaletti2014structure, masuda2017random, dearruda2018fundamentals}.

The sheer size of network science precludes us from overviewing the field comprehensively (let alone exhaustively). Our purpose is instead specific, that is, to examine the question of what constitutes a community in networks. We are particularly interested in decomposing and understanding the community structure of networks through the prism of generative models (as opposed to heuristic community-detection methods). For readers interested in the topic of community detection more broadly, modern developments and the current state of the art from the standpoint of physics can be found in Refs.~\cite{lancichinetti2009community, fortunato2010community, lancichinetti2011finding, hric2014community, fortunato2016community, schaub2017many, rosvall2019different}. The specific direction that we are singling out from the breadth of network science leads to at least two questions. What makes network-community structure so important? And why should one look favourably at generative models? We shall endeavour to answer these questions in the next section.

\subsection{Community detection: contexts and methods}

Detecting communities in networks is an intuitively appealing task. Thinking of a network as a means to consistently simplify the picture of a complex system, while retaining the ability to see how the system is stitched together into one whole, we may gain deeper insights into the functioning of the system by identifying the network's more basic constituents. Community detection can furthermore be considered a natural unsupervised-learning task (see Section~\ref{SS:AIfundamentals}). Indeed, when clustering is used to probe the internal structure of a dataset, among the first steps is to define a similarity or distance function between two data points (in essence forming a graph representation of the dataset), but networks already come with links that indicate relations between nodes. One underlying motivation behind community detection is therefore to leverage the preexisting information on network topology, and thus node-node relations, in order to learn about networks, much like an unsupervised learner learns about datasets. Given that communities as more basic constituents come together to form a network, it is of little surprise that the network's divisibility into communities shapes the dynamics that unfolds in networks. We have already mentioned this in the case of the evolutionary dynamics of cooperation~\cite{jiang2013spreading, gomez2012evolutionary}, but the same goes for, among others, epidemiological dynamics~\cite{chung2014generalized, valdez2020epidemic} and related decision making~\cite{masuda2009immunization, gross2020epidemic}. Arguably one of the most important features of community detection is the ability to make informed decisions about errors in measuring network structure. Doing so, however, demands having a `standard' or a `blueprint' that tells us whether we should expect a link where there is none or expect no links where there is one. Such a standard or blueprint is provided by generative models.

The importance of generative models is analogous to the importance of mechanistic (i.e., process-based) models in the context of dynamical systems. If we measure, say, the growth of a city, and the number of hospitals or schools necessary to sustain the city, we may gain the ability to plan for the future. If the number of hospitals increases sublinearly or linearly with city size, then the city growth is likely to be manageable. If, however, the number of schools increases supralinearly, then the city growth is at some point likely to deplete the resources needed for building and operating more schools. Although the ability to plan for the future is very much desirable, without a mechanistic model of city growth, we are in the dark as to what causes the number of hospitals to increase sublinearly (which is manageable) and the number of schools supralinearly (which is unmanageable). Actionable insights may be gained by pinpointing the processes behind the sublinear increase in one case and the supralinear increase in the other. In a similar vein, a generative model that fits an empirical network very well may strongly favour the presence of a link when there is none in the data. Such a situation would give us much confidence that the link is missing due to a measurement error. We may even uncover the process that creates this particular link. Surely, the best known generative models in network science are Erd\H{o}s-R\'{e}nyi~\cite{erodos1959random}, Watts-Strogatz~\cite{watts1998collective}, and Barab\'{a}si-Albert~\cite{albert2002statistical} (in which the processes of growth and preferential attachment decide network topology). When it comes to generating modular networks, stochastic block models have become a staple~\cite{holland1983stochastic}. We shall rely heavily on this model type moving henceforward.

\begin{figure}[!t]
\centering\includegraphics[scale=1.0]{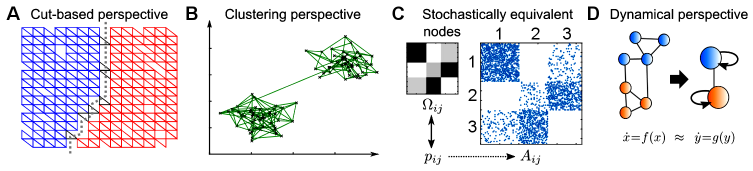}
\caption{Four main community-detection contexts and method classes. \textbf{A,} Cut-based methods seek to find network partitioning that minimises the number of between-community links without imposing dense within-community linking. \textbf{B,} Clustering methods embody the intuitive idea that links within communities are dense and across communities sparse. \textbf{C,} Methods seeking stochastically equivalent nodes posit that two nodes from the same community link to nodes from other communities with exactly the same probability. In the shown example, there are three communities identified by the block structure of the adjacency matrix. Community 1 has dense internal links, barely any links with community 2, and moderately dense links with community 3. \textbf{D,} Dynamic methods emphasise system behaviour over system topology. In particular, the role dynamics is considered crucial.\newline
Source: Reprinted figure from Ref.~\cite{schaub2017many} under the Creative Commons Attribution 4.0 International (CC BY 4.0).}
\label{fig:commdetectmethods}
\end{figure}

Before we introduce and define stochastic block models, and see how they are used for statistical inference, it is useful to look at four main community-detection contexts and method classes (Fig.~\ref{fig:commdetectmethods}). The reason for this is to show that the problem of community detection has no single correct formulation, let alone a single correct solution. For example, when designing a distributed computing system spread over several locations, the best partitioning is the one that minimises expensive long-distance links. Such a problem is best tackled using cut-based methods (Fig.~\ref{fig:commdetectmethods}A). If, by contrast, the aim is to understand the structure of large organisations based on social interactions, a collection of strongly interacting, and thus densely interconnected, individuals is likely to act as a functional group within the organisation. Clearly, this is a problem for clustering methods (Fig.~\ref{fig:commdetectmethods}B). When competing interests drive social interactions, strongly interacting, and thus densely interconnected, individuals may be opponents who belong to different teams. This is a type of problem for methods seeking stochastically equivalent nodes (Fig.~\ref{fig:commdetectmethods}C). Finally, if we aim to identify groups of individuals threatened by an epidemic, then interconnectedness comes secondary to epidemiological dynamics. Dynamic methods are expected to yield the best results (Fig.~\ref{fig:commdetectmethods}D).

The examples outlined above show that in community detection, context dictates methods. And yet, some methods are more heuristic and phenomenological, whereas others are more rigorous and fundamental. The latter is especially true of methods seeking stochastically equivalent nodes. These methods are founded on generative models and statistical inference. A prime example is stochastic block models to which we turn next.

\subsection{Introducing stochastic block models}

\begin{figure}[!t]
\centering\includegraphics[scale=1.0]{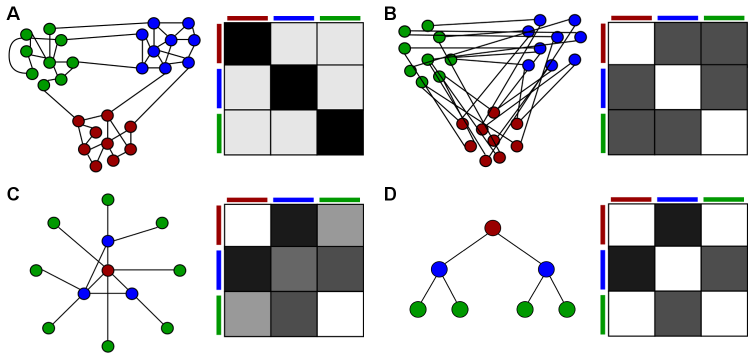}
\caption{Representing various network topologies using stochastic block models. Each of the four networks has an adjacency matrix divided into blocks, where the grayscale indicates link probabilities (white=0 and black=1). \textbf{A,} Assortative network structure in which within-community links are dense, but between-community links are sparse. \textbf{B,} Disassortative network structure in which within-community links are sparse, but between-community links are dense. \textbf{C,} Core-periphery network structure. \textbf{D,} Hierarchical network structure.\newline
Source: Reprinted figure from Ref.~\cite{funke2019stochastic} under the Creative Commons Attribution 4.0 International (CC BY 4.0).}
\label{fig:SBMstructures}
\end{figure}

As our discussion has already indicated, stochastic block models (abbreviated SBMs) power one of the most popular techniques for community detection that comes from the domain of statistical inference. The technique is based on the construction of an SBM that is fitted to network data~\cite{holland1983stochastic, nowicki2001estimation}. The model parameters are estimated by maximising likelihood, and once this is done, they provide information not only about the network structure, but also about the within-network node relationships, thus forming a flexible modelling tool for analysis and prediction. An additional advantage of SBMs over other methods is that SBMs are not limited to assortative mono-layered networks; it is conceptually straightforward to generalise the technique to a wide range of topological and dynamical network constructs (Fig.~\ref{fig:SBMstructures}). Mixed memberships and overlapping communities \cite{airoldi2008mixed, fan2016copula, pal2019scalable}, weighted networks~\cite{aicher2015learning, peixoto2018nonparametric}, multilayer networks~\cite{peixoto2015inferring, stanley2016clustering, paul2016consistent, valles2016multilayer, debacco2017community}, temporal networks~\cite{fu2009dynamic, xu2014dynamic, peel2015detecting, corneli2016exact, ghasemian2016detectability, matias2017statistical, peixoto2017modelling, zhang2017random}, networks that possess node attributes~\cite{peel2015active} or are annotated with metadata~\cite{hric2016network, newman2016structure} pose no problems to SBM-based community detection~\cite{peixoto2019bayesian}.

The first SBM algorithms were developed in social sciences to detect communities of `approximately equivalent' nodes. The algorithms were deterministic and were based on permuting the adjacency matrix to reveal the block structure and relationships among community members~\cite{white1976social, arabie1978constructing, rosvall2019different}. Shortly afterwards, pioneering works formalised the generative model and established a stochastic formulation of node equivalence in a community in such a way that equivalent nodes were associated with equivalent probabilities \cite{holland1983stochastic, wasserman1987stochastic, nowicki2001estimation}. During the 1990s, computer power started to grow tremendously, which opened up the doors to new opportunities for network analysis in the digital environment. Consequently, SBM methods and algorithms underwent a rapid and diverse development over the last 25 years. SBMs are nowadays applied in a large number of natural and social scientific, and engineering fields (see Table I in Ref.~\cite{peixoto2017nonparametric}).

One of the most significant contributions to the development of community-detection methods based on statistical inference comes from physicists Karrer and Newman, who adapted the standard SBM model to take into account node-degree heterogeneity~\cite{karrer2011stochastic}. Thus obtained degree-corrected SBM enabled the application of SBMs to real-world networks and paved the way for the development of multiple model variants \cite{gulikers2017spectral, chen2018convexified}. Newman, also the creator of one of the most popular heuristic methods based on modularity optimisation~\cite{newman2004finding}, in a recent study~\cite{newman2016equivalence} showed that there is an equivalence between likelihood maximisation for SBMs and generalised modularity maximisation for planted-partition model with simplified community structure. The planted-partition model is a reduced version of a standard SBM in which links materialise with two probabilities, one for within communities and the other for between communities. Equivalence results such as Newman's, and the extent to which they hold~\cite{zhang2020statistical}, are of much interest because they show that methods developed with very different motivations and intuitions in mind may end up serving the same purpose.

Another major contribution from statistical physics comes from Peixoto who developed a microcanonical view of SBMs by which the traditional (i.e., canonical), probabilistic definition of link formation between nodes belonging to two separate communities is replaced by a precise number of links~\cite{peixoto2012entropy, peixoto2019bayesian}. The canonical and microcanonical definitions are in accordance with the jargon of statistical physics; in the canonical generative model constraints on node degrees and the number of links are imposed on average, whereas in the microcanonical model these constraints are exact. One of the chief results in recent years is the development of a non-parametric microcanonical model using Bayesian inference that does not require prior knowledge of the number of communities, as well as the nested versions of this model for community detection~\cite{peixoto2017nonparametric, peixoto2019bayesian}. The result is a culmination of previous developments in the context of microcanonical SBMs, including the work on making use of minimum description length (MDL)~\cite{rissanen1978modeling, rosvall2007information} to determine the number of communities~\cite{peixoto2013parsimonious}, devising efficient Markov chain Monte Carlo inference algorithms~\cite{peixoto2014efficient}, proposing nonparametric nested models~\cite{peixoto2014hierarchical}, and incorporating model selection~\cite{peixoto2015model}. For a hands-on experience, most of the present SBM variants are implemented in \textit{graph-tool}, a Python module for manipulation and statistical analysis of networks (available at \url{https://graph-tool.skewed.de/}).

In the rest of this introductory review, our focus will be on the above-mentioned concepts, whereas detailed reviews of the development of other SBM variants are given in Refs.~\cite{funke2019stochastic, lee2019review, jin2021survey}. Of note is that the development of new SBM variants is in many cases motivated by the specifics of real-world networks. The theoretical results related to establishment of the fundamental limits for community detection in the SBM, both with respect to information-theoretic and computational thresholds, are extensively reviewed in Ref.~\cite{abbe2017community}.

\subsection{Defining stochastic block models}

The construction of SBMs as a generative network tool is based on the idea that network nodes are divided into communities, and that the existence of a link between two nodes is determined by the communities to which these nodes belong. These considerations impose conditions for generating ensembles that in a statistical sense represent a network. There are two main approaches to defining an SBM: (i) canonical, in which conditions are imposed on average, and (ii) microcanonical, in which conditions are imposed exactly.

\paragraph*{Canonical form} The traditional formulation of SBMs is in canonical form. SBMs in canonical form are parameterised by two parameters, $\mathbf{b}$ and $\mathbf{W}$, as follows; $N$ nodes are distributed in $K$ communities, and the affiliation of nodes to communities is expressed by the vector $\mathbf{b}=[b_i]$ of dimension $N$, where the value $b_i=r\in\{1,2,\ldots,K\}$ denotes the affiliation of the $i$th node to the $r$th community. The number of nodes in each community can be read from the vector $\mathbf{b}$. We denote the size of the $r$th community by $n_r$ and form a vector of community sizes $\mathbf{n}=[n_r]$ of dimension $K$. The matrix $\mathbf{W}=[w_{rs}]$ of dimension $K\times K$ specifies the probabilities $w_{rs}$ that a link is formed between any two nodes belonging to the communities $r$ and $s$, that is, $P(i\Leftrightarrow j)=w_{rs}$, where $b_i=r$ and $b_j=s$.

The matrix $\mathbf{W}$ can be specified in multiple ways. When the values $w_{rs}\sim\mathcal{B}(p_{rs})$ follow a Bernoulli distribution with the parameter $p_{rs}$, then the probability that there is a link between nodes $i$ and $j$ is $P(i\Leftrightarrow j)=p_{rs}$, and the probability that there is no link is $P(i\nLeftrightarrow j)=1-p_{rs} $. In this case, the probability that a generative procedure, taking a node division $\mathbf{b}$ as a parameter, creates an undirected and unweighted network with the adjacency matrix $\mathbf{A}=[A_{ij}]$ equals to
\begin{linenomath}
\begin{equation}
P(\mathbf{A}|\mathbf{p},\mathbf{b})=\prod_{i<j}p_{b_i,b_j}^{A_{ij}} (1-p_{b_i,b_j})^{1-A_{ij}},
\end{equation}
\end{linenomath}
where $\mathbf{W}=\mathbf{p}=[p_{rs}]$ is matrix of dimension $K\times K$ comprising Bernoulli parameters.

A commonly used distribution for ease of calculation is the Poisson distribution. In this case, referred to as the standard SBM, the values $w_{rs}\sim\mathcal{P}(\lambda_{rs})$ are determined by the parameters $\lambda_{rs}$, which in turn define the probability $P(i\Leftrightarrow j; k)=\lambda_{rs}^k e^{-\lambda_{rs}}/k!$ that there is a link of order $k$ between nodes $i$ and $j$. A network created using a Poisson distribution, unlike a Bernoulli one, has multiple links between nodes, which is convenient in the case of generating or analysing value networks. The Poisson distribution can, however, be used in conjunction with networks having just one link between two nodes. This is achieved by recognising that the probability of multiple links decreases with $1/N$ in sparse networks in which the total number of links is proportional to $N$, so the existence of multiple links can be ignored when $N$ is large~\cite{peixoto2019bayesian}, or simply multiple links are compressed into a single one. The probability that the standard SBM generates a network with the adjacency matrix $\mathbf{A}=[A_{ij}]$ equals to
\begin{linenomath}
\begin{equation}
P(\mathbf{A}|\mathbf{\Lambda},\mathbf{b)}= \prod_{i<j} \frac{\lambda_{b_i,b_j}^{A_{ij}}}{A_{ij}!}e^{-\lambda_{b_i,b_j}} \times \prod_{i} \frac{(\lambda_{b_i,b_i}/2)^{A_{ii}}}{(A_{ii}/2)!}e^{-\frac{\lambda_{b_i,b_i}}{2}},
\end{equation}
\end{linenomath}
where $\mathbf{W}=\mathbf{\Lambda}=[\lambda_{rs}]$ is matrix of dimension $K\times K$ comprising Poisson parameters.

\paragraph*{Degree-corrected SBM} The assumption behind the standard SBM is that nodes in the same community are statistically equivalent, that is, they all have the same number of links, on average. This is seldom true for real networks whose node-degree heterogeneity may span orders of magnitude~\cite{newman2010networks}. To accommodate such node-degree heterogeneity, Ref.~\cite{karrer2011stochastic} proposes a modified model, called the degree-corrected SBM, in which each node is assigned the parameter $\theta_i$ controlling the node's expected degree $\langle k_{i}\rangle$, regardless of community affiliation. This means that in addition to the vector $\mathbf{b}$ and the matrix $\mathbf{\Lambda}$, another model parameter $\boldsymbol{\theta}=[\theta_i]$, taking the form of an $N$-vector, is required to define the model. The parameter $\boldsymbol{\theta}$ generates heterogeneity within communities, where $P(i\Leftrightarrow j; k)=(\theta_i\theta_j\lambda_{rs})^k e^{-\theta_i\theta_j\lambda_{rs}}/k!$ is the probability that there is a link of multiplicity $k$ between nodes $i\in b_i=r$ and $j\in b_j=s$. Given the additional parameter $\boldsymbol{\theta}$, a network with the adjacency matrix $\mathbf{A}=[A_{ij}]$ is generated with the probability
\begin{linenomath}
\begin{align}
P(\mathbf{A}|\mathbf{\Lambda},\boldsymbol{\theta},\mathbf{b)}&=
\prod_{i<j} \frac{(\theta_i\theta_j\lambda_{b_i,b_j})^{A_{ij}}}{A_{ij}!}e^{-\theta_i\theta_j\lambda_{b_i,b_j}}\nonumber\\
&\times\prod_{i} \frac{(\theta_i^2\lambda_{b_i,b_i}/2)^{A_{ii}}}{(A_{ii}/2)!}e^{-\frac{\theta_i^2\lambda_{b_i,b_i}}{2}}.\label{eqn:dcsmb}
\end{align}
\end{linenomath}

\paragraph*{Microcanonical form} In canonical form, restrictions on node degrees and the number of links between communities $r$ and $s$ are `soft', expressed via expected values. This means that across various model realisations, node degrees and the number of links fluctuate around the mean values. In microcanonical form, the conditions are `hard' in the sense that node degrees and the number of links are strictly determined for each realisation. Specifically, let the vector $\mathbf{k}=[k_i]$ of dimension $N$ define node degrees, and furthermore let $\mathbf{e}=[e_{rs}]$ be a matrix of dimension $K\times K$ whose values $e_{rs}$ define the number of links between communities $r$ and $s$. For convenience, the diagonal elements $e_{rr}$ are defined as double the number of links within community $r$. The generative process~\cite{fosdick2018configuring} assigns $k_i$ semi-links (called `stubs') to each node $i$, whereupon the stubs are randomly joined together until reaching the condition that between communities $r$ and $s$ there are exactly $e_{rs}$ links. Connecting the stubs (i.e., wiring the network) to satisfy said condition can be done in $\Omega(\mathbf{e})$ ways
\begin{linenomath}
\begin{equation}
\Omega(\mathbf{e})=\frac{\prod_r e_r!}{\prod_{r<s}e_{rs}!\prod_{r}e_{rr}!!},
\label{eqn:micro1}
\end{equation}
\end{linenomath}
where $e_r=\sum_s e_{rs}$ and $(2m)!!=2^mm!$. However, not every wiring produces a unique network. Given the adjacency matrix $\mathbf{A}$, the number of different stub wirings, $\Xi(\mathbf{A})$, that produce the same network is given by~\cite{fosdick2018configuring, peixoto2019bayesian}
\begin{linenomath}
\begin{equation}
\Xi(\mathbf{A})=\frac{\prod_i k_i!}{\prod_{i<j}A_{ij}!\prod_{i}A_{ii}!!}.
\label{eqn:micro2}
\end{equation}
\end{linenomath}
This implies that the network with the adjacency matrix $\mathbf{A}$ is generated with the probability
\begin{linenomath}
\begin{equation}
P(\mathbf{A}|\mathbf{k},\mathbf{e},\mathbf{b})= \frac{\Xi(\mathbf{A})}{\Omega(\mathbf{e})}.
\label{eqn:micro3}
\end{equation}
\end{linenomath}
The last relationship is valid under the `hard' constraints
\begin{linenomath}
\begin{align}
k_i&=\sum_{j} A_{ij},\nonumber\\
e_{rs}&=\sum_{ij}A_ {ij}\delta_{b_i,r}\delta_{b_j,s}.\label{eqn:hard}
\end{align}
\end{linenomath}
If these constraints are not satisfied, then $P(\mathbf{A}|\mathbf{k},\mathbf{e},\mathbf{b})=0$.

\paragraph*{Relating canonical and microcanonical forms} As briefly mentioned before, equivalence results are important because they reveal how different motivations and intuitions may serve the same purpose. Even if full equivalence cannot be established, understanding conditions under which separate mathematical constructs exhibit similar behaviour is of great interest.

Canonical and microcanonical forms generate the same networks in the asymptotic sense if node degrees and the number of links between communities are large enough numbers. Namely, by expanding the relation in Eq.~(\ref{eqn:dcsmb}), we get
\begin{linenomath}
\begin{align}
P(\mathbf{A}|\mathbf{\Lambda},\boldsymbol{\theta})&=\prod_{r<s}\lambda_{rs}^{e_rs}e^{-\hat{\theta}_r\hat{\theta}_s\lambda_{rs}} \prod_{r}\lambda_{rr}^{\frac{e_{rr}}{2}}e^{-\frac{\hat{\theta}_r^{2}\lambda_{rr}}{2}} \nonumber\\
&\times \frac{\prod_i \theta_i^{k_i}}{\prod_{i<j}A_{ij}!\prod_{i}(A_{ii}/2)!}, \label{eqn:dcsbm2}
\end{align}
\end{linenomath}
where $\hat{\theta}_r=\sum_i \theta_i \delta_{b_i,r}$. The parameters $\theta_i$ and $\lambda_{rs}$ form a product in the expression for the probability $P(\mathbf{A}|\mathbf{\Lambda},\boldsymbol{\theta})$, so their individual values can be re-scaled provided that the product remains the same. If we choose parameterisation
\begin{linenomath}
\begin{equation}
\hat{\theta}_r=\sum_i \theta_i\delta_{b_i,r}=1
\label{eqn:seeit}
\end{equation}
\end{linenomath}
for each community $r$, then $\lambda_{rs}=\langle e_{rs}\rangle$ is the expected number of links between communities $r$ and $s$, and \smash{$\theta_{i}=\frac{\langle k_i \rangle}{\sum_s \lambda_{b_i,s}}$} is proportional to the expected node degree~\cite{peixoto2019bayesian}. If furthermore the Stirling's factorial approximation $\ln(m!)\approx m\ln(m)-m$ is applied to Eqs.~(\ref{eqn:micro1}) and (\ref{eqn:micro2}), it can be shown that the microcanonical likelihood $P(\mathbf{A}|\mathbf{k},\mathbf{e},\mathbf{b})$ in Eq.~(\ref{eqn:micro3}) approaches asymptotically the likelihood $P(\mathbf{A}|\mathbf{\Lambda},\boldsymbol{\theta})$ in Eq.~(\ref{eqn:dcsbm2}), for large enough $k_i$ and $e_{rs}$. For small or sparse networks, however, the differences between canonical and microcanonical forms can be substantial~\cite{bianconi2009entropy, squartini2015breaking}.

As we have just seen, there is no exact equivalence between canonical and microcanonical forms. From the viewpoint of inference, however, such a lack of equivalence is immaterial because the models are unidentifiable anyway; it is impossible to tell from a single network realisation whether the network came from the canonical or the microcanonical model. Bayesian inference offers yet another perspective on the relationship between the two models, which will be discussed shortly.

\subsection{Statistical inference of communities}

Having defined the most common SBMs, it is now time to put them to good use, that is, use them for statistical inference. Given an observed network with the adjacency matrix $\mathbf{A}=[A_{ij}]$, $i,j\in\{1,2,...,N\}$, statistical inference using SBMs consists of finding the model parameters that generate the observed network. More specifically, the problem is to find the node partition $\mathbf{b}$ that maximises the log-likelihood function $\mathcal{L}=P(\mathbf{A}|\mathbf{b})$. Karrer and Newman~\cite{karrer2011stochastic} derived an unnormalised log-likelihood function for the standard model
\begin{subequations}
\begin{linenomath}
\begin{equation}
\mathcal{L}^\mathrm{KN} = \ln P(\mathbf{A}|\mathbf{b})=\sum_{r,s} e_{rs} \ln\frac{e_{rs}}{n_{r} n_{s}},
\label{eqn:LKN1}
\end{equation}
\end{linenomath}
where, as before, $e_{rs}=\sum_{ij} A_{ij}\delta_{b_{i},r}\delta_{b_{j},s}$ is the total number of links between communities $r$ and $s$, or if $r=s$, double the number of links in community $r$. Peixoto~\cite{peixoto2012entropy} derived a similar result for the microcanonical SBM
\begin{linenomath}
\begin{align}
\mathcal{L}^\mathrm{P} &= \frac{1}{2} \sum_{r,s} n_r n_s H(\frac{e_{rs}}{n_{r}n_{s}}) \nonumber\\
&\overset{\langle k\rangle\ll N}{\approx} -E+\frac{1}{2} \sum_{r,s} e_{rs} \ln\frac{e_{rs}}{n_{r} n_{s}}, \label{eqn:LP2}
\end{align}
\end{linenomath}
where $H(x)=-x\ln x-(1-x)\ln(1-x)$ is the binary entropy function. In the case of the degree-corrected SBM, the analogous log-likelihood relations for $\mathbf{b}$ are (see Refs.~\cite{karrer2011stochastic, peixoto2012entropy} for details)
\begin{linenomath}
\begin{align}
\mathcal{L}_\mathrm{d.c.}^\mathrm{KN} &= \sum_{r,s} e_{rs} \ln\frac{e_{rs}}{e_{r} e_{s}}, \label{eqn:LP3}\\
\mathcal{L}_\mathrm{d.c.}^\mathrm{P}  &= -E + \sum_k N_k \ln(k!) + \frac{1}{2}\sum_{r,s} e_{rs} \ln\frac{e_{rs}}{e_{r} e_{s}}, \label{eqn:LP4}
\end{align}
\end{linenomath}
\end{subequations}
where $e_r=\sum_s e_{rs}$ is the total number of links with nodes affiliated with community $r$, and $N_k$ is the number of nodes with the degree $k$. The likelihood functions in Eqs.~(\ref{eqn:LKN1})--(\ref{eqn:LP4}), despite showing a way forward, suffer from a serious drawback. Naively minimising them would result in communities with only one node. It is therefore necessary to either know the number of communities $K$ in advance or somehow estimate the value of $K$.

The methods for inferring the number of communities are diverse~\cite{lee2019review}, but there are generally two dominant directions: (i) SBMs are fitted using various values of the parameter $K$, and the optimal parameter value is determined by some measure or criterion; and (ii) $K$ is not considered an external parameter, but is determined internally by the inference algorithm. The former approaches are called parametric, whereas the latter are called, somewhat anticlimactically, non-parametric. Parametric approaches oftentimes add to the likelihood function a part that acts as a penalty for the increasing number of communities. An example of this is the use of the Bayesian information criterion (BIC)~\cite{fu2009dynamic, xing2010state}, the minimum-description-length (MDL) principle~\cite{peixoto2013parsimonious}, or some other variation in terms of how to estimate likelihood or modify penalty~\cite{yan2016bayesian, wang2017likelihood, hu2020corrected}.

Examples of integrating procedures for determining the number of  communities $K$ into inference algorithms are due to Newman and Reinert~\cite{newman2016estimating}, who obtained a closed-form likelihood expression for the degree-corrected SBM, as well as C\^{o}me and Latouche~\cite{come2015model} who used the exact integrated complete likelihood. Nowadays, however, among the most commonly used versions is the non-parametric microcanonical formulation due to Peixoto~\cite{peixoto2017nonparametric, peixoto2019bayesian}.

\begin{figure}[!t]
\centering\includegraphics[scale=1.0]{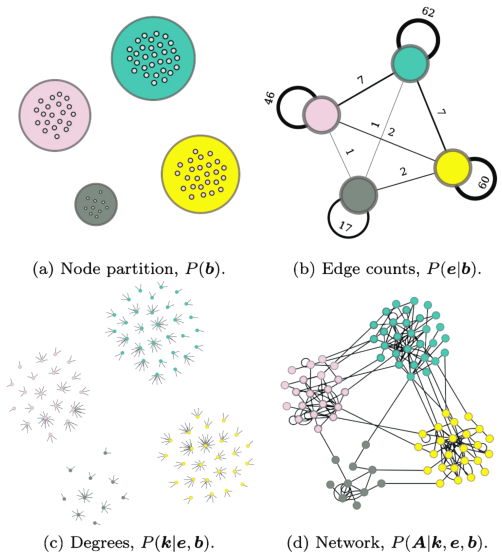}
\caption{Non-parametric microcanonical stochastic block model and the corresponding generative process. Panel (a) illustrates the sampling of node partitions from the distribution $P(\mathbf{b})$. Panel (b) illustrates the sampling of edge counts from the distribution $P(\mathbf{e}|\mathbf{b})$. Panel (c) illustrates the sampling of node degrees from the distribution $P(\mathbf{k}|\mathbf{e},\mathbf{b})$. Nodes are accompanied with semi-links or stubs that are yet to be wired into a network. Finally, panel (d) illustrates the sampling of the network from the distribution $P(\mathbf{A}|\mathbf{k},\mathbf{e},\mathbf{b})$.\newline
Source: Reprinted figure from Ref.~\cite{peixoto2017nonparametric}.}
\label{fig:SBMgeneration}
\end{figure}

\paragraph*{Non-parametric microcanonical SBM} The total joint distribution for data and model parameters in microcanonical form is
\begin{linenomath}
\begin{equation}
P(\mathbf{A},\mathbf{k},\mathbf{e},\mathbf{b}) = P(\mathbf{A}|\mathbf{k},\mathbf{e},\mathbf{b}) P(\mathbf{k}|\mathbf{e},\mathbf{b}) P(\mathbf{e}|\mathbf{b}) P(\mathbf{b}),
\label{eqn:bayes}
\end{equation}
\end{linenomath}
where $P(\mathbf{A}|\mathbf{k},\mathbf{e},\mathbf{b})$ is defined in Eq.~(\ref{eqn:micro3}), whereas $P(\mathbf{k}|\mathbf{e},\mathbf{b})$, $P(\mathbf{e}|\mathbf{b})$, and $P(\mathbf{b})$ are prior distributions (Fig.~\ref{fig:SBMgeneration}). According to Bayes' theorem, the posterior distribution of a network partitioning into communities is
\begin{linenomath}
\begin{equation}
P(\mathbf{b}|\mathbf{A}) = \frac{P(\mathbf{A}|\mathbf{b})P(\mathbf{b})}{P(\mathbf{A})} = \frac{P(\mathbf{A},\mathbf{b})}{P(\mathbf{A})} = \frac{P(\mathbf{A},\mathbf{b})}{\sum_b P(\mathbf{A},\mathbf{b})},
\label{eqn:bayestheorem}
\end{equation}
\end{linenomath}
where $P(\mathbf{A},\mathbf{b})$ is the marginal joint distribution after integrating out the parameters $\mathbf{k}$ and $\mathbf{e}$
\begin{linenomath}
\begin{equation}
P(\mathbf{A},\mathbf{b}) = \sum_{\mathbf{k}, \mathbf{e}} P(\mathbf{A},\mathbf{k},\mathbf{e},\mathbf{b})= P(\mathbf{A},\mathbf{k'},\mathbf{e'},\mathbf{b}) ,
\label{eqn:bayestheoremB}
\end{equation}
\end{linenomath}
where $\mathbf{k'}=\mathbf{k'}(\mathbf{A},\mathbf{b})$ and $\mathbf{e'}=\mathbf{e'}(\mathbf{A},\mathbf{b})$ are compliant with Eq.~(\ref{eqn:hard}), that is, the hard constraints on node degrees and the number of links. Eqs.~(\ref{eqn:bayes})--(\ref{eqn:bayestheoremB}) offer conceptual guidance as to what we want to achieve through Bayesian statistical inference. Specifically, we want to maximise the probability $P(\mathbf{b}|\mathbf{A})$ of partitioning $\mathbf{b}$ conditional on observing the adjacency matrix $\mathbf{A}$. This turns out to be equivalent to maximising the joint distribution $P(\mathbf{A},\mathbf{b})$ because the denominator in Eq.~(\ref{eqn:bayestheorem}) acts just as a normalisation constant. The joint probability $P(\mathbf{A},\mathbf{b})$, however, is knowable only via Eq.~(\ref{eqn:bayes}), implying that inference is a multi-part procedure. We need to specify all the prior distributions, and then maximise the resulting joint distribution.

The prior distributions are key ingredients of the inference procedure. Because in most cases there is no empirical information about priors, the prior selection is purposely kept uninformative. This prevents introducing bias to the posterior distribution, and allows the data to guide the partitioning of networks into communities. The following relations respectively define the priors for the parameters $\mathbf{b}$ and $\mathbf{e}$ (see Refs.~\cite {peixoto2017nonparametric, peixoto2019bayesian} for details)
\begin{linenomath}
\begin{align}
P(\mathbf{b}) &= P(\mathbf{b}|\mathbf{n})P(\mathbf{n}|K)P(K)=\frac{\prod_r n_r!}{N!}\binom{N-1}{K-1}^{-1}\frac{1}{N},
\label{eqn:prior1}\\
P(\mathbf{e}|\mathbf{b}) &= \left(\binom{K(K+1)/2}{E}\right)^{-1},
\label{eqn:prior2}
\end{align}
\end{linenomath}
where $((\cdot))$ is the multiset binomial coefficient, and $E$ is the total number of links. The number of communities $K$ is in this context called a hyperparameter. The corresponding distribution $P(K)$ is rather fittingly called a hyperprior. The prior distribution for the parameter $\mathbf{k}$ is specified in one of two ways (see Ref.~\cite {peixoto2017nonparametric} for details)
\begin{linenomath}
\begin{subequations}
\begin{align}
P_\mathrm{u}(\mathbf{k}|\mathbf{e},\mathbf{b}) &= \prod_r\left(\binom{n_r}{e_r}\right)^{-1},
\label{eqn:prior3}\\
P_\mathrm{h}(\mathbf{k}|\mathbf{e},\mathbf{b}) &= \prod_r \frac{\prod_k N_k^r!}{n_r!}q(e_r,n_r)^{-1}\,,
\end{align}
\end{subequations}
\end{linenomath}
where $N_k^r$ is the number of degree-$k$ nodes in community $r$, and $q(m,n)=q(m,n-1)+q(m-n,n)$ with the boundary conditions $q(m,1)=1$ for $m>1$ and $q(m,n)=0$ for $m\leq0$ or $n\leq0$. Selecting the uniform prior $P_\mathrm{u}(\mathbf{k}|\mathbf{e},\mathbf{b})$ may cause that most nodes have similar degrees, in which case incorporating the heterogeneous prior $P_h(\mathbf{k}|\mathbf{e},\mathbf{b})$ may help.

\paragraph*{Bayesian equivalence} As already discussed, we fall short of achieving the exact equivalence between canonical and microcanonical forms, but the Bayesian framework offers another perspective on the subject. We start by marginalising the likelihood $P(\mathbf{A}|\mathbf{\Lambda},\boldsymbol{\theta})$ from Eq.~(\ref{eqn:dcsbm2}) as follows~\cite{peixoto2017nonparametric}
\begin{linenomath}
\begin{equation}
P(\mathbf{A}|\mathbf{b}) = \int P(\mathbf{A}|\mathbf{\Lambda},\boldsymbol{\theta})P(\mathbf{\Lambda})P(\boldsymbol{\theta}|\mathbf{b})\dup{\mathbf{\lambda}}\dup{\boldsymbol{\theta}}.
\label{eqn:int}
\end{equation}
\end{linenomath}
We have previously shown that, in the case of $\hat{\theta}_r=1$ in Eq.~(\ref{eqn:seeit}), the values of $\lambda_{rs}$ represent the expected number of nodes between communities $r$ and $s$. If we choose a non-informative prior for $\lambda$ in the form of an exponential distribution whose expectation is $\bar{\lambda}=2E/K(K+1)$
\begin{linenomath}
\begin{equation}
P(\lambda_{rs})=
\begin{cases}
\frac{1}{\bar{\lambda}}e^{-\frac{\lambda_{rs}}{\bar{\lambda}}}, & \text{if}~r\ne s,\\
\frac{1}{2\bar{\lambda}}e^{-\frac{\lambda_{rr}}{2\bar{\lambda}}}, & \text{if}~r=s,
\end{cases}
\end{equation}
\end{linenomath}
and for $\boldsymbol{\theta}$ we choose another non-informative prior defined by
\begin{linenomath}
\begin{equation}
P(\boldsymbol{\theta}|\mathbf{b}) = \prod_r (n_r-1)!\delta(\hat{\theta}_r-1),
\end{equation}
\end{linenomath}
then from Eq.~(\ref{eqn:int}) it follows that~\cite{peixoto2017nonparametric}
\begin{linenomath}
\begin{align}
P(\mathbf{A}|\mathbf{b}) &=  \frac{\prod_{r<s}e_{rs}!\prod_{r}e_{rr}!!}{\prod_{i<j}A_{ij}!\prod_{i}A_{ii}!!} \nonumber\\
&\times \prod_r\frac{(n_r-1)!}{(e_r+n_r-1)!} \prod_i k_i! \nonumber\\
&\times \frac{\bar{\lambda}^E}{(\bar{\lambda}+1)^{E+K(K+1)/2}} \nonumber\\
&= P(\mathbf{A}|\mathbf{k},\mathbf{e},\mathbf{b})P(\mathbf{k}|\mathbf{e},\mathbf{b})P(\mathbf{e}).
\end{align}
\end{linenomath}
The quantity $P(\mathbf{A}|\mathbf{k},\mathbf{e},\mathbf{b})$ is precisely the microcanonical likelihood from Eq.~(\ref{eqn:micro3}), the quantity $P(\mathbf{k}|\mathbf{e},\mathbf{b})$ is precisely the uniform prior from Eq.~(\ref{eqn:prior3}), whereas the quantity
\begin{linenomath}
\begin{equation}
P(\mathbf{e}) = \frac{\bar{\lambda}^E}{(\bar{\lambda}+1)^{E+K(K+1)/2}}
\end{equation}
\end{linenomath}
differs from the microcanonical prior in Eq.~(\ref{eqn:prior2}) only in the sense that the total number of links may fluctuate, but in expectation the canonical and microcanonical forms are equivalent to one another~\cite{peixoto2017nonparametric}.

\paragraph*{Minimum-description-length interpretation} The non-parametric microcanonical SBM can be reinterpreted from an information-theoretical perspective, which offers an intuitive explanation of why this model is robust to overfitting. Namely, we can write $P(\mathbf{A},\mathbf{k},\mathbf{e},\mathbf{b}) = 2^{-\Sigma}$, where by taking the logarithm of both sides, we get a value called the description length of data~\cite{rissanen1978modeling}
\begin{linenomath}
\begin{align}
\Sigma &= -\log_2 P(\mathbf{A},\mathbf{k},\mathbf{e},\mathbf{b}) \nonumber\\
&= -\log_2( P(\mathbf{A}|\mathbf{k},\mathbf{e},\mathbf{b}) P(\mathbf{k},\mathbf{e},\mathbf{b}) ) \nonumber\\
&= \mathcal{S}+\mathcal{L}.
\end{align}
\end{linenomath}
The quantity $\mathcal{S} = -\log_2 P(\mathbf{A}|\mathbf{k},\mathbf{e},\mathbf{b})$ equals the number of bits needed to describe a network when the model parameters are known, and the quantity $\mathcal{L} = -\log_2 P(\mathbf{k},\mathbf{e},\mathbf{b})$ equals the number of bits needed to describe the model. By maximising the joint probability distribution in Eq.~(\ref{eqn:bayes}), a set of parameters is automatically obtained that gives the minimum description length. Moreover, the increasing value of $\mathcal{L}$ acts as a penalty on the number of parameters, thus also limiting the number of communities. Without such a penalty the increasing number of communities tends to decrease the value of $\mathcal{S}$, meaning that ultimately each node would comprise its own community. In sum, the MDL interpretation shows that the non-parametric microcanonical SBM is a formal implementation of Occam's razor, according to which the simplest model with a sufficient significance level is to be preferred.

\paragraph*{Resolution limit and the nested SBM} Community detection by means of SBMs comes with a resolution limit, meaning that communities below the resolution minimum will not be assigned sufficient statistical significance. Instead, such communities will be merged into larger ones. A well-known example is a network of 64 10-node cliques (i.e., complete subgraphs) in which the cliques are mutually disconnected (i.e., there's no links between any two cliques). Fitting the microcanonical SBM yields 32 communities, each comprising two cliques~\cite{peixoto2019bayesian}. The model thus suffers from undefitting that manifests as an inability to detect communities below the resolution limit, which in turn scales with $\mathcal{O}(\sqrt{N})$~\cite{peixoto2013parsimonious}.

\begin{figure}[!t]
\centering\includegraphics[scale=1.0]{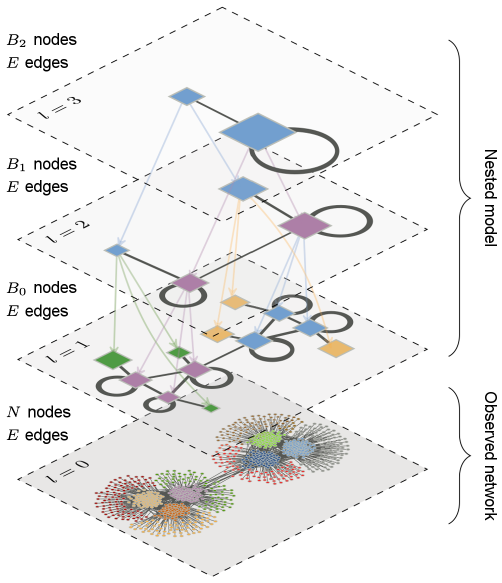}
\caption{Nested stochastic block model in action. Stacked on top of the observed network are the corresponding three levels of multigraph representation ($l=1$ to $l=3$). To break the resolution limit, a higher level serves as a prior for the next lower level.\newline
Source: Reprinted figure from Ref.~\cite{peixoto2014hierarchical} under the Creative Commons Attribution 3.0 Unported (CC BY 3.0).}
\label{fig:SBMhierarchical}
\end{figure}

Peixoto~\cite{peixoto2014hierarchical} offered a solution to the resolution-limit problem in the form of a nested SBM. This type of SBM is based on a simple idea that the communities and the number of links between them, as determined by fitting an SBM, form a new multigraph (i.e., a network in which any two nodes may be linked multiple times, including closed loops). According to this idea, the communities represent the nodes of the new multigraph, the number of links between any two communities represents link multiplicity, and the number of links within a community represents loop multiplicity (Fig.~\ref{fig:SBMhierarchical}). It is then possible to fit the SBM to the new multigraph again, producing yet another multigraph. By repeating the procedure recursively, we get a smaller and smaller number of communities until we finally reach the multigraph with one community. The reason why this method improves the resolution limit is that a higher-level multigraph serves as the information prior for the next lower level. The method generalises the `flat' model described above, and is applicable to large networks, be they assortative or disassortative.

\paragraph*{Inference algorithms} Algorithms that effectively infer community affiliations often come from the Monte-Carlo class of methods used in statistical physics. Although for different SBM variants exact expressions for the posterior probability can be derived in accordance with Eq.~(\ref{eqn:bayestheorem}), up to the normalisation constant, the distributions are in most cases quite complicated. Therefore, Markov-chain Monte-Carlo (MCMC) methods for sampling from complicated distributions are proving to be an efficient and easy-to-implement inferential tool.

Examples of MCMC algorithms are Metropolis or Metropoplis-Hasting algorithms~\cite{hastings1970monte, chib1995understanding}. In the Metropolis algorithm, to sample from our target distribution $P(\mathbf{b}|\mathbf{A})$, we initialise the algorithm at an arbitrary position $\mathbf{b}=\mathbf{b}_0$. Next, a candidate replacement $\mathbf{b}'$ is sampled from a symmetric, but otherwise arbitrary, distribution; if this distribution is normal and centralised around $\mathbf{b}$, choosing the candidate replacement $\mathbf{b}'$ amounts to making a random-walk step. We then calculate the ratio $f(\mathbf{b}')/f(\mathbf{b})$, where the function $f$ needs only to be proportional to our target distribution $P(\mathbf{b}|\mathbf{A})$. This proportionality requirement is, in fact, one of the main strengths of the Metropolis algorithm because calculating the normalisation constant for the target distribution is often a non-trivial task. If said ratio is greater-or-equal than unity, i.e., $f(\mathbf{b}')/f(\mathbf{b})\geq 1$, then the replacement candidate $\mathbf{b}'$ comes from a region that is more probable under $P(\mathbf{b}|\mathbf{A})$ than the region around $\mathbf{b}$; the candidate replacement becomes the new current sample. If, by contrast, the ratio is lower than unity, i.e., $f(\mathbf{b}')/f(\mathbf{b})<1$, then the replacement candidate can still be accepted as the new current sample, but the probability of doing so decreases as the ratio gets closer to zero. Occasional acceptance of candidate replacements that are less probable under $P(\mathbf{b}|\mathbf{A})$ is necessary because `less probable' is still possible. Once a sufficiently large sample is obtained, the algorithm can be stopped, but of note is that there is no natural termination criterion. The Metropolis-Hastings algorithm relaxes the condition that candidate replacements need to be drawn from a symmetric distribution. Beside MCMC algorithms, frequently used are variational~\cite{gopalan2012scalable, kim2013efficient, hayashi2016tractable, matias2018semiparametric} and greedy methods~\cite{karrer2011stochastic, come2015model, yan2016bayesian}, which have been extensively reviewed in Ref.~\cite{lee2019review}.

\subsection{Future outlook}

Community detection is a fast-paced research domain whose explosive development over the past 20 years or so stems from mathematical foundations that were laid decades ago. We started our overview of research on community detection with four main community-detection contexts, meanwhile emphasising the importance of generative models for consistent statistical inference of communities from network data. We singled out SBMs as a poster child for tremendous progress that had been made. Here, we recognise that as the SBM-based theory has been maturing, new methodological advances started taking roots and shaping the field's future~\cite{jin2021survey}.

Much effort has been put into reformulating the problem of community detection to conform to the format of some machine-learning technique. Examples of this are community detection using topic models or matrix factorisation~\cite{jin2021survey}. Topic models originate in machine learning and natural-language processing, and are based on an idea that documents, as word collections, refer to a limited number of topics. A topic in this approach is a cluster of similar words. The main task of a specific topic model is to map a set of documents into word-use statistics, and from there make two inferences. One inference is the collection of topics that the documents collectively cover. The other inference is how much a certain topic is represented in a given document. Community detection based on topic modelling thus implies packaging nodes and links as words and documents for the model to process. The returned topics are then interpreted as communities. In a similar fashion, matrix factorisation is a class of collaborative-filtering algorithms devised for recommender systems with the goal to learn low-dimensional representations of users and items that can be used to predict how users rate items (e.g., how subscribers rate shows on a streaming service). Community detection based on matrix factorisation treats the adjacency matrix as ratings, while low-dimensional representations to be learned are those of links outgoing from and incoming into a fixed number of communities. In practice this means that an $N\times N$ adjacency matrix is decomposed into a product of an $N\times K$ matrix of outgoing links and $K\times N$ matrix of incoming links. The latter two matrices respectively give probabilities that the $i$th node generates an outgoing link from community $r$ and that the $j$th node receives an incoming link into community $r$, which ultimately determines the community structure.

With the recent rise in popularity of neural networks and deep learning (see Section~\ref{SS:deeplearning}), it is unsurprising that the problem of community detection has also been cast in the form suitable for deep neural networks~\cite{jin2021survey}. The main idea in this context is to learn node representation, that is, extract important features that set nodes apart, including their community affiliations. In deep neural networks, features are encoded by hidden network layers, but the whole business of feature extraction is perhaps easier to present by referring to a more `manual' approach in which network nodes are embedded into a low-dimensional vector space. Such an embedding is achieved by first defining a node-similarity measure on networks. A well-known example is unbiased, fixed-length random walks that measure similarity in terms of the probability of visiting the node $v$ during a random walk starting from the node $u$~\cite{perozzi2014deepwalk}. Second, a map (i.e., an embedding) is defined that associates network nodes with vectors in the vector space. Lastly, the parameters of this embedding are optimised in such a way that the similarity measure on networks is well approximated by (some function of) the scalar product in the vector space. This exact problem has, in fact, been rigorously treated by mathematicians with the diffusion distance being the similarity measure defined on networks, diffusion maps serving as the embedding, and the Euclidean distance approximating the diffusion distance in the low-dimensional vector space~\cite{coifman2005geometric}. In practice, however, measuring node similarity is expensive and therefore done locally. The embedding is optimised using only measured (as opposed to all) node similarities. Once this is achieved, community detection takes place by clustering vectors, corresponding to network nodes, in the low-dimensional vector space. Ref.~\cite{grover2016node2vec} demonstrates the approach in action.

Among the briefly described approaches to community detection inspired by machine learning, deep neural networks in particular shed the connection to generative processes and statistical inference. This, however, is unlikely to slow down further proliferation of such methods. If anything, the most practical methods in terms of the ability to handle large network sizes (e.g., billions of nodes) and various network types (e.g., multilayer, dynamic, and incomplete) are likely to prosper in the future. We nonetheless expect community-detection methods founded on generative processes and statistical inference to continue going strong due to fundamental advantages and unrivalled rigour.

\FloatBarrier

\section{Human-machine networks}
\label{S:HMN}

Human society is currently experiencing the impact of a digital transition by which data about human behaviour has evolved from a limited and unused resource to a manifold of permanently growing real-time data streams called \textit{big data}. Today, big data is being pervasively generated, collected, analysed, and utilised within various smart systems to support enjoyable and comfortable living and working conditions. We are, in fact, witnessing a transition from \textit{information} to \textit{knowledge society} that has prompted the emergence of technologies whose potential to unravel both individual and collective behavioural phenomena is unprecedented. Examples of such technologies include:
\begin{itemize}
\item Google Knowledge Graph augmenting web navigation of Internet users~\cite{singhal2012introducing},
\item Facebook Social Graph revealing users' personal relations~\cite{ugander2011anatomy},
\item LinkedIn Economic Graph digitally mapping every member of the workforce~\cite{madey2014linkedin},
\item Unacast Real World Graph explaining how people move around the planet~\cite{walle2020introducing}, and
\item Pinterest Taste Graph visually exploring what people like and what inspires them~\cite{milinovich2017introducing}.
\end{itemize}

Within the ongoing digital transformation, the ambition of ensuring smooth progress along the data-information-knowledge-wisdom hierarchy~\cite{sharma2009dikw} undoubtedly stays one of the major challenges given the ever-growing size, diversity, and frequency of generated data. The vehicle for navigating such a knowledge hierarchy is \textit{Data Science}, an interdisciplinary field dealing with various methodologies and technologies (i.e., algorithms and systems) to automatically derive knowledge from data. Understanding and augmenting human intelligence and human-decision making processes is the necessary next step to harness data science in the pursuit of \textit{actionable knowledge} for improving human lives.

\begin{figure}[!t]
\makebox[\textwidth][c]{\includegraphics[scale=1.0]{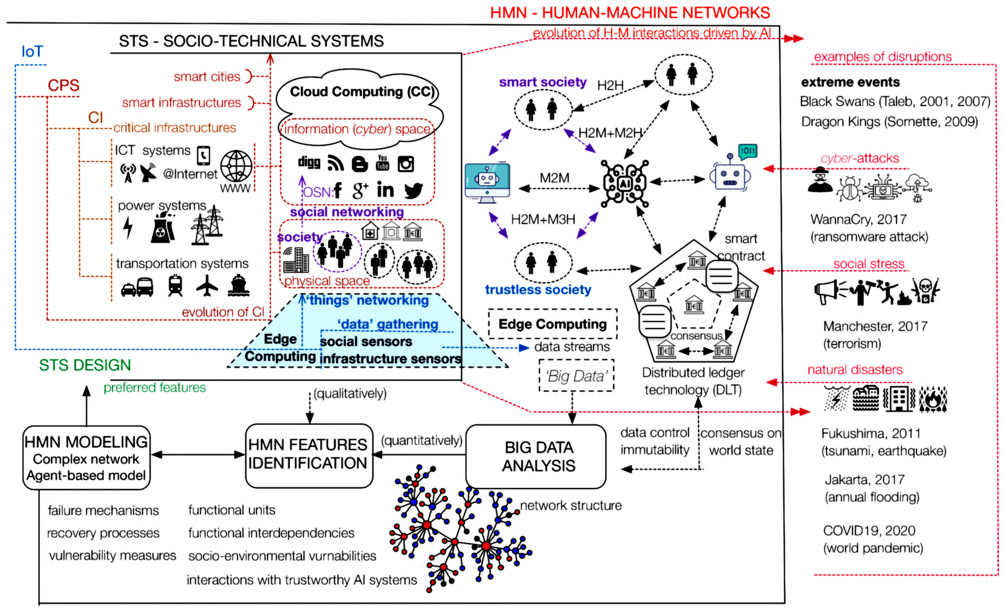}}
\caption{Conceptual illustration of the evolution of socio-technical systems (STS) and human-machine networks (HMN) driven by emerging technologies. Although the interconnection of structures and functionalities ensures scalable development, there is also the risk of a single local failure in one subsystem provoking a cascade of failures throughout other subsystems. For example, a failure of the power grid can cause the failure of information and communication technology systems, upon which financial, healthcare, or security services may depend down the line. Further vulnerabilities include the spread of disinformation or cyber-attacks. Ensuring not only scalability, but also robustness and security of human-machine networks therefore constitutes a critical task within the framework of social-technical systems.}
\label{fig:HMN}
\end{figure}

Due to their influence on human lives, the evolution of data-science methodologies and technologies is intertwined with social and organisational structures in what is known as \textit{socio-technical systems}. From this viewpoint, novel technologies emerge to satisfy societal needs, all the while society adapts to better accommodate those technologies (Fig.~\ref{fig:HMN}). For example, Internet as a global packet data network and the World Wide Web as an information system provide critical services for modern-day society, yet neither were originally designed with such broad purposes in mind.

Some of the emerging and evolving technologies within the framework of socio-technical systems are Distributed Ledger Technology (e.g., blockchain)~\cite{ballandies2018decrypting, wang2019survey}, Cyber-Physical Systems~\cite{zhou2019cyber}, Internet of Things~\cite{pticek2016beyond}, Cloud, and post-Cloud paradigms Fog, Edge, and Dew Computing~\cite{skala2015scalable}. The post-Cloud paradigms in particular strive to relocate computing resources closer to end users to mitigate cloud-related issues of highly centralised computation~\cite{guberovic2021dew}. All these technologies, together with advances in the research of Artificial Intelligence (AI)~\cite{wang2020convergence, deng2020edge, lim2020federated} already influence everyday lives~\cite{wang2020deep} to the point that humans and machines are entangled into elaborate \textit{human-machine networks}~\cite{tsvetkova2017understanding}. Because of their increasing complexity and relevance in modern societies, human-machine networks are among the most challenging, and yet, most important environments to study human and machine co-behaviour~\cite{rahwan2019machine}.

The need to understand human behaviour in conjunction with big data collected from a variety of human-machine networks has given rise to a new discipline called \textit{Computational Social Science}~\cite{lazer2009social, holme2015mechanistic}. This discipline encompasses modern trends in social-physics research~\cite{schweitzer2018sociophysics, bhattacharya2019social} based on the joint use of computational big-data analyses on the one hand, and models from physical sciences on the other hand. Specifically, methods are borrowed from behavioural economics and social psychology, network science, data science, and machine learning, as well as game theory and the theory of critical phenomena. Discoveries are thus possible on three distinct levels:
\begin{enumerate}
\item \textit{Data analyses} generate insights directly from collected data,
\item \textit{Modelling} attempts to capture plausible governing mechanisms and processes, and
\item \textit{Simulations} yield system-wide or component-wise predictions of behaviour within the human-machine network of interest.
\end{enumerate}
The goal is to develop expressive-yet-simple models that can be calibrated and validated against real-word data as opposed to using phenomenological models that are limited to generic insights into governing mechanisms and processes. This type of data-driven modelling overcomes the drawbacks of black-box machine-learning algorithms in which the underlying physical principles of social systems remain entirely neglected.

Data-driven modelling as described above opens the door to deep insights into human behavioural patterns and better decisions in response to critical social problems. Areas of potential betterment include monitoring socio-economic deprivation of individuals and countries~\cite{blumenstock2015predicting},
increasing public wealth and health~\cite{zhang2016support, jebb2018happiness},
controlling safety and crime~\cite{bogomolov2015moves},
mapping epidemics~\cite{ginsberg2009detecting, wesolowski2012quantifying}, managing natural disasters~\cite{wilson2016rapid, ofli2016combining},
and securing social inclusion~\cite{hobbs2016online, podobnik2017predicting}. Multiple blooming research directions have caused much excitement, culminating in the idea of \textit{social-good algorithms}~\cite{lepri2017tyranny, abebe2018mechanism, shi2020artificial, vinuesa2020role} that should guide all aspects of sustainable development~\cite{vinuesa2020role}, decision-making, and resource optimisation~\cite{abebe2018mechanism, shi2020artificial}.

Due to their supposed influence over so many aspects of modern-day life, social-good algorithms are undoubtedly powerful. Yet, with great power comes great responsibility. Concerns have already been raised around a range of social, ethical, and legal issues, including privacy and security~\cite{demontjoye2013unique, demontjoye2015unique}, transparency and accountability~\cite{pasquale2015black}, and discrimination and bias~\cite{ruths2014social}. The fact that people remain largely unaware of how algorithms utilise their data and affect their lives has been encapsulated in the expression \textit{black-box society}~\cite{pasquale2015black}. Especially today's Internet-enabled human-machine networks such as online social networks, search engines, or other cloud-based platforms are highly centralised, exposing users' personal data to potential commercial or political misuses. Adding to the mix AI systems built upon complex deep-learning models, infamous for their lack of transparency, accentuates the need to tread carefully in the near future.

A key challenge for researchers and policymakers, in order to avoid the pitfalls of a black-box society, is to ensure maximum transparency in ever-growing interactions between humans and machines. Attempts to rise to this challenge have culminated in the framework of \textit{Trustworthy Artificial Intelligence}, at the core of which lies recognition that AI systems must be more easily interpretable in general, and explainable to various user groups in particular~\cite{xie2020explainable}. Further requirements in this context are alignment with fundamental human rights and legal practices, as well as service to societal common good~\cite{floridi2019establishing, theodorou2020towards}. The European Union (EU) through the European Commission's High-Level Expert Group on AI has taken an early lead in setting the pathway towards trustworthy AI; for example, the document ``Ethics Guidelines for Trustworthy AI''~\cite{alapietila2019ethics}, published in April 2019, provides concrete guidance on how to operationalise the above-stated requirements in future human-machine networks.

Trustworthy AI, with its three key characteristics (lawful, ethical, and robust) and seven key requirements (human agency and oversight, technical robustness and safety, privacy and data governance, transparency, diversity, non-discrimination and fairness, environmental and societal well-being and accountability) clearly sets the research-and-development direction for emerging AI-driven human-machine networks. Especially the focus on human-understandable algorithms that manipulate various types of unstructured, semi-structured, or structured data is sowing the seeds of an indispensable tool in the toolbox~\cite{zdeborova2017machine} of computational branches of social physics. These promising developments and a growing interest in AI research notwithstanding, existing studies about collective human~\cite{bhattacharya2019social} and machine behaviour~\cite{rahwan2019machine}, and their symbiotic interdependence~\cite{tsvetkova2017understanding, shirado2017locally, crandall2018cooperating}, are highly fragmented. The fragments, comprising Human-Imitative AI, Intelligence Augmentation, and Intelligent Infrastructure, are complementary to one another and should fuse together in future studies of large-scale human behaviour.

The aim of this chapter is to help aspiring social physicists (i) to navigate, at times chaotic, advances in the domain of AI-driven human-machine networks, and (ii) to identify research directions where future breakthroughs may lie. To this end, the following sections start with an up-to-date review of studies at the interface between human~\cite{bhattacharya2019social} and machine~\cite{rahwan2019machine} behaviour. Thereafter, the methodological fundamentals of AI are laid out in a form condensed for easy understanding. After a brief mention of exemplary uses in social-good and sustainable-development contexts, the AI methodology is exemplified in detail via the use of AI agents in game theory, and especially for the purpose of promoting cooperation. The chapter concludes with an outlook for the future.

\subsection{Literature walkthrough}

The use of machine learning in general, and deep learning in particular, to understand large-scale human behavioural phenomena has traditionally been scattered among research communities. Human and machine behaviour have thus been researched independently for the most part. Only recently an interface between these research topics has started to emerge. To gradually zero in on this interface, we categorise the literature into (i) general AI, machine-learning, or deep-learning techniques, (ii) AI, machine-learning, or deep-learning techniques to address societal challenges, (iii) AI, machine-learning, or deep-learning techniques in social physics, and (iv) collective human-machine behaviour.

There exists a large number of machine-learning and deep-learning surveys tailored to the needs of both specific~\cite{wang2020convergence, lim2020federated, mohammadi2018deep, zhang2019deep, raghu2020survey, mehta2019high} or general audiences~\cite{bengio2009learning, bengio2013representation, deng2014deep, lecun2015deep, schmidhuber2015deep, goodfellow2016deep, liu2017survey, fan2019selective}. Prioritising the latter, Ref.~\cite{lecun2015deep} presents an overview of most popular models and provides a long-term outlook for the field. Ref.~\cite{schmidhuber2015deep} offers a comprehensive historical overview of the relevant work that deep learning builds on. Ref.~\cite{deng2014deep} is notable for presenting deep-learning applications to a variety of information processing tasks. The topic of AI for social good, aiming at advancing and employing AI, machine learning, or deep learning to address societal challenges, is covered in Ref.~\cite{shi2020artificial}. Considering the ongoing Covid-19 pandemic, Ref.~\cite{bullock2020mapping} overviews recent studies that utilise machine learning to tackle aspects of the pandemic at different scales, including molecular, clinical, and societal. Yet other recent studies have investigated economic~\cite{barua2020understanding, baker2020covid}, social~\cite{nicola2020socio}, as well as psychological and mental impacts~\cite{zhang2020monitoring} of the drastic life changes brought about by the pandemic. Lastly, Ref.~\cite{zhang2019deep} surveys recent developments in deep learning for recommender systems, which are currently one of the most established AI, machine-learning, or deep-learning applications within human-machine networks, having an important role in many online services and mobile apps.

Depending on whether machine learning is conducted with or without labelled input-output example pairs, we respectively distinguish between supervised and unsupervised learning. An intermediate approach, called semi-supervised learning, is useful in applications when unlabelled data is readily available or easy to acquire, while labelled data is often expensive or otherwise difficult to collect~\cite{oliver2018realistic}. Ref.~\cite{ouali2020overview} is a comprehensive overview of recent advances in the domain of semi-supervised deep-learning techniques. Meanwhile, in the domain of unsupervised deep learning, progress has materialised in the form of generative models such as Variational Autoencoders and Generative Adversarial Networks. The use of the former in deep-learning contexts is covered in Ref.~\cite{kingma2019introduction}, while the latter, with applications, is covered in Ref.~\cite{gui2020review}.

\textit{Deep reinforcement learning} is a core AI research direction aimed at solving complex sequential decision-making tasks with potentially wide-ranging applications such as robotics, smart infrastructure, healthcare, finance, and others. Ref.~\cite{franccois2018introduction} offers an introduction to deep reinforcement-learning techniques, models, and applications, while Refs.~\cite{li2017deep, arulkumaran2017deep} represent more comprehensive guides into the field. Furthermore, there exist numerous resources that allow for a more hands-on approach:
\begin{itemize}
\item ``Spinning Up in Deep RL'' is a practical introduction~\cite{achiam2018spinning},
\item OpenAI Gym is a collection of benchmark problems (i.e., environments) for comparing deep reinforcement-learning algorithms~\cite{brockman2016openai},
\item OpenAI Baselines is a set of baseline implementations of deep reinforcement-learning algorithms~\cite{dhariwal2017openai}, and
\item rlpyt is an open-source repository of modular and parallelised implementations of various deep reinforcement-learning algorithms~\cite{stooke2019rlpyt}.
\end{itemize}

More recently, research on deep reinforcement learning has taken a turn from single-agent to multi-agent scenarios~\cite{hernandez2019survey, hernandez2020very}. Four topics of interest have crystallised in this context: (i) \textit{the analysis of emergent behaviours} pertains to evaluating single-agent deep reinforcement-learning algorithms in multi-agent scenarios, for example, cooperative, competitive, and mixed; (ii) \textit{learning communication} pertains to agents learning both through actions and messages; (iii) \textit{learning cooperation} pertains to agents learning to cooperate using only actions and (local) observations; and (iv) \textit{agents modelling agents} pertains to agents reasoning about other agents to fulfil a task, for example, cooperative or competitive. Cooperative tasks between actors in human-machine networks are of particular interest to the social-physics agenda and will, therefore, be discussed in more detail later.

Turning to the specifics of the AI, machine-learning, or deep-learning use in science, Ref.~\cite{raghu2020survey} overviews the techniques for applying deep-learning models in conjunction with limited data (self-supervision, semi-supervised learning, and data augmentation), as well as the techniques for interpretability and representation analyses. Ref.~\cite{roscher2020explainable} discusses the natural-science applications of explainable machine learning via three core concepts: \textit{transparency}, \textit{interpretability}, and \textit{explainability}. Finally, Ref.~\cite{xie2020explainable} is a guide into the explainable deep learning aimed at researchers just entering the field.

Alongside many other branches of science, physics has not been immune to adopting the machine-learning methodology~\cite{zdeborova2017machine, iten2020discovering}. Statistical physics has, in fact, inspired the exploration and development of machine-learning models~\cite{zdeborova2020understanding, udrescu2020ai, wu2019toward, choudhary2019physics, koch2018mutual}. A recent comprehensive review introduces the key concepts and tools of machine learning in a physicist-friendly manner accompanied with a set of Python Jupyter notebooks that demonstrate the application of modern machine-learning and deep-learning packages on physics-inspired datasets~\cite{mehta2019high}. Refs.~\cite{carleo2019machine, bahri2020statistical} cover, from a theoretical perspective, the intersection between the foundational machine-learning and deep-learning concepts and statistical mechanics. The methods from statistical mechanics have furthermore begun to provide conceptual insights into deep-learning models regarding the model expressivity~\cite{mhaskar2016learning, lin2017does, lin2018generalization, decelle2019learning}, the shape of the model loss landscape~\cite{pennington2017geometry}, model training and information propagation dynamics~\cite{saxe2019information, goldfeld2020information, piran2020dual}, model generalisation capabilities~\cite{achille2019information, hafez2019compressed}, and the model ability to `imagine', that is, build deep generative models of data~\cite{kingma2019introduction, dangelo2020learning}.

To conclude this literature walkthrough, there is an immensely rich body of literature on the AI, machine-learning, and deep-learning techniques, and the contribution of physicists to this richness has been non-negligible to say the least. But where should an aspiring social physicist look for potential breakthroughs? Ref.~\cite{rahwan2019machine} proposes a new field of scientific study called \textit{machine behaviour} to better understand how AI agents might affect society, culture, the economy, and politics. The four primary motivations for the study of machine behaviour are the ever-increasing ubiquity of AI algorithms in human daily activities, but also, their complexity, opacity, and a lack of explainability. This `black-box' nature of AI algorithms poses substantial challenges to predicting the effects of such algorithms, whether positive or negative, on humanity and society. There are three scales at which to study machine behaviour: individual machines, machine networks, and human-machine networks. Ref.~\cite{tsvetkova2017understanding} surveys the state-of-the-art developments on the third scale, identifying eight different types of human-machine networks depending on structure and interactions. These eight types are public-resource computing, crowdsourcing, web-search engines, crowdsensing, online markets, social media, multiplayer online games and virtual worlds, and mass collaborations. Nowadays, however, the omnipresence and usability of human-machine networks is causing novel trends to emerge by which the limits between the eight listed types are beginning to blur, while hybrid types keep cropping up. This state of affairs suggests a long and winding road, and thereby plenty of opportunities, towards the ultimate goal of the aforementioned social-good algorithms.

\subsection{Fundamentals of artificial intelligence}
\label{SS:AIfundamentals}

Here, we introduce some of the most fundamental terms and concepts of AI, machine learning, and deep learning. Readers who wish a first-hand experience with the methods and techniques that arise from these concepts may want to consult Ref.~\cite{boucher2020montrealai} as a form of a getting-started tutorial.

Historically, the term `artificial intelligence' was introduced in the late 1950s to refer to the aim of making intelligent machines that have the high-level cognitive capability to act, reason, think, and learn like humans, that is, to mimic human intelligence~\cite{russell2002artificial, goodfellow2016deep, jordan2019artificial}. The evolution towards this aim of \textit{human-imitative AI} further led to the emergence of AI systems that \textit{augment human intelligence}, as well as those that constitute \textit{intelligent infrastructure} in order to make living and working environments safer and more supportive of human needs~\cite{jordan2019artificial}. Distilled to these three complementary research activities, that is, human-imitative AI, intelligence augmentation, and intelligent infrastructure, the research on AI comprises a very broad and diverse set of techniques for building and integrating intelligent agents into software solutions and hardware platforms.

\begin{figure}[!t]
\centering
\includegraphics[scale=1.0]{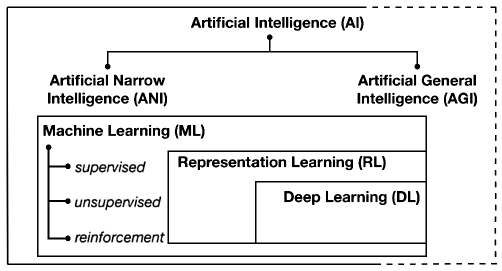}
\caption{A Venn diagram showing overview and composition of AI technology. Deep learning is a part of machine learning, which in turn is a part of AI. Machine learning is commonly divided into supervised learning, unsupervised learning, and reinforcement learning. Deep learning for the most part inherits this division. The major benefit of deep learning over traditional machine learning is the automatic feature extraction that circumvents expensive feature engineering by hand.}
\label{fig:AI_overview}
\end{figure}

Because human intelligence is general, the aim of achieving complete human-imitative AI is often called artificial general intelligence or the `strong' AI~\cite{searle1980minds}. Achieving general intelligence that would encompass all, or most, of human cognitive processes is beyond the reach of current AI research~\cite{fjelland2020general}, meaning that presently only the `weak' AI systems are realisable. These weak AIs are also referred to as artificial narrow intelligence. A further implication is that machines learn to perform good on a specific, well-defined task, but cannot augment humans outside of a limited domain in which the machine learned to operate. Limitations notwithstanding, modern-day artificial narrow intelligence is widely used in various domains, ranging from science to business to health care and more. Interestingly, most systems that classify as AI today are, in fact, based on the machine-learning and deep-learning techniques (Fig.~\ref{fig:AI_overview}).

Machine learning as a subfield of AI enables artificial agents (i.e., machines) to automatically learn from data, make decisions and predictions by themselves, and help in human decision making without being explicitly programmed with expert knowledge~\cite{shalev2014understanding, mehta2019high}. This ability of AI systems to automatically extract new knowledge from data is an extension of knowledge-based approach to AI by which knowledge about the world is to be hard-coded using formal languages, while machine reasoning should follow logical inference rules on statements formulated in such languages. In practice, a successful machine-learning algorithm recognises important features in a `training' dataset in order to make inductive inferences or predictions about data samples unseen during training. The machine-learning algorithm must thus `generalise' beyond data in the training dataset; the goal is not to minimise an evaluation objective on the training dataset, but rather on new, previously unseen samples.

\begin{figure}[!t]
\centering
\includegraphics[scale=1.0]{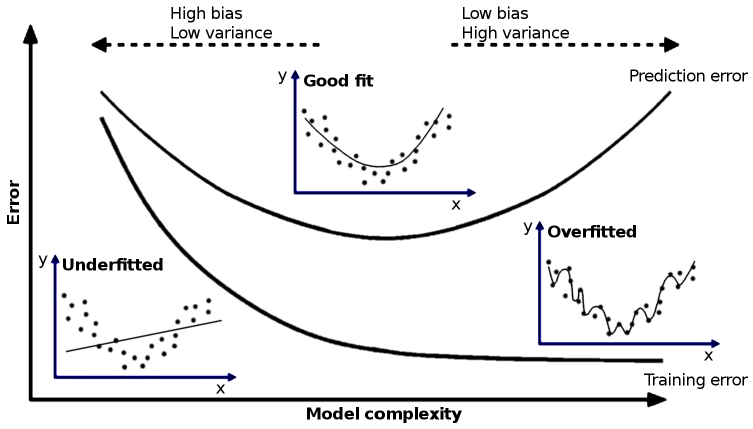}
\caption{Bias-variance tradeoff or how to strike a balance between the hypothesis-set complexity and the dataset size. As the model becomes more and more complex, the training error always decreases. The prediction error, however, decreases only up to a point, and then starts to increase again. The inability of an overly simple hypothesis set to represent the knowledge contained in a dataset creates a bias in predictions, whereas the ability of an overly complex hypothesis set to fit any data perfectly represents noise (i.e., a specific realisation thereof) more than knowledge. In this case, predictions have a large variance.}
\label{fig:bias_variance_tradeoff}
\end{figure}

In addition to a training dataset, every machine-learning algorithm needs a hypothesis set, an error function (also called objective function or cost function), and an optimisation procedure. The algorithm searches the hypothesis set to find the hypothesis that best represents knowledge contained in the dataset, which in practice means relying on the optimisation procedure to minimise an estimate of the prediction (i.e., out-of-sample) error. The key here is to strike a balance between the dataset size and the hypothesis-set complexity (Fig.~\ref{fig:bias_variance_tradeoff}). An overly simple hypothesis set contains no single hypothesis that can represent the knowledge that is contained in the data. An overly complex hypothesis set, by contrast, always contains a hypothesis that fits the data perfectly. In doing so, however, the seemingly perfect hypothesis represents not only knowledge, but also the specific realisation of noise, which could be due to genuine stochasticity or measurement errors, or alternatively, deterministic in origin~\cite{abumostafa2012learning}.

The performance of an inductively learned model degrades for one of the following three reasons~\cite{domingos2012few}. First, the hypothesis set may not contain a suitable representation of reality. In the case of a classification task, for example, a classifier that is outside of the hypothesis set cannot be learned, although the best hypothesis may still yield a reasonable approximation. Second, the error function may have many local optima over the hypothesis set in which case even representable reality may be hard to learn. Finite data, time, and memory enable searches through only a tiny subset of all possibilities. Finally, the choice of the optimisation method also determines the scope of search, where methods that try out more hypotheses reduce bias but increase variance and vice versa. Oftentimes it is advantageous to reduce a learning problem to a well-known optimisation problem by transforming the objective function or introducing additional constraints or relaxations. Ref.~\cite{bennett2006interplay} describes in detail the desirable properties of an optimisation procedure for machine learning. Such properties are good generalisation, scalability to large datasets, good performance in terms of execution times and memory requirements, simple and easy implementation of algorithm, exploitation of problem structure, fast convergence to an approximate solution, robustness and numerical stability for the chosen class of machine learners, and theoretically known convergence and complexity.

Based on feedback available during the learning phase, it is possible to distinguish between three machine-learning contexts: supervised, unsupervised, and reinforcement. In \textit{supervised learning}, the training dataset contains samples that are labelled with an additional `ground truth'. A machine learner attempts to learn a target map from data samples to the ground-truth value. If the ground truth is a discrete class from some finite set of classes, then the learner is facing a \textit{classification} task. Sometimes the classification task is such that a probability distribution over the set of classes is more accessible than the direct classifier. If, by contrast, the ground truth is continuous, then the learner is facing a \textit{regression} task.

Among practical obstacles to supervised learning is that obtaining a complete set of labels for all data samples is often difficult and expensive. Learning then uses both a smaller, labelled subset of the full dataset, as well as a larger, unlabelled subset. This type of learning context is refereed to as \textit{semi-supervised learning}. The two dominant paradigms in semi-supervised learning are transductive learning and inductive learning~\cite{oliver2018realistic, ouali2020overview}. The former does not concern itself with generalisation, but instead attempts to infer the correct labels for the unlabelled subset. This is achieved by assigning labels to unlabelled samples such that ultimately a given optimisation criterion is satisfied across all data, that is, originally labelled and unlabelled subsets taken together. Among the popular techniques are those for learning node representation on networks (i.e., graphs) such as node2vec~\cite{grover2016node2vec} or DeepWalk~\cite{perozzi2014deepwalk}. The goal of node-representation learning is to find a low-dimensional space of features with a scalar product that approximates some measure of node similarity in the original network. A commonly used measure of node similarity is random walks. The learned feature space can be exploited for network community detection, link prediction, and other network-science problems. A more common paradigm, however, is that of inductive learning by which the learner attempts to infer the correct (i.e., generalisable) target map from samples to labels. Irrespective of the employed paradigm, semi-supervised learning is designed to fill the gap between supervised and unsupervised learning, just as the name would suggest.

In \textit{unsupervised learning}, a machine learner is only concerned with extracting insightful patterns from a dataset without relying on any direct feedback. There is no access to supervision signals in the form of discrete or continuous labels. Tasks commonly associated with this machine-learning contexts are partitioning of the dataset into clusters of similar data instances, anomaly detection in the dataset, blind source separation (e.g., picking a single conversation out of many), density estimation of the underlying probability distribution, or learning latent representations of the data by dimensionality reduction techniques. Unsupervised learning is especially useful in exploratory data analysis because of the ability to identify structure in the dataset on its own.

\begin{figure}[!t]
\centering\includegraphics[scale=1.0]{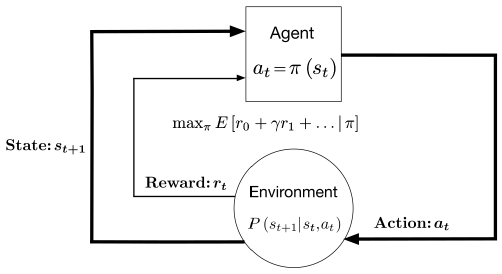}
\caption{Constituents of a Markov decision process in the reinforcement learning setting. The agent needs to make decisions in discrete rounds $t=0,1,\ldots,n$. Decisions are made based on a policy $a_t=\pi(s_t)$, which prescribes the action $a_t$ to be taken given the current state $s_t$. The action triggers a reward signal $r_t$, while the agent transitions into a new state $s_{t+1}$ according to the transition probability $P(s_{t+1}|s_t, a_t)$. The primary objective is for the agent is to find the policy that maximises the total expected reward $E[r_0+\gamma r_1+\ldots|\pi]$ received over the long run, where $\gamma$ is a discounting factor. All relevant information about the past is contained in the current state $s_t$, which encodes aspects of the environment that the agent can sense or influence.}
\label{fig:RL}
\end{figure}

In \textit{reinforcement learning}, a machine learner, often termed an agent, learns how to achieve long-term goals in a complex, uncertain environment. A theoretical underpinning of reinforcement learning is given in the form of Markov decision processes, that is, stochastic decision-making models comprising a set $S$ of agent's states in the environment, a set $A$ of actions available to the agent, and the state-transition probabilities $P(s'|s,a)$ from the state $s\in S$ to a state $s'\in S$ via the action $a\in A$ (Fig.~\ref{fig:RL}). A reward signal $r$ is generated when the agent transitions between states. The objective is to find a policy $\pi(s)$ that prescribes which action should be taken in a given state in order to maximise the cumulative reward obtained over the agent's lifetime. Reinforcement learning comes into play when the environment is so complex or uncertain that the optimal policy is unknowable a priori. The agent learns a model of the environment through execution and simulation, continuously using feedback from past decisions to reinforce good strategies. The learning process is fraught with danger of focusing too much on immediate rewards, thus preventing the discovery of better alternatives, especially those whose rewards are delayed. This is the essence of a trade-off between exploitation and exploration; the agent strives to exploit what is already known in order to accumulate rewards, but at the same time the agent should go exploring to find action selections that pay off more further down the line.

The performance of machine learning approaches discussed heretofore depends heavily on the representation of data. Much effort is therefore put into the feature-engineering process, during which raw data is rendered in a form suitable for modelling. Automating this process is one of the holy grails of machine learning, and will be discussed next.

\subsection{Deep learning}
\label{SS:deeplearning}

\begin{figure}[!t]
\centering\includegraphics[scale=1.0]{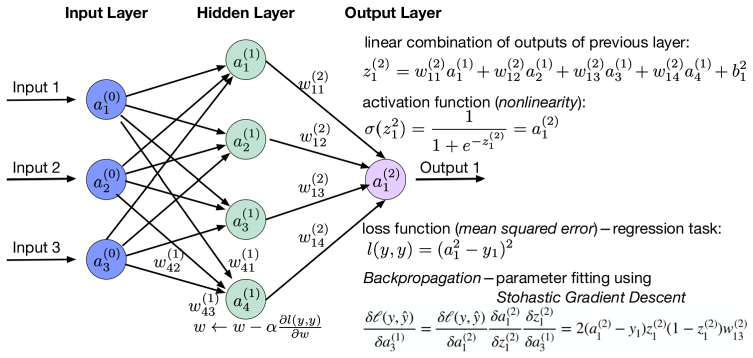}
\caption{Conceptual illustration of a simple artificial neural network. Generally, artificial neural networks are graph structures comprising multiple layers that perform a number of linear and non-linear transformations on input data. Layers between the first (i.e., input) and the last (i.e., output) layer are called hidden layers. Each layer consists of neurons that receive information from preceding layers across weighted edges. Artificial neural networks propagate information forward to calculate the final output, but also backward to perform weight estimation. \textit{Backpropagation} is the key algorithm that makes weight estimation, and thus the training of deep models, computationally tractable and highly efficient; the algorithm amounts to a shrewd application of the chain rule for derivatives. The gradient of the loss function with respect to each weight is calculated one layer at a time, and then weights are updated in the direction in which the loss function decreases the most. The calculation is iterated until the algorithm converges to accurate outputs on the training dataset.}
\label{fig:ANN_illustration2}
\end{figure}

Current efforts to automate the feature-engineering process rely on the idea to generate large numbers of candidate features and then select the best feature subsets with respect to a given learning task. This is done while taking into account that features that look irrelevant in isolation may be relevant in combination with other features~\cite{domingos2012few}. Recently, replacing traditional domain expertise and human engineering to hand craft feature extractors has become possible through the development of deep learning~\cite{goodfellow2016deep}. As a subfield of AI and machine learning, deep learning uses multilayer neural networks (hence the term `deep') to exploit the many layers of non-linear information processing for automatic (supervised or unsupervised) feature extraction, and subsequently for pattern analysis and classification (Fig.~\ref{fig:ANN_illustration2}). In deep neural networks, in particular, activation functions are used at the end of hidden units to introduce non-linear complexities to the model. The most common activation functions are
\begin{linenomath}
\begin{subequations}
\begin{align}
&\mathrm{Rectified~linear~unit}& &f(x)=\max\{0, x\},\\
&\mathrm{Sigmoid}& &f(x)=\sigma(x)=\frac{1}{1+e^{-x}},\\
&\mathrm{Swish}& &f(x)=x \sigma(\beta x),\\
&\mathrm{Hyperbolic~tangent}& &f(x)=\tanh (x)=\frac{e^{x}-e^{-x}}{e^{x}+e^{-x}},\\
&\mathrm{Softmax}& &f(\mathbf{x})_{i}= \operatorname{softmax}(\mathbf{x})_{i} = \frac{\exp \left(\mathbf{x}_{i}\right)}{\sum_{j=1}^{n} \exp \left(\mathbf{x}_{j}\right)},
\end{align}
\end{subequations}
\end{linenomath}
where softmax is often used for normalising the output layer of a neural network.

Deep learning usually requires large datasets to eliminate the need for manual feature extraction. Because a machine learner is fed with raw data to learn its own representations, deep learning is a form of \textit{representation learning}. Learned representations are contained in the multiple layers of the neural network~\cite{esteva2019guide} and encode data by means of a sparse, latent structure with far fewer features than at the beginning. Such elimination of redundant features makes downstream data processing and the final learning task far less intensive. Consequently, most of the recent AI success comes from the utilisation of representation learning with end-to-end trained deep neural-network models in tasks such as image, text, and speech recognition or strategic board and video games. Through enabling the automatic feature engineering, deep learning substantially reduces the reliance on domain-expert knowledge, outperforming in the process traditional methods based on hand-crafted feature engineering, and achieving the performance that equals or supersedes that of humans.

\begin{figure}[!t]
\centering\includegraphics[scale=1.0]{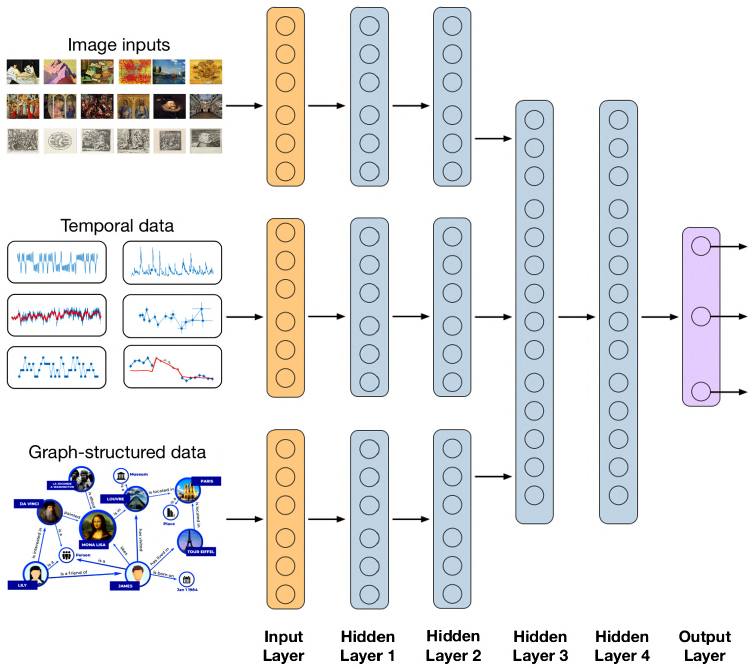}
\caption{Illustrative example of a large-scale deep neural network. The network accepts as inputs a variety of data types---images, time-series, or graph-structured data---and then in its lower-level hidden layers learns useful representations for each data type.
}
\label{fig:DNN_illustration2}
\end{figure}

Variants of deep neural networks are designed to improve performance in specific problem domains (Fig.~\ref{fig:DNN_illustration2}). \textit{Convolutional neural networks} thus excel in computer-vision tasks, while \textit{recurrent neural networks} with special gated mechanisms (such as long short-term memory~\cite{hochreiter1997long} or gated recurrent unit~\cite{chung2014empirical}) resolve issues of a vanishing gradient when learning long-term dependencies. Types of encoder-decoder architectures, combined with an attention mechanism~\cite{vaswani2017attention}, are furthermore naturally suited for modelling time series~\cite{qin2017dual} and sequential data, offering state of the art performance in the space of natural-language processing. Finally, the development of graph neural networks further accelerated progress in developing general AI architectures for handling unstructured and non-Euclidean data~\cite{ward2020practical}.

Despite recent advances in deep learning~\cite{stone2016artificial, turkina2018importance}, many obstacles remain to be overcome. The most common drawback is that popular deep-learning techniques need large amounts of data samples in order to generalise and make predictions on unseen inputs, thus being extremely data inefficient. In supervised learning, data inefficiency translates into the need to label, often manually, thousands of data samples; doing so is time-consuming, cumbersome, expensive, and ultimately unreliable. Likewise reinforcement learning demands access to a large number of training trajectories, which in turn must be obtained via human-machine interactions in the real world that are hard to set up. Attempts to resolve the described issues therefore deserve some attention.

\subsection{Learning to learn}

Attempts to improve the data efficiency of deep learning have shown a couple of promising ways forward. \textit{Transfer learning}~\cite{cetinic2018fine, petangoda2020foliated}, for instance, relies on the idea that knowledge can be transferred from existing to new models. This approach, inspired by how humans as life-long learning entities use own experiences, exploits structural similarities between learning tasks. To exemplify, an image-recognition model may consist of two parts, a feature-extractor part and a classifier part that is made up of fully connected layers and the output layer. If the model is pretrained for a specific task, the feature-extracting part of the model could still be used for a different task, while fully connected layers of the classifier part are replaced and retrained. Such retraining requires much less data because only a fraction of weights of the original model must be estimated.

A further step forward towards increased data efficiency, and more fundamentally artificial general intelligence, is \textit{meta learning}. The goal of this subfield of machine learning is to mimic human learning of new concepts~\cite{battaglia2016interaction, lake2017building, lake2019omniglot}, which often happens quickly and with only a few examples provided. Meta learning is also known as `learning to learn' thanks to efficiently exploiting previous learning experiences when optimising algorithms to generalise to novel tasks~\cite{hospedales2020meta}. Such previous experiences include properties of the learning problem, algorithm properties, or patterns already derived from the data, which in turn make it possible to select, alter, or combine elements of learning algorithms to perform well in a previously unseen context.

A common use of meta learning is in the context of supervised few-shot learning~\cite{wang2020generalizing} which consists of a series of training tasks followed by a series of testing tasks. In a single training task, a dataset \smash{$\mathcal{D}=\left\{\left(\mathbf{x}_i, y_i\right)\right\}$} containing data instances $\mathbf{x}_i$ and their corresponding labels $y_i$, is divided into a support set $\mathcal{S}$ for learning the task and a query set $\mathcal{Q}$ for determining the classification performance, that is, evaluating the error function. The model parameters are updated based on this performance. Because the support set in each training task contains $N$ different classes with $K$ examples per class, this approach is known as $N$-way-$K$-shot classification. Key is that classes differ from one training task to another; to exemplify, let us consider a computer-vision model to distinguish animals of different species. The number of classes may be $N=3$. One training task for such a model may have the support set with $K$ instances of lions, tunas, and turtles, but the next training task may have the support set with $K$ instances of mice, elephants, and seals. The model tries to use the information in the support sets to classify animals in the query sets. Once the training is complete, testing proceeds on tasks with previously unseen classes, say, cats, dogs, and spiders. The point is that the model learns to discriminate data classes in general (i.e., one animal species from another), rather than a particular subset of classes (e.g., cats from dogs).

Meta-learning approaches can be metric-based, model-based, and optimisation based. Metric-based approaches predict the probability of class $y$ conditional on a data instance $\mathbf{x}$ and the support set $\mathcal{S}$, which is achieved by means of a weighted sum of labels $y_i\in\mathcal{S}$, where weights are given by a kernel function that measures the distance between the data instance $\mathbf{x}$ and instances $\mathbf{x}_i\in\mathcal{S}$. Well-known examples in this context include Siamese neural networks~\cite{koch2015siamese}, matching networks~\cite{vinyals2016matching}, prototypical networks~\cite{snell2017prototypical}, and relation networks~\cite{sung2018learning}.

Model-based approaches make no assumptions about the conditional probability of class $y$; rather the idea is to design a model for fast learning that updates its parameters over the course of a few training steps. For example, external memory can be used to expedite the neural-network learning process. In the basic setup, a controller neural network receives inputs and generates outputs while reading from and writing to a memory matrix. It is appropriate to think of the controller as a CPU of a computer and of the memory matrix as RAM with a benefit that the whole system can learn to use memory for various tasks instead of sticking to a fixed set of procedures on data. To be usable in a meta-learning context, the described controller-network plus memory-matrix system needs to be trained such that memory encodes information about novel tasks fast and that any stored representation is promptly accessible. Ref.~\cite{santoro2016meta} prescribes a training technique that forces memory to hold current inputs for later use. This enables successful classification when a novel instance $\mathbf{x}$ from an already-seen class $y$ is presented at an arbitrary point in time.

Optimisation-based approaches recognise that deep learning models use backpropagation of gradients to learn, and yet the gradient-based optimisation has never been designed to work with a few training samples, nor to converge after a few optimisation steps. To overcome these problems, optimisation itself can be treated as a model to be learned~\cite{ravi2016optimization}. In the popular model-agnostic meta-learning~\cite{finn2017model, raghu2019rapid}, what is learned is a shared set of model parameter values for initialising optimisation. This shared set leads to quick specialisation on wide variety of tasks, which is achieved by a training procedure that first optimises one shared set of model parameter values to specialise on a batch of tasks, but then uses the results to find an updated shared set that is better at learning with fewer examples.

The described developments have yielded meta-learning methods capable of achieving human and superhuman performance in simple tasks such as one-shot classification. This is just an initial step though. Hopes are that meta-learning approaches will serve an important role in the future discovery of artificial general intelligence.

\subsection{AI agents for promoting cooperation}

Here, the focus is put on cooperation and AI research aiming to promote cooperation between human and artificial agents in human-machine networks. Besides human-human (H2H) cooperation already discussed in Section~\ref{S:Coop}, we differentiate between machine-machine (M2M) cooperation, human-machine cooperation facilitated by the former (H2M), and human-machine cooperation facilitated by the latter (M2H).

Artificial learners in human-machine networks are expected to take an active part in society, interacting with both humans and other artificial learners in a complex environment of competition and conflict. What may promote cooperation in such an environment is some form of reciprocity, which underpins the demand for learning algorithms that ensure the emergence of reciprocity in human-machine networks. Evolutionary game theory offers a methodological framework for studying the evolution of cooperation in multi-agent systems in which individual agents must choose between selfish interests and common good. Ref.~\cite{liang2019survey} in particular covers game-theoretical methods in the contexts characteristic of human-machine networks such as crowdsourcing, Internet of Things, and blockchain.

\paragraph*{M2M cooperation} In a social-dilemma setting, how can reciprocity, usually observed as a tit-for-tat strategy, emerge in a network of self-interested, reward-maximising reinforcement learners? Ref.~\cite{foerster2017learning} shows that naive and commonly defecting reinforcement learners start to cooperate when they incorporate in their own learning process the awareness of their opponent's learning. Appropriately dubbed learning with opponent-learning awareness or LOLA, the approach leads to the emergence of tit-for-tat and consequent cooperation in the iterated prisoners' dilemma. LOLA agents exemplify the AI design based on the `theory of mind'~\cite{press2012iterated, stewart2012extortion, rabinowitz2018machine, glynatsi2020using}, that is, the ability to know the opponent's behaviour and correspondingly alter own behaviour using only human-like, high-level models of other agents rather than the underlying physical mechanisms. Interestingly, agents with a theory of mind about their opponents have a way of dealing with extortionate zero-determinant strategies by being deliberately hurtful until the extortionist opponent becomes fairer~\cite{press2012iterated}.

Axelrod's influential study on the evolution of cooperation~\cite{axelrod1981evolution}, involving a round-robin tournament in which strategy entries submitted by game theorists competed in a 200-move iterated prisoner's dilemma (and which was won by the tit-for-tat strategy), still inspires research today. In fact, there exists a whole Axelrod library~\cite{knight2016open} of strategies that has been used to organise tournaments similar to Axelrod's original. Ref.~\cite{harper2017reinforcement} conducted such a tournament with a twist of introducing 5\,\% noise, that is, a chance that an action is flipped by a random shock. The purpose was to compare the performance and robustness of 176 available strategies for the iterated prisoner's dilemma. The Axelrod library contains a variety of machine-learning strategies most of which use many rounds of memory, and perform extremely well in tournaments. These strategies encode a variety of other strategies, including the classics such as tit-for-tat, handshake, and grudging. For example, the LookerUp strategy, which does the best in the standard tournament, is a lookup table encoding a set of deterministic responses based on the opponent's first $n_1$ moves, the opponent's last $m_1$ moves, and the LookerUp agent's own last $m_2$ moves. LookerUp is an archetype that can be used to train deterministic memory-$n$ strategies with parameters $n_1=0$ and $m_1=m_2=n$, which for $n=1$ cooperate if the last round was mutual cooperation and defect otherwise (known as Grim or Grudger). Interestingly, in this particular tournament, the tit-for-tat strategy could not win any matches.

Pretrained strategies are generally better than human-designed strategies at maximising payoff against a diverse set of opponents. Furthermore, strategies trained using reinforcement learning and evolutionary algorithms with the objective of maximising the payoff difference (rather than own total payoff) resemble zero-determinant strategies, which are generally cooperative and do not defect first, although their performance declines in the presence of noise. Good performance in both standard and noisy tournaments is exhibited by single-layer neural networks albeit their downside is utilising handcrafted features based on the history of play. One of the best-performing strategies in terms of the overall average score is the Desired Belief Strategy~\cite{au2006accident}, which actively analyses the opponent and responds depending on whether the opponent's action is perceived as noise or a genuine behavioural change. Ultimately, an inescapable conclusions is that reinforcement learning is an effective means to construct strong strategies for various iterated social-dilemma situations~\cite{foerster2017learning, harper2017reinforcement, leibo2017multi, peysakhovich2018towards}.

Newer studies go beyond pairwise interactions in the iterated prisoner's dilemma to examine whether multiple agents cooperate effectively as a team against another team of agents. This requires learning to cooperate with teammates through communication while competing with the opposing team~\cite{lowe2017multi, park2019multi, bachrach2020negotiating}. Ref.~\cite{lowe2017multi} proposed an approach called multi-agent deep deterministic policy gradients (MADDPG) that performs well in a number of mixed competitive-cooperative environments. MADDPG is an extension of actor-critic algorithms in reinforcement learning~\cite{degris2012off, haarnoja2018soft}. These algorithms fuse the strengths of actor-only and critic-only methods~\cite{konda2000actor}. The former methods focus on a parameterised family of policies such that the performance gradient is estimated by simulation, upon which a parameter update is made in a direction of improvement. Among the drawbacks of actor-only methods is that new gradient estimates are independent of past estimates, precluding accumulation and consolidation of previously learned knowledge. Critic-only methods, by contrast, try to estimate the value function (i.e., the total expected reward $E[r_0+\gamma r_1+\ldots|\pi]$) in order to infer a near-optimal policy from there. A drawback is that there are no guarantees whether the inferred policy will indeed be near-optimal. In MADDPG, the actor is used to select actions, while a central critic evaluates those actions by observing the joint state and actions of all agents. In this sense, MADDPG follows the centralised learning with decentralised execution paradigm~\cite{kraemer2016multi, oroojlooyjadid2019review, zhang2019multi}, which assumes unrestricted communication bandwidth during training, as well as the central controller's ability to receive and process all agents' information. To relax these assumptions, Flexible Fully-decentralised Approximate Actor-critic (F2A2) algorithm~\cite{li2020f2a2} was proposed as a variant of multi-agent reinforcement learning based on decentralised training with decentralised execution. A strong suit of the F2A2 algorithm is its ability to handle competitive-cooperative partially observable stochastic games~\cite{hansen2004dynamic}. Here, the term `partially observable' refers to situations in which agents only know the probability of making an observation conditional on the current state as opposed to directly determining the current state. `Stochastic games' furthermore refer to situations in which multiple decision makers interact with one another, while the environment changes in response to decisions made.

\paragraph*{Hybrid H2M and M2H cooperation} Beside utilising the AI techniques, especially multi-agent reinforcement learning, to learn cooperativeness among machine learners, a growing number of studies investigate hybrid human-machine systems~\cite{shirado2017locally, crandall2018cooperating, rovatsos2019we, ishowo2019behavioural}. A key issue in this context is that, in order to interact with humans in social-dilemma situations, machine learners must understand and incorporate moral, trusting, and cooperative human intuitions~\cite{yoeli2013powering, lerer2017maintaining, peysakhovich2018towards, bonnefon2020machine}.

Ref.~\cite{crandall2018cooperating} examines how to build machine learners that can cooperate with people and other machines at levels that rival human cooperativeness in two-player repeated stochastic games with perfect information. The study identifies three key properties that algorithms should posses to be successful: (i) \textit{generality} in terms of superior performance in many scenarios rather than a specific one, (ii) \textit{flexibility} to both deter potentially exploitative opponent behaviours and elicit cooperation in hesitant opponents, and (iii) \textit{learning speed} sufficient to learn effective behaviours after only a few interactions with people. An algorithm displaying these desirable properties is the new simple rule-based expert algorithm termed S++ that uses a version of aspiration learning~\cite{karandikar1998evolving} to select which strategy to follow from a finite set of expert strategies.

An interesting question in the context of hybrid H2M and M2H cooperation is whether humans remain willing to cooperate with machine learners once the true nature of the latter is revealed to the former. In an iterated prisoner's dilemma in which the actions of machine learners were driven by the S++ algorithm~\cite{rovatsos2019we, ishowo2019behavioural} shows that cooperativeness goes down when humans assume a non-human opponent. The same happens even in contact with zealous non-human opponents whose behaviour is constant. The results thus point to a transparency-efficiency tradeoff by which being transparent about the true nature of the system is likely to harm efficiency. A possible way around this and similar problems is to combine behavioural- and computer-science expertise to make algorithmic decision-making interpretable by many stakeholders, which in turn would allow people to exercise agency and build trust~\cite{stoyanovich2020imperative}.

Ref.~\cite{shirado2017locally} is a particularly nice demonstration that machines can help humans work towards a common goal. In a classic colour coordination game~\cite{kearns2006experimental} embedded into an artificially-constrained social network, non-human agents proved useful in achieving the collective aim of colouring nodes with one of three colours in such a way that each node's colour differs from the colour of every neighbouring node. Non-human agents used a local optimal colouring strategy with occasional random-colour choices, thus introducing a certain level of noise. Low-noise non-human agents placed centrally in the network improved the resolution of colour conflicts, boosted success rates, and increased the speed with which the problem was solved by nudging humans to occasionally deviate and open up to possibilities. Non-human agents thus facilitated human-human interactions at distant network position in effect helping humans to help each other. Although this illustrative example demonstrates how non-human agents can positively facilitate human cooperation and coordination, there are also examples showing that technology-mediated interactions between humans in online social networks and social-media ecosystems can be used to deceive or manipulate~\cite{ferrara2017disinformation, lazer2018science, stella2018bots, stewart2019information}. A task for ongoing and future research is therefore to understand the dynamics of both positive and negative human-machine interactions, and more importantly offer human-centric solutions the foster the former and avoid the latter.

\subsection{Future outlook}

AI is expected to enable people to collaborate with machines in an efficient manner for the purpose of solving complex problems. This is further expected to drive the emergence of ever newer and more widespread kinds of human-machine networks. To make the integration of such networks into society as seamless as possible, research heretofore indicates that it will be critical to ensure good mutual communication, trust, clarity, and understanding between humans and machines, that is, the AI technology will have to be human-centric and explainable. Success in integrating human-machine networks into society will then open the doors to a continued massive assimilation of unstructured, semi-structured, or structured data that should be put to good use by addressing a wide spectrum of social-good problems.

Concerns about the black-box nature and opaqueness of deep-learning systems have hampered more widespread AI applications~\cite{crawford2016there}. To address the problem, a strong case is being made for \textit{Human-centric and explainable AI} as a framework towards human-understandable interpretations of algorithmic behaviour. In this way, human operators should be put back into the driver's seat to continually improve the robustness, fairness, accountability, transparency, and explainability~\cite{xie2020explainable} of AI technologies.

Furthermore, agreed upon methods to assess the sustained effects of AI on human populations in social, cultural, and political contexts are currently non-existent~\cite{crawford2016there}. Such methods are much needed if AI technologies are to fulfil their promise as an enabler for tackling societal issues or improving human well being. Concrete examples here would be helping to attain sustainable development goals~\cite{vinuesa2020role} or to alleviate the effects of the Covid-19 pandemic through molecular and clinical breakthroughs~\cite{bullock2020mapping}. More broadly, but under the condition that the above-stated concerns are resolved, we expect AI to permeate a myriad of research domains and topics, including environmental sustainability (e.g., climate, resource, and biodiversity conservation), social-media abuses (e.g., fake-news, hate-speech, and fraud detection), public safety (e.g., disaster and crime prevention), and more.

\FloatBarrier

\section{Criminology}
\label{S:Crim}

\begin{figure}[!t]
\centering\includegraphics[scale=1.0]{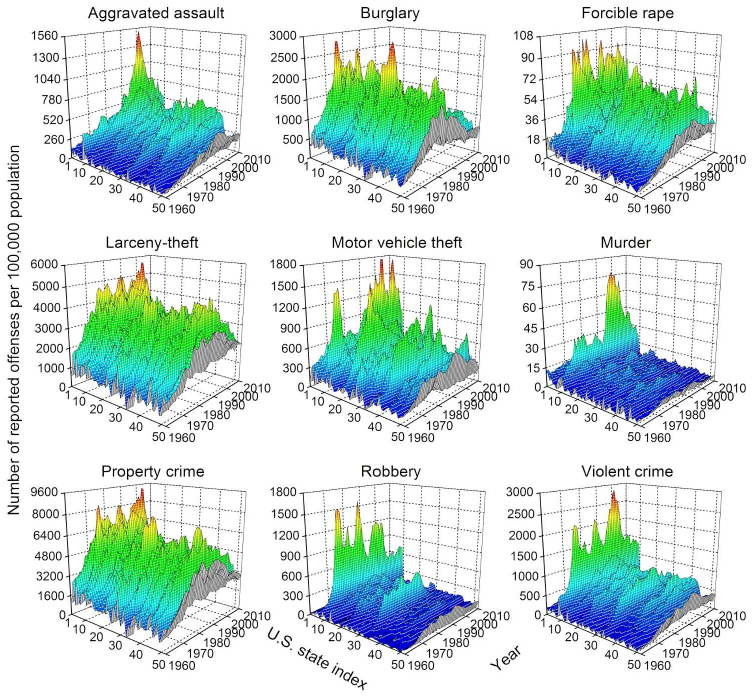}
\caption{The recurrent nature of crime through lens of data from the Federal Bureau of Investigation. Regardless of type and severity, crime is remarkably recurrent despite our best prevention and punishment efforts. While positive and negative trends are inferrable, crime events (measured as number of offenses per 100,000 population) between 1960 and 2010 fluctuate more or less persistently. More importantly, there is no trend inferable to suggest that crime rates are going down, let alone that crime is vanishing. The U.S. state index is alphabetical, including the District of Columbia being 9th, and the U.S. total being 52nd.\newline
Source: Reprinted figure from Ref.~\cite{dorsogna2015statistical}.}
\label{fig:data}
\end{figure}

Containing the spreading of crime remains a major challenge across human societies. Empirical data show consistently that crime is recurrent and proliferates, even more so if it is left unchecked. Fig.~\ref{fig:data} shows data provided by the United States Federal Bureau of Investigation, indicating that even in strongly monitored and policed states, crime deterrence approaches do not have the desired impact. Indeed, eradicating crime culture is a steep uphill battle, especially in underprivileged social circumstances that do not foster the sense of shared social responsibility. Crime is also problematic in power-driven environments, where greed often overrides the moral compass.

In the realm of physics research, crime is considered as a complex phenomenon, where non-linear feedback loops and self-organisation create conditions that are difficult to foretell, control, and often also difficult to understand~\cite{dorsogna2015statistical, ball2012why, helbing2015saving}. Complexity science in general contends with models in which a large number of relatively simple agents exhibits complex, counterintuitive, and often unexpected behaviours, and models of crime are in this regard no exception.

Although our understanding of the emergence and diffusion of crime is an ongoing learning experience, recent research shows that methods of statistical physics can significantly contribute to a better understanding of criminal activity. Herein, we review different approaches aimed at modelling and improving our understanding of crime, focusing in particular on the mathematical description of crime hotspots with partial differential equations, on the self-exciting point process and agent-based modelling, adversarial evolutionary games, and the network science behind the formation of gangs and large-scale organised crime. As we hope we will succeed in showing, physics can relevantly inform the design of successful crime prevention strategies, as well as improve the accuracy of expectations about how different policing interventions should impact malicious human activity that deviates from social norms.

\subsection{The broken windows theory}

The 1982 seminal paper by Wilson and Kelling~\cite{wilson1982broken} contains many examples and stories that bring the `broken windows theory' to life. For example, how an unattended broken window invites by-passers to behave mischievously or disorderly. Or how a subway graffiti points to an unkempt environment that people can desecrate, signalling also that more egregious behaviour might be tolerated. Or how drunks, addicts, prostitutes, and loiterers are more likely to frequent neglected subway stations than orderly and carefully patrolled ones. Thus, on first glance unimportant and petty signals of disorder invite antisocial behaviour and, over time, serious crime---i.e., one broken window soon becomes many.

To physicists, this broken windows theory may be reminiscent of complexity science and self-organised criticality~\cite{bak1996how}, where seemingly small and irrelevant changes at one point in time might have significant and often unexpected and unwanted consequences later on. Moreover, feedback loops, bifurcations, and catastrophes~\cite{kuznetsov2004elements}, as well as phase transitions~\cite{stanley1971introduction}, are commonly associated with emergent phenomena in complex social systems~\cite{castellano2009statistical}.

Besides the `broken windows theory', there exist other theories of criminal behaviour. According to `routine activity theory'~\cite{cohen1979social}, for example, most criminal acts are born out of the convergence of three factors, namely the presence of likely offenders, the presence of suitable targets, and the absence of guardians to protect against the attempted crime. Residential burglary, armed robberies, pick-pocketing, and rape are all examples of such criminal acts.

While intuitively the above three factors are relatively straightforward conditions that obviously favour criminal activity, mathematically they allow us to model the dynamics of criminal offences as deviations from simple random walks. This is due to built-in heterogeneities in target selection that may drive criminal activity towards preferred locations and away from less desired ones. The degree of target attractiveness may change in time and depend on mundane factors such as the day of the week or weather conditions, or on the more sophisticated interplay between landscape, criminal activity, and law enforcement responses. Crime dynamics may also include learning mechanisms or feedback loops. These elements ultimately lead to the emergence of non-trivial patterns such as spatially localised crime hotspots~\cite{short2008statistical} and repeat and near-repeat victimisation~\cite{johnson1997new, townsley2000repeat, johnson2004stability, short2009measuring}, wherein the odds of a second victimisation of the original target or a target in its vicinity are greatly enhanced.

The complexity of crime dynamics that stems from the above-described fundamental considerations has as a consequence the fact that the mitigation and displacement of crime is a highly non-trivial task---a task which, based for example on data shown in Fig.~\ref{fig:data}, we often fail at~\cite{green1995cleaning, braga2001effects, braga2014effects, weisburd2006does, taniguchi2009size}.

In this light, it is important to note that straightforward gain-loss principles that underlie rational choice theories are likely too simple and naive. Anticipating that stronger punishment would just lead to less crime is simply not aligned with the reality, not in empirical data, and not from mathematical models that at least to some degree attempt to capture the complexity of crime~\cite{becker1968crime, doob2003sentence}.

\subsection{Crime hotspots}

\begin{figure}[!t]
\makebox[\textwidth][c]{\includegraphics[scale=1.0]{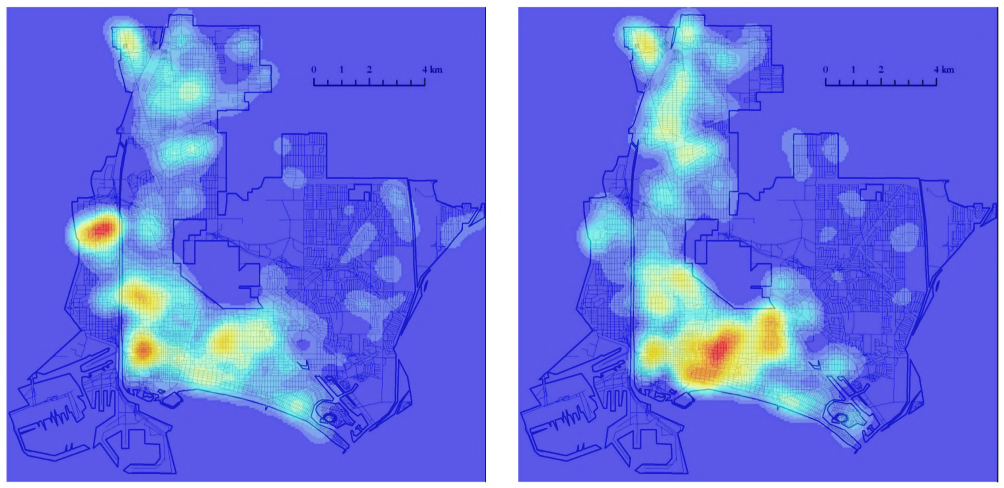}}
\caption{Dynamic changes in residential burglary hotspots in Long Beach California, as observed for two consecutive three-month periods, starting in June 2011. The emergence of different burglary patterns is related to how offenders move within their environments and how they respond to the successes and failures of their illicit activities. Returns to previously victimised locations or locations in their vicinities are common and in agreement with the `routine activity theory'~\cite{cohen1979social}.\newline
Source: Reprinted figure from Ref.~\cite{short2008statistical}.}
\label{fig:hot}
\end{figure}

Presented empirical observations of spatio-temporal clusters of crime in urban areas (Fig.~\ref{fig:hot}) motivated the development of a statistical model of criminal behaviour~\cite{short2008statistical}. The model was developed to study residential burglary, where target sites are stationary, which is not that case in crimes where offenders and targets are mobile, as in assault or pick-pocketing. The model builds on the assumption that burglars are opportunistic, and that they thus victimise areas that are sufficiently close to where they live, and where possibly they have committed crimes before~\cite{johnson2007space}. Another important assumption is that the distances that criminals are willing to travel to engage in criminal acts are best described by monotonically decreasing functions~\cite{rengert1999distance}. The movement of offenders is usually described as a biased random walk, whereby the bias is twofold as follows. In the first place, a given home may be intrinsically more attractive to a burglar due to its perceived wealth, the ease of access, or the predictable routine of its residents. Secondly, there may be learned elements that bias the burglar towards a specific location, for example to a previously victimised home where a successful break-in was once already possible.

To quantify the bias towards any given location and to determine the subsequent rate of burglary, the hotspot crime model includes a dynamically changing attractiveness field~\cite{short2008statistical}. Moreover, the tendency for repeat victimisation is included in the model by temporarily increasing the attractiveness field in response to past crimes~\cite{bernasco2003effects, bernasco2005residential}. Since residential burglary entails non-moving crime targets, and for simplicity, it is convenient to start with a discrete model on a square lattice with periodic boundary conditions. Each lattice site $s=(i,j)$ is a real estate with attractiveness $A_s(t)$ and the number of criminals $n_s(t)$. The higher the value of $A_s(t)$, the higher the bias towards site $s$ and the more likely it will be the subject of crime. Moreover, once site $s$ has been victimised, its attractiveness further increases. The following decomposition is introduced
\begin{linenomath}
\begin{equation}
A_s(t)=A^{0}_s+B_s(t),
\label{eq:field}
\end{equation}
\end{linenomath}
where $A^{0}_s$ is the static, though possibly spatially varying, component of the attractiveness field, and $B_s(t)$ represents the dynamic component associated with repeat and near-repeat victimisation. More precisely, $B_s(t+1)=B_s(t)(1-\omega)+E_s(t)$, where $\omega$ sets a time scale over which repeat victimisations are most likely to occur, while $E_s(t)$ is the number of events that occurred at site $s$ between $t$ and $t+1$. To take into account the broken windows theory~\cite{wilson1982broken}, we let $B_s(t)$ spread locally from each site $s$ towards its nearest neighbours $s'$ according to
\begin{linenomath}
\begin{equation}
B_s(t+1)=\left[(1-\eta)B_s(t)+\frac{\eta}{z}\sum\limits_{s'}B_{s'}(t)\right](1-\omega)+E_s(t)
\label{eq:discreteB}
\end{equation}
\end{linenomath}
where the sum runs over the nearest neighbour sites associated to site $s$, $z$ is the coordination number of the lattice, and $\eta$ is a parameter between zero and one that determines the significance of neighbourhood effects. Higher values of $\eta$ lead to a greater degree of spreading of the attractiveness generated by a given burglary event, and vice-versa for lower values. For simplicity, we can further assume that the spacing between sites $\ell$ and the discrete time unit $\delta t$ over which criminal actions occur are both equal to one, and that every time a site $s$ is subject to crime its dynamic attractiveness $B_s(t)$ increases by one. Interaction networks other than the square lattice, which better describe the city grid or social networks can be easily accommodated as well.

Criminal activity is included in the model by allowing individuals to perform one of two actions at every time step. A criminal may either burglarise the site they currently occupy, or move to a neighbouring one. Burglaries are modelled as random events occurring with probability $p_s(t)=1-\exp[-A_s(t)]$. Whenever site $s$ is subject to crime, the corresponding criminal is removed from the lattice, representing the tendency of actual burglars to flee the location of their crime. To balance this removal, new criminal agents are generated at a rate $\Gamma$ uniformly at random on the lattice. If site $s$ is not burglarised, the criminal will move to one of its neighbouring sites with probability $1- p_s(t)=\exp[-A_s(t)]$. The movement is thus modelled as a biased random walk so that site $s'$ is visited with probability
\begin{linenomath}
\begin{equation}
q_{s \to s'}(t)=\frac{A_{s'}(t)}{\sum\limits_{s'}A_{s'}(t)},
\end{equation}
\end{linenomath}
where the sum runs over all neighbouring sites of $s$. The position of the criminals and the biasing attractiveness field in Eqs.~(\ref{eq:field}) and (\ref{eq:discreteB}) create non-linear feedback loops which give rise to complex patterns of aggregation that are reminiscent of actual crime hotspots, similar to those depicted in Fig.~\ref{fig:hot}. The model actually displays four different regimes of $A_s(t)$, as shown in Fig.~3 of Ref.~\cite{short2008statistical}, all of which apply to different realities of residential burglary.

A continuum version of the above-described discrete model has also been introduced~\cite{short2010dissipation, short2010nonlinear}, the bifurcation analysis of which can also outline suggestions for crime hotspot suppression and policing. According to~\cite{short2008statistical}, the continuum version of the dynamics of the attractiveness field takes the form
\begin{linenomath}
\begin{equation}
\frac{\partial B}{\partial t}=\frac{\eta D}{z} \nabla^2B-\omega B + \epsilon D \rho A,
\label{eq:continB}
\end{equation}
\end{linenomath}
where $D=\ell^2/\delta t$, $\epsilon= \delta t$, and $\rho(s,t)=n_s(t)/\ell^2$. The continuum equation for criminal number density, denoted as $\rho$, is given by
\begin{linenomath}
\begin{equation}
\frac{\partial \rho}{\partial t}=\frac{D}{z} \vec{\nabla} \left[\vec{\nabla} \rho - \frac{2 \rho}{A} \vec{\nabla}A  \right] -\rho A +\gamma,
\label{eq:continR}
\end{equation}
\end{linenomath}
where offenders exit the system at a rate $\rho A$, and are reintroduced at a constant rate per unit area $\gamma = \Gamma / \ell^2$. Eqs.~(\ref{eq:continB}) and (\ref{eq:continR}) are coupled partial differential equations that describe the spatio-temporal evolution of the attractiveness $B$ and the offender population $\rho$, and they belong to the general class of reaction-diffusion equations that frequently exhibit spatial pattern formation~\cite{cross1993pattern}.

\begin{figure}[!t]
\centering{\includegraphics[scale=1.0]{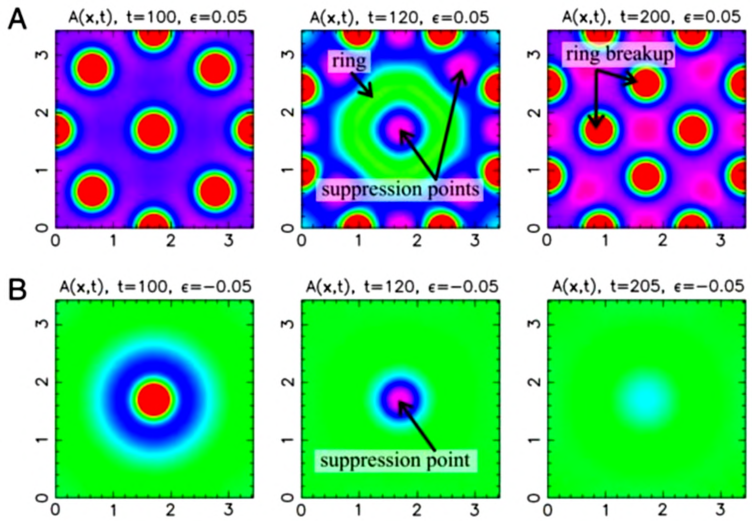}}
\caption{Failure and success of crime hotspot suppression. In the upper row, crime hotspots emerge as a supercritical bifurcation. When subjected to suppression, they simply displace but never vanish completely. New hotspots always emerge in positions adjacent to the original ones. In the lower row, crime hotspots emerge via a subcritical bifurcation. When subjected to suppression, the hotspot gradually vanishes without giving rise to new hotspots in nearby locations. The colour maps encode the time evolution of the attractiveness field $B$. We refer to Ref.~\cite{short2010dissipation} for further details.\newline
Source: Reprinted figure from Ref.~\cite{short2010dissipation}.}
\label{fig:suppress}
\end{figure}

In order to study the effects of police intervention, the crime rate $\rho A$ in Eq.~(\ref{eq:continR}) is set to zero at given hotspot locations and for a given time frame~\cite{short2010dissipation}. Calculations then show that only subcritical crime hotspots may be permanently eradicated by means of a suitable suppression mechanism, while supercritical hotspots are only displaced but never fully removed from the population (Fig.~\ref{fig:suppress}).

The mathematical models describing the nucleation and diffusion of crime hotspots can be upgraded to include spatial disorder, as well as approaches to dynamically adapt suppression measures to evolving crime patterns, or to choose from different deployment strategies and more rigorous analysis~\cite{rodriguez2010local, jones2010statistical, cantrell2012global, berestycki2013traveling, zipkin2014cops}. Along similar lines, related research includes the consideration of dynamical systems that take into account the competition between citizens, criminals, and guards~\cite{nuno2008triangle},
the effects of socio-economic classes and changes in police efficiency and resources allocated to them~\cite{nuno2011mathematical}, the impact of imprisonment and recidivism~\cite{mcmillon2014modeling}, and the possibility of self-defence of communities against crime~\cite{sooknanan2012criminals}.

\subsection{Crime as a self-exciting point process}

Certain types of crime exhibit similar space and time clustering as earthquake activity. Examples include burglary, gang violence, and property crime. Just like clustering patterns observed by seismologists indicate that the occurrence of an earthquake is likely to induce a series of aftershocks near the location of the initial event, so are these types of crime likely to reoccur near initially victimised spots, thus leading to crime swarms and clusters, and lending themselves to the application of seismology methods to model criminal activity. The self-exciting point processes is one such method~\cite{mohler2011self}.

A space-time point process is defined by a collection of points with location $(x,y)$ at time $t$, where a certain event took place. This event can be an earthquake, a lightning strike, or a criminal act. The process is then described by a conditional rate $\lambda(x,y,t)$, which gives the occurrence rate at location $(x,y)$ in dependence on the history $H(t)$ of the point process up to time $t$~\cite{daley2003introduction}. There is typically an initial or parent crime, akin to a parent earthquake, which generates several follow-up or offspring crimes, akin to aftershocks. The follow-up crimes are described by a triggering function $g(x,y,t)$, which depends on previous criminal activity, but with an amplitude that decreases with increasing spatio-temporal distance from it. In modelling crime, a multiplicative factor $\nu(t)$ for the background activity is also needed, which takes into account fluctuations due to weather, seasonality, or day time. Decades of research in seismology have lead to well-defined forms for above functions. In crime, on the other hand, non-parametric methods and calibrations using data are necessary for their estimation. For details we refer to the seminal work by Mohler et al.~\cite{mohler2011self}.

The self-exciting point process has been applied and tested on urban crime using residential burglary data from the Los Angeles Police Department~\cite{mohler2011self}. Traditionally, crime hotspot maps were generated by means of a pre-assigned fixed kernel, using previous crime occurrences as input~\cite{bowers2004prospective}. However, the point process methodology has been found to yield better results, and this also for types of crime where near-repeat effects do not play such a prominent role, like robberies and car theft. The main source of this superiority has been attributed to a better balance between exogenous and endogenous contributions to crime rates and to the method relying on direct inference from data, rather than on an imposition of hotspot maps using a pre-assigned fixed kernel.

Self-exciting point processes have also been used to analyse temporal patterns of civilian death reports in Iraq~\cite{lewis2012self}. Similarly to urban crimes, the rate of violent events has been partitioned into the sum of a Poisson background rate and a self-exciting component in which previous bombings or other episodes of violence generate a sequence of offspring events according to a Poisson distribution. The study showed that point processes are well suited for modelling the temporal dynamics of violence in Iraq.

The geographic profiling of criminal offenders can also be made using self-exciting point processes in order to estimate the probability density for the home base of a criminal who has committed a given set of spatially distributed crimes. Target selection from a hypothetical home base is informed by geographic inhomogeneities such as housing types, parks, freeways or other physical barriers, as well as directional bias and preferred distances to crime~\cite{mohler2012geographic}. These techniques have also been used to model intra-gang violence that results from retaliation after an initial attack~\cite{zipkin2014cops}.

Future research along this line could be aimed at further refining point process models towards crime type and local geography. Doing so would facilitate the application of this promising methodology.

\subsection{Social dilemmas of crime}

The prisoner's dilemma game is amongst the most frequently employed theoretical frameworks to study pairwise social dilemmas~\cite{axelrod1984evolution}. In the prisoner's dilemma game, two players should simultaneously decide whether to cooperator or defect, and then based on their choices receive payoffs accordingly. A social dilemma arises because mutual cooperation yields the highest collective payoff, but the payoff for a defector is higher if the opponent decides to cooperate. Mutual defection is therefore the only rational outcome if we assume that both players act in self-interest so as to maximise their individual payoffs. In the long run this leads to the proliferation of defection and ultimately to the `tragedy of the commons'~\cite{hardin1968tragedy, perc2017statistical}, where common resources are lost to societies due to overexploitation and lack of shared social responsibility.

While criminal behaviour does not map directly to the prisoner's dilemma game, the framework of evolutionary games, and evolutionary social dilemmas in particular~\cite{nowak2006five, perc2010coevolutionary}, lends itself very well to modelling crime. In this context, social order can be considered as the public good that is threatened by criminal activity, with competition arising between criminals and those trying to prevent crime. However, committing crimes is not necessarily equivalent to defection, because unlike defectors, criminals may actively seek to harm others. Likewise, fighting crime is often more than just cooperating, in particular because it may involve risk that goes beyond contributing some fraction of one's wealth into a common pool. Thus, a more deliberate formulation of competing strategies may elevate, and is in fact needed for, the accuracy of the modelling approach.

\begin{figure}[!t]
\centering{\includegraphics[scale=1.0]{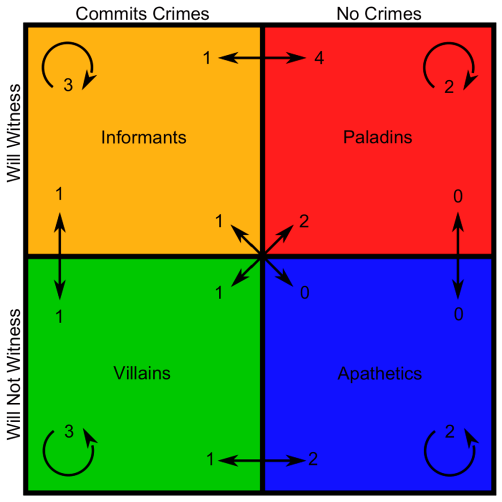}}
\caption{Crime as a four-strategy evolutionary game, comprising informants, paladins, villains, and apathetics. The four strategies are defined by their propensities to commit crimes and serve as witnesses in criminal investigations. Arrows between strategies indicate the number of possible game pairings and outcomes in which the update step leads to a strategy change. For example, there are two ways by means of which a villain can be converted into a paladin. Circular arrows within each strategy quadrant indicate updates such that player strategies remain unchanged.\newline
Source: Reprinted figure from Ref.~\cite{short2010cooperation}.}
\label{fig:strats}
\end{figure}

With these considerations in mind, an adversarial evolutionary game with four competing strategies, as shown in Fig.~\ref{fig:strats}, has been proposed in Ref.~\cite{short2010cooperation}. The game entails informants ($I$) and villains ($V$) as those who commit crimes, as well as paladins ($P$) and apathetics ($A$) as those who do not. Informants and paladins actively contribute to crime abatement by collaborating with authorities whenever asked to do so. All players may witness crimes or be the victims of crime, in agreement with victimisation surveys~\cite{lynch2006understanding}. Thus, paladins are model citizens that do not commit crimes and collaborate with authorities. At the other end of the spectrum we have the villains, who commit crimes and do not report them. Somewhere in between we have informants who report on other offenders while still committing crimes, and apathetics who neither commit crimes nor report crimes of others. The lack of active cooperation in apathetics may be due to inherent apathy, fear of retaliation, or ostracism from the community at large. Apathetics are similar to second-order free riders in the context of the public goods game with punishment \cite{fehr2004dont, szolnoki2017second}, in that they cooperate at first order by contributing to the public goods as in not committing crimes, but defect at second order by not punishing offenders.

At each round of the game a criminal is selected randomly from the $V+I$ pool together with a potential victim from the $N-1$ remainder of the population. The two selected players begin the game with a unitary payoff. After a crime occurs, the criminal player increases their payoff by $\delta$, while the victim looses $\delta$. If the victim is either an apathetic or a villain, the crime is not reported to the authorities and therefore successful: the victim's payoff is decreased to $1-\delta$ and the victimiser's is increased to $1+\delta$. If, on the other hand, the victim is a paladin or an informant, the crime is reported to the authorities and an investigation begins. For this, a subset $M$ of the $N-2$ remaining players is drawn, and the victimiser is convicted with probability $w=(m_P+m_I)/M$, where $m_P$ and $m_I$ are the number of paladins and informants within $M$. In case of a conviction, the victim is refunded $\delta$, and the payoff of the criminal becomes $1-\theta$, where $\theta$ determines the punishment fine. With probability $1-w$ the crime is left unpunished, in which case the criminal retains $1+\delta$, while the victim's payoff is further decreased to $1-\delta-\epsilon$, where $\epsilon$ is due to retaliation of the accused who, having escaped punishment, feels empowered in their revenge. Other interpretations of $\epsilon$ may be damages to personal image or credibility, or loss of faith in the system after making an accusation that is unsubstantiated by the community. Notably, in the latter case, the choice of reporting one's victimisation to authorities may be even more detrimental to the witness than the original criminal act ($\epsilon>\delta$), which is common in societies that are heavily marred by war, by mafia, or drug cartels, where very few people will serve as witnesses to crimes.

Parameter values of $\delta$, $\theta$, and $\epsilon$ are always used such that all payoffs remain positive. At the end of each round of the game, the player with the smaller payoff changes its strategy according to proportional imitation~\cite{schlag1998imitate}. In particular, if the victimiser is emulated, the loser simply adopts the victimiser's strategy and ends the update as either a villain or an informant. If the victim is emulated, the loser mimics the victim's propensity to serve as a witness but adopts a noncriminal strategy regardless of the victim's. In this case, the update results with the loser becoming either a paladin or an apathetic (see Fig.~\ref{fig:strats} for details).

\begin{figure}[!t]
\centering{\includegraphics[scale=1.0]{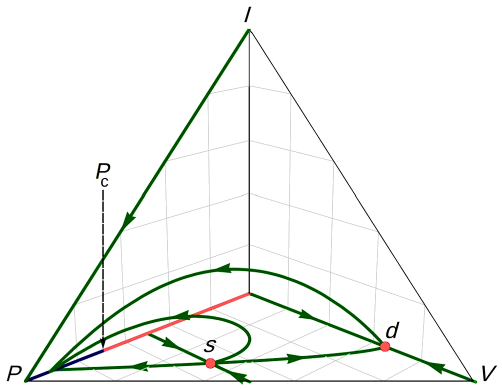}}
\caption{The emergence of utopia in a society with informants. All trajectories with $I_0>0$ evolve towards a crime-free state. The ternary diagram shows unstable fixed points in light red, unstable fixed lines in thick light red, stable fixed lines in thick dark blue, and trajectories beginning (or ending) along various eigenvectors as thick green arrows. The dystopian fixed point $d$ and the saddle point $s$ are unstable to increases in $I$, so that the only attracting final states for $I_0>0$ are those utopias with $P>P_\mathrm{c}$. These results were obtained with $\delta=0.3$, $\theta=0.6$, and $\epsilon=0.2$.\newline
Source: Reprinted figure from Ref.~\cite{short2010cooperation}.}
\label{fig:ternary}
\end{figure}

Simulations of the four-strategy evolutionary game described above reveal that informants are key to the emergence of utopia---a crime-free society. Indeed, a crime-dominated society can become crime-free by imposing an optimal number of informants $I_0$ at the onset of the game. The dynamics depends on the chosen parameter values. A utopia may be elusive in extremely adversarial societies in which initially we have high numbers of villains and apathetics. However, by deriving a deterministic version of the above described game~\cite{short2010cooperation}, it is possible to show that if there are at least some informants initially present in the population ($I_0 > 0$), the final state is always utopia regardless of $\delta$, $\theta$, and $\epsilon$ (Fig.~\ref{fig:ternary}).

While beneficial, the presence of informants may come at a cost, either in training an undercover informant, or in convincing a criminal to collaborate with authorities, or in tolerating the criminal acts that informants will keep committing. One may thus consider an optimal control problem~\cite{short2013external} to investigate the active recruitment of informants from the general population in terms of associated costs and benefits. Higher recruitment levels may be the most beneficial in abating crime, but they may also be prohibitively expensive. The optimal control problem was expressed via three control functions subject to a system of delay differential equations. The research showed that optimal recruitment strategies change drastically as parameters and resource constraints vary, and moreover, that more information about individual player strategies leads only to marginally lower costs~\cite{short2013external}.

The important role of informants within the reviewed adversarial evolutionary game~\cite{short2010cooperation} has also been studied by means of human experiments in~\cite{dorsogna2013criminal}. The goal was to test whether, and if yes to what degree, informants are actually critical for crime abatement as predicted by theory. Remarkably good agreements between simulations and laboratory experiments have been observed for different parameterisations of the game, thus lending full support to the approach.

In addition to social dilemmas, the evolution of crime can also be studied by means of the inspection game~\cite{tsebelis1990penalty}. Rational choice theories predict that increasing fines should diminish crime~\cite{becker1968crime}. However, a three strategy inspection game in which, in addition to criminals ($C$) and punishing inspectors ($P$), ordinary individuals ($O$) are present leads to significantly different and counterintuitive outcomes \cite{perc2013understanding, perc2015double}. The $O$ players neither commit crimes nor do they participate in inspection activities. They represent the masses that catalyse rewards for criminals and costs for inspectors. Ordinary individuals receive no bonus payoffs upon encountering inspectors or their peers. Only when paired with criminals do they suffer the consequences of crime in form of a negative payoff $-g \le0$. Criminals, on the other hand, gain the reward $g\ge 0$ for committing a crime. When paired with inspectors, criminals receive a payoff $g-f$, where $f\ge 0$ is a punishment fine. When two criminals are paired none of them receive any benefits. Inspectors, on the other hand, always have the cost of inspection, $c \ge 0$, but when confronted with a criminal, an inspector receives the reward $r\ge 0$ for a successful apprehension. This game was studied via Monte Carlo simulations on a square lattice with periodic boundary conditions where each lattice site is occupied either by a criminal, a punishing inspector, or an ordinary citizen. The game evolves by first randomly selecting player $s$ to play the inspection game with their four nearest neighbours, yielding the payoff $P_{s}$. One of the nearest neighbours of player $s$, $s'$, is then chosen randomly to play the game with their four nearest neighbours, leading to $P_{s'}$. Finally, player $s'$ imitates the strategy of player $s$ with probability
\begin{linenomath}
\begin{equation}
q=\frac{1}{1+\exp\left(\frac{P_{s'}-P_{s}}{K}\right)},
\label{eq:MNL}
\end{equation}
\end{linenomath}
where $K$ determines the level of uncertainty in the strategy adoption process. The chosen form in Eq.~(\ref{eq:MNL}) corresponds to the empirically supported multinomial logit model~\cite{mcfadden1974conditional}, which for two decision alternatives becomes the Fermi function~\cite{szabo2007evolutionary, szolnoki2009topology}. A finite value of $K$ accounts for the fact that better performing players are readily imitated, although it is possible to adopt a strategy by player who is performing worse, for example due to imperfect information or errors in decision making.

Monte Carlo simulations reveal that the collective behaviour of the three-strategy spatial inspection game is indeed complex and counterintuitive, with both continuous and discontinuous transitions between different phases. Here a phase is either a single-strategy or a multi-strategy stable state that is uninvadable by any other combination of strategies or a single strategy. Usually, evolutionary games with more than two competing strategies require the stability of subsystem solutions be performed for the accurate and correct determination of phase transitions~\cite{perc2017stability}. A subsystem solution can be formed by any subset of all possible strategies. The winner between two subsystem solutions can be determined by the average moving direction of the invasion front that separates them, yet it is crucial that the competing subsystem solutions are characterised by a proper composition and spatio-temporal structure before the competition starts. In this way the three-strategy inspection game also yields a cyclic dominance phase~\cite{szolnoki2014cyclic}, which emerges spontaneously due to pattern formation and is robust against initial condition variations, and in which all three competing strategies coexist.

Taken together, these results indicate that crime should be viewed not only as the result of offending actions committed by certain individuals, but also as the result of social interactions between people who adjust their behaviour in response to societal cues and imitative interactions. The emergence of crime thus should not be ascribed merely to the `criminal nature' of particular individuals, but also to the social context, the systems of reward and punishment, the level of engagement of the community, as well as to the interactions between individuals. This more comprehensive view of crime may have relevant implications for policies and law enforcement.

\subsection{Criminal networks}

The goal of this section is to review how methods of physics, and in particular of network science~\cite{albert2002statistical, dorogovtsev2002evolution, boccaletti2006complex, barthelemy2011spatial, estrada2012structure, holme2012temporal, boccaletti2014structure, kivela2014multilayer}, can contribute to better understanding organised crime~\cite{mallory2011understanding}, such as drug cartels, the formation of gangs, or political corruption networks~\cite{ribeiro2018dynamical}.

Criminal structures like the Italian Mafia~\cite{gambetta1998trust}, street gangs, or drug cartels~\cite{beittel2009mexico} often emerge when fear and despair become so ingrained within a society that the social norm is simply to accept crime. In such a case, witnesses and even victims of crime often choose not to cooperate with law enforcement in the prosecution of criminals. Instead, people sometimes try to fit in, although acquiescence and acceptance are usually slippery slopes towards later forms of active engagement. Ultimately this thus leads to the growth of a criminal network.

Criminological research has identified a number of factors that may promote the regional development of crime, including unemployment~\cite{raphael2001identifying, lin2008does}, economic deprivation~\cite{lafree1999declining}, untoward youth culture~\cite{curtis1997improbable}, failing social institutions~\cite{lafree1998losing}, issues with political legitimacy~\cite{lafree1999declining}, as well as lenient local law enforcement strategies~\cite{corsaro2009testing, mcgarrell2010project}, to name the most prominent examples. Policies aimed at reducing recruitment into organised crime have also been incorporated in agent-based models with a multiplex-network structure to capture the effects of household, kinship, school, work, friends, and co-offending social relations~\cite{calderoni2021recruitment}. Recent work on declining criminal behaviour in the U.S. suggests that trends in the levels of crime may be best understood as arising from a complex interplay of a rich myriad of said factors~\cite{gomez2007decomposing, zimring2006thegreat}, while most recent empirical data indicate that social networks of criminals have a particularly strong impact on the occurrence of crime---the more the criminals are connected into networks, the higher the crime rate~\cite{papachristos2012social, papachristos2013corner}.

\begin{figure}[!t]
\centering\includegraphics[scale=1.0]{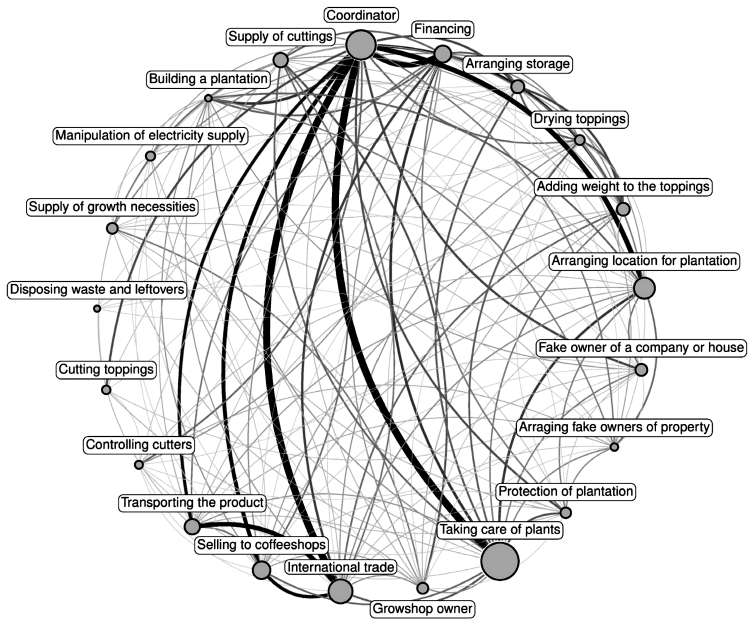}
\caption{A criminal network in the Netherlands involved in cannabis cultivation. Nodes represent the actors that are needed for successful production and distribution of cannabis. The network is highly resilient to targeted disruption strategies. Even worse, research shows that perturbations may lead to reorganisation towards an even more robust and resilient network. Node sizes represent the number of actors fulfilling the associated role, and link thickness corresponds to the total number of links between actor groups.\newline
Source: Reprinted figure from Ref.~\cite{duijn2014relative}.}
\label{fig:cannabis}
\end{figure}

The assumption that there is a network structure behind organised crime invites the idea that removing the leader, or the most important hubs of the network~\cite{albert2000error}, will disrupt the organisation sufficiently to hinder or at least heavily disrupt criminal activity. Law enforcement agencies thus often attempt to identify and arrest the `ring leader' of an identified criminal organisation. But even if successful, such operations rarely have the desired effect. A recent study analysing cannabis production and distribution networks in the Netherlands shows that this strategy may in fact be fundamentally flawed~\cite{duijn2014relative}. All attempts towards network disruption analysed in the study proved to be at best unsuccessful (Fig.~\ref{fig:cannabis}). At worst, they had the opposite effect in that they have increased the efficiency of the network. The latter was achieved by means of nifty reorganisations, such that the attack ultimately made these networks stronger and more resilient to future such attempts. By combining computational modelling and social network analysis with unique criminal network intelligence data from the Dutch Police, Duijn et al.~\cite{duijn2014relative} have concluded that criminal network interventions are likely to be effective only if applied at the very early stages of network growth, before the network gets a chance to organise, or to reorganise to maximum resilience.

\begin{figure}[!t]
\centering\includegraphics[scale=1.0]{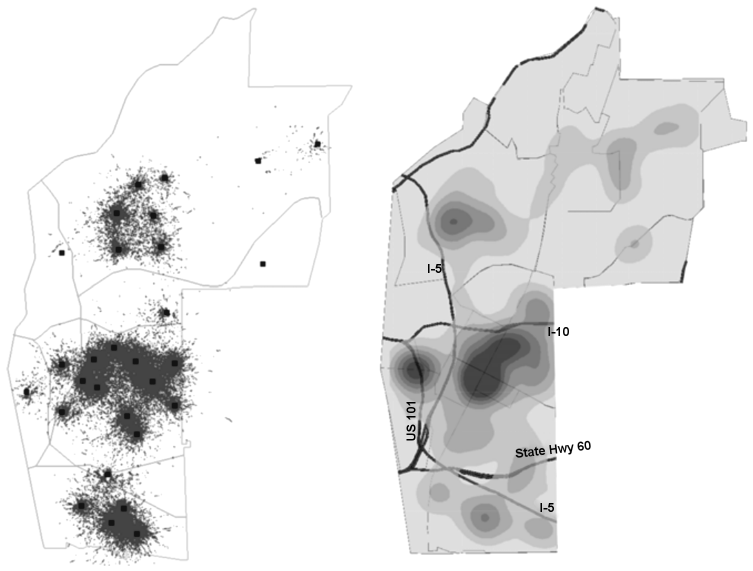}
\caption{Reconstructing a gang network from data with the biased L\'evy walk network method. Interactions between agents simulated using the biased L\'evy walk network method are shown left, while the actual density map of gang-related violent crimes in Hollenbeck between 1998 and 2000 is shown right. Thick lines represent major freeways crossing the city.\newline
Source: Reprinted figure from Ref.~\cite{hegemann2011geographical}.}
\label{fig:SBLN}
\end{figure}

Gang rivalries have been studied by means of agent-based simulations in conjunction with
data from the Hollenbeck policing division of the Los Angeles Police Department~\cite{hegemann2011geographical}. The details of the model were as follows. Each agent is part of an evolving rivalry network that includes past interactions between gang members. Individuals perform random walks where the jump length is drawn from a truncated L\'evy distribution and where bias in the direction of rivals is included. Gang home bases, historical turfs, and geographic details that may limit movement such as freeways, rivers, and parks were all taken into account in the simulated biased L\'evy walk network. Typical gang behaviour, as inferred from the criminology literature, has also been considered. Using metrics from graph theory, it was possible to show that simulated biased L\'evy walk network modelling is in fact the most accurate in replicating actual gang networks. In Fig.~\ref{fig:SBLN}, we reproduce a picture from~\cite{hegemann2011geographical}, showing simulated results and an actual map of violent crimes in Hollenbeck, which are indeed in very good agreement. This approach can also be used to infer unknown rivalry interactions, in particular because the simulated biased L\'evy walk network converges to stable long-term configurations. The authors of Ref.~\cite{hegemann2011geographical} have also noted that the method is portable and can be applied to other geographical locations, offering insight into gang rivalry distributions even in the absence of data. The method may also be extended to test sociological concepts related to gang interactions such as territoriality and allegiances within gangs.

Police department field interview cards were later used to study the behavioural patterns of roughly 748 suspected gang members who were stopped and questioned in Hollenbeck~\cite{vangennip2013community}. The goal was to identify any social communities among street gang members by creating a fully-connected ad hoc network where individuals represent nodes and links encode geographical and social data. Individuals stopped together were assumed to share a friendly or social link and the distance $d_{i,j}$ between stop locations of individuals was recorded. This information was used to determine the affinity matrix $W_{i,j}$ associated with the network. Its entries are composed of a term that decays as a function of $d_{i,j}$, representing geographical information, and of an adjacency matrix whose entries are zero or one depending on whether individuals were stopped together or not. The latter represents social information. By using spectral clustering methods, distinct groups were then identified and interpreted as distinct social communities among Hollenbeck gang members. These clustered communities were finally matched with actual gang affiliations recorded from the police field interview cards.

To evaluate the quality of identified clusters, the authors of Ref.~\cite{vangennip2013community} used a purity measure, defined as the number of correctly identified gang members in each cluster divided by the total number of gang members. Results showed that using geographical information alone leads to clustering purity of about 56\,\% with respect to the true affiliations of the 748 individuals taken in consideration. Adding social data may improve purity levels, especially if this data is used in conjunction with other information, such as friendship or rivalry networks. These results may be used as a practical tool for law enforcement in providing useful starting points when trying to identify possible culprits of a gang attack.

An interesting physics-inspired approach to modelling gang aggregation and territory formation by means of an Ising-like model has also been proposed in Ref.~\cite{barbaro2013territorial}. In particular, otherwise indistinguishable agents were allowed to aggregate within two distinct gangs and to lay graffiti on the sites they occupy. Interactions among individuals were indirect and occurred only via the graffiti markings present on-site and on nearest-neighbour sites. Within this model, gang clustering and territory formation may arise under specific parameter choices, and a phase transition may occur between well-mixed and well separated, clustered configurations. In the mean-field version of the model, parameter regimes were identified where the transition is first or second order. In all cases, however, these clustering transitions were driven by gang-to-graffiti couplings since direct gang-to-gang interactions were not included in the model. The role of graffiti and vandalism has been reviewed also by Thompson et al.~\cite{thompson2012broken}, who analysed the urban rail industry, where graffiti markings have significant impact on expenditure, timely operation of services, and on passenger perception of safety.

Methods of network science are also well suited to study a more subtle form of crime, namely political corruption. Indeed, corrupt behaviour in politics limits economic growth~\cite{rose1975economics, shleifer1993corruption, mauro1995corruption, bardhan1997corruption, shao2007quantitative}, embezzles public funds~\cite{haque2008public}, and promotes socio-economic inequality in modern democracies~\cite{mauro1995corruption, gupta2002does}. The World Bank estimates that the annual cost of corruption exceeds 5\,\% of the global Gross Domestic Product, which amounts to \$2.6 trillion USD, with \$$1$ trillion USD being paid in bribes around the world. In another estimation by the non-governmental organisation Transparency International, corrupt officials receive as much as \$$40$ billion USD bribes per year in developing countries, and nearly two out of five business executives have to pay bribes when dealing with public institutions. Despite the difficulties in trying to estimate the cost of global corruption, there is consensus that massive financial resources are lost every year to this cause, leading to devastating consequences for companies, countries, and the society as a whole.

The shortage of studies aimed at understanding the finer details of corruption processes is in considerable part due to the difficulties in finding reliable and representative data about people who are involved~\cite{xu2005criminal}. On the one hand, this is certainly also because those who are involved do their best to remain undetected, but also because information that does leak into the public is often spread over different media outlets offering conflicting points of view. In short, lack of information and misinformation~\cite{delvicario2016spreading} both act to prevent in-depth research.

\begin{figure}[!t]
\centering{\includegraphics[scale=1.0]{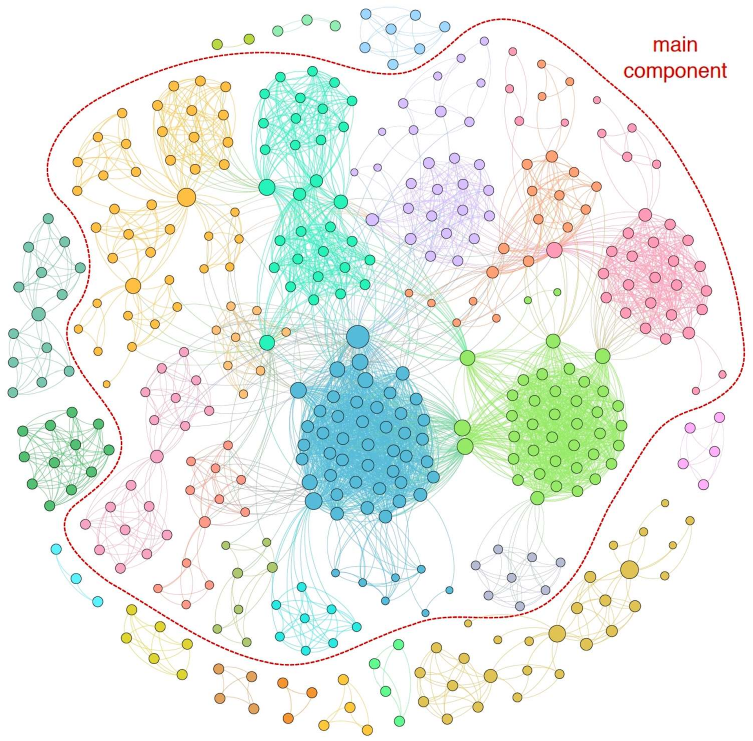}}
\caption{Network representation of people involved in corruption scandals in Brazil from 1987 to 2014 (from Ref.~\cite{ribeiro2018dynamical}). Each vertex represents a person and the edges among them occur when two individuals appear at least once in the same corruption scandal. Node sizes are proportional to their degrees and the colour code refers to the modular structure of the network, as obtained with the network-cartography approach~\cite{guimera2005cartography}. There are 27 significant modules, and 14 of them are within the giant component indicated by the red dashed loop.\newline
Source: Reprinted figure from Ref.~\cite{ribeiro2018dynamical}.}
\label{fig:cornet}
\end{figure}

To overcome these challenges, Refs.~\cite{ribeiro2018dynamical, luna2020corruption, nicolas2021conspiracy} have employed datasets that allow in-depth insights into corruption scandals in Brazil and Mexico. The Brazilian dataset~\cite{ribeiro2018dynamical} in particular provides details of political-corruption activities of 404 people who were from 1987 to 2014 involved in 65 important and well-documented scandals. Notably, Brazil has been ranked 79th in the Corruption Perceptions Index, which surveyed 176 countries in its 2016 edition, which places it behind African countries such as Suriname (64th) and Ghana (70th), and way behind its neighbouring countries such as Uruguay (21th) and Chile (24th). Methods of time series analysis and network science have been applied to reveal the dynamical organisation of political corruption networks in Brazil, which in turn reveals fascinating details about individual involvement in particular scandals, and it allows the prediction of future corruption partners with useful accuracy~\cite{ribeiro2018dynamical}. Research showed that the number of people involved in corruption cases is exponentially distributed, and that the time series of the yearly number of people involved in corruption has a correlation function that oscillates with a four-year period. This indicates a strong relationship with the changes in political power due to the four-year election cycle. By linking together people that were involved in the same political scandal in a given year, it was also possible to create a network representation of people that took part in corruption scandals (Fig.~\ref{fig:cornet}). The network has an exponential degree distributions with plateaus that follow abrupt variations in years associated with important changes in the political powers governing Brazil. By maximising the modularity of the latest stage of the corruption network, we can observe statistically significant modular structures that do not coincide with corruption cases but usually merge more than one case.

Based on this research, it is also possible to apply different algorithms for predicting missing links in corruption networks. By using a snapshot of the network in a given year, Ribeiro et al.~\cite{ribeiro2018dynamical} have tested the ability of these algorithms to predict missing links that appear in future iterations of the corruption network. Obtained results show that some of these algorithms have a significant predictive power in that they can correctly predict missing links between individuals in the corruption network, which could be used effectively in prosecution and mitigation of future corruption scandals.

Lastly, we mention promising efforts to detect criminal organisations~\cite{ferrara2014detecting} and to predict crime~\cite{bogomolov2014once} based on demographics and mobile data. It is known that the usage of communication media such as mobile phones and online social networks leaves digital traces, and research shows that this data can be used successfully for detecting and characterising criminal organisations. We hope that this section shows that, with the help of network science and community detection~\cite{fortunato2010community, fortunato2016community}, law enforcement agencies could better understand hierarchies within criminal organisations, more reliably identify members who play central roles in them, as well as obtain valuable information on connections among different sub-groups and their respective responsibilities in the illicit undertakings.

\subsection{Rehabilitation and recidivism}

The final, and perhaps even the most important, stage in treating crime is the rehabilitation of past offenders. Only if past offenders acknowledge and understand their wrongdoing, and only if after paying their dues they can be integrated successfully back into society, can we consider the problem solved. Otherwise, we are patching the problem of crime with temporary solutions that in the long run do not actually lead to better societal outcomes. Sadly, the available data indicate that it is in these later stages of crime abatement where societies fail the most, and in particular fail to give offenders a second chance at a new start in life, which often pushes them, or at least provides a nudge, into recidivism.

The dilemma that commonly shows in such cases is often referred to as the `stick versus carrot' dilemma. In other words, should rehabilitation programs focus on punishing wrongdoing (stick), or should they focus on generously rewarding steps of progress along the way (carrot)? There is ample research in evolutionary game theory that addresses this dilemma in the context of cooperation in the public goods game~\cite{perc2017statistical}. Notably, there is no simple conclusion or resolution of the dilemma, due to the fact that the outcome depends substantially on the context and other circumstances that are taken into account in the model. Theoretically at least, punishment seems to be more promising simply because it can stop once the target behaviour is achieved. Rewarding, on the other hand, often creates a self-enforcing loop in that the more progress is achieved, the higher the rewards that are expected to uphold the good trend. However, research on punishment also emphasises the very negative consequences of antisocial punishment and with it related concerns to use sanctions as a means to promote collaborative efforts and to raise social welfare~\cite{herrmann2008antisocial, rand2009positive}. Evidence suggesting that rewards may be as effective as punishment and lead to higher total earnings without potential damage to reputation~\cite{milinski2002reputation} or fear from retaliation~\cite{dreber2008winners} has also been mounting. Moreover, Rand and Nowak~\cite{rand2011evolution} provide firm evidence that antisocial punishment renders the concept of sanctioning ineffective, and argue further that healthy levels of cooperation are likelier to be achieved through less destructive means.

Regardless of whether the burden is placed on punishment~\cite{gachter2008long, boyd2010coordinated, perc2012self, szolnoki2017second} or reward \cite{hilbe2010incentives, szolnoki2010reward, szolnoki2012evolutionary, szolnoki2015antisocial, fang2019synergistic, hu2020rewarding} or both~\cite{szolnoki2013correlation}, the problem with both actions is that they are costly. Cooperators who abstain from either punishing or rewarding therefore become second-order free riders, and they can seriously challenge the success of sanctioning as well as rewarding. In the context of rehabilitating criminals, the question is how much punishment for the crime and how much reward for eschewing wrongdoing in the future is in order for optimal results, as well as whether these efforts should be placed on individuals or institutions~\cite{sigmund2010social, szolnoki2011phase}, all the while also assuming of course that the resources are limited~\cite{perc2012sustainable, chen2014optimal}.

To improve our understanding of these important considerations, Berenji et al.~\cite{berenji2014recidivism} have introduced an evolutionary game to study the effects of carrot and stick intervention programs on criminal recidivism. Their model assumes that each player may commit crimes and may be arrested after a criminal offence. In the case of a conviction, a criminal is punished and later given resources for rehabilitation in order to prevent recidivism. After their release into society, players may choose to continue committing crimes or to become paladins ($P$). The later option is an optimal outcome, indicating they have been permanently reformed. Players are given $r$ chances to become paladins. If after the $r$-th arrest and rehabilitation an individual relapses into crime it is marked as an unreformable ($U$). States $P$ and $U$ are sinks, meaning they mark the end of the evolutionary process for each particular individual. The $P/U$ ratio is therefore a natural order parameter, such that societies with a lot of crime are characterised by $P/U \to 0$ while crime-free societies are characterised by $P/U \to \infty$. The main parameters of the game are the resources allocated for rehabilitation $h$, the duration of the rehabilitation $\tau$, and the severity of punishment $\theta$.

\begin{figure}[!t]
\centering{\includegraphics[scale=1.0]{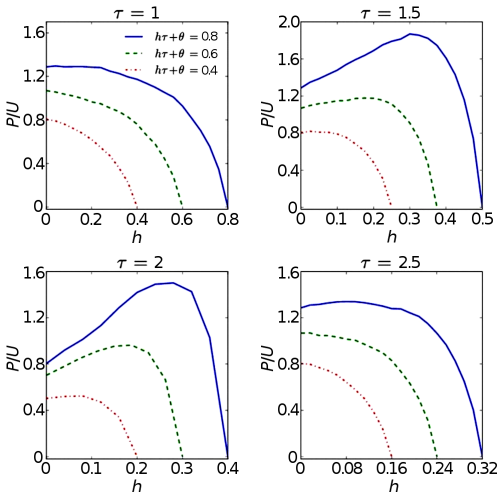}}
\caption{Minimising recidivism requires carefully balanced rehabilitation programs, where both punishment and reward play a crucial role. Either neglecting punishment in favour of generous rehabilitation or vice versa will ultimately fail in successfully reintegrating offenders into society. Depicted is the ratio between paladins and unreformables $P/U$ in dependence on the amount of resources for rehabilitation $h$, as obtained for different values of the duration of intervention $\tau$ (see top of individual graphs). In all cases the severity of punishment $\theta$ is adjusted so that $h\tau+\theta=C$ (see legend in the top left graph), taking into account the fact that available resources are finite. The upper right graph reveals that the optimal parameter values are $h=0.3$, $\tau=1.5$, and $\theta=0.35$, which indicates that the most successful strategy is to allocate the limited resources so that after being punished, criminals experience impactful intervention programs, especially during the first stages of their return to society.\newline
Source: Reprinted figure from Ref.~\cite{berenji2014recidivism} under the Creative Commons Attribution 4.0 International (CC BY 4.0).}
\label{fig:optimal}
\end{figure}

Simulations of this model have been performed which include the constraint $h \tau + \theta = C$, where $C$ is the total amount of available resources, and where $h \tau$ is the part of these resources that are spent on rehabilitation---the carrots---while $\theta$ is the remainder, spent on punishment---the sticks. Because $C$ is finite, increasing one effort decreases the other, hence the `stick versus carrot' dilemma. As $C$ increases, the ratio $P/U$ will increase as well (Fig.~\ref{fig:optimal}). This means that with more general resources available, the conversion to paladins becomes more efficient. For a given value of $C$, the most successful strategy in reducing crime, warranting the highest $P/U$ ratio, is to optimally allocate resources so that after being punished, criminals experience impactful intervention programs, especially during the first stages of their return to society. Indeed, the upper right panel of Fig.~\ref{fig:optimal} reveals that for the case of $N=400$ players, the optimal parameter values are $h=0.3$, $\tau=1.5$ and $\theta=0.35$, which indicates that the available resources $C$ need to be balanced so that there is enough stick (a sufficiently high $\theta$) and enough carrots (a sufficiently high $h$) for a long enough time (a sufficiently high $\tau$). Within this model, excessively harsh or lenient punishment is less effective than when the two are well-balanced. In the first case, there are not enough resources for rehabilitation left, in the second, punishment was not strong enough to discourage criminals from committing further crimes upon release to society.

The findings reviewed in this section have important sociological implications, and they provide useful guidance on how to minimise recidivism while maximising social reintegration of criminal offenders. At the same time, we note that research dedicated specifically to rehabilitation and recidivism at the interface of physics and criminology is rather sparse, so that this is certainly an avenue worth exploring more prominently in the future, especially given its importance in assuring long-term success of prior crime prevention strategies.

\subsection{Rosy outlooks for less crime}

The physics of crime is a developing and vibrant field, with ample opportunities for novel discoveries and improvements of existing models and theory. The model of crime hotspots, for example, could be upgraded to account for the distribution of real estate that better reflects the layout of an actual city. It would then be interesting to learn whether and how the introduced heterogeneity in the interaction network affects the emergence and diffusion of hotspots. If the crime is no longer residential burglary but crime that involves moving targets, further extensions towards social networks whose structure varies over time also become viable. If crime is treated as an evolutionary game the possibilities for upgrades range from increased strategic complexity to the integration of more realistic, possibly co-evolving, interaction networks that describe human interactions. In the realm of adversarial evolutionary games, it would be interesting to study the impact of different strategy adoption rules, in particular because imitation-based rules are frequently contested with best-response dynamics in the realm of human behaviour. In addition to the outlined extensions and upgrades of existing models, it is also possible to envisage new classes of models, especially such that would build more on self-organisation and growth from first principles to eventually arrive at model societies with varying levels of crime. Here the hierarchical growth of criminal networks involving persuasion to join an organisation and fidelity to either committing or not committing crimes could be fertile starting grounds.

As we hope this section shows, the physics of crime can provide useful insights into the emergence of criminal behaviour, as well as indicate effective policies for crime mitigation. We also hope the reviewed results may be useful to police and other security agencies in developing better and more cost-effective crime mitigation schemes while optimising the use of their limited resources. Indeed, the physics of crime has far-reaching implications, and we emphasise that the time is ripe for these insights to be used in synergy with traditional crime-related research to yield more effective crime mitigation policies. Many examples of ineffective policies clearly highlight that an insufficient understanding of the complex dynamical interactions underlying criminal activity may cause adverse effects from well-intended deterrence strategies. A new way of thinking, maybe even a new kind of science for deterring crime is thus needed---in particular one that takes into account not just the obvious and similarly linear relations between various factors, but one that also looks at the interdependence and interactions of each individual and its social environment. One then finds that this gives rise to strongly counterintuitive results that can only be understood as the outcome of emergent, collective dynamics. This is why physics can make important and substantial contributions to the better understanding and containment of crime.

The aim here is to highlight valuable theoretical resources that can help us bridge the gaps between data and models of criminal activity. Employing these resources should certainly contribute to rosier outlooks for human societies with less crime.

\FloatBarrier

\section{Migrations}
\label{S:Mig}

Incoming and outgoing migrations, respectively called immigration and emigration, pose substantial challenges to society. Unsuccessful integration of immigrants, for instance, often leads to cultural and socio-economic segregation that, if unchecked, may trigger unrest and ethnic clashes. The explosiveness of such situations reveals that interethnic tolerance is subject to cascades and tipping points as the harbingers of radical political transformations. Once a tipping point is crossed, tolerance typically evaporates fast on its own, but the transformation can further be catalysed by shocks in the form of economic crises or pandemics. Fortunately, concepts and methods studied in statistical physics and its relatives, complexity and network sciences, can help develop our quantitative understanding of large-scale social dynamics. Examples in this context include, but are not limited to, tipping points and phase transitions~\cite{stanley1971introduction, dorogovtsev2008critical, scheffer2012anticipating}, cascade failures~\cite{buldyrev2010catastrophic, yu2016system}, resilience or robustness~\cite{albert2000error, cohen2000resilience}, and recoveries or repairs~\cite{majdandzic2016multiple, majdandzic2014spontaneous}.

\begin{figure}[!t]
\centering{\includegraphics[scale=1.0]{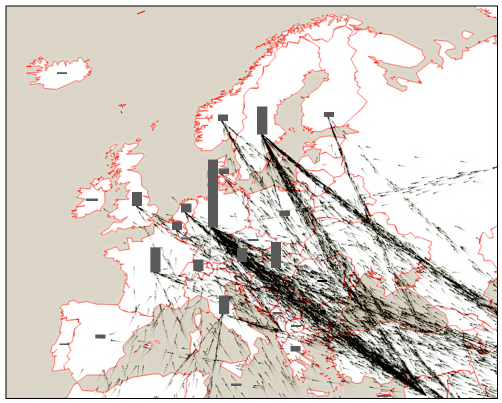}}
\caption{Main corridors and destination countries during the European migrant crisis. The crisis started in 2014, peaked in 2015, and was declared over in 2019. Germany received by far the largest number of asylum requests in 2015, followed by Hungary, Sweden, Austria, Italy, and France (vertical bars).\newline
Source: Reprinted figure from Ref.~\cite{podobnik2019emergence}.}
\label{fig:migrantcrisis}
\end{figure}

When migrations take place, numbers matter. The European Union (EU), for example, handled in an orderly manner about 300,000 asylum seekers yearly up until 2014, but as that number quadrupled in two short years (Fig.~\ref{fig:migrantcrisis}), a deep crisis emerged, prompting prominent political figures to prognosticate an end to the EU as a political project~\cite{greenhill2016open}. Some members of the Schengen Zone responded to the crisis by invoking the 'exceptional circumstances' provision of the Article 26 of the Borders Code to unilaterally reinstate internal border controls, while others chose to erect barbed-wire fences along borders with their non‐Schengen neighbours. Without a clear solution in sight, the debate on the subject polarised around two 'ideological blackmails'; one side argued that the EU's borders must stay fully opened to refugees, whereas the other side argued that the borders must be swiftly and completely closed~\cite{vzivzek2016against}. The crisis furthermore served as a platform for the rise of right-wing political populism throughout Europe~\cite{podobnik2017predicting}. That a large increase in migrant inflows can provoke such a knee-jerk reaction in terms of strengthening border controls and embracing populist policies suggests conditional---albeit heterogeneous---tolerance levels towards diversity of peoples, values, lifestyles, etc. Indeed, the general population seems to heed Karl Popper's maxim that unconditional tolerance must lead to the demise of tolerance~\cite{popper2013open}.

In the language of thermodynamics, and as evidenced by the existence of international borders, human society is far from an equilibrium state. Maintaining a steady non-equilibrium state requires a costly investment of energy, which is precisely the cost that the European countries attempted to avoid by forming the EU and the Schengen Area. The recent migrant crisis, however, is a stark reminder that freeing trade and the movement of people demands utmost care to harmonise relations not only between political elites who orchestrate agreements, but also on the microscopic scale of individual interactions. It is in this latter context that the economist Durlauf argued that statistical mechanics can inform research in social sciences~\cite{durlauf1999howcan}. The basic idea of statistical mechanics---that every atom is influenced by other atoms even beyond just the immediate neighbours---is similar to the ideas of social science that an individual's decisions depend upon the decisions of others. This, in turn, has led to an intriguing possibility that a common mathematical formalism underlies natural and social phenomena~\cite{weidlich1991physics, weidlich2006sociodynamics, galam2008sociophysics}.

When people migrate, be it for personal safety (refugees) or in search of better life (economic migrants), they are forced to establish new contacts and friendships, as well as acclimatise to a new language and culture as a part of an integration process~\cite{bansak2018improving}. Individual socio-economic interactions that take place during this process often pose social dilemmas that involve balancing selfish interests and common good~\cite{dawes1980social, vanlange2013psychology}. Evolutionary game theory offers a formal framework to resolve social dilemmas, bringing into focus five mechanisms of social viscosity that help `lubricate' interactions between individuals. These mechanisms include three types of reciprocities, direct, indirect, and network, and two types of selection, kin and group~\cite{nowak2006five}. Intriguingly, evolutionary game theory has maintained a close tie to statistical physics ever since the discovery of game-driven spatial chaos in structured populations~\cite{nowak1993spatial, hauert2005game}. Later, this has led to an even stronger tie to network science, especially via the emergence of network reciprocity~\cite{ohtsuki2006replicator} and subsequent discoveries that put social networks at the forefront of resolving key social dilemmas~\cite{santos2005scale, santos2008social}.

We have, heretofore, identified statistical physics, complexity and network sciences, and evolutionary game theory as some of the fields that could help model large-scale social dynamics due to migrations. Incidentally, the concept of phase transitions has exerted tremendous influence on all these fields~\cite{newman1999renormalization, hinrichsen2000nonequi, szabo2002phase, perc2016phase}, begging the question as to what causes such widespread fascination with this concept. It has become increasingly evident that, aside from physical systems such as water and ferromagnetic materials, many dynamical complex systems also possess critical thresholds---called tipping or crossover points---near which the system undergoes a swift transition from one state to another~\cite{scheffer2009early}. Among the famous examples of this are ecosystems~\cite{scheffer2003catastrophic}, but similar arguments have been made for stock-market crashes and political revolutions~\cite{levy2005social, levy2008stock}.

In the wake of the EU's migrant crisis, and in view of momentary flirting with border controls and populist policies during the crisis, it is worthwhile examining if the EU members approached a tipping point sometime in 2015 or 2016. Although a devastating phase transition that would put a stop to the EU political project has been averted for now, some signs of a system transitioning from one state to another have remained in public records. For instance, by the second half of 2015, an estimated 58\,\% of the EU citizens harboured concerns about immigration, more than a double compared to a year before (24\,\%) and more than a triple compared to two years before (16\,\%)~\cite{unknown2016standard}. A similar rise of a single overwhelming concern had previously been seen in the second half of 2011, when economic situation preoccupied an estimated 59\,\% of the EU citizens, and thus paved the way for a landmark surrender of the European Central Bank to ultra-low interest-rate policies. More important, however, is the fact that both in 2011 and 2015, a normally multidimensional space of a population's concerns shrank to the point when one dimension dominated all. This shared-concern phenomenon has many parallels with long-range order in physical systems by which a system's remote constituents exhibit highly correlated behaviour~\cite{white1979long}. The ordered state is often established via symmetry breaking upon a phase transition from a disordered state, with a famous example of this being the spontaneous magnetisation of ferromagnetic materials below the Curie temperature~\cite{bertotti1998hysteresis}. Aside from physical systems, long-range order in terms of the cross-correlation between remote constituents has been observed in biological systems, concretely, the healthy operation of cellulo-social collectives~\cite{podobnik2020beta}. Natural and social sciences have thereafter made progress by generalising the described ideas to the notion of self-organisation. This notion has proven relevant to human endeavours on all scales ranging from economics~\cite{witt1997self} to robotics~\cite{pfeifer2007self} to traffic flows~\cite{kerner1998experimental} and many others~\cite{helbing2012social}.

Returning to the problem of migrations, the EU crisis had in the meantime abated, especially when the focus shifted to the ongoing Covid-19 pandemic~\cite{nicola2020socio}. The underlying causes of migrations, however, have not been resolved (e.g., Middle Eastern turmoils and African armed conflicts~\cite{park2015europe}). Adding to this state of affairs the predicted consequences of climate change, future migrant waves are to be expected~\cite{mueller2014heat}. The question is then is the current world order ready?

\subsection{Tolerance}

A way to define the notion of tolerance, or toleration, is mutual acceptance of conflicting worldviews without resorting to suppressive violence. The worldview of others is in this notion seen in a negative light, yet there are vindicating circumstances that outweigh the negatives~\cite{forst2003toleration, sullivan1993political}. Immigration squarely fits into this `tolerance dichotomy' because immigrants are commonly seen as a threat to job security and a source of increased competition in the job market, especially during economic downturns, yet immigrants also serve as a much needed workforce in labour-deficient sectors, especially during the periods of economic growth~\cite{kunovich2013labor}. Aside from the economic dimension, the space of factors that shape attitudes towards immigrants has multiple other dimensions. These can roughly be classified as individual or collective~\cite{paas2012attitudes}. The former include educational attainment, cultural conflicts, political involvement, interpersonal trust, and public safety, whereas the latter include immigrant abundance and foreign investments.

\begin{figure}[!t]
\makebox[\textwidth][c]{\includegraphics[scale=1.0]{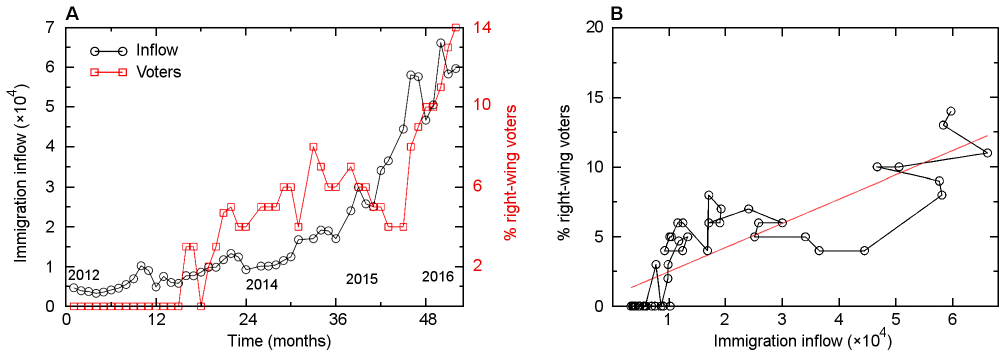}}
\caption{Right-wing populist parties gain popularity when the immigrant inflow increases. \textbf{A,} In Germany, the increasing inflow of immigrants rather clearly coincided with the increasing support for a right-wing populist party. \textbf{B,} Significant regression emerges when the German case is presented as a scatter plot between the immigrant inflow and the percentage of right-wing populist voters.\newline
Source: Reprinted figure from Ref.~\cite{podobnik2017predicting} under the Creative Commons Attribution 4.0 International (CC BY 4.0).}
\label{fig:inflow}
\end{figure}

That there exists a plethora of factors affecting attitudes towards immigrants goes a long way in confirming that tolerance is conditional rather than unconditional. In Germany, for example, a sudden increase in the immigrant inflow at the onset of the migrant crisis tightly correlated with the increase in the popularity of an anti-immigrant right-wing populist party (Fig.~\ref{fig:inflow}). An even starker example is that in nine out of 15 polled Eastern European countries in 2016, more than half the population expressed views that their country should refuse any Syrian refugees~\cite{esipova2017syrian}. Interestingly, none of these nine countries have been the main refugee destination, and yet their appetite for immigration was minute compared to that of their Western European neighbours, suggesting a strong cultural divide in tolerance. The described conditional nature of tolerance in the real world is in sharp contrast with primary legislation, specifically the European Convention on Human Rights, which envisions unconditional tolerance by stating that any refugee or displaced person has the right to EU protection if an asylum is claimed within the EU. It is this latter notion of tolerance that Karl Popper challenged~\cite{popper2013open}, as do simple models of the evolution of human cooperation. In the model of Nowak and Sigmund~\cite{nowak1998evolution}, cooperation evolves via indirect reciprocity as practised by conditional cooperators, whereas unconditional cooperators inadvertently undermine cooperation because they help others regardless of how those others behave, thus giving defectors an edge to invade and prevail. Similarly in the tag-based model of cooperation by Riolo, Cohen, and Axelrod~\cite{riolo2001evolution}, an overly tolerant population is vulnerable to mutants who rarely give help to others.

The conditional nature of tolerance, and especially its manifestations such as anti-immigrant sentiments and right-wing populism~\cite{semyonov2006rise, unknown2016standard, gorodzeisky2016not, podobnik2017predicting, hooghe2018explaining}, indicate that we need a better quantitative understanding of the interplay between the processes of immigration and integration. To this end, Ref.~\cite{podobnik2017fear} studies the balance between immigration and integration rates in relation to the tolerance of the local population for newcomers. Combining the elements of evolutionary game theory and network science, an artificial society is envisioned to form a social network that grows not just intrinsically, but also by attracting newcomers from the outside. Newcomers are attracted to a benefit differential gained by cooperating with the locals. All cooperative interactions, represented by social-network links, yield a per-capita fitness $\Phi_1$ for locals and $\Phi_2$ for newcomers such that the growth of the two sub-populations is given by $N_1(t+1)=\Phi_1(t) N_1(t)$ and $N_2(t+1)=\Phi_2(t) N_2(t)$. Denoting the fraction of newcomers in the population with $f_\mathrm{n}=N_2/N$, where $N=N_1+N_2$ is the population size, we have that $f_\mathrm{n}(t+1)=R(t)f_\mathrm{n}(t)$. Here, $R=\Phi_2/\overline{\Phi}$ is the newcomer-to-population fitness ratio and \smash{$\overline{\Phi} = (1-f_\mathrm{n})\Phi_1+f_\mathrm{n}\Phi_2$} is the average population fitness. As long as the benefit differential is positive, it holds that $R>1$, causing the fraction of newcomers in the social network to grow. The benefit differential is a government policy that cannot be changed by individuals which is why the following individual-scale processes take place. First, as locals get increasingly surrounded by newcomers, the tolerance of the former for the latter gradually saturates. Second, tolerance saturation causes a local to either radicalise and stop cooperating with newcomers altogether, or to remain cooperative but tacitly support radicals. Third, depending on their surroundings, newcomers can pick up the cultural patterns of locals, thus becoming integrated and no longer exerting pressure on the tolerance of locals. The social dynamics arising from the described setup, as will be explained next, distinguishes between a successful and an unsuccessful immigration policy.

\begin{figure}[!t]
\centering{\includegraphics[scale=1.0]{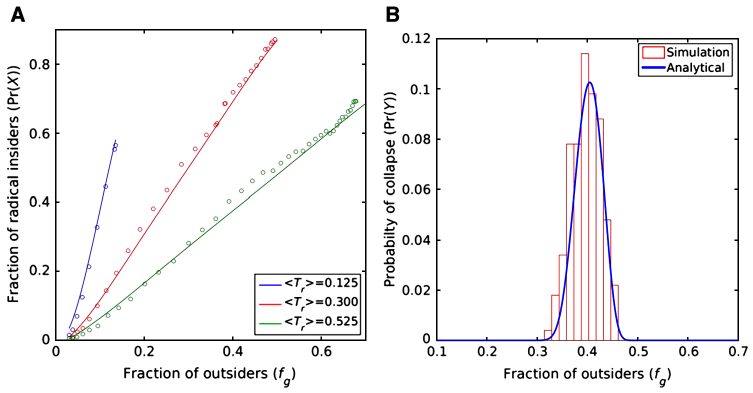}}
\caption{Outcomes of immigration policies. \textbf{A,} Probability of a randomly selected local being radicalised is a monotonically increasing function of the fraction of newcomers. The tolerance of the local population for newcomers controls how strongly the former react to the latter, as well as the maximum abundance of the latter. \textbf{B,} Probability of reaching the critical number of radicals and their tacit supporters to attain a majority need not always increase with the fraction of newcomers. If the tolerance of the local population is sufficient, tolerant locals together with newcomers comprise the majority.\newline
Source: Reprinted figure from Ref.~\cite{podobnik2017fear}.}
\label{fig:prXprY}
\end{figure}

Outcomes of a given immigration policy can be predicted analytically using a mean-field approximation or estimated numerically using computer simulations. Examples of predictable outcomes are the probability $\Pr(X)$ that a randomly chosen local is radicalised or the probability $\Pr(Y)$ that the number of radicals and their tacit supporters exceeds a critical threshold ($N_1^\mathrm{c}$) necessary to initiate oppressive action against newcomers. In a democracy, this might be electing a right-wing populist party to power. The mean-field analysis shows that
\begin{linenomath}
\begin{equation}
\Pr(X)=\frac{1}{k_{\mathrm{max}}-k_{\mathrm{min}}+1} \sum\limits_{l=k_{\mathrm{min}}}^{k_{\mathrm{max}}} \sum\limits_{k=l}^{\lra{k}}\binom{\lra{k}}{k} f_\mathrm{n}^k \left(1-f_\mathrm{n}\right)^{\lra{k} - k},
\label{eq:probX}
\end{equation}
\end{linenomath}
where $k_{\mathrm{min}}$ (respectively, $k_{\mathrm{max}}$) is the number of newcomer neighbours at which the least (resp., the most) tolerant local radicalises, whereas $\lra{k}$ is the average number of neighbours (i.e., the average node degree). The mean-field analysis further shows that
\begin{linenomath}
\begin{equation}
\Pr(Y)=e^{-\lambda} \sum\limits_{l\geq N_1^\mathrm{c}}\frac{\lambda^l}{l!}\approx 1-F_\mathrm{norm}\left(\frac{N_1^\mathrm{c}}{N_1};\frac{\lambda}{N_1};\frac{\sqrt{\lambda}}{N_1}\right),
\label{eq:probY}
\end{equation}
\end{linenomath}
where $\lambda=N_1 f_\mathrm{n} (1-g_\mathrm{r})$ and $g_\mathrm{r}$ is the fraction of radicalised locals. The function $F_\mathrm{norm}$ is the cumulative distribution function of a normal random variable with the mean $\mu=\lambda/N_1$ and the standard deviation $\sigma=\sqrt{\lambda}/N_1$. The mean-field and numerical results are in good agreement (Fig.~\ref{fig:prXprY}).

\begin{figure}[!t]
\makebox[\textwidth][c]{\includegraphics[scale=1.0]{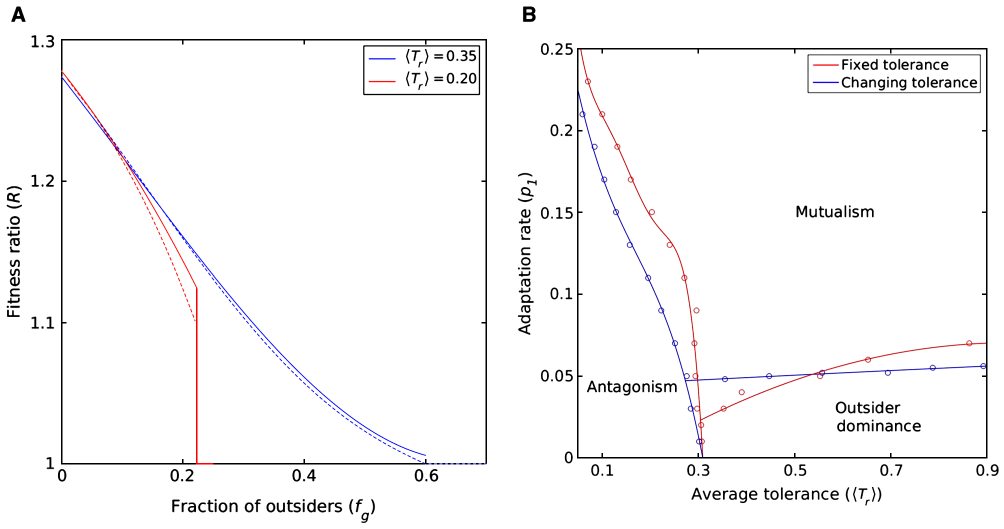}}
\caption{Immigration-policy outcomes. \textbf{A,} Newcomer-to-population fitness ratio $R$ decreases because abundant newcomers saturate the tolerance of locals who then stop being cooperative. If locals are relatively tolerant, the decrease in $R$ is continuous. By contrast, if locals are relatively intolerant, the decrease in $R$ may be discontinuous, signifying a sudden collapse of cooperation between newcomers and locals. Solid curves are simulations, whereas dotted curves are a mean-field approximation. \textbf{B,} Success of immigration policies depends on the tolerance of locals and the integration rate of newcomers. Tolerant locals and integrable newcomers are likely to interact in a mutually beneficial manner. Tolerant locals, however, may lose the majority to non-integrable newcomers. Finally, intolerant locals and non-integrable newcomers make for an explosive combination.\newline
Source: Reprinted figure from Ref.~\cite{podobnik2017fear}.}
\label{fig:phase}
\end{figure}

Ultimately, the phase space of immigration policies can be explored using numerical simulations. By doing so, three types of outcomes reveal themselves:
\begin{itemize}
\item \textit{Mutualism} is a set of equilibrium states reached by a smooth reduction of the fitness ratio $R$ to a level at which locals maintain a sustainable majority (continuous blue curve in Fig.~\ref{fig:phase}A).
\item \textit{Newcomer dominance} is a set of equilibrium states, also reached by a smooth reduction of the fitness ratio $R$, but to a level at which newcomers form a majority (continuous blue curve in Fig.~\ref{fig:phase}A).
\item \textit{Antagonism} is a set of non-equilibrium, absorbing states due to a complete breakdown of cooperation between the two sub-populations (discontinuous red curve in Fig.~\ref{fig:phase}A).
\end{itemize}
If society is sufficiently tolerant and integrates newcomers at a reasonable rate, it is likely for local and newcomer populations to interact in a mutually beneficial manner (Fig.~\ref{fig:phase}B). If, however, integration is too slow, a peaceful transition to the newcomer majority may take place (Fig.~\ref{fig:phase}B). Finally, peace may succumb to turmoil and violence if slow integration is matched with low tolerance (Fig.~\ref{fig:phase}B). These outcomes show that a successful immigration policy must be carefully though out, taking into account (i.e., measuring and monitoring) the tolerance of locals and the integrability of newcomers.

\subsection{Integration and culture}

Defining an exact criterion that would mark the end of the integration process is difficult. To partly circumvent the problem and quantitatively analyse the situation, economists often look at economic integration that compares the earnings of immigrants to those of natives. Early cross-sectional analysis showed that at the time of arrival in the United States immigrants earn 17\,\% less than natives, but the difference disappears in 15 years, whereas after 30 years immigrants even earn 11\,\% more than natives~\cite{chiswick1978effect}. Borjas, however, argued that the cross-sectional picture fails to account for a gradual decline in skills of immigrant cohorts after 1965, concluding that almost all immigrants since the second half of the sixties experience ``the same sluggish relative earnings growth'' and that earnings parity between immigrants and natives is ``extremely unlikely''~\cite{borjas1994economics}.

Cultural integration is particularly difficult to express in quantitative terms. Some anecdotal evidence paints a picture that integration is facilitated by `cultural similarity' between native and immigrant populations. Largely successful immigration policies in this context are considered to be those of Australia and Canada, whereas within the EU borders, there are Luxembourg, Portugal, and Spain. In Luxembourg, for example, 45\,\% of the total population are foreign nationals, who over 90\,\% originate from other EU countries, meaning that the cultural similarity between native and immigrant populations is relatively large. If Luxembourg is too small and too rich to be taken as a representative example, there is the case of Portugal, where somewhat over 3\,\% of the population is foreign and dominated by the Brazilians (who are `cultural neighbours' of the Portuguese). The 3\,\% number, however, is deceptively small because Portugal has naturalised many of its immigrants, illustrating the importance of successful integration of the migrant population. Furthermore, in Spain, slightly under 10\,\% of the population comprises foreign nationals, again dominated by cultural neighbours from Latin American countries. Interestingly, there is also a sizeable proportion of culturally more distant minorities, and Spain has even been targeted by terrorist attacks, yet the right-wing populist movement has never gained a foothold. The Spanish case thus illustrates the importance of a balanced approach which secures peaceful and prosperous co-existence of a highly diverse population.

It is instructive to contrast the above-mentioned successful cases with the situation in France. The French population comprises about 8.5\,\% of foreign nationals, but also an additional 10.5\,\% direct descendants of immigrants. About half of these are of culturally distant origin, such as Arab-Berber, Sub-Saharan, and Turkish. Difficulties in absorbing such a large and culturally distant migrant population prompted a series of terrorist attacks across France to which the native majority responded by increasingly swinging towards the political right. Ultimately, the French case is a sort of antithesis to the Spanish one, and illustrates just how hard it may be to strike the much needed balance when dealing with an inflow of migrants.

The cultural-similarity hypothesis finds some empirical support in the data on economic integration as well. Being of non-EU origins and living in a non-mixed household (i.e., having a non-native spouse) had a significant negative impact on immigrant earnings in a range of EU countries~\cite{buchel2005immigrants}. This negative effect was smaller for immigrants with non-EU origins who lived in a mixed household (i.e., with a native spouse) or for EU-born immigrants who lived in a non-mixed household. Finally, the effect was insignificant for EU-born immigrants who lived in a mixed household. From a theoretical perspective, the concept of cultural similarity was introduced by Axelrod in the context of his seminal model of social influence and cultural change~\cite{axelrod1997dissemination}.

Axelrod's model of social influence is based on three principles:
\begin{enumerate}
\item \textit{Agent-based modelling} means that mechanisms are first specified at the individual scale, and then the consequences of such mechanisms are examined at the population scale to discover the collective or emergent properties of the system.
\item \textit{The lack of a central authority} means that cultural change occurs in a bottom-up manner without coordination from a global overseer.
\item \textit{Adaptive instead of rational actors} means that local circumstances dictate how influencing or being influenced takes place (cf.\ Ref.~\cite{onnela2010spontaneous}). There is no cost-benefit analysis nor forward-looking strategic evaluation.
\end{enumerate}
Culture in Axelrod's model is a multidimensional trait set. Each cultural dimension (say, formal wear) is accompanied by its own traits (say, morning dress, dress suit, ceremonial dress, uniform, religious clothing, national costumes, or frock coats). In an abstract form, a trait is represented by an ordinal number, implying that culture itself is a list of trait numbers. If two actors have the same culture, then all their corresponding trait numbers are equal. Cultural similarity is the percentage of cultural dimensions that share the same trait values. An interaction between two neighbours happens with a probability proportional to their cultural similarity such that the focal actor adopts one trait value from the neighbouring actor.

Axelrod's model is set up to ensure a local convergence of culture, and yet, global polarisation may emerge from the model~\cite{axelrod1997dissemination}. Such an outcome is more likely with fewer cultural dimensions, but lots of traits per dimension. In the context of immigrant integration, these results imply that without a concerted effort from the central government, social influence by natives may actually push immigrants aside rather then integrate them. Crises are likely to exacerbate the problem because they tend to narrow the number of cultural dimensions that pervade political discourse as exemplified in the introduction. The shrinkage in the number of relevant cultural dimensions is consistent with the mathematical formalism of phase transitions in the vicinity of tipping points, suggesting a way for quantitative surveys to measure how far a population is from the tipping point at which radical societal and political changes become probable.

Speaking of phase transitions, physicists have extended Axelrod's social-influence model and subjected it to intensive study in order to understand the model's dynamics~\cite{castellano2000nonequilibrium, weisbuch2002meet, klemm2003nonequilibrium, klemm2003global, weisbuch2004bounded}. Ref.~\cite{castellano2000nonequilibrium}, in particular, discusses a phase transition in Axelrod's model separating an ordered from a disordered phase. In the ordered phase, a dominant cultural region spans a large fraction of the whole system, whereas in the disordered phase, the system's state is fragmented into many cultural regions whose sizes are distributed in a non-trivial manner. The transition turns from continuous to discontinuous as the number of cultural dimensions increases. In relation to immigration policies, these results imply that seemingly similar conditions may lead to greatly different outcomes; there could be a dominant cultural region enveloping most of society, or society could get shattered into many isolated cultural regions. It is questionable whether this latter outcome is compatible with modern-day nation states.

Unlike Axelrod's model in which local convergence is encouraged, there are models of social dynamics that favour divergent individuals~\cite{lee2017modeling, juul2019hipsters, touboul2019hipster}. A typical example is the seceder model~\cite{dittrich2000spontaneous, dittrich2001survival, gronlund2004networking}. This model leads to complex group formation such that, at random times, new groups split from old ones or existing groups go extinct. In medium-sized populations, the distance between two groups that are furthest apart tends to saturate, but in large-sized populations, this distance seems to increase linearly forever. The model thus mimics how subcultures pop in and out of existence~\cite{youth2005modeling}, but may also offer insights into how immigrant communities develop while seeking to preserve cultural heritage and uniqueness. Conditions that lead to radicalisation, or the prevention thereof, could perhaps be better understood by drawing inspiration from the seceder model.

\subsection{Populism and polarisation}

Since its rise in the 1980s, populism has become a tool wielded by parties across the whole political spectrum~\cite{mudde2004populist}. In the context of migrations, however, it is right-wing populism that is of most interest. This particular type of populism is widely seen as pathological and pseudo-democratic in the sense that it is accompanied by a radically xenophobic and authoritarian political programme~\cite{betz1993two}.

Examining the determinants of right-wing populism, Ref.~\cite{oesch2008explaining} suggests that economic parameters play a smaller role than it is often assumed. An analysis of the results of the European Social Survey points in the direction that the electorates of the right-wing populist parties are more afraid of the negative influence of immigrants on a country's culture and heritage than on the country's economy. Data, in fact, suggests that low unemployment rates provide a fertile soil for the growth of right-wing populism~\cite{arzheimer2006political, bjorklund2007unemployment}. Ref.~\cite{smith2010does} furthermore finds that right-wing populist parties benefit from more crime, especially by linking crime to more immigration.

Analysing poll data from a group of EU countries affected by the recent migrant crisis, Ref.~\cite{podobnik2017predicting} finds that over the three-year period from 2014 to 2016 (i.e., in the midst of the migrant crisis), the percentage of right-wing populist voters in a given country depended on the prevalence of immigrants in this country's population and the total immigration inflow into the entire EU. The latter was likely due to the perception that the EU functions as a supranational state in which a lack of inner borders means that `someone else's problem' can easily become `my problem'. When the annual immigrant inflow exceeded 0.4\,\% of a country's population, it invariably led to an annual increase in the right-wing populist voters anywhere between 1\,\% and 5\,\%, implying that a prolonged large inflow could eventually cause right-wing populism to prevail.

Ref.~\cite{podobnik2017predicting} proceeds to mechanistically describe the rise of right-wing populism using a network-science model that accounts for the existence of tipping points in social dynamics. The model is constructed by placing a constant number of native `insider' agents in a random network of social contacts. Immigrant `outsider' agents subsequently enter the network. Each insider notices the percentage of outsiders in their neighbourhood and based on this percentage decides whether or not to turn to right-wing populism. Such a decision is based on local information, but non-local information can also affect decisions, as can misinformation. Three assumptions formalise these ideas:
\begin{enumerate}
\item \textit{Global influences}, such as unfavourable socioeconomics and media reports, are assumed to induce a small negative bias in the decision making of any insider agent anywhere in the network. Direct contact with immigrants is unnecessary for anti-immigrant sentiments as evidenced by the BREXIT referendum in which low-immigrant areas mainly voted Leave~\cite{lawton2016hard}.
\item \textit{Opinion contagion}, complementing global influences, allows the seeds of populism to take root and turn into a full fledged right-wing populist movement. The contagion happens because of connections between insider agents. Namely, when an insider agent is surrounded by a critical number of right-wing populist-supporting neighbours, the agent's decision making becomes negatively biased.
\item \textit{Local influences}, specifically the perceived abundance of outsiders, is assumed to negatively bias the decision making of insider agents whose tolerance threshold has been exceeded by the number of outsider neighbours. For example, in local elections in Greece in November 2010, the far-right Golden Dawn party received only 5.3\,\% of the vote, but in some neighbourhoods of Athens with large immigrant communities the party won nearly 20\,\%.
\end{enumerate}

\begin{figure}[!t]
\makebox[\textwidth][c]{\includegraphics[scale=1.0]{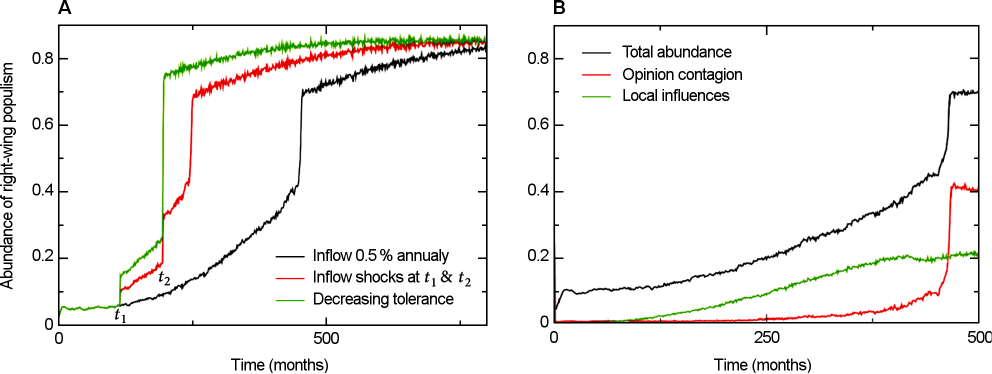}}
\caption{Rise of right-wing populism in a finitely tolerant population. \textbf{A,} Simple assumptions about the interactions of insider and outsider agents in a social network lead to a non-linear dynamics and discontinuous jumps in the abundance of right-wing populist supporters. \textbf{B,} Breakdown of the causes behind right-wing populism reveals that relatively far from the tipping point, the abundance of right-wing populist supporters responds to local influences. As the network approaches its tipping point, however, opinion contagion takes over and accelerates the transition to society dominated by right-wing populism.\newline
Source: Reprinted figure from Ref.~\cite{podobnik2017predicting} under the Creative Commons Attribution 4.0 International (CC BY 4.0).}
\label{fig:populdyn}
\end{figure}

Simulation runs with an annual outsider inflow of 0.5\,\% of the total population show that a tipping point starts manifesting itself as the fraction of outsiders approaches the tolerance threshold of insiders. The abundance of right-wing populist supporters increases non-linearly and eventually undergoes a sudden, discontinuous jump at about 37 years (450 months) into the simulation (black curve in Fig.~\ref{fig:populdyn}A). The jump occurs much earlier if there are inflow shocks. Such shocks happen at times $t_1$ and $t_2$, and cause inflows that are equivalent to about 5\,\% of the total population. The closer the system to the tipping point, the effect of exactly the same shock becomes disproportionately larger (red curve in Fig.~\ref{fig:populdyn}A).

Examining contributions to the rise of right-wing populism, it is evident that global influences can seed populist ideas here and there, but such ideas cannot be sustained without other processes. At first it is local influences that drive the increase in the abundance of right-wing populist supporters in the network (Fig.~\ref{fig:populdyn}B). Opinion contagion remains a relatively small contributor until the tipping point is approached (Fig.~\ref{fig:populdyn}B). Near the tipping point, however, opinion contagion is explosive and overtakes local influences as the main source of right-wing populism. Thereafter right-wing populist supporters dominate society.

Right-wing populist policies are deeply divisive with a large potential to polarise societies. Similarly to how models of social influence helped us to gain insights into cultural integration in the preceding section, here we rely on models of opinion dynamics to gain insights into social polarisation. Ref.~\cite{deffuant2000mixing} is an early and influential work in this context, examining the dynamics of continuous opinions in well-mixed and lattice-structured populations. Continuous opinions nicely correspond to a variety of possible positions on the political spectrum.

The opinion-dynamics model in Ref.~\cite{deffuant2000mixing} has a very simple structure that in some aspects resembles the structure of previously discussed Axelrod's model of social influence. Each agent holds an opinion $x_i\in [0,1]$ that is initially drawn at random from the uniform distribution. In simulations, a pair of agents meet either by chance (well-mixed case) or because they are neighbours in the social network (lattice-structured case). If their opinions are closer than a threshold $d$, $|x_i-x_j|<d$, the two agents' opinions approach one another at a convergence rate $\mu$; otherwise, the opinions remain unchanged. Just as two cultures interact only if they share some common points, so do two opinions interact only if they are close enough to begin with.

\begin{figure}[!t]
\makebox[\textwidth][c]{\includegraphics[scale=1.0]{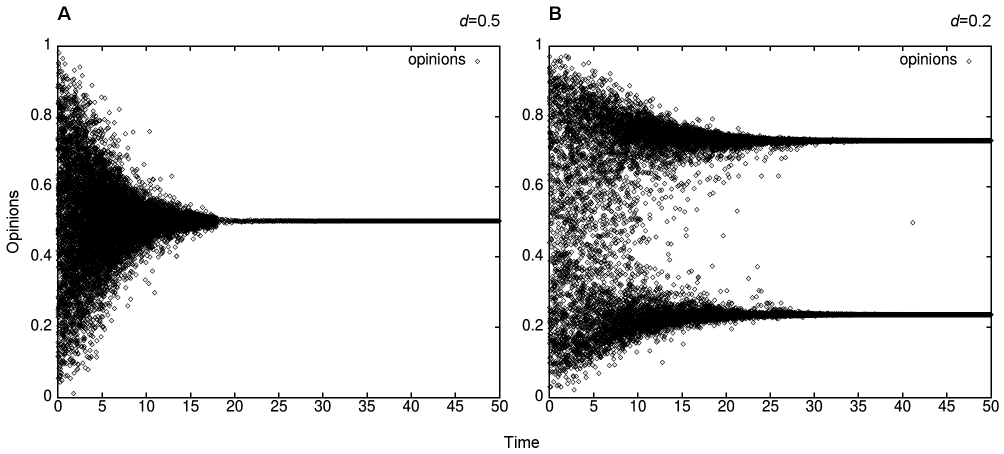}}
\caption{Opinion dynamics may lead to a consensus or polarise society. \textbf{A,} Assuming that individuals tolerate opinions that differ from their own opinion by at most a distance $d$, sufficiently large $d$ values lead to consensus. \textbf{B,} As the $d$ value decreases, society gets polarised between two competing opinions. Even smaller $d$ values cause more than two competing opinions to persevere.\newline
Source: Reprinted figure from Ref.~\cite{deffuant2000mixing}.}
\label{fig:consensus}
\end{figure}

Simulation results reveal a critical role of the threshold $d$. Opinions converge by the very nature of the model, but consensus is observed only for large enough $d$ values (Fig.~\ref{fig:consensus}A). For smaller $d$ values the population polarises, first forming two distinct opinions whose gap cannot be closed (Fig.~\ref{fig:consensus}B), and then even more, with the approximate number of distinct opinions being $\lfloor 2d \rfloor$. In a lattice-structured population, the results are similar, although not without some additional interesting properties. For large enough $d$ values, one percolating opinion dominates the lattice, while a small number of isolated opinions get randomly scattered around. As the $d$ values decrease, it is still possible to observe one percolating opinion, but also a few sizeable clusters of distinct opinions. Interestingly, opinions within any single non-percolating cluster are quite similar but not entirely identical.

Interpreting the described results in the context of migrations, when less tolerant natives start resorting to right-wing populism, many of the right-wing populist policies are unpalatable to more tolerant natives, thus creating a rift in the population. If communication across the rift is very limited, opinion dynamics suggests that society is likely to polarise. Aside from immigration, evidence shows that a similar, catalytic role is played by partisanship, religious orientation, and even geographical differences~\cite{abramowitz2008polarization,saavedra2007identifying}. Polarisation, furthermore, leads to more political participation~\cite{abramowitz2008polarization}, hinting that once a rift forms, each sub-population doubles down on its own opinion. Evidence that rifts across which limited communication occur is found in the phenomenon of echo chambers, that is, groups of like-minded individuals who subscribe to, and mutually reinforce, a certain narrative~\cite{jasny2015empirical, williams2015network, boutyline2017social, usher2018twitter, cinelli2021echo}. Of note here is that the extent to which online echo chambers impact society is a matter of debate~\cite{flaxman2016filter}.

Similar to Axelrod's model of social influence, the opinion-dynamics model of Deffuant et al.~\cite{deffuant2000mixing} has attracted a lot of attention among physicists~\cite{ben2003bifurcations, amblard2004role, stauffer2004simulation, ben2005opinion, holme2006nonequilibrium}. In adaptive networks~\cite{gross2008adaptive}, for instance, agents can sever links with those agents who harbour very different opinions, while preferentially linking with those agents who harbour similar opinions. Such adaptability reduces the chances of reaching both the consensus state and any of the highly fragmented states, thus in effect promoting polarisation~\cite{kozma2008consensus}. Extending the opinion-dynamics model even further, a recent study aims at explaining the above-mentioned phenomenon of echo chambers and polarisation on social media~\cite{baumann2020modeling}. This same body of research also encompasses voting phenomena and the emergence of language. The voter model is concerned with capturing opinion dynamics under a strong influence of an individual's social context~\cite{suchecki2005voter, mobilia2007role, masuda2010heterogeneous}, including statistical regularities of real-world electoral processes~\cite{fernandez2014voter, redner2019reality}. The theory of language emergence is concerned with communication strategies employed by individual speakers, and how the systematic use of such strategies leads to a consensus by which one word conveys the same idea to everyone~\cite{steels1995self, baronchelli2006sharp, loreto2007emergence, baronchelli2008depth, lu2009naming, degiuli}. All this goes to show that the research on opinion dynamics has an enormous scope and breadth, with many important contributions yet to come.

\subsection{Future outlook}

Although it could be argued that the geopolitical situation leading to the migrant crisis that hit Europe in recent years is a one-off event, neither the EU nor other destination countries have the luxury of ignoring large-scale migrations in the near- to mid-term future. Climate change, for example, is expected to displace millions of people over the next few decades~\cite{mueller2014heat}. Worse yet, the recent crisis thought us that large-scale migrations can be accompanied with immeasurable suffering and the loss of human life. For these reasons, we see the need for an interdisciplinary research agenda whose main goal would be attaining a quantitative understanding of migrations and the underlying social dilemmas. The necessary scientific tools that could aid sociological and economic research are available in the form of network science (collective phenomena), statistical physics (tipping points and phase transitions), evolutionary game theory (cooperative Nash equilibria), and others. We furthermore believe that modern approaches should be data driven, and therefore rely on the methods of Bayesian statistics, econometrics, and machine learning.

\paragraph*{Fast and slow migrations} Current research indicates that serious imbalances between (i) the inflow of migrants, (ii) the willingness of the native population to accept them, and (iii) migrant integration causes a knee-jerk political response by which right-wing populists are voted into offices. Subsequently, discriminatory laws and regulations targeting migrants or underrepresented minorities are often proposed and even enacted. It thus becomes essential to consider the role of speed at which processes in social dynamics unravel. If the inflow of migrants is moderate, the native population should feel secure and adapt, but if the inflow is extremely high, the natives may perceive migrants as a threat and---in a sort of Popperian twist---respond antagonistically. These two limits suggest that tolerance is a relative category depending on the magnitude of immigrant inflow, which is an effect that must be included in a genuine model of large-scale migrations. Otherwise, rising right-wing populism may lead to a demise of values such as free thought, liberty, democracy, human rights, and even peace.

\paragraph*{Migration dilemma} Countries today face a migration dilemma that has two complementary dimensions. The first dimension is faced by countries with a positive net inflow of migrants (i.e., immigrants). Such countries, comprising mostly `old democracies', must decide whether to decline or accept immigrants, and in the latter case, decide the rate of acceptance before the integration capacity is fully utilised. Therein lies the dilemma for old democracies. The second dimension is faced by countries with a positive net outflow of migrants (i.e., emigrants), usually economic ones, who seek better life elsewhere. Such countries, comprising mostly `new democracies', must decide whether to replenish their populations with immigrants or face socio-economic instabilities in the future. Therein lies the dilemma for new democracies. Irrespective of how individual countries choose to resolve the dilemma, it is important to recognise that the consequences will likely be non-local because of strong socio-economic interdependencies in a globalised world. This finally returns us to network science, statistical physics, and evolutionary game theory as theoretical frameworks equipped to handle interdependent complex systems~\cite{buldyrev2010catastrophic}, tipping points~\cite{vespignani2012modelling}, and incentives for human behaviour~\cite{arefin2020social}. Intertwining these frameworks with sociological and economic knowledge might provide us with a detailed enough picture of how to resolve the migration dilemma for the benefit of all.

\FloatBarrier

\section{Contagion phenomena}
\label{S:CP}

At the time of writing, the Covid-19 pandemic has been raging around the world for nearly two years, while the cause of this pandemic, the SARS-CoV-2 virus, has infected over 200 million people. At this scale, the pandemic is of course a global public health issue of which general population is acutely aware. Viral outbreaks, however, oftentimes fall short of such widespread awareness despite their recurrent occurrences and deadly potential. To understand just how frequent and dangerous viral outbreaks are, it is illustrative to consider the following examples. Influenza A(H1N1) outbreak started in Mexico in 2009 and subsequently reached 214 countries and regions, causing 18,500 deaths~\cite{neumann2009emergence}. Avian influenza A(H7N9) outbreak started in mainland China in 2013, giving rise to only 419 cases because of no sustained human-human transmission, but still causing 127 deaths due to high virulence in humans~\cite{liu2013origin}. Ebola virus disease emerged in West Africa in 2014 and went on to infect more than 25,000 people, resulting in over 11,000 known fatalities although the true case fatality rate is suspected to be above 70\,\%~\cite{kucharski2014case}. Finally, the MERS-CoV outbreak started in the Middle East in 2012 and proceeded to cause over 1,350 human infections with a death toll of more than 500 people from 26 countries~\cite{corti2015prophylactic}.

For nearly 100 years, the bread and butter of research into disease outbreaks have been compartmental epidemiological models. It was Kermack and McKendrick who in 1927 put forth a mathematical framework for compartmental models in epidemiology~\cite{kermack1927contribution}. Furthermore, the work of Reed and Frost from about the same time, but exposed later by others~\cite{abbey1952examination}, had been the first to introduce a chain-binomial model with the recognisable susceptible-infectious-recovered (SIR) structure. In SIR models, a population exposed to an active pathogen is divided into three compartments. Healthy individuals are susceptible to the pathogen ($S$), infectious individuals transmit the pathogen ($I$), while recovered individuals no longer respond to the pathogen due to acquired immunity ($R$). The simplest SIR model is therefore given by
\begin{subequations}
\begin{linenomath}
\begin{align}
\dd{S}{t} &= -\beta SI,\\
\dd{I}{t} &= \beta SI - \gamma I,\\
\dd{R}{t} &= \gamma I,
\end{align}
\end{linenomath}
\end{subequations}
where $\beta$ and $\gamma$ are infection and recovery rates, respectively. The pathogen cannot spread if $\dd{I}{t}\leq0$, which leads to the condition \smash{$R_0=\frac{\beta S(0)}{\gamma}\leq1$}. The quantity $R_0$ is called the basic reproductive number, and stands for the expected number of cases caused by a single case in a completely susceptible population. Once the epidemic takes off, its decline is subject to the condition \smash{$R_\mathrm{e}=\frac{\beta S(t)}{\gamma}$}, in which case we are talking about an effective reproductive number because the population is no longer completely susceptible. Among the common extensions of the model is to add a population growth rate $\Lambda$, as well as the different mortality rates for each compartment $d_i$, $i\in\{S,I,R\}$. The model then takes the shape
\begin{subequations}
\begin{linenomath}
\begin{align}
\dd{S}{t} &= \Lambda -\beta SI - d_\mathrm{S}S,\\
\dd{I}{t} &= \beta SI - \gamma I - d_\mathrm{I}I,\\
\dd{R}{t} &= \gamma I - d_\mathrm{R}R.
\end{align}
\end{linenomath}
\end{subequations}

Another straightforward extension of the SIR model is to consider a network of populations (e.g., cities). The model then becomes
\begin{subequations}
\begin{linenomath}
\begin{align}
\dd{S_i}{t} &= \Lambda_i -\beta_i S_i I_i - d_i^\mathrm{S}S_i + \sum\limits_{i=1}^n a_{ij} S_j - S_i \sum\limits_{i=1}^n a_{ji},\\
\dd{I_i}{t} &= \beta_i S_i I_i - \gamma_i I_i - d_i^\mathrm{I}I_i + \sum\limits_{i=1}^n b_{ij} I_j - I_i \sum\limits_{i=1}^n b_{ji},\\
\dd{R_i}{t} &= \gamma I_i - d_i^\mathrm{R}R_i + \sum\limits_{i=1}^n c_{ij} R_j - R_i \sum\limits_{i=1}^n c_{ji},
\end{align}
\end{linenomath}
\end{subequations}
where the matrices $A=[a_{ij}]$, $B=[b_{ij}]$, and $C=[c_{ij}]$ quantify mobility rates between populations $i$ and $j$~\cite{li2009global}. Models of this type are often referred to as metapopulation epidemiological models. Since the 1980s, metapopulation models built
around data on the worldwide air-transportation network~\cite{guimera2005worldwide} have become one of the main tools for studying the global spread of emerging epidemics. This is also where our focus lies in the subsequent sections of this chapter.

\subsection{Data-driven metapopulation models}

Metapopulation epidemiological models are built as a network of populations such that transmission dynamics occurs on two scales. The smaller scale describes the local disease spread within each population, while the larger scale describes the disease spread between between populations due to the movements of infectious individuals.

\paragraph*{Disease spread within a population} Let us consider a disease that can be described by an SIR compartmental model, such as pandemic influenza or measles. Susceptible individuals ($S$) become infectious ($I$) at a rate $\beta$ after encountering an infectious individual, whereas each infectious individual becomes recovered or dies at a rate $\mu$. Assuming that the contacts between susceptible and infectious individuals are frequency dependent, we can model transitions in population $i$ from susceptible to infectious individuals and from infectious to recovered individuals over the unit time period $\Delta t$ using the following equations
\begin{linenomath}
\begin{subequations}
\begin{align}
S_i(t + \Delta t) & = S_i(t) - \Delta S_i(t),\\
I_i(t + \Delta t) & = I_i(t) + \Delta S_i(t) - \Delta I_i(t),\\
R_i(t + \Delta t) & = R_i(t) + \Delta I_i(t) \\
\Delta S_i(t) & \sim \mathcal{B}(S_i(t), \beta \frac{I_i(t)}{N_i} \Delta t)\\
\Delta I_i(t) & \sim \mathcal{B}(I_i(t), \mu \Delta t)
\end{align}
\end{subequations}
\end{linenomath}
where $\mathcal{B}(n, p)$ denotes a binomial random variable with the parameters $n$ for the number of trials and $p$ for the probability of success, whereas $N_i = S_i(t) + I_i(t) + R_i(t)$ denotes the size of the $i$th population. For pathogens with a relatively short doubling time (e.g., SARS-CoV-2, MERS-CoV, influenza A(H1N1), Ebola, measles, etc.), it is appropriate to set a short unit time period (e.g., $\Delta t= 0.05$\,d).

\paragraph*{Disease spread between populations} In metapopulation models, individuals are imagined to jump between populations using a transportation network. The simplest scenario is to consider that individual movement is stochastic, which approximates well the international spread of infectious diseases~\cite{hufnagel2004PNAS, colizza2006PNAS, bajardi2011human, lin2018NC}. To account for more complex patterns of human mobility, metapopulation models have also been extended to address the memory effects of individual mobility (e.g., daily commuting)~\cite{balcan2009PNAS, poletto2012SR, vespignani2020science1}, differential social mixing patterns due to socioeconomic stratification~\cite{soriano2018PRX}, and age-related factors~\cite{bedford2015nature}. Using as an example the simplest metapopulation epidemiological model with $G$ populations, the spread of epidemics from each population $i$ to downstream populations that are directly connected to population $i$ can be described using
\begin{linenomath}
\begin{subequations}
\begin{align}
X_i(t) & = \{X_{i1}(t),\dots,X_{iG}(t)\} \sim \mathcal{M}(S_i(t), w_{i1}\Delta t,\dots, w_{iG}\Delta t),\\
Y_i(t) & = \{Y_{i1}(t),\dots,Y_{iG}(t)\} \sim \mathcal{M}(I_i(t), w_{i1}\Delta t,\dots, w_{iG}\Delta t),\\
Z_i(t) & = \{Z_{i1}(t),\dots,Z_{iG}(t)\} \sim \mathcal{M}(R_i(t), w_{i1}\Delta t,\dots, w_{iG}\Delta t),
\end{align}
\end{subequations}
\end{linenomath}
where $\mathcal{M}$ denotes a multinomial random variable, $w_{ij}$ denotes the mobility rate between populations $i$ and $j$, $X_{ij}(t)$, $Y_{ij}(t)$, and $Z_{ij}(t)$ respectively denote the number of susceptible, infectious, and recovered individuals who travel from population $i$ to population $j$ between the times $t$ and $t + \Delta t$.

\paragraph*{The need for data} To build metapopulation epidemiological models for studying real-world epidemics, it is essential to parameterise the connectivity of the underlying transportation network, as well as population flows along this network. This has been made possible by recent advancements in digital data collection and storage (see Chapter~\ref{S:HMN}). Examples of suitable data sources include:
\begin{itemize}
\item \textit{Official Aviation Guide}, a data-subscription service covering the worldwide flight booking database that has previously been used for building state-of-the-art global epidemic simulators (e.g., GLEAM)~\cite{vespignani2020science1}.
\item \textit{mobile service providers} (e.g., the Telenor Group or Orange) whose data-subscription services offer access to anonymous call detail records (CDRs), which allow the quantification of aggregated population movements between cities~\cite{gonzalez2008nature, wesolowski2012science}.
\item \textit{search-engine and social-media companies} (e.g., Google, Baidu, Facebook, Tencent, etc.) whose open-access data portals or data-subscription services offer population-movement information derived from mobile location-based services (LBS)~\cite{lai2020effect, castro2021using}.
\end{itemize}
Compared to the CDR data, the LBS data allows the stratification of population flows by points of interests or modes of transportation. For example, Google's Covid-19 Community Mobility Reports provide six data-streams called `grocery and pharmacy', `parks', `transit stations', `retail and recreation', `residential', and `workplaces'~\cite{nouvellet2021NC}. Tencent's migration-data portal separates travel by aeroplanes, railways, or highways~\cite{xu2021JTM}. Overall, modern sources of digital data usable in epidemiology have become so rich that the term `digital epidemiology' is now being used to describe the extent to which epidemiologists have come to rely on such sources. This is the topic of our next section.

\subsection{Digital epidemiology}

\paragraph*{Definition} Relative to traditional epidemiology that uses data generated by the public health system to understand the incidence, distribution, and possible control of diseases and other health-related factors, digital epidemiology uses digital data generated elsewhere~\cite{salathe2018digital}. Among the main reasons for this is that data collected by professionals in clinics, hospitals, and laboratories is accurate but very costly in terms of logistics, time, labour, and materials. Tapping into alternative and cost-efficient sources therefore seems like a natural way to move forward~\cite{cervellin2017google}.

Internet and mobile-phone uses result in billions of digital-communication records that document health-related behaviours such as symptom reports or attitudes towards vaccines. For example, Google searches provide estimates of influenza activity in a near real-time manner~\cite{ginsberg2009detecting}, whereas mobile-phone data tracks population movements during disease outbreaks~\cite{bengtsson2011improved}. Such capabilities have led to the applications of digital epidemiology in surveilling and predicting the spread of air-borne and vector-borne viruses, parasites, and other pathogens.

\paragraph*{Data sources} Google Trends is a free service providing normalised trends in search activity depending on user-specified geographical regions and time frames. Refs.~\cite{polgreen2008using, ginsberg2009detecting} found that the relative frequency of certain queries about influenza is closely correlated with influenza cases reported by public health agencies in the United States. These pioneering works thus opened the door to using search queries to detect near real-time influenza epidemics.


The original algorithm in Refs.~\cite{polgreen2008using, ginsberg2009detecting} has later been shown to suffer from several major limitations that lead to inaccurate estimates \cite{lazer2014parable, santillana2014can}. The algorithm is static and fails to account for time-series properties such as the seasonality of influenza activity, whereas aggregating multiple query terms into a single variable ignores changes in internet-search behaviour over time. In response, Ref.~\cite{yang2015accurate} proposed the AutoRegression with Google search data (ARGO) model to address the above-said shortcomings. Based on the ARGO model, Ref.~\cite {lu2019improved} further improved influenza prediction by incorporating spatial and temporal synchronicities seen in historical flu activity at the state-level in the US.

In addition to success stories from developed countries such as the data-rich United States, Google Trends has proven valuable in developing countries with poorer data environments for infectious disease nowcasting and forecasting. For example, Ref.~\cite{clemente2019improved} has demonstrated the power of the ARGO model to improve influenza predictions for 8 different Latin-American countries: Argentina, Bolivia, Brazil, Chile, Mexico, Paraguay, Peru, and Uruguay.

In addition to monitoring influenza-like illnesses, internet search queries have also been applied to vector-borne diseases such as dengue, malaria, and Chagas disease in tropical and temperate low- to middle-income countries~\cite{althouse2011prediction, chan2011using, alasaad2013war, ocampo2013using, milinovich2014using}. Because of an increasing internet-access availability, but relatively limited traditional epidemiological data-collection capabilities~\cite{messina2014global}, internet-based surveillance methods have the potential to greatly complement the work of public-health agencies on preparedness against vector-born diseases. Two independent studies~ \cite{althouse2011prediction, chan2011using} have shown that web search query data has a high correlation ranging from 0.82 to 0.99 with dengue activity in Bolivia, Brazil, India, Indonesia, and Singapore. Afterwards, Google developed a prediction tool called Google Dengue Trends to provide timely information to public health agencies from Mexico~\cite{gluskin2014evaluation}, Taiwan~\cite{yang2017advances}, Venezuela~\cite{strauss2017google}, and the Philippines~\cite{ho2018using}.

Twitter is another big-data source of interest to public-health researchers because of the real-time nature of content, precise geotagged locations, and publicly available information. Tweet-based disease surveillance and prediction exploits time-series trends in health-related keyword volumes as a predictor variable in regression models, which is conceptually similar to how Google Trends is used. However, because most tweets employ everyday language to describe a combination of symptoms rather than a diagnosis, there is a need for natural-language processing in order to identify relevant information. Ref.~\cite{gesualdo2013influenza} found a large correlation coefficient between the trends of influenza mentions on Twitter and influenza-like illnesses identified by traditional surveillance systems in the US. Ref.~\cite{broniatowski2013national} further recorded a high prediction accuracy (85\,\%) in relation to the weekly change of influenza prevalence on both national and city scales (e.g., in the New York city).

When it comes to vector-borne diseases, Twitter has proven useful in surveying and predicting dengue in Brazil on national, regional, and city scales~\cite{dealmeida2017dengue}. The best performance has been observed on the national scale, on which the correlation coefficient between predictions and dengue activity was as large as 0.98.

Twitter is also an important source of data to assess health-related behaviours such as concerns about disease outbreaks and attitudes to public-health measures~\cite{deiner2019facebook, hu2021revealing}. Ref.~\cite{salathe2011assessing} relied on publicly available Twitter data to resolve the spatio-temporal sentiment towards a new influenza A (H1N1) vaccine in 2009. The study identified a strong correlation between the online sentiment and estimated vaccination rates by region. Subsequent data-driven simulations of disease spread found that the clusters of a negative vaccine sentiment tend to coincide with an increased likelihood of disease outbreaks. Ref.~\cite{salathe2013dynamics} additionally found that sentiments themselves are contagious. Negative sentiments in particular are more contagious than positive ones.

To exhaust the potential of Twitter data, the following two applications must also be considered. Tweets provide an opportunity to discuss medications, which perhaps could be used to detect drug-related adverse events and thus improve pharmacovigilance. Ref.~\cite{bian2012towards} showcases an analysis of 2 billion tweets in search of adverse events related to 5 different cancer drugs. The analysis identified 239 potential drug users. Each potential case was then examined by two experts, which led to 72 definite positives. From these positives, 27 drug-related adverse event were detected, thus providing a proof-of-concept solution towards improving pharmacovigilance based on Twitter data. The other application arises from the fact that tweets are geo-referenced. This implies the possibility to capture human-movement patterns in order to track and control infectious diseases. Twitter stores geographic coordinates that offer insights into movements on various temporal (from daily onward) and spatial (from local to national to international) scales~\cite{jurdak2015understanding, blanford2015geo}.

Facebook is among the most visited websites in the world, but has not been used as much in public-health contexts because of limited data access in the past. Facebook Data for Good is a recent project aimed at broadening access for social-welfare purposes~\cite{unknown2020facebook}. One of the major advantages of Facebook data is the information on social connections. Albeit these connections are established in an online environment, there is a strong correlation with the geography of health-related activities. Ref.~\cite{kuchler2021jue} thus found that Covid-19 tends to spread between regions with more social-network connections as indicated by Facebook. This showcases that data from online social networks could be used to forecast the spread of air-borne diseases based on proximity indicators derived from digital interactions.

Integrating the information on human movements into epidemiological models leads to insights into the disease spread, as well as the optimal resource allocation to contain the spread. Ref.~\cite{spelta2020after} is an attempt to do so using Facebook movement data while evaluating the economic consequences of alternative lockdown-lifting scenarios in various Italian districts. The results show that there is a tradeoff between disease transmission and worker mobility such that a given economic loss on the national scale induces heterogeneous regional losses. Furthermore, humanitarian organisations need to know where to allocate resources to help people who are most affected by a disease outbreak or other disasters. Ref.~\cite{maas2019facebook} shows how aggregating Facebook usage in areas impacted by such events can be used to produce disaster maps outlining population evacuations and long-term displacements.

\paragraph*{Ensemble estimates} Despite the apparent success of relying on digital-data sources for surveilling and predicting contagions, this methodology has been criticised and concerns have been raised~\cite{cook2011assessing, butler2013google, olson2013reassessing, lazer2014parable, santillana2014can}. Ensemble models have been developed in response to such criticisms, following the idea that combining multiple digital-data sources circumvents the weaknesses that any individual source might have.

Ref.~\cite{santillana2015combining} outlines an ensemble machine-learning model to predict influenza activity in the US by leveraging Google Trends, Twitter data, Flu Near You~\cite{smolinski2015flu}, and the CDC data on influenza-like illnesses. The results demonstrate that combining information from multiple data sources improves real-time predictions up to four weeks ahead. Encouraging results have also been obtained in middle-income countries from Latin America where available data is scarcer~\cite{santillana2016perspectives}. Similarly, Ref.~\cite{mcgough2017forecasting} combines the information on Zika virus disease from Google searches, Twitter, HealthMap~\cite{brownstein2008surveillance}, and suspected cases during the 2015-2016 Latin American outbreak to predict new weekly cases up to three weeks ahead.

\begin{figure}[p]
\makebox[\textwidth][c]{\includegraphics[scale=1.0]{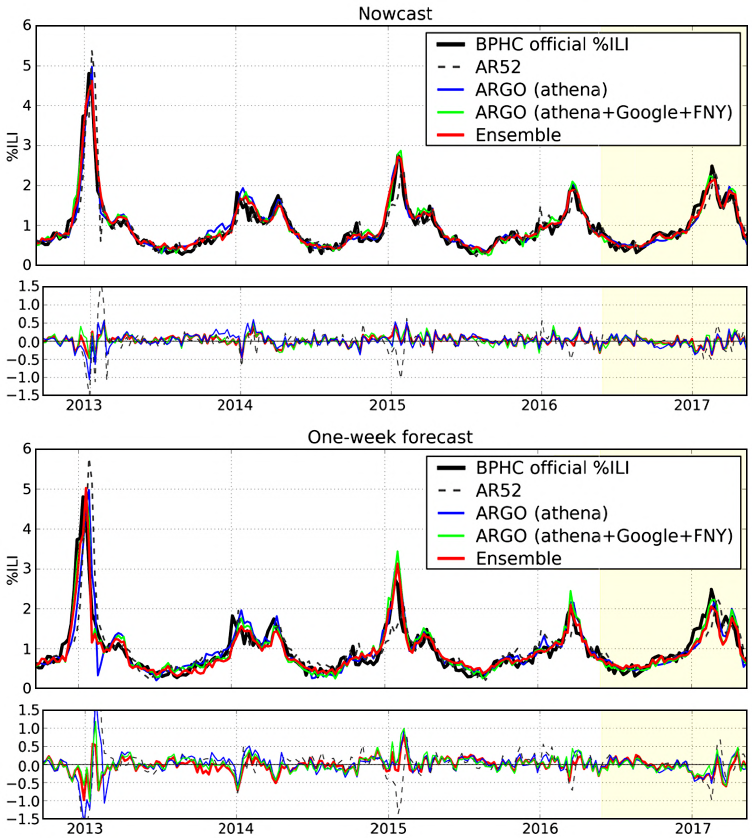}}
\caption{Nowcasts and one-week forecasts (with errors) of influenza-like illnesses in the Boston metropolitan area from September 2012 to May 2017. BPHC stands for Boston Public Health Commission. AR52 stands for an autoregressive baseline model using 52 weeks of past data to make predictions. ARGO stands for autoregression with general online information with the sources of online information being athenahealth (athena), Goolge Trends (Google), and Flu Near You (FNY).\newline
Source: Reprinted figure from Ref.~\cite{lu2018accurate} under the Creative Commons Attribution 4.0 International (CC BY 4.0).}
\label{fig:multipleSources}
\end{figure}

Refining spatial resolution decreases the correlation between predictions by digital-surveillance systems and estimates by public-health systems~\cite{baltrusaitis2018comparison}. Ensemble models are a promising way forward in this context. Ref.~\cite{liu2020real}, for example, combines official health reports, internet searches for Covid-19 on Baidu, news media, and results from an agent-based epidemiological model to produce accurate forecasts two days ahead on the provincial scale in China. Another similar example is Ref.~\cite{lu2018accurate} which combines Google searches, Twitter data, electronic health records, and Flu Near You~\cite{smolinski2015flu} to predict influenza outbreaks in the Boston metropolitan area one week ahead (Fig.~\ref{fig:multipleSources}).

This brief overview by no means exhausts all possible sources of digital data that can be used in epidemiology. What is more, the variety of such sources is bound to increase in the short- to mid-term future. A key development, however, will be to couple them with models that offer mechanistic insights into epidemics. This, of course, includes state-of-the-art metapopulation models discussed in detail in this chapter.

\subsection{Analytical results from metapopulation models}

From the viewpoint of statistical physics, the epidemic threshold is an important concept arising inspired by the studies of critical phenomena and phase transitions~\cite{pastor2015RMP}. The epidemic threshold delineates where in the parameter space an epidemic wanes and where in the parameter space the epidemic intensifies. Interestingly, studying the spread of computer viruses on the Internet has shown that the epidemic threshold can be negligibly small if the node-degree distribution of a network is strongly heterogeneous, as is the case for scale-free networks~\cite{pastor2001PRL, castellano2010PRL}. This was a surprising finding at the time, motivating various searches for novel public-health policies. For example, subsequent attempts to formulate effective vaccination strategies in networks have largely focused on vaccinating hub nodes, that is, those nodes that possess the most connections~\cite{cohen2003PRL}.

The research on epidemic threshold has eventually been extended to metapopulation epidemiological models. Refs.~\cite{colizza2007PRL, colizza2008JTB}, for example, describe a mean-field derivation leading to a closed-form expression for the epidemic threshold of a multiple-population network in which individuals randomly move between populations. The threshold was found to decrease as the network becomes more node-degree heterogeneous. Refs.~\cite{balcan2011NP, balcan2012JTB} extend this line of work to account for recurrent mobility patterns.

A concept of interest in addition to the epidemic threshold is the epidemic arrival time (EAT). After a disease outbreak in a city of origin, for example, Wuhan in case of the Covid-19 pandemic, the disease can spread to other cities through the travel of individuals. The EAT for each downstream city $j$ is the time at which the first infected case is imported into this city. The EAT measures the spreading velocity of the disease and encodes relatively reliable information for inferring key epidemiological parameters in the early phases of novel epidemics~\cite{fraser2009science, lin2018NC, du2020EIDrisk, wu2020lancet}.

Although several seminal studies~\cite{gautreau2008global, tomba2008simple, brockmann2013science} have explored the potential for developing a simple summary statistics to approximate the EAT, a general analytical framework leading to a closed-form expression for the probability distribution of the EAT has remained elusive. To fill this knowledge gap, Ref.~\cite{lin2018NC} derives the probability distribution of the EAT in three metapopulation models with increasingly complex network structure: (i) the simplest two-population model, (ii) the shortest-path tree of the worldwide air-transportation network, and (iii) the whole worldwide air-transportation network.

\begin{figure}[!t]
\makebox[\textwidth][c]{\includegraphics[scale=1.0]{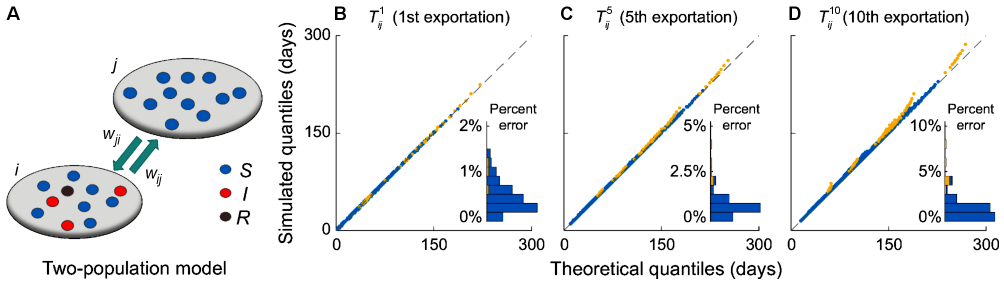}}
\caption{Two-population epidemiological model. \textbf{A,} Model schematics showing the situation in which origin population $i$ connects only to population $j$. \textbf{B--D,} Q-Q plots showing the analytical and simulated quantiles of the random variables $T^1_{ij}$, $T^5_{ij}$, and $T^{10}_{ij}$. Insets show the corresponding histograms of the percent error in $E[T^n_{ij}]$. Simulations entail 100 epidemic scenarios sampled using the Latin-hypercube sampling from the following parameter space. Doubling and generation times both ranged between 3 and 30 days, the seed size $s_i$ ranged between 1 and 100, the mobility rate $w_{ij}$ ranged between $10^{-6}$ and $10^{-3}$, and the population size $N_i$ ranged between 0.1 and 10 million. The latter two parameters were respectively chosen according to the Official Airline Guide (OAG) air-traffic data and the Gridded Population of the World dataset (Version 4). Simulated quantiles for each of these 100 scenarios were computed from 10,000 stochastic realisations. Points on the diagonal indicate that analytically calculated and numerically simulated arrival-time quantiles are equal. Blue and yellow points distinguish scenarios in which $P(X_{ij} \geq n) = 1$ from those in which $P(X_{ij} \geq n) < 1$, where $X_{ij}$ is the number of exportations.\newline
Source: Reprinted figure from Ref.~\cite{lin2018NC} under the Creative Commons Attribution 4.0 International (CC BY 4.0).}
\label{fig:2pop}
\end{figure}

\paragraph*{Two-population model} The simplest problem for which the EAT distribution can be found analytically is that of two populations. An infectious diseases is assumed to originate from population $i$ which is in turn connected to population $j$ (Fig.~\ref{fig:2pop}A). This situation corresponds to initial stages of an epidemic when new infections emerge in the origin population, while all the other populations are aggregated into a single population momentarily unaffected by the disease~\cite{bajardi2011human}. Two key mathematical assumptions~\cite{lin2018NC, gautreau2008global, tomba2008simple} made at this point are:
\begin{enumerate}
\item Exportation of infections from population $i$ to $j$ is a non-homogeneous Poisson process~\cite{ross1996stochastic} with the intensity function (i.e., the expected number of
infections exported from population $i$ to population $j$ at time $t$) given by $w_{ij} I_i(t)$, where $I_i(t)$ is the number of infectious people (i.e., disease prevalence) in population $i$ at time $t$, and $w_{ij}$ is the per-capita mobility rate from population $i$ to population $j$.
\item After the new epidemic establishes itself in origin population $i$, the first few exportations from population $i$ to population $j$ occur while disease prevalence is still growing exponentially in the origin, that is, $I_i(t) = s_i \exp(\lambda_i t)$, where $s_i$ is t initial seed size in origin population $i$ at time $t=0$, and $\lambda_i$ is the local epidemic growth rate.
\end{enumerate}

With the above-stated assumptions, the $n$th EAT in population $j$, denoted $T^n_{ij}$, is a random variable whose probability density function can be expressed in closed form
\begin{linenomath}
\begin{equation}
f_n(t|\lambda_i,\alpha_{ij}) = \left(\frac{e^{\lambda_i t} - 1}{\lambda_i}\right)^{n-1} \frac{\alpha^n_{ij}}{(n-1)!}\exp\left(\lambda_i t - \alpha_{ij}\frac{e^{\lambda_i t} - 1}{\lambda_i}\right),
\label{eq:EQ2pop}
\end{equation}
\end{linenomath}
where $\alpha_{ij} = s_i w_{ij}$ is the adjusted mobility rate. The last expression can be validated numerically. To this end, analytical and simulated EATs were compared over a wide range of epidemic scenarios (Fig.~\ref{fig:2pop}). Among others, doubling and generation times varied between 3 and 30 days, which was enough to cover many present-day infectious diseases. Covid-19 has a doubling time of 5-7 days, whereas Ebola has a doubling time of more than 20 days.

Eq.~(\ref{eq:EQ2pop}) can be used to derive a number of corollaries:
\begin{enumerate}
\item Exportation of the first $n$ infections is a non-homogeneous Poisson process with the intensity function $\alpha_{ij} \exp(\lambda_i t)$.
\item The cumulative distribution function of the $n$th EAT is
\begin{linenomath}
\begin{equation}
F_n(t|\lambda_i,\alpha_{ij}) =
\Gamma\left[n, \frac{\alpha_{ij}}{\lambda_i}\left(e^{\lambda_i t} - 1\right) \right],
\label{eq:EQ2popCDF}
\end{equation}
\end{linenomath}
where $\Gamma$ denotes the lower incomplete Gamma function.
\item The expected first EAT is
\begin{linenomath}
\begin{equation}
E[T^1_{ij}] = \frac{1}{\lambda_i}\exp\left(\frac{\alpha_{ij}}{\lambda_i}\right)\mathbf{E}_1\left(\frac{\alpha_{ij}}{\lambda_i}\right),
\label{eq:EQ2popMean}
\end{equation}
\end{linenomath}
where $\mathbf{E}_m(x) = x^{m-1} \int^{\infty}_x \mu^{-m} e^{-\mu} \dup{\mu} $ is the exponential integral.
\item If $\alpha_{ij} \ll \lambda_i$ and $\gamma$ denotes the Euler-Mascheroni constant, then the expected first EAT can be approximated using
\begin{linenomath}
\begin{equation}
E[T^1_{ij}] = \frac{1}{\lambda_i}\left[ \ln\left( \frac{\lambda_i}{\alpha_{ij}} \right) - \gamma \right],
\label{eq:EQ2popMean2}
\end{equation}
\end{linenomath}
which is equivalent to the first-EAT statistic in Ref.~\cite{gautreau2008global}.
\item The expected $n$th EAT is
\begin{linenomath}
\begin{equation}
E[T^n_{ij}] = \frac{1}{\lambda_i}\exp\left( \frac{\alpha_{ij}}{\lambda_i} \right) \sum^n_{m=1} \mathbf{E}_m\left( \frac{\alpha_{ij}}{\lambda_i} \right).
\label{eq:EQ2popMean3}
\end{equation}
\end{linenomath}
\item For any positive integers $m$ and $n$, $m < n$, the probability density function of $T^n_{ij}-T^m_{ij}$ conditional on $T^m_{ij}$ is
\begin{linenomath}
\begin{equation}
f_{n-m}\left(t|\lambda_i,\alpha_{ij}e^{\lambda_i T^m_{ij}}\right),
\end{equation}
\end{linenomath}
which can be reinterpreted as the probability density function of the $(n-m)$th EAT with the seed size $s_i \exp(\lambda_i T^m_{ij})$. Using this relation recursively, we deduce that the joint probability density function of $T^1_{ij}=t_1, \ldots, T^n_{ij}=t_n$ is
\begin{linenomath}
\begin{equation}
\prod^n_{m=1} f_1\left(t_m|\lambda_i,\alpha_{ij}e^{\lambda_i t_{m-1}}\right),
\label{eq:EQ2popprod}
\end{equation}
\end{linenomath}
for all $0=t_0 < t_1 < t_2 < \ldots < t_{n-1} < t_n$.
\item The expected $(n-1)$th EAT given an epidemic that starts at time $T^1_{ij}$ with the seed size $s_i\exp(\lambda_i T^1_{ij})$ is
\begin{linenomath}
\begin{equation}
E[T^n_{ij}|T^1_{ij}] = T^1_{ij} + \frac{1}{\lambda_i}\exp\left( \frac{\alpha_{ij}}{\lambda_i}e^{\lambda_i T^1_{ij}} \right)\sum^{n-1}_{m=1}\mathbf{E}_m\left( \frac{\alpha_{ij}}{\lambda_i}e^{\lambda_i T^1_{ij}} \right).
\label{eq:CondExp}
\end{equation}
\end{linenomath}
\end{enumerate}
These corollaries will prove useful in extending the two-population model to the analysis of the epidemic arrival process in the shortest-path tree of the worldwide air-transportation network, and the whole worldwide air-transportation network.

\begin{figure}[!t]
\makebox[\textwidth][c]{\includegraphics[scale=1.0]{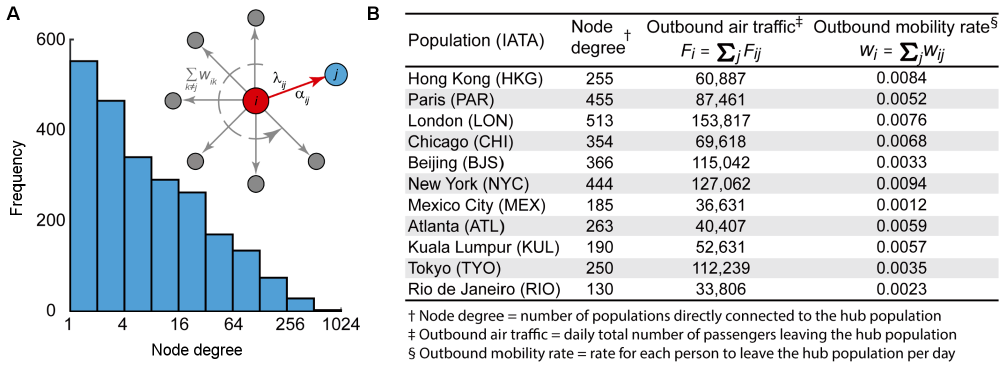}}
\caption{Network properties of hub populations. \textbf{A,} Histogram shows the distribution of node degrees for all populations in the worldwide air-transportation network. The node degree of a given population equals the number of populations to which this population is directly connected. The inset illustrates that travel-hub population $i$ is connected to multiple populations, including population $j$. \textbf{B,} List of several major hub cities situated all over the world. Shown are the node degree, the daily outbound-traffic volume, and the daily outbound per-capita mobility rate.}
\label{fig:HubPop}
\end{figure}

\begin{figure}[!t]
\centering{\includegraphics[scale=1.0]{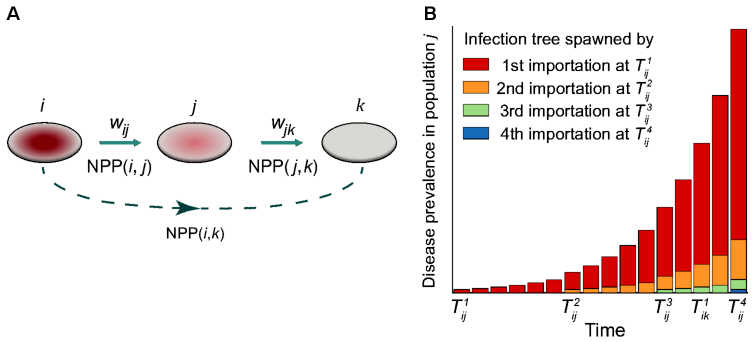}}
\caption{Continuous-seeding effect. \textbf{A,} Schematic of the epidemic arrival process (mathematically, a non-homogenous Poisson process) over an acyclic path connecting origin population $i$ to population $k$ via population $j$ (i.e., $\psi: i \rightarrow j \rightarrow k $). \textbf{B,} In this example, the epidemic arrives in population $k$ after population $j$ has imported three infections from the origin, that is, $T^3_{ij} < T^1_{ik} < T^4_{ij}$. In the absence of continuous seeding adjustment, infection
trees spawned by the second and subsequent importations in population $j$ are ignored.\newline
Source: Reprinted figure from Ref.~\cite{lin2018NC} under the Creative Commons Attribution 4.0 International (CC BY 4.0).}
\label{fig:contiSeed}
\end{figure}

\paragraph*{Modelling the shortest-path tree} A dominant sub-network of the worldwide air-transportation network is its shortest-path tree. In this sub-network, each downstream population is connected to the epidemic origin via only one path. Ref.~\cite{brockmann2013science} suggests that an emerging epidemic spreads from the origin population to other populations mainly through the shortest-path tree, that is, the infrastructure of the shortest-path tree drives the global spread of the disease across the worldwide air-transportation network. Ref.~\cite{lin2018NC} demonstrates that the $n$th EAT $T^n_{ik}$ from origin population $i$ to any population $k$ in the shortest-path tree is accurately characterised by Eq.~(\ref{eq:EQ2pop}), but the local epidemic growth rate $\lambda_i$ and the adjusted mobility rate $\alpha_{ij}$ need to be re-parameterised to account for what is called \textit{the hub effect} (Fig.~\ref{fig:HubPop}) and \textit{the continuous-seeding effect} (Fig. \ref{fig:contiSeed}).

Travel hubs such as Hong Kong, London, and Paris are characterised by direct flights to many locations. In network-science terminology, the node degree of travel hubs is much larger than unity. This creates many opportunities for infection exportation, perhaps to the point that the local growth of the disease prevalence, $I_i(t)$, is noticeably reduced. If indeed a noticeable proportion of infections travel outward as the epidemic unfolds, the local epidemic growth rate, $\lambda_i$, needs an adjustment.

\begin{figure}[!t]
\makebox[\textwidth][c]{\includegraphics[scale=1.0]{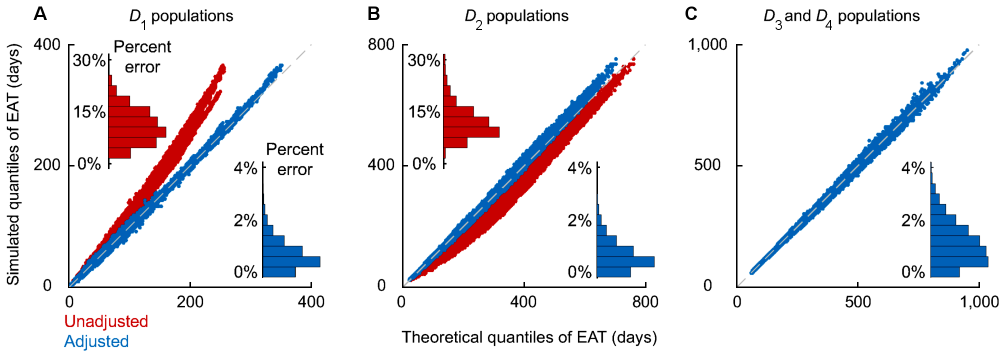}}
\caption{Numerical validation of analytics for the shortest-path tree of the worldwide air-transportation network. Shown are the Q-Q plots comparing the analytical and simulated quantiles of various EATs for downstream populations in the shortest-path tree. Insets show the corresponding histograms of the percent error in the expected EATs. The origin of the epidemic was assumed to be in Hong Kong. The same 100 epidemic scenarios as in Fig.~\ref{fig:2pop} were used. The symbol $D_c$ stands for the set of all populations that are separated by $c$ degrees of separation from the epidemic origin. \textbf{A,} The results for the set $D_1$ before (red) and after (blue) adjusting for the hub effect. \textbf{B,} The results for the set $D_2$ before (red) and after (blue) adjusting for the hub effect and continuous seeding. \textbf{C,} The results for the sets $D_3$ and $D_4$ after adjusting for the hub effect and continuous seeding, and performing path reduction.\newline
Source: Reprinted figure from Ref.~\cite{lin2018NC} under the Creative Commons Attribution 4.0 International (CC BY 4.0).}
\label{fig:WANSPT}
\end{figure}

Suppose that hub population $i$ is directly connected to two or more populations, one of which is population $j$ (Fig.~\ref{fig:HubPop}A). In the shortest-path tree, all infectious individuals who disperse from population $i$ to populations other than $j$ no longer contribute to disease exportations to population $j$. A consequence is that the probability density function of the $n$th EAT is still given by Eq.~(\ref{eq:EQ2pop}), but the local epidemic growth rate in hub population $i$ from the perspective of population $j$ must be adjusted to
\begin{linenomath}
\begin{equation}
\lambda_{ij} = \lambda_i - \sum_{k\neq j} w_{ik}.
\label{eq:hubEffect}
\end{equation}
\end{linenomath}
The random variable $T^n_{ij}$ therefore has the probability density function $f_n(t|\lambda_{ij}, \alpha_{ij})$, implying that the disease prevalence in hub population $i$ grows exponentially at the effective growth rate $\lambda_{ij}$, while the number of exported infections from population $i$ to population $j$ at time $t$ remains the same as before, that is, $w_{ij} I_i(t)$. The need for the described adjustment can be numerically validated in a similar manner as the two-population model (Fig.~\ref{fig:WANSPT}A).

Although a single seeding event seeds the origin population with the disease, all other populations in the shortest-path tree are continually seeded by infections exported from upstream populations (Fig.~\ref{fig:contiSeed}). Such continuous seeding has been documented in the case of Zika virus in Florida coming from the Caribbean~\cite{grubaugh2017nature} and SARS-CoV-2 in the UK coming from Europe~\cite{filipe2020natmicrob}.

Let $D_c$ be the set of populations that are separated by $c$ degrees of separation from the origin population in the shortest-path tree. Let furthermore population $k$ in $D_2$ be connected to origin population $i$ via population $j$ along the path $\psi: i \rightarrow j \rightarrow k$. After the epidemic arrives in population $j$ at time $T^1_{ij}$, population $i$ continues to export infections to population $j$ before the epidemic arrives in population $k$ at time $T^1_{ik}$ (Fig.~\ref{fig:contiSeed}). Based on the two-population model, each imported infection that arrives in population $j$ at times $T^1_{ij}, T^2_{ij}, \ldots$ causes new exponential spreading at the hub-adjusted rate $\lambda_{jk}$. The overall disease prevalence, $I_j(t)$, in population $j$ at time $t$ is therefore the sum of disease prevalence over all exponential spreadings
\begin{linenomath}
\begin{equation}
I_j(t) = \sum^{\infty}_{m=1} \mathbf{I}\{t > T^m_{ij}\} e^{\lambda_{jk} (t - T^m_{ij})},
\end{equation}
\end{linenomath}
where $T^m_{ij}$ is the $m$th EAT in population $j$ and $\mathbf{I}\{\cdot\}$ is the indicator function.

Based on the two-population model, the exportation of infections from population $j$ to population $k$ is a non-homogeneous Poisson process with the intensity function $w_{ij} I_j(t)$, which itself is a complex stochastic process due to its dependence on the random variables $T^1_{ij}, T^2_{ij}, \ldots$. This leads to the probability density function of the random variable $T^n_{ik}$ (for $n=1,2,\ldots$), conditional on the prevalence $I_j(t)$ (and hence $T^1_{ij}, T^2_{ij}, \ldots$), in the form
\begin{linenomath}
\begin{equation}
g_n(t|w_{jk} I_j) = f_\mathrm{Poisson}\left[n-1, w_{jk}\int\limits^t_0 I_j(u) \dup{u} \right] w_{jk} I_j(t),
\end{equation}
\end{linenomath}
where $f_\mathrm{Poisson}(\cdot,\mu)$ is the probability mass function of a Poisson random variable with the mean $\mu$. Consequently, the unconditional probability density function of the random variable $T^n_{ik}$ is
\begin{linenomath}
\begin{equation}
E_{T^1_{ij},T^2_{ij},\ldots}\left[ g_n(t|w_{jk} I_j) \right],
\end{equation}
\end{linenomath}
where integration proceeds over the joint probability density function of $T^1_{ij}=t_1,T^2_{ij}=t_2,\ldots$, which in turn is given by Eq.~(\ref{eq:EQ2popprod}) after replacing $\lambda_i$ with $\lambda_{ij}$ to account for the hub effect.

Ref.~\cite{lin2018NC} proceeds to demonstrate that the complex dependence on $T^1_{ij}=t_1,T^2_{ij}=t_2,\ldots$ can be simplified with a little loss of accuracy. To this end, the following certainty equivalent approximation (CEA) is made; $T^m_{ij}\approx E[T^m_{ij}|T^1_{ij}]$ for all $m>1$. An intuitive interpretation is that most of uncertainty in the $m$th EAT in population $j$ is due to uncertainty in the first EAT in this population, where the latter uncertainty is characterised by the probability density function in Eq.~(\ref{eq:EQ2pop}) after inserting $n=1$ and replacing $\lambda_i$ with $\lambda_{ij}$ to account for the hub effect. The prevalence $I_j(t)$ becomes
\begin{linenomath}
\begin{align}
I^\mathrm{CEA}_j(t) &  = \sum^{\infty}_{m=1} \mathbf{I}\left\{ t > E\left[T^m_{ij}|T^1_{ij}\right] \right\}e^{\lambda_{jk}\left(t - E[T^m_{ij}|T^1_{ij}]\right)},\nonumber\\
& = \sum^{\infty}_{m=1} \mathbf{I}\left\{ t > T^1_{ij} + \Delta T^m_{ij} \right\} e^{\lambda_{jk} \left( t - T^1_{ij} - \Delta T^m_{ij}  \right)},
\end{align}
\end{linenomath}
where from Eq.~(\ref{eq:CondExp}) it follows
\begin{linenomath}
\begin{align}
\Delta T^m_{ij} & = E\left[T^m_{ij}|T^1_{ij}\right] - T^1_{ij},\nonumber\\
& = \frac{1}{\lambda_{ij}} \exp\left( \frac{\alpha_{ij}}{\lambda_{ij}} e^{\lambda_{ij} T^1_{ij}} \right)\sum^{m-1}_{q=1}\mathbf{E}_q \left( \frac{\alpha_{ij}}{\lambda_{ij}} e^{\lambda_{ij} T^1_{ij}} \right).
\end{align}
\end{linenomath}
Finally, the unconditional probability density function of $T^n_{ik}$ is given by
\begin{linenomath}
\begin{equation}
E_{T^1_{ij}} \left[ g_n \left( t | w_{jk} I^\mathrm{CEA}_j \right) \right].
\label{eq:eq2CEAPDF}
\end{equation}
\end{linenomath}

Although the last expression is perfectly suitable to handle EATs for all populations in the set $D_2$, additional approximations are necessary to handle populations in sets $D_c$, $c\geq 3$. One such approximation is \textit{path reduction}~\cite{lin2018NC} by which the path $\psi: i\rightarrow j \rightarrow k$ is treated as the direct path $\psi': i\rightarrow k$. This allows us, for $n=1$, to replace the probability density function in Eq.~(\ref{eq:eq2CEAPDF}) with the probability density function in Eq.~(\ref{eq:EQ2pop}), but with suitably corrected parameters $f_1(t|\lambda_\psi,\alpha_\psi)$. The corrected parameters $\lambda_\psi$ and $\alpha_\psi$ are obtained by minimising the Kullback-Leibler divergence (i.e., the relative entropy)~\cite{cover2006elements, lin2018NC} for the first EAT through the path $\psi$
\begin{linenomath}
\begin{equation}
D_\mathrm{KL}=\int\limits_0^\infty E_{T^1_{ij}} \left[ g_1 \left( t | w_{jk} I^\mathrm{CEA}_j \right) \right] \ln\frac{ E_{T^1_{ij}} \left[ g_1 \left( t | w_{jk} I^\mathrm{CEA}_j \right) \right]}{f_1(t|\lambda_\psi,\alpha_\psi)}.
\end{equation}
\end{linenomath}
The quantity $D_\mathrm{KL}$ can be understood as a measure of how much one probability distribution differs from another, reference probability distribution. By minimising the quantity $D_\mathrm{KL}$, we therefore reduce the two leg path $\psi$ to the one-leg path $\psi'$ such that the probability distribution of the first EAT in population $k$ remains unaffected. Epidemic spread from the origin population $i$ to any population $k$ in $D_2$ is thus regarded as a two-population problem, but with the local epidemic growth rate $\lambda_\psi$ and the adjusted mobility rate $\alpha_\psi$. If we now have an even longer path $\varphi: i\rightarrow j \rightarrow k \rightarrow m$ (i.e., $\varphi\in D_3$), we first apply path reduction to the two-leg part $i\rightarrow j \rightarrow k$, and then treat the remainder with the methods developed for the set $D_2$ (Fig.~\ref{fig:WANSPT}B,\,C).

\begin{figure}[!t]
\centering{\includegraphics[scale=1.0]{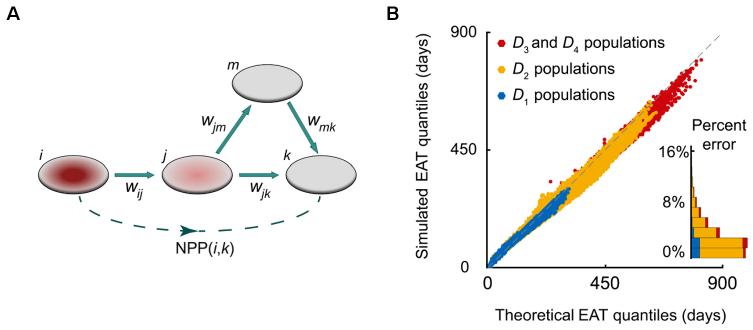}}
\caption{Numerical validation of analytics for the whole worldwide air-transportation network. \textbf{A,} Multiple acyclic paths may connect downstream population $k$ to origin population $i$. \textbf{B,} Q-Q plots comparing the analytical and simulated quantiles of various EATs for downstream populations in the whole worldwide air-transportation network. Insets show the corresponding histograms of the percent error in the expected EATs. The origin of the epidemic was again assumed to be in Hong Kong. The same 100 epidemic scenarios as in Fig.~\ref{fig:2pop} were used. Data points are coloured in blue for 255 $D_1$ populations, yellow for 1839 $D_2$ populations, and red for 207 $D_3$ and 7 $D_4$ populations.\newline
Source: Reprinted figure from Ref.~\cite{lin2018NC} under the Creative Commons Attribution 4.0 International (CC BY 4.0).}
\label{fig:WANFull}
\end{figure}

\paragraph*{Modelling the whole worldwide air-transportation network} To find EATs for the whole worldwide air-transportation network, it is necessary to account for the fact that each downstream population $k$ can be connected to origin population $i$ via multiple paths (Fig.~\ref{fig:WANFull}A). Furthermore, paths may include cycles and may intersect one another, implying that there is some degree of dependence between them. It is intuitive to assume that cycles introduce considerable delays, which makes path with cycles largely irrelevant relative to acyclic paths. If it also holds that dependence between acyclic paths is sufficiently weak to treat them as almost independent, then the following calculation becomes plausible. First, all paths that connect origin population $i$ to downstream population $k$ should be decomposed into a set $\Psi_{ik}$ of `independent' acyclic paths. Then, all these pseudo-independent paths are fully reduced until they are characterised by the parameters $\lambda_\psi$ and $\alpha_\psi$. Finally, EATs for population $k$ can be approximated by the superposition of non-homogeneous Poisson processes~\cite{ross1996stochastic} such that the intensity function of the superpositioned process is $\sum_{\psi\in\Psi_{ik}} \alpha_\psi \exp(\lambda_\psi t)$. Numerical results show that the entire analytical framework combining the two-population analytics, adjustment for the hub effect, adjustment for continuous seeding, path reduction, and path superposition accurately estimates EATs for almost all populations in the worldwide air-transportation network (Fig.~\ref{fig:WANFull}B).

\subsection{Bayesian inference}

Bayesian inference is a class of statistical methods for data analysis and parameter estimation based on Bayes' theorem~\cite{vandeschoot2021bayesian}. Let $P(A)$ and $P(B)$ be the probabilities of observing events $A$ and $B$, respectively. Let $P(A|B)$ be the conditional probability of observing event $A$ given the observation of event $B$, and $P(B|A)$ the conditional probability of observing event $B$ given the observation of event $A$. Bayes’ theorem says that these two conditional probabilities are related by
\begin{linenomath}
\begin{equation}
P(A|B) = \frac{P(B|A) P(A)}{P(B)}.
\label{eq.bayes}
\end{equation}
\end{linenomath}
An analogous relationship links data to model parameters. With $D$ denoting the observed data and $\theta$ denoting the model parameters of the data generating process, Eq.~(\ref{eq.bayes}) can be rewritten as
\begin{linenomath}
\begin{equation}
P(\theta | D) = \frac{P(D|\theta) P(\theta)}{\int P(D|\theta) P(\theta) d\theta}.
\label{eq.bayesRule}
\end{equation}
\end{linenomath}
The prior probability $P(\theta)$ represents initial beliefs about the model parameters before any data analysis. The likelihood function $P(D|\theta)$ represents the conditional probability of observing the data $D$ given the parameters $\theta$. The posterior probability $P(\theta|D)$ summarises the updated knowledge about the parameters upon synthesising the prior knowledge with the observed data.

The denominator of Eq.~(\ref{eq.bayesRule}) requires integration over all model parameters $\theta$, which is often very complex and analytically intractable. Numerical integration may also become computationally prohibitive as the number of parameters increases. However, the denominator only plays the role of a normalisation constant. Sufficient for inference is the proportionality
\begin{linenomath}
\begin{equation}
P(\theta | D) \propto P(D|\theta) P(\theta).
\label{eq.posterior}
\end{equation}
\end{linenomath}
Accordingly, Bayesian inference mainly comprises two steps; (i) formalising the prior distribution of each model parameter using background knowledge and literature reviews and (ii) designing the likelihood function by using probabilistic models to account for the underlying data-generating process.

Specifying prior distributions is a nontrivial task~\cite{vandeschoot2021bayesian}. Existing studies in the fields of infectious diseases modelling, network theory, bioinformatics, and statistical physics tend to use the simplest uninformative flat or diffuse prior~\cite{lambert2018student}. The main aim of such a simplification is to assess the capacity of the likelihood model in fitting the observed data. Despite being useful in resolving low-dimensional problems with a few parameters to infer, the flat or diffuse prior cannot be regarded as a universal tool for fitting. In particular, for high-dimensional problems with many parameters to infer, the usage of flat or diffuse priors can lead to biased estimations or convergence failures~\cite{vandeschoot2021bayesian}. Recent progress in the field suggests that even weakly informative priors may be a better option. How to setup prior distributions is explained in detail in Refs.~\cite{gelman2017prior, lemoine2019moving}.

To estimate model parameters, Bayesian methods often formulate the likelihood function using probabilistic models. The purpose of such models is to describe the data-generating process behind observed data. In this section, we proceed by outlining two epidemiological case studies to explain how to develop the likelihood function using probabilistic models for Bayesian inference.

\paragraph*{Inferring the basic reproductive number $R_0$ for the 2009 influenza A(H1N1) in Greater Mexico City} As briefly discussed at the beginning of this section, the basic reproductive number $R_0$ is an important epidemiological quantity, giving the expected number of secondary infections induced by an infectious person in a fully susceptible population. Estimating $R_0$ during the early stage of an outbreak is key to understanding the potential of the disease for interpersonal transmissions, the requirements for a vaccine to achieve herd immunity, and the extent of non-pharmaceutical interventions to control the outbreak.

Here, we look at the estimation of $R_0$ in the case of the 2009 influenza A (H1N1) epidemic in Greater Mexico City. Ref.~\cite{balcan2009seasonal} is well-known for presenting a maximum likelihood method that was used to estimate $R_0$ in this particular case. The method runs a large number of computer simulations to explore the parameter space, which is computationally so intensive that it requires the use of a supercomputer. Ref.~\cite{lin2018NC} lessens the computational burden by using Bayesian inference in conjunction with disease-exportation records from Mexico to the first 12 countries as summarised in Ref.~\cite{balcan2009seasonal}. This inference combines the two-population model in Eq.~(\ref{eq:EQ2pop}) with adjustments for the hub effect in Eq.~(\ref{eq:hubEffect}).

\begin{figure}[!t]
\centering{\includegraphics[scale=1.0]{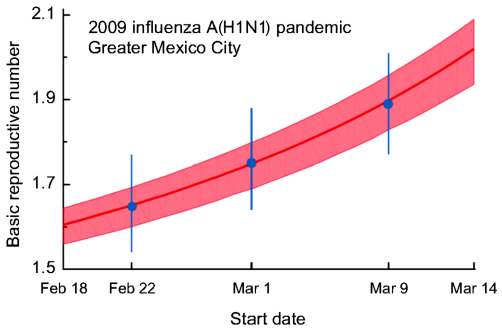}}
\caption{Inference of the basic reproductive number $R_0$ for the 2009 influenza A(H1N1) pandemic in Greater Mexico City. The value of $R_0$ is inferred using EATs for the first 12 countries seeded by Mexico, as documented in Ref.~\cite{balcan2009seasonal}. The red curve and the red-shaded area respectively indicate the posterior medians and the 95\,\% credible intervals. The blue dots and error bars respectively show the mean and the 95\,\% confidence intervals from Ref.~\cite{balcan2009seasonal} depending on whether the influenza A(H1N1) pandemic started in La Gloria, Mexico on 11, 18, or 25 February 2009.\newline
Source: Reprinted figure from Ref.~\cite{lin2018NC} under the Creative Commons Attribution 4.0 International (CC BY 4.0).}
\label{fig:AH1N1}
\end{figure}

The estimates of the basic reproductive number $R_0$ for the 2009 influenza A(H1N1) using the GLEAM simulator powered by high-performance computing~\cite{balcan2009seasonal} equal 1.65 with the 95\,\%-confidence interval (CI) [1.54, 1.77], 1.75 with the 95\,\% CI [1.64, 1.88], or 1.89 with the 95\,\% CI [1.77, 2.01] depending on whether the outbreak in La Gloria, Mexico, started on 11, 18, or 25 February 2009, respectively. Bayesian inference, by contrasts, rests on a likelihood function that can be written in a closed form that only depends on the basic reproductive number $R_0$
\begin{linenomath}
\begin{equation}
L(R_0) = \prod_{j\in A} f_1(t_j|\lambda_{ij},\alpha_{ij}) \prod_{j\in B}F_1(t_j|\lambda_{ij}, \alpha_{ij}),
\label{eq.AH1N1.exp}
\end{equation}
\end{linenomath}
where population $i$ denotes the Greater Mexico City as the epidemic origin, $t_j$ denotes the observed EAT for population $j$ which can be exact (set A) or left-censored (set B), $\lambda_{ij}=\lambda_i - \sum_{k\neq j}w_{ik}$ denotes the hub-adjusted epidemic growth rate, and $\alpha_{ij}$ denotes the adjusted mobility rate. It holds that $R_0=1+\lambda_i Z$, where $Z$ is the mean infectious period. This likelihood function leads to the same estimates of $R_0$ as the GLEAM simulator but without relying on high-performance computing (Fig.~\ref{fig:AH1N1}). Such a substantial reduction in computational complexity and resource requirements is expected to greatly improve the efficiency and timeliness of pandemic forecasting in the future.

\paragraph*{Estimating the transportation risk of Covid-19 from Wuhan to other cities in China} Covid-19 is caused by the SARS-CoV-2 virus. Due to the rapid global expansion of this virus, rising death numbers, unknown animal reservoir, and the increasing evidence of interpersonal transmissions~\cite{cowling2020NEJM}, the World Health Organization (WHO) declared a public-health emergency of international concern on 30 January 2020. An important concern at the time was the risk of new cases spreading from Wuhan to other locations. This concern was addressed in Ref.~\cite{du2020EIDrisk}, a quick case study performed during early 2020 using the methods proposed in Ref.~\cite{lin2018NC}.

Estimating the risk of new cases spreading from an origin population outwards begins with an epidemiological model. Let $\Delta I_\mathrm{W}(t)$ be the daily number of new infections in Wuhan from 1 December 2019 through 22 January 2020. Based on the epidemiological data from the first 425 Covid-19 cases confirmed in Wuhan by 22 January 2020~\cite{cowling2020NEJM}, the epidemic was assumed to grow exponentially
\begin{linenomath}
\begin{equation}
\Delta I_\mathrm{W}(t) = i_0 \exp(\lambda t),
\label{eq.exp.growth.EID2019}
\end{equation}
\end{linenomath}
where $i_0$ is the number of initial cases on 1 December 2019, and $\lambda$ is the local epidemic growth rate between 1 December 2019 and 22 January 2020. New Covid-19 cases were typically detected with a mean delay of $D=10$ days~\cite{imai2020Report2}, which included an incubation period of 5-6 days~\cite{cowling2020NEJM} and a delay from symptom
onset to detection of 4-5 days. With this in mind, the number of infectious cases at time $t$ is given by
\begin{linenomath}
\begin{equation}
I_\mathrm{W}(t) = \sum_{u=t-D}^t \Delta I_\mathrm{W}(u),
\end{equation}
\end{linenomath}
and the prevalence of infectious cases is
\begin{linenomath}
\begin{equation}
\eta(t) = I_\mathrm{W}(t) / N_\mathrm{W},
\end{equation}
\end{linenomath}
with $N_\mathrm{W}=11.1$ million denoting the population size of Wuhan.

The next step in estimating the risk of new cases spreading from an origin population outwards entails specifying a model of mobility. Assuming that the visitors to Wuhan and the residents of Wuhan share the same daily risk of infection, a non-homogenous Poisson process can be used to estimate the risk for exporting Covid-19 infections from Wuhan~\cite{gautreau2008global, tomba2008simple, lin2018NC}. Let $W_j(t)$ be the number of Wuhan residents travelling to city $j$ on day $t$, and $M_j(t)$ the number of travellers from city $j$ travelling back from Wuhan on the same day. The intensity function of the non-homogeneous Poisson process is then $\eta(t) \left[ W_j(t)+M_j(t) \right]$, and the probability of introducing at least one Covid-19 case from Wuhan to city $j$ by time $t$ is
\begin{linenomath}
\begin{equation}
1 - \exp\left( -\int\limits^t_{t_0}\eta(u) \left[ W_j(u)+M_j(u) \right]du\right),
\label{DU2020EIDEqExport}
\end{equation}
\end{linenomath}
where $t_0$ is the start of the study period, that is, 1 December 2019.

To estimate the unknown parameters, such as the number of initial cases $i_0$ and the local epidemic growth rate $\lambda$, a likelihood function is needed. This function can incorporate diverse information, including the information on disease exportations outside of China even if risk is to be quantified solely for Chinese cities. Ref.~\cite{du2020EIDrisk} in particular used the data on EATs due to 19 Wuhan residents who travelled to 11 cities outside of China before 22 January 2020.

Let $N_j$ be the number of infectious Wuhan residents detected at location $j$ outside of China, and $t_j^i$ the time at which the $i$th detection occurs. Furthermore, let $t_j^0$ denote 1 January 2020, which is the date on which international surveillance for infected travellers from Wuhan began. Finally, let $t_\mathrm{e}$ denote 22 January 2020 which is the end of the study period. As mentioned above, the rate at which infected residents of Wuhan arrive at location $j$ at time $t$ is $\eta(t)W_j(t)$. Accordingly, the likelihood function for observing the EATs due to 19 Wuhan residents travelling outside of China by 22 January 2020 is
\begin{linenomath}
\begin{equation}
\prod^{11}_{j=0}\exp\left(-\int\limits^{t_\mathrm{e}}_{t_j^{N_j}} \eta(t)W_j(t) dt \right) \prod^{N_j}_{i=1} \eta(t_j^i)W_j(t_j^i) \exp\left(-\int\limits^{t_j^i}_{t_j^{i-1}} \eta(t)W_j(t) dt \right).
\label{DU2020EID}
\end{equation}
\end{linenomath}
The first product quantifies the probability of not seeing any exportations at location $j$ between the time of the $N_j$th exportation and the study end. The second product quantifies the probability density of seeing the $i$th exportation at location $j$ at time $t_j^i$. All cities included in the study but without observed Covid-19 cases before 22 January 2020 were treated as a single location indexed by $j=0$.

\begin{figure}[p]
\makebox[\textwidth][c]{\includegraphics[scale=1.0]{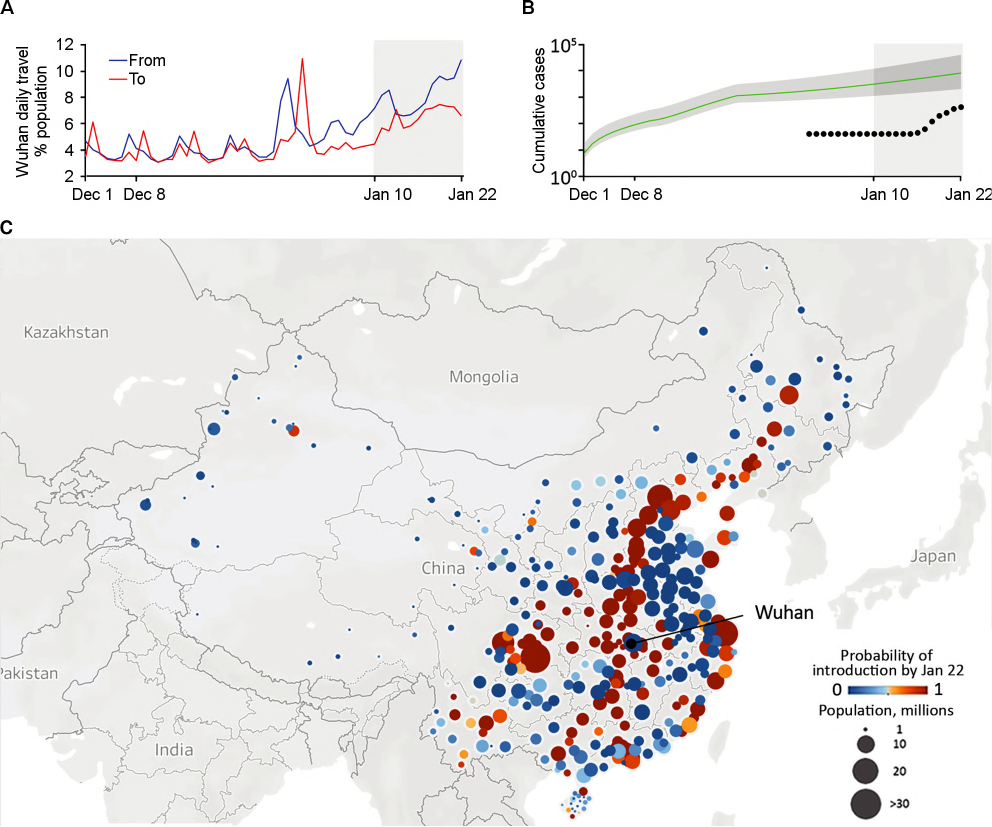}}
\caption{Risks of exporting Covid-19 from Wuhan, China, before the lockdown on 23 January 2020. \textbf{A,} Daily travel volume to and from Wuhan. \textbf{B,} Estimated and confirmed cumulative Covid-19 cases in Wuhan. Green line and grey shaded area indicate the mean and the 95\,\% credible interval (CrI) of the estimated cumulative true infections since 1 December 2019. Black dots indicate the cumulative confirmed-case counts during 1-22 January 2020. January 10 marks the beginning of the Spring Festival travel season in China. \textbf{C,} Probabilities that Chinese cities import more than one Covid-19 case from Wuhan by 22 January 2020. 131 cities (orange and red circles) exhibited a high risk of more than 50\,\%.\newline
Source: Reprinted figure from Ref.~\cite{du2020EIDrisk} under the Creative Commons Attribution 4.0 International (CC BY 4.0).}
\label{fig:EIDRISK}
\end{figure}

With the likelihood function in Eq.~(\ref{DU2020EID}), it becomes possible to estimate the number of initial cases $i_0$ on 1 December 2019 and the local epidemic growth rate $\lambda$ between 1 December 2019 and 22 January 2020. Ref.~\cite{du2020EIDrisk} used the Markov Chain Monte Carlo method with the Hamiltonian Monte Carlo sampling and non-informative flat priors. By assuming that the incubation period is exponentially distributed with the mean $L$ of 3-6 days and the infectious period is also exponentially distributed with the mean $Z$ of 2-7 days, the basic reproduction number is $R_0 = (1+\lambda L)(1 + \lambda Z)$. With all the parameters estimated, Eq.~(\ref{DU2020EIDEqExport}) calculates the risk of transporting at least one case from Wuhan to a downstream city $j$ before the lockdown of Wuhan on 23 January 2020 (Fig.~\ref{fig:EIDRISK}).

Ref.~\cite{du2020EIDrisk} estimates the Covid-19 doubling time ($=\ln(2)/\lambda$) at 7.31 days with the 95\,\% credible interval (CrI) [6.26, 9.66] days. Other studies using similar methods have yielded
congruent results; for example, Ref.~\cite{wu2020lancet} estimates the doubling time at 6.4 days with the 95\,\% CrI [5.8, 7.1] days and the basic reproductive number $R_0$ at 2.7 with the 95\,\% CrI [2.5, 2.9]. Ref.~\cite{li2020science} presents a metapopulation network model covering 375 Chinese cities and employs Tencent migration data to capture population movements during the 2021 Spring Festival period in China. By fitting the model to the reported 801 Covid-19 cases throughout China after the lockdown of Wuhan on 23 January 2020, the basic reproductive number $R_0$ is estimated at 2.38 with the 95\,\% CrI [2.03, 2.77]. Interestingly, Ref.~\cite{sanche2020high} estimates the basic reproduction number during the first wave of Covid-19 in mainland China at $R_0=5.7$ with the 95\,\% confidence interval [3.8, 8.9], which is inconsistent with other studies. This extreme result, however, may be due to an improper assumption of a single infection occurring at the initial time.

\subsection{Challenges and future work}

We have taken a look at the state of the art in epidemiological modelling and how it relates to the budding field of digital epidemiology. Here, we outline some pressing issues and ideas for further progress.

The metapopulation approach described herein considers mixing in each of the populations (i.e., cities) to be homogeneous. However, looking at mobility patterns only on the intercity scale hides away important epidemiological phenomena that happen on the intracity scale~\cite{luo2018large, yin2021data}. Ref.~\cite{chang2021nature}, for instance, describes a metapopulation susceptible-exposed-infectious-removed (SEIR) model that incorporates intracity mobility to explore the spread dynamic of Covid-19 in ten metropolitan areas in the US. Doing so has enabled identifying higher infection rates among disadvantaged racial and socioeconomic groups because of the differences in mobility. Specifically, disadvantaged groups have relatively little control over reducing their mobility and consequent exposure to infectious diseases.

In addition to mobility patterns, the epidemiological importance of contact patterns is impossible to overlook. These latter patterns have been shown to be highly assortative with age; especially school children and young adults tend to mix with similarly aged people~\cite{mossong2008social, prem2017projecting, mistry2021inferring}. This leaves younger populations potentially more vulnerable to infectious diseases unless there are attenuating biological circumstances, as is the case with (the early variants of) SARS-CoV-2~\cite{odriscoll2021age}.

Another key factor in epidemiology is individual heterogeneity in the ability to transmit an infectious disease. Superspreaders, for example, cause disproportionate number of secondary cases during the outbreaks of measles, influenza, rubella, smallpox, Ebola, monkeypox, SARS, and Covid-19. In the case of Covid-19, about 19\,\% of infectious individuals seed 80\,\% of all local transmissions~\cite{adam2020clustering}. SARS-CoV and SARS-CoV-2 viruses have, in fact, recently been shown to cause the number of secondary infections that follows a fat-tailed distribution, thus emphasising a large heterogeneity in transmission ability among individuals~\cite{wong2020evidence}. Accounting for this heterogeneity in epidemiological models generates the results that differ substantially from average-based approaches such that outbreaks are rarer but more explosive~\cite{lloyd2005superspreading}. Future models should therefore account not only for heterogeneity in contact networks, but also heterogeneity in infectiousness and possibly other epidemiological parameters.

Since the onset of the Covid-19 pandemic in December 2019, the amount of research on non-pharmaceutical interventions has exploded~\cite{perra2021non}. A big reason for such an explosion of interest is that non-pharmaceutical interventions are the only means of reducing the spread of a novel pathogen. They are also effective. For example, social distancing alone has proven sufficient to control Covid-19 in China~\cite{zhang2020changes}. Fast isolation of infectious individuals has furthermore reduced the time between successive disease onsets in a transmission chain (i.e., the serial interval), thus signalling effectiveness in preventing multiple secondary infections that would have arisen without isolation~\cite{ali2020serial}. These examples show that determining the optimal combinations of non-pharmaceutical interventions in given circumstances may save many lives, and should therefore be a major component of epidemiological modelling.

This chapter has hopefully demonstrated just how much data-hungry epidemiology is. A major limitation in the use of digital-data sources, such as Google Trends, Twitter, and Facebook, is that that none of them have been constructed with epidemiology in mind. This limitation can be overcome by establishing data standards and specialised systems akin to HealthMap~\cite{brownstein2008surveillance} and Influenzanet~\cite{paolotti2014web}, but also proving the usefulness of these systems to public-health agencies. Achieving so is no small feat because complicated data-analysis and modelling methodologies are often too demanding for otherwise busy public-health officials. In this context, relying on easy-to-use interactive interfaces called visual analytics~\cite{thomas2006visual} may help. Ref.~\cite{luo2016visual}, for example, demonstrates how to effectively couple data mining and agent-based epidemic modelling with a visual-analytics environment to facilitate human decision making in controlling infectious diseases.

\FloatBarrier

\section{Environment}
\label{S:Env}

Anthropogenic impact on the geological record is such that the 20th century saw a start of a new geological epoch---Anthropocene~\cite{waters2016anthropocene}. While climate change may be the (politically) most prominent issue today, our civilisation has additional profound impacts on every single ecosystem through pollution and habitat loss. Surprisingly, these are ultimately social issues because only through societal consensus on the need for action can they be controlled~\citep{pellow1999framing}.

Building social consensus against environmental degradation is extremely difficult. Preventing environmental degradation incurs unwanted costs either through the need for direct investment into unprofitable infrastructure and processes (e.g., water treatment facilities, carbon capture, safe waste disposal), or through opportunity costs (mostly land use restrictions due to ecosystem conservation, e.g., nature protected areas). Hence, to stay competitive, economies generally allow externalisation of environmental degradation costs until it is proven that costs of environmental degradation outweigh their economic benefits. Even once proven, building the social consensus can be extremely difficult; climate change and plastic pollution are just the two most prominent current cases that highlight the difficulties.

Despite the difficulties, change is possible if sufficiently strong scientific case can be made, and costs of degradation can be quantified. For example, theory on the adverse effect of at least some chlorofluorocarbons (CFCs) on the protective ozone layer was settled in 1974~\cite{molina1974stratospheric}, and experimentally confirmed in 1985~\cite{farman1985large}. It took only two years following the indisputable evidence for international ratification of a treaty to phase out the use of ozone-depleting substances~\cite{united2006handbook}, and the ozone layer is recovering~\cite{wmo2018scientific}. Tetraethyllead in gasoline followed a similar path from suspicion in mid-1920s~\cite{hoffman1927deaths} to ban in 1990s and 2000s as scientific evidence pioneered by Patterson~\cite{patterson1965contaminated} accumulated. Due to largely local nature of the problem, nations were able to take regulatory steps independently, with Japan banning low-octane lead gasoline in as early as 1975~\cite{yoshinaga2012lead}, and some countries continuing the use to date. Similar stories can be told of numerous other chemicals ranging from mercury, to dichlorodiphenyltrichloroethane (DDT), to polychlorinated biphenyls (PCBs) and polybrominated diphenyl ethers (PBDEs).

History---and the present---clearly demonstrate that the burden of proof that an externality is overly damaging squarely lies with environmental science, declaratory proclamations of the precautionary principle notwithstanding. Furthermore, the weight of the evidence required to elicit science-based activism that may eventually lead to societal consensus and (ultimately) solutions is extremely high. To provide sufficiently strong evidence, environmental science had to evolve, mostly towards physics. Because climate change will be--- due to its importance---discussed in a separate section (Section~\ref{S:GCC}), this section will focus on pollution and physical habitat loss. Historically, physical habitat loss due to competition of wildlife for natural resources with humans has been the chief driver of extinctions; pollution as an environmental problem, however, is a relatively new phenomenon. Before the industrial age, humans simply could not (and had no motivation to) produce toxic chemicals at a scale that could severely impact wildlife.

\subsection{Pollution}

There are many types of pollution, all traceable to introduction of materials or energy into the environment. The materials range from simple to complex chemicals, to particles, or even displaced natural materials. Energy pollution can also be diverse, ranging from noise, to non-ionising electromagnetic radiation (including light), to ionising radiation. Judging impact of pollution can be challenging because the judgement depends on both the point of view, and on the existing knowledge. For example radioactive debris from the Chernobyl nuclear plant explosion were not negative from the point of view of wildlife; in fact, wildlife is thriving despite the moderately toxic environment~\cite{deryabina2015long} simply because anthropogenic influence before evacuation was even worse. Focusing on the organismal level helps minimise such ambiguities.

Traditionally, toxicity has been estimated by exposing model organisms to varying concentrations and doses of chemicals in food or the environment while tracking endpoints like survivorship or mortality, fertility, cognitive ability, and---more recently---biomolecular targets like gene expression, protein levels, and reactive oxygen species. Such data yields dose-response curves used to predict the no-effect concentration, that is, the environmental concentration at which no negative effects are expected. Some calculations of the concentration simply divide by 1000 the concentration at which 50\,\% of organisms in short-term experiments showed a response, such as death or some other endpoint~\cite{debruijn2003technical}.

Considering the consequences of getting it wrong, the inadequacy of testing and observations as bases for regulation is staggering:
\begin{itemize}
\item Species or even individuals have different sensitivities to toxic exposure; the precautionary principle demands that generally the most sensitive model organisms should be used~\cite{debruijn2003technical}, but the choice of test organism is limited because in general only animals thriving in laboratory environments can realistically be utilised, and endangered species cannot be used at all.
\item Exposures in the lab are standardised, and typically last only a small fraction of organisms' life span~\cite{parasuraman2011toxicological}. Consequently, multigenerational studies are extremely rare and effects of long-term or cross-generational exposure are rarely captured.
\item Environmental conditions such as food availability, temperature, and humidity change exposure and its effects~\cite{witeska2003effects}. Relevance of tests in standardised laboratory environments to effects in highly variable natural environments is, therefore, limited.
\item Ecological feedbacks that concentrate toxicants, such as biotransformation and bioaccumulation, can result in much higher exposure for some species than indicated by experimental observations~\citep{wang1987factors}.
\item Even though chemicals mix in the environment, and mixtures can exacerbate toxicity~\cite{norwood2003effects}, most chemicals are tested---and their legal limits set---independently.
\item Only a small fraction of new chemical compounds are tested for toxicity. More than 20 million substances were reported as of January 2017, with about a million compounds added annually at an exponentially  growing rate~\cite{llanos2019exploration}; number of tested chemicals is measured in thousands.
\end{itemize}
Accordingly, relying on testing for legislation is highly impractical at best, misleading or wrong at worst. Consequences of underrating dangers may be dire---as previous examples show, it is extremely difficult to eliminate impacts of a chemical once it permeates the environment. Microplastics~\cite{galloway2016marine} and whatever is causing the decline of polinators~\cite{vanbergen2013threats} are shaping up to be the next major problems, but each new compound presents a new risk by itself or in synergy with other pollutants. High-throughput screening could help identify the most dangerous chemicals and direct preventive research in the future~\cite{barrick2017role}, but process-based modelling is leading the way in predictive ecotoxicology.

Process-based models rely on physical principles to explain observations, predict effects of hitherto unobserved exposures, and capture interactions between organisms and the environment. Although process-based models cover the whole range of scales, from molecular to landscape and ecosystem, the scales are not sufficiently linked~\cite{murphy2018incorporating}.

For example, quantitative structure-activity relationship (QSAR) models link characteristics of chemicals to their physicochemical, biological, and environmental properties. While essentially correlative in nature, QSARs use mechanistic descriptors derived from physico-chemical properties of chemicals. Types of the properties used to derive the descriptors determine QSAR dimension: chemical formula (0D), sub-structural fragment (1D), graph theory (2D), spatial geometry (3D), conformation/orientation/protonation state (4D), induced fit on ligand-based virtual or pseudo-receptor model (5D), 5D plus other solvation conditions (6D), and real target receptor model data (7D). In particular, descriptors in 3D QSARs involve a number of concepts from physics such as energy minimisation, classical approaches to molecular mechanics, quantum mechanics implemented using Born-Oppenheimer or Hartree-Fock approximations, or density functional theory to reduce computational loads and reduce descriptors to quantitative variables suitable for the correlative step in QSAR analysis. The correlative step, that is, relating descriptors to outcomes, can utilise any of the number of statistical methods ranging from multiple regression to artificial neural networks.

Despite their proven track record, QSARs have limitations. Due to the statistical nature of development and rigorous validation requirements, QSARs are still very data-intensive. Furthermore, when used to predict high-level responses (such as mortality), QSARs offer limited insight into metabolic pathways of toxic action. Understanding these, however, can be crucial for considering effects of chemicals in untested systems and mixtures. To overcome these shortcomings, the Organisation for Economic Co-operation and Development (OECD) actively supports development of a modular framework of metabolic cause-effect transfer functions: adverse-outcome-pathway (AOP) framework~\cite{ankley2010adverse}.

The AOP framework captures effects of exposure by modelling a sequence of all relevant molecular and cellular events. The framework consists (as of  January 2021) of more than 2,000 \textit{Key Events} (KEs). Exposure to a stressor triggers a KE that assumes the status of a \textit{Molecular Initiating Event} (MIE), which initiates a chain of KEs through a series of \textit{KE Relationships} (KER) akin to if$\rightarrow$then prescriptions. The chain terminates with an \textit{Adverse Outcome} (AO). For example, in AOP~\#15 any one of 10 stressors recognised by the framework can cause DNA alkylation (MIE), which through KER~\#24 leads to inadequate DNA repair (KE~\#155). KE~\#155 then through KER~\#164 leads to KE~\#185 (increase in mutations). The increase, through KER~\#202, leads to a heritable increase of mutations in offspring (AO~\#336). Elements of AOP~\#15 relate to seven other AOPs. Each element in the AOP chain has detailed documentation, a rigorous scientific background, and has been reviewed by experts.

Due to stringent background and review requirements, only 16 of more than 300 existing pathways have been endorsed across the OECD, but the framework is expected to expand exponentially as new pathways reuse old events and relationships, thus requiring fewer new ones. Currently, the framework focuses on humans, but similarity of organisms on bio-molecular level offers hope that eventually other organisms could benefit as well. High-throughput screening could provide a fast and affordable way to determine new key events and relationships. Currently, AOPs are not suitable for environmental ecotoxicology, making Dynamic Energy Budget (DEB) model the tool of choice for advanced links between environmental exposure and ecologically relevant organism-level endpoints.

DEB theory~\cite{kooijman2010dynamic} is, essentially, an application of the laws of thermodynamics to all three types of macrochemical reactions of a heterotrophic aerobe: assimilation, growth, and dissipation~\cite{jusup2017physics}. Focusing on (i) the four building blocks constituting 99\,\% of living biomass (C, H, O, and N), and (ii) effects of a fairly limited number of `hub' metabolites crucial for metabolic-network function that are markedly similar between species on macromolecular and cellular levels, the theory makes a number of simplifications leading to a robust theoretical framework able to capture ontogeny of living organisms and make testable predictions. Standardised DEB models are aggregated in the AmP database~\cite{marques2018amp}, which at the moment of writing contains entries for over 2,800 species. Detailed derivation of the standard DEB for the physics-minded reader is given in Ref.~\cite{jusup2017physics}.

\begin{figure}[!t]
\centering\includegraphics[scale=1.0]{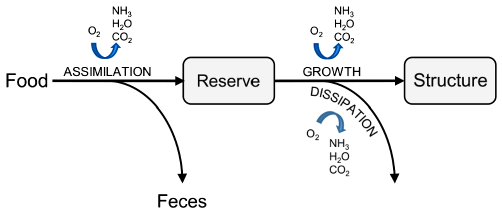}
\caption{Basic metabolic processes of heterotrophic aerobes according to standard DEB theory. Food is assimilated into reserve. In turn, the reserve is (i) converted into structure representing growth, (ii) committed to reproduction when possible, and (iii) used to power various dissipative processes such as maturation, maintenance, and metabolic inefficiencies (i.e., overheads) of growth, assimilation, and reproduction. Non-assimilated food (faeces) and other excess metabolites such as carbon dioxide, water, and nitrogenous waste are excreted into the environment. }
\label{fig:ENV_stdDEB}
\end{figure}

Starting from the first law of thermodynamics applied to an organism, the rate of change in internal energy $U$ is:
\begin{linenomath}
\begin{equation}
\label{eq:ENV_energy_bal}
\frac{\dup{U}}{\dup{t}} = \dot{Q} + \dot{W} + {\sum_i \bar h_i \left. \frac{\dup{M_i}}{\dup{t}} \right|_\mathrm{in}} - {\sum_i \bar h_i \left. \frac{\dup{M_i}}{\dup{t}} \right|_\mathrm{out}},
\end{equation}
\end{linenomath}
where $\dot{Q}$ and $\dot{W}$ respectively represent heat-transfer rate and mechanical power, $i \in \lbrace X, P \rbrace$ stands for organic substances in food and faeces, $i \in \lbrace C, H, O, N \rbrace$ stands for metabolites, $\bar h_i$ are molar enthalpies, and $M_i$ is the amount of substance $i$ in the organism. Next, using the mass, energy, and entropy balances, it can be shown that the total organismal Gibbs free energy is a sum of Gibbs free energies in compartments of the organism, assuming homeostasis of each compartment, that is, the chemical composition of each compartment remains constant throughout the life of the organism. The standard DEB theory recognises two compartments: energy reserve and structure, with an additional compartment tracking energy committed to maturation and reproduction (Fig.~\ref{fig:ENV_stdDEB}). If the organism is isomorphic (i.e., of constant shape), energy reserve and structure in a standard DEB model can be described by a simple set of coupled ordinary differential equations, especially when scaled
\begin{linenomath}
\begin{subequations}
\begin{align}
\frac{\dup{e}}{\dup{\tau}} &= g\frac{f - e}{l}, \label{eq:ENV_stdDEBScaled1}\\
\frac{\dup{l}}{\dup{\tau}} &= \frac{g}{3}\frac{e - l - l_T}{e + g}, \label{eq:ENV_stdDEBScaled2}
\end{align}
\end{subequations}
\end{linenomath}
where $0\leq f\leq 1$ is the scaled functional response representing surface-specific assimilation rate relative to the maximum, $e$ is the energy density of reserve relative to the maximum, $l$ is the length of the organism relative to the maximum possible length for $f=1$, $g$ is a compound parameter called the energy investment ratio, and $l_T$ is the heating length, scaled parameter accounting for energy spent on maintaining target body temperature in endotherms. In constant environments, the standard DEB model converges to the von Bertalanffy growth equation~\cite{vonbertalanffy1957quantitative}, which is still the most widely used energy-based equation for organismal growth. The von Bertalanffy growth equation is, however, a demand-side model; it can help estimate energy required for observed growth, but cannot predict how growth (or any other processes) would respond to changes in the environment.

\begin{figure}[!t]
\centering\includegraphics[scale=1.0]{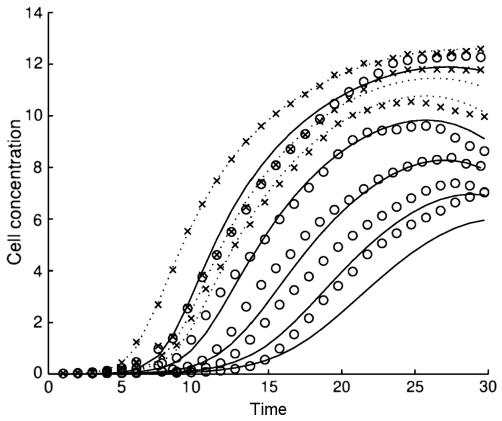}
\caption{Cadmium-ion toxicity. DEB model for Cd-ion toxicity predicts bacterial population dynamics for exposures of up to 150\,mg(Cd)/L (solid curves) with a single common parameter set fitted only using observations at 0, 10, and 20\,mg(Cd)/L (dotted curves).\newline
Source: Reprinted figure from Ref.~\cite{klanjscek2012modeling} under the Creative Commons Attribution 4.0 International (CC BY 4.0).}
\label{fig:ENV_cadmium}
\end{figure}

DEB models, on the other hand, make a quantum leap in that they causally link environmental energy and material availability to organismal growth and reproduction whilst preserving mass and energy balances. Additionally, the standard DEB theory also enables tracking of metabolism- and stress-related hazard rate, that is, the risk of death due to accumulated (cellular) damage. These features make some extraordinary feats possible. For example, the ability to track material fluxes in the context of interactions between the environment and the organism enabled a revolution in toxicokinetic (the distribution of toxicants) and toxicodynamic (toxicant effects) modelling by (i) resolving a number of long-standing issues of empirical dose-response curves~\cite{jager2011some}, and (ii) enabling prediction of bacterial population dynamics under exposure seven times greater than used for model fitting~\cite{klanjscek2012modeling} (Fig.~\ref{fig:ENV_cadmium}), as well as identification of nano-toxicity and its mechanisms~\cite{klanjscek2013dynamic}.

\begin{figure}[!t]
\centering\includegraphics[scale=1.0]{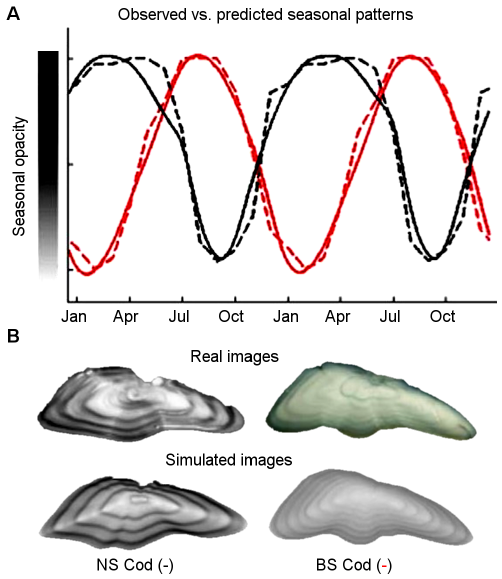}
\caption{Comparison of real and simulated otoliths. \textbf{A,} Seasonal variability in opacity patterns. Observed (dashed) and simulated (solid) variability in opacity of southern North Sea (NS, black) and Barents Sea (BS, red) anchovy otoliths. \textbf{B,} Opacity images: actual (top left) and simulated (bottom left) North Sea anchovy otoliths, and actual (top right) and simulated (bottom right) Barents Sea anchovy otoliths. Of note is that only environmental forcing (temperature and food) differ between the two populations; the model and parameter values are equal in both simulations.\newline
Source: Reprinted figure from Ref.~\cite{fablet2011shedding} under the Creative Commons Attribution 4.0 International (CC BY 4.0).}
\label{fig:ENV_otolith}
\end{figure}

\begin{figure}[!t]
\centering\includegraphics[scale=1.0]{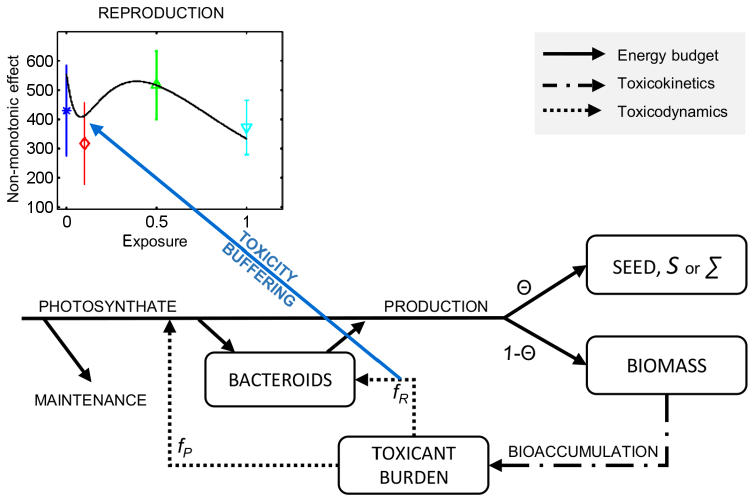}
\caption{Ecological interactions affect the outcome of pollution. Despite clear soybean plant cellular damage caused by exposure to CeO$_2$ naoparticles and negative effects of low exposure to growth and yield, higher exposures surprisingly improve plant growth and yield (inset). DEB model of coupled plant-bacteroid dynamics provides an explanation. Photosynthate (energy) is utilised for (i) maintenance of the plant, (ii) growth of bacteroids; remaining energy is used for (iii) seed production with proportion $\Theta$, and plant growth with proportion $1-\Theta$. Bioaccumulated toxicant affects both the bacteroids ($f_R$) and the plant ($f_P$). In nitrogen-poor soil, the bacteroids provide nitrogen needed for plant growth, but in nitrogen-rich soil, the bacteroids reduce energy available to the plant without providing any benefits. Small exposure affects the plant without killing bacteroids, causing a depression in growth and yield. Higher exposures, however, kill off the bacteroids but not the plant. This leaves more energy for the plant, thus improving growth and yield. See Ref.~\cite{klanjscek2017host} for further details.}
\label{fig:ENV_soybean}
\end{figure}

\begin{figure}[!t]
\centering\includegraphics[scale=1.0]{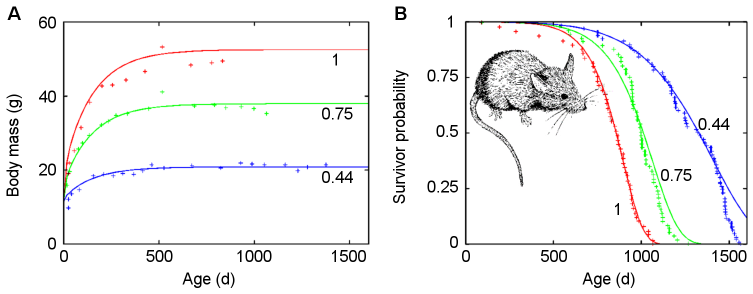}
\caption{Longevity of mice depends on food abundance. \textbf{A,} Growth of mice. \textbf{B,} Survival probability. Solid curves (model simulations) and data (`+') are shown for food abundances of 0.44 (blue), 0.75 (green), and 1 (red) relative to the maximum. Mice at restricted diets have a smaller metabolic activity that produces less damage-inducing compounds, thus accumulating less damage per unit of time, and living longer. See Ref.~\cite{kooijman2010dynamic} for further details.\newline
Source: Courtesy of Sebastiaan A. L. M. Kooijman.}
\label{fig:ENV_hazard}
\end{figure}

Further exemplary successes achieved by mechanistic modelling advocated by DEB theory include reconstructing otolith growth in anchovy~\cite{fablet2011shedding, pecquerie2012reconstructing} that enables tracking of historic environmental conditions and therefore determination of organism's population range (Fig.~\ref{fig:ENV_otolith}). Ecological interactions can explain why a level of exposure that measurably harms a plant can increase its growth and yield (Fig.~\ref{fig:ENV_soybean}). Finally, the concept of hazard rate unifies the Wiebull (allometric) and Gompertz (exponential) models of ageing, and is able to explain why hungry mice live longer (Fig.~\ref{fig:ENV_hazard}).

DEB makes strides in understanding toxic effects of exposure on individual and population levels, but falls short of the ultimate goal of ecotoxicology: understanding multi-generational effects of toxicants in environmental settings where multiple toxicants combine and interact with other stressors, and where complex ecological interactions could greatly affect outcomes. Reaching the goal requires modelling multiple populations exposed to a variety of toxicants in heterogeneous (spatially explicit) environments.

Many building blocks necessary to reach said ultimate goal of ecotoxicology already exist, and tools reduce barriers to entry. A conceptual framework has been developed that links the sub-cellular and sub-organismal AOP framework to individual-level DEB models~\cite{murphy2018incorporating}. \textit{General Unified Threshold model for Survival} (GUTS) is a modelling framework for toxicity test analysis in which `survival' is the endpoint~\cite{jager2011general}, with open source software toolbox available at \url{openguts.info}. The framework has been recognised in the Organisation for Economic Co-operation and Development (OECD) guidance for toxicity testing since 2006~\cite{oecd2003current}. Effects of toxicant mixtures have successfully been estimated from effects of each toxicant alone~\cite{jager2010biology, jager2014dynamic, margerit2016dynamic}.

\textit{DEBKiss} is a simplified version of the standard DEB model that, at the expense of generality, substantially reduces the barrier to entry to DEB modelling whilst preserving much of the utility, especially for specific questions for which cross-species comparison is not of primary importance~\cite{jager2013debkiss}. Older simple energy budget models exist, most notably net-assimilation and net-production models~\cite{ledder2004dynamic}. These too have been proven useful in specific circumstances, but lack scientific rigour and generality of the standard DEB family of models.

\textit{DEBTool} is a set of Matlab scripts to estimate parameters of a DEB model and run it. The tool is open source. Accompanying this tool is \textit{Add-my-Pet} (AmP) database of DEB model parameters containing parameters and data for over 2800 species and growing. Ref.~\cite{marques2018amp} provides a good overview of the database functionality.

DEB models are naturally suited for \textit{individual based modelling} (IBM) in which each organism is modelled separately; Ref.~\cite{martin2012dynamic} provides a Graphical User Interface (GUI) in NetLogo for simple IBM-based population modelling. The standard DEB model can be run directly in the GUI, but more complex features---including spatial heterogeneity---require additional programming. Population dynamics can be modelled in a number of ways if IBM approaches are impractical, including the Euler-Lotka equation~\cite{deroos2008demographic, beekman2019thermodynamic}, matrix population models~\cite{klanjscek2006integrating, ijima2019effects}, continuous-time physiologically structured~\cite{deroos1992studying, deroos2001physiologically}, and integral-projection models~\cite{smallegange2017mechanistic}. Of those, matrix population models have a particularly low barrier to entry and can also separate the population into patches~\cite{hunter2005use}, include predation, and other ecological interactions. Escalator Boxcar Train tool (EBTtool) is a GUI-based environment for implementation and analysis of physiologically structured models based on any physiological model of an individual including DEB.

\textit{Maxent}~\cite{phillips2006maximum} is a Java application that utilises data on species occurrences and environmental conditions to predict the geographic distribution of species using a maximum entropy approach. The software, appearing in thousands of publications, has been open source since 2017~\cite{phillips2017opening}. \textit{NicheMapR} is an R package available at \url{mrke.github.io/} that serves roughly the same purpose as Maxent, but is substantially more advanced. While Maxent integrates only observation data, NicheMapR is able to integrate metabolic models such as DEB, account for heat and water exchange using principles of biophysical ecology including behaviour, as well as for microclimate---using micrometereology, soil physics, and hydrology---to which the organism is exposed~\cite{kearney2020nichemapr}.

Therefore, frameworks and tools linking scales of biological organisation already exist, with physics providing the necessary glue. Crucially, frameworks make predictions that can be falsified, thus respecting the scientific process. Due to their mechanistic origins, the tools and frameworks are largely modular and can therefore be used to investigate effects of other anthropogenic pressures, including climate change and habitat loss.

\subsection{Physical habitat loss}

Physical habitat loss happens when a domestic species can no longer inhabit an area. Anthropogenic pressures, including climate change, can drive habitat loss through change of local environmental conditions that can facilitate exclusion of a domestic species by introducing---or merely making the local environment better for---an invasive species. Most habitat loss to date, however, has been due to land and sea use, that is, direct competition between wildlife and humans for space and related natural resources such as water and, in the case of predators, prey biomass. Measurable human influence on the environment may have started millions of years ago~\cite{faurby2020brain}; today, 95\,\% of Earth's land masses show human influence, with the remaining 5\,\% being mostly in inhospitable areas, including ice~\cite{kennedy2019managing}.

Unlike human-driven extinctions, the idea that we might need to preserve at least some of the natural biodiversity is relatively new. First legal environmental protections focusing on ecosystem preservation sprung up in the second half of the 18$^\mathrm{th}$ century, and significantly proliferated only in the 20$^\mathrm{th}$. The idea that environmental impact should be assessed at least for large-scale projects is even newer. The USA was in 1969 the first to require environmental impact assessments (EIA) for large-scale projects. To date, protected areas and the assessments remain the most effective tools for habitat preservation and biodiversity conservation; the number of protected areas has rapidly grown in the past decades, and the range of projects required to have an environmental impact assessment has been greatly expanded in the developed world.

To date, 15.4\,\% of terrestrial areas and 7.6\,\% of marine areas are classified as protected. These include protected areas ranging from strict nature preserves to airports~\cite{belle2018protected}. Strict nature preserves where no human activities are allowed except scientific monitoring and management interventions are rare. The vast majority of protections allow at least some activities. Therefore, management of protected areas involves balancing human activities with nature preservation.

More often than not, such management is focused on maximising human activities while keeping indicators of environmental damage within acceptable limits. Hence, protected areas are managed to a level of human activities just shy of environmental destruction, that is, to maximum damage levels that the environment can sustain in the long term. Clearly, this is not ideal even if the managing authority has the greenest of intentions. What we may perceive as the safe level of damage based on current data may reduce the resilience of the protected ecosystem to the point at which it succumbs to new pressures, such as the climate change. Tourism in national parks is a poster child for such a balancing act. Because nature is the main product offered to tourists, managing authorities have a vested interest in preserving the environment. Nevertheless, visitations are often maximised at the expense of nature~\cite{holloway2019business}.

Similar considerations hold for fisheries, with the added complication that the fishing industry has to contend with both the regulatory and natural uncertainties. For example, EU sets total allowable catch on (bi)annual basis to preserve fish stock and biodiversity. Without public subsidies, a particularly low value could decimate the fishing fleet because the industry may not be able to absorb the reduction in earnings. Such a reduction carries negative social and economic implications, but also makes the industry unable to respond to a higher allowed catch in the future. Hence, the regulatory authority has to balance the industry's need for income stability with conservation goals, often sacrificing one for the other.

Likewise, environmental impact studies have to balance conservation with the industry's interest in externalising as many costs as possible. Typically, a study has to show that the planned project meets environmental standards set by relevant authorities in terms of environmental indicators (e.g., water purity) and risk mitigation of catastrophic environmental incidents (e.g., oil spills). Because mitigation of environmental impact is expensive, companies have large incentives to aim towards the minimums required by the standards. Therefore, as in the case of activities in protected areas, environmental impact of industry is effectively managed towards the maximum acceptable environmental degradation.

Defining acceptability is not exclusively in the domain of environmental sciences. Lower environmental standards attract investments, thus incentivising legislators to be as lax with the standards as the local population will allow. The willingness of the locals to tolerate environmental destruction partly depends on their awareness of environmental and health issues, but more so on their standards of living and needs for employment. Therefore, the balance between conservation and environmental degradation is set at the intersection of environmental sciences, economics, social sciences, health sciences, and  politics. In countries where the population can afford to worry about environmental and health issues, red lines are drawn at ecological and health tipping points, that is, at environmental damage levels that, if increased, are all but guaranteed to cause unacceptable irreparable long-term damage to ecosystems or have clearly measurable effects on human health.

\subsection{Tipping points}

\begin{figure}[!t]
\centering\includegraphics[scale=1.0]{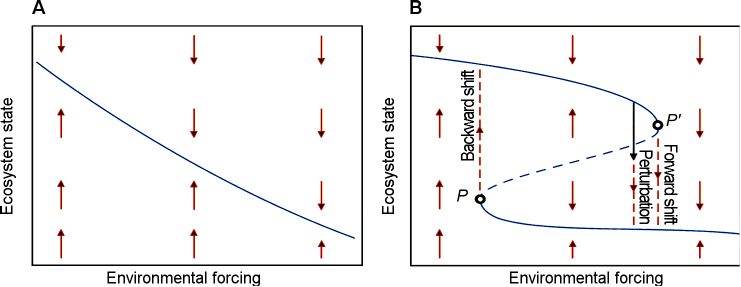}
\caption{Example of a tipping point in ecosystems. Curves show an ecosystem's equilibrium state as a function of environmental forcings such as nutrient abundance, temperature, salinity, precipitation, human exploitation etc. Arrows show the direction in which the ecosystem's non-equilibrium state shifts over time given the conditions. \textbf{A,} Ecosystem equilibrium is a monotonic, slightly convex function of conditions. If the ecosystem state is initially below (above) the equilibrium curve, an upward (downward) shift is to be expected over time. \textbf{B,} Non-linear feedbacks may lead to pitchfork bifurcation where, for a given condition, multiple equilibria exist (here, three equilibria exist between points $P$ and $P'$). This transforms the ecosystem's convergence dramatically. If conditions in the tipping point $P'$ change ever so slightly, the ecosystem undergoes a large forward shift. Even if conditions remain constant, a small perturbation close to the tipping point $P'$ could induce the shift. This is because the unstable part of the equilibrium curve (dashed) indicates unstable equilibria around which the direction of convergence changes abruptly. Finally, akin to hysteresis in ferromagnetic materials, recovery following a forward shift at $P'$ requires return of the forcing variable (conditions) all the way to the tipping point $P$. See Refs.~\cite{scheffer2000socioeconomic, scheffer2001catastrophic, scheffer2003catastrophic, scheffer2009early} for further details.}
\label{fig:tippingpts}
\end{figure}

Tipping point is a point in parameter space at which a small perturbation induces a significant change in the state of the system (Fig.~\ref{fig:tippingpts}). Such points are a common feature of complex systems, and affect ecology at all scales---from molecular through organismal, individual and population, to ecosystem level.

\begin{figure}[!t]
\centering\includegraphics[scale=1.0]{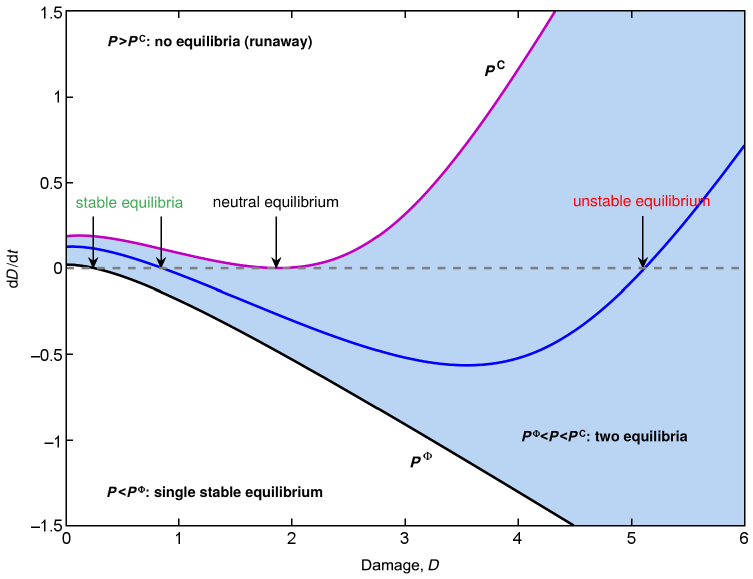}
\caption{Example of a molecular tipping point turning to an individual-level tipping point. Solid curves delineate areas of distinct damage dynamics (\smash{$\frac{\dup{D}}{\dup{t}}$}) as a function of damage ($D$); the dashed line represents an equilibrium (\smash{$\frac{\dup{D}}{\dup{t}}=0$}). As damage production due to exposure ($P$) increases from zero, the system has a single stable equilibrium, and damage levels are small. When $P$ reaches a critical value of $P^\Phi$ (black solid curve), further increase in exposure results in two distinct equilibria, a stable and an unstable one (shaded area with a blue solid curve for illustration); damage levels are controlled unless initial damage levels are too high ($D$ greater than the unstable equilibrium point for the given $P$). Further increase of $P$ past $P^\mathrm{C}$ leads to a saddle-node bifurcation in that the two equilibria combine into one neutral equilibrium and disappear, thus initiating runaway dynamics for any $D$; damage is not controlled, and the organism eventually dies from exposure. See Ref.~\cite{klanjscek2016feedbacks} for further details.}
\label{fig:molecularTippingPoints}
\end{figure}

Molecular-level tipping points in ecology typically occur when a negative feedback maintaining homeostasis is overcome by a forcing variable such as exposure to toxicants (Fig.~\ref{fig:molecularTippingPoints}). For low levels of exposure, the cell is able to maintain homeostasis by up-regulating molecular defence and repair mechanisms, even for large amounts of cellular damage. At the exposure level that exceeds the capacity of biomolecular control, even a small additional exposure will initiate a positive feedback loop of damage creation; additional cellular damage reduces the cell's ability to defend against exposure and increase in damage, thus accelerating (runaway) damage accumulation. Therefore, runaway damage accumulation initiated by a molecular tipping point (breakdown of molecular control) leads to death of the individual cell or organism, thus creating an individual-level tipping point. For complex systems, there could be intermediate states in which, for example, the organism is able to prevent runaway damage creation if damage levels are low, but not if the initial damage levels are high (Fig.~\ref{fig:molecularTippingPoints}). In such cases, any preexisting damage or additional damage-creating stress would result in death from exposure levels that would not harm an otherwise unstressed organism. This is particularly important to note when using experimental exposure data to interpret real-world situations in which organisms are rarely subject to a single source of stress.

\begin{figure}[!t]
\centering\includegraphics[scale=1.0]{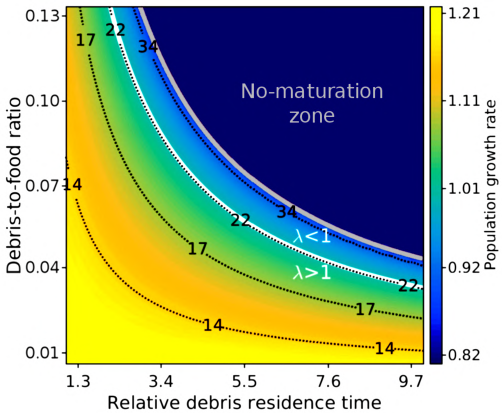}
\caption{Individual- and population-level tipping points caused by plastics. Effects of plastic debris on loggerhead turtles were investigated using DEB. Black curves represent maturation age, while the colour on the graph represents population growth rate depending on the ratio of debris to food ($y$-axis), and residence time of debris in the gut ($x$-axis). If both are high enough, individuals can neither maturate nor reproduce (the dark blue zone; individual-level tipping point). In this zone one would not observe reproducing individuals at all. If the ratio and the residence time are low enough, population is growing, and reproducing individuals can be observed. However, there is a zone (light blue between the white and grey curves) in which healthy reproducing individuals are observed, but the population is nevertheless going extinct (population-level tipping point). For loggerhead turtles this is a realistic prospect given that the ratio of debris to food required has already been observed. Process-based modelling expedites identification of such subtle but extremely important tipping point.\newline
Source: Reprinted figure from Ref.~\cite{marn2020quantifying}.}
\label{fig:ENV_turtles}
\end{figure}

Food abundance is a major forcing in ecosystems that can initiate a number of ecological tipping points. For example, when food abundance reduces below corresponding tipping-point levels, ontogeny changes dramatically as individuals cannot grow, mature, or survive. Interestingly, however, even if an individual seems to be thriving, a population might have crossed a tipping point leading to species extinction~(Fig.~\ref{fig:ENV_turtles}).

There is ample evidence of ecological tipping points, that is, tipping points at the level of whole ecosystems. Among the best-known examples is overfishing-induced ecosystem regime shift off the coast of Newfoundland in the early 1990s~\cite{hamilton2001outport, hamilton2004above}. The northern cod fishery that operated in the area was a product of systematic cod exploitation from the 16$^\mathrm{th}$ century onwards. Since the 1880s, the fishery yielded over 200,000 tonnes of cod annually, but introduction of new technologies in the 1960s led to an explosion in yields which peaked at 810,000 tonnes in 1968. Technologies in question included more powerful trawlers equipped with radars, electronic navigation, and sonars. This enabled fishing longer, over larger areas, and at greater depths. Considerable amounts of non-commercial bycatch was extracted from the sea, including cod's prey, the capelin.

The marine ecosystem off the coast of Newfoundland finally gave in to enormous pressure in the early 1990s. In 1991, the fishery still landed about 129,000 tonnes of cod, which was roughly 69\,\% of the plan for 1992. This plan, however, was entirely unrealistic, and in the same year, authorities announced a two-year moratorium on cod fishing in response to a dire state of fish stocks~\cite{myers1997fish}. The moratorium was eventually extended to a full decade, but even in 2002, there was no sign of cod recovery. Instead the ecosystem was dominated by invertebrates, crabs and shrimps.

First signs of recovery were reported in 2011~\cite{frank2011transient}, suggesting that prey-fish stocks (such as the aforementioned capelin) had exploded in the wake of the cod-fishery collapse. This in turn had put enormous pressure on cod eggs and larvae. By 2005, however, prey-fish stocks themselves went into a downward spiral, providing a window of opportunity for the cod. An optimistic assessment of the situation was reiterated in 2015~\cite{rose2015northern}. The reported recovery proposes a dynamically intriguing possibility; namely, a predator species (cod) historically held a prey species (capelin) in check, but upon the collapse of the predator, the prey multiplied enough to exert pressure on the predator's offspring, thus preventing a recovery. Such dynamics will be examined next through the prism of mathematical modelling.

Ref.~\cite{deroos2002size} presents a food-chain model consisting of an unstructured predator population feeding on a structured prey species, which in turn feeds on an unstructured resource (i.e., zooplankton) species. The term `structured' designates that in the prey population there is an explicit distinction between juvenile and adult individuals. This is not the case in `unstructured' populations. In the model, the ontogeny of prey individuals is specified in terms of three functions dependent on the resource abundance, $R$, and the individual's body size, $l$. \textit{The ingestion function} is
\begin{linenomath}
\begin{equation}
I(R,l)=I_\mathrm{m}l^2\frac{R}{R_\mathrm{h}+R},
\end{equation}
\end{linenomath}
where $I_\mathrm{m}$ is the maximum surface-area-specific mass ingestion rate and $R_\mathrm{h}$ is the half-saturation constant, that is, $I(R_\mathrm{h},l)=\frac{1}{2}I_\mathrm{m}l^2$. \textit{The growth function} is
\begin{linenomath}
\begin{equation}
g(R,l)=\gamma\left[\frac{I(R,l)}{I_\mathrm{m}l^2}l_\mathrm{m}-l\right],
\end{equation}
\end{linenomath}
where $\gamma$ is the maximum growth rate and $l_\mathrm{m}$ is the maximum body size achieved when the resource is abundant, that is, $R\rightarrow\infty$. \textit{The reproduction function} is
\begin{linenomath}
\begin{equation}
b(R,l)=
\begin{cases}
0, & \mathrm{for}~l<l_\mathrm{j}\\
r_\mathrm{m}\frac{I(R,l)}{I_\mathrm{m}}, & \mathrm{for}~l\geq l_\mathrm{j}
\end{cases},
\end{equation}
\end{linenomath}
where $l_\mathrm{j}$ is the length at sexual maturation and $r_\mathrm{m}$ is the maximum surface-area-specific reproduction rate. Additionally, the prey species is subject to a natural-mortality rate $\mu$ and a predator-induced mortality
\begin{linenomath}
\begin{equation}
d(P)=
\begin{cases}
0, & \mathrm{for}~l<l_\mathrm{b}\\
\frac{aP}{1+aT_\mathrm{h}B}, & \mathrm{for}~l_\mathrm{b}\leq l< l_\mathrm{v}\\
0, & \mathrm{for}~l\geq l_\mathrm{v}
\end{cases},
\end{equation}
\end{linenomath}
where $l_\mathrm{b}$ and $l_\mathrm{v}$ specify the range of body sizes in which the prey species is vulnerable to predation, $a$ is the maximum predator ingestion rate, and $T_\mathrm{h}$ is the mass-specific prey-processing time of the predator. The quantity $P$ stands for the predator abundance, whereas the quantity $B$ is the prey biomass accounting only for individuals with body sizes in the predation-vulnerability range, $l_\mathrm{b}\leq l\leq l_\mathrm{v}$.

The biomass $B$ is not a state variable directly tracked by the model. Instead, the model keeps track of the prey density $c(t,l)$ of body size $l$, from which we have
\begin{linenomath}
\begin{equation}
B(t)=\int\limits_{l_\mathrm{b}}^{l_\mathrm{v}}\beta l^3 c(t,l)\dup{l},
\end{equation}
\end{linenomath}
where a typical weight-length relationship is used with the proportionality constant $\beta$. The dynamics of the state variable $c(t,l)$ is given in terms of a partial differential equation
\begin{subequations}
\begin{linenomath}
\begin{equation}
\pp{c(t,l)}{t}+\pp{}{t}\left[g(R,l)c(t,l)\right]=- \left[\mu+d(P)\right]c(t,l).
\end{equation}
\end{linenomath}
To be solvable, this equation needs a boundary condition, which is given in terms of the reproduction function
\begin{linenomath}
\begin{equation}
g(R,l)c(t,l)=\int\limits_{l_\mathrm{j}}^{l_\mathrm{m}}b(R,l)c(t,l)\dup{l}.
\end{equation}
\end{linenomath}
The dynamics of the state variable $R$ is given in terms of an integro-differential equation
\begin{linenomath}
\begin{equation}
\dd{R}{t}=\rho(K-R)-\int\limits_{l_\mathrm{b}}^{l_\mathrm{m}}I(R,l)c(t,l)\dup{l},
\end{equation}
\end{linenomath}
where $\rho$ is the zooplankton inflow rate and $K$ is the corresponding carrying capacity. The model is fully specified with a differential equation describing the predator dynamics
\begin{linenomath}
\begin{equation}
\dd{P}{t}=\epsilon d(P)B -\delta P,
\end{equation}
\end{linenomath}
\end{subequations}
where $\epsilon$ is the growth efficiency of predator on prey and $\delta$ is the predator's natural mortality.

\begin{figure}[!t]
\centering\includegraphics[scale=1.0]{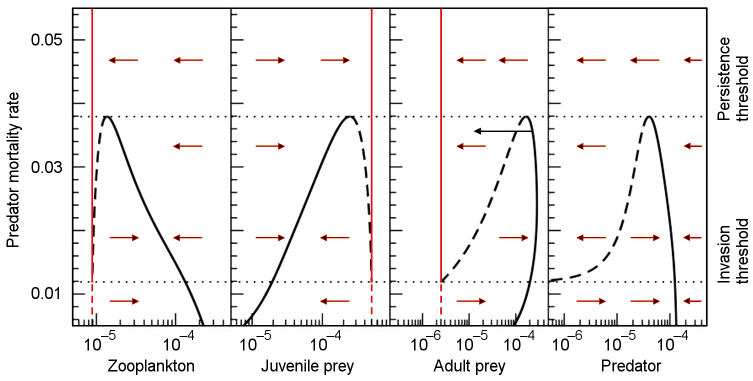}
\caption{Zooplankton, prey, and predator dynamics in a marine ecosystem subjected to human exploitation of the predator species. Below the invasion threshold, zooplankton, adult prey, and predator are abundant, while juvenile prey is rare. Above the persistence threshold zooplankton and adult prey are rare, juvenile prey is abundant, while predator goes extinct. Between the two thresholds zooplankton and juvenile prey are inversely related, while adult prey will be abundant if predator is abundant too or adult prey will be rare if predator goes extinct. Red vertical lines (black curves) denote stable equilibria without (with) predator. Red arrows indicate the direction of the model's convergence, whereas the black arrow represents a perturbation affecting the abundance of adult prey. See Ref.~\cite{deroos2002size} for further details.\newline
Source: Courtesy of Andre M. de Roos.}
\label{fig:deroosmodel}
\end{figure}

The described model generates extremely rich dynamics. Here, we are interested in the case of increasing human exploitation of the predator species, which increases the predator's mortality rate. Depending on this mortality rate, there are two key thresholds, the invasion threshold and the persistence threshold (Fig.~\ref{fig:deroosmodel}). Below the former threshold, predator is abundant, as are adult prey and zooplankton, while juvenile prey is rare. Between the two thresholds, the predator abundance decreases with the mortality rate, allowing adult prey to reach its peak, which in turn means more offspring (i.e., juvenile prey) and a consequent zooplankton reduction. Above the latter threshold, however, predator suddenly goes extinct, adult prey and zooplankton become rare, while juvenile prey abounds. This is the ecosystem's tipping point.

Close to the tipping point various perturbations may turn the ecosystem's state upside down even if the predator mortality rate is below the persistence threshold. For example, an adverse event affecting adult prey would be sufficient to change the model's convergence in such a way that adult prey becomes rare and predator ultimately collapses (Fig.~\ref{fig:deroosmodel}). As a consequence, the ecosystem would be dominated by juvenile prey, just as was the case off the coast of Newfoundland where capelin and other prey-fish stocks became abundant after the cod fisher had gone bust.

\subsection{Future outlook}

Arguably, the primary socially relevant objective of modern environmental sciences should be to identify and improve our understanding of tipping points where irreparable environmental damage becomes inevitable. Indeed, incorporation of physics into environmental sciences leads to an unprecedented improvement in our understanding of such tipping points. Simultaneously, incorporation of economics through the concept of ecosystem services has raised awareness of trade-offs between short-term monetary gain from allowing externalisation of environmental costs, and long-term losses from reduced services provided by the ecosystem~\cite{king2015trade}. Now more than ever, therefore, environmental sciences can help the decision-making process by (i) calculating positive and negative externalities, (ii) providing easily available large-scale longitudinal monitoring, and (iii) conducting risk assessment of ecological scenarios. Three major issues remain, though, implying new environmental research and management opportunities.

\paragraph*{The mindset of acceptable change} Thinking that acceptable impacts do not have negative effects is referred to as the mindset of acceptable change. For example, even process-based models in ecotoxicology rely on the concept of no-effect concentration. Such a concentration, however, does not actually exist---even traces of a toxicant elicit responses at the biomolecular level, and can up-regulate a number of cellular processes~\cite{okayama2016general}. We do not observe effects of small concentrations on ontogenetic endpoints (e.g., reproduction or growth) only because of homeostatic regulatory mechanisms. These regulatory mechanisms are complex dynamical systems that utilise feedback loops to buffer against negative change, and maintain cellular or organismal homeostasis. Hence, detectable effect on an ontogenetic endpoint can be interpreted as a failure of the homeostatic buffering mechanism, and the no-effect concentration can be viewed as the concentration at which at least one homeostatic mechanism is about to fail. Is such a concentration safe, or can the up-regulation of the homeostatic mechanisms lead to harm in the long term? Does the homeostatic mechanism maintain its effectiveness when environmental conditions change? Answering such questions requires models of toxic effects that incorporate regulatory dynamics at the organismal level, but we are aware of only one such model~\cite{klanjscek2016feedbacks}. The same considerations are valid when factoring acceptable environmental impact of human activities relevant in environmental impact studies and protected area management.

As discussed in the previous section, the mindset of acceptable change tends to result in management towards maximum environmental impact shy of tipping the environment into unwanted states. Perhaps it is therefore not surprising that, guided by funding, environmental sciences have been researching the overarching question of the acceptable change mindset: ``How far can we stretch the environment?'' We propose to change the question into something more constructive.

For example, in national park management, the question could be: ``How can we improve the environment?'' Surprisingly, the answer is not to ban tourism. National parks require expensive monitoring, management, scientific research, and active conservation efforts. Tourism provides the necessary funds either directly through entrance fees, or indirectly through local taxes collected from economic activities related to tourism. Hence, some tourism activity may be necessary for preservation, especially in poor regions where funding conservation is not high on the list of priorities. Tourism then contributes to the ability of a region to fund conservation but, even more importantly, employment opportunities in tourism provide motivation for conservation at all societal levels. Therefore, the new question does not affect whether tourists should be allowed to visit a protected area; it may, however, seek to limit the number of tourists, and affect their spatio-temporal distribution.

The difference in environmental impact between the two approaches can be huge. For example, while still in the mindset of acceptable change, two major national parks in Croatia (National Park Plitvice and National Park Krka) used to fight overcrowding by spreading the visitors throughout the park. This approach resulted in newly cleared paths, further habitat fragmentation, new conflicts between visitors and wildlife, and other negative environmental impacts. Due to a number of research projects, the mindset changed. Now, the two parks (i) streamline visitor experience to reduce perception of crowding, (ii) use  monetary incentives to reduce peaks in demand, (iii) offer tourist attractions outside of the protected area, and (iv) limit simultaneous number of visitors in the area to safeguard visitor experience. These approaches enabled the parks to concentrate a greater number of tourists into a smaller area without sacrificing visitor experience. The higher concentrations of tourists made advanced environmental impact mitigation measures practical, thus drastically reducing the impact per visitor.

Re-visiting the management goals, therefore, reduced environmental degradation whilst increasing the number of visitors, their happiness, and income from entrance fees. Getting to the solutions, however, required interdisciplinary research and collaboration between social science, environmental science, and economics. We suggest goals of environmental impact studies, and environmental standards, should be revisited similarly.

\paragraph*{The Big Picture} Ecosystems, responsible for the very air we breathe, depend on photosynthesis---processes on atomic levels that turn light into organic matter (but see Ref.~\cite{baross1985submarine}). This leads to a local decrease in entropy essentially fuelled by nuclear reactions in the sun. All further interactions increase local entropy, hence our biosphere is limited by that initial step. Transferring the organic matter through the food chain is extremely inefficient; about 90\,\% of energy is lost at each step~\cite{lindeman1942trophic}, thus the number of levels in the food chain is also limited. Therefore, as humans co-opt an increasing proportion of space and the photosynthetic production, ecosystems are put under increasing strain. Because ecosystems are complex dynamic systems, long-term consequences of changing the forcing of the system (and directly influencing all of its components) is extremely hard to predict.

Observations can provide only limited insights into small components of the ecosystem. Even large-scale monitoring can inform us about the past and the present, but not about the future. We can try to extrapolate from experiments, but they are of limited scope by definition, and therefore applicable to a limited set of environmental conditions and interactions. Furthermore, investigating organisms informs us about the individual, not the population over multiple generations in an uncontrolled environment.

To overcome the limitations of observation, we need to continue developing mechanistic models. Ideally, models would link gene expression to organismal ontogeny, to population status, to ecosystem dynamics. Then, we could truly explore optimal solutions to environmental and a host of societal and economic problems. Modelling must, however, be tempered by reality.

\paragraph*{Virtual reality} Tools for linking sub-cellular processes to ecosystem-scale effects are being developed at an accelerating rate. Further development of these tools will create an unprecedented \textit{in-silico} test-bed for risk assessment and testing of potential ecological scenarios to minimise effects of anthropogenic pressures on the environment. However, increasing complexity of the underlying theory and physics in these tools makes it difficult for any one person to understand the background and assumptions of all the linkages. Simultaneously, the tools are increasingly easy to use, and their interfaces require less understanding of the underlying processes. The increasing complexity and ease-of-use could then combine to yield 'virtual reality' quasi-scientific consensus where model outputs drive policy even when divorced from reality. This effect can already be seen in fisheries where model-driven quota settings repeatedly fail to attain predicted results, and yet neither the procedures nor the modelling have been affected.

Dangers of succumbing to the virtual reality should be avoided by aggressively pursuing integration of data into modelling. Fortunately, relevant data collection is increasing exponentially. Biophysics generates knowledge on the (sub-)molecular level; remote sensing through satellites and drones, in combination with in-situ sensors creates snapshots of the environment going back decades; citizen science initiatives and obligatory reporting required by western governments contains a wealth of information on the environment. These sources are, however, not cross-linked and integrated, nor will they be without sufficiently accurate underlying models spanning relevant levels of biological organisation. Environmental science has only recently started the iterative cycle of mutually improving data collection and modelling, a process that brought physics to the forefront of natural sciences. The transition may require a change in attitude as well as competencies.

Ultimately, long-term solutions to environmental problems will be found at the intersection of environmental sciences, social sciences, economics, and politics. Environmental science can only identify problems and offer potential solutions. Choosing and implementing those solutions always has been a matter of policy and governance, which require social consensus; getting to the consensus is for social sciences to (re)solve. Economics clearly affects the social dialog, and ultimately makes implementation of solutions possible by finding ways to make environmentally sound choices in line with monetary incentives of the modern society. The high degree of ecosystem complexity implies that the best our civilisation can currently do is to aim at precautionary adaptive management fuelled by rigorous interaction between modelling and data. To achieve the needed breakthrough, a new generation of environmental scientists versed in mathematics, machine learning, and physics may be required.

\FloatBarrier

\section{Global climate change}
\label{S:GCC}

\begin{figure}[p]
\makebox[\textwidth][c]{\includegraphics[scale=1.0]{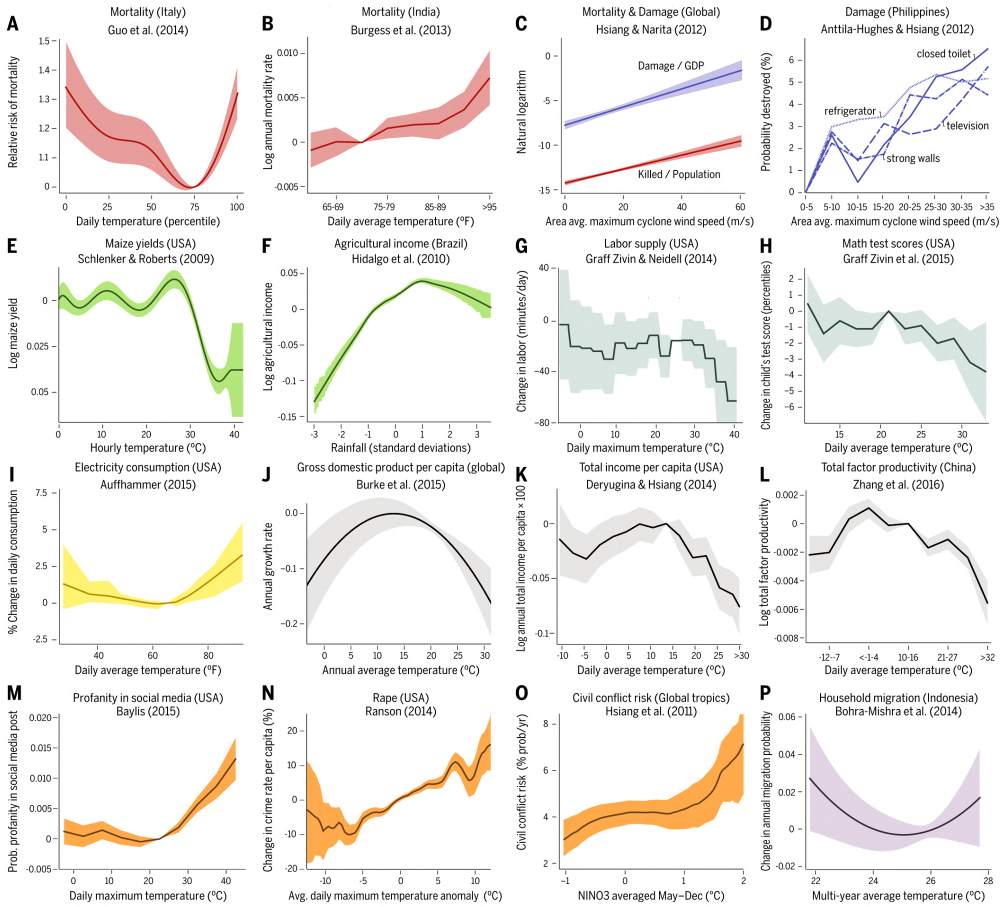}}
\caption{Social consequences of climate variables. The causal effect of climatological events on various social outcomes is described by a dose-response function. Colours indicate categories of outcome variables: \textbf{A,\,B,} red, mortality~\cite{guo2014global, burgess2013unequal}; \textbf{C,\,D,} blue, cyclone damage~\cite{hsiang2012adaptation, anttila2013destruction}; \textbf{E,\,F,} green, agriculture~\cite{schlenker2009nonlinear, hidalgo2010economic}; \textbf{G,\,H,} teal, labour productivity~\cite{graff2014temperature, graff2018temperature}; \textbf{I,} yellow, electricity consumption~~\cite{auffhammer2018climate}; \textbf{J--L,} grey, aggregate economic indicators~\cite{burke2015global, deryugina2014does, zhang2018temperature}; \textbf{M--O,} orange, aggression, violence, and conflict~\cite{baylis2020temperature, ranson2014crime, hsiang2011civil}; \textbf{P,} purple, migration~\cite{bohra2014nonlinear}. Shaded areas are confidence intervals.\newline
Source: Reprinted figure from Ref.~\cite{carleton2016social}.}
\label{fig:Fig13_1}
\end{figure}

Climate change is among the greatest challenges that humanity has to face. Overwhelming scientific evidence indicates that a changing climate has tremendous influence on societies, both past and present, with serious consequences for the future~\cite{pachauri2014climate}. Recent advances in quantitative empirical research have illuminated the key connections in the coupled climate-human system \cite{holland1986statistics}. Numerous statistical analyses have addressed the causal effects between specific climatic conditions and social outcomes, such as agriculture, economics, conflict, migration, and health~\cite{carleton2016social}. Fig.~\ref{fig:Fig13_1} highlights a number of empirical studies that demonstrate how climatological events affect various social outcomes on both regional and global scales.

Many societal effects of climate have been evidenced. For example, temperatures around the 75th percentile of available records in Italy are associated with the lowest mortality there~\cite{guo2014global}. An increase of one standard deviation in high-temperature days in India increases annual mortality among rural populations by 7.3\,\%~\cite{burgess2013unequal}. More intense tropical cyclones lead to more destruction in a global cross-section of countries~\cite{hsiang2012adaptation}. Non-linear temperature effects signal severe damages to the U.S. maize yields~\cite{schlenker2009nonlinear}.
Insufficient rainfall in Brazil causes steep drops in agricultural income~\cite{hidalgo2010economic}. Labour productivity in the U.S. is strongly determined by the daily average temperature~\cite{graff2014temperature, graff2018temperature}, as is residential electricity consumption in California~\cite{auffhammer2018climate}, total income per capita in the U.S.~\cite{deryugina2014does}, and total factor productivity in China~\cite{zhang2018temperature}. The annual average temperature, furthermore, plays a key role in the growth of gross domestic product per capita~\cite{burke2015global}. Interpersonal aggression, both petty and criminal, increases with temperature and sometimes decreases with rainfall; examples include the use of profanity on social media~\cite{baylis2020temperature} on the one end, and rape~\cite{ranson2014crime} on the other end. Even civil conflicts escalate in the tropics in response to El Ni\~{n}o-type warming that takes place in the tropical central and eastern Pacific Ocean~\cite{hsiang2011civil}. Finally, changes in the multi-year average temperature have a greater effect on permanent outward migration of households in Indonesia than other natural disasters~\cite{bohra2014nonlinear}. All these quantitative empirical examples reveal that climate is indeed a major factor affecting social outcomes, often with first-order consequences. Even more important, however, is that understanding the relationship between climate and society offers insights into how modern society can best respond to the current climatic events, and how future climate trajectories may impact humanity.

When considering the climate-human relationship, a crucial realisation is that, while climate drives social outcomes, human activities and emissions also impact the dynamics of climate change. According to the Intergovernmental Panel on Climate Change's (IPCC) fifth assessment report:
\begin{quote}
It is extremely likely that more than half of the observed increase in global average surface temperature from 1951 to 2010 was caused by the anthropogenic increase in greenhouse gases concentrations and other anthropogenic forcings together.
\end{quote}
The underlying mechanism of human contributions to global warming is clear. The combustion of fossil fuels like coal and oil emits into the atmosphere greenhouse gases (GHGs), primarily carbon dioxide CO$_2$. GHGs block convective heat from escaping into space, which ultimately manifests as the rise in temperature.

Given the interdependence between climate and society, mitigating risks due to climate change requires an integrated perspective that goes beyond just the grasp of physical facts. It is also necessary to mobilise human action, which is, as a research problem, in the domain of multiple disciplines such as behavioural economics, social psychology, and evolutionary game theory. Scientists are attempting to rise to the challenge through the ongoing integration of climate science, social sciences, and humanities, giving rise to a new `science of the Earth' called \textit{Earth System Science}~\cite{steffen2020emergence}. The aim is to build a
unified comprehension of the Earth and its human population.

\paragraph*{Climate extreme events and global warming} Among the most visible consequences of climate change are increases in the intensity and frequency of extreme weather and climate events. These include heat waves, droughts, wildfires, floods, and hurricanes, to name a few. Such extreme events endanger not only human lives, but livelihood as well, as evidenced by, for example, fresh-water shortages and reduced food production. An extreme event is said to occur when the value of a climatic variable moves beyond the corresponding critical threshold. Of note, however, is that extreme events may be due to natural climate variability that is unrelated to anthropogenic forcings.

Extreme weather and climate events are grouped into three categories~\cite{seneviratne2012changes}: (i) extremes of atmospheric weather and climate variables, including temperature, precipitation, and wind; (ii) weather and climate phenomena that influence the occurrence of extremes in weather or climatic variables, or represent extremes themselves, including monsoons, tropical cyclones, and extratropical cyclones; (iii) impacts on the natural physical environment, such as the aforementioned heat waves, droughts, floods and more. The distinction between these categories, while intuitive, is also somewhat blurred, and the categories are highly correlated.

The changes in the frequency, intensity, spatial extent, duration, and timing of weather and climate extremes may precipitate unprecedented risks and disasters for both natural physical environment and human society. Strengthening resilience against disruptive weather phenomena and climate change at national, regional, and local levels is therefore of vital importance. The ability to anticipate and predict extreme events would greatly aid the efforts to strengthen the resilience of human communities, but non-linear feedbacks, coupled interactions, and complex structure of the climate system pose formidable challenges. The situation is even more dire when attempts are made to account for the coupled climate-human system.

\paragraph*{Tipping points in climate and social systems} The concept of a tipping point commonly refers to ``a critical threshold at which a tiny perturbation can qualitatively alter the state or development of a system''~\cite{lenton2008tipping}. The term `tipping point' was popularised by the writer Malcolm Gladwell who used it to describe intriguing sociological events during which little things can make a big difference~\cite{gladwell2006tipping}. Many complex systems, including climate~\cite{rahmstorf2002ocean, dakos2008slowing, caesar2018observed} and human society~\cite{doyle2016social, centola2010spread}, have tipping points at which an abrupt shift to a contrasting dynamical regime may occur. In relation to the climate system, Ref.~\cite{lenton2008tipping} introduced the term `tipping element' to describe large-scale components of the climate system that may approach or exceed a tipping point. Some examples are the ice-loss acceleration in Greenland, the diminishing sea-ice area in the Arctic, novel pests and fire patterns in boreal forest, the slowdown of the Atlantic thermohaline circulation, intense droughts in the Amazon rainforest, the large-scale die-offs of coral reefs, the decay of the Antarctic ice sheet, and others. Climate scientists have long suspected that by the present time, many tipping points of the climate system will have been exceeded, pushing the Earth ever closer to a~\textit{global tipping point}. Exceeding the global tipping point would constitute a downright existential threat to civilisation~\cite{lenton2019climate}. It is, therefore, high time for international action, such as reducing the emissions of GHGs, to improve the planet's resilience against extreme climate.

Recently, Ref.~\cite{otto2020social} proposed a framework of social tipping dynamics for stabilising Earth's climate such that the planet is put back on track to halve global emissions by 2030 and tip the scales to net zero emissions by 2050. The study extends the idea of tipping elements from the components of the climate system to the subdomains of the planetary socioeconomic system, calling such subdomains \textit{social tipping elements}, ``where the required disruptive change may take place and lead to a sufficiently fast reduction in anthropogenic greenhouse gas emissions''. In this context, \textit{social tipping interventions} have the potential to set social tipping elements on the path of change, for example, by (i) highlighting the moral implications of fossil fuels, (ii) strengthening climate education and engagement, and (iii) disclosing information on greenhouse gas emissions. By doing so, social tipping dynamics could be harnessed to foster climate-change mitigation.

Anticipating and predicting tipping points before they are breached would yield substantial socioeconomic benefits. Many techniques have been developed to this end based on the theory of early-warning signals for critical transitions~\cite{scheffer2009early}. The phenomenon of \textit{critical slowing down} is considered as one of the most important clues that a dynamic system has lost resilience and is fast-approaching a tipping point~\cite{scheffer2012anticipating}. Critical slowing down is recognised by an increase in the auto-correlation and the variance of the system's state variables. For example, Ref.~\cite{dakos2008slowing} analysed eight abrupt shifts in ancient climate and found that significant increases in the lag-1 auto-correlation had preceded these shifts. Ref.~\cite{carpenter2006rising} examined a lake model in the vicinity of a bifurcation point and found an increasing variance of lake‐water phosphorus about a decade prior to the shift to a new, nutrient-rich state. Other quantities and techniques have also been proposed as early-warning signals. Examples include the detrended fluctuation analysis~\cite{livina2007modified}, power spectra~\cite{prettyman2018novel}, flickering before transitions~\cite{wang2012flickering}, skewness and kurtosis \cite{guttal2008changing, biggs2009turning}, and others. See Ref.~\cite{dakos2012methods} for a more detailed review on the subject of tipping points and early-warning signals.

\subsection{Modelling the climate system}

\paragraph*{What is a climate model?} There are two main tools that supported the development of climate science: (i) observations of a changing Earth system and (ii) computer modelling and simulations. The term `observations' in the context of climate science usually refers to instrumental data (i.e., meteorological stations or satellites), reanalyses data (e.g., ECMWF, NCEP-NCAR, and JRA), and proxy data (e.g., coral records, tree rings and ice-core records, etc.). `Computer modelling' and climate models, by contrast, refer to attempts to simulate physical, chemical, and biological processes that take place in the atmosphere, cryosphere, land, ocean, and lithosphere and collectively produce climate. A climate model comprises a series of equations that describe said processes, and is typically implemented in numerical form that is suitable for processing on powerful computers. Crucially, scientists use climate models to project how climate may change over the course of the predictable future.

The history of climate modelling by means of numerical methods likely begins with Richardson's work in the 1920s~\cite{richardson1922weather}, in which he proposed a novel idea to forecast weather using differential equations while viewing the atmosphere as a network of gridded cells. In 1938, Callendar published a seminal paper~\cite{callendar1938artificial} describing a one-dimensional radiative transfer model to show that rising CO$_2$ levels are warming the atmosphere. The first computerised, regional-weather forecast was tested in 1950 on the electronic numerical integrator and computer (ENIAC). The first three-dimensional general circulation model of the global atmosphere that could realistically depict seasonal patterns in the troposphere was released by Phillips in 1956. This was followed by the establishment of the National Center for Atmospheric Research (NCAR) in 1960, which soon thereafter became the leading climate modelling centre. The 1967 study by Manabe and Wetherald~\cite{manabe1967thermal} introduced an influential 1D radiative-convective model to generate the first credible prediction of the surface temperature in response to the CO$_2$ content of the atmosphere. NASA's Nimbus III satellite was launched in 1969 with the specific task of taking measurements of Earth. The National Oceanic and Atmospheric Administration (NOAA) was created in 1970, and similar to NCAR soon thereafter became the world's leading centre for climate-change research. The Met Office's first general circulation model released in 1972 rounds up early developments in the field.

Rising awareness of climate change led to the establishment of the IPCC in 1988 with the aim to ``provide the world with a clear scientific view on the current state of knowledge in climate change and its potential environmental and socio-economic impacts.'' The first IPCC assessment report~\cite{ipcc1990policymakers} was published two years later with a summary stating that ``under the IPCC Business-as-Usual emissions of greenhouse gases, the average rate of increase of global mean temperature during the next century is estimated to be about 0.3\,$^{\circ}$C per decade.'' With the more widespread development of coupled atmosphere-ocean global circulation models, a need arose for standardising their outputs, which resulted in the launch of Coupled Model Intercomparison Project in 1995. By the late 2000s climate models could be used in conjunction with paleoclimate data to explore climatic tipping elements~\cite{lenton2008tipping}. The biophysical understanding of Earth, including the climate system, was integrated with policy and governance in the planetary boundaries framework~\cite{rockstrom2009safe}. Interestingly, between about 1998 and 2012, Earth seemed hardly to warm, which became known as the \textit{global warming hiatus}, prompting some to question previous conclusions. The newer findings, however, reconciled models and data, leading the authors of Ref.~\cite{medhaug2017reconciling} to conclude that ``we are now more confident than ever that human influence is dominant in long-term warming.''

A variety of climate models to date, from simple energy-balance models to elaborate general circulation models, differ in their complexity and relative advantages. Especially the general circulation models are highly reliant on the most advanced supercomputers in order to assimilate all the required data. Despite the rapid advancement of the field since its inception, there is no single, comprehensive model that could encapsulate all the non-linear interactions between climate-determining subsystems. Consequently, the longer-term predictive skills regarding climate variability and climate change remain limited and dependent on the precise initial conditions.

\paragraph*{Hierarchy of climate models} Earth's climate is a complex system influenced by many factors, for example, solar radiation, clouds, winds, ocean currents, and many others. The system is furthermore subdivided into subsystems---the atmosphere, the ocean, the cryosphere, the biosphere, the pedosphere, and the lithosphere---that interact at the interfaces such as air-ocean, air-ice, ice-ocean, as well as land-air and land-ocean. Over the years, climate modelling has benefited from a variety of approaches to constructing climate models that integrate, to a larger or lesser degree, said components and interactions. In particular, much has been learned from models focused on specific aspects of the climate system (e.g., El Ni\~{n}o events and monsoons), while abandoning the pretence that full complexity can be accounted for. This line of thinking and climate model development is now known as a \textit{hierarchical modelling approach}~\cite{neelin2010climate}.

We distinguish four model categories based on their complexity:
\begin{enumerate}
    \item\textit{Energy-balance models} estimate the changes in the climate system by analysing Earth's energy budget, that is, by balancing the incoming solar radiation and the outgoing terrestrial radiation.
    \item\textit{Radiative-convective models} simulate the vertical profile of atmospheric temperature and the associated transfer of energy under the assumption of radiative-convective equilibrium.
    \item\textit{Statistical-dynamical models} combine the features of energy-balance and radiative-convective models in order to study horizontal energy flows and processes that disrupt such flows.
    \item\textit{General circulation models} attempt to capture the fundamental physics and chemistry of the climate system, including the exchange of energy and materials between the components of this system.
\end{enumerate}

An energy-balance model can take one of two simple forms, the zero-dimensional model such that Earth is a single compartment with a global mean effective temperature or the one-dimensional model such that temperature is latitudinally resolved. In the one-dimensional model, each latitudinal zone is described by the following equation
\begin{linenomath}
\begin{equation}
(\text {Shortwave in}) =(\text {Transport out})+(\text {Longwave out}),
\label{eq:EQ13_1}
\end{equation}
\end{linenomath}
or more formally
\begin{linenomath}
\begin{equation}
S(\phi)\{1-\alpha(\phi)\}=c \{T(\phi)-\bar{T}\}+\{A+B T(\phi)\},
\label{eq:EQ13_2}
\end{equation}
\end{linenomath}
where $S(\phi)$ is the mean annual radiation incident at latitude $\phi$, $\alpha(\phi)$ is the albedo at latitude $\phi$ (0.62 for $T<-10$\,$^\circ$C and 0.3 otherwise), $c$ is the horizontal heat-transport coefficient (3.81\,W\,m$^{-2}$\,$^\circ$C$^{-1}$), $T(\phi)$ is the surface temperature at latitude $\phi$, $\bar{T}$ stands for the mean global surface temperature, and $A$ and $B$ are constants governing the longwave radiation loss ($A=204.0$\,W\,m$^{-2}$ and $B=2.17$\,W\,m$^{-2}$\,$^\circ$C$^{-1}$). Of note is that some implementations of energy-balance models also simulate energy transfers between the atmosphere and the ocean.

Radiative-convective models add complexity relative to the energy balance models. Thus, one-dimensional radiative-convective models account for the vertical dimension, while two-dimensional models additionally account for one horizontal dimension. Using such models it possible to predict how GHGs modify effective emissivity and surface temperature. A radiative-convective model has the following mathematical form
\begin{linenomath}
\begin{subequations}
\begin{align}
&S=\alpha_{c} S+\alpha_{g}\left(1-a_{c}\right)^{2}\left(1-\alpha_{c}\right) S+\varepsilon \sigma T_{c}^{4}+(1-\varepsilon) \sigma T_{g}^{4},\\
&a_{c}\left(1-\alpha_{c}\right) S+a_{c} \alpha_{g}\left(1-a_{c}\right)\left(1-\alpha_{c}\right) S+\varepsilon \sigma T_{g}^{4}=2 \varepsilon \sigma T_{c}^{4},\\
&\left(1-\alpha_{g}\right)\left(1-a_{c}\right)\left(1-\alpha_{c}\right) S+\varepsilon \sigma T_{c}^{4}=\sigma T_{g}^{4}.
\end{align}
\end{subequations}
\end{linenomath}
These equations respectively represent the energy balances at (i) the top of the atmosphere, (ii) the cloud level, and (iii) the surface. The equations are directly solvable upon setting the values for the cloud shortwave absorption, $a_{c}$, the cloud albedo, $\alpha_{c}$, infrared emissivity, $\varepsilon$, and the surface albedo, $\alpha_{g}$. The parameter $\sigma$ is the Stefan-Boltzmann constant. Radiative-convective models, as seen here, incorporate information about radiation fluxes throughout the atmosphere, including the fluxes of solar radiation, cloud cover, and land.

While statistical-dynamical models make more of a practical leap, general circulation models mark the next true conceptual leap in climate modelling. General circulation models are the most complex and `complete' model type used in climate-change science. These three-dimensional models are constructed by discretising the differential equations that express the conservation of momentum, mass, and energy. The model closure is achieved by adding an equation of state for the atmosphere. Expressed in a mathematical form, we have:
\begin{enumerate}
    \item\textit{Conservation of momentum}
    \begin{linenomath}
    \begin{equation}
    \DD{\mathbf{v}}{t}=-2 \mathbf{\Omega} \times \mathbf{v}-\rho^{-1} \nabla p+\mathbf{g}+\mathbf{F}.
    \label{eq:EQ13_4}
    \end{equation}
    \end{linenomath}
    \item\textit{Conservation of mass}
    \begin{linenomath}
    \begin{equation}
    \DD{\rho}{t}=-\rho \nabla \cdot \mathbf{v}+C-E.
    \label{eq:EQ13_5}
    \end{equation}
    \end{linenomath}
    \item\textit{Conservation of energy}
    \begin{linenomath}
    \begin{equation}
    \DD{I}{t}=-p \DD{}{t}\rho^{-1}+Q.
    \label{eq:EQ13_6}
    \end{equation}
    \end{linenomath}
    \item\textit{Ideal gas law}
    \begin{linenomath}
    \begin{equation}
    p=\rho R T.
    \label{eq:EQ13_7}
    \end{equation}
    \end{linenomath}
\end{enumerate}
The physical meanings of the symbols are as follows:
\begin{itemize}[topsep=0pt,parsep=0pt,itemsep=0pt]
    \item $\mathbf{v}=$velocity relative to Earth,
    \item $t=$time,
    \item $\DD{}{t}=\frac{\partial}{\partial t}+\mathbf{v} \cdot \nabla=$total time derivative,
    \item $\Omega=$Earth's angular velocity vector,
    \item $\rho=$atmospheric density,
    \item $p=$atmospheric pressure,
    \item $\mathbf{g}=$apparent gravitational acceleration,
    \item $\mathbf{F}=$force (other than gravity) per unit mass,
    \item $C=$creation rate of atmospheric constituents,
    \item $E=$destruction rate of atmospheric constituents,
    \item $I=c_{p}T=$internal energy per unit mass,
    \item $T=$temperature,
    \item $Q=$heating rate per unit mass,
    \item $R=$gas constant, and
    \item $c_p=$specific heat of air at constant pressure.
\end{itemize}

Aside from the aforementioned model types, there are other classes of climate models that attempt to capture specific aspects of the climate system, but in a simplified way. These are known as \textit{intermediate complexity models}. A representative example is the Cane-Zebiak model~\cite{zebiak1987model}, developed to simulate El Ni\~{n}o events, as well as conduct experimental climate predictions.

Notably, the physical models of the climate system lack any form of human dynamics, treating instead Earth's global population as an outside force. Attempts to fill this gap produced integrated assessment models in which human dynamics plays a prominent part~\cite{parson1997integrated, vanvuuren2011well}. These models aim to link socioeconomics with the biosphere and the atmosphere into one modelling framework for the purpose of simulating costs of specific climate-stabilisation policies.

Integrated assessment models ``represent many of the most important interactions among technologies, relevant human systems (e.g., energy, agriculture, the economic system), and associated greenhouse gas emissions in a single integrated framework''~\cite{pachauri2014climate}. This means not only an integrated representation of the physical laws driving natural systems, but also the changing preferences that drive human society. Typically, there are two main types of integrated assessment models---simple and complex. Simple models are run in a spreadsheet by utilising simplified equations, while detailed relationships between the economy, energy, and Earth systems are left out~\cite{hope1993policy}. These models are commonly used to evaluate the `social cost of carbon'. By contrast, complex integrated assessment models account for energy technologies and uses, changes in land use, and societal trends. Separate modules represent the global economy and the climate system. The basic structure of an integrated assessment model can be broken down as follows:
\begin{itemize}[topsep=0pt,parsep=0pt,itemsep=0pt]
\item\textit{Model inputs}---assumptions about how the world works and changes, such as the GDP, populations, policies, and so on;
\item\textit{Model itself}---modules that represent economic, energy, land, and climate systems;
\item\textit{Model outputs}---quantitative predictions about the economy, land-use changes, greenhouse gas emissions and energy-use pathways, and future human development.
\end{itemize}
Despite their comprehensiveness, complex models are imperfect, often failing to capture nuanced social mores, habits, and behaviours, both present and future. Such imperfections notwithstanding, complex integrated assessment models are a valuable tool in exploring the contributions of social factors to climate change. The key questions that can be addressed in this way include, for instance, how to avoid global warming of more than 1.5\,$^\circ$C at the lowest cost or what the implications are of current national pledges to reduce the emissions of GHGs. Taken together, global circulation, integrated assessment, and other climate models offer powerful means to explore Earth system dynamics at a range of spatial and temporal scales, all the while incorporating both physical and social mechanisms and processes.

\subsection{Climate networks}

In recent years, network science has emerged as a novel framework to study climate phenomena such as El Ni\~{n}o--Southern Oscillation (ENSO), extreme-rainfall patterns, and air-pollution variability~\cite{holme2021ncc}. It is, of course, worthwhile to review this topic in its own right, but even more so given that network science interfaces physics with so many other disciplines.

\paragraph*{Basic concepts} Networks have proven to be a versatile tool to explore the structural and dynamical properties of complex systems beyond physics, for example, in biological, ecological, and social sciences~\cite{newman2010networks}. A particular strength of the network representation of a complex system is the ability to map out the system's topological features. Climate, as mentioned previously, is a quintessential example of a complex system comprising many non-linearly coupled subsystems with multiple forcings and feedbacks. The desire to model climate's complexity has, therefore, led to the birth of the idea of \textit{climate networks} in which geographical locations on a longitude-latitude grid become network nodes, while the degree of similarity, or `connectedness', between the climate records obtained at two different locations determines whether there is a network link between these locations~\cite{tsonis2004architecture, tsonis2006networks}. The climate-network framework has been applied successfully to analyse, model, and predict various climate phenomena, such as ENSO~\cite{tsonis2008topology, yamasaki2008climate, gozolchiani2008pattern, gozolchiani2011emergence, ludescher2013improved, meng2017percolation, fan2017network, meng2018forecasting}, extreme rainfall~\cite{boers2014prediction, boers2019complex}, Indian summer monsoon~\cite{stolbova2016tipping}, Atlantic meridional overturning circulation~\cite{vandermheen2013interaction, feng2014deep},
Atlantic multidecadal oscillation~\cite{feng2014north}, teleconnection paths \cite{zhou2015teleconnection}, the impacts of CO$_2$~\cite{ying2020rossby}, and others.

\begin{figure}[!t]
\centering\includegraphics[scale=1.0]{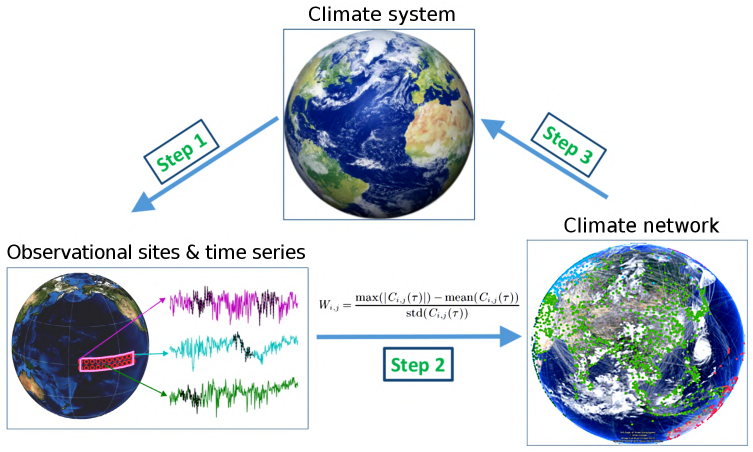}
\caption{Constructing and utilising a climate network. In step 1, a spatial grid of nodes where climatological time series have been observed is defined. In step 2, the cross-correlation between time series at two node locations is calculated. When such a cross-correlation is strong, the node locations are deemed to be linked. In step 3, the topology of the climate network is analysed using the methods of network science to reveal the properties of the climate system.}
\label{fig:Fig13_3}
\end{figure}

Climate networks are constructed and utilised in three steps (Fig.~\ref{fig:Fig13_3}):
\begin{enumerate}
    \item\textit{Step 1.} Define a spatial grid of nodes containing the climatological variable of interest (e.g., temperature, geopotential height, precipitation, etc.).
    \item\textit{Step 2.} Build links between nodes pairs based on statistical correlations between the time series recorded at the two node locations.
    \item\textit{Step 3.} Interpret the dynamical processes of the climate system (e.g., winds, ocean currents, atmospheric circulation, Rossby waves, etc.) via the structural properties of the climate network.
\end{enumerate}
A detailed overview of methodology for constructing and analysing climate networks can be found, for example, in Ref.~\cite{donges2015unified}.

Building links between nodes is of central importance in constructing climate networks. Among the most direct ways to decide whether, or how strongly, two nodes are connected is the \textit{Pearson correlation}. Let us suppose that a climate observable $T$ (e.g., the sea-surface temperature anomaly) is measured at a number of fixed stations. At station $i$, which is to be identified with the $i$th node in the climate network, measuring the climate observable yields a time series $T_i(t)$. If the time series is subdivided into, for instance, calendar years, months, or days, this is further indexed with the index $y$ producing $T_i^y(t)$. Then, the time-delayed Pearson cross-correlation function between nodes $i$ and $j$ is~\cite{meng2018forecasting},
\begin{linenomath}
\begin{equation}
C^{y}_{i,j}(-\tau)=\frac{\langle T_i^{y}(t) T_j^{y}(t-\tau) \rangle-\langle T_i^{y}(t)\rangle \langle T_j^{y}(t-\tau) \rangle }{\sqrt{\langle (T_i^{y}(t)-\langle T_i^{y}(t)\rangle)^2\rangle}\cdot\sqrt{\langle (T_j^{y}(t-\tau)-\langle T_j^{y}(t-\tau)\rangle)^2\rangle}},
\label{eq:EQ13_8}
\end{equation}
\end{linenomath}
and
\begin{linenomath}
\begin{equation}
C^{y}_{i,j}(\tau)=\frac{\langle T_i^{y}(t-\tau) T_j^{y}(t) \rangle-\langle T_i^{y}(t-\tau)\rangle \langle T_j^{y}(t) \rangle }{\sqrt{\langle (T_i^{y}(t-\tau)-\langle T_i^{y}(t-\tau)\rangle)^2\rangle}\cdot\sqrt{\langle (T_j^{y}(t)-\langle T_j^{y}(t)\rangle)^2\rangle}},
\label{eq:EQ13_9}
\end{equation}
\end{linenomath}
where $0\leq\tau\leq\tau_\mathrm{max}$ is the time lag and $\langle\cdot\rangle$ denotes averaging over the variable $t$. Next, it is possible to define the positive and negative link strengths of between nodes $i$ and $j$ as~\cite{fan2018climate}
\begin{linenomath}
\begin{equation}
W^{+,y}_{i,j} = \frac{\max(C^{y}_{i,j}) - {\rm mean}(C^{y}_{i,j})}{{\rm std}(C^{y}_{i,j})},
\label{eq:EQ13_10}
\end{equation}
\end{linenomath}
and
\begin{linenomath}
\begin{equation}
W^{-,y}_{i,j} = \frac{\min(C^{y}_{i,j}) - {\rm mean}(C^{y}_{i,j})}{{\rm std}(C^{y}_{i,j})},
\label{eq:EQ13_11}
\end{equation}
\end{linenomath}
where max, min, mean, and std respectively denote the maximum, minimum, mean, and the standard deviations of the cross-correlation function over the time lag $\tau$. The direction of the link is taken to be from node $i$ to node $j$ if $W^{+,y}_{i,j}>{}\left|W^{-,y}_{i,j}\right|$ and from node $j$ to node $i$ otherwise~\cite{fan2017network}.

An alternative method to construct climate networks is the \textit{event synchronisation} method, which was originally developed to measure synchronisation and infer the direction of time delay between signals~\cite{quiroga2002event}. The method is based on the relative timings of events in a time series, where an event is defined as, for instance, reaching a threshold or a local maximum. Supposing two time series, $X(t)$ and $Y(t)$, an event $l$ seen in $X$ at time $t_{l}^{x}$ is considered to be synchronised with an event $m$ seen in $Y$ at time $t_{m}^{y}$, if $0<\left|t_{l}^{x}-t_{m}^{y}\right|<\tau_{l m}^{x y}$, where
\begin{linenomath}
\begin{equation}
\tau_{l m}^{x y} =\frac{1}{2} \min \left\{t_{l+1}^{x}-t_{l}^{x}, t_{l}^{x}-t_{l-1}^{x}, t_{m+1}^{y}-t_{m}^{y}, t_{m}^{y}-t_{m-1}^{y}\right\}.
\label{eq:EQ13_12}
\end{equation}
\end{linenomath}
The quantity $\tau_{l m}^{x y}$ represents a minimum time lag between two consecutive events of the same type occurring in one of the two time series. Synchronisation thus requires that when the event of interest is seen in one time series, say, $X$, the same event should occur in the other time series $Y$ too before being seen again in the time series $X$. If, furthermore, $e_{x}$ and $e_{y}$ denote the number of events in $X$ and $Y$, respectively, then $l = 1, 2, \ldots, e_{x}$ and $m = 1, 2, \ldots, e_{y}$. A counter of instances when the event happens in $X$ shortly after happening in $Y$ is
\begin{linenomath}
\begin{equation}
c(x | y)=\sum_{l=1}^{e_{x}} \sum_{m=1}^{e_{y}} J_{x y}^{l m}
\label{eq:EQ13_13}
\end{equation}
\end{linenomath}
with
\begin{linenomath}
\begin{equation}
J_{x y}^{l m}=\left\{\begin{array}{ll}
1, & \text { if } 0<t_{l}^{x}-t_{m}^{y} \leq \tau_{l m}^{x y}, \\
1 / 2, & \text { if } t_{l}^{x}=t_{m}^{y}, \\
0, & \text { otherwise. }
\end{array}\right.
\label{eq:EQ13_14}
\end{equation}
\end{linenomath}
Analogous reasoning is used to define $c(y | x)$. Finally, the symmetrical and anti-symmetrical combinations of these counters are
\begin{linenomath}
\begin{equation}
Q_{xy}=\frac{c(y | x)+c(x | y)}{\sqrt{e_{x} e_{y}}},
q_{xy}=\frac{c(y | x)-c(x | y)}{\sqrt{e_{x} e_{y}}}.
\label{eq:EQ13_15}
\end{equation}
\end{linenomath}
Here, $Q_{xy}$ measures the strength of event synchronisation, while $q_{xy}$ estimates the direction of time delay. In the construction of climate networks, the former (latter) quantity determines the link strength (direction).

Other methods to quantify time-series, and thus node, similarity exist and can be used in the construction of climate networks. Examples are the \textit{mutual information} method~\cite{feng2014north} and the  \textit{$\epsilon$-recurrence} method~\cite{donner2010recurrence}. Whichever method of quantifying similarity is adopted, it is common to discard weak links by applying a thresholding criterion. The network can, in fact, be made unweighted and undirected by defining the adjacency matrix as
\begin{linenomath}
\begin{equation}
A_{i,j}=\mathcal{H}\left(W_{i,j}-W_\mathrm{c}\right).
\label{eq:EQ13_16}
\end{equation}
\end{linenomath}
where $\mathcal{H}$ is the Heaviside function, $W_{i,j}$ are link weights calculated using, say, the Pearson cross-correlation, and $W_\mathrm{c}$ is a threshold. Once the climate network is constructed, it can be subjected to structural analyses using the methods of network science. Several examples of such analyses are outlined next, showing how the climate-network framework reveals new knowledge about climatic events of great societal relevance, for example, ENSO, extreme rainfall, and air pollution.

\paragraph*{El Ni\~{n}o-Southern Oscillation (ENSO) forecasting} ENSO is among the most prominent phenomena of climate variability on the interannual time scale~\cite{dijkstra2005nonlinear, clarke2008introduction}. The term refers to fluctuations between anomalous warm El Ni\~{n}o and cold La Ni\~{n}a conditions in the eastern Pacific Ocean. The occurrence of an El Ni\~{n}o event can trigger numerous disruptions around the globe, causing climate-related disasters, such as droughts, floods, fishery declines, famines, plagues, and even political and social unrest. To adequately prepare for these potential disruptions, it is pivotal to develop reliable prediction skills for when and where climate may turn extreme. After the first forecasting model from the 1980s, that is, the aforementioned Cane-Zebiak model, a number of dynamical and statistical models have been proposed to predict the  El Ni\~{n}o events. International Research Institute for Climate and Society, for example, offers some 20 climate models for ENSO forecasts. Model richness notwithstanding, early and reliable ENSO forecasting remains a substantial challenge. Good prediction skill is generally limited to about 6 months ahead, due to the presence of the so-called `\textit{spring predictability barrier}', which greatly amplifies errors arising from the coupling and feedbacks in the equatorial atmosphere-ocean system \cite{duan2013spring}.

\begin{figure}[!t]
\centering
\includegraphics[scale=1.0]{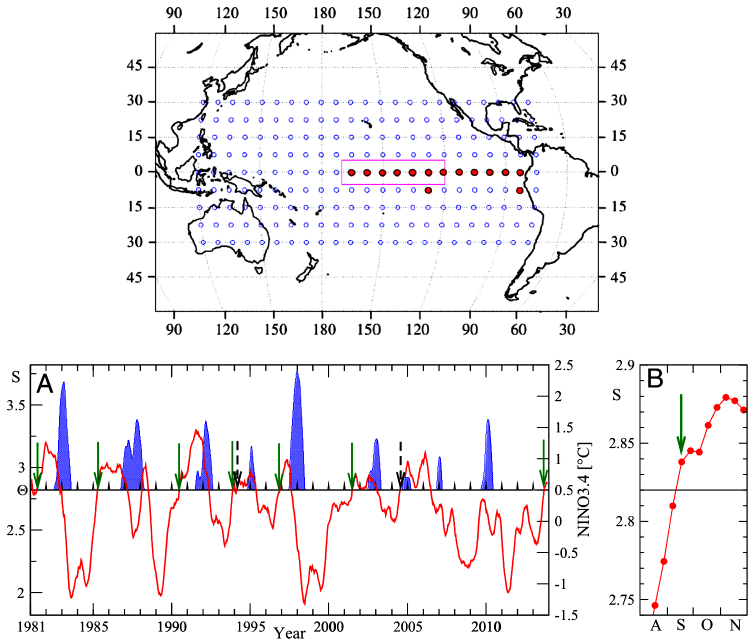}
\caption{Climate network scheme for forecasting El Ni\~{n}o events. The upper panel shows the geographical locations of grid points (i.e., nodes). The network consists of 14 nodes in the El Ni\~{n}o basin (solid red symbols) and 193 nodes outside this domain (open symbols). The red rectangle denotes the NINO3.4 region (5\,$^\circ$S--5\,$^\circ$N, 170\,$^\circ$W--120\,$^\circ$W). In the lower left panel, the red curve is the average link weight $W$ of the climate network as it changes through time, the horizontal line indicates the decision threshold $\Theta = 2.82$, and the blue areas show the El Ni\~{n}o events. When $W$ crosses the threshold from below, an alarm is sounded indicating that there is an impending El Ni\~{n}o event in the following calendar year. Correct predictions are marked by solid green arrows and false alarms by dashed black arrows. The lower right panel is a magnification for August (A), September (S), October (O), and November (N) of 2013.\newline
Source: Reprinted figure from Ref.~\cite{ludescher2014very}.}
\label{fig:Fig13_4}
\end{figure}

To improve the El Ni\~{n}o forecasting skill, especially beyond the spring predictability barrier, Ref.~\cite{ludescher2014very} resorted to an approach based on climate networks that yields reliable predictions about one year in advance. Nodes for the construction of the climate network were mainly located in the tropical Pacific (Fig.~\ref{fig:Fig13_4}, upper panel), with only a minority of nodes, denoted in red, inside the El Ni\~{n}o basin. Link weights were calculated using the Pearson cross-correlation method (see Eqs.~\ref{eq:EQ13_8} and \ref{eq:EQ13_10}). To obtain the mean strength of dynamical teleconnections in the climate network, link strength was averaged across all links
\begin{linenomath}
\begin{equation}
W^{y} =\frac{1}{n_1 n_2} \sum_{i=1}^{n_1}\sum_{j=1}^{n_2} W^{+,y}_{i,j}.
\label{eq:EQ13_17}
\end{equation}
\end{linenomath}
Here, $n_1$ and $n_2$ respectively stand for the number of red and blue nodes in the upper panel of Fig.~\ref{fig:Fig13_4}. The quantity $W^{y}$ was then compared with a decision threshold $\Theta = 2.82$, which had to be crossed from below in order to consider signalling an alarm. Additionally, the NINO3.4 index, that is, NOAA's primary indicator for monitoring El Ni\~{n}o and La Ni\~{n}a events, needed to be below 0.5\,$^\circ$C. If both conditions were satisfied, then the alarm would be signalled for the following calendar year (Fig.~\ref{fig:Fig13_4}, lower left panel). The prediction accuracy of the climate network, as measured by a type of receiver-operating-characteristic analysis, turned out to be much higher than that of the state-of-the-art climate models, for example, Kirtman~\cite{kirtman2003cola} and Chen-Cane \cite{chen2004predictability} models. The approach, in fact, successfully predicted in 2013 the onset of the 2014--2016 strong El Ni\~{n}o event (Fig.~\ref{fig:Fig13_4}C, lower right panel).

Afterwards, Ref.~\cite{meng2017percolation} proposed another framework for predicting the onset of El Ni\~{n}o events that combined a time-evolving, weighted climate network with the elements of percolation theory. In this approach, nodes near-homogeneously covered the entire globe rather than just the tropical Pacific. The climate network was shown to undergo abrupt percolation transitions usually about one year before an El Ni\~{n}o event, thus providing a reliable early-warning indicator. These research efforts were followed by yet another approach based on climate networks that successfully predicted one year in advance the onset of the 2018--2019 El Ni\~{n}o event~\cite{meng2018forecasting}. In this last approach the climate network was located entirely in the El Ni\~{n}o basin.

ENSO greatly affects atmospheric circulation patterns and exhibits strong regional and remote influences on weather. To investigate the global impacts of ENSO, Ref.~\cite{fan2017network} resorted to constructing a series of directed and weighted climate networks based on the near-surface air temperature. Regions that are characterised by larger positive or negative network links correlated more with the NINO3.4 index, thus becoming warmer (cooler) during El Ni\~{n}o (La Ni\~{n}a) periods. Although regions affected by ENSO vary from one event to another, and are difficult to predict, the climate network analysis offered a new perspective on the problem with much potential for further successes.

ENSO, as referred to heretofore, is sometimes called Eastern Pacific ENSO to distinguish it from a temperature anomaly that arises in the central Pacific~\cite{larkin2005definition}, which is called Central Pacific ENSO~\cite{kao2009contrasting} or ENSO Modoki~\cite{ashok2007nino} (Japanese `modoki' roughly translates as `pseudo'). ENSO Modoki has distinct teleconnections and affects many parts of the world, yet distinguishing and predicting the type of ENSO in practice remains a challenge. Climate networks may help, as evidenced by a novel method to predict the type of El Ni\~{n}o events, as well as estimate their impacts in advance~\cite{lu2020impacts}.

\paragraph*{Extreme-precipitation patterns} Precipitation, at its extremes, poses a threat to society, resulting in the loss of life and property in floods and landslides. Flooding due to extreme precipitation in India, for example, affected over 800 million people in the period between 1950--2015, leaving 17 million without homes and causing 69,000 deaths~\cite{roxy2017threefold}. In early 2017 in coastal Peru, a series of extreme precipitation events caused severe floods, killing 114 people, displacing 184,000 people, and creating damages in excess of \$3 billion USD~\cite{son2020climate}.

Even more disconcerting than historical records is the fact that the intensity of extreme weather events is expected to strengthen under global warming. The basic mechanism of how a temperature rise fuels extreme rainfall is clear, (i) warmer ocean waters carry energy more easily to the atmosphere when storms form, and (ii) for every degree of surface-temperature warming, the atmosphere holds about 7\,\% more water vapour~\cite{fischer2016observed}. Accordingly, climate models also predict intensification in the annual maximum precipitation, although they possibly underestimate the true future state of affairs~\cite{borodina2017models}, thus exposing gaps in our understanding of the factors involved. Such gaps, for example, include limited knowledge of global and regional teleconnection patterns associated with extreme rainfall. Recent progress in this context has relied heavily on climate networks, not only by mapping extreme-rainfall teleconnections, but also suggesting the underlying mechanisms behind the observed phenomena.

Ref.~\cite{boers2014prediction} offered a new conceptual route to study the spatial characteristics of the synchronicity of extreme rainfall in South America during the monsoon seasons. First, the study defined extreme-rainfall events as those above the 99th percentile over the spatial domain covering 40\,$^\circ$S--15\,$^\circ$N and 30\,$^\circ$W--85\,$^\circ$W at a resolution of 0.25\,$^\circ$, and the temporal domain extending from 1998 to 2012 at a resolution of 3 hours. A climate network was then constructed using the aforementioned event-synchronisation method. The synchronisation strength into $S_{i}^\mathrm{in}$ (out of $S_{i}^\mathrm{out}$) a climate-network node was defined as the sum of weights of all links pointing to (from) this node. The network divergence $\Delta S$ was then introduced to spatially resolve the temporal order of extreme-rainfall events,
\begin{linenomath}
\begin{equation}
\Delta S_{i}=S_{i}^\mathrm{in}-S_{i}^\mathrm{out}=\sum_{j=1}^{N} A_{i j}-\sum_{j=1}^{N} A_{j i},
\label{eq:EQ13_18}
\end{equation}
\end{linenomath}
where $A_{i j}$ is the adjacency matrix of the climate network. The positive (negative) values of $\Delta S_{i}$ indicated sink (source) nodes, that is, locations where extreme events occur shortly, within two days, after (before) occurring at many other locations. Typical propagation pathways of extreme events could thus be identified along which extreme events have high predictability. The method was applied to the real-time satellite-derived rainfall data to successfully predict more than 60\,\% of extreme-rainfall events in the Central Andes of South America, with the success rate going above 90\,\% during El Ni\~{n}o conditions.

\begin{figure}[p]
\makebox[\textwidth][c]{\includegraphics[scale=1.0]{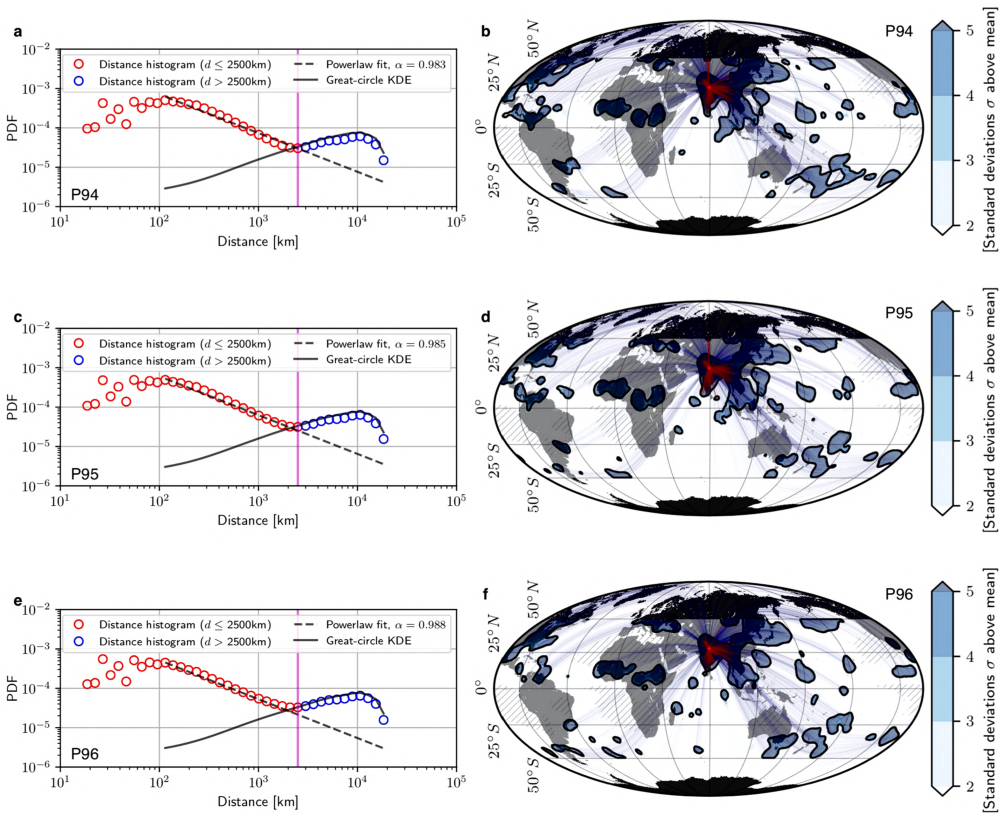}}
\caption{Distance distribution and teleconnection pattern in south-central Asia for different extreme-event percentiles. Left-column panels show the probability density function of the significant link distances (red and blue circles), the power-law fit over the range 100--2,500\,km (dashed line), and the kernel-density estimate (KDE) of the distribution of all possible great-circle distances (solid black line) for extreme-event percentiles $\alpha = 0.94$, $0.95$, and $0.96$, respectively. Right-column panels show link bundles attached to south-central Asia for extreme-event percentiles $\alpha = 0.94$, $0.95$, and $0.96$, respectively. Links shorter (longer) than 2,500\,km are denoted in red (blue).\newline
Source: Reprinted figure from Ref.~\cite{boers2019complex}.}
\label{fig:Fig13_5}
\end{figure}

A similar methodology based on climate networks, and specifically on the event-synchronisation method, has been applied to investigate the spatial configuration of synchronisation between extreme-rainfall events around the globe. Ref.~\cite{boers2019complex}, using a network with 576,000 nodes, found that the distribution of distances between significant links (p-value $p<0.005$) decays according to a power law with a coefficient $\approx$1 up to distances of about 2,500\,km, while the probability of significant longer-distance links is much larger than expected from the power law. The relative underabundance of shorter-distance links is due to regional weather systems, yet the relative overabundance of longer-distance links, which form a global rainfall teleconnection pattern, is probably dominated by the \textit{Rossby waves}. The described picture is robust to the choice of the extreme-event percentile (Fig.~\ref{fig:Fig13_5}, left-column panels). Furthermore, climate networks revealed that the extreme-rainfall events in the monsoon systems of south-central Asia, east Asia, and Africa are strongly synchronised (Fig.~\ref{fig:Fig13_5}, right-column panels). The use of climate networks thus made inroads towards the global predictability of natural hazards associated with extreme rainfall.

\subsection{Impact of Rossby waves on air pollution}

Air pollution is a major health concern worldwide. According to the World Health Organization (WHO)~\cite{guterres2019stressing}:
\begin{quote}
    An estimated 9 out of 10 people worldwide are exposed to air pollutants that exceed World Health Organization (WHO) air quality guidelines. [P]olluted air kills some 7 million people each year, causes long-term health problems, such as asthma, and reduces children’s cognitive development. According to the World Bank, air pollution costs societies more than \$5 trillion every year.
\end{quote}
Climate change and air pollution are closely related. For example, the main sources of CO$_2$ emissions are also a major source of air pollutants. Conversely, many air pollutants, such as particulate matter, ozone, nitrogen dioxide, etc. contribute to climate change by affecting the amount of incoming radiation that is reflected or absorbed by the atmosphere. The exact feedbacks between weather and climate dynamics at different pressure levels, on the one hand, and the fluctuations in air pollution, on the other hand, is a subject of intense study.

\begin{figure}[!t]
\centering\includegraphics[scale=1.0]{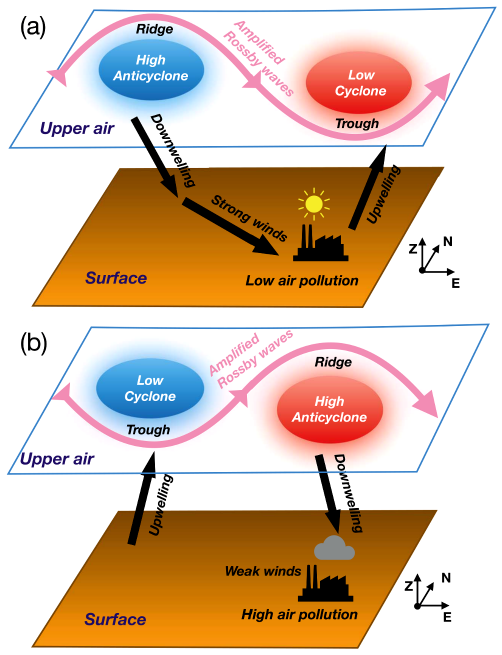}
\caption{Schematic representation of Rossby waves influencing air pollution. Panel (a) shows low-pollution conditions. Panel (b) shows high-pollution conditions. Blue and red colours respectively represent the negative and positive out‐degree clusters of the climate network.\newline
Source: Reprinted figure from Ref.~\cite{zhang2019significant}.}
\label{fig:Fig13_6}
\end{figure}

To study air-pollution spreading and diffusion patterns, Ref.~\cite{zhang2019significant} employed a multilayer and multivariable network analysis designed to delineate the influence of the upper air dynamics (at 500\,hPa geopotential height) on the temporal variability of the surface air pollution ($\mathrm{PM}_{2.5}$) in China and U.S. Two multilayer networks were considered, one with dominant negative-correlated interlinks and the other with positive-correlated interlinks, corresponding to negative, Eq.~(\ref{eq:EQ13_11}), and positive, Eq.~(\ref{eq:EQ13_10}), weights, respectively. Only the links for which $\left|W\right|>W_\mathrm{c} = 4.5$ were selected based on shuffled data-significance tests. Applying this methodology showed that the upper air critical regimes substantially influence the surface air pollution. Specifically, Rossby waves influence the air-pollution fluctuations through the development of cyclone and anticyclone systems that control local winds and air stability (Fig.~\ref{fig:Fig13_6}). High-pressure anticyclones form on the ridges, while low-pressure cyclones form on the troughs of Rossby waves. The former, identified by negative out-degree clusters in climate networks, cause the air to downwell. The latter, identified by positive out-degree clusters in climate networks, cause the air to upwell. The described downwelling and upwelling pattern induces strong winds that keep air pollution low. As Rossby waves travel, upwelling replaces downwelling and vice versa, which weakens the winds and leads to subsequent accumulation of pollution in the air. Recognising the outlined mechanism behind the air-pollution fluctuations helps to improve the prediction of extreme pollution events, and once again highlights the potential of climate networks to unveil intricate interactions and feedbacks in the climate system.

\subsection{Critical phenomena in the climate system}

\paragraph*{Critical points, exponents, and universality} The concept of critical phenomena is most commonly associated with physical systems that undergo phase transitions at a \textit{critical point}. Examples include the vapour-to-liquid-to-solid transitions of substances at their critical points characterised by a specific value of temperature and pressure, or the ferromagnetism-to-paramagnetism transition of some solids at their Curie point under zero magnetic field. Critical phenomena are present in both nature (lakes, oceans, terrestrial ecosystems, etc.) and society~\cite{scheffer2009critical}, including the climate system itself. Ref.~\cite{dorogovtsev2008critical} offers a general overview of the subject.

Curiously, critical phenomena seem to be independent of the details of the physical system at hand. Instead, only the system's general features seem to matter~\cite{stanley1971phase}, such as (i) the spatial dimensionality (e.g., the system's arrangement in a two-dimensional, three-dimensional, or more-dimensional lattice), (ii) the dimensionality of the order parameter (e.g., the system's spin dimensionality), and (iii) the range of microscopic interactions (e.g., only first neighbours interact). Many physical quantities that describe the system's state near a critical point have a power-law form whose main feature is the \textit{critical exponent}. The ubiquity of power laws is often referred to as \textit{universality}---different systems with the same values of the critical exponent are said to belong to the same universality class. In what follows, we focus on some of the critical phenomena specific to the Earth climate system.

\paragraph*{Critical phenomena in atmospheric precipitation} Earth's atmosphere is a fluid in complex motion that dynamically varies in space and time. Despite its dynamic complexity, from a meteorological perspective, the atmosphere is driven by slow large-scale forcing (moisture convergence, evaporation, and radiative cooling) and rapid convective-buoyancy release
(small-scale convection). Convection intensifies above a critical point in the water vapour, causing the onset of heavy precipitation. This intensification is reflected in the average precipitation rate as a function of the water vapour, which exhibits a relatively simple power-law behaviour as predicted by the theory of critical phenomena~\cite{peters2006critical}.

\begin{figure}[!t]
\makebox[\textwidth][c]{\includegraphics[scale=1.0]{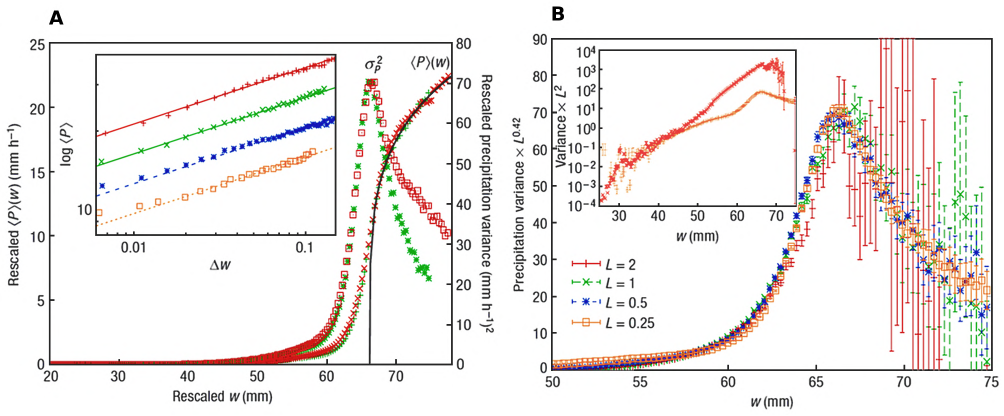}}
\caption{Critical phenomena in atmospheric precipitation. \textbf{A,} The average precipitation rates $\lra{P}(w)$ (order parameter) and their variances $\sigma_{p}^{2}(w)$ (susceptibility) are shown as a function of the water vapour (tuning parameter) for the eastern (red, 170\,$^\circ$W--70\,$^\circ$W) and the western (green, 120\,$^\circ$E--170\,$^\circ$W) Pacific Ocean. The solid curve stands for a power-law fit above the critical point. The inset shows, using double-logarithmic scales, $\lra{P}(w)$ as a function of the reduced water vapour, $\Delta w \equiv\left(w-w_{\mathrm{c}}\right) / w_{\mathrm{c}}$, for the western Pacific (green, 120\,$^\circ$E--170\,$^\circ$W), the eastern Pacific (red, 170\,$^\circ$W--70\,$^\circ$W), the Atlantic (blue, 70\,$^\circ$W--20\,$^\circ$E), and the Indian Ocean (pink, 30\,$^\circ$E--120\,$^\circ$E). Note the same slope irrespective of the climatic region. \textbf{B,} Finite-size scaling of the variance $\sigma_{p}^{2}(w; L)$ of the order parameter in the western Pacific. Near the critical point, $w>57$\,mm, the collapse of the curves is good, indicating $\sigma_{p}^{2}(w_\mathrm{c}; L)\propto{}L^{-0.42}$, as expected from the theory of critical phenomena. The inset shows that relatively far away from the critical point, $w<40$\,mm, trivial scaling $\sigma_{p}^{2}(w; L)\propto{}L^{-2}$ works adequately. The error bars represent standard errors.\newline
Source: Reprinted figure from Ref.~\cite{peters2006critical}.}
\label{fig:Fig13_7}
\end{figure}

Using satellite data from the Tropical Rainfall Measuring Mission, Ref.~\cite{peters2006critical} analysed the relationship between the precipitation rate, $P$, and the water vapour, $w$. Various major ocean basins were covered by oceanic grid points between 20\,$^\circ$S--20\,$^\circ$N. The precipitation and water-vapour data were collected at 0.25\,$^\circ$ latitude-longitude resolution. From a statistical physics perspective, quantities $P$ and $w$ were regarded as the \textit{order parameter} and \textit{tuning parameter}, respectively. It was found that, when the tuning parameter crosses its critical value, $w_\mathrm{c}$, the order parameter can be well approximated by a power-law of the form
\begin{linenomath}
\begin{equation}
\lra{P}(w)=a\left(w-w_{\mathrm{c}}\right)^{\beta},
\label{eq:EQ13_19}
\end{equation}
\end{linenomath}
where $a$ is a system-dependent constant and $\beta$ is a critical exponent. The operator $\lra{\cdot}$ refers to averaging over all observations in a given region. The same power-law fits the data irrespective of the climatic region (Fig.~\ref{fig:Fig13_7}A). Similarly, the critical exponent is universal and independent of the climatic region, with the value of $0.215 \pm 0.02$ (inset in Fig.~\ref{fig:Fig13_7}A).

The susceptibility of the system, $\chi(w; L)$, was defined by means of the variance of the order parameter such that
\begin{linenomath}
\begin{equation}
\chi(w; L)=L^{d} \sigma_{p}^{2}(w; L),
\label{eq:EQ13_20}
\end{equation}
\end{linenomath}
where $d$ stands for the system's dimensionality and $L$ for the spatial resolution. Near the critical point $w_\mathrm{c}$, however, the theory of critical phenomena suggests that~\cite{privman1990finite}
\begin{linenomath}
\begin{equation}
\chi(w; L)=L^{\gamma / \nu} \tilde{\chi}\left(\Delta w L^{1 / \nu}\right),
\label{fig:EQ13_21}
\end{equation}
\end{linenomath}
where $\gamma$ and $\nu$ are the standard critical exponents, $\Delta w \equiv\left(w-w_{\mathrm{c}}\right) / w_{\mathrm{c}}$ is the reduced water vapour, and $\tilde{\chi}(x)$ is the usual finite-size scaling function. When $\Delta w=0$ (i.e., $w = w_c$), $\tilde{\chi}(0)$ is constant, implying the scaling relationship $\sigma_{P}^{2}(w_\mathrm{c}; L) \propto L^{\gamma / \nu-d}$. Data confirm this relationship with the critical exponent ratio of $\gamma/\nu=1.58$ (Fig.~\ref{fig:Fig13_7}B). For $w>57$\,mm, all data indeed collapse into a single function. Furthermore, relatively far from the critical point, $w<40$\,mm, scaling is expectedly trivial (inset in Fig.~\ref{fig:Fig13_7}B). These results show that the balance between slow large-scale forcing via moisture convergence, evaporation, and radiative cooling and rapid convective buoyancy release via small-scale convection leads to continuous (i.e., second order) phase transition such that below the critical point, there is very little precipitation, but once the critical point is crossed, precipitation rapidly increases with the water vapour. Interestingly, the balance between forcing and buoyancy release is stable, further suggesting that the described atmospheric criticality is in fact self-organised.

\paragraph*{Hadley cell and percolation} The Hadley cell is a global-scale three-dimensional tropical atmospheric circulation that transports energy and angular momentum poleward. This circulation enables, among others, the trade winds, hurricanes, and the jet streams. The locations of the subtropical dry zones and the major tropical and subtropical deserts are strongly associated with the subsiding branches of the Hadley cell~\cite{held2006robust}. Therefore, understanding how structure and intensity of the Hadley cell may change under global warming has attracted widespread attention. For example, an analysis of satellite observations indicated a poleward expansion, by $\approx$2\,$^\circ$, of the Hadley cell over the period from 1979 to 2005~\cite{fu2006enhanced}. A physical mechanism for the expansion of the Hadley cell was proposed shortly afterwards~\cite{lu2007expansion}, followed by a discovery of a robust weakening of the Hadley cell in the 21st century through the analysis of 30 different CMIP5 coupled model simulations~\cite{seo2014mechanism}. Observations, theory, and climate models are thus coming together to suggest the poleward expansion and weakening of the Hadley cell under global warming.

A standard approach to determining the strength of the Hadley cell is to
compute the observed zonal-mean mass-stream function, $\Psi$. This function relates to the zonal-mean meridional wind velocity $V$ via
\begin{linenomath}
\begin{equation}
[\overline{V}] =  \frac{g}{2\pi R \cos\phi} \frac{\partial \Psi}{\partial p},
\label{eq:EQ13_22}
\end{equation}
\end{linenomath}
where the operators $\bar{\cdot}$ and $[\cdot]$ stand for temporal and zonal averaging, respectively. The quantity $g$ is the gravitational acceleration, $R$ is the mean Earth radius, $\phi$ is the latitude, and $p$ designates pressure coordinates. When calculating the $\Psi$ field, it is common to assume $\Psi = 0$ at the top of the atmosphere. Based on Eq.~(\ref{eq:EQ13_22}), the edges of the Hadley cell are identified as the first latitude poleward of the maximum of $\Psi_{500}$ at which $\Psi_{500}=0$, where the index 500 indicates the value of the stream function at 500\,hPa~\cite{lu2007expansion}. Although this conventional analysis has been applied to investigate the structure and intensity of the Hadley cell, there are some important limitations: (i) the latitude-longitude structure of the Hadley cell is not fully resolved because Eq.~(\ref{eq:EQ13_22}) only accounts for the latitudinal direction, and (ii) in contrast to theory and models that predict the decreasing intensity of the Hadley cell, the reanalysis datasets point to an increasing intensity~\cite{mitas2005has}.

To circumvent the limitations of the conventional approach, Ref.~\cite{fan2018climate} analysed the structure and intensity of the Hadley cell using climate networks and \textit{percolation} theory. The main question of percolation theory~\cite{aharony2003introduction} can be posed in several different ways, but in the context of network science, one seeks the probability $q$ of node failure such that after $q$\,\% of nodes do fail, the network changes from being connected to being disconnected. It turns out that there exist a critical probability $q_\mathrm{c}$ below which (i.e., for $q<q_\mathrm{c}$) the network remains connected with probability one, but above which (i.e., for $q>q_\mathrm{c}$) the network gets disconnected with probability one. This criticality strictly holds only for infinite networks, but is, in fact, easily observed in practice in networks with $\mathcal{O}(100)$ nodes. In Ref.~\cite{fan2018climate}, the near-surface atmosphere was represented with a two-dimensional grid of points that turned into a lattice by adding links between nearest neighbours. These links were added as follows. First, the strength of each link, $W_{i,j}$, was calculated based on Eq.~(\ref{eq:EQ13_10}). Link strengths were then sorted in descending order. The strongest link was the first one to be added, then the second strongest, the third strongest, and so on. The resulting lattice-shaped climate network was found to undergo an abrupt phase transition in the order parameter $G_1$, defined as the largest connected network component. Because the original grid points were embedded into the spherical Earth surface, the right expression for the order parameter was
\begin{linenomath}
\begin{equation}
G_1(M) = \frac{\max \left[\sum\limits_{i\in S_1 (M)} \cos(\phi_i),\cdots, \sum\limits_{i\in S_m (M)} \cos(\phi_i),\cdots,\right]}{\sum\limits_{i=1}^{N} \cos(\phi_i)},
\label{eq:EQ13_23}
\end{equation}
\end{linenomath}
where $M$ is the number of added links, $\phi_i$ is the latitude of grid point $i$, and $S_j$ is the $j$th connected network component in terms of the number of nodes. The percolation threshold at $M=M_\mathrm{c}$ was determined by recording jumps in $G_1(M)$ with each added link, and singling out the largest jump to mark the value $G_\mathrm{c}=G_1(M_\mathrm{c})$ and the corresponding critical link weight $W_\mathrm{c}$~\cite{fan2020universal}. By altering the resolution of grid points, the theory of critical phenomena was used to confirm that the order parameter is indeed discontinuous at the percolation threshold and thus consistent with a first-order phase transition. The largest connected component of the climate network at the percolation threshold, obtained upon applying the described methodology, is located in the tropics, as expected from an analogue of the Hadley cell.

\begin{figure}[!t]
\makebox[\textwidth][c]{\includegraphics[scale=1.0]{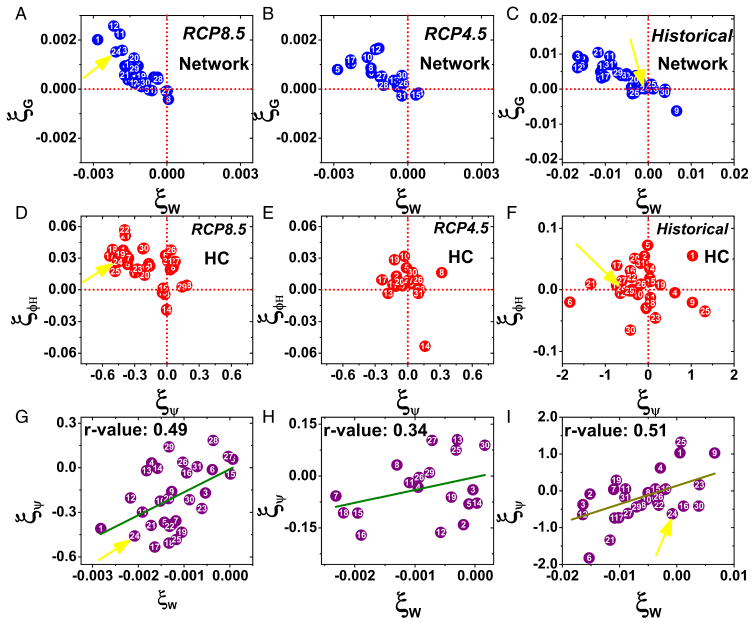}}
\caption{Comparing the rates of change of the Hadley cell and its climate-network analogue under three climate-change scenarios. The climate network analogue of the Hadley cell is the largest connected network component at the percolation threshold. Panels A--C show that this component is getting larger over time, while the critical link weight is decreasing under all scenarios for most of the CMIP5 climate models. Panels D--F show that the size of the Hadley cell and its intensity exhibit qualitatively similar behaviour as their climate-network analogues, although there is somewhat more ambiguity in the results, especially in the Historical scenario. Panels G--I show that the rates of change obtained via the climate-network analysis and the conventional approach show significant correlation. Numbering in the circles indexes the 31 CMIP5 climate models.\newline
Source: Reprinted figure from Ref.~\cite{fan2018climate}.}
\label{fig:Fig13_8}
\end{figure}

The purpose here was more than just finding a climate-network analogue of the Hadley cell. To determine the temporal evolution of the quantities $G_\mathrm{c}$ and $W_\mathrm{c}$, a sequence of climate networks was constructed using successive and non-overlapping temporal windows with the length of 60\,mos. The results could be fitted adequately with simple linear relationships
\begin{linenomath}
\begin{subequations}
\begin{align}
G_\mathrm{c}(t) &= a + \xi_G t,\\
W_\mathrm{c}(t) &= b + \xi_W t,
\end{align}
\end{subequations}
\end{linenomath}
where $a$ and $b$ are constants, while $\xi_G$ and $\xi_W$ are the rates of change of the quantities $G_\mathrm{c}$ and $W_\mathrm{c}$. Denoting the analogous rates of change for the Hadley cell, obtained via the conventional approach, with $\xi_{\phi_{H}}$ and $\xi_\Psi$, the results show a consistent expansion from the tropics poleward of the largest connected network component at the percolation threshold, as well as the weakening of the corresponding critical link weight. The same holds for the 31 CMIP5 21st century climate models and the reanalysis data (ERA-Interim and ERA-40). Put more quantitatively, most of the CMIP5 models exhibit $\xi_G>0$ and $\xi_W<0$, and similarly $\xi_{\phi_{H}}>0$ and $\xi_{\Psi}<0$ irrespective of the the climate scenario (Fig.~\ref{fig:Fig13_8}, cf. panels A--C and D--F). The results obtained via climate-network analysis and using the conventional approach are highly correlated (Fig.~\ref{fig:Fig13_8}, panels G--I).

The poleward expansion of the Hadley cell may result in (i) a drier future in some tropical or subtropical regions~\cite{lu2007expansion} and (ii) a poleward migration of the location of the maximum tropical-cyclone intensity~\cite{kossin2014poleward}. The climate-network analysis described herein may therefore help to identify regions that are more probable to
experience precipitation decline or hurricane intensification. Among the prime candidate regions to be affected by the Hadley-cell expansion are northern India, southern Africa, and western Australia. Local governments in these and other potentially exposed regions should keep a close eye on climate science and take risk-mitigating actions until there is still time.

\subsection{Future outlook}

This chapter started by outlining the social consequences of global climate change and briefly venturing into the history of climate modelling. Thereafter, two research topics central to the current climate-change discourse---climate networks and critical phenomena---were introduced. Both of these topics originate from statistical physics, thus sharing their origins with many other themes in the present review that fall more squarely into the domain of social physics. The reason for the shared origins of the climate-change research and the social-physics research in the stricter sense is that the climate system is an epitome of complexity as much as human society is. Despite the fast-paced progress seen over the past two decades, there is still a lot of work ahead, especially in the context of integrating climate predictions into social dynamics.

Because climate networks are constructed by applying similarity measures to observational data, the underlying physical mechanisms and processes often remain hidden or unclear. Shedding light on such mechanisms and processes may, however, substantially impact our understanding of climate change and subsequently improve the predictive power of numerical climate models. A promising methodology that has emerged in recent years and could play an instrumental role in demystifying climate change is machine learning and AI~\cite{dijkstra2019application}. Considering the inherent `black box' structure of climate systems, the integration of AI and visual analytics provides a potential solution~\cite{luo2017impact}.

Another key issue is the question of analysing the climate resilience of ecosystems and economies. Currently, for example, there is a lack of appropriate models to fully understand and predict the effects of cascading failures~\cite{buldyrev2010catastrophic}, triggered by extreme climate and weather events, on critical interdependent infrastructures. Closing this knowledge gap is a crucial step towards climate-resilient society.

\FloatBarrier

\section{Epilogue: Keeping the dialogue open}
\label{S:Epi}

We hope this review has given the reader an overview of physicists' contributions to multidisciplinary social science. To make the story contiguous, we had to sacrifice some topics, like the physics of art (music~\cite{berezovsky2019structure}, painting~\cite{kim2014large}, dance~\cite{laws2002physics}, etc.), agriculture~\cite{gandica2021bali}, gastronomy~\cite{ahn2011flavor}, ethnology (how ethnic groups remember their shared history)~\cite{lee2010emergence}, civil unrest~\cite{braha2012global}, etc. Conversely, we imagine some readers finding our definition of social physics too generous, especially at the border between physics and engineering, artificial intelligence, and climate modelling. Perhaps a better title would be \textit{Human physics}---`human' as in `topics affecting humans'.

We started our expos\'e arguing that physics has played a fundamental role in the modern movement towards multidisciplinarity. However, physicists entering multidisciplinary research have a bad reputation for their imperious attitude: ``Step aside! We'll show you how it's done.'' When this happens, even if unintentional, it threatens mutual respect and understanding between collaborators and jeopardises the overall success of collaborative interactions. But what better way to ensure mutual respect and understanding than to keep the dialogue open.

To illustrate what we have in mind, a physicist's strength lies in putting quantitative methods to good use, be it rigorous data analyses or complex numerical simulations. The use of quantitative methods, however, is preceded by formulating research hypotheses of interest or model assumptions of relevance to the problem at hand. Seeking inputs from experts is absolutely crucial in this stage because intuition and common sense cannot replace expert knowledge, and may easily lead to simplistic and naive hypotheses or assumptions. Accordingly, before quantitative methods are employed, physicists for the most part need to be on the receiving end of the dialogue with their multidisciplinary collaborators.

Another strength that is rather unique to physicists is seeing the big picture and consequently making approximations that simplify the problem, but still account for the main processes at play. It is important to recognise that such approximations and subsequent simplifications go against the training received by researchers from many other disciplines. Among ecologists, for example, the focus on biodiversity is so prevalent that general patterns often come secondary to exceptions. Accordingly, when quantitative methods are employed, physicists for the most part need to be on the transmitting end of the dialogue with their multidisciplinary collaborators.

Many more situations are bound to arise in practice in which keeping the dialogue open will be crucial to success. They demand patience and care, but when resolved satisfactorily, they lead to insightful and impactful research that is so much needed to ensure the continued prosperity of humankind.

\section*{Acknowledgements}
We are thankful to Sheikh Taslim Ali, Yosef Ashkenazy, Simon Cauchemez, Xiaosong Chen, Benjamin J. Cowling, Zhanwei Du, Shlomo Havlin, Juergen Kurths, Josef Ludescher, Jun Meng, Lauren A. Meyers, Henrik Salje, Hans Joachim Schellnhuber, Maximilian Schich, Joseph T. Wu, and Xiaoke Xu. M.\,J. was supported by the Japan Society for the Promotion of Science (JSPS) KAKENHI grants-in-aid for scientific research JP20H04288, JP21K04545, and JP21H03625. P.\,H. was supported by JSPS KAKENHI JP21H04595. K.\,K. was supported by the Japan Science and Technology Agency (JST), PRESTO grant JPMJPR20M2. Z.\,W. was supported by the National Natural Science Foundation for Distinguished Young Scholars (grant no.\ 62025602), the National Natural Science Foundation of China (grants nos.\ U1803263, 11931015, 81961138010), the Fok Ying-Tong Education Foundation of China (grant no.\ 171105), and the Key Technology Research-and-Development Program of the Science and Technology-Scientific and Technological Innovation Team of Shaanxi Province (grant no.\ 2020TD-013). S.\,G. and T.\,K. were supported by the Croatian Science Foundation project 2018-01-3150 AqADAPT. B.\,P. was supported by the Slovenian Research Agency (grant no.\ J7-3156). L.\,W. was supported by the European Research Council (grant no.\ 804744), the Engineering and Physical Sciences Research Council (EPSRC) Impact Acceleration Grant RG90413, Health and Medical Research Fund, Food and Health Bureau, Government of the Hong Kong Special Administrative Region (grant no.\ 20190712), National Natural Science Foundation of China (grant no.\ 62173065 and 11975025), and the International COVID-19 Data Alliance (ICODA), an initiative funded by the COVID-19 Therapeutics Accelerator and convened by Health Data Research UK. W.\,L. was supported by The National University of Singapore Start-up and Ministry of Education Tier 1 grants (grant no.\ WBS R-109-000-270-133). J.\,F. was supported by the East Africa Peru India Climate Capacities---EPICC project, which is a part of the International Climate Initiative (IKI) sponsored by the Federal Ministry for the Environment, Nature Conservation, and Nuclear Safety (BMU) on the basis of a decision adopted by the German Bundestag. The Potsdam Institute for Climate Impact Research (PIK) is leading the execution of the project together with project partners The Energy and Resources Institute (TERI) and the German Meteorological Service, Deutscher Wetterdienst (DWD). M.\,P. was supported by the Slovenian Research Agency (grant nos.\ P1-0403, J1-2457, and J1-9112).

\clearpage







\begin{thebibliography}{1161}
\expandafter\ifx\csname natexlab\endcsname\relax\def\natexlab#1{#1}\fi
\providecommand{\url}[1]{\texttt{#1}}
\providecommand{\href}[2]{#2}
\providecommand{\path}[1]{#1}
\providecommand{\DOIprefix}{doi:}
\providecommand{\ArXivprefix}{arXiv:}
\providecommand{\URLprefix}{URL: }
\providecommand{\Pubmedprefix}{pmid:}
\providecommand{\doi}[1]{\href{http://dx.doi.org/#1}{\path{#1}}}
\providecommand{\Pubmed}[1]{\href{pmid:#1}{\path{#1}}}
\providecommand{\bibinfo}[2]{#2}
\ifx\xfnm\relax \def\xfnm[#1]{\unskip,\space#1}\fi
\bibitem[{Ball(2002)}]{ball2002physical}
\bibinfo{author}{P.~Ball},
\newblock \bibinfo{title}{The physical modelling of society: a historical
  perspective},
\newblock \bibinfo{journal}{Physica A} \bibinfo{volume}{314}
  (\bibinfo{year}{2002}) \bibinfo{pages}{1--14}.
\bibitem[{Sinatra et~al.(2015)Sinatra, Deville, Szell, Wang, and
  Barab{\'a}si}]{sinatra2015century}
\bibinfo{author}{R.~Sinatra}, \bibinfo{author}{P.~Deville},
  \bibinfo{author}{M.~Szell}, \bibinfo{author}{D.~Wang}, \bibinfo{author}{A.-L.
  Barab{\'a}si},
\newblock \bibinfo{title}{A century of physics},
\newblock \bibinfo{journal}{Nat. Phys.} \bibinfo{volume}{11}
  (\bibinfo{year}{2015}) \bibinfo{pages}{791--796}.
\bibitem[{Bromley(2002)}]{bromley2002century}
\bibinfo{author}{D.~A. Bromley}, \bibinfo{title}{A century of physics},
  \bibinfo{publisher}{Springer Science \& Business Media},
  \bibinfo{year}{2002}.
\bibitem[{Gabrielse et~al.(2006)Gabrielse, Hanneke, Kinoshita, Nio, and
  Odom}]{gabrielse2006new}
\bibinfo{author}{G.~Gabrielse}, \bibinfo{author}{D.~Hanneke},
  \bibinfo{author}{T.~Kinoshita}, \bibinfo{author}{M.~Nio},
  \bibinfo{author}{B.~Odom},
\newblock \bibinfo{title}{New determination of the fine structure constant from
  the electron g value and {QED}},
\newblock \bibinfo{journal}{Phys. Rev. Lett.} \bibinfo{volume}{97}
  (\bibinfo{year}{2006}) \bibinfo{pages}{030802}.
\bibitem[{Everitt et~al.(2011)Everitt, DeBra, Parkinson, Turneaure, Conklin,
  Heifetz, Keiser, Silbergleit, Holmes, Kolodziejczak
  et~al.}]{everitt2011gravity}
\bibinfo{author}{C.~W.~F. Everitt}, \bibinfo{author}{D.~B. DeBra},
  \bibinfo{author}{B.~W. Parkinson}, \bibinfo{author}{J.~P. Turneaure},
  \bibinfo{author}{J.~W. Conklin}, \bibinfo{author}{M.~I. Heifetz},
  \bibinfo{author}{G.~M. Keiser}, \bibinfo{author}{A.~S. Silbergleit},
  \bibinfo{author}{T.~Holmes}, \bibinfo{author}{J.~Kolodziejczak}, et~al.,
\newblock \bibinfo{title}{Gravity probe {B}: final results of a space
  experiment to test general relativity},
\newblock \bibinfo{journal}{Phys. Rev. Lett.} \bibinfo{volume}{106}
  (\bibinfo{year}{2011}) \bibinfo{pages}{221101}.
\bibitem[{Nisbet et~al.(2000)Nisbet, Muller, Lika, and
  Kooijman}]{nisbet2000molecules}
\bibinfo{author}{R.~M. Nisbet}, \bibinfo{author}{E.~B. Muller},
  \bibinfo{author}{K.~Lika}, \bibinfo{author}{S.~A. L.~M. Kooijman},
\newblock \bibinfo{title}{From molecules to ecosystems through dynamic energy
  budget models},
\newblock \bibinfo{journal}{J. Anim. Ecol.} \bibinfo{volume}{69}
  (\bibinfo{year}{2000}) \bibinfo{pages}{913--926}.
\bibitem[{Brown et~al.(2004)Brown, Gillooly, Allen, Savage, and
  West}]{brown2004toward}
\bibinfo{author}{J.~H. Brown}, \bibinfo{author}{J.~F. Gillooly},
  \bibinfo{author}{A.~P. Allen}, \bibinfo{author}{V.~M. Savage},
  \bibinfo{author}{G.~B. West},
\newblock \bibinfo{title}{Toward a metabolic theory of ecology},
\newblock \bibinfo{journal}{Ecology} \bibinfo{volume}{85}
  (\bibinfo{year}{2004}) \bibinfo{pages}{1771--1789}.
\bibitem[{Kooijman(2010)}]{kooijman2010dynamic}
\bibinfo{author}{S.~A. L.~M. Kooijman}, \bibinfo{title}{Dynamic energy budget
  theory for metabolic organisation}, \bibinfo{publisher}{Cambridge University
  Press}, \bibinfo{year}{2010}.
\bibitem[{Jusup et~al.(2017)Jusup, Sousa, Domingos, Labinac, Marn, Wang, and
  Klanj{\v{s}}{\v{c}}ek}]{jusup2017physics}
\bibinfo{author}{M.~Jusup}, \bibinfo{author}{T.~Sousa},
  \bibinfo{author}{T.~Domingos}, \bibinfo{author}{V.~Labinac},
  \bibinfo{author}{N.~Marn}, \bibinfo{author}{Z.~Wang},
  \bibinfo{author}{T.~Klanj{\v{s}}{\v{c}}ek},
\newblock \bibinfo{title}{Physics of metabolic organization},
\newblock \bibinfo{journal}{Phys. Life Rev.} \bibinfo{volume}{20}
  (\bibinfo{year}{2017}) \bibinfo{pages}{1--39}.
\bibitem[{Kearney and Porter(2009)}]{kearney2009mechanistic}
\bibinfo{author}{M.~Kearney}, \bibinfo{author}{W.~Porter},
\newblock \bibinfo{title}{Mechanistic niche modelling: combining physiological
  and spatial data to predict species' ranges},
\newblock \bibinfo{journal}{Ecol. Lett.} \bibinfo{volume}{12}
  (\bibinfo{year}{2009}) \bibinfo{pages}{334--350}.
\bibitem[{Kearney and Porter(2020)}]{kearney2020nichemapr}
\bibinfo{author}{M.~R. Kearney}, \bibinfo{author}{W.~P. Porter},
\newblock \bibinfo{title}{{NicheMapR}--an {R} package for biophysical
  modelling: the ectotherm and {D}ynamic {E}nergy {B}udget models},
\newblock \bibinfo{journal}{Ecography} \bibinfo{volume}{43}
  (\bibinfo{year}{2020}) \bibinfo{pages}{85--96}.
\bibitem[{Grassly and Fraser(2008)}]{grassly2008mathematical}
\bibinfo{author}{N.~C. Grassly}, \bibinfo{author}{C.~Fraser},
\newblock \bibinfo{title}{Mathematical models of infectious disease
  transmission},
\newblock \bibinfo{journal}{Nat. Rev. Microbiol.} \bibinfo{volume}{6}
  (\bibinfo{year}{2008}) \bibinfo{pages}{477--487}.
\bibitem[{Sun et~al.(2016)Sun, Jusup, Jin, Wang, and Wang}]{sun2016pattern}
\bibinfo{author}{G.-Q. Sun}, \bibinfo{author}{M.~Jusup},
  \bibinfo{author}{Z.~Jin}, \bibinfo{author}{Y.~Wang},
  \bibinfo{author}{Z.~Wang},
\newblock \bibinfo{title}{Pattern transitions in spatial epidemics:
  {M}echanisms and emergent properties},
\newblock \bibinfo{journal}{Phys. Life Rev.} \bibinfo{volume}{19}
  (\bibinfo{year}{2016}) \bibinfo{pages}{43--73}.
\bibitem[{Wodarz and Nowak(2002)}]{wodarz2002mathematical}
\bibinfo{author}{D.~Wodarz}, \bibinfo{author}{M.~A. Nowak},
\newblock \bibinfo{title}{Mathematical models of {HIV} pathogenesis and
  treatment},
\newblock \bibinfo{journal}{BioEssays} \bibinfo{volume}{24}
  (\bibinfo{year}{2002}) \bibinfo{pages}{1178--1187}.
\bibitem[{Beauchemin and Handel(2011)}]{beauchemin2011review}
\bibinfo{author}{C.~A. Beauchemin}, \bibinfo{author}{A.~Handel},
\newblock \bibinfo{title}{A review of mathematical models of influenza {A}
  infections within a host or cell culture: lessons learned and challenges
  ahead},
\newblock \bibinfo{journal}{BMC Public Health} \bibinfo{volume}{11}
  (\bibinfo{year}{2011}) \bibinfo{pages}{1--15}.
\bibitem[{Kaplan and Craver(2011)}]{kaplan2011explanatory}
\bibinfo{author}{D.~M. Kaplan}, \bibinfo{author}{C.~F. Craver},
\newblock \bibinfo{title}{The explanatory force of dynamical and mathematical
  models in neuroscience: {A} mechanistic perspective},
\newblock \bibinfo{journal}{Philos. Sci.} \bibinfo{volume}{78}
  (\bibinfo{year}{2011}) \bibinfo{pages}{601--627}.
\bibitem[{Bassett et~al.(2018)Bassett, Zurn, and Gold}]{bassett2018nature}
\bibinfo{author}{D.~S. Bassett}, \bibinfo{author}{P.~Zurn},
  \bibinfo{author}{J.~I. Gold},
\newblock \bibinfo{title}{On the nature and use of models in network
  neuroscience},
\newblock \bibinfo{journal}{Nat. Rev. Neurosci.} \bibinfo{volume}{19}
  (\bibinfo{year}{2018}) \bibinfo{pages}{566--578}.
\bibitem[{Michor et~al.(2004)Michor, Iwasa, and Nowak}]{michor2004dynamics}
\bibinfo{author}{F.~Michor}, \bibinfo{author}{Y.~Iwasa}, \bibinfo{author}{M.~A.
  Nowak},
\newblock \bibinfo{title}{Dynamics of cancer progression},
\newblock \bibinfo{journal}{Nat. Rev. Cancer} \bibinfo{volume}{4}
  (\bibinfo{year}{2004}) \bibinfo{pages}{197--205}.
\bibitem[{Eftimie et~al.(2016)Eftimie, Gillard, and
  Cantrell}]{eftimie2016mathematical}
\bibinfo{author}{R.~Eftimie}, \bibinfo{author}{J.~J. Gillard},
  \bibinfo{author}{D.~A. Cantrell},
\newblock \bibinfo{title}{Mathematical models for immunology: current state of
  the art and future research directions},
\newblock \bibinfo{journal}{Bull. Math. Biol.} \bibinfo{volume}{78}
  (\bibinfo{year}{2016}) \bibinfo{pages}{2091--2134}.
\bibitem[{Abelson(1967)}]{abelson1967mathematical}
\bibinfo{author}{R.~P. Abelson},
\newblock \bibinfo{title}{Mathematical models in social psychology},
\newblock in: \bibinfo{editor}{L.~Berkowitz} (Ed.),
  \bibinfo{booktitle}{Advances in experimental social psychology},
  volume~\bibinfo{volume}{3}, \bibinfo{publisher}{Elsevier},
  \bibinfo{year}{1967}, pp. \bibinfo{pages}{1--54}.
\bibitem[{Rodgers(2010)}]{rodgers2010epistemology}
\bibinfo{author}{J.~L. Rodgers},
\newblock \bibinfo{title}{The epistemology of mathematical and statistical
  modeling: a quiet methodological revolution.},
\newblock \bibinfo{journal}{Am. Psychol.} \bibinfo{volume}{65}
  (\bibinfo{year}{2010}) \bibinfo{pages}{1--12}.
\bibitem[{S{\o}rensen(1978)}]{sorensen1978mathematical}
\bibinfo{author}{A.~B. S{\o}rensen},
\newblock \bibinfo{title}{Mathematical models in sociology},
\newblock \bibinfo{journal}{Annu. Rev. Sociol.} \bibinfo{volume}{4}
  (\bibinfo{year}{1978}) \bibinfo{pages}{345--371}.
\bibitem[{Carrington et~al.(2005)Carrington, Scott, and
  Wasserman}]{carrington2005models}
\bibinfo{author}{P.~J. Carrington}, \bibinfo{author}{J.~Scott},
  \bibinfo{author}{S.~Wasserman}, \bibinfo{title}{Models and methods in social
  network analysis}, \bibinfo{publisher}{Cambridge University Press},
  \bibinfo{year}{2005}.
\bibitem[{Schweber(1993)}]{schweber1993physics}
\bibinfo{author}{S.~S. Schweber},
\newblock \bibinfo{title}{Physics, community and the crisis in physical
  theory},
\newblock \bibinfo{journal}{Phys. Today} \bibinfo{volume}{46}
  (\bibinfo{year}{1993}) \bibinfo{pages}{34--34}.
\bibitem[{Lykken and Spiropulu(2014)}]{lykken2014supersymmetry}
\bibinfo{author}{J.~Lykken}, \bibinfo{author}{M.~Spiropulu},
\newblock \bibinfo{title}{Supersymmetry and the crisis in physics},
\newblock \bibinfo{journal}{Sci. Am.} \bibinfo{volume}{310}
  (\bibinfo{year}{2014}) \bibinfo{pages}{34--39}.
\bibitem[{Smolin(2007)}]{smolin2007trouble}
\bibinfo{author}{L.~Smolin}, \bibinfo{title}{The trouble with physics: the rise
  of string theory, the fall of a science, and what comes next},
  \bibinfo{publisher}{Houghton Mifflin Harcourt}, \bibinfo{year}{2007}.
\bibitem[{Hossenfelder(2019)}]{hossenfelder2018screams}
\bibinfo{author}{S.~Hossenfelder}, \bibinfo{title}{Screams for explanation:
  finetuning and naturalness in the foundations of physics},
  \bibinfo{year}{2019}. \bibinfo{note}{{e}-print arXiv:1801.02176}.
\bibitem[{Alves et~al.(2012)Alves, Arkani-Hamed, Arora, Bai, Baumgart, Berger,
  Buckley, Butler, Chang, Cheng et~al.}]{alves2012simplified}
\bibinfo{author}{D.~Alves}, \bibinfo{author}{N.~Arkani-Hamed},
  \bibinfo{author}{S.~Arora}, \bibinfo{author}{Y.~Bai},
  \bibinfo{author}{M.~Baumgart}, \bibinfo{author}{J.~Berger},
  \bibinfo{author}{M.~Buckley}, \bibinfo{author}{B.~Butler},
  \bibinfo{author}{S.~Chang}, \bibinfo{author}{H.-C. Cheng}, et~al.,
\newblock \bibinfo{title}{Simplified models for {LHC} new physics searches},
\newblock \bibinfo{journal}{J. Phys. G} \bibinfo{volume}{39}
  (\bibinfo{year}{2012}) \bibinfo{pages}{105005}.
\bibitem[{Safronova et~al.(2018)Safronova, Budker, DeMille, Kimball,
  Derevianko, and Clark}]{safronova2018search}
\bibinfo{author}{M.~Safronova}, \bibinfo{author}{D.~Budker},
  \bibinfo{author}{D.~DeMille}, \bibinfo{author}{D.~F.~J. Kimball},
  \bibinfo{author}{A.~Derevianko}, \bibinfo{author}{C.~W. Clark},
\newblock \bibinfo{title}{Search for new physics with atoms and molecules},
\newblock \bibinfo{journal}{Rev. Mod. Phys.} \bibinfo{volume}{90}
  (\bibinfo{year}{2018}) \bibinfo{pages}{025008}.
\bibitem[{Farmer(2000)}]{farmer2000physicists}
\bibinfo{author}{J.~D. Farmer},
\newblock \bibinfo{title}{Physicists attempt to scale the ivory towers of
  finance},
\newblock \bibinfo{journal}{Int. J. Theor. Appl. Finance} \bibinfo{volume}{3}
  (\bibinfo{year}{2000}) \bibinfo{pages}{311--333}.
\bibitem[{{United Nations, Department of Economic and Social Affairs,
  Population Division}(2018)}]{un2018world}
\bibinfo{author}{{United Nations, Department of Economic and Social Affairs,
  Population Division}}, \bibinfo{title}{{World Urbanization Prospects: {T}he
  2018 Revision}}, \bibinfo{year}{2018}. \bibinfo{note}{Online edition.
  Available at: \url{https://population.un.org/wup/Download/}}.
\bibitem[{Caves(2005)}]{caves2005encyclopedia}
\bibinfo{author}{R.~W. Caves}, \bibinfo{title}{Encyclopedia of the City},
  \bibinfo{publisher}{Taylor \& Francis}, \bibinfo{year}{2005}.
\bibitem[{Cohen(2011)}]{cohen2011cities}
\bibinfo{author}{M.~P. Cohen}, \bibinfo{title}{Cities in times of crisis. {T}he
  response of local governments in light of the global economic crisis: the
  role of the formation of human capital, urban innovation and strategic
  planning}, \bibinfo{year}{2011}. \bibinfo{note}{Institute of Urban and
  Regional Development (IURD) Working Paper No. 2011-01. Available at:
  \url{https://escholarship.org/uc/item/3432p4rb}}.
\bibitem[{Perc et~al.(2017)Perc, Jordan, Rand, Wang, Boccaletti, and
  Szolnoki}]{perc2017statistical}
\bibinfo{author}{M.~Perc}, \bibinfo{author}{J.~J. Jordan},
  \bibinfo{author}{D.~G. Rand}, \bibinfo{author}{Z.~Wang},
  \bibinfo{author}{S.~Boccaletti}, \bibinfo{author}{A.~Szolnoki},
\newblock \bibinfo{title}{Statistical physics of human cooperation},
\newblock \bibinfo{journal}{Phys. Rep.} \bibinfo{volume}{687}
  (\bibinfo{year}{2017}) \bibinfo{pages}{1--51}.
\bibitem[{Harush and Barzel(2017)}]{harush2017dynamic}
\bibinfo{author}{U.~Harush}, \bibinfo{author}{B.~Barzel},
\newblock \bibinfo{title}{Dynamic patterns of information flow in complex
  networks},
\newblock \bibinfo{journal}{Nat. Commun.} \bibinfo{volume}{8}
  (\bibinfo{year}{2017}) \bibinfo{pages}{2181}.
\bibitem[{Van~Dijk(2007)}]{vandijk2007mafia}
\bibinfo{author}{J.~Van~Dijk},
\newblock \bibinfo{title}{Mafia markers: assessing organized crime and its
  impact upon societies},
\newblock \bibinfo{journal}{Trends Organ. Crime} \bibinfo{volume}{10}
  (\bibinfo{year}{2007}) \bibinfo{pages}{39--56}.
\bibitem[{Taylor(1995)}]{taylor1995impact}
\bibinfo{author}{R.~B. Taylor},
\newblock \bibinfo{title}{The impact of crime on communities},
\newblock \bibinfo{journal}{Ann. Am. Acad. Political Soc. Sci.}
  \bibinfo{volume}{539} (\bibinfo{year}{1995}) \bibinfo{pages}{28--45}.
\bibitem[{Bijak et~al.(2008)Bijak, Kupiszewska, and
  Kupiszewski}]{bijak2008replacement}
\bibinfo{author}{J.~Bijak}, \bibinfo{author}{D.~Kupiszewska},
  \bibinfo{author}{M.~Kupiszewski},
\newblock \bibinfo{title}{Replacement migration revisited: {S}imulations of the
  effects of selected population and labor market strategies for the aging
  {E}urope, 2002--2052},
\newblock \bibinfo{journal}{Popul. Res. Policy Rev.} \bibinfo{volume}{27}
  (\bibinfo{year}{2008}) \bibinfo{pages}{321--342}.
\bibitem[{Williamson(2013)}]{williamson2013demographic}
\bibinfo{author}{J.~G. Williamson},
\newblock \bibinfo{title}{Demographic dividends revisited},
\newblock \bibinfo{journal}{Asian Dev. Rev.} \bibinfo{volume}{30}
  (\bibinfo{year}{2013}) \bibinfo{pages}{1--25}.
\bibitem[{Postelnicescu(2016)}]{postelnicescu2016europe}
\bibinfo{author}{C.~Postelnicescu},
\newblock \bibinfo{title}{Europe's new identity: {T}he refugee crisis and the
  rise of nationalism},
\newblock \bibinfo{journal}{Eur. J. Psychol} \bibinfo{volume}{12}
  (\bibinfo{year}{2016}) \bibinfo{pages}{203--209}.
\bibitem[{Holmes and Casta{\~n}eda(2016)}]{holmes2016representing}
\bibinfo{author}{S.~M. Holmes}, \bibinfo{author}{H.~Casta{\~n}eda},
\newblock \bibinfo{title}{Representing the ``{E}uropean refugee crisis'' in
  {G}ermany and beyond: {D}eservingness and difference, life and death},
\newblock \bibinfo{journal}{Am. Ethnol.} \bibinfo{volume}{43}
  (\bibinfo{year}{2016}) \bibinfo{pages}{12--24}.
\bibitem[{Fineberg(2014)}]{fineberg2014pandemic}
\bibinfo{author}{H.~V. Fineberg},
\newblock \bibinfo{title}{Pandemic preparedness and response---lessons from the
  {H1N1} influenza of 2009},
\newblock \bibinfo{journal}{N. Engl. J. Med.} \bibinfo{volume}{370}
  (\bibinfo{year}{2014}) \bibinfo{pages}{1335--1342}.
\bibitem[{Eggen et~al.(2004)Eggen, Behra, Burkhardt-Holm, Escher, and
  Schweigert}]{eggen2004challenges}
\bibinfo{author}{R.~I.~L. Eggen}, \bibinfo{author}{R.~Behra},
  \bibinfo{author}{P.~Burkhardt-Holm}, \bibinfo{author}{B.~I. Escher},
  \bibinfo{author}{N.~Schweigert},
\newblock \bibinfo{title}{Challenges in ecotoxicology},
\newblock \bibinfo{journal}{Environ. Sci. Technol.} \bibinfo{volume}{38}
  (\bibinfo{year}{2004}) \bibinfo{pages}{58A--64A}.
\bibitem[{Kirilenko and Sedjo(2007)}]{kirilenko2007climate}
\bibinfo{author}{A.~P. Kirilenko}, \bibinfo{author}{R.~A. Sedjo},
\newblock \bibinfo{title}{Climate change impacts on forestry},
\newblock \bibinfo{journal}{Proc. Natl. Acad. Sci. USA} \bibinfo{volume}{104}
  (\bibinfo{year}{2007}) \bibinfo{pages}{19697--19702}.
\bibitem[{Louis and Hess(2008)}]{louis2008climate}
\bibinfo{author}{M.~E.~S. Louis}, \bibinfo{author}{J.~J. Hess},
\newblock \bibinfo{title}{Climate change: impacts on and implications for
  global health},
\newblock \bibinfo{journal}{Am. J. Prev. Med.} \bibinfo{volume}{35}
  (\bibinfo{year}{2008}) \bibinfo{pages}{527--538}.
\bibitem[{Wheeler and Von~Braun(2013)}]{wheeler2013climate}
\bibinfo{author}{T.~Wheeler}, \bibinfo{author}{J.~Von~Braun},
\newblock \bibinfo{title}{Climate change impacts on global food security},
\newblock \bibinfo{journal}{Science} \bibinfo{volume}{341}
  (\bibinfo{year}{2013}) \bibinfo{pages}{508--513}.
\bibitem[{Batty(2008)}]{batty2008size}
\bibinfo{author}{M.~Batty},
\newblock \bibinfo{title}{The size, scale, and shape of cities},
\newblock \bibinfo{journal}{Science} \bibinfo{volume}{319}
  (\bibinfo{year}{2008}) \bibinfo{pages}{769--771}.
\bibitem[{Batty(2013)}]{batty2013new}
\bibinfo{author}{M.~Batty}, \bibinfo{title}{The new science of cities},
  \bibinfo{publisher}{The MIT Press}, \bibinfo{year}{2013}.
\bibitem[{Barthelemy(2019)}]{barthelemy2019statistical}
\bibinfo{author}{M.~Barthelemy},
\newblock \bibinfo{title}{The statistical physics of cities},
\newblock \bibinfo{journal}{Nat. Rev. Phys.} \bibinfo{volume}{1}
  (\bibinfo{year}{2019}) \bibinfo{pages}{406--415}.
\bibitem[{Arcaute et~al.(2015)Arcaute, Hatna, Ferguson, Youn, Johansson, and
  Batty}]{arcaute2015constructing}
\bibinfo{author}{E.~Arcaute}, \bibinfo{author}{E.~Hatna},
  \bibinfo{author}{P.~Ferguson}, \bibinfo{author}{H.~Youn},
  \bibinfo{author}{A.~Johansson}, \bibinfo{author}{M.~Batty},
\newblock \bibinfo{title}{Constructing cities, deconstructing scaling laws},
\newblock \bibinfo{journal}{J. R. Soc. Interface} \bibinfo{volume}{12}
  (\bibinfo{year}{2015}) \bibinfo{pages}{20140745}.
\bibitem[{Rozenfeld et~al.(2008)Rozenfeld, Rybski, Andrade, Batty, Stanley, and
  Makse}]{rozenfeld2008laws}
\bibinfo{author}{H.~D. Rozenfeld}, \bibinfo{author}{D.~Rybski},
  \bibinfo{author}{J.~S. Andrade}, \bibinfo{author}{M.~Batty},
  \bibinfo{author}{H.~E. Stanley}, \bibinfo{author}{H.~A. Makse},
\newblock \bibinfo{title}{Laws of population growth},
\newblock \bibinfo{journal}{Proc. Natl. Acad. Sci. USA} \bibinfo{volume}{105}
  (\bibinfo{year}{2008}) \bibinfo{pages}{18702--18707}.
\bibitem[{Zipf(1949)}]{zipf1949human}
\bibinfo{author}{G.~K. Zipf}, \bibinfo{title}{Human behavior and the principle
  of least effort}, \bibinfo{publisher}{Addison-Wesley Press},
  \bibinfo{year}{1949}.
\bibitem[{Zanette and Manrubia(1997)}]{zanette1997role}
\bibinfo{author}{D.~H. Zanette}, \bibinfo{author}{S.~C. Manrubia},
\newblock \bibinfo{title}{Role of intermittency in urban development: a model
  of large-scale city formation},
\newblock \bibinfo{journal}{Phys. Rev. Lett.} \bibinfo{volume}{79}
  (\bibinfo{year}{1997}) \bibinfo{pages}{523}.
\bibitem[{Soo(2005)}]{soo2005zipf}
\bibinfo{author}{K.~T. Soo},
\newblock \bibinfo{title}{Zipf's law for cities: a cross-country
  investigation},
\newblock \bibinfo{journal}{Reg. Sci. Urban. Econ.} \bibinfo{volume}{35}
  (\bibinfo{year}{2005}) \bibinfo{pages}{239--263}.
\bibitem[{Malevergne et~al.(2011)Malevergne, Pisarenko, and
  Sornette}]{malevergne2011testing}
\bibinfo{author}{Y.~Malevergne}, \bibinfo{author}{V.~Pisarenko},
  \bibinfo{author}{D.~Sornette},
\newblock \bibinfo{title}{Testing the pareto against the lognormal
  distributions with the uniformly most powerful unbiased test applied to the
  distribution of cities},
\newblock \bibinfo{journal}{Phys. Rev. E} \bibinfo{volume}{83}
  (\bibinfo{year}{2011}) \bibinfo{pages}{036111}.
\bibitem[{Marsili and Zhang(1998)}]{marsili1998interacting}
\bibinfo{author}{M.~Marsili}, \bibinfo{author}{Y.-C. Zhang},
\newblock \bibinfo{title}{Interacting individuals leading to {Z}ipf's law},
\newblock \bibinfo{journal}{Phys. Rev. Lett.} \bibinfo{volume}{80}
  (\bibinfo{year}{1998}) \bibinfo{pages}{2741}.
\bibitem[{Simon(1955)}]{simon1955class}
\bibinfo{author}{H.~A. Simon},
\newblock \bibinfo{title}{On a class of skew distribution functions},
\newblock \bibinfo{journal}{Biometrika} \bibinfo{volume}{42}
  (\bibinfo{year}{1955}) \bibinfo{pages}{425--440}.
\bibitem[{Merton(1968)}]{merton1968matthew}
\bibinfo{author}{R.~K. Merton},
\newblock \bibinfo{title}{The {M}atthew effect in science: {T}he reward and
  communication systems of science are considered},
\newblock \bibinfo{journal}{Science} \bibinfo{volume}{159}
  (\bibinfo{year}{1968}) \bibinfo{pages}{56--63}.
\bibitem[{Price(1976)}]{cumulative_advantage}
\bibinfo{author}{D.~S. Price},
\newblock \bibinfo{title}{A general theory of bibliometric and other cumulative
  advantage processes},
\newblock \bibinfo{journal}{J. Am. Soc. Inf. Sci.} \bibinfo{volume}{27}
  (\bibinfo{year}{1976}) \bibinfo{pages}{292--306}.
\bibitem[{Gabaix(1999)}]{gabaix1999zipf}
\bibinfo{author}{X.~Gabaix},
\newblock \bibinfo{title}{Zipf's law for cities: an explanation},
\newblock \bibinfo{journal}{Q. J. Econ.} \bibinfo{volume}{114}
  (\bibinfo{year}{1999}) \bibinfo{pages}{739--767}.
\bibitem[{Newman(2005)}]{newman2005power}
\bibinfo{author}{M.~E.~J. Newman},
\newblock \bibinfo{title}{Power laws, {P}areto distributions and {Z}ipf's law},
\newblock \bibinfo{journal}{Contemp. Phys.} \bibinfo{volume}{46}
  (\bibinfo{year}{2005}) \bibinfo{pages}{323--351}.
\bibitem[{Decker et~al.(2007)Decker, Kerkhoff, and Moses}]{decker2007global}
\bibinfo{author}{E.~H. Decker}, \bibinfo{author}{A.~J. Kerkhoff},
  \bibinfo{author}{M.~E. Moses},
\newblock \bibinfo{title}{Global patterns of city size distributions and their
  fundamental drivers},
\newblock \bibinfo{journal}{PLOS ONE} \bibinfo{volume}{2}
  (\bibinfo{year}{2007}) \bibinfo{pages}{e934}.
\bibitem[{Manrubia et~al.(1999)Manrubia, Zanette, and
  Sole}]{manrubia1999transient}
\bibinfo{author}{S.~C. Manrubia}, \bibinfo{author}{D.~H. Zanette},
  \bibinfo{author}{R.~V. Sole},
\newblock \bibinfo{title}{Transient dynamics and scaling phenomena in urban
  growth},
\newblock \bibinfo{journal}{Fractals} \bibinfo{volume}{7}
  (\bibinfo{year}{1999}) \bibinfo{pages}{1--8}.
\bibitem[{Witten and Sander(1981)}]{witten1981diffusion}
\bibinfo{author}{T.~A. Witten}, \bibinfo{author}{L.~M. Sander},
\newblock \bibinfo{title}{Diffusion-limited aggregation, a kinetic critical
  phenomenon},
\newblock \bibinfo{journal}{Phys. Rev. Lett.} \bibinfo{volume}{47}
  (\bibinfo{year}{1981}) \bibinfo{pages}{1400--1403}.
\bibitem[{Andersson et~al.(2002)Andersson, Lindgren, Rasmussen, and
  White}]{andersson2002urban}
\bibinfo{author}{C.~Andersson}, \bibinfo{author}{K.~Lindgren},
  \bibinfo{author}{S.~Rasmussen}, \bibinfo{author}{R.~White},
\newblock \bibinfo{title}{Urban growth simulation from ``first principles''},
\newblock \bibinfo{journal}{Phys. Rev. E} \bibinfo{volume}{66}
  (\bibinfo{year}{2002}) \bibinfo{pages}{026204}.
\bibitem[{Makse et~al.(1995)Makse, Havlin, and Stanley}]{makse1995modelling}
\bibinfo{author}{H.~A. Makse}, \bibinfo{author}{S.~Havlin},
  \bibinfo{author}{H.~E. Stanley},
\newblock \bibinfo{title}{Modelling urban growth patterns},
\newblock \bibinfo{journal}{Nature} \bibinfo{volume}{377}
  (\bibinfo{year}{1995}) \bibinfo{pages}{608--612}.
\bibitem[{Seto et~al.(2012)Seto, G{\"u}neralp, and Hutyra}]{seto2012global}
\bibinfo{author}{K.~C. Seto}, \bibinfo{author}{B.~G{\"u}neralp},
  \bibinfo{author}{L.~R. Hutyra},
\newblock \bibinfo{title}{Global forecasts of urban expansion to 2030 and
  direct impacts on biodiversity and carbon pools},
\newblock \bibinfo{journal}{Proc. Natl. Acad. Sci. USA} \bibinfo{volume}{109}
  (\bibinfo{year}{2012}) \bibinfo{pages}{16083--16088}.
\bibitem[{Doxiadis(1963)}]{doxiadis1963ekistics}
\bibinfo{author}{C.~A. Doxiadis},
\newblock \bibinfo{title}{Ekistics and traffic},
\newblock \bibinfo{journal}{Traffic Q.} \bibinfo{volume}{17}
  (\bibinfo{year}{1963}) \bibinfo{pages}{439--457}.
\bibitem[{Abler et~al.(1971)Abler, Adams, and Gould}]{abler1971spatial}
\bibinfo{author}{R.~Abler}, \bibinfo{author}{J.~S. Adams},
  \bibinfo{author}{P.~Gould}, \bibinfo{title}{Spatial organization: the
  geographer's view of the world}, \bibinfo{publisher}{Prentice-Hall},
  \bibinfo{year}{1971}.
\bibitem[{Christaller(1966)}]{christaller1966central}
\bibinfo{author}{W.~Christaller}, \bibinfo{title}{Central Places in Southern
  Germany}, \bibinfo{publisher}{Prentice-Hall}, \bibinfo{year}{1966}.
\bibitem[{Losch(1954)}]{losch1954economics}
\bibinfo{author}{A.~Losch}, \bibinfo{title}{The Economics of Location},
  \bibinfo{publisher}{Yale University Press}, \bibinfo{year}{1954}.
\bibitem[{Parr(1973)}]{parr1973structure}
\bibinfo{author}{J.~B. Parr},
\newblock \bibinfo{title}{Structure and size in the urban system of
  {L{\"o}sch}},
\newblock \bibinfo{journal}{Econ. Geogr.} \bibinfo{volume}{49}
  (\bibinfo{year}{1973}) \bibinfo{pages}{185--212}.
\bibitem[{Barth{\'e}lemy and Flammini(2008)}]{barthelemy2008modeling}
\bibinfo{author}{M.~Barth{\'e}lemy}, \bibinfo{author}{A.~Flammini},
\newblock \bibinfo{title}{Modeling urban street patterns},
\newblock \bibinfo{journal}{Phys. Rev. Lett.} \bibinfo{volume}{100}
  (\bibinfo{year}{2008}) \bibinfo{pages}{138702}.
\bibitem[{Barth\'elemy(2016)}]{barthelemy2016structure}
\bibinfo{author}{M.~Barth\'elemy}, \bibinfo{title}{The Structure and Dynamics
  of Cities: Urban Data Analysis and Theoretical Modeling},
  \bibinfo{publisher}{Cambridge University Press}, \bibinfo{year}{2016}.
\bibitem[{Bettencourt(2021)}]{bettencourt:book}
\bibinfo{author}{L.~M.~A. Bettencourt}, \bibinfo{title}{Introduction to Urban
  Science: Evidence and Theory of Cities as Complex Systems},
  \bibinfo{publisher}{MIT Press}, \bibinfo{address}{Cambridge MA},
  \bibinfo{year}{2021}.
\bibitem[{Barth\'elemy(2018)}]{barthelemy2018morphogenesis}
\bibinfo{author}{M.~Barth\'elemy}, \bibinfo{title}{Morphogenesis of spatial
  networks}, \bibinfo{publisher}{Springer}, \bibinfo{year}{2018}.
\bibitem[{Gastner and Newman(2006)}]{gastner2006shape}
\bibinfo{author}{M.~T. Gastner}, \bibinfo{author}{M.~E.~J. Newman},
\newblock \bibinfo{title}{Shape and efficiency in spatial distribution
  networks},
\newblock \bibinfo{journal}{J. Stat. Mech.} \bibinfo{volume}{2006}
  (\bibinfo{year}{2006}) \bibinfo{pages}{P01015}.
\bibitem[{Fabrikant et~al.(2002)Fabrikant, Koutsoupias, and
  Papadimitriou}]{fabrikant2002heuristically}
\bibinfo{author}{A.~Fabrikant}, \bibinfo{author}{E.~Koutsoupias},
  \bibinfo{author}{C.~H. Papadimitriou},
\newblock \bibinfo{title}{Heuristically optimized trade-offs: {A} new paradigm
  for power laws in the {I}nternet},
\newblock in: \bibinfo{editor}{P.~Widmayer}, \bibinfo{editor}{S.~Eidenbenz},
  \bibinfo{editor}{F.~Triguero}, \bibinfo{editor}{R.~Morales},
  \bibinfo{editor}{R.~Conejo}, \bibinfo{editor}{M.~Hennessy} (Eds.),
  \bibinfo{booktitle}{International Colloquium on Automata, Languages, and
  Programming}, volume \bibinfo{volume}{2380} of
  \textit{\bibinfo{series}{Lecture Notes in Computer Science}},
  \bibinfo{publisher}{Springer}, \bibinfo{year}{2002}, pp.
  \bibinfo{pages}{110--122}.
\bibitem[{Schelling(1971)}]{schelling1971dynamic}
\bibinfo{author}{T.~C. Schelling},
\newblock \bibinfo{title}{Dynamic models of segregation},
\newblock \bibinfo{journal}{J. Math. Sociol.} \bibinfo{volume}{1}
  (\bibinfo{year}{1971}) \bibinfo{pages}{143--186}.
\bibitem[{Vinkovi{\'c} and Kirman(2006)}]{vinkovic2006physical}
\bibinfo{author}{D.~Vinkovi{\'c}}, \bibinfo{author}{A.~Kirman},
\newblock \bibinfo{title}{A physical analogue of the {S}chelling model},
\newblock \bibinfo{journal}{Proc. Natl. Acad. Sci. USA} \bibinfo{volume}{103}
  (\bibinfo{year}{2006}) \bibinfo{pages}{19261--19265}.
\bibitem[{Dall'Asta et~al.(2008)Dall'Asta, Castellano, and
  Marsili}]{dallasta2008statistical}
\bibinfo{author}{L.~Dall'Asta}, \bibinfo{author}{C.~Castellano},
  \bibinfo{author}{M.~Marsili},
\newblock \bibinfo{title}{Statistical physics of the {S}chelling model of
  segregation},
\newblock \bibinfo{journal}{J. Stat. Mech.} \bibinfo{volume}{2008}
  (\bibinfo{year}{2008}) \bibinfo{pages}{L07002}.
\bibitem[{Stauffer(2013)}]{stauffer2013biased}
\bibinfo{author}{D.~Stauffer},
\newblock \bibinfo{title}{A biased review of sociophysics},
\newblock \bibinfo{journal}{J. Stat. Phys.} \bibinfo{volume}{151}
  (\bibinfo{year}{2013}) \bibinfo{pages}{9--20}.
\bibitem[{Bak(1996)}]{bak1996how}
\bibinfo{author}{P.~Bak}, \bibinfo{title}{How Nature Works: The Science of
  Self-Organized Criticality}, \bibinfo{publisher}{Copernicus},
  \bibinfo{year}{1996}.
\bibitem[{West(2017)}]{west2017scale}
\bibinfo{author}{G.~B. West}, \bibinfo{title}{Scale: The Universal Laws of Life
  and Death in Organisms, Cities and Companies}, \bibinfo{publisher}{Weidenfeld
  \& Nicolson}, \bibinfo{year}{2017}.
\bibitem[{West et~al.(1997)West, Brown, and Enquist}]{west1997general}
\bibinfo{author}{G.~B. West}, \bibinfo{author}{J.~H. Brown},
  \bibinfo{author}{B.~J. Enquist},
\newblock \bibinfo{title}{A general model for the origin of allometric scaling
  laws in biology},
\newblock \bibinfo{journal}{Science} \bibinfo{volume}{276}
  (\bibinfo{year}{1997}) \bibinfo{pages}{122--126}.
\bibitem[{Bettencourt et~al.(2007)Bettencourt, Lobo, Helbing, K{\"u}hnert, and
  West}]{bettencourt2007growth}
\bibinfo{author}{L.~M. Bettencourt}, \bibinfo{author}{J.~Lobo},
  \bibinfo{author}{D.~Helbing}, \bibinfo{author}{C.~K{\"u}hnert},
  \bibinfo{author}{G.~B. West},
\newblock \bibinfo{title}{Growth, innovation, scaling, and the pace of life in
  cities},
\newblock \bibinfo{journal}{Proc. Natl. Acad. Sci. USA} \bibinfo{volume}{104}
  (\bibinfo{year}{2007}) \bibinfo{pages}{7301--7306}.
\bibitem[{Um et~al.(2009)Um, Son, Lee, Jeong, and Kim}]{um2009scaling}
\bibinfo{author}{J.~Um}, \bibinfo{author}{S.-W. Son}, \bibinfo{author}{S.-I.
  Lee}, \bibinfo{author}{H.~Jeong}, \bibinfo{author}{B.~J. Kim},
\newblock \bibinfo{title}{Scaling laws between population and facility
  densities},
\newblock \bibinfo{journal}{Proc. Natl. Acad. Sci. USA} \bibinfo{volume}{106}
  (\bibinfo{year}{2009}) \bibinfo{pages}{14236--14240}.
\bibitem[{Bettencourt(2013)}]{bettencourt2013origins}
\bibinfo{author}{L.~M.~A. Bettencourt},
\newblock \bibinfo{title}{The origins of scaling in cities},
\newblock \bibinfo{journal}{Science} \bibinfo{volume}{340}
  (\bibinfo{year}{2013}) \bibinfo{pages}{1438=--1441}.
\bibitem[{Ravenstein(1885)}]{ravenstein1885laws}
\bibinfo{author}{E.~G. Ravenstein},
\newblock \bibinfo{title}{The laws of migration},
\newblock \bibinfo{journal}{J. Stat. Soc. Lond.} \bibinfo{volume}{48}
  (\bibinfo{year}{1885}) \bibinfo{pages}{167--235}.
\bibitem[{Zipf(1946)}]{zipf1946hypothesis}
\bibinfo{author}{G.~K. Zipf},
\newblock \bibinfo{title}{The {P1 P2 / D} hypothesis: {O}n the intercity
  movement of persons},
\newblock \bibinfo{journal}{Am. Soc. Rev.} \bibinfo{volume}{11}
  (\bibinfo{year}{1946}) \bibinfo{pages}{677--686}.
\bibitem[{Chen and Huang(2018)}]{chen2018scaling}
\bibinfo{author}{Y.~Chen}, \bibinfo{author}{L.~Huang},
\newblock \bibinfo{title}{A scaling approach to evaluating the distance
  exponent of the urban gravity model},
\newblock \bibinfo{journal}{Chaos Solitons Fractals} \bibinfo{volume}{109}
  (\bibinfo{year}{2018}) \bibinfo{pages}{303--313}.
\bibitem[{Lee and Holme(2015)}]{lee2015relating}
\bibinfo{author}{M.~Lee}, \bibinfo{author}{P.~Holme},
\newblock \bibinfo{title}{Relating land use and human intra-city mobility},
\newblock \bibinfo{journal}{PLOS ONE} \bibinfo{volume}{10}
  (\bibinfo{year}{2015}) \bibinfo{pages}{e0140152}.
\bibitem[{Stouffer(1940)}]{stouffer1940intervening}
\bibinfo{author}{S.~A. Stouffer},
\newblock \bibinfo{title}{Intervening opportunities: {A} theory relating
  mobility and distance},
\newblock \bibinfo{journal}{Am. Soc. Rev.} \bibinfo{volume}{5}
  (\bibinfo{year}{1940}) \bibinfo{pages}{845--867}.
\bibitem[{Gonzalez et~al.(2008)Gonzalez, Hidalgo, and
  Barab\'asi}]{gonzalez2008understanding}
\bibinfo{author}{M.~C. Gonzalez}, \bibinfo{author}{C.~A. Hidalgo},
  \bibinfo{author}{A.-L. Barab\'asi},
\newblock \bibinfo{title}{Understanding individual human mobility patterns},
\newblock \bibinfo{journal}{Nature} \bibinfo{volume}{453}
  (\bibinfo{year}{2008}) \bibinfo{pages}{779--782}.
\bibitem[{Park et~al.(2010)Park, Lee, and Gonz{\'a}lez}]{park2010eigenmode}
\bibinfo{author}{J.~Park}, \bibinfo{author}{D.-S. Lee}, \bibinfo{author}{M.~C.
  Gonz{\'a}lez},
\newblock \bibinfo{title}{The eigenmode analysis of human motion},
\newblock \bibinfo{journal}{J. Stat. Mech.} \bibinfo{volume}{2010}
  (\bibinfo{year}{2010}) \bibinfo{pages}{P11021}.
\bibitem[{Song et~al.(2010)Song, Qu, Blumm, and Barab{\'a}si}]{song2010limits}
\bibinfo{author}{C.~Song}, \bibinfo{author}{Z.~Qu}, \bibinfo{author}{N.~Blumm},
  \bibinfo{author}{A.-L. Barab{\'a}si},
\newblock \bibinfo{title}{Limits of predictability in human mobility},
\newblock \bibinfo{journal}{Science} \bibinfo{volume}{327}
  (\bibinfo{year}{2010}) \bibinfo{pages}{1018--1021}.
\bibitem[{Lu et~al.(2012)Lu, Bengtsson, and Holme}]{lu2012predictability}
\bibinfo{author}{X.~Lu}, \bibinfo{author}{L.~Bengtsson},
  \bibinfo{author}{P.~Holme},
\newblock \bibinfo{title}{Predictability of population displacement after the
  2010 {H}aiti earthquake},
\newblock \bibinfo{journal}{Proc. Natl. Acad. Sci. USA} \bibinfo{volume}{109}
  (\bibinfo{year}{2012}) \bibinfo{pages}{11576--11581}.
\bibitem[{Louf et~al.(2013)Louf, Jensen, and Barthelemy}]{louf2013emergence}
\bibinfo{author}{R.~Louf}, \bibinfo{author}{P.~Jensen},
  \bibinfo{author}{M.~Barthelemy},
\newblock \bibinfo{title}{Emergence of hierarchy in cost-driven growth of
  spatial networks},
\newblock \bibinfo{journal}{Proc. Natl. Acad. Sci. USA} \bibinfo{volume}{110}
  (\bibinfo{year}{2013}) \bibinfo{pages}{8824--8829}.
\bibitem[{Lee et~al.(2018)Lee, Cheon, and Lee}]{minjin2018imbalance}
\bibinfo{author}{M.~Lee}, \bibinfo{author}{S.~Cheon}, \bibinfo{author}{S.~Lee},
  \bibinfo{title}{Imbalance of pairwise efficiency in urban street network},
  \bibinfo{year}{2018}. \bibinfo{note}{{e}-print arXiv:1808.05844}.
\bibitem[{Lee et~al.(2017)Lee, Barbosa, Youn, Holme, and
  Ghoshal}]{lee2017morphology}
\bibinfo{author}{M.~Lee}, \bibinfo{author}{H.~Barbosa},
  \bibinfo{author}{H.~Youn}, \bibinfo{author}{P.~Holme},
  \bibinfo{author}{G.~Ghoshal},
\newblock \bibinfo{title}{Morphology of travel routes and the organization of
  cities},
\newblock \bibinfo{journal}{Nat. Commun.} \bibinfo{volume}{8}
  (\bibinfo{year}{2017}) \bibinfo{pages}{2229}.
\bibitem[{{de Berg} et~al.(1997){de Berg}, {van Kreveld}, Overmars, and
  Schwarzkopf}]{deberg1997visibility}
\bibinfo{author}{M.~{de Berg}}, \bibinfo{author}{M.~{van Kreveld}},
  \bibinfo{author}{M.~Overmars}, \bibinfo{author}{O.~Schwarzkopf},
  \bibinfo{title}{Computational Geometry: Algorithms and Applications},
  \bibinfo{publisher}{Springer}, \bibinfo{year}{1997}, pp.
  \bibinfo{pages}{305--315}.
\bibitem[{Lee and Holme(2012)}]{lee2012exploring}
\bibinfo{author}{S.~H. Lee}, \bibinfo{author}{P.~Holme},
\newblock \bibinfo{title}{Exploring maps with greedy navigators},
\newblock \bibinfo{journal}{Phys. Rev. Lett.} \bibinfo{volume}{108}
  (\bibinfo{year}{2012}) \bibinfo{pages}{128701}.
\bibitem[{Wolbers and Hegarty(2010)}]{wolbers2010determines}
\bibinfo{author}{T.~Wolbers}, \bibinfo{author}{M.~Hegarty},
\newblock \bibinfo{title}{What determines our navigational abilities?},
\newblock \bibinfo{journal}{Trends Cogn. Sci.} \bibinfo{volume}{14}
  (\bibinfo{year}{2010}) \bibinfo{pages}{138--146}.
\bibitem[{Luo et~al.(2019)Luo, Wang, Liu, and Gao}]{luo2019cities}
\bibinfo{author}{W.~Luo}, \bibinfo{author}{Y.~Wang}, \bibinfo{author}{X.~Liu},
  \bibinfo{author}{S.~Gao},
\newblock \bibinfo{title}{Cities as spatial and social networks: towards a
  spatio-socio-semantic analysis framework},
\newblock in: \bibinfo{editor}{X.~Ye}, \bibinfo{editor}{X.~Liu} (Eds.),
  \bibinfo{booktitle}{Cities as Spatial and Social Networks},
  \bibinfo{publisher}{Springer}, \bibinfo{year}{2019}, pp.
  \bibinfo{pages}{21--37}.
\bibitem[{Ni(2015)}]{ni2015traffic}
\bibinfo{author}{D.~Ni}, \bibinfo{title}{Traffic Flow Theory: Characteristics,
  Experimental Methods, and Numerical Techniques},
  \bibinfo{publisher}{Butterworth Heinemann}, \bibinfo{year}{2015}.
\bibitem[{Kessels(1999)}]{kessels2019traffic}
\bibinfo{author}{F.~Kessels}, \bibinfo{title}{Traffic Flow Modelling:
  Introduction to Traffic Flow Theory Through a Genealogy of Models},
  \bibinfo{publisher}{Springer}, \bibinfo{year}{1999}.
\bibitem[{Greenshields et~al.(1935)Greenshields, Bibbins, Channing, and
  Miller}]{greenshields1935study}
\bibinfo{author}{B.~D. Greenshields}, \bibinfo{author}{J.~R. Bibbins},
  \bibinfo{author}{W.~S. Channing}, \bibinfo{author}{H.~H. Miller},
\newblock \bibinfo{title}{A study of traffic capacity},
\newblock in: \bibinfo{booktitle}{Highway Research Board Proceedings 14},
  \bibinfo{publisher}{Highway Research Board}, \bibinfo{year}{1935}, pp.
  \bibinfo{pages}{448--477}.
\bibitem[{Chowdhury et~al.(2000)Chowdhury, Santen, and
  Schadschneider}]{chowdhury2000statistical}
\bibinfo{author}{D.~Chowdhury}, \bibinfo{author}{L.~Santen},
  \bibinfo{author}{A.~Schadschneider},
\newblock \bibinfo{title}{Statistical physics of vehicular traffic and some
  related systems},
\newblock \bibinfo{journal}{Phys. Rep.} \bibinfo{volume}{329}
  (\bibinfo{year}{2000}) \bibinfo{pages}{199--329}.
\bibitem[{Helbing(2001)}]{helbing2001traffic}
\bibinfo{author}{D.~Helbing},
\newblock \bibinfo{title}{Traffic and related self-driven many-particle
  systems},
\newblock \bibinfo{journal}{Rev. Mod. Phys.} \bibinfo{volume}{73}
  (\bibinfo{year}{2001}) \bibinfo{pages}{1067--1141}.
\bibitem[{Nagatani(2002)}]{nagatani2002physics}
\bibinfo{author}{T.~Nagatani},
\newblock \bibinfo{title}{The physics of traffic jams},
\newblock \bibinfo{journal}{Rep. Prog. Phys.} \bibinfo{volume}{65}
  (\bibinfo{year}{2002}) \bibinfo{pages}{1331--1386}.
\bibitem[{Mihaita et~al.(2019)Mihaita, Li, He, and
  Rizoiu}]{mihaita2019motorway}
\bibinfo{author}{A.~S. Mihaita}, \bibinfo{author}{H.~Li},
  \bibinfo{author}{Z.~He}, \bibinfo{author}{M.-A. Rizoiu},
\newblock \bibinfo{title}{Motorway traffic flow prediction using advanced deep
  learning},
\newblock in: \bibinfo{booktitle}{Proceedings of the 2019 IEEE Intelligent
  Transportation Systems Conference (ITSC)}, \bibinfo{publisher}{IEEE},
  \bibinfo{year}{2019}, pp. \bibinfo{pages}{1683--1690}.
\bibitem[{{van Wageningen-Kessels} et~al.(2015){van Wageningen-Kessels}, {van
  Lint}, Vuik, and Hoogendoorn}]{fvwk2015genealogy}
\bibinfo{author}{F.~{van Wageningen-Kessels}}, \bibinfo{author}{H.~{van Lint}},
  \bibinfo{author}{K.~Vuik}, \bibinfo{author}{S.~Hoogendoorn},
\newblock \bibinfo{title}{Genealogy of traffic flow models},
\newblock \bibinfo{journal}{EURO J. Transp. Logist.} \bibinfo{volume}{4}
  (\bibinfo{year}{2015}) \bibinfo{pages}{445--473}.
\bibitem[{Lighthill and Whitham(1955)}]{lighthill1955kinematic}
\bibinfo{author}{M.~J. Lighthill}, \bibinfo{author}{G.~B. Whitham},
\newblock \bibinfo{title}{On kinematic waves {II}. {A} theory of traffic flow
  on long crowded roads},
\newblock \bibinfo{journal}{Proc. R. Soc. A} \bibinfo{volume}{229}
  (\bibinfo{year}{1955}) \bibinfo{pages}{317--345}.
\bibitem[{Reuschel(1950)}]{reuschel1950fahrzeugbewegungen}
\bibinfo{author}{A.~Reuschel},
\newblock \bibinfo{title}{Fahrzeugbewegungen in der {Kolonne}},
\newblock \bibinfo{journal}{\"{O}sterreichisches Ingenieur Archiv}
  \bibinfo{volume}{4} (\bibinfo{year}{1950}) \bibinfo{pages}{193--215}.
\bibitem[{Nagel and Schreckenberg(1992)}]{nagel1992cellular}
\bibinfo{author}{K.~Nagel}, \bibinfo{author}{M.~Schreckenberg},
\newblock \bibinfo{title}{A cellular automaton model for freeway traffic},
\newblock \bibinfo{journal}{J. Phys. I} \bibinfo{volume}{2}
  (\bibinfo{year}{1992}) \bibinfo{pages}{2221--2229}.
\bibitem[{Hall et~al.(1986)Hall, Allen, and Gunter}]{hall1986empirical}
\bibinfo{author}{F.~L. Hall}, \bibinfo{author}{B.~L. Allen},
  \bibinfo{author}{M.~A. Gunter},
\newblock \bibinfo{title}{Empirical analysis of freeway flow-density
  relationships},
\newblock \bibinfo{journal}{Transp. Res. A} \bibinfo{volume}{20}
  (\bibinfo{year}{1986}) \bibinfo{pages}{197--210}.
\bibitem[{Kerner(2017)}]{kerner2017breakdown}
\bibinfo{author}{B.~S. Kerner}, \bibinfo{title}{Breakdown in Traffic Networks:
  Fundamentals of Traffic Science}, \bibinfo{publisher}{Springer},
  \bibinfo{year}{2017}.
\bibitem[{Kerner and Rehborn(1996)}]{kerner1996experimental}
\bibinfo{author}{B.~S. Kerner}, \bibinfo{author}{H.~Rehborn},
\newblock \bibinfo{title}{Experimental properties of complexity in traffic
  flow},
\newblock \bibinfo{journal}{Phys. Rev. E} \bibinfo{volume}{53}
  (\bibinfo{year}{1996}) \bibinfo{pages}{R4275--R4278}.
\bibitem[{Tadaki et~al.(2013)Tadaki, Kikuchi, Fukui, Nakayama, Nishinari,
  Shibata, Sugiyama, Yosida, and Yukawa}]{tadaki2013phase}
\bibinfo{author}{S.-i. Tadaki}, \bibinfo{author}{M.~Kikuchi},
  \bibinfo{author}{M.~Fukui}, \bibinfo{author}{A.~Nakayama},
  \bibinfo{author}{K.~Nishinari}, \bibinfo{author}{A.~Shibata},
  \bibinfo{author}{Y.~Sugiyama}, \bibinfo{author}{T.~Yosida},
  \bibinfo{author}{S.~Yukawa},
\newblock \bibinfo{title}{Phase transition in traffic jam experiment on a
  circuit},
\newblock \bibinfo{journal}{New J. Phys.} \bibinfo{volume}{15}
  (\bibinfo{year}{2013}) \bibinfo{pages}{103034}.
\bibitem[{Treiterer and Myers(1974)}]{treiterer1974hysteresis}
\bibinfo{author}{J.~Treiterer}, \bibinfo{author}{J.~A. Myers},
\newblock \bibinfo{title}{The hysteresis phenomenon in traffic flow},
\newblock in: \bibinfo{editor}{D.~Buckley} (Ed.),
  \bibinfo{booktitle}{Proceedings of the 6th International Symposium on
  Transportation and Traffic Theory}, \bibinfo{publisher}{Reed},
  \bibinfo{year}{1974}, pp. \bibinfo{pages}{13--38}.
\bibitem[{Leutzbach(1988)}]{leutzbach1988intro}
\bibinfo{author}{W.~Leutzbach}, \bibinfo{title}{Introduction to the Theory of
  Traffic Flow}, \bibinfo{publisher}{Springer}, \bibinfo{year}{1988}.
\bibitem[{Sugiyama et~al.(2008)Sugiyama, Fukui, Kikuchi, Hasebe, Nakayama,
  Nishinari, Tadaki, and Yukawa}]{sugiyama2008traffic}
\bibinfo{author}{Y.~Sugiyama}, \bibinfo{author}{M.~Fukui},
  \bibinfo{author}{M.~Kikuchi}, \bibinfo{author}{K.~Hasebe},
  \bibinfo{author}{A.~Nakayama}, \bibinfo{author}{K.~Nishinari},
  \bibinfo{author}{S.~Tadaki}, \bibinfo{author}{S.~Yukawa},
\newblock \bibinfo{title}{Traffic jams without bottlenecks: {E}xperimental
  evidence for the physical mechanism of the formation of a jam},
\newblock \bibinfo{journal}{New J. Phys.} \bibinfo{volume}{10}
  (\bibinfo{year}{2008}) \bibinfo{pages}{033001}.
\bibitem[{Kesting et~al.(2018)Kesting, Treiber, and
  Helbing}]{kesting2008agents}
\bibinfo{author}{A.~Kesting}, \bibinfo{author}{M.~Treiber},
  \bibinfo{author}{D.~Helbing},
\newblock \bibinfo{title}{Agents for traffic simulation},
\newblock in: \bibinfo{editor}{A.~M. Uhrmacher}, \bibinfo{editor}{D.~Weyns}
  (Eds.), \bibinfo{booktitle}{Multi-agent systems: Simulation and
  applications}, \bibinfo{publisher}{CRC Press}, \bibinfo{year}{2018}, pp.
  \bibinfo{pages}{325--356}.
\bibitem[{Kerner(1998)}]{kerner1998experimental}
\bibinfo{author}{B.~S. Kerner},
\newblock \bibinfo{title}{Experimental features of self-organization in traffic
  flow},
\newblock \bibinfo{journal}{Phys. Rev. Lett.} \bibinfo{volume}{81}
  (\bibinfo{year}{1998}) \bibinfo{pages}{3797--3800}.
\bibitem[{Prigogine and Andrews(1960)}]{prigogine1960boltzmann}
\bibinfo{author}{I.~Prigogine}, \bibinfo{author}{F.~C. Andrews},
\newblock \bibinfo{title}{A {B}oltzmann-like approach for traffic flow},
\newblock \bibinfo{journal}{Oper. Res.} \bibinfo{volume}{8}
  (\bibinfo{year}{1960}) \bibinfo{pages}{789--797}.
\bibitem[{Paveri-Fontana(1975)}]{paveri1975boltzmann}
\bibinfo{author}{S.~Paveri-Fontana},
\newblock \bibinfo{title}{On {B}oltzmann-like treatments for traffic flow: a
  critical review of the basic model and an alternative proposal for dilute
  traffic analysis},
\newblock \bibinfo{journal}{Transport. Res.} \bibinfo{volume}{9}
  (\bibinfo{year}{1975}) \bibinfo{pages}{225--235}.
\bibitem[{Yukawa and Kikuchi(1995)}]{yukawa1995coupled}
\bibinfo{author}{S.~Yukawa}, \bibinfo{author}{M.~Kikuchi},
\newblock \bibinfo{title}{Coupled-map modeling of one-dimensional traffic
  flow},
\newblock \bibinfo{journal}{J. Phys. Soc. Jpn.} \bibinfo{volume}{64}
  (\bibinfo{year}{1995}) \bibinfo{pages}{35--38}.
\bibitem[{Krau{\ss} et~al.(1996)Krau{\ss}, Wagner, and
  Gawron}]{krauss1996continuous}
\bibinfo{author}{S.~Krau{\ss}}, \bibinfo{author}{P.~Wagner},
  \bibinfo{author}{C.~Gawron},
\newblock \bibinfo{title}{Continuous limit of the {Nagel-Schreckenberg} model},
\newblock \bibinfo{journal}{Phys. Rev. E} \bibinfo{volume}{54}
  (\bibinfo{year}{1996}) \bibinfo{pages}{3707--3712}.
\bibitem[{Horiguchi and Sakakibara(1998)}]{horiguchi1998numerical}
\bibinfo{author}{T.~Horiguchi}, \bibinfo{author}{T.~Sakakibara},
\newblock \bibinfo{title}{Numerical simulations for traffic-flow models on a
  decorated square lattice},
\newblock \bibinfo{journal}{Physica A} \bibinfo{volume}{252}
  (\bibinfo{year}{1998}) \bibinfo{pages}{388--404}.
\bibitem[{Helbing and Huberman(1998)}]{helbing1998coherent}
\bibinfo{author}{D.~Helbing}, \bibinfo{author}{B.~A. Huberman},
\newblock \bibinfo{title}{Coherent moving states in highway traffic},
\newblock \bibinfo{journal}{Nature} \bibinfo{volume}{396}
  (\bibinfo{year}{1998}) \bibinfo{pages}{738--740}.
\bibitem[{Takayasu and Takayasu(1993)}]{takayasu1993noise}
\bibinfo{author}{M.~Takayasu}, \bibinfo{author}{H.~Takayasu},
\newblock \bibinfo{title}{1/f noise in a traffic model},
\newblock \bibinfo{journal}{Fractals} \bibinfo{volume}{1}
  (\bibinfo{year}{1993}) \bibinfo{pages}{860--866}.
\bibitem[{MacDonald et~al.(1968)MacDonald, Gibbs, and
  Pipkin}]{macdonald1968kinetics}
\bibinfo{author}{C.~T. MacDonald}, \bibinfo{author}{J.~H. Gibbs},
  \bibinfo{author}{A.~C. Pipkin},
\newblock \bibinfo{title}{Kinetics of biopolymerization on nucleic acid
  templates},
\newblock \bibinfo{journal}{Biopolymers} \bibinfo{volume}{6}
  (\bibinfo{year}{1968}) \bibinfo{pages}{1--25}.
\bibitem[{Burgers(1948)}]{burgers1948mathematical}
\bibinfo{author}{J.~M. Burgers},
\newblock \bibinfo{title}{A mathematical model illustrating the theory of
  turbulence},
\newblock in: \bibinfo{editor}{R.~{von Mises}}, \bibinfo{editor}{T.~{von
  Karman}} (Eds.), \bibinfo{booktitle}{Advances in Applied Mechanics},
  volume~\bibinfo{volume}{1}, \bibinfo{publisher}{Academic Press},
  \bibinfo{year}{1948}, pp. \bibinfo{pages}{171--199}.
\bibitem[{Forster et~al.(1976)Forster, Nelson, and Stephen}]{forster1976long}
\bibinfo{author}{D.~Forster}, \bibinfo{author}{D.~R. Nelson},
  \bibinfo{author}{M.~J. Stephen},
\newblock \bibinfo{title}{Long-time tails and the large-eddy behavior of a
  randomly stirred fluid},
\newblock \bibinfo{journal}{Phys. Rev. Lett.} \bibinfo{volume}{36}
  (\bibinfo{year}{1976}) \bibinfo{pages}{867}.
\bibitem[{Biham et~al.(1992)Biham, Middleton, and Levine}]{biham1992self}
\bibinfo{author}{O.~Biham}, \bibinfo{author}{A.~A. Middleton},
  \bibinfo{author}{D.~Levine},
\newblock \bibinfo{title}{Self-organization and a dynamical transition in
  traffic-flow models},
\newblock \bibinfo{journal}{Phys. Rev. A} \bibinfo{volume}{46}
  (\bibinfo{year}{1992}) \bibinfo{pages}{R6124--R6127}.
\bibitem[{Helbing et~al.(1997)Helbing, Keltsch, and
  Molnar}]{helbing1997modelling}
\bibinfo{author}{D.~Helbing}, \bibinfo{author}{J.~Keltsch},
  \bibinfo{author}{P.~Molnar},
\newblock \bibinfo{title}{Modelling the evolution of human trail systems},
\newblock \bibinfo{journal}{Nature} \bibinfo{volume}{388}
  (\bibinfo{year}{1997}) \bibinfo{pages}{47--50}.
\bibitem[{Feliciani and Nishinari(2016)}]{feliciani2016empirical}
\bibinfo{author}{C.~Feliciani}, \bibinfo{author}{K.~Nishinari},
\newblock \bibinfo{title}{Empirical analysis of the lane formation process in
  bidirectional pedestrian flow},
\newblock \bibinfo{journal}{Phys. Rev. E} \bibinfo{volume}{94}
  (\bibinfo{year}{2016}) \bibinfo{pages}{032304}.
\bibitem[{Helbing et~al.(2000)Helbing, Farkas, and
  Vicsek}]{helbing2000simulating}
\bibinfo{author}{D.~Helbing}, \bibinfo{author}{I.~Farkas},
  \bibinfo{author}{T.~Vicsek},
\newblock \bibinfo{title}{Simulating dynamical features of escape panic},
\newblock \bibinfo{journal}{Nature} \bibinfo{volume}{407}
  (\bibinfo{year}{2000}) \bibinfo{pages}{487--490}.
\bibitem[{Henderson(1971)}]{henderson1971statistics}
\bibinfo{author}{L.~F. Henderson},
\newblock \bibinfo{title}{The statistics of crowd fluids},
\newblock \bibinfo{journal}{Nature} \bibinfo{volume}{229}
  (\bibinfo{year}{1971}) \bibinfo{pages}{381--383}.
\bibitem[{Krajewski et~al.(2018)Krajewski, Bock, Kloeker, and
  Eckstein}]{krajewski2018highd}
\bibinfo{author}{R.~Krajewski}, \bibinfo{author}{J.~Bock},
  \bibinfo{author}{L.~Kloeker}, \bibinfo{author}{L.~Eckstein},
\newblock \bibinfo{title}{The {highD} dataset: {A} drone dataset of
  naturalistic vehicle trajectories on {G}erman highways for validation of
  highly automated driving systems},
\newblock in: \bibinfo{booktitle}{2018 21st International Conference on
  Intelligent Transportation Systems (ITSC)}, \bibinfo{publisher}{IEEE},
  \bibinfo{year}{2018}, pp. \bibinfo{pages}{2118--2125}.
\bibitem[{Halkias and Colyar(2006)}]{halkias2006ngsim}
\bibinfo{author}{J.~Halkias}, \bibinfo{author}{J.~Colyar},
  \bibinfo{title}{{NGSIM---Interstate 80 Freeway Dataset}},
  \bibinfo{year}{2006}. \bibinfo{note}{Official website:
  \url{https://www.fhwa.dot.gov/publications/research/operations/07030/index.cfm}.
  Archived at: \url{https://doi.org/10.17605/OSF.IO/N7G9X}}.
\bibitem[{Andersen and Nowak(2013)}]{andersen2013financial}
\bibinfo{author}{J.~V. Andersen}, \bibinfo{author}{A.~Nowak},
\newblock \bibinfo{title}{Financial markets as interacting individuals: {P}rice
  formation from models of complexity},
\newblock in: \bibinfo{booktitle}{An Introduction to Socio-Finance},
  \bibinfo{publisher}{Springer}, \bibinfo{year}{2013}, pp.
  \bibinfo{pages}{59--76}.
\bibitem[{Ormerod(2016)}]{ormerod2016ten}
\bibinfo{author}{P.~Ormerod},
\newblock \bibinfo{title}{Ten years after ``{W}orrying trends in
  econophysics'': developments and current challenges},
\newblock \bibinfo{journal}{Eur. Phys. J. Spec. Top.} \bibinfo{volume}{225}
  (\bibinfo{year}{2016}) \bibinfo{pages}{3281--3291}.
\bibitem[{Dragulescu and Yakovenko(2000)}]{dragulescu2000statistical}
\bibinfo{author}{A.~Dragulescu}, \bibinfo{author}{V.~M. Yakovenko},
\newblock \bibinfo{title}{Statistical mechanics of money},
\newblock \bibinfo{journal}{Eur. Phys. J. B} \bibinfo{volume}{17}
  (\bibinfo{year}{2000}) \bibinfo{pages}{723--729}.
\bibitem[{Aoyama et~al.(2010)Aoyama, Fujiwara, Ikeda, Iyetomi, and
  Souma}]{aoyama2010econophysics}
\bibinfo{author}{H.~Aoyama}, \bibinfo{author}{Y.~Fujiwara},
  \bibinfo{author}{Y.~Ikeda}, \bibinfo{author}{H.~Iyetomi},
  \bibinfo{author}{W.~Souma}, \bibinfo{title}{Econophysics and companies:
  statistical life and death in complex business networks},
  \bibinfo{publisher}{Cambridge University Press}, \bibinfo{year}{2010}.
\bibitem[{Braha et~al.(2011)Braha, Stacey, and Bar-Yam}]{braha2011corporate}
\bibinfo{author}{D.~Braha}, \bibinfo{author}{B.~Stacey},
  \bibinfo{author}{Y.~Bar-Yam},
\newblock \bibinfo{title}{Corporate competition: {A} self-organized network},
\newblock \bibinfo{journal}{Soc. Netw.} \bibinfo{volume}{33}
  (\bibinfo{year}{2011}) \bibinfo{pages}{219--230}.
\bibitem[{Luo et~al.(2014)Luo, Yin, Di, Hardisty, and
  MacEachren}]{luo2014geovisual}
\bibinfo{author}{W.~Luo}, \bibinfo{author}{P.~Yin}, \bibinfo{author}{Q.~Di},
  \bibinfo{author}{F.~Hardisty}, \bibinfo{author}{A.~M. MacEachren},
\newblock \bibinfo{title}{A geovisual analytic approach to understanding
  geo-social relationships in the international trade network},
\newblock \bibinfo{journal}{PLOS ONE} \bibinfo{volume}{9}
  (\bibinfo{year}{2014}) \bibinfo{pages}{e88666}.
\bibitem[{Harmon et~al.(2015)Harmon, Lagi, de~Aguiar, Chinellato, Braha,
  Epstein, and Bar-Yam}]{harmon2015anticipating}
\bibinfo{author}{D.~Harmon}, \bibinfo{author}{M.~Lagi}, \bibinfo{author}{M.~A.
  de~Aguiar}, \bibinfo{author}{D.~D. Chinellato}, \bibinfo{author}{D.~Braha},
  \bibinfo{author}{I.~R. Epstein}, \bibinfo{author}{Y.~Bar-Yam},
\newblock \bibinfo{title}{Anticipating economic market crises using measures of
  collective panic},
\newblock \bibinfo{journal}{PLOS ONE} \bibinfo{volume}{10}
  (\bibinfo{year}{2015}) \bibinfo{pages}{e0131871}.
\bibitem[{Abergel et~al.(2017)Abergel, Aoyama, Chakrabarti, Chakraborti, Deo,
  Raina, and Vodenska}]{abergel2017econophysics}
\bibinfo{author}{F.~Abergel}, \bibinfo{author}{H.~Aoyama},
  \bibinfo{author}{B.~K. Chakrabarti}, \bibinfo{author}{A.~Chakraborti},
  \bibinfo{author}{N.~Deo}, \bibinfo{author}{D.~Raina},
  \bibinfo{author}{I.~Vodenska}, \bibinfo{title}{Econophysics and sociophysics:
  Recent progress and future directions}, \bibinfo{publisher}{Springer},
  \bibinfo{year}{2017}.
\bibitem[{Wachs and Kert{\'e}sz(2019)}]{wachs2019network}
\bibinfo{author}{J.~Wachs}, \bibinfo{author}{J.~Kert{\'e}sz},
\newblock \bibinfo{title}{A network approach to cartel detection in public
  auction markets},
\newblock \bibinfo{journal}{Sci. Rep.} \bibinfo{volume}{9}
  (\bibinfo{year}{2019}) \bibinfo{pages}{10818}.
\bibitem[{Mantegna(1991)}]{mantegna1991levy}
\bibinfo{author}{R.~N. Mantegna},
\newblock \bibinfo{title}{L{\'e}vy walks and enhanced diffusion in {M}ilan
  stock exchange},
\newblock \bibinfo{journal}{Physica A} \bibinfo{volume}{179}
  (\bibinfo{year}{1991}) \bibinfo{pages}{232--242}.
\bibitem[{Mandelbrot(1982)}]{mandelbrot1983fractal}
\bibinfo{author}{B.~B. Mandelbrot}, \bibinfo{title}{The fractal geometry of
  nature}, \bibinfo{publisher}{Times Books}, \bibinfo{year}{1982}.
\bibitem[{Vicsek(1992)}]{vicsek1992fractal}
\bibinfo{author}{T.~Vicsek}, \bibinfo{title}{Fractal growth phenomena},
  \bibinfo{publisher}{World scientific}, \bibinfo{year}{1992}.
\bibitem[{Takayasu(1990)}]{takayasu1990fractals}
\bibinfo{author}{H.~Takayasu}, \bibinfo{title}{Fractals in the physical
  sciences}, \bibinfo{publisher}{Manchester University Press},
  \bibinfo{year}{1990}.
\bibitem[{Bak et~al.(1987)Bak, Tang, and Wiesenfeld}]{bak1987self}
\bibinfo{author}{P.~Bak}, \bibinfo{author}{C.~Tang},
  \bibinfo{author}{K.~Wiesenfeld},
\newblock \bibinfo{title}{Self-organized criticality: {A}n explanation of 1/f
  noise},
\newblock \bibinfo{journal}{Phys. Rev. Lett.} \bibinfo{volume}{59}
  (\bibinfo{year}{1987}) \bibinfo{pages}{381}.
\bibitem[{Takayasu et~al.(1992)Takayasu, Miura, Hirabayashi, and
  Hamada}]{takayasu1992statistical}
\bibinfo{author}{H.~Takayasu}, \bibinfo{author}{H.~Miura},
  \bibinfo{author}{T.~Hirabayashi}, \bibinfo{author}{K.~Hamada},
\newblock \bibinfo{title}{Statistical properties of deterministic threshold
  elements---the case of market price},
\newblock \bibinfo{journal}{Physica A} \bibinfo{volume}{184}
  (\bibinfo{year}{1992}) \bibinfo{pages}{127--134}.
\bibitem[{Mantegna and Stanley(1995)}]{mantegna1995scaling}
\bibinfo{author}{R.~N. Mantegna}, \bibinfo{author}{H.~E. Stanley},
\newblock \bibinfo{title}{Scaling behaviour in the dynamics of an economic
  index},
\newblock \bibinfo{journal}{Nature} \bibinfo{volume}{376}
  (\bibinfo{year}{1995}) \bibinfo{pages}{46--49}.
\bibitem[{Takayasu et~al.(1997)Takayasu, Sato, and
  Takayasu}]{takayasu1997stable}
\bibinfo{author}{H.~Takayasu}, \bibinfo{author}{A.-H. Sato},
  \bibinfo{author}{M.~Takayasu},
\newblock \bibinfo{title}{Stable infinite variance fluctuations in randomly
  amplified {L}angevin systems},
\newblock \bibinfo{journal}{Phys. Rev. Lett.} \bibinfo{volume}{79}
  (\bibinfo{year}{1997}) \bibinfo{pages}{966}.
\bibitem[{Mantegna and Stanley(1999)}]{mantegna1999introduction}
\bibinfo{author}{R.~Mantegna}, \bibinfo{author}{H.~Stanley}, \bibinfo{title}{An
  Introduction to Econophysics: Correlations and Complexity in Finance},
  \bibinfo{publisher}{Cambridge University Press}, \bibinfo{year}{1999}.
\bibitem[{Sornette(2003)}]{sornette2003stock}
\bibinfo{author}{D.~Sornette}, \bibinfo{title}{Why stock markets crash:
  critical events in complex financial systems}, \bibinfo{publisher}{Princeton
  University Press}, \bibinfo{year}{2003}.
\bibitem[{Bouchaud and Potters(2003)}]{bouchaud2003theory}
\bibinfo{author}{J.-P. Bouchaud}, \bibinfo{author}{M.~Potters},
  \bibinfo{title}{Theory of financial risk and derivative pricing: from
  statistical physics to risk management}, \bibinfo{publisher}{Cambridge
  University Press}, \bibinfo{year}{2003}.
\bibitem[{Takayasu(2013)}]{takayasu2013empirical}
\bibinfo{author}{H.~Takayasu}, \bibinfo{title}{Empirical science of financial
  fluctuations: The advent of econophysics}, \bibinfo{publisher}{Springer},
  \bibinfo{year}{2013}.
\bibitem[{Takayasu(2012)}]{takayasu2012application}
\bibinfo{author}{H.~Takayasu}, \bibinfo{title}{The Application of Econophysics:
  Proceedings of the Second {Nikkei} Econophysics Symposium},
  \bibinfo{publisher}{Springer}, \bibinfo{year}{2012}.
\bibitem[{Takayasu et~al.(2010)Takayasu, Watanabe, and
  Takayasu}]{takayasu2010econophysics}
\bibinfo{author}{M.~Takayasu}, \bibinfo{author}{T.~Watanabe},
  \bibinfo{author}{H.~Takayasu}, \bibinfo{title}{Econophysics approaches to
  large-scale business data and financial crisis},
  \bibinfo{publisher}{Springer}, \bibinfo{year}{2010}.
\bibitem[{Hirabayashi et~al.(1993)Hirabayashi, Takayasu, Miura, and
  Hamada}]{hirabayashi1993behavior}
\bibinfo{author}{T.~Hirabayashi}, \bibinfo{author}{H.~Takayasu},
  \bibinfo{author}{H.~Miura}, \bibinfo{author}{K.~Hamada},
\newblock \bibinfo{title}{The behavior of a threshold model of market price in
  stock exchange},
\newblock \bibinfo{journal}{Fractals} \bibinfo{volume}{1}
  (\bibinfo{year}{1993}) \bibinfo{pages}{29--40}.
\bibitem[{Yamada et~al.(2007)Yamada, Takayasu, and
  Takayasu}]{yamada2007characterization}
\bibinfo{author}{K.~Yamada}, \bibinfo{author}{H.~Takayasu},
  \bibinfo{author}{M.~Takayasu},
\newblock \bibinfo{title}{Characterization of foreign exchange market using the
  threshold-dealer-model},
\newblock \bibinfo{journal}{Physica A} \bibinfo{volume}{382}
  (\bibinfo{year}{2007}) \bibinfo{pages}{340--346}.
\bibitem[{Sato and Takayasu(1998)}]{sato1998dynamic}
\bibinfo{author}{A.-H. Sato}, \bibinfo{author}{H.~Takayasu},
\newblock \bibinfo{title}{Dynamic numerical models of stock market price: from
  microscopic determinism to macroscopic randomness},
\newblock \bibinfo{journal}{Physica A} \bibinfo{volume}{250}
  (\bibinfo{year}{1998}) \bibinfo{pages}{231--252}.
\bibitem[{Engle(1982)}]{engle1982autoregressive}
\bibinfo{author}{R.~F. Engle},
\newblock \bibinfo{title}{Autoregressive conditional heteroscedasticity with
  estimates of the variance of {U}nited {K}ingdom inflation},
\newblock \bibinfo{journal}{Econometrica} \bibinfo{volume}{50}
  (\bibinfo{year}{1982}) \bibinfo{pages}{987--1007}.
\bibitem[{Sato and Takayasu(2002)}]{sato2002derivation}
\bibinfo{author}{A.-H. Sato}, \bibinfo{author}{H.~Takayasu},
\newblock \bibinfo{title}{Derivation of {ARCH(1)} process from market price
  changes based on deterministic microscopic multi-agent},
\newblock in: \bibinfo{editor}{H.~Takayasu} (Ed.),
  \bibinfo{booktitle}{Empirical Science of Financial Fluctuations},
  \bibinfo{publisher}{Springer}, \bibinfo{year}{2002}, pp.
  \bibinfo{pages}{171--178}.
\bibitem[{Yamada et~al.(2009)Yamada, Takayasu, Ito, and
  Takayasu}]{yamada2009solvable}
\bibinfo{author}{K.~Yamada}, \bibinfo{author}{H.~Takayasu},
  \bibinfo{author}{T.~Ito}, \bibinfo{author}{M.~Takayasu},
\newblock \bibinfo{title}{Solvable stochastic dealer models for financial
  markets},
\newblock \bibinfo{journal}{Phys. Rev. E} \bibinfo{volume}{79}
  (\bibinfo{year}{2009}) \bibinfo{pages}{051120}.
\bibitem[{Matsunaga et~al.(2012)Matsunaga, Yamada, Takayasu, and
  Takayasu}]{matsunaga2012construction}
\bibinfo{author}{K.~Matsunaga}, \bibinfo{author}{K.~Yamada},
  \bibinfo{author}{H.~Takayasu}, \bibinfo{author}{M.~Takayasu},
\newblock \bibinfo{title}{Construction of the spread dealer model and its
  application (in japanese)},
\newblock \bibinfo{journal}{Trans. Jpn. Soc. Artif. Intell.}
  \bibinfo{volume}{27} (\bibinfo{year}{2012}) \bibinfo{pages}{365--375}.
\bibitem[{Ciacci et~al.(2020)Ciacci, Sueshige, Takayasu, Christensen, and
  Takayasu}]{ciacci2020microscopic}
\bibinfo{author}{A.~Ciacci}, \bibinfo{author}{T.~Sueshige},
  \bibinfo{author}{H.~Takayasu}, \bibinfo{author}{K.~Christensen},
  \bibinfo{author}{M.~Takayasu},
\newblock \bibinfo{title}{The microscopic relationships between triangular
  arbitrage and cross-currency correlations in a simple agent based model of
  foreign exchange markets},
\newblock \bibinfo{journal}{PLOS ONE} \bibinfo{volume}{15}
  (\bibinfo{year}{2020}) \bibinfo{pages}{e0234709}.
\bibitem[{Aiba et~al.(2002)Aiba, Hatano, Takayasu, Marumo, and
  Shimizu}]{aiba2002triangular}
\bibinfo{author}{Y.~Aiba}, \bibinfo{author}{N.~Hatano},
  \bibinfo{author}{H.~Takayasu}, \bibinfo{author}{K.~Marumo},
  \bibinfo{author}{T.~Shimizu},
\newblock \bibinfo{title}{Triangular arbitrage as an interaction among foreign
  exchange rates},
\newblock \bibinfo{journal}{Physica A} \bibinfo{volume}{310}
  (\bibinfo{year}{2002}) \bibinfo{pages}{467--479}.
\bibitem[{Ito et~al.(2020)Ito, Yamada, Takayasu, and
  Takayasu}]{ito2020execution}
\bibinfo{author}{T.~Ito}, \bibinfo{author}{K.~Yamada},
  \bibinfo{author}{M.~Takayasu}, \bibinfo{author}{H.~Takayasu},
  \bibinfo{title}{Execution risk and arbitrage opportunities in the foreign
  exchange markets}, \bibinfo{year}{2020}. \bibinfo{note}{National Bureau of
  Economic Research (NBER) Working Paper No. 26706. Available at:
  \url{https://www.nber.org/papers/w26706}}.
\bibitem[{Takayasu et~al.(2002)Takayasu, Takayasu, and
  Okazaki}]{takayasu2002transaction}
\bibinfo{author}{M.~Takayasu}, \bibinfo{author}{H.~Takayasu},
  \bibinfo{author}{M.~P. Okazaki},
\newblock \bibinfo{title}{Transaction interval analysis of high resolution
  foreign exchange data},
\newblock in: \bibinfo{editor}{H.~Takayasu} (Ed.),
  \bibinfo{booktitle}{Empirical Science of Financial Fluctuations},
  \bibinfo{publisher}{Springer}, \bibinfo{year}{2002}, pp.
  \bibinfo{pages}{18--25}.
\bibitem[{Takayasu and Takayasu(2003)}]{takayasu2003self}
\bibinfo{author}{M.~Takayasu}, \bibinfo{author}{H.~Takayasu},
\newblock \bibinfo{title}{Self-modulation processes and resulting generic $1/f$
  fluctuations},
\newblock \bibinfo{journal}{Physica A} \bibinfo{volume}{324}
  (\bibinfo{year}{2003}) \bibinfo{pages}{101--107}.
\bibitem[{Takayasu et~al.(2006{\natexlab{a}})Takayasu, Mizuno, Ohnishi, and
  Takayasu}]{takayasu2006temporal}
\bibinfo{author}{M.~Takayasu}, \bibinfo{author}{T.~Mizuno},
  \bibinfo{author}{T.~Ohnishi}, \bibinfo{author}{H.~Takayasu},
\newblock \bibinfo{title}{Temporal characteristics of moving average of foreign
  exchange markets},
\newblock in: \bibinfo{editor}{H.~Takayasu} (Ed.),
  \bibinfo{booktitle}{Practical Fruits of Econophysics},
  \bibinfo{publisher}{Springer}, \bibinfo{year}{2006}{\natexlab{a}}, pp.
  \bibinfo{pages}{29--32}.
\bibitem[{Takayasu et~al.(2006{\natexlab{b}})Takayasu, Mizuno, and
  Takayasu}]{takayasu2006potential}
\bibinfo{author}{M.~Takayasu}, \bibinfo{author}{T.~Mizuno},
  \bibinfo{author}{H.~Takayasu},
\newblock \bibinfo{title}{Potential force observed in market dynamics},
\newblock \bibinfo{journal}{Physica A} \bibinfo{volume}{370}
  (\bibinfo{year}{2006}{\natexlab{b}}) \bibinfo{pages}{91--97}.
\bibitem[{Takayasu et~al.(2010)Takayasu, Watanabe, Mizuno, and
  Takayasu}]{takayasu2010theoretical}
\bibinfo{author}{M.~Takayasu}, \bibinfo{author}{K.~Watanabe},
  \bibinfo{author}{T.~Mizuno}, \bibinfo{author}{H.~Takayasu},
\newblock \bibinfo{title}{Theoretical base of the {PUCK}-model with application
  to foreign exchange markets},
\newblock in: \bibinfo{editor}{M.~Takayasu}, \bibinfo{editor}{T.~Watanabe},
  \bibinfo{editor}{H.~Takayasu} (Eds.), \bibinfo{booktitle}{Econophysics
  Approaches to Large-Scale Business Data and Financial Crisis},
  \bibinfo{publisher}{Springer}, \bibinfo{year}{2010}, pp.
  \bibinfo{pages}{79--98}.
\bibitem[{Takayasu and Takayasu(2009)}]{takayasu2009continuum}
\bibinfo{author}{M.~Takayasu}, \bibinfo{author}{H.~Takayasu},
\newblock \bibinfo{title}{Continuum limit and renormalization of market price
  dynamics based on {PUCK} model},
\newblock \bibinfo{journal}{Prog. Theor. Phys. Supp.} \bibinfo{volume}{179}
  (\bibinfo{year}{2009}) \bibinfo{pages}{1--7}.
\bibitem[{Watanabe et~al.(2009{\natexlab{a}})Watanabe, Takayasu, and
  Takayasu}]{watanabe2009observation}
\bibinfo{author}{K.~Watanabe}, \bibinfo{author}{H.~Takayasu},
  \bibinfo{author}{M.~Takayasu},
\newblock \bibinfo{title}{Observation of two types of behaviors of financial
  bubbles and the related higher-order potential forces},
\newblock \bibinfo{journal}{Prog. Theor. Phys. Supp.} \bibinfo{volume}{179}
  (\bibinfo{year}{2009}{\natexlab{a}}) \bibinfo{pages}{8--16}.
\bibitem[{Watanabe et~al.(2009{\natexlab{b}})Watanabe, Takayasu, and
  Takayasu}]{watanabe2009random}
\bibinfo{author}{K.~Watanabe}, \bibinfo{author}{H.~Takayasu},
  \bibinfo{author}{M.~Takayasu},
\newblock \bibinfo{title}{Random walker in temporally deforming higher-order
  potential forces observed in a financial crisis},
\newblock \bibinfo{journal}{Phys. Rev. E} \bibinfo{volume}{80}
  (\bibinfo{year}{2009}{\natexlab{b}}) \bibinfo{pages}{056110}.
\bibitem[{Maslov(2000)}]{maslov2000simple}
\bibinfo{author}{S.~Maslov},
\newblock \bibinfo{title}{Simple model of a limit order-driven market},
\newblock \bibinfo{journal}{Physica A} \bibinfo{volume}{278}
  (\bibinfo{year}{2000}) \bibinfo{pages}{571--578}.
\bibitem[{Bouchaud et~al.(2002)Bouchaud, M{\'e}zard, and
  Potters}]{bouchaud2002statistical}
\bibinfo{author}{J.-P. Bouchaud}, \bibinfo{author}{M.~M{\'e}zard},
  \bibinfo{author}{M.~Potters},
\newblock \bibinfo{title}{Statistical properties of stock order books:
  empirical results and models},
\newblock \bibinfo{journal}{Quant. Finance} \bibinfo{volume}{2}
  (\bibinfo{year}{2002}) \bibinfo{pages}{251--256}.
\bibitem[{Yura et~al.(2014)Yura, Takayasu, Sornette, and
  Takayasu}]{yura2014financial}
\bibinfo{author}{Y.~Yura}, \bibinfo{author}{H.~Takayasu},
  \bibinfo{author}{D.~Sornette}, \bibinfo{author}{M.~Takayasu},
\newblock \bibinfo{title}{Financial brownian particle in the layered order-book
  fluid and fluctuation-dissipation relations},
\newblock \bibinfo{journal}{Phys. Rev. Lett,} \bibinfo{volume}{112}
  (\bibinfo{year}{2014}) \bibinfo{pages}{098703}.
\bibitem[{Yura et~al.(2015)Yura, Takayasu, Sornette, and
  Takayasu}]{yura2015financial}
\bibinfo{author}{Y.~Yura}, \bibinfo{author}{H.~Takayasu},
  \bibinfo{author}{D.~Sornette}, \bibinfo{author}{M.~Takayasu},
\newblock \bibinfo{title}{Financial {K}nudsen number: {B}reakdown of continuous
  price dynamics and asymmetric buy-and-sell structures confirmed by
  high-precision order-book information},
\newblock \bibinfo{journal}{Phys. Rev. E} \bibinfo{volume}{92}
  (\bibinfo{year}{2015}) \bibinfo{pages}{042811}.
\bibitem[{Daniels et~al.(2003)Daniels, Farmer, Gillemot, Iori, and
  Smith}]{daniels2003quantitative}
\bibinfo{author}{M.~G. Daniels}, \bibinfo{author}{J.~D. Farmer},
  \bibinfo{author}{L.~Gillemot}, \bibinfo{author}{G.~Iori},
  \bibinfo{author}{E.~Smith},
\newblock \bibinfo{title}{Quantitative model of price diffusion and market
  friction based on trading as a mechanistic random process},
\newblock \bibinfo{journal}{Phys. Rev. Lett.} \bibinfo{volume}{90}
  (\bibinfo{year}{2003}) \bibinfo{pages}{108102}.
\bibitem[{Smith et~al.(2003)Smith, Farmer, Gillemot, and
  Krishnamurthy}]{smith2003statistical}
\bibinfo{author}{E.~Smith}, \bibinfo{author}{J.~D. Farmer},
  \bibinfo{author}{L.~Gillemot}, \bibinfo{author}{S.~Krishnamurthy},
\newblock \bibinfo{title}{Statistical theory of the continuous double auction},
\newblock \bibinfo{journal}{Quant. Finance} \bibinfo{volume}{3}
  (\bibinfo{year}{2003}) \bibinfo{pages}{481--514}.
\bibitem[{Slanina(2013)}]{slanina2013essentials}
\bibinfo{author}{F.~Slanina}, \bibinfo{title}{Essentials of econophysics
  modelling}, \bibinfo{publisher}{Oxford University Press},
  \bibinfo{year}{2013}.
\bibitem[{Bouchaud et~al.(2018)Bouchaud, Bonart, Donier, and
  Gould}]{bouchaud2018trades}
\bibinfo{author}{J.-P. Bouchaud}, \bibinfo{author}{J.~Bonart},
  \bibinfo{author}{J.~Donier}, \bibinfo{author}{M.~Gould},
  \bibinfo{title}{Trades, quotes and prices: financial markets under the
  microscope}, \bibinfo{publisher}{Cambridge University Press},
  \bibinfo{year}{2018}.
\bibitem[{Kanazawa et~al.(2018{\natexlab{a}})Kanazawa, Sueshige, Takayasu, and
  Takayasu}]{kanazawa2018derivation}
\bibinfo{author}{K.~Kanazawa}, \bibinfo{author}{T.~Sueshige},
  \bibinfo{author}{H.~Takayasu}, \bibinfo{author}{M.~Takayasu},
\newblock \bibinfo{title}{Derivation of the {B}oltzmann equation for financial
  {B}rownian motion: {D}irect observation of the collective motion of
  high-frequency traders},
\newblock \bibinfo{journal}{Phys. Rev. Lett.} \bibinfo{volume}{120}
  (\bibinfo{year}{2018}{\natexlab{a}}) \bibinfo{pages}{138301}.
\bibitem[{Kanazawa et~al.(2018{\natexlab{b}})Kanazawa, Sueshige, Takayasu, and
  Takayasu}]{kanazawa2018kinetic}
\bibinfo{author}{K.~Kanazawa}, \bibinfo{author}{T.~Sueshige},
  \bibinfo{author}{H.~Takayasu}, \bibinfo{author}{M.~Takayasu},
\newblock \bibinfo{title}{Kinetic theory for financial {B}rownian motion from
  microscopic dynamics},
\newblock \bibinfo{journal}{Phys. Rev. E} \bibinfo{volume}{98}
  (\bibinfo{year}{2018}{\natexlab{b}}) \bibinfo{pages}{052317}.
\bibitem[{Sueshige et~al.(2018)Sueshige, Kanazawa, Takayasu, and
  Takayasu}]{sueshige2018ecology}
\bibinfo{author}{T.~Sueshige}, \bibinfo{author}{K.~Kanazawa},
  \bibinfo{author}{H.~Takayasu}, \bibinfo{author}{M.~Takayasu},
\newblock \bibinfo{title}{Ecology of trading strategies in a forex market for
  limit and market orders},
\newblock \bibinfo{journal}{PLOS ONE} \bibinfo{volume}{13}
  (\bibinfo{year}{2018}) \bibinfo{pages}{e0208332}.
\bibitem[{Schmidt(2012)}]{schmidt2012ecology}
\bibinfo{author}{A.~B. Schmidt}, \bibinfo{title}{Ecology of the modern
  institutional spot {FX}: The {EBS} market in 2011}, \bibinfo{year}{2012}.
  \bibinfo{note}{Social Science Research Network (SSRN) Paper No. 1984070.
  Available at: \url{https://ssrn.com/abstract=1984070}}.
\bibitem[{Poirer(2012)}]{poirer2012high}
\bibinfo{author}{I.~Poirer},
\newblock \bibinfo{title}{High-frequency trading and the flash crash:
  structural weaknesses in the securities markets and proposed regulatory
  responses},
\newblock \bibinfo{journal}{Hastings Bus. Law J.} \bibinfo{volume}{8}
  (\bibinfo{year}{2012}) \bibinfo{pages}{445--472}.
\bibitem[{(2020)}]{unknown2016ebs}
\bibinfo{title}{{EBS} dealing rules---appendix {EBS} market},
  \bibinfo{year}{2020}. \bibinfo{note}{Available at:
  \url{https://www.cmegroup.com/trading/market-tech-and-data-services/files/ebs-dealing-rules-appendix-ebs-market.pdf}.
  Archived at: \url{https://doi.org/10.17605/OSF.IO/N7G9X}}.
\bibitem[{Gardiner(1985)}]{gardiner1985handbook}
\bibinfo{author}{C.~W. Gardiner}, \bibinfo{title}{Handbook of stochastic
  methods}, \bibinfo{publisher}{Springer}, \bibinfo{year}{1985}.
\bibitem[{Haken(2013)}]{haken2013synergetics}
\bibinfo{author}{H.~Haken}, \bibinfo{title}{Synergetics: {I}ntroduction and
  advanced topics}, \bibinfo{publisher}{Springer Science \& Business Media},
  \bibinfo{year}{2013}.
\bibitem[{Landau and Lifshitz(1976)}]{landau1976course}
\bibinfo{author}{L.~D. Landau}, \bibinfo{author}{E.~M. Lifshitz},
  \bibinfo{title}{Course of theoretical physics volume 1: {M}echanics},
  \bibinfo{publisher}{Elsevier}, \bibinfo{year}{1976}.
\bibitem[{Evans and Morriss(2008)}]{evans2008statistical}
\bibinfo{author}{D.~J. Evans}, \bibinfo{author}{G.~Morriss},
  \bibinfo{title}{Statistical mechanics of nonequilbrium liquids},
  \bibinfo{publisher}{Cambridge University Press}, \bibinfo{year}{2008}.
\bibitem[{Hansen and McDonald(1990)}]{hansen1990theory}
\bibinfo{author}{J.-P. Hansen}, \bibinfo{author}{I.~R. McDonald},
  \bibinfo{title}{Theory of simple liquids}, \bibinfo{publisher}{Elsevier},
  \bibinfo{year}{1990}.
\bibitem[{Van~Kampen(1992)}]{vankampen1992stochastic}
\bibinfo{author}{N.~G. Van~Kampen}, \bibinfo{title}{Stochastic processes in
  physics and chemistry}, volume~\bibinfo{volume}{1},
  \bibinfo{publisher}{Elsevier}, \bibinfo{year}{1992}.
\bibitem[{Spohn(1980)}]{spohn1980kinetic}
\bibinfo{author}{H.~Spohn},
\newblock \bibinfo{title}{Kinetic equations from hamiltonian dynamics:
  {M}arkovian limits},
\newblock \bibinfo{journal}{Rev. Mod. Phys.} \bibinfo{volume}{52}
  (\bibinfo{year}{1980}) \bibinfo{pages}{569}.
\bibitem[{Lux(1996)}]{lux1996stable}
\bibinfo{author}{T.~Lux},
\newblock \bibinfo{title}{The stable {P}aretian hypothesis and the frequency of
  large returns: an examination of major {G}erman stocks},
\newblock \bibinfo{journal}{Appl. Financ. Econ.} \bibinfo{volume}{6}
  (\bibinfo{year}{1996}) \bibinfo{pages}{463--475}.
\bibitem[{Longin(1996)}]{longin1996asymptotic}
\bibinfo{author}{F.~M. Longin},
\newblock \bibinfo{title}{The asymptotic distribution of extreme stock market
  returns},
\newblock \bibinfo{journal}{J. Business} \bibinfo{volume}{69}
  (\bibinfo{year}{1996}) \bibinfo{pages}{383--408}.
\bibitem[{Guillaume et~al.(1997)Guillaume, Dacorogna, Dav{\'e}, M{\"u}ller,
  Olsen, and Pictet}]{guillaume1997bird}
\bibinfo{author}{D.~M. Guillaume}, \bibinfo{author}{M.~M. Dacorogna},
  \bibinfo{author}{R.~R. Dav{\'e}}, \bibinfo{author}{U.~A. M{\"u}ller},
  \bibinfo{author}{R.~B. Olsen}, \bibinfo{author}{O.~V. Pictet},
\newblock \bibinfo{title}{From the bird's eye to the microscope: {A} survey of
  new stylized facts of the intra-daily foreign exchange markets},
\newblock \bibinfo{journal}{Finance Stoch.} \bibinfo{volume}{1}
  (\bibinfo{year}{1997}) \bibinfo{pages}{95--129}.
\bibitem[{Plerou et~al.(1999)Plerou, Gopikrishnan, Amaral, Meyer, and
  Stanley}]{plerou1999scaling}
\bibinfo{author}{V.~Plerou}, \bibinfo{author}{P.~Gopikrishnan},
  \bibinfo{author}{L.~A.~N. Amaral}, \bibinfo{author}{M.~Meyer},
  \bibinfo{author}{H.~E. Stanley},
\newblock \bibinfo{title}{Scaling of the distribution of price fluctuations of
  individual companies},
\newblock \bibinfo{journal}{Phys. Rev. E} \bibinfo{volume}{60}
  (\bibinfo{year}{1999}) \bibinfo{pages}{6519}.
\bibitem[{Sueshige et~al.(2019)Sueshige, Sornette, Takayasu, and
  Takayasu}]{sueshige2019classification}
\bibinfo{author}{T.~Sueshige}, \bibinfo{author}{D.~Sornette},
  \bibinfo{author}{H.~Takayasu}, \bibinfo{author}{M.~Takayasu},
\newblock \bibinfo{title}{Classification of position management strategies at
  the order-book level and their influences on future market-price formation},
\newblock \bibinfo{journal}{PLOS ONE} \bibinfo{volume}{14}
  (\bibinfo{year}{2019}) \bibinfo{pages}{e0220645}.
\bibitem[{Farmer(2002)}]{farmer2002market}
\bibinfo{author}{J.~D. Farmer},
\newblock \bibinfo{title}{Market force, ecology and evolution},
\newblock \bibinfo{journal}{Ind. Corp. Chang.} \bibinfo{volume}{11}
  (\bibinfo{year}{2002}) \bibinfo{pages}{895--953}.
\bibitem[{Farmer and Skouras(2013)}]{farmer2013ecological}
\bibinfo{author}{J.~D. Farmer}, \bibinfo{author}{S.~Skouras},
\newblock \bibinfo{title}{An ecological perspective on the future of computer
  trading},
\newblock \bibinfo{journal}{Quant. Finance} \bibinfo{volume}{13}
  (\bibinfo{year}{2013}) \bibinfo{pages}{325--346}.
\bibitem[{Odean(1998)}]{odean1998investors}
\bibinfo{author}{T.~Odean},
\newblock \bibinfo{title}{Are investors reluctant to realize their losses?},
\newblock \bibinfo{journal}{J. Finance} \bibinfo{volume}{53}
  (\bibinfo{year}{1998}) \bibinfo{pages}{1775--1798}.
\bibitem[{Grinblatt and Keloharju(2000)}]{grinblatt2000investment}
\bibinfo{author}{M.~Grinblatt}, \bibinfo{author}{M.~Keloharju},
\newblock \bibinfo{title}{The investment behavior and performance of various
  investor types: a study of {F}inland's unique data set},
\newblock \bibinfo{journal}{J. Financ. Econ.} \bibinfo{volume}{55}
  (\bibinfo{year}{2000}) \bibinfo{pages}{43--67}.
\bibitem[{Lee et~al.(2004)Lee, Liu, Roll, and Subrahmanyam}]{lee2004order}
\bibinfo{author}{Y.-T. Lee}, \bibinfo{author}{Y.-J. Liu},
  \bibinfo{author}{R.~Roll}, \bibinfo{author}{A.~Subrahmanyam},
\newblock \bibinfo{title}{Order imbalances and market efficiency: {E}vidence
  from the {Taiwan Stock Exchange}},
\newblock \bibinfo{journal}{J. Financ. Quant. Anal.} \bibinfo{volume}{39}
  (\bibinfo{year}{2004}) \bibinfo{pages}{327--341}.
\bibitem[{de~Lachapelle and Challet(2010)}]{delachapelle2010turnover}
\bibinfo{author}{D.~M. de~Lachapelle}, \bibinfo{author}{D.~Challet},
\newblock \bibinfo{title}{Turnover, account value and diversification of real
  traders: evidence of collective portfolio optimizing behavior},
\newblock \bibinfo{journal}{New J. Phys.} \bibinfo{volume}{12}
  (\bibinfo{year}{2010}) \bibinfo{pages}{075039}.
\bibitem[{Toth et~al.(2012)Toth, Eisler, Lillo, Kockelkoren, Bouchaud, and
  Farmer}]{toth2012does}
\bibinfo{author}{B.~Toth}, \bibinfo{author}{Z.~Eisler},
  \bibinfo{author}{F.~Lillo}, \bibinfo{author}{J.~Kockelkoren},
  \bibinfo{author}{J.-P. Bouchaud}, \bibinfo{author}{J.~D. Farmer},
\newblock \bibinfo{title}{How does the market react to your order flow?},
\newblock \bibinfo{journal}{Quant. Finance} \bibinfo{volume}{12}
  (\bibinfo{year}{2012}) \bibinfo{pages}{1015--1024}.
\bibitem[{Tumminello et~al.(2012)Tumminello, Lillo, Piilo, and
  Mantegna}]{tumminello2012identification}
\bibinfo{author}{M.~Tumminello}, \bibinfo{author}{F.~Lillo},
  \bibinfo{author}{J.~Piilo}, \bibinfo{author}{R.~N. Mantegna},
\newblock \bibinfo{title}{Identification of clusters of investors from their
  real trading activity in a financial market},
\newblock \bibinfo{journal}{New J. Phys.} \bibinfo{volume}{14}
  (\bibinfo{year}{2012}) \bibinfo{pages}{013041}.
\bibitem[{Lillo et~al.(2015)Lillo, Micciche, Tumminello, Piilo, and
  Mantegna}]{lillo2015news}
\bibinfo{author}{F.~Lillo}, \bibinfo{author}{S.~Micciche},
  \bibinfo{author}{M.~Tumminello}, \bibinfo{author}{J.~Piilo},
  \bibinfo{author}{R.~N. Mantegna},
\newblock \bibinfo{title}{How news affects the trading behaviour of different
  categories of investors in a financial market},
\newblock \bibinfo{journal}{Quant. Finance} \bibinfo{volume}{15}
  (\bibinfo{year}{2015}) \bibinfo{pages}{213--229}.
\bibitem[{Musciotto et~al.(2018)Musciotto, Marotta, Piilo, and
  Mantegna}]{musciotto2018long}
\bibinfo{author}{F.~Musciotto}, \bibinfo{author}{L.~Marotta},
  \bibinfo{author}{J.~Piilo}, \bibinfo{author}{R.~N. Mantegna},
\newblock \bibinfo{title}{Long-term ecology of investors in a financial
  market},
\newblock \bibinfo{journal}{Palgrave Commun.} \bibinfo{volume}{4}
  (\bibinfo{year}{2018}) \bibinfo{pages}{92}.
\bibitem[{Frank(2007)}]{frank2007life}
\bibinfo{author}{S.~A. Frank},
\newblock \bibinfo{title}{All of life is social},
\newblock \bibinfo{journal}{Curr. Biol.} \bibinfo{volume}{17}
  (\bibinfo{year}{2007}) \bibinfo{pages}{R648--R650}.
\bibitem[{Smith(1982)}]{smith1982evolution}
\bibinfo{author}{J.~M. Smith}, \bibinfo{title}{Evolution and the Theory of
  Games}, \bibinfo{publisher}{Cambridge University Press},
  \bibinfo{year}{1982}.
\bibitem[{Sachs et~al.(2004)Sachs, Mueller, Wilcox, and
  Bull}]{sachs2004evolution}
\bibinfo{author}{J.~L. Sachs}, \bibinfo{author}{U.~G. Mueller},
  \bibinfo{author}{T.~P. Wilcox}, \bibinfo{author}{J.~J. Bull},
\newblock \bibinfo{title}{The evolution of cooperation},
\newblock \bibinfo{journal}{Q. Rev. Biol.} \bibinfo{volume}{79}
  (\bibinfo{year}{2004}) \bibinfo{pages}{135--160}.
\bibitem[{Cooper and Kagel(2016)}]{cooper2016other}
\bibinfo{author}{D.~J. Cooper}, \bibinfo{author}{J.~H. Kagel},
\newblock \bibinfo{title}{Other-regarding preferences},
\newblock in: \bibinfo{editor}{J.~H. Kagel}, \bibinfo{editor}{A.~E. Roth}
  (Eds.), \bibinfo{booktitle}{The Handbook of Experimental Economics, Volume
  2}, \bibinfo{publisher}{Princeton University Press}, \bibinfo{year}{2016},
  pp. \bibinfo{pages}{217--289}.
\bibitem[{Henrich and Muthukrishna(2021)}]{henrich2021origins}
\bibinfo{author}{J.~Henrich}, \bibinfo{author}{M.~Muthukrishna},
\newblock \bibinfo{title}{The origins and psychology of human cooperation},
\newblock \bibinfo{journal}{Annual Review of Psychology} \bibinfo{volume}{72}
  (\bibinfo{year}{2021}) \bibinfo{pages}{207--240}.
\bibitem[{Ak\c{c}ay(2020)}]{akcay2020deconstructing}
\bibinfo{author}{E.~Ak\c{c}ay},
\newblock \bibinfo{title}{Deconstructing evolutionary game theory: coevolution
  of social behaviors with their evolutionary setting},
\newblock \bibinfo{journal}{Am. Nat.} \bibinfo{volume}{195}
  (\bibinfo{year}{2020}) \bibinfo{pages}{315--330}.
\bibitem[{Dawes(1980)}]{dawes1980social}
\bibinfo{author}{R.~M. Dawes},
\newblock \bibinfo{title}{Social dilemmas},
\newblock \bibinfo{journal}{Annu. Rev. Psychol.} \bibinfo{volume}{31}
  (\bibinfo{year}{1980}) \bibinfo{pages}{169--193}.
\bibitem[{Kuhn(2019)}]{kuhn2019prisoner}
\bibinfo{author}{S.~Kuhn},
\newblock \bibinfo{title}{Prisoner's {D}ilemma},
\newblock in: \bibinfo{editor}{E.~N. Zalta} (Ed.), \bibinfo{booktitle}{The
  Stanford Encyclopedia of Philosophy (Winter 2019 Edition)},
  \bibinfo{publisher}{Metaphysics Research Lab, Stanford University},
  \bibinfo{year}{2019}, pp. \bibinfo{pages}{---}. \bibinfo{note}{Available at:
  \url{https://plato.stanford.edu/archives/win2019/entries/prisoner-dilemma/}}.
\bibitem[{Weibull(1997)}]{weibull1997evolutionary}
\bibinfo{author}{J.~W. Weibull}, \bibinfo{title}{Evolutionary game theory},
  \bibinfo{publisher}{MIT Press}, \bibinfo{year}{1997}.
\bibitem[{Tanimoto(2015)}]{tanimoto2015fundamentals}
\bibinfo{author}{J.~Tanimoto}, \bibinfo{title}{Fundamentals of evolutionary
  game theory and its applications}, \bibinfo{publisher}{Springer},
  \bibinfo{year}{2015}.
\bibitem[{Taylor and Jonker(1978)}]{taylor1978evolutionary}
\bibinfo{author}{P.~D. Taylor}, \bibinfo{author}{L.~B. Jonker},
\newblock \bibinfo{title}{Evolutionary stable strategies and game dynamics},
\newblock \bibinfo{journal}{Math. Biosci.} \bibinfo{volume}{40}
  (\bibinfo{year}{1978}) \bibinfo{pages}{145--156}.
\bibitem[{Doebeli and Hauert(2005)}]{doebeli2005models}
\bibinfo{author}{M.~Doebeli}, \bibinfo{author}{C.~Hauert},
\newblock \bibinfo{title}{Models of cooperation based on the {P}risoner's
  {D}ilemma and the {S}nowdrift game},
\newblock \bibinfo{journal}{Ecol. Lett.} \bibinfo{volume}{8}
  (\bibinfo{year}{2005}) \bibinfo{pages}{748--766}.
\bibitem[{Wang et~al.(2015)Wang, Jusup, Kokubo, and
  Tanimoto}]{wang2015universal}
\bibinfo{author}{Z.~Wang}, \bibinfo{author}{M.~Jusup},
  \bibinfo{author}{S.~Kokubo}, \bibinfo{author}{J.~Tanimoto},
\newblock \bibinfo{title}{Universal scaling for the dilemma strength in
  evolutionary games},
\newblock \bibinfo{journal}{Phys. Life Rev.} \bibinfo{volume}{14}
  (\bibinfo{year}{2015}) \bibinfo{pages}{1--30}.
\bibitem[{Ito and Tanimoto(2018)}]{ito2018scaling}
\bibinfo{author}{H.~Ito}, \bibinfo{author}{J.~Tanimoto},
\newblock \bibinfo{title}{Scaling the phase-planes of social dilemma strengths
  shows game-class changes in the five rules governing the evolution of
  cooperation},
\newblock \bibinfo{journal}{R. Soc. Open Sci.} \bibinfo{volume}{5}
  (\bibinfo{year}{2018}) \bibinfo{pages}{181085}.
\bibitem[{Arefin et~al.(2020)Arefin, Kabir, Jusup, Ito, and
  Tanimoto}]{arefin2020social}
\bibinfo{author}{M.~R. Arefin}, \bibinfo{author}{K.~M.~A. Kabir},
  \bibinfo{author}{M.~Jusup}, \bibinfo{author}{H.~Ito},
  \bibinfo{author}{J.~Tanimoto},
\newblock \bibinfo{title}{Social efficiency deficit deciphers social dilemmas},
\newblock \bibinfo{journal}{Sci. Rep.} \bibinfo{volume}{10}
  (\bibinfo{year}{2020}) \bibinfo{pages}{16092}.
\bibitem[{Nowak(2006)}]{nowak2006five}
\bibinfo{author}{M.~A. Nowak},
\newblock \bibinfo{title}{Five rules for the evolution of cooperation},
\newblock \bibinfo{journal}{Science} \bibinfo{volume}{314}
  (\bibinfo{year}{2006}) \bibinfo{pages}{1560--1563}.
\bibitem[{Taylor and Nowak(2007)}]{taylor2007transforming}
\bibinfo{author}{C.~Taylor}, \bibinfo{author}{M.~A. Nowak},
\newblock \bibinfo{title}{Transforming the dilemma},
\newblock \bibinfo{journal}{Evolution} \bibinfo{volume}{61}
  (\bibinfo{year}{2007}) \bibinfo{pages}{2281--2292}.
\bibitem[{Eshel and Cavalli-Sforza(1982)}]{eshel1982assortment}
\bibinfo{author}{I.~Eshel}, \bibinfo{author}{L.~L. Cavalli-Sforza},
\newblock \bibinfo{title}{Assortment of encounters and evolution of
  cooperativeness},
\newblock \bibinfo{journal}{Proc. Natl. Acad. Sci. USA} \bibinfo{volume}{79}
  (\bibinfo{year}{1982}) \bibinfo{pages}{1331--1335}.
\bibitem[{Newton(2018)}]{newton2018evolutionary}
\bibinfo{author}{J.~Newton},
\newblock \bibinfo{title}{Evolutionary game theory: {A} renaissance},
\newblock \bibinfo{journal}{Games} \bibinfo{volume}{9} (\bibinfo{year}{2018})
  \bibinfo{pages}{31}.
\bibitem[{Kay et~al.(2020)Kay, Keller, and Lehmann}]{kay2020evolution}
\bibinfo{author}{T.~Kay}, \bibinfo{author}{L.~Keller},
  \bibinfo{author}{L.~Lehmann},
\newblock \bibinfo{title}{The evolution of altruism and the serial rediscovery
  of the role of relatedness},
\newblock \bibinfo{journal}{Proc. Natl. Acad. Sci. USA} \bibinfo{volume}{117}
  (\bibinfo{year}{2020}) \bibinfo{pages}{28894--28898}.
\bibitem[{Tudge and Brede(2015)}]{tudge2015tale}
\bibinfo{author}{S.~J. Tudge}, \bibinfo{author}{M.~Brede},
\newblock \bibinfo{title}{A tale of two theorems: {C}omment on ``universal
  scaling for the dilemma strength in evolutionary games'' by {Z. Wang} et
  al.},
\newblock \bibinfo{journal}{Phys. Life Rev.} \bibinfo{volume}{14}
  (\bibinfo{year}{2015}) \bibinfo{pages}{49--51}.
\bibitem[{Ohtsuki et~al.(2006)Ohtsuki, Hauert, Lieberman, and
  Nowak}]{ohtsuki2006simple}
\bibinfo{author}{H.~Ohtsuki}, \bibinfo{author}{C.~Hauert},
  \bibinfo{author}{E.~Lieberman}, \bibinfo{author}{M.~A. Nowak},
\newblock \bibinfo{title}{A simple rule for the evolution of cooperation on
  graphs and social networks},
\newblock \bibinfo{journal}{Nature} \bibinfo{volume}{441}
  (\bibinfo{year}{2006}) \bibinfo{pages}{502--505}.
\bibitem[{Ohtsuki and Nowak(2006{\natexlab{a}})}]{ohtsuki2006evolutionary}
\bibinfo{author}{H.~Ohtsuki}, \bibinfo{author}{M.~A. Nowak},
\newblock \bibinfo{title}{Evolutionary games on cycles},
\newblock \bibinfo{journal}{Proc. R. Soc. B} \bibinfo{volume}{273}
  (\bibinfo{year}{2006}{\natexlab{a}}) \bibinfo{pages}{2249--2256}.
\bibitem[{Ohtsuki and Nowak(2006{\natexlab{b}})}]{ohtsuki2006replicator}
\bibinfo{author}{H.~Ohtsuki}, \bibinfo{author}{M.~A. Nowak},
\newblock \bibinfo{title}{The replicator equation on graphs},
\newblock \bibinfo{journal}{J. Theor. Biol.} \bibinfo{volume}{243}
  (\bibinfo{year}{2006}{\natexlab{b}}) \bibinfo{pages}{86--97}.
\bibitem[{Nowak and May(1992)}]{nowak1992evolutionary}
\bibinfo{author}{M.~A. Nowak}, \bibinfo{author}{R.~M. May},
\newblock \bibinfo{title}{Evolutionary games and spatial chaos},
\newblock \bibinfo{journal}{Nature} \bibinfo{volume}{359}
  (\bibinfo{year}{1992}) \bibinfo{pages}{826--829}.
\bibitem[{Nowak and May(1993)}]{nowak1993spatial}
\bibinfo{author}{M.~A. Nowak}, \bibinfo{author}{R.~M. May},
\newblock \bibinfo{title}{The spatial dilemmas of evolution},
\newblock \bibinfo{journal}{Int. J. Bifurc. Chaos} \bibinfo{volume}{3}
  (\bibinfo{year}{1993}) \bibinfo{pages}{35--78}.
\bibitem[{G{\'o}mez-Gardenes et~al.(2007)G{\'o}mez-Gardenes, Campillo,
  Flor{\'\i}a, and Moreno}]{gomez2007dynamical}
\bibinfo{author}{J.~G{\'o}mez-Gardenes}, \bibinfo{author}{M.~Campillo},
  \bibinfo{author}{L.~M. Flor{\'\i}a}, \bibinfo{author}{Y.~Moreno},
\newblock \bibinfo{title}{Dynamical organization of cooperation in complex
  topologies},
\newblock \bibinfo{journal}{Phys. Rev. Lett.} \bibinfo{volume}{98}
  (\bibinfo{year}{2007}) \bibinfo{pages}{108103}.
\bibitem[{Poncela et~al.(2007)Poncela, G{\'o}mez-Gardenes, Flor{\'\i}a, and
  Moreno}]{poncela2007robustness}
\bibinfo{author}{J.~Poncela}, \bibinfo{author}{J.~G{\'o}mez-Gardenes},
  \bibinfo{author}{L.~M. Flor{\'\i}a}, \bibinfo{author}{Y.~Moreno},
\newblock \bibinfo{title}{Robustness of cooperation in the evolutionary
  prisoner's dilemma on complex networks},
\newblock \bibinfo{journal}{New J. Physics} \bibinfo{volume}{9}
  (\bibinfo{year}{2007}) \bibinfo{pages}{184}.
\bibitem[{G{\'o}mez-Garde{\~n}es et~al.(2008)G{\'o}mez-Garde{\~n}es, Poncela,
  Flor{\'\i}a, and Moreno}]{gomez2008natural}
\bibinfo{author}{J.~G{\'o}mez-Garde{\~n}es}, \bibinfo{author}{J.~Poncela},
  \bibinfo{author}{L.~M. Flor{\'\i}a}, \bibinfo{author}{Y.~Moreno},
\newblock \bibinfo{title}{Natural selection of cooperation and degree hierarchy
  in heterogeneous populations},
\newblock \bibinfo{journal}{J. Theor. Biol.} \bibinfo{volume}{253}
  (\bibinfo{year}{2008}) \bibinfo{pages}{296--301}.
\bibitem[{Devlin and Treloar(2009)}]{devlin2009evolution}
\bibinfo{author}{S.~Devlin}, \bibinfo{author}{T.~Treloar},
\newblock \bibinfo{title}{Evolution of cooperation through the heterogeneity of
  random networks},
\newblock \bibinfo{journal}{Phys. Rev. E} \bibinfo{volume}{79}
  (\bibinfo{year}{2009}) \bibinfo{pages}{016107}.
\bibitem[{Poncela et~al.(2009)Poncela, G{\'o}mez-Garde{\~n}es, Flor{\'\i}a,
  Moreno, and S{\'a}nchez}]{poncela2009cooperative}
\bibinfo{author}{J.~Poncela}, \bibinfo{author}{J.~G{\'o}mez-Garde{\~n}es},
  \bibinfo{author}{L.~M. Flor{\'\i}a}, \bibinfo{author}{Y.~Moreno},
  \bibinfo{author}{A.~S{\'a}nchez},
\newblock \bibinfo{title}{Cooperative scale-free networks despite the presence
  of defector hubs},
\newblock \bibinfo{journal}{EPL (Europhys. Lett.)} \bibinfo{volume}{88}
  (\bibinfo{year}{2009}) \bibinfo{pages}{38003}.
\bibitem[{Santos and Pacheco(2005)}]{santos2005scale}
\bibinfo{author}{F.~C. Santos}, \bibinfo{author}{J.~M. Pacheco},
\newblock \bibinfo{title}{Scale-free networks provide a unifying framework for
  the emergence of cooperation},
\newblock \bibinfo{journal}{Phys. Rev. Lett.} \bibinfo{volume}{95}
  (\bibinfo{year}{2005}) \bibinfo{pages}{098104}.
\bibitem[{Szolnoki et~al.(2008)Szolnoki, Perc, and Danku}]{szolnoki2008towards}
\bibinfo{author}{A.~Szolnoki}, \bibinfo{author}{M.~Perc},
  \bibinfo{author}{Z.~Danku},
\newblock \bibinfo{title}{Towards effective payoffs in the prisoner's dilemma
  game on scale-free networks},
\newblock \bibinfo{journal}{Physica A} \bibinfo{volume}{387}
  (\bibinfo{year}{2008}) \bibinfo{pages}{2075--2082}.
\bibitem[{Yamauchi et~al.(2010)Yamauchi, Tanimoto, and
  Hagishima}]{yamauchi2010controls}
\bibinfo{author}{A.~Yamauchi}, \bibinfo{author}{J.~Tanimoto},
  \bibinfo{author}{A.~Hagishima},
\newblock \bibinfo{title}{What controls network reciprocity in the {P}risoner's
  {D}ilemma game?},
\newblock \bibinfo{journal}{BioSystems} \bibinfo{volume}{102}
  (\bibinfo{year}{2010}) \bibinfo{pages}{82--87}.
\bibitem[{Nowak et~al.(2010)Nowak, Tarnita, and Antal}]{nowak2010evolutionary}
\bibinfo{author}{M.~A. Nowak}, \bibinfo{author}{C.~E. Tarnita},
  \bibinfo{author}{T.~Antal},
\newblock \bibinfo{title}{Evolutionary dynamics in structured populations},
\newblock \bibinfo{journal}{Philos. Trans. R. Soc. B} \bibinfo{volume}{365}
  (\bibinfo{year}{2010}) \bibinfo{pages}{19--30}.
\bibitem[{Allen et~al.(2017)Allen, Lippner, Chen, Fotouhi, Momeni, Yau, and
  Nowak}]{allen2017evolutionary}
\bibinfo{author}{B.~Allen}, \bibinfo{author}{G.~Lippner},
  \bibinfo{author}{Y.-T. Chen}, \bibinfo{author}{B.~Fotouhi},
  \bibinfo{author}{N.~Momeni}, \bibinfo{author}{S.-T. Yau},
  \bibinfo{author}{M.~A. Nowak},
\newblock \bibinfo{title}{Evolutionary dynamics on any population structure},
\newblock \bibinfo{journal}{Nature} \bibinfo{volume}{544}
  (\bibinfo{year}{2017}) \bibinfo{pages}{227--230}.
\bibitem[{Su et~al.(2016)Su, Li, Zhou, and Wang}]{su2016interactive}
\bibinfo{author}{Q.~Su}, \bibinfo{author}{A.~Li}, \bibinfo{author}{L.~Zhou},
  \bibinfo{author}{L.~Wang},
\newblock \bibinfo{title}{Interactive diversity promotes the evolution of
  cooperation in structured populations},
\newblock \bibinfo{journal}{New J. Physics} \bibinfo{volume}{18}
  (\bibinfo{year}{2016}) \bibinfo{pages}{103007}.
\bibitem[{Sendi{\~n}a-Nadal et~al.(2020)Sendi{\~n}a-Nadal, Leyva, Perc, Papo,
  Jusup, Wang, Almendral, Manshour, and Boccaletti}]{sendina2020diverse}
\bibinfo{author}{I.~Sendi{\~n}a-Nadal}, \bibinfo{author}{I.~Leyva},
  \bibinfo{author}{M.~Perc}, \bibinfo{author}{D.~Papo},
  \bibinfo{author}{M.~Jusup}, \bibinfo{author}{Z.~Wang}, \bibinfo{author}{J.~A.
  Almendral}, \bibinfo{author}{P.~Manshour}, \bibinfo{author}{S.~Boccaletti},
\newblock \bibinfo{title}{Diverse strategic identities induce dynamical states
  in evolutionary games},
\newblock \bibinfo{journal}{Phys. Rev. Res.} \bibinfo{volume}{2}
  (\bibinfo{year}{2020}) \bibinfo{pages}{043168}.
\bibitem[{Su et~al.(2018)Su, Li, and Wang}]{su2018evolution}
\bibinfo{author}{Q.~Su}, \bibinfo{author}{A.~Li}, \bibinfo{author}{L.~Wang},
\newblock \bibinfo{title}{Evolution of cooperation with interactive identity
  and diversity},
\newblock \bibinfo{journal}{J. Theor. Biol.} \bibinfo{volume}{442}
  (\bibinfo{year}{2018}) \bibinfo{pages}{149--157}.
\bibitem[{Jia et~al.(2020)Jia, Wang, Song, Romi{\'c}, Li, Jusup, and
  Wang}]{jia2020evolutionary}
\bibinfo{author}{D.~Jia}, \bibinfo{author}{X.~Wang}, \bibinfo{author}{Z.~Song},
  \bibinfo{author}{I.~Romi{\'c}}, \bibinfo{author}{X.~Li},
  \bibinfo{author}{M.~Jusup}, \bibinfo{author}{Z.~Wang},
\newblock \bibinfo{title}{Evolutionary dynamics drives role specialization in a
  community of players},
\newblock \bibinfo{journal}{J. R. Soc. Interface} \bibinfo{volume}{17}
  (\bibinfo{year}{2020}) \bibinfo{pages}{20200174}.
\bibitem[{Szolnoki et~al.(2014)Szolnoki, Mobilia, Jiang, Szczesny, Rucklidge,
  and Perc}]{szolnoki2014cyclic}
\bibinfo{author}{A.~Szolnoki}, \bibinfo{author}{M.~Mobilia},
  \bibinfo{author}{L.-L. Jiang}, \bibinfo{author}{B.~Szczesny},
  \bibinfo{author}{A.~M. Rucklidge}, \bibinfo{author}{M.~Perc},
\newblock \bibinfo{title}{Cyclic dominance in evolutionary games: a review},
\newblock \bibinfo{journal}{J. R. Soc. Interface} \bibinfo{volume}{11}
  (\bibinfo{year}{2014}) \bibinfo{pages}{20140735}.
\bibitem[{Szab{\'o} and Hauert(2002)}]{szabo2002evolutionary}
\bibinfo{author}{G.~Szab{\'o}}, \bibinfo{author}{C.~Hauert},
\newblock \bibinfo{title}{Evolutionary prisoner's dilemma games with voluntary
  participation},
\newblock \bibinfo{journal}{Phys. Rev. E} \bibinfo{volume}{66}
  (\bibinfo{year}{2002}) \bibinfo{pages}{062903}.
\bibitem[{Shen et~al.(2021)Shen, Jusup, Shi, Wang, Perc, and
  Holme}]{shen2021exit}
\bibinfo{author}{C.~Shen}, \bibinfo{author}{M.~Jusup},
  \bibinfo{author}{L.~Shi}, \bibinfo{author}{Z.~Wang},
  \bibinfo{author}{M.~Perc}, \bibinfo{author}{P.~Holme},
\newblock \bibinfo{title}{Exit rights open complex pathways to cooperation},
\newblock \bibinfo{journal}{J. R. Soc. Interface} \bibinfo{volume}{18}
  (\bibinfo{year}{2021}) \bibinfo{pages}{20200777}.
\bibitem[{Guo et~al.(2020)Guo, Song, Ge{\v{c}}ek, Li, Jusup, Perc, Moreno,
  Boccaletti, and Wang}]{guo2020novel}
\bibinfo{author}{H.~Guo}, \bibinfo{author}{Z.~Song},
  \bibinfo{author}{S.~Ge{\v{c}}ek}, \bibinfo{author}{X.~Li},
  \bibinfo{author}{M.~Jusup}, \bibinfo{author}{M.~Perc},
  \bibinfo{author}{Y.~Moreno}, \bibinfo{author}{S.~Boccaletti},
  \bibinfo{author}{Z.~Wang},
\newblock \bibinfo{title}{A novel route to cyclic dominance in voluntary social
  dilemmas},
\newblock \bibinfo{journal}{J. R. Soc. Interface} \bibinfo{volume}{17}
  (\bibinfo{year}{2020}) \bibinfo{pages}{20190789}.
\bibitem[{Hauert and Doebeli(2004)}]{hauert2004spatial}
\bibinfo{author}{C.~Hauert}, \bibinfo{author}{M.~Doebeli},
\newblock \bibinfo{title}{Spatial structure often inhibits the evolution of
  cooperation in the snowdrift game},
\newblock \bibinfo{journal}{Nature} \bibinfo{volume}{428}
  (\bibinfo{year}{2004}) \bibinfo{pages}{643--646}.
\bibitem[{Gosak et~al.(2018)Gosak, Markovi{\v{c}}, Dolen{\v{s}}ek, Rupnik,
  Marhl, Sto{\v{z}}er, and Perc}]{gosak2018network}
\bibinfo{author}{M.~Gosak}, \bibinfo{author}{R.~Markovi{\v{c}}},
  \bibinfo{author}{J.~Dolen{\v{s}}ek}, \bibinfo{author}{M.~S. Rupnik},
  \bibinfo{author}{M.~Marhl}, \bibinfo{author}{A.~Sto{\v{z}}er},
  \bibinfo{author}{M.~Perc},
\newblock \bibinfo{title}{Network science of biological systems at different
  scales: {A} review},
\newblock \bibinfo{journal}{Phys. Life Rev.} \bibinfo{volume}{24}
  (\bibinfo{year}{2018}) \bibinfo{pages}{118--135}.
\bibitem[{Pilosof et~al.(2017)Pilosof, Porter, Pascual, and
  K{\'e}fi}]{pilosof2017multilayer}
\bibinfo{author}{S.~Pilosof}, \bibinfo{author}{M.~A. Porter},
  \bibinfo{author}{M.~Pascual}, \bibinfo{author}{S.~K{\'e}fi},
\newblock \bibinfo{title}{The multilayer nature of ecological networks},
\newblock \bibinfo{journal}{Nat. Ecol. Evol.} \bibinfo{volume}{1}
  (\bibinfo{year}{2017}) \bibinfo{pages}{0101}.
\bibitem[{Buldyrev et~al.(2010)Buldyrev, Parshani, Paul, Stanley, and
  Havlin}]{buldyrev2010catastrophic}
\bibinfo{author}{S.~V. Buldyrev}, \bibinfo{author}{R.~Parshani},
  \bibinfo{author}{G.~Paul}, \bibinfo{author}{H.~E. Stanley},
  \bibinfo{author}{S.~Havlin},
\newblock \bibinfo{title}{Catastrophic cascade of failures in interdependent
  networks},
\newblock \bibinfo{journal}{Nature} \bibinfo{volume}{464}
  (\bibinfo{year}{2010}) \bibinfo{pages}{1025--1028}.
\bibitem[{Gao et~al.(2011)Gao, Buldyrev, Havlin, and
  Stanley}]{gao2011robustness}
\bibinfo{author}{J.~Gao}, \bibinfo{author}{S.~V. Buldyrev},
  \bibinfo{author}{S.~Havlin}, \bibinfo{author}{H.~E. Stanley},
\newblock \bibinfo{title}{Robustness of a network of networks},
\newblock \bibinfo{journal}{Phys. Rev. Lett.} \bibinfo{volume}{107}
  (\bibinfo{year}{2011}) \bibinfo{pages}{195701}.
\bibitem[{Pocock et~al.(2012)Pocock, Evans, and Memmott}]{pocock2012robustness}
\bibinfo{author}{M.~J. Pocock}, \bibinfo{author}{D.~M. Evans},
  \bibinfo{author}{J.~Memmott},
\newblock \bibinfo{title}{The robustness and restoration of a network of
  ecological networks},
\newblock \bibinfo{journal}{Science} \bibinfo{volume}{335}
  (\bibinfo{year}{2012}) \bibinfo{pages}{973--977}.
\bibitem[{Dong et~al.(2013)Dong, Gao, Du, Tian, Stanley, and
  Havlin}]{dong2013robustness}
\bibinfo{author}{G.~Dong}, \bibinfo{author}{J.~Gao}, \bibinfo{author}{R.~Du},
  \bibinfo{author}{L.~Tian}, \bibinfo{author}{H.~E. Stanley},
  \bibinfo{author}{S.~Havlin},
\newblock \bibinfo{title}{Robustness of network of networks under targeted
  attack},
\newblock \bibinfo{journal}{Phys. Rev. E} \bibinfo{volume}{87}
  (\bibinfo{year}{2013}) \bibinfo{pages}{052804}.
\bibitem[{Evans et~al.(2013)Evans, Pocock, and Memmott}]{evans2013robustness}
\bibinfo{author}{D.~M. Evans}, \bibinfo{author}{M.~J. Pocock},
  \bibinfo{author}{J.~Memmott},
\newblock \bibinfo{title}{The robustness of a network of ecological networks to
  habitat loss},
\newblock \bibinfo{journal}{Ecol. Lett.} \bibinfo{volume}{16}
  (\bibinfo{year}{2013}) \bibinfo{pages}{844--852}.
\bibitem[{Wang et~al.(2015)Wang, Wang, Szolnoki, and
  Perc}]{wang2015evolutionary}
\bibinfo{author}{Z.~Wang}, \bibinfo{author}{L.~Wang},
  \bibinfo{author}{A.~Szolnoki}, \bibinfo{author}{M.~Perc},
\newblock \bibinfo{title}{Evolutionary games on multilayer networks: a
  colloquium},
\newblock \bibinfo{journal}{Eur. Phys. J. B} \bibinfo{volume}{88}
  (\bibinfo{year}{2015}) \bibinfo{pages}{1--15}.
\bibitem[{De~Domenico et~al.(2013)De~Domenico, Sol{\'e}-Ribalta, Cozzo,
  Kivel{\"a}, Moreno, Porter, G{\'o}mez, and
  Arenas}]{dedomenico2013mathematical}
\bibinfo{author}{M.~De~Domenico}, \bibinfo{author}{A.~Sol{\'e}-Ribalta},
  \bibinfo{author}{E.~Cozzo}, \bibinfo{author}{M.~Kivel{\"a}},
  \bibinfo{author}{Y.~Moreno}, \bibinfo{author}{M.~A. Porter},
  \bibinfo{author}{S.~G{\'o}mez}, \bibinfo{author}{A.~Arenas},
\newblock \bibinfo{title}{Mathematical formulation of multilayer networks},
\newblock \bibinfo{journal}{Phys. Rev. X} \bibinfo{volume}{3}
  (\bibinfo{year}{2013}) \bibinfo{pages}{041022}.
\bibitem[{Kivel{\"a} et~al.(2014)Kivel{\"a}, Arenas, Barthelemy, Gleeson,
  Moreno, and Porter}]{kivela2014multilayer}
\bibinfo{author}{M.~Kivel{\"a}}, \bibinfo{author}{A.~Arenas},
  \bibinfo{author}{M.~Barthelemy}, \bibinfo{author}{J.~P. Gleeson},
  \bibinfo{author}{Y.~Moreno}, \bibinfo{author}{M.~A. Porter},
\newblock \bibinfo{title}{Multilayer networks},
\newblock \bibinfo{journal}{J. Complex Netw.} \bibinfo{volume}{2}
  (\bibinfo{year}{2014}) \bibinfo{pages}{203--271}.
\bibitem[{G{\'o}mez-Gardenes et~al.(2012)G{\'o}mez-Gardenes, Reinares, Arenas,
  and Flor{\'\i}a}]{gomez2012evolution}
\bibinfo{author}{J.~G{\'o}mez-Gardenes}, \bibinfo{author}{I.~Reinares},
  \bibinfo{author}{A.~Arenas}, \bibinfo{author}{L.~M. Flor{\'\i}a},
\newblock \bibinfo{title}{Evolution of cooperation in multiplex networks},
\newblock \bibinfo{journal}{Sci. Rep.} \bibinfo{volume}{2}
  (\bibinfo{year}{2012}) \bibinfo{pages}{620}.
\bibitem[{Jiang and Perc(2013)}]{jiang2013spreading}
\bibinfo{author}{L.-L. Jiang}, \bibinfo{author}{M.~Perc},
\newblock \bibinfo{title}{Spreading of cooperative behaviour across
  interdependent groups},
\newblock \bibinfo{journal}{Sci. Rep.} \bibinfo{volume}{3}
  (\bibinfo{year}{2013}) \bibinfo{pages}{2483}.
\bibitem[{G{\'o}mez-Gardenes et~al.(2012)G{\'o}mez-Gardenes, Gracia-L{\'a}zaro,
  Floria, and Moreno}]{gomez2012evolutionary}
\bibinfo{author}{J.~G{\'o}mez-Gardenes},
  \bibinfo{author}{C.~Gracia-L{\'a}zaro}, \bibinfo{author}{L.~M. Floria},
  \bibinfo{author}{Y.~Moreno},
\newblock \bibinfo{title}{Evolutionary dynamics on interdependent populations},
\newblock \bibinfo{journal}{Phys. Rev. E} \bibinfo{volume}{86}
  (\bibinfo{year}{2012}) \bibinfo{pages}{056113}.
\bibitem[{Wang et~al.(2012)Wang, Szolnoki, and Perc}]{wang2012evolution}
\bibinfo{author}{Z.~Wang}, \bibinfo{author}{A.~Szolnoki},
  \bibinfo{author}{M.~Perc},
\newblock \bibinfo{title}{Evolution of public cooperation on interdependent
  networks: {T}he impact of biased utility functions},
\newblock \bibinfo{journal}{EPL (Europhys. Lett.)} \bibinfo{volume}{97}
  (\bibinfo{year}{2012}) \bibinfo{pages}{48001}.
\bibitem[{Shen et~al.(2018)Shen, Chu, Shi, Jusup, Perc, and
  Wang}]{shen2018coevolutionary}
\bibinfo{author}{C.~Shen}, \bibinfo{author}{C.~Chu}, \bibinfo{author}{L.~Shi},
  \bibinfo{author}{M.~Jusup}, \bibinfo{author}{M.~Perc},
  \bibinfo{author}{Z.~Wang},
\newblock \bibinfo{title}{Coevolutionary resolution of the public goods dilemma
  in interdependent structured populations},
\newblock \bibinfo{journal}{EPL (Europhys. Lett.)} \bibinfo{volume}{124}
  (\bibinfo{year}{2018}) \bibinfo{pages}{48003}.
\bibitem[{Barabasi(2005)}]{barabasi2005origin}
\bibinfo{author}{A.-L. Barabasi},
\newblock \bibinfo{title}{The origin of bursts and heavy tails in human
  dynamics},
\newblock \bibinfo{journal}{Nature} \bibinfo{volume}{435}
  (\bibinfo{year}{2005}) \bibinfo{pages}{207--211}.
\bibitem[{Vazquez(2005)}]{vazquez2005exact}
\bibinfo{author}{A.~Vazquez},
\newblock \bibinfo{title}{Exact results for the {B}arab{\'a}si model of human
  dynamics},
\newblock \bibinfo{journal}{Physi. Rev. Lett.} \bibinfo{volume}{95}
  (\bibinfo{year}{2005}) \bibinfo{pages}{248701}.
\bibitem[{Starnini et~al.(2013)Starnini, Baronchelli, and
  Pastor-Satorras}]{starnini2013modeling}
\bibinfo{author}{M.~Starnini}, \bibinfo{author}{A.~Baronchelli},
  \bibinfo{author}{R.~Pastor-Satorras},
\newblock \bibinfo{title}{Modeling human dynamics of face-to-face interaction
  networks},
\newblock \bibinfo{journal}{Phys. Rev, Lett.} \bibinfo{volume}{110}
  (\bibinfo{year}{2013}) \bibinfo{pages}{168701}.
\bibitem[{Holme and Saram{\"a}ki(2012)}]{holme2012temporal}
\bibinfo{author}{P.~Holme}, \bibinfo{author}{J.~Saram{\"a}ki},
\newblock \bibinfo{title}{Temporal networks},
\newblock \bibinfo{journal}{Phys. Rep.} \bibinfo{volume}{519}
  (\bibinfo{year}{2012}) \bibinfo{pages}{97--125}.
\bibitem[{Holme(2015)}]{holme2015modern}
\bibinfo{author}{P.~Holme},
\newblock \bibinfo{title}{Modern temporal network theory: a colloquium},
\newblock \bibinfo{journal}{Eur. Phys. J. B} \bibinfo{volume}{88}
  (\bibinfo{year}{2015}) \bibinfo{pages}{1--30}.
\bibitem[{Masuda and Lambiotte(2016)}]{masuda2016guide}
\bibinfo{author}{N.~Masuda}, \bibinfo{author}{R.~Lambiotte}, \bibinfo{title}{A
  guide to temporal networks}, \bibinfo{publisher}{World Scientific},
  \bibinfo{year}{2016}.
\bibitem[{Masuda et~al.(2013)Masuda, Klemm, and
  Egu{\'\i}luz}]{masuda2013temporal}
\bibinfo{author}{N.~Masuda}, \bibinfo{author}{K.~Klemm}, \bibinfo{author}{V.~M.
  Egu{\'\i}luz},
\newblock \bibinfo{title}{Temporal networks: slowing down diffusion by long
  lasting interactions},
\newblock \bibinfo{journal}{Phys. Rev. Lett.} \bibinfo{volume}{111}
  (\bibinfo{year}{2013}) \bibinfo{pages}{188701}.
\bibitem[{Scholtes et~al.(2014)Scholtes, Wider, Pfitzner, Garas, Tessone, and
  Schweitzer}]{scholtes2014causality}
\bibinfo{author}{I.~Scholtes}, \bibinfo{author}{N.~Wider},
  \bibinfo{author}{R.~Pfitzner}, \bibinfo{author}{A.~Garas},
  \bibinfo{author}{C.~J. Tessone}, \bibinfo{author}{F.~Schweitzer},
\newblock \bibinfo{title}{Causality-driven slow-down and speed-up of diffusion
  in non-{M}arkovian temporal networks},
\newblock \bibinfo{journal}{Nat. Commun.} \bibinfo{volume}{5}
  (\bibinfo{year}{2014}) \bibinfo{pages}{1--9}.
\bibitem[{Masuda et~al.(2017)Masuda, Porter, and Lambiotte}]{masuda2017random}
\bibinfo{author}{N.~Masuda}, \bibinfo{author}{M.~A. Porter},
  \bibinfo{author}{R.~Lambiotte},
\newblock \bibinfo{title}{Random walks and diffusion on networks},
\newblock \bibinfo{journal}{Phys. Rep.} \bibinfo{volume}{716}
  (\bibinfo{year}{2017}) \bibinfo{pages}{1--58}.
\bibitem[{Valdano et~al.(2015)Valdano, Ferreri, Poletto, and
  Colizza}]{valdano2015analytical}
\bibinfo{author}{E.~Valdano}, \bibinfo{author}{L.~Ferreri},
  \bibinfo{author}{C.~Poletto}, \bibinfo{author}{V.~Colizza},
\newblock \bibinfo{title}{Analytical computation of the epidemic threshold on
  temporal networks},
\newblock \bibinfo{journal}{Phys. Rev. X} \bibinfo{volume}{5}
  (\bibinfo{year}{2015}) \bibinfo{pages}{021005}.
\bibitem[{Rocha et~al.(2011)Rocha, Liljeros, and Holme}]{rocha2011simulated}
\bibinfo{author}{L.~E. Rocha}, \bibinfo{author}{F.~Liljeros},
  \bibinfo{author}{P.~Holme},
\newblock \bibinfo{title}{Simulated epidemics in an empirical spatiotemporal
  network of 50,185 sexual contacts},
\newblock \bibinfo{journal}{PLOS Comput. Biol.} \bibinfo{volume}{7}
  (\bibinfo{year}{2011}) \bibinfo{pages}{e1001109}.
\bibitem[{Li et~al.(2017)Li, Cornelius, Liu, Wang, and
  Barab{\'a}si}]{li2017fundamental}
\bibinfo{author}{A.~Li}, \bibinfo{author}{S.~P. Cornelius},
  \bibinfo{author}{Y.-Y. Liu}, \bibinfo{author}{L.~Wang},
  \bibinfo{author}{A.-L. Barab{\'a}si},
\newblock \bibinfo{title}{The fundamental advantages of temporal networks},
\newblock \bibinfo{journal}{Science} \bibinfo{volume}{358}
  (\bibinfo{year}{2017}) \bibinfo{pages}{1042--1046}.
\bibitem[{Perc and Szolnoki(2010)}]{perc2010coevolutionary}
\bibinfo{author}{M.~Perc}, \bibinfo{author}{A.~Szolnoki},
\newblock \bibinfo{title}{Coevolutionary games---{A} mini review},
\newblock \bibinfo{journal}{BioSystems} \bibinfo{volume}{99}
  (\bibinfo{year}{2010}) \bibinfo{pages}{109--125}.
\bibitem[{Zimmermann et~al.(2004)Zimmermann, Egu{\'\i}luz, and
  San~Miguel}]{zimmermann2004coevolution}
\bibinfo{author}{M.~G. Zimmermann}, \bibinfo{author}{V.~M. Egu{\'\i}luz},
  \bibinfo{author}{M.~San~Miguel},
\newblock \bibinfo{title}{Coevolution of dynamical states and interactions in
  dynamic networks},
\newblock \bibinfo{journal}{Phys. Rev. E} \bibinfo{volume}{69}
  (\bibinfo{year}{2004}) \bibinfo{pages}{065102}.
\bibitem[{Egu{\'\i}luz et~al.(2005)Egu{\'\i}luz, Zimmermann, Cela-Conde, and
  Miguel}]{eguiluz2005cooperation}
\bibinfo{author}{V.~M. Egu{\'\i}luz}, \bibinfo{author}{M.~G. Zimmermann},
  \bibinfo{author}{C.~J. Cela-Conde}, \bibinfo{author}{M.~S. Miguel},
\newblock \bibinfo{title}{Cooperation and the emergence of role differentiation
  in the dynamics of social networks},
\newblock \bibinfo{journal}{Am. J. Sociol.} \bibinfo{volume}{110}
  (\bibinfo{year}{2005}) \bibinfo{pages}{977--1008}.
\bibitem[{Fu et~al.(2009)Fu, Wu, and Wang}]{fu2009partner}
\bibinfo{author}{F.~Fu}, \bibinfo{author}{T.~Wu}, \bibinfo{author}{L.~Wang},
\newblock \bibinfo{title}{Partner switching stabilizes cooperation in
  coevolutionary prisoner’s dilemma},
\newblock \bibinfo{journal}{Phys. Rev. E} \bibinfo{volume}{79}
  (\bibinfo{year}{2009}) \bibinfo{pages}{036101}.
\bibitem[{Du and Fu(2011)}]{du2011partner}
\bibinfo{author}{F.~Du}, \bibinfo{author}{F.~Fu},
\newblock \bibinfo{title}{Partner selection shapes the strategic and
  topological evolution of cooperation},
\newblock \bibinfo{journal}{Dyn. Games Appl.} \bibinfo{volume}{1}
  (\bibinfo{year}{2011}) \bibinfo{pages}{354--369}.
\bibitem[{Melamed et~al.(2020)Melamed, Sweitzer, Simpson, Abernathy, Harrell,
  and Munn}]{melamed2020homophily}
\bibinfo{author}{D.~Melamed}, \bibinfo{author}{M.~Sweitzer},
  \bibinfo{author}{B.~Simpson}, \bibinfo{author}{J.~Z. Abernathy},
  \bibinfo{author}{A.~Harrell}, \bibinfo{author}{C.~W. Munn},
\newblock \bibinfo{title}{Homophily and segregation in cooperative networks},
\newblock \bibinfo{journal}{Am. J. Sociol.} \bibinfo{volume}{125}
  (\bibinfo{year}{2020}) \bibinfo{pages}{1084--1127}.
\bibitem[{Cardillo et~al.(2014)Cardillo, Petri, Nicosia, Sinatra,
  G{\'o}mez-Gardenes, and Latora}]{cardillo2014evolutionary}
\bibinfo{author}{A.~Cardillo}, \bibinfo{author}{G.~Petri},
  \bibinfo{author}{V.~Nicosia}, \bibinfo{author}{R.~Sinatra},
  \bibinfo{author}{J.~G{\'o}mez-Gardenes}, \bibinfo{author}{V.~Latora},
\newblock \bibinfo{title}{Evolutionary dynamics of time-resolved social
  interactions},
\newblock \bibinfo{journal}{Phys. Rev. E} \bibinfo{volume}{90}
  (\bibinfo{year}{2014}) \bibinfo{pages}{052825}.
\bibitem[{Li et~al.(2020)Li, Zhou, Su, Cornelius, Liu, Wang, and
  Levin}]{li2020evolution}
\bibinfo{author}{A.~Li}, \bibinfo{author}{L.~Zhou}, \bibinfo{author}{Q.~Su},
  \bibinfo{author}{S.~P. Cornelius}, \bibinfo{author}{Y.-Y. Liu},
  \bibinfo{author}{L.~Wang}, \bibinfo{author}{S.~A. Levin},
\newblock \bibinfo{title}{Evolution of cooperation on temporal networks},
\newblock \bibinfo{journal}{Nat. Commun.} \bibinfo{volume}{11}
  (\bibinfo{year}{2020}) \bibinfo{pages}{2259}.
\bibitem[{Perra et~al.(2012)Perra, Gon{\c{c}}alves, Pastor-Satorras, and
  Vespignani}]{perra2012activity}
\bibinfo{author}{N.~Perra}, \bibinfo{author}{B.~Gon{\c{c}}alves},
  \bibinfo{author}{R.~Pastor-Satorras}, \bibinfo{author}{A.~Vespignani},
\newblock \bibinfo{title}{Activity driven modeling of time varying networks},
\newblock \bibinfo{journal}{Sci. Rep.} \bibinfo{volume}{2}
  (\bibinfo{year}{2012}) \bibinfo{pages}{469}.
\bibitem[{Alvarez-Rodriguez et~al.(2021)Alvarez-Rodriguez, Battiston,
  de~Arruda, Moreno, Perc, and Latora}]{alvarez2021evolutionary}
\bibinfo{author}{U.~Alvarez-Rodriguez}, \bibinfo{author}{F.~Battiston},
  \bibinfo{author}{G.~F. de~Arruda}, \bibinfo{author}{Y.~Moreno},
  \bibinfo{author}{M.~Perc}, \bibinfo{author}{V.~Latora},
\newblock \bibinfo{title}{Evolutionary dynamics of higher-order interactions in
  social networks},
\newblock \bibinfo{journal}{Nat. Hum. Behav.} \bibinfo{volume}{5}
  (\bibinfo{year}{2021}) \bibinfo{pages}{586--595}.
\bibitem[{Santos et~al.(2008)Santos, Santos, and Pacheco}]{santos2008social}
\bibinfo{author}{F.~C. Santos}, \bibinfo{author}{M.~D. Santos},
  \bibinfo{author}{J.~M. Pacheco},
\newblock \bibinfo{title}{Social diversity promotes the emergence of
  cooperation in public goods games},
\newblock \bibinfo{journal}{Nature} \bibinfo{volume}{454}
  (\bibinfo{year}{2008}) \bibinfo{pages}{213--216}.
\bibitem[{Szolnoki et~al.(2009)Szolnoki, Perc, and
  Szab{\'o}}]{szolnoki2009topology}
\bibinfo{author}{A.~Szolnoki}, \bibinfo{author}{M.~Perc},
  \bibinfo{author}{G.~Szab{\'o}},
\newblock \bibinfo{title}{Topology-independent impact of noise on cooperation
  in spatial public goods games},
\newblock \bibinfo{journal}{Phys. Rev. E} \bibinfo{volume}{80}
  (\bibinfo{year}{2009}) \bibinfo{pages}{056109}.
\bibitem[{Battiston et~al.(2020)Battiston, Cencetti, Iacopini, Latora, Lucas,
  Patania, Young, and Petri}]{battiston2020networks}
\bibinfo{author}{F.~Battiston}, \bibinfo{author}{G.~Cencetti},
  \bibinfo{author}{I.~Iacopini}, \bibinfo{author}{V.~Latora},
  \bibinfo{author}{M.~Lucas}, \bibinfo{author}{A.~Patania},
  \bibinfo{author}{J.-G. Young}, \bibinfo{author}{G.~Petri},
\newblock \bibinfo{title}{Networks beyond pairwise interactions: structure and
  dynamics},
\newblock \bibinfo{journal}{Phys. Rep.} \bibinfo{volume}{874}
  (\bibinfo{year}{2020}) \bibinfo{pages}{1--92}.
\bibitem[{Hauert and Szabo(2003)}]{hauert2003prisoner}
\bibinfo{author}{C.~Hauert}, \bibinfo{author}{G.~Szabo},
\newblock \bibinfo{title}{Prisoner's dilemma and public goods games in
  different geometries: compulsory versus voluntary interactions},
\newblock \bibinfo{journal}{Complexity} \bibinfo{volume}{8}
  (\bibinfo{year}{2003}) \bibinfo{pages}{31--38}.
\bibitem[{Wu et~al.(2019)Wu, Wang, and Evans}]{wu2019large}
\bibinfo{author}{L.~Wu}, \bibinfo{author}{D.~Wang}, \bibinfo{author}{J.~A.
  Evans},
\newblock \bibinfo{title}{Large teams develop and small teams disrupt science
  and technology},
\newblock \bibinfo{journal}{Nature} \bibinfo{volume}{566}
  (\bibinfo{year}{2019}) \bibinfo{pages}{378--382}.
\bibitem[{Roca et~al.(2009)Roca, Cuesta, and
  S{\'a}nchez}]{roca2009evolutionary}
\bibinfo{author}{C.~P. Roca}, \bibinfo{author}{J.~A. Cuesta},
  \bibinfo{author}{A.~S{\'a}nchez},
\newblock \bibinfo{title}{Evolutionary game theory: {T}emporal and spatial
  effects beyond replicator dynamics},
\newblock \bibinfo{journal}{Phys. Life Rev.} \bibinfo{volume}{6}
  (\bibinfo{year}{2009}) \bibinfo{pages}{208--249}.
\bibitem[{Nowak et~al.(2004)Nowak, Sasaki, Taylor, and
  Fudenberg}]{nowak2004emergence}
\bibinfo{author}{M.~A. Nowak}, \bibinfo{author}{A.~Sasaki},
  \bibinfo{author}{C.~Taylor}, \bibinfo{author}{D.~Fudenberg},
\newblock \bibinfo{title}{Emergence of cooperation and evolutionary stability
  in finite populations},
\newblock \bibinfo{journal}{Nature} \bibinfo{volume}{428}
  (\bibinfo{year}{2004}) \bibinfo{pages}{646--650}.
\bibitem[{Hindersin et~al.(2019)Hindersin, Wu, Traulsen, and
  Garc{\'\i}a}]{hindersin2019computation}
\bibinfo{author}{L.~Hindersin}, \bibinfo{author}{B.~Wu},
  \bibinfo{author}{A.~Traulsen}, \bibinfo{author}{J.~Garc{\'\i}a},
\newblock \bibinfo{title}{Computation and simulation of evolutionary game
  dynamics in finite populations},
\newblock \bibinfo{journal}{Sci. Rep.} \bibinfo{volume}{9}
  (\bibinfo{year}{2019}) \bibinfo{pages}{6946}.
\bibitem[{Lieberman et~al.(2005)Lieberman, Hauert, and
  Nowak}]{lieberman2005evolutionary}
\bibinfo{author}{E.~Lieberman}, \bibinfo{author}{C.~Hauert},
  \bibinfo{author}{M.~A. Nowak},
\newblock \bibinfo{title}{Evolutionary dynamics on graphs},
\newblock \bibinfo{journal}{Nature} \bibinfo{volume}{433}
  (\bibinfo{year}{2005}) \bibinfo{pages}{312--316}.
\bibitem[{Santos et~al.(2006)Santos, Pacheco, and
  Lenaerts}]{santos2006evolutionary}
\bibinfo{author}{F.~C. Santos}, \bibinfo{author}{J.~M. Pacheco},
  \bibinfo{author}{T.~Lenaerts},
\newblock \bibinfo{title}{Evolutionary dynamics of social dilemmas in
  structured heterogeneous populations},
\newblock \bibinfo{journal}{Proc. Natl. Acad. Sci. USA} \bibinfo{volume}{103}
  (\bibinfo{year}{2006}) \bibinfo{pages}{3490--3494}.
\bibitem[{Maciejewski et~al.(2014)Maciejewski, Fu, and
  Hauert}]{maciejewski2014evolutionary}
\bibinfo{author}{W.~Maciejewski}, \bibinfo{author}{F.~Fu},
  \bibinfo{author}{C.~Hauert},
\newblock \bibinfo{title}{Evolutionary game dynamics in populations with
  heterogenous structures},
\newblock \bibinfo{journal}{PLOS Comput. Biol.} \bibinfo{volume}{10}
  (\bibinfo{year}{2014}) \bibinfo{pages}{e1003567}.
\bibitem[{Wu et~al.(2013)Wu, Gokhale, van Veelen, Wang, and
  Traulsen}]{wu2013interpretations}
\bibinfo{author}{B.~Wu}, \bibinfo{author}{C.~S. Gokhale},
  \bibinfo{author}{M.~van Veelen}, \bibinfo{author}{L.~Wang},
  \bibinfo{author}{A.~Traulsen},
\newblock \bibinfo{title}{Interpretations arising from {W}rightian and
  {M}althusian fitness under strong frequency dependent selection},
\newblock \bibinfo{journal}{Ecol. Evol.} \bibinfo{volume}{3}
  (\bibinfo{year}{2013}) \bibinfo{pages}{1276--1280}.
\bibitem[{Doebeli et~al.(2017)Doebeli, Ispolatov, and Simon}]{doebeli2017point}
\bibinfo{author}{M.~Doebeli}, \bibinfo{author}{Y.~Ispolatov},
  \bibinfo{author}{B.~Simon},
\newblock \bibinfo{title}{Point of view: Towards a mechanistic foundation of
  evolutionary theory},
\newblock \bibinfo{journal}{eLife} \bibinfo{volume}{6} (\bibinfo{year}{2017})
  \bibinfo{pages}{e23804}.
\bibitem[{Wu et~al.(2013)Wu, Garc{\'\i}a, Hauert, and
  Traulsen}]{wu2013extrapolating}
\bibinfo{author}{B.~Wu}, \bibinfo{author}{J.~Garc{\'\i}a},
  \bibinfo{author}{C.~Hauert}, \bibinfo{author}{A.~Traulsen},
\newblock \bibinfo{title}{Extrapolating weak selection in evolutionary games},
\newblock \bibinfo{journal}{PLOS Comput. Biol.} \bibinfo{volume}{9}
  (\bibinfo{year}{2013}) \bibinfo{pages}{e1003381}.
\bibitem[{Wu et~al.(2015)Wu, Bauer, Galla, and Traulsen}]{wu2015fitness}
\bibinfo{author}{B.~Wu}, \bibinfo{author}{B.~Bauer},
  \bibinfo{author}{T.~Galla}, \bibinfo{author}{A.~Traulsen},
\newblock \bibinfo{title}{Fitness-based models and pairwise comparison models
  of evolutionary games are typically different---even in unstructured
  populations},
\newblock \bibinfo{journal}{New J. Phys.} \bibinfo{volume}{17}
  (\bibinfo{year}{2015}) \bibinfo{pages}{023043}.
\bibitem[{S{\'a}nchez(2015)}]{sanchez2015theory}
\bibinfo{author}{A.~S{\'a}nchez},
\newblock \bibinfo{title}{Theory must be informed by experiments (and back):
  comment on ``{U}niversal scaling for the dilemma strength in evolutionary
  games'', by {Z. Wang} et al.},
\newblock \bibinfo{journal}{Phys. Life Rev.} \bibinfo{volume}{14}
  (\bibinfo{year}{2015}) \bibinfo{pages}{52--53}.
\bibitem[{S{\'a}nchez(2018)}]{sanchez2018physics}
\bibinfo{author}{A.~S{\'a}nchez},
\newblock \bibinfo{title}{Physics of human cooperation: experimental evidence
  and theoretical models},
\newblock \bibinfo{journal}{J. Stat. Mech. Theory Exp.} \bibinfo{volume}{2018}
  (\bibinfo{year}{2018}) \bibinfo{pages}{024001}.
\bibitem[{Dreber et~al.(2008)Dreber, Rand, Fudenberg, and
  Nowak}]{dreber2008winners}
\bibinfo{author}{A.~Dreber}, \bibinfo{author}{D.~G. Rand},
  \bibinfo{author}{D.~Fudenberg}, \bibinfo{author}{M.~A. Nowak},
\newblock \bibinfo{title}{Winners don't punish},
\newblock \bibinfo{journal}{Nature} \bibinfo{volume}{452}
  (\bibinfo{year}{2008}) \bibinfo{pages}{348--351}.
\bibitem[{Wu et~al.(2009)Wu, Zhang, Zhou, He, Zheng, Cressman, and
  Tao}]{wu2009costly}
\bibinfo{author}{J.-J. Wu}, \bibinfo{author}{B.-Y. Zhang},
  \bibinfo{author}{Z.-X. Zhou}, \bibinfo{author}{Q.-Q. He},
  \bibinfo{author}{X.-D. Zheng}, \bibinfo{author}{R.~Cressman},
  \bibinfo{author}{Y.~Tao},
\newblock \bibinfo{title}{Costly punishment does not always increase
  cooperation},
\newblock \bibinfo{journal}{Proc. Natl. Acad. Sci. USA} \bibinfo{volume}{106}
  (\bibinfo{year}{2009}) \bibinfo{pages}{17448--17451}.
\bibitem[{Li et~al.(2018)Li, Jusup, Wang, Li, Shi, Podobnik, Stanley, Havlin,
  and Boccaletti}]{li2018punishment}
\bibinfo{author}{X.~Li}, \bibinfo{author}{M.~Jusup}, \bibinfo{author}{Z.~Wang},
  \bibinfo{author}{H.~Li}, \bibinfo{author}{L.~Shi},
  \bibinfo{author}{B.~Podobnik}, \bibinfo{author}{H.~E. Stanley},
  \bibinfo{author}{S.~Havlin}, \bibinfo{author}{S.~Boccaletti},
\newblock \bibinfo{title}{Punishment diminishes the benefits of network
  reciprocity in social dilemma experiments},
\newblock \bibinfo{journal}{Proc. Natl. Acad. Sci. USA} \bibinfo{volume}{115}
  (\bibinfo{year}{2018}) \bibinfo{pages}{30--35}.
\bibitem[{Gracia-L{\'a}zaro et~al.(2012)Gracia-L{\'a}zaro, Ferrer, Ruiz,
  Taranc{\'o}n, Cuesta, S{\'a}nchez, and Moreno}]{gracia2012heterogeneous}
\bibinfo{author}{C.~Gracia-L{\'a}zaro}, \bibinfo{author}{A.~Ferrer},
  \bibinfo{author}{G.~Ruiz}, \bibinfo{author}{A.~Taranc{\'o}n},
  \bibinfo{author}{J.~A. Cuesta}, \bibinfo{author}{A.~S{\'a}nchez},
  \bibinfo{author}{Y.~Moreno},
\newblock \bibinfo{title}{Heterogeneous networks do not promote cooperation
  when humans play a {P}risoner's {D}ilemma},
\newblock \bibinfo{journal}{Proc. Natl. Acad. Sci. USA} \bibinfo{volume}{109}
  (\bibinfo{year}{2012}) \bibinfo{pages}{12922--12926}.
\bibitem[{Gruji{\'c} et~al.(2014)Gruji{\'c}, Gracia-L{\'a}zaro, Milinski,
  Semmann, Traulsen, Cuesta, Moreno, and S{\'a}nchez}]{grujic2014comparative}
\bibinfo{author}{J.~Gruji{\'c}}, \bibinfo{author}{C.~Gracia-L{\'a}zaro},
  \bibinfo{author}{M.~Milinski}, \bibinfo{author}{D.~Semmann},
  \bibinfo{author}{A.~Traulsen}, \bibinfo{author}{J.~A. Cuesta},
  \bibinfo{author}{Y.~Moreno}, \bibinfo{author}{A.~S{\'a}nchez},
\newblock \bibinfo{title}{A comparative analysis of spatial {P}risoner's
  {D}ilemma experiments: {C}onditional cooperation and payoff irrelevance},
\newblock \bibinfo{journal}{Sci. Rep.} \bibinfo{volume}{4}
  (\bibinfo{year}{2014}) \bibinfo{pages}{4615}.
\bibitem[{Rand et~al.(2014)Rand, Nowak, Fowler, and
  Christakis}]{rand2014static}
\bibinfo{author}{D.~G. Rand}, \bibinfo{author}{M.~A. Nowak},
  \bibinfo{author}{J.~H. Fowler}, \bibinfo{author}{N.~A. Christakis},
\newblock \bibinfo{title}{Static network structure can stabilize human
  cooperation},
\newblock \bibinfo{journal}{Proc. Natl. Acad. Sci. USA} \bibinfo{volume}{111}
  (\bibinfo{year}{2014}) \bibinfo{pages}{17093--17098}.
\bibitem[{Szab{\'o} and T{\H{o}}ke(1998)}]{szabo1998evolutionary}
\bibinfo{author}{G.~Szab{\'o}}, \bibinfo{author}{C.~T{\H{o}}ke},
\newblock \bibinfo{title}{Evolutionary prisoner’s dilemma game on a square
  lattice},
\newblock \bibinfo{journal}{Phys. Rev. E} \bibinfo{volume}{58}
  (\bibinfo{year}{1998}) \bibinfo{pages}{69}.
\bibitem[{Traulsen et~al.(2006)Traulsen, Nowak, and
  Pacheco}]{traulsen2006stochastic}
\bibinfo{author}{A.~Traulsen}, \bibinfo{author}{M.~A. Nowak},
  \bibinfo{author}{J.~M. Pacheco},
\newblock \bibinfo{title}{Stochastic dynamics of invasion and fixation},
\newblock \bibinfo{journal}{Phys. Rev. E} \bibinfo{volume}{74}
  (\bibinfo{year}{2006}) \bibinfo{pages}{011909}.
\bibitem[{Apesteguia et~al.(2007)Apesteguia, Huck, and
  Oechssler}]{apesteguia2007imitation}
\bibinfo{author}{J.~Apesteguia}, \bibinfo{author}{S.~Huck},
  \bibinfo{author}{J.~Oechssler},
\newblock \bibinfo{title}{Imitation---theory and experimental evidence},
\newblock \bibinfo{journal}{J. Econ. Theory} \bibinfo{volume}{136}
  (\bibinfo{year}{2007}) \bibinfo{pages}{217--235}.
\bibitem[{Traulsen et~al.(2010)Traulsen, Semmann, Sommerfeld, Krambeck, and
  Milinski}]{traulsen2010human}
\bibinfo{author}{A.~Traulsen}, \bibinfo{author}{D.~Semmann},
  \bibinfo{author}{R.~D. Sommerfeld}, \bibinfo{author}{H.-J. Krambeck},
  \bibinfo{author}{M.~Milinski},
\newblock \bibinfo{title}{Human strategy updating in evolutionary games},
\newblock \bibinfo{journal}{Proc. Natl. Acad. Sci. USA} \bibinfo{volume}{107}
  (\bibinfo{year}{2010}) \bibinfo{pages}{2962--2966}.
\bibitem[{Gruji{\'c} and Lenaerts(2020)}]{grujic2020people}
\bibinfo{author}{J.~Gruji{\'c}}, \bibinfo{author}{T.~Lenaerts},
\newblock \bibinfo{title}{Do people imitate when making decisions? {E}vidence
  from a spatial {P}risoner’s {D}ilemma experiment},
\newblock \bibinfo{journal}{R. Soc. Open Sci.} \bibinfo{volume}{7}
  (\bibinfo{year}{2020}) \bibinfo{pages}{200618}.
\bibitem[{Poncela-Casasnovas et~al.(2016)Poncela-Casasnovas,
  Guti{\'e}rrez-Roig, Gracia-L{\'a}zaro, Vicens, G{\'o}mez-Garde{\~n}es,
  Perell{\'o}, Moreno, Duch, and S{\'a}nchez}]{poncela2016humans}
\bibinfo{author}{J.~Poncela-Casasnovas},
  \bibinfo{author}{M.~Guti{\'e}rrez-Roig},
  \bibinfo{author}{C.~Gracia-L{\'a}zaro}, \bibinfo{author}{J.~Vicens},
  \bibinfo{author}{J.~G{\'o}mez-Garde{\~n}es},
  \bibinfo{author}{J.~Perell{\'o}}, \bibinfo{author}{Y.~Moreno},
  \bibinfo{author}{J.~Duch}, \bibinfo{author}{A.~S{\'a}nchez},
\newblock \bibinfo{title}{Humans display a reduced set of consistent behavioral
  phenotypes in dyadic games},
\newblock \bibinfo{journal}{Sci. Adv.} \bibinfo{volume}{2}
  (\bibinfo{year}{2016}) \bibinfo{pages}{e1600451}.
\bibitem[{Peysakhovich et~al.(2014)Peysakhovich, Nowak, and
  Rand}]{peysakhovich2014humans}
\bibinfo{author}{A.~Peysakhovich}, \bibinfo{author}{M.~A. Nowak},
  \bibinfo{author}{D.~G. Rand},
\newblock \bibinfo{title}{Humans display a `cooperative phenotype' that is
  domain general and temporally stable},
\newblock \bibinfo{journal}{Nat. Commun.} \bibinfo{volume}{5}
  (\bibinfo{year}{2014}) \bibinfo{pages}{4939}.
\bibitem[{Axelrod and Hamilton(1981)}]{axelrod1981evolution}
\bibinfo{author}{R.~Axelrod}, \bibinfo{author}{W.~D. Hamilton},
\newblock \bibinfo{title}{The evolution of cooperation},
\newblock \bibinfo{journal}{Science} \bibinfo{volume}{211}
  (\bibinfo{year}{1981}) \bibinfo{pages}{1390--1396}.
\bibitem[{Bendor and Swistak(1997)}]{bendor1997evolutionary}
\bibinfo{author}{J.~Bendor}, \bibinfo{author}{P.~Swistak},
\newblock \bibinfo{title}{The evolutionary stability of cooperation},
\newblock \bibinfo{journal}{Am. Political Sci. Rev.} \bibinfo{volume}{91}
  (\bibinfo{year}{1997}) \bibinfo{pages}{290--307}.
\bibitem[{Keser and Van~Winden(2000)}]{keser2000conditional}
\bibinfo{author}{C.~Keser}, \bibinfo{author}{F.~Van~Winden},
\newblock \bibinfo{title}{Conditional cooperation and voluntary contributions
  to public goods},
\newblock \bibinfo{journal}{Scand. J. Econ.} \bibinfo{volume}{102}
  (\bibinfo{year}{2000}) \bibinfo{pages}{23--39}.
\bibitem[{Fischbacher et~al.(2001)Fischbacher, G{\"a}chter, and
  Fehr}]{fischbacher2001people}
\bibinfo{author}{U.~Fischbacher}, \bibinfo{author}{S.~G{\"a}chter},
  \bibinfo{author}{E.~Fehr},
\newblock \bibinfo{title}{Are people conditionally cooperative? {E}vidence from
  a public goods experiment},
\newblock \bibinfo{journal}{Econ. Lett.} \bibinfo{volume}{71}
  (\bibinfo{year}{2001}) \bibinfo{pages}{397--404}.
\bibitem[{Gruji{\'c} et~al.(2010)Gruji{\'c}, Fosco, Araujo, Cuesta, and
  S{\'a}nchez}]{grujic2010social}
\bibinfo{author}{J.~Gruji{\'c}}, \bibinfo{author}{C.~Fosco},
  \bibinfo{author}{L.~Araujo}, \bibinfo{author}{J.~A. Cuesta},
  \bibinfo{author}{A.~S{\'a}nchez},
\newblock \bibinfo{title}{Social experiments in the mesoscale: {H}umans playing
  a spatial prisoner's dilemma},
\newblock \bibinfo{journal}{PLOS ONE} \bibinfo{volume}{5}
  (\bibinfo{year}{2010}) \bibinfo{pages}{e13749}.
\bibitem[{Shi et~al.(2020)Shi, Romi\'{c}, Ma, Wang, Podobnik, Stanley, Holme,
  and Jusup}]{shi2020freedom}
\bibinfo{author}{L.~Shi}, \bibinfo{author}{I.~Romi\'{c}},
  \bibinfo{author}{Y.~Ma}, \bibinfo{author}{Z.~Wang},
  \bibinfo{author}{B.~Podobnik}, \bibinfo{author}{H.~E. Stanley},
  \bibinfo{author}{P.~Holme}, \bibinfo{author}{M.~Jusup},
\newblock \bibinfo{title}{Freedom of choice adds value to public goods},
\newblock \bibinfo{journal}{Proc. Natl. Acad. Sci. USA} \bibinfo{volume}{117}
  (\bibinfo{year}{2020}) \bibinfo{pages}{17516--17521}.
\bibitem[{Guti{\'e}rrez-Roig et~al.(2014)Guti{\'e}rrez-Roig, Gracia-L{\'a}zaro,
  Perell{\'o}, Moreno, and S{\'a}nchez}]{gutierrez2014transition}
\bibinfo{author}{M.~Guti{\'e}rrez-Roig},
  \bibinfo{author}{C.~Gracia-L{\'a}zaro}, \bibinfo{author}{J.~Perell{\'o}},
  \bibinfo{author}{Y.~Moreno}, \bibinfo{author}{A.~S{\'a}nchez},
\newblock \bibinfo{title}{Transition from reciprocal cooperation to persistent
  behaviour in social dilemmas at the end of adolescence},
\newblock \bibinfo{journal}{Nat. Commun.} \bibinfo{volume}{5}
  (\bibinfo{year}{2014}) \bibinfo{pages}{4362}.
\bibitem[{Horita et~al.(2017)Horita, Takezawa, Inukai, Kita, and
  Masuda}]{horita2017reinforcement}
\bibinfo{author}{Y.~Horita}, \bibinfo{author}{M.~Takezawa},
  \bibinfo{author}{K.~Inukai}, \bibinfo{author}{T.~Kita},
  \bibinfo{author}{N.~Masuda},
\newblock \bibinfo{title}{Reinforcement learning accounts for moody conditional
  cooperation behavior: experimental results},
\newblock \bibinfo{journal}{Sci. Rep.} \bibinfo{volume}{7}
  (\bibinfo{year}{2017}) \bibinfo{pages}{1--10}.
\bibitem[{Tversky(1972)}]{tversky1972elimination}
\bibinfo{author}{A.~Tversky},
\newblock \bibinfo{title}{Elimination by aspects: {A} theory of choice.},
\newblock \bibinfo{journal}{Psychol. Rev.} \bibinfo{volume}{79}
  (\bibinfo{year}{1972}) \bibinfo{pages}{281--299}.
\bibitem[{Kahneman and Tversky(1979)}]{kahneman1979prospect}
\bibinfo{author}{D.~Kahneman}, \bibinfo{author}{A.~Tversky},
\newblock \bibinfo{title}{Prospect theory: {A}n analysis of decision under
  risk},
\newblock \bibinfo{journal}{Econometrica} \bibinfo{volume}{47}
  (\bibinfo{year}{1979}) \bibinfo{pages}{363--391}.
\bibitem[{Schoemaker(1982)}]{schoemaker1982expected}
\bibinfo{author}{P.~J. Schoemaker},
\newblock \bibinfo{title}{The expected utility model: Its variants, purposes,
  evidence and limitations},
\newblock \bibinfo{journal}{J. Econ. Lit.} \bibinfo{volume}{20}
  (\bibinfo{year}{1982}) \bibinfo{pages}{529--563}.
\bibitem[{Wang et~al.(2017)Wang, Jusup, Wang, Shi, Iwasa, Moreno, and
  Kurths}]{wang2017onymity}
\bibinfo{author}{Z.~Wang}, \bibinfo{author}{M.~Jusup}, \bibinfo{author}{R.-W.
  Wang}, \bibinfo{author}{L.~Shi}, \bibinfo{author}{Y.~Iwasa},
  \bibinfo{author}{Y.~Moreno}, \bibinfo{author}{J.~Kurths},
\newblock \bibinfo{title}{Onymity promotes cooperation in social dilemma
  experiments},
\newblock \bibinfo{journal}{Sci. Adv.} \bibinfo{volume}{3}
  (\bibinfo{year}{2017}) \bibinfo{pages}{e1601444}.
\bibitem[{Hoffman et~al.(1996)Hoffman, McCabe, and Smith}]{hoffman1996social}
\bibinfo{author}{E.~Hoffman}, \bibinfo{author}{K.~McCabe},
  \bibinfo{author}{V.~L. Smith},
\newblock \bibinfo{title}{Social distance and other-regarding behavior in
  dictator games},
\newblock \bibinfo{journal}{Am. Econ. Rev.} \bibinfo{volume}{86}
  (\bibinfo{year}{1996}) \bibinfo{pages}{653--660}.
\bibitem[{Wang et~al.(2018)Wang, Jusup, Shi, Lee, Iwasa, and
  Boccaletti}]{wang2018exploiting}
\bibinfo{author}{Z.~Wang}, \bibinfo{author}{M.~Jusup},
  \bibinfo{author}{L.~Shi}, \bibinfo{author}{J.-H. Lee},
  \bibinfo{author}{Y.~Iwasa}, \bibinfo{author}{S.~Boccaletti},
\newblock \bibinfo{title}{Exploiting a cognitive bias promotes cooperation in
  social dilemma experiments},
\newblock \bibinfo{journal}{Nat. Commun.} \bibinfo{volume}{9}
  (\bibinfo{year}{2018}) \bibinfo{pages}{2954}.
\bibitem[{Pettibone and Wedell(2000)}]{pettibone2000examining}
\bibinfo{author}{J.~C. Pettibone}, \bibinfo{author}{D.~H. Wedell},
\newblock \bibinfo{title}{Examining models of nondominated decoy effects across
  judgment and choice},
\newblock \bibinfo{journal}{Organ. Behav. Hum. Decis. Process.}
  \bibinfo{volume}{81} (\bibinfo{year}{2000}) \bibinfo{pages}{300--328}.
\bibitem[{Group et~al.(2014)Group, Fawcett, Fallenstein, Higginson, Houston,
  Mallpress, Trimmer, and McNamara}]{group2014evolution}
\bibinfo{author}{T.~M. A.~D. Group}, \bibinfo{author}{T.~W. Fawcett},
  \bibinfo{author}{B.~Fallenstein}, \bibinfo{author}{A.~D. Higginson},
  \bibinfo{author}{A.~I. Houston}, \bibinfo{author}{D.~E. Mallpress},
  \bibinfo{author}{P.~C. Trimmer}, \bibinfo{author}{J.~M. McNamara},
\newblock \bibinfo{title}{The evolution of decision rules in complex
  environments},
\newblock \bibinfo{journal}{Trends Cogn. Sci.} \bibinfo{volume}{18}
  (\bibinfo{year}{2014}) \bibinfo{pages}{153--161}.
\bibitem[{Rand et~al.(2011)Rand, Arbesman, and Christakis}]{rand2011dynamic}
\bibinfo{author}{D.~G. Rand}, \bibinfo{author}{S.~Arbesman},
  \bibinfo{author}{N.~A. Christakis},
\newblock \bibinfo{title}{Dynamic social networks promote cooperation in
  experiments with humans},
\newblock \bibinfo{journal}{Proc. Natl. Acad. Sci. USA} \bibinfo{volume}{108}
  (\bibinfo{year}{2011}) \bibinfo{pages}{19193--19198}.
\bibitem[{Gardner and West(2004)}]{gardner2004cooperation}
\bibinfo{author}{A.~Gardner}, \bibinfo{author}{S.~A. West},
\newblock \bibinfo{title}{Cooperation and punishment, especially in humans},
\newblock \bibinfo{journal}{Am. Nat.} \bibinfo{volume}{164}
  (\bibinfo{year}{2004}) \bibinfo{pages}{753--764}.
\bibitem[{Hauert et~al.(2007)Hauert, Traulsen, Brandt, Nowak, and
  Sigmund}]{hauert2007via}
\bibinfo{author}{C.~Hauert}, \bibinfo{author}{A.~Traulsen},
  \bibinfo{author}{H.~Brandt}, \bibinfo{author}{M.~A. Nowak},
  \bibinfo{author}{K.~Sigmund},
\newblock \bibinfo{title}{Via freedom to coercion: the emergence of costly
  punishment},
\newblock \bibinfo{journal}{Science} \bibinfo{volume}{316}
  (\bibinfo{year}{2007}) \bibinfo{pages}{1905--1907}.
\bibitem[{Melis and Semmann(2010)}]{melis2010human}
\bibinfo{author}{A.~P. Melis}, \bibinfo{author}{D.~Semmann},
\newblock \bibinfo{title}{How is human cooperation different?},
\newblock \bibinfo{journal}{Philos. Trans. R. Soc. B} \bibinfo{volume}{365}
  (\bibinfo{year}{2010}) \bibinfo{pages}{2663--2674}.
\bibitem[{Nikiforakis(2008)}]{nikiforakis2008punishment}
\bibinfo{author}{N.~Nikiforakis},
\newblock \bibinfo{title}{Punishment and counter-punishment in public good
  games: {C}an we really govern ourselves?},
\newblock \bibinfo{journal}{J. Public Econ.} \bibinfo{volume}{92}
  (\bibinfo{year}{2008}) \bibinfo{pages}{91--112}.
\bibitem[{Nikiforakis and Engelmann(2011)}]{nikiforakis2011altruistic}
\bibinfo{author}{N.~Nikiforakis}, \bibinfo{author}{D.~Engelmann},
\newblock \bibinfo{title}{Altruistic punishment and the threat of feuds},
\newblock \bibinfo{journal}{J. Econ. Behav. Organ.} \bibinfo{volume}{78}
  (\bibinfo{year}{2011}) \bibinfo{pages}{319--332}.
\bibitem[{Balafoutas et~al.(2014)Balafoutas, Grechenig, and
  Nikiforakis}]{balafoutas2014third}
\bibinfo{author}{L.~Balafoutas}, \bibinfo{author}{K.~Grechenig},
  \bibinfo{author}{N.~Nikiforakis},
\newblock \bibinfo{title}{Third-party punishment and counter-punishment in
  one-shot interactions},
\newblock \bibinfo{journal}{Econ. Lett.} \bibinfo{volume}{122}
  (\bibinfo{year}{2014}) \bibinfo{pages}{308--310}.
\bibitem[{Fehl et~al.(2011)Fehl, van~der Post, and
  Semmann}]{fehl2011coevolution}
\bibinfo{author}{K.~Fehl}, \bibinfo{author}{D.~J. van~der Post},
  \bibinfo{author}{D.~Semmann},
\newblock \bibinfo{title}{Co-evolution of behaviour and social network
  structure promotes human cooperation},
\newblock \bibinfo{journal}{Ecol. Lett.} \bibinfo{volume}{14}
  (\bibinfo{year}{2011}) \bibinfo{pages}{546--551}.
\bibitem[{Pepper and Smuts(2002)}]{pepper2002mechanism}
\bibinfo{author}{J.~W. Pepper}, \bibinfo{author}{B.~B. Smuts},
\newblock \bibinfo{title}{A mechanism for the evolution of altruism among
  nonkin: positive assortment through environmental feedback},
\newblock \bibinfo{journal}{Am. Nat.} \bibinfo{volume}{160}
  (\bibinfo{year}{2002}) \bibinfo{pages}{205--213}.
\bibitem[{Fletcher and Zwick(2006)}]{fletcher2006unifying}
\bibinfo{author}{J.~A. Fletcher}, \bibinfo{author}{M.~Zwick},
\newblock \bibinfo{title}{Unifying the theories of inclusive fitness and
  reciprocal altruism},
\newblock \bibinfo{journal}{Am. Nat.} \bibinfo{volume}{168}
  (\bibinfo{year}{2006}) \bibinfo{pages}{252--262}.
\bibitem[{Fletcher and Doebeli(2009)}]{fletcher2009simple}
\bibinfo{author}{J.~A. Fletcher}, \bibinfo{author}{M.~Doebeli},
\newblock \bibinfo{title}{A simple and general explanation for the evolution of
  altruism},
\newblock \bibinfo{journal}{Proc. R. Soc. B} \bibinfo{volume}{276}
  (\bibinfo{year}{2009}) \bibinfo{pages}{13--19}.
\bibitem[{Momeni et~al.(2013)Momeni, Waite, and Shou}]{momeni2013spatial}
\bibinfo{author}{B.~Momeni}, \bibinfo{author}{A.~J. Waite},
  \bibinfo{author}{W.~Shou},
\newblock \bibinfo{title}{Spatial self-organization favors heterotypic
  cooperation over cheating},
\newblock \bibinfo{journal}{eLife} \bibinfo{volume}{2} (\bibinfo{year}{2013})
  \bibinfo{pages}{e00960}.
\bibitem[{Klein et~al.(2014)Klein, Ratliff, Vianello, Adams~Jr, Bahn{\'\i}k,
  Bernstein, Bocian, Brandt, Brooks, Brumbaugh et~al.}]{klein2014investigating}
\bibinfo{author}{R.~A. Klein}, \bibinfo{author}{K.~A. Ratliff},
  \bibinfo{author}{M.~Vianello}, \bibinfo{author}{R.~B. Adams~Jr},
  \bibinfo{author}{{\v{S}}.~Bahn{\'\i}k}, \bibinfo{author}{M.~J. Bernstein},
  \bibinfo{author}{K.~Bocian}, \bibinfo{author}{M.~J. Brandt},
  \bibinfo{author}{B.~Brooks}, \bibinfo{author}{C.~C. Brumbaugh}, et~al.,
\newblock \bibinfo{title}{Investigating variation in replicability},
\newblock \bibinfo{journal}{Soc. Psychol.} \bibinfo{volume}{45}
  (\bibinfo{year}{2014}) \bibinfo{pages}{142--152}.
\bibitem[{Collaboration(2015)}]{open2015estimating}
\bibinfo{author}{O.~S. Collaboration},
\newblock \bibinfo{title}{Estimating the reproducibility of psychological
  science},
\newblock \bibinfo{journal}{Science} \bibinfo{volume}{349}
  (\bibinfo{year}{2015}) \bibinfo{pages}{aac4716}.
\bibitem[{Camerer et~al.(2018)Camerer, Dreber, Holzmeister, Ho, Huber,
  Johannesson, Kirchler, Nave, Nosek, Pfeiffer et~al.}]{camerer2018evaluating}
\bibinfo{author}{C.~F. Camerer}, \bibinfo{author}{A.~Dreber},
  \bibinfo{author}{F.~Holzmeister}, \bibinfo{author}{T.-H. Ho},
  \bibinfo{author}{J.~Huber}, \bibinfo{author}{M.~Johannesson},
  \bibinfo{author}{M.~Kirchler}, \bibinfo{author}{G.~Nave},
  \bibinfo{author}{B.~A. Nosek}, \bibinfo{author}{T.~Pfeiffer}, et~al.,
\newblock \bibinfo{title}{Evaluating the replicability of social science
  experiments in {N}ature and {S}cience between 2010 and 2015},
\newblock \bibinfo{journal}{Nat. Hum. Behav.} \bibinfo{volume}{2}
  (\bibinfo{year}{2018}) \bibinfo{pages}{637--644}.
\bibitem[{Klein et~al.(2018)Klein, Vianello, Hasselman, Adams, Adams~Jr, Alper,
  Aveyard, Axt, Babalola, Bahn{\'\i}k et~al.}]{klein2018many}
\bibinfo{author}{R.~A. Klein}, \bibinfo{author}{M.~Vianello},
  \bibinfo{author}{F.~Hasselman}, \bibinfo{author}{B.~G. Adams},
  \bibinfo{author}{R.~B. Adams~Jr}, \bibinfo{author}{S.~Alper},
  \bibinfo{author}{M.~Aveyard}, \bibinfo{author}{J.~R. Axt},
  \bibinfo{author}{M.~T. Babalola}, \bibinfo{author}{{\v{S}}.~Bahn{\'\i}k},
  et~al.,
\newblock \bibinfo{title}{Many {L}abs 2: {I}nvestigating variation in
  replicability across samples and settings},
\newblock \bibinfo{journal}{Adv. Methods Pract. Psychol. Sci.}
  \bibinfo{volume}{1} (\bibinfo{year}{2018}) \bibinfo{pages}{443--490}.
\bibitem[{Serra-Garcia and Gneezy(2021)}]{serra2021nonreplicable}
\bibinfo{author}{M.~Serra-Garcia}, \bibinfo{author}{U.~Gneezy},
\newblock \bibinfo{title}{Nonreplicable publications are cited more than
  replicable ones},
\newblock \bibinfo{journal}{Sci. Adv.} \bibinfo{volume}{7}
  (\bibinfo{year}{2021}) \bibinfo{pages}{eabd1705}.
\bibitem[{Nosek et~al.(2015)Nosek, Alter, Banks, Borsboom, Bowman, Breckler,
  Buck, Chambers, Chin, Christensen et~al.}]{nosek2015promoting}
\bibinfo{author}{B.~A. Nosek}, \bibinfo{author}{G.~Alter},
  \bibinfo{author}{G.~C. Banks}, \bibinfo{author}{D.~Borsboom},
  \bibinfo{author}{S.~D. Bowman}, \bibinfo{author}{S.~J. Breckler},
  \bibinfo{author}{S.~Buck}, \bibinfo{author}{C.~D. Chambers},
  \bibinfo{author}{G.~Chin}, \bibinfo{author}{G.~Christensen}, et~al.,
\newblock \bibinfo{title}{Promoting an open research culture},
\newblock \bibinfo{journal}{Science} \bibinfo{volume}{348}
  (\bibinfo{year}{2015}) \bibinfo{pages}{1422--1425}.
\bibitem[{Munaf{\`o} et~al.(2017)Munaf{\`o}, Nosek, Bishop, Button, Chambers,
  Du~Sert, Simonsohn, Wagenmakers, Ware, and Ioannidis}]{munafo2017manifesto}
\bibinfo{author}{M.~R. Munaf{\`o}}, \bibinfo{author}{B.~A. Nosek},
  \bibinfo{author}{D.~V.~M. Bishop}, \bibinfo{author}{K.~S. Button},
  \bibinfo{author}{C.~D. Chambers}, \bibinfo{author}{N.~P. Du~Sert},
  \bibinfo{author}{U.~Simonsohn}, \bibinfo{author}{E.-J. Wagenmakers},
  \bibinfo{author}{J.~J. Ware}, \bibinfo{author}{J.~P.~A. Ioannidis},
\newblock \bibinfo{title}{A manifesto for reproducible science},
\newblock \bibinfo{journal}{Nat. Hum. Behav} \bibinfo{volume}{1}
  (\bibinfo{year}{2017}) \bibinfo{pages}{0021}.
\bibitem[{Shrout and Rodgers(2018)}]{shrout2018psychology}
\bibinfo{author}{P.~E. Shrout}, \bibinfo{author}{J.~L. Rodgers},
\newblock \bibinfo{title}{Psychology, science, and knowledge construction:
  {B}roadening perspectives from the replication crisis},
\newblock \bibinfo{journal}{Annu. Rev. Psychol.} \bibinfo{volume}{69}
  (\bibinfo{year}{2018}) \bibinfo{pages}{487--510}.
\bibitem[{Benjamin et~al.(2018)Benjamin, Berger, Johannesson, Nosek,
  Wagenmakers, Berk, Bollen, Brembs, Brown, Camerer
  et~al.}]{benjamin2018redefine}
\bibinfo{author}{D.~J. Benjamin}, \bibinfo{author}{J.~O. Berger},
  \bibinfo{author}{M.~Johannesson}, \bibinfo{author}{B.~A. Nosek},
  \bibinfo{author}{E.-J. Wagenmakers}, \bibinfo{author}{R.~Berk},
  \bibinfo{author}{K.~A. Bollen}, \bibinfo{author}{B.~Brembs},
  \bibinfo{author}{L.~Brown}, \bibinfo{author}{C.~Camerer}, et~al.,
\newblock \bibinfo{title}{Redefine statistical significance},
\newblock \bibinfo{journal}{Nat. Hum. Behav.} \bibinfo{volume}{2}
  (\bibinfo{year}{2018}) \bibinfo{pages}{6--10}.
\bibitem[{Brandt et~al.(2014)Brandt, IJzerman, Dijksterhuis, Farach, Geller,
  Giner-Sorolla, Grange, Perugini, Spies, and
  Van't~Veer}]{brandt2014replication}
\bibinfo{author}{M.~J. Brandt}, \bibinfo{author}{H.~IJzerman},
  \bibinfo{author}{A.~Dijksterhuis}, \bibinfo{author}{F.~J. Farach},
  \bibinfo{author}{J.~Geller}, \bibinfo{author}{R.~Giner-Sorolla},
  \bibinfo{author}{J.~A. Grange}, \bibinfo{author}{M.~Perugini},
  \bibinfo{author}{J.~R. Spies}, \bibinfo{author}{A.~Van't~Veer},
\newblock \bibinfo{title}{The replication recipe: {W}hat makes for a convincing
  replication?},
\newblock \bibinfo{journal}{J. Exp. Soc. Psychol.} \bibinfo{volume}{50}
  (\bibinfo{year}{2014}) \bibinfo{pages}{217--224}.
\bibitem[{Maxwell et~al.(2015)Maxwell, Lau, and Howard}]{maxwell2015psychology}
\bibinfo{author}{S.~E. Maxwell}, \bibinfo{author}{M.~Y. Lau},
  \bibinfo{author}{G.~S. Howard},
\newblock \bibinfo{title}{Is psychology suffering from a replication crisis?
  {W}hat does ``failure to replicate'' really mean?},
\newblock \bibinfo{journal}{Am Psychol.} \bibinfo{volume}{70}
  (\bibinfo{year}{2015}) \bibinfo{pages}{487--498}.
\bibitem[{Nosek et~al.(2018)Nosek, Ebersole, DeHaven, and
  Mellor}]{nosek2018preregistration}
\bibinfo{author}{B.~A. Nosek}, \bibinfo{author}{C.~R. Ebersole},
  \bibinfo{author}{A.~C. DeHaven}, \bibinfo{author}{D.~T. Mellor},
\newblock \bibinfo{title}{The preregistration revolution},
\newblock \bibinfo{journal}{Proc. Natl. Acad. Sci. USA} \bibinfo{volume}{115}
  (\bibinfo{year}{2018}) \bibinfo{pages}{2600--2606}.
\bibitem[{Nosek et~al.(2019)Nosek, Beck, Campbell, Flake, Hardwicke, Mellor,
  van't Veer, and Vazire}]{nosek2019preregistration}
\bibinfo{author}{B.~A. Nosek}, \bibinfo{author}{E.~D. Beck},
  \bibinfo{author}{L.~Campbell}, \bibinfo{author}{J.~K. Flake},
  \bibinfo{author}{T.~E. Hardwicke}, \bibinfo{author}{D.~T. Mellor},
  \bibinfo{author}{A.~E. van't Veer}, \bibinfo{author}{S.~Vazire},
\newblock \bibinfo{title}{Preregistration is hard, and worthwhile},
\newblock \bibinfo{journal}{Trends Cogn. Sci.} \bibinfo{volume}{23}
  (\bibinfo{year}{2019}) \bibinfo{pages}{815--818}.
\bibitem[{Soderberg et~al.(2021)Soderberg, Errington, Schiavone, Bottesini,
  Thorn, Vazire, Esterling, and Nosek}]{soderberg2021initial}
\bibinfo{author}{C.~K. Soderberg}, \bibinfo{author}{T.~M. Errington},
  \bibinfo{author}{S.~R. Schiavone}, \bibinfo{author}{J.~Bottesini},
  \bibinfo{author}{F.~S. Thorn}, \bibinfo{author}{S.~Vazire},
  \bibinfo{author}{K.~M. Esterling}, \bibinfo{author}{B.~A. Nosek},
\newblock \bibinfo{title}{Initial evidence of research quality of registered
  reports compared with the standard publishing model},
\newblock \bibinfo{journal}{Nat. Hum. Behav.} \bibinfo{volume}{5}
  (\bibinfo{year}{2021}) \bibinfo{pages}{990--997}.
\bibitem[{Muthukrishna and Henrich(2019)}]{muthukrishna2019problem}
\bibinfo{author}{M.~Muthukrishna}, \bibinfo{author}{J.~Henrich},
\newblock \bibinfo{title}{A problem in theory},
\newblock \bibinfo{journal}{Nat. Hum. Behav.} \bibinfo{volume}{3}
  (\bibinfo{year}{2019}) \bibinfo{pages}{221--229}.
\bibitem[{Brumfiel(2011)}]{brumfiel2011particles}
\bibinfo{author}{G.~Brumfiel}, \bibinfo{title}{Particles break light-speed
  limit}, \bibinfo{year}{2011}. \bibinfo{note}{Available at:
  \url{https://www.nature.com/articles/news.2011.554}}.
\bibitem[{Adam et~al.(2012)Adam, Agafonova, Aleksandrov, Altinok, Sanchez,
  Anokhina, Aoki, Ariga, Ariga, Autiero et~al.}]{adam2012measurement}
\bibinfo{author}{T.~Adam}, \bibinfo{author}{N.~Agafonova},
  \bibinfo{author}{A.~Aleksandrov}, \bibinfo{author}{O.~Altinok},
  \bibinfo{author}{P.~A. Sanchez}, \bibinfo{author}{A.~Anokhina},
  \bibinfo{author}{S.~Aoki}, \bibinfo{author}{A.~Ariga},
  \bibinfo{author}{T.~Ariga}, \bibinfo{author}{D.~Autiero}, et~al.,
\newblock \bibinfo{title}{Measurement of the neutrino velocity with the {OPERA}
  detector in the {CNGS} beam},
\newblock \bibinfo{journal}{J. High Energy Phys.} \bibinfo{volume}{2012}
  (\bibinfo{year}{2012}) \bibinfo{pages}{1--37}.
\bibitem[{Lee et~al.(2015)Lee, Sigmund, Dieckmann, and Iwasa}]{lee2015games}
\bibinfo{author}{J.-H. Lee}, \bibinfo{author}{K.~Sigmund},
  \bibinfo{author}{U.~Dieckmann}, \bibinfo{author}{Y.~Iwasa},
\newblock \bibinfo{title}{Games of corruption: {H}ow to suppress illegal
  logging},
\newblock \bibinfo{journal}{J. Theor. Biol.} \bibinfo{volume}{367}
  (\bibinfo{year}{2015}) \bibinfo{pages}{1--13}.
\bibitem[{Lee et~al.(2019)Lee, Iwasa, Dieckmann, and Sigmund}]{lee2019social}
\bibinfo{author}{J.-H. Lee}, \bibinfo{author}{Y.~Iwasa},
  \bibinfo{author}{U.~Dieckmann}, \bibinfo{author}{K.~Sigmund},
\newblock \bibinfo{title}{Social evolution leads to persistent corruption},
\newblock \bibinfo{journal}{Proc. Natl. Acad. Sci.} \bibinfo{volume}{116}
  (\bibinfo{year}{2019}) \bibinfo{pages}{13276--13281}.
\bibitem[{Arefin et~al.(2019)Arefin, Masaki, Kabir, and
  Tanimoto}]{arefin2019interplay}
\bibinfo{author}{M.~R. Arefin}, \bibinfo{author}{T.~Masaki},
  \bibinfo{author}{K.~A. Kabir}, \bibinfo{author}{J.~Tanimoto},
\newblock \bibinfo{title}{Interplay between cost and effectiveness in influenza
  vaccine uptake: a vaccination game approach},
\newblock \bibinfo{journal}{Proc. R. Soc. A} \bibinfo{volume}{475}
  (\bibinfo{year}{2019}) \bibinfo{pages}{20190608}.
\bibitem[{Kabir et~al.(2019)Kabir, Jusup, and Tanimoto}]{kabir2019behavioral}
\bibinfo{author}{K.~A. Kabir}, \bibinfo{author}{M.~Jusup},
  \bibinfo{author}{J.~Tanimoto},
\newblock \bibinfo{title}{Behavioral incentives in a vaccination-dilemma
  setting with optional treatment},
\newblock \bibinfo{journal}{Phys. Rev. E} \bibinfo{volume}{100}
  (\bibinfo{year}{2019}) \bibinfo{pages}{062402}.
\bibitem[{Nowak and Sigmund(1993)}]{nowak1993strategy}
\bibinfo{author}{M.~Nowak}, \bibinfo{author}{K.~Sigmund},
\newblock \bibinfo{title}{A strategy of win-stay, lose-shift that outperforms
  tit-for-tat in the {P}risoner's {D}ilemma game},
\newblock \bibinfo{journal}{Nature} \bibinfo{volume}{364}
  (\bibinfo{year}{1993}) \bibinfo{pages}{56--58}.
\bibitem[{Ohtsuki and Iwasa(2006)}]{ohtsuki2006leading}
\bibinfo{author}{H.~Ohtsuki}, \bibinfo{author}{Y.~Iwasa},
\newblock \bibinfo{title}{The leading eight: social norms that can maintain
  cooperation by indirect reciprocity},
\newblock \bibinfo{journal}{J. Theor. Biol.} \bibinfo{volume}{239}
  (\bibinfo{year}{2006}) \bibinfo{pages}{435--444}.
\bibitem[{Ohtsuki and Iwasa(2007)}]{ohtsuki2007global}
\bibinfo{author}{H.~Ohtsuki}, \bibinfo{author}{Y.~Iwasa},
\newblock \bibinfo{title}{Global analyses of evolutionary dynamics and
  exhaustive search for social norms that maintain cooperation by reputation},
\newblock \bibinfo{journal}{J. Theor. Biol.} \bibinfo{volume}{244}
  (\bibinfo{year}{2007}) \bibinfo{pages}{518--531}.
\bibitem[{Holme(2003)}]{holme2003network}
\bibinfo{author}{P.~Holme},
\newblock \bibinfo{title}{Network dynamics of ongoing social relationships},
\newblock \bibinfo{journal}{EPL (Europhys. Lett.)} \bibinfo{volume}{64}
  (\bibinfo{year}{2003}) \bibinfo{pages}{427}.
\bibitem[{Newman and Park(2003)}]{newman2003social}
\bibinfo{author}{M.~E. Newman}, \bibinfo{author}{J.~Park},
\newblock \bibinfo{title}{Why social networks are different from other types of
  networks},
\newblock \bibinfo{journal}{Phys. Rev. E} \bibinfo{volume}{68}
  (\bibinfo{year}{2003}) \bibinfo{pages}{036122}.
\bibitem[{Holme et~al.(2004)Holme, Edling, and Liljeros}]{holme2004structure}
\bibinfo{author}{P.~Holme}, \bibinfo{author}{C.~R. Edling},
  \bibinfo{author}{F.~Liljeros},
\newblock \bibinfo{title}{Structure and time evolution of an {I}nternet dating
  community},
\newblock \bibinfo{journal}{Soc. Netw.} \bibinfo{volume}{26}
  (\bibinfo{year}{2004}) \bibinfo{pages}{155--174}.
\bibitem[{Rocha et~al.(2010)Rocha, Liljeros, and Holme}]{rocha2010information}
\bibinfo{author}{L.~E. Rocha}, \bibinfo{author}{F.~Liljeros},
  \bibinfo{author}{P.~Holme},
\newblock \bibinfo{title}{Information dynamics shape the sexual networks of
  {I}nternet-mediated prostitution},
\newblock \bibinfo{journal}{Proc. Natl. Acad. Sci. USA} \bibinfo{volume}{107}
  (\bibinfo{year}{2010}) \bibinfo{pages}{5706--5711}.
\bibitem[{Pinter-Wollman et~al.(2014)Pinter-Wollman, Hobson, Smith, Edelman,
  Shizuka, De~Silva, Waters, Prager, Sasaki, Wittemyer
  et~al.}]{pinter2014dynamics}
\bibinfo{author}{N.~Pinter-Wollman}, \bibinfo{author}{E.~A. Hobson},
  \bibinfo{author}{J.~E. Smith}, \bibinfo{author}{A.~J. Edelman},
  \bibinfo{author}{D.~Shizuka}, \bibinfo{author}{S.~De~Silva},
  \bibinfo{author}{J.~S. Waters}, \bibinfo{author}{S.~D. Prager},
  \bibinfo{author}{T.~Sasaki}, \bibinfo{author}{G.~Wittemyer}, et~al.,
\newblock \bibinfo{title}{The dynamics of animal social networks: analytical,
  conceptual, and theoretical advances},
\newblock \bibinfo{journal}{Behav. Ecol.} \bibinfo{volume}{25}
  (\bibinfo{year}{2014}) \bibinfo{pages}{242--255}.
\bibitem[{Jalili et~al.(2017)Jalili, Orouskhani, Asgari, Alipourfard, and
  Perc}]{jalili2017link}
\bibinfo{author}{M.~Jalili}, \bibinfo{author}{Y.~Orouskhani},
  \bibinfo{author}{M.~Asgari}, \bibinfo{author}{N.~Alipourfard},
  \bibinfo{author}{M.~Perc},
\newblock \bibinfo{title}{Link prediction in multiplex online social networks},
\newblock \bibinfo{journal}{R. Soc. Open Sci.} \bibinfo{volume}{4}
  (\bibinfo{year}{2017}) \bibinfo{pages}{160863}.
\bibitem[{Battiston et~al.(2012)Battiston, Puliga, Kaushik, Tasca, and
  Caldarelli}]{battiston2012debtrank}
\bibinfo{author}{S.~Battiston}, \bibinfo{author}{M.~Puliga},
  \bibinfo{author}{R.~Kaushik}, \bibinfo{author}{P.~Tasca},
  \bibinfo{author}{G.~Caldarelli},
\newblock \bibinfo{title}{Debtrank: {T}oo central to fail? {F}inancial
  networks, the {FED} and systemic risk},
\newblock \bibinfo{journal}{Sci. Rep.} \bibinfo{volume}{2}
  (\bibinfo{year}{2012}) \bibinfo{pages}{541}.
\bibitem[{Rocha and Holme(2010)}]{rocha2010network}
\bibinfo{author}{L.~E.~C. Rocha}, \bibinfo{author}{P.~Holme},
\newblock \bibinfo{title}{The network organisation of consumer complaints},
\newblock \bibinfo{journal}{EPL (Europhys. Lett.)} \bibinfo{volume}{91}
  (\bibinfo{year}{2010}) \bibinfo{pages}{28005}.
\bibitem[{Kito et~al.(2014)Kito, Brintrup, New, and
  Reed-Tsochas}]{kito2014structure}
\bibinfo{author}{T.~Kito}, \bibinfo{author}{A.~Brintrup},
  \bibinfo{author}{S.~New}, \bibinfo{author}{F.~Reed-Tsochas},
  \bibinfo{title}{The structure of the {T}oyota supply network: an empirical
  analysis}, \bibinfo{year}{2014}. \bibinfo{note}{Social Science Research
  Network (SSRN) Paper No. 2412512. Available at:
  \url{https://ssrn.com/abstract=2412512}}.
\bibitem[{Inoue and Todo(2019)}]{inoue2019firm}
\bibinfo{author}{H.~Inoue}, \bibinfo{author}{Y.~Todo},
\newblock \bibinfo{title}{Firm-level propagation of shocks through supply-chain
  networks},
\newblock \bibinfo{journal}{Nat. Sustain.} \bibinfo{volume}{2}
  (\bibinfo{year}{2019}) \bibinfo{pages}{841--847}.
\bibitem[{Holme(2003)}]{holme2003congestion}
\bibinfo{author}{P.~Holme},
\newblock \bibinfo{title}{Congestion and centrality in traffic flow on complex
  networks},
\newblock \bibinfo{journal}{Adv. Complex Syst.} \bibinfo{volume}{6}
  (\bibinfo{year}{2003}) \bibinfo{pages}{163--176}.
\bibitem[{DeLaurentis et~al.(2008)DeLaurentis, Han, and
  Kotegawa}]{delaurentis2008network}
\bibinfo{author}{D.~DeLaurentis}, \bibinfo{author}{E.-P. Han},
  \bibinfo{author}{T.~Kotegawa},
\newblock \bibinfo{title}{Network-theoretic approach for analyzing connectivity
  in air transportation networks},
\newblock \bibinfo{journal}{J. Aircr.} \bibinfo{volume}{45}
  (\bibinfo{year}{2008}) \bibinfo{pages}{1669--1679}.
\bibitem[{Hu and Zhu(2009)}]{hu2009empirical}
\bibinfo{author}{Y.~Hu}, \bibinfo{author}{D.~Zhu},
\newblock \bibinfo{title}{Empirical analysis of the worldwide maritime
  transportation network},
\newblock \bibinfo{journal}{Physica A} \bibinfo{volume}{388}
  (\bibinfo{year}{2009}) \bibinfo{pages}{2061--2071}.
\bibitem[{Xie and Levinson(2009)}]{xie2009modeling}
\bibinfo{author}{F.~Xie}, \bibinfo{author}{D.~Levinson},
\newblock \bibinfo{title}{Modeling the growth of transportation networks: {A}
  comprehensive review},
\newblock \bibinfo{journal}{Netw. Spat. Econ.} \bibinfo{volume}{9}
  (\bibinfo{year}{2009}) \bibinfo{pages}{291--307}.
\bibitem[{Derrible and Kennedy(2011)}]{derrible2011applications}
\bibinfo{author}{S.~Derrible}, \bibinfo{author}{C.~Kennedy},
\newblock \bibinfo{title}{Applications of graph theory and network science to
  transit network design},
\newblock \bibinfo{journal}{Transp. Rev.} \bibinfo{volume}{31}
  (\bibinfo{year}{2011}) \bibinfo{pages}{495--519}.
\bibitem[{Du et~al.(2016)Du, Zhou, Jusup, and Wang}]{du2016physics}
\bibinfo{author}{W.-B. Du}, \bibinfo{author}{X.-L. Zhou},
  \bibinfo{author}{M.~Jusup}, \bibinfo{author}{Z.~Wang},
\newblock \bibinfo{title}{Physics of transportation: {T}owards optimal capacity
  using the multilayer network framework},
\newblock \bibinfo{journal}{Sci. Rep.} \bibinfo{volume}{6}
  (\bibinfo{year}{2016}) \bibinfo{pages}{19059}.
\bibitem[{Rosas-Casals et~al.(2007)Rosas-Casals, Valverde, and
  Sol{\'e}}]{rosas2007topological}
\bibinfo{author}{M.~Rosas-Casals}, \bibinfo{author}{S.~Valverde},
  \bibinfo{author}{R.~V. Sol{\'e}},
\newblock \bibinfo{title}{Topological vulnerability of the {E}uropean power
  grid under errors and attacks},
\newblock \bibinfo{journal}{Int. J. Bifurc. Chaos} \bibinfo{volume}{17}
  (\bibinfo{year}{2007}) \bibinfo{pages}{2465--2475}.
\bibitem[{Pagani and Aiello(2013)}]{pagani2013power}
\bibinfo{author}{G.~A. Pagani}, \bibinfo{author}{M.~Aiello},
\newblock \bibinfo{title}{The power grid as a complex network: a survey},
\newblock \bibinfo{journal}{Physica A} \bibinfo{volume}{392}
  (\bibinfo{year}{2013}) \bibinfo{pages}{2688--2700}.
\bibitem[{Kim et~al.(2015)Kim, Lee, and Holme}]{kim2015community}
\bibinfo{author}{H.~Kim}, \bibinfo{author}{S.~H. Lee},
  \bibinfo{author}{P.~Holme},
\newblock \bibinfo{title}{Community consistency determines the stability
  transition window of power-grid nodes},
\newblock \bibinfo{journal}{New J. Phys.} \bibinfo{volume}{17}
  (\bibinfo{year}{2015}) \bibinfo{pages}{113005}.
\bibitem[{Donges et~al.(2009)Donges, Zou, Marwan, and
  Kurths}]{donges2009complex}
\bibinfo{author}{J.~F. Donges}, \bibinfo{author}{Y.~Zou},
  \bibinfo{author}{N.~Marwan}, \bibinfo{author}{J.~Kurths},
\newblock \bibinfo{title}{Complex networks in climate dynamics},
\newblock \bibinfo{journal}{Eur. Phys. J. Spec. Top.} \bibinfo{volume}{174}
  (\bibinfo{year}{2009}) \bibinfo{pages}{157--179}.
\bibitem[{Fan et~al.(2021)Fan, Meng, Ludescher, Chen, Ashkenazy, Kurths,
  Havlin, and Schellnhuber}]{fan2021statistical}
\bibinfo{author}{J.~Fan}, \bibinfo{author}{J.~Meng},
  \bibinfo{author}{J.~Ludescher}, \bibinfo{author}{X.~Chen},
  \bibinfo{author}{Y.~Ashkenazy}, \bibinfo{author}{J.~Kurths},
  \bibinfo{author}{S.~Havlin}, \bibinfo{author}{H.~J. Schellnhuber},
\newblock \bibinfo{title}{Statistical physics approaches to the complex earth
  system},
\newblock \bibinfo{journal}{Phys. Rep.} \bibinfo{volume}{869}
  (\bibinfo{year}{2021}) \bibinfo{pages}{1--84}.
\bibitem[{Barab{\'a}si(2007)}]{barabasi2007network}
\bibinfo{author}{A.-L. Barab{\'a}si},
\newblock \bibinfo{title}{Network medicine---from obesity to the
  ``diseasome''},
\newblock \bibinfo{journal}{N. Engl. J. Med.} \bibinfo{volume}{357}
  (\bibinfo{year}{2007}) \bibinfo{pages}{404--407}.
\bibitem[{Hofmann and Curtiss(2018)}]{hofmann2018complex}
\bibinfo{author}{S.~G. Hofmann}, \bibinfo{author}{J.~Curtiss},
\newblock \bibinfo{title}{A complex network approach to clinical science},
\newblock \bibinfo{journal}{Eur. J. Clin. Invest.} \bibinfo{volume}{48}
  (\bibinfo{year}{2018}) \bibinfo{pages}{e12986}.
\bibitem[{Barab{\'a}si et~al.(2020)Barab{\'a}si, Menichetti, and
  Loscalzo}]{barabasi2020unmapped}
\bibinfo{author}{A.-L. Barab{\'a}si}, \bibinfo{author}{G.~Menichetti},
  \bibinfo{author}{J.~Loscalzo},
\newblock \bibinfo{title}{The unmapped chemical complexity of our diet},
\newblock \bibinfo{journal}{Nat. Food} \bibinfo{volume}{1}
  (\bibinfo{year}{2020}) \bibinfo{pages}{33--37}.
\bibitem[{Herrera(2020)}]{herrera2020contribution}
\bibinfo{author}{J.~C.~S. Herrera},
\newblock \bibinfo{title}{The contribution of network science to the study of
  food recipes. {A} review paper},
\newblock \bibinfo{journal}{Appetite} \bibinfo{volume}{159}
  (\bibinfo{year}{2020}) \bibinfo{pages}{105048}.
\bibitem[{Buld{\'u} et~al.(2018)Buld{\'u}, Busquets, Mart{\'\i}nez,
  Herrera-Diestra, Echegoyen, Galeano, and Luque}]{buldu2018using}
\bibinfo{author}{J.~M. Buld{\'u}}, \bibinfo{author}{J.~Busquets},
  \bibinfo{author}{J.~H. Mart{\'\i}nez}, \bibinfo{author}{J.~L.
  Herrera-Diestra}, \bibinfo{author}{I.~Echegoyen},
  \bibinfo{author}{J.~Galeano}, \bibinfo{author}{J.~Luque},
\newblock \bibinfo{title}{Using network science to analyse football passing
  networks: {D}ynamics, space, time, and the multilayer nature of the game},
\newblock \bibinfo{journal}{Front. Psychol.} \bibinfo{volume}{9}
  (\bibinfo{year}{2018}) \bibinfo{pages}{1900}.
\bibitem[{Buld{\'u} et~al.(2019)Buld{\'u}, Busquets, Echegoyen
  et~al.}]{buldu2019defining}
\bibinfo{author}{J.~Buld{\'u}}, \bibinfo{author}{J.~Busquets},
  \bibinfo{author}{I.~Echegoyen}, et~al.,
\newblock \bibinfo{title}{Defining a historic football team: {U}sing {N}etwork
  {S}cience to analyze {G}uardiola’s {FC} {B}arcelona},
\newblock \bibinfo{journal}{Sci. Rep.} \bibinfo{volume}{9}
  (\bibinfo{year}{2019}) \bibinfo{pages}{13602}.
\bibitem[{Noh and Rieger(2004)}]{noh2004random}
\bibinfo{author}{J.~D. Noh}, \bibinfo{author}{H.~Rieger},
\newblock \bibinfo{title}{Random walks on complex networks},
\newblock \bibinfo{journal}{Phys. Rev. Lett.} \bibinfo{volume}{92}
  (\bibinfo{year}{2004}) \bibinfo{pages}{118701}.
\bibitem[{Perkins et~al.(2014)Perkins, Foxall, Glass, and
  Edwards}]{perkins2014scaling}
\bibinfo{author}{T.~J. Perkins}, \bibinfo{author}{E.~Foxall},
  \bibinfo{author}{L.~Glass}, \bibinfo{author}{R.~Edwards},
\newblock \bibinfo{title}{A scaling law for random walks on networks},
\newblock \bibinfo{journal}{Nat. Commun.} \bibinfo{volume}{5}
  (\bibinfo{year}{2014}) \bibinfo{pages}{5121}.
\bibitem[{Wu(2005)}]{wu2005synchronization}
\bibinfo{author}{C.~W. Wu},
\newblock \bibinfo{title}{Synchronization in networks of nonlinear dynamical
  systems coupled via a directed graph},
\newblock \bibinfo{journal}{Nonlinearity} \bibinfo{volume}{18}
  (\bibinfo{year}{2005}) \bibinfo{pages}{1057}.
\bibitem[{Scardovi and Sepulchre(2009)}]{scardovi2009synchronization}
\bibinfo{author}{L.~Scardovi}, \bibinfo{author}{R.~Sepulchre},
\newblock \bibinfo{title}{Synchronization in networks of identical linear
  systems},
\newblock \bibinfo{journal}{Automatica} \bibinfo{volume}{45}
  (\bibinfo{year}{2009}) \bibinfo{pages}{2557--2562}.
\bibitem[{Del~Genio et~al.(2016)Del~Genio, G{\'o}mez-Garde{\~n}es, Bonamassa,
  and Boccaletti}]{delgenio2016synchronization}
\bibinfo{author}{C.~I. Del~Genio}, \bibinfo{author}{J.~G{\'o}mez-Garde{\~n}es},
  \bibinfo{author}{I.~Bonamassa}, \bibinfo{author}{S.~Boccaletti},
\newblock \bibinfo{title}{Synchronization in networks with multiple interaction
  layers},
\newblock \bibinfo{journal}{Sci. Adv.} \bibinfo{volume}{2}
  (\bibinfo{year}{2016}) \bibinfo{pages}{e1601679}.
\bibitem[{Bansal et~al.(2007)Bansal, Grenfell, and
  Meyers}]{bansal2007individual}
\bibinfo{author}{S.~Bansal}, \bibinfo{author}{B.~T. Grenfell},
  \bibinfo{author}{L.~A. Meyers},
\newblock \bibinfo{title}{When individual behaviour matters: homogeneous and
  network models in epidemiology},
\newblock \bibinfo{journal}{J. R. Soc. Interface} \bibinfo{volume}{4}
  (\bibinfo{year}{2007}) \bibinfo{pages}{879--891}.
\bibitem[{Danon et~al.(2011)Danon, Ford, House, Jewell, Keeling, Roberts, Ross,
  and Vernon}]{danon2011networks}
\bibinfo{author}{L.~Danon}, \bibinfo{author}{A.~P. Ford},
  \bibinfo{author}{T.~House}, \bibinfo{author}{C.~P. Jewell},
  \bibinfo{author}{M.~J. Keeling}, \bibinfo{author}{G.~O. Roberts},
  \bibinfo{author}{J.~V. Ross}, \bibinfo{author}{M.~C. Vernon},
\newblock \bibinfo{title}{Networks and the epidemiology of infectious disease},
\newblock \bibinfo{journal}{Interdiscip. Perspect. Infect. Dis.}
  \bibinfo{volume}{2011} (\bibinfo{year}{2011}) \bibinfo{pages}{284909}.
\bibitem[{Brockmann and Helbing(2013)}]{brockmann2013hidden}
\bibinfo{author}{D.~Brockmann}, \bibinfo{author}{D.~Helbing},
\newblock \bibinfo{title}{The hidden geometry of complex, network-driven
  contagion phenomena},
\newblock \bibinfo{journal}{science} \bibinfo{volume}{342}
  (\bibinfo{year}{2013}) \bibinfo{pages}{1337--1342}.
\bibitem[{Hummon and Doreian(2003)}]{hummon2003some}
\bibinfo{author}{N.~P. Hummon}, \bibinfo{author}{P.~Doreian},
\newblock \bibinfo{title}{Some dynamics of social balance processes: bringing
  {H}eider back into balance theory},
\newblock \bibinfo{journal}{Soc. Netw.} \bibinfo{volume}{25}
  (\bibinfo{year}{2003}) \bibinfo{pages}{17--49}.
\bibitem[{Antal et~al.(2005)Antal, Krapivsky, and Redner}]{antal2005dynamics}
\bibinfo{author}{T.~Antal}, \bibinfo{author}{P.~L. Krapivsky},
  \bibinfo{author}{S.~Redner},
\newblock \bibinfo{title}{Dynamics of social balance on networks},
\newblock \bibinfo{journal}{Phys. Rev. E} \bibinfo{volume}{72}
  (\bibinfo{year}{2005}) \bibinfo{pages}{036121}.
\bibitem[{Antal et~al.(2006)Antal, Krapivsky, and Redner}]{antal2006social}
\bibinfo{author}{T.~Antal}, \bibinfo{author}{P.~L. Krapivsky},
  \bibinfo{author}{S.~Redner},
\newblock \bibinfo{title}{Social balance on networks: {T}he dynamics of
  friendship and enmity},
\newblock \bibinfo{journal}{Physica D} \bibinfo{volume}{224}
  (\bibinfo{year}{2006}) \bibinfo{pages}{130--136}.
\bibitem[{Iacopini et~al.(2018)Iacopini, Milojevi{\'c}, and
  Latora}]{iacopini2018network}
\bibinfo{author}{I.~Iacopini}, \bibinfo{author}{S.~Milojevi{\'c}},
  \bibinfo{author}{V.~Latora},
\newblock \bibinfo{title}{Network dynamics of innovation processes},
\newblock \bibinfo{journal}{Phys. Rev. Lett.} \bibinfo{volume}{120}
  (\bibinfo{year}{2018}) \bibinfo{pages}{048301}.
\bibitem[{Boccaletti et~al.(2006)Boccaletti, Latora, Moreno, Chavez, and
  Hwang}]{boccaletti2006complex}
\bibinfo{author}{S.~Boccaletti}, \bibinfo{author}{V.~Latora},
  \bibinfo{author}{Y.~Moreno}, \bibinfo{author}{M.~Chavez},
  \bibinfo{author}{D.-U. Hwang},
\newblock \bibinfo{title}{Complex networks: {S}tructure and dynamics},
\newblock \bibinfo{journal}{Phys. Rep.} \bibinfo{volume}{424}
  (\bibinfo{year}{2006}) \bibinfo{pages}{175--308}.
\bibitem[{Fortunato(2010)}]{fortunato2010community}
\bibinfo{author}{S.~Fortunato},
\newblock \bibinfo{title}{Community detection in graphs},
\newblock \bibinfo{journal}{Phys Rep.} \bibinfo{volume}{486}
  (\bibinfo{year}{2010}) \bibinfo{pages}{75--174}.
\bibitem[{Fortunato and Hric(2016)}]{fortunato2016community}
\bibinfo{author}{S.~Fortunato}, \bibinfo{author}{D.~Hric},
\newblock \bibinfo{title}{Community detection in networks: {A} user guide},
\newblock \bibinfo{journal}{Phys. Rep.} \bibinfo{volume}{659}
  (\bibinfo{year}{2016}) \bibinfo{pages}{1--44}.
\bibitem[{Barth{\'e}lemy(2011)}]{barthelemy2011spatial}
\bibinfo{author}{M.~Barth{\'e}lemy},
\newblock \bibinfo{title}{Spatial networks},
\newblock \bibinfo{journal}{Phys. Rep.} \bibinfo{volume}{499}
  (\bibinfo{year}{2011}) \bibinfo{pages}{1--101}.
\bibitem[{Boccaletti et~al.(2014)Boccaletti, Bianconi, Criado, Del~Genio,
  G{\'o}mez-Gardenes, Romance, Sendina-Nadal, Wang, and
  Zanin}]{boccaletti2014structure}
\bibinfo{author}{S.~Boccaletti}, \bibinfo{author}{G.~Bianconi},
  \bibinfo{author}{R.~Criado}, \bibinfo{author}{C.~I. Del~Genio},
  \bibinfo{author}{J.~G{\'o}mez-Gardenes}, \bibinfo{author}{M.~Romance},
  \bibinfo{author}{I.~Sendina-Nadal}, \bibinfo{author}{Z.~Wang},
  \bibinfo{author}{M.~Zanin},
\newblock \bibinfo{title}{The structure and dynamics of multilayer networks},
\newblock \bibinfo{journal}{Phys. Rep.} \bibinfo{volume}{544}
  (\bibinfo{year}{2014}) \bibinfo{pages}{1--122}.
\bibitem[{de~Arruda et~al.(2018)de~Arruda, Rodrigues, and
  Moreno}]{dearruda2018fundamentals}
\bibinfo{author}{G.~F. de~Arruda}, \bibinfo{author}{F.~A. Rodrigues},
  \bibinfo{author}{Y.~Moreno},
\newblock \bibinfo{title}{Fundamentals of spreading processes in single and
  multilayer complex networks},
\newblock \bibinfo{journal}{Phys. Rep.} \bibinfo{volume}{756}
  (\bibinfo{year}{2018}) \bibinfo{pages}{1--59}.
\bibitem[{Lancichinetti and Fortunato(2009)}]{lancichinetti2009community}
\bibinfo{author}{A.~Lancichinetti}, \bibinfo{author}{S.~Fortunato},
\newblock \bibinfo{title}{Community detection algorithms: a comparative
  analysis},
\newblock \bibinfo{journal}{Phys. Rev. E} \bibinfo{volume}{80}
  (\bibinfo{year}{2009}) \bibinfo{pages}{056117}.
\bibitem[{Lancichinetti et~al.(2011)Lancichinetti, Radicchi, Ramasco, and
  Fortunato}]{lancichinetti2011finding}
\bibinfo{author}{A.~Lancichinetti}, \bibinfo{author}{F.~Radicchi},
  \bibinfo{author}{J.~J. Ramasco}, \bibinfo{author}{S.~Fortunato},
\newblock \bibinfo{title}{Finding statistically significant communities in
  networks},
\newblock \bibinfo{journal}{PLOS ONE} \bibinfo{volume}{6}
  (\bibinfo{year}{2011}) \bibinfo{pages}{e18961}.
\bibitem[{Hric et~al.(2014)Hric, Darst, and Fortunato}]{hric2014community}
\bibinfo{author}{D.~Hric}, \bibinfo{author}{R.~K. Darst},
  \bibinfo{author}{S.~Fortunato},
\newblock \bibinfo{title}{Community detection in networks: {S}tructural
  communities versus ground truth},
\newblock \bibinfo{journal}{Phys. Rev. E} \bibinfo{volume}{90}
  (\bibinfo{year}{2014}) \bibinfo{pages}{062805}.
\bibitem[{Schaub et~al.(2017)Schaub, Delvenne, Rosvall, and
  Lambiotte}]{schaub2017many}
\bibinfo{author}{M.~T. Schaub}, \bibinfo{author}{J.-C. Delvenne},
  \bibinfo{author}{M.~Rosvall}, \bibinfo{author}{R.~Lambiotte},
\newblock \bibinfo{title}{The many facets of community detection in complex
  networks},
\newblock \bibinfo{journal}{Appl. Netw. Sci.} \bibinfo{volume}{2}
  (\bibinfo{year}{2017}) \bibinfo{pages}{4}.
\bibitem[{Rosvall et~al.(2019)Rosvall, Delvenne, Schaub, and
  Lambiotte}]{rosvall2019different}
\bibinfo{author}{M.~Rosvall}, \bibinfo{author}{J.-C. Delvenne},
  \bibinfo{author}{M.~T. Schaub}, \bibinfo{author}{R.~Lambiotte},
\newblock \bibinfo{title}{Different approaches to community detection},
\newblock in: \bibinfo{editor}{P.~Doreian}, \bibinfo{editor}{V.~Batagelj},
  \bibinfo{editor}{A.~Ferligoj} (Eds.), \bibinfo{booktitle}{Advances in network
  clustering and blockmodeling}, \bibinfo{publisher}{Wiley Online Library},
  \bibinfo{year}{2019}, pp. \bibinfo{pages}{105--119}.
\bibitem[{Chung et~al.(2014)Chung, Baek, Kim, Ha, and
  Jeong}]{chung2014generalized}
\bibinfo{author}{K.~Chung}, \bibinfo{author}{Y.~Baek},
  \bibinfo{author}{D.~Kim}, \bibinfo{author}{M.~Ha},
  \bibinfo{author}{H.~Jeong},
\newblock \bibinfo{title}{Generalized epidemic process on modular networks},
\newblock \bibinfo{journal}{Phys. Rev. E} \bibinfo{volume}{89}
  (\bibinfo{year}{2014}) \bibinfo{pages}{052811}.
\bibitem[{Valdez et~al.(2020)Valdez, Braunstein, and
  Havlin}]{valdez2020epidemic}
\bibinfo{author}{L.~D. Valdez}, \bibinfo{author}{L.~A. Braunstein},
  \bibinfo{author}{S.~Havlin},
\newblock \bibinfo{title}{Epidemic spreading on modular networks: {T}he fear to
  declare a pandemic},
\newblock \bibinfo{journal}{Phys. Rev. E} \bibinfo{volume}{101}
  (\bibinfo{year}{2020}) \bibinfo{pages}{032309}.
\bibitem[{Masuda(2009)}]{masuda2009immunization}
\bibinfo{author}{N.~Masuda},
\newblock \bibinfo{title}{Immunization of networks with community structure},
\newblock \bibinfo{journal}{New J. Phys.} \bibinfo{volume}{11}
  (\bibinfo{year}{2009}) \bibinfo{pages}{123018}.
\bibitem[{Gross and Havlin(2020)}]{gross2020epidemic}
\bibinfo{author}{B.~Gross}, \bibinfo{author}{S.~Havlin},
\newblock \bibinfo{title}{Epidemic spreading and control strategies in spatial
  modular network},
\newblock \bibinfo{journal}{Appl. Netw. Sci.} \bibinfo{volume}{5}
  (\bibinfo{year}{2020}) \bibinfo{pages}{95}.
\bibitem[{Erodos(1959)}]{erodos1959random}
\bibinfo{author}{P.~Erodos},
\newblock \bibinfo{title}{On random graphs i},
\newblock \bibinfo{journal}{Publ. Math. Debr.} \bibinfo{volume}{6}
  (\bibinfo{year}{1959}) \bibinfo{pages}{290--297}.
\bibitem[{Watts and Strogatz(1998)}]{watts1998collective}
\bibinfo{author}{D.~J. Watts}, \bibinfo{author}{S.~H. Strogatz},
\newblock \bibinfo{title}{Collective dynamics of `small-world' networks},
\newblock \bibinfo{journal}{Nature} \bibinfo{volume}{393}
  (\bibinfo{year}{1998}) \bibinfo{pages}{440--442}.
\bibitem[{Albert and Barab{\'a}si(2002)}]{albert2002statistical}
\bibinfo{author}{R.~Albert}, \bibinfo{author}{A.-L. Barab{\'a}si},
\newblock \bibinfo{title}{Statistical mechanics of complex networks},
\newblock \bibinfo{journal}{Rev. Mod. Phys.} \bibinfo{volume}{74}
  (\bibinfo{year}{2002}) \bibinfo{pages}{47}.
\bibitem[{Holland et~al.(1983)Holland, Laskey, and
  Leinhardt}]{holland1983stochastic}
\bibinfo{author}{P.~W. Holland}, \bibinfo{author}{K.~B. Laskey},
  \bibinfo{author}{S.~Leinhardt},
\newblock \bibinfo{title}{Stochastic blockmodels: {F}irst steps},
\newblock \bibinfo{journal}{Soc. Netw.} \bibinfo{volume}{5}
  (\bibinfo{year}{1983}) \bibinfo{pages}{109--137}.
\bibitem[{Nowicki and Snijders(2001)}]{nowicki2001estimation}
\bibinfo{author}{K.~Nowicki}, \bibinfo{author}{T.~A.~B. Snijders},
\newblock \bibinfo{title}{Estimation and prediction for stochastic
  blockstructures},
\newblock \bibinfo{journal}{J. Am. Stat. Assoc.} \bibinfo{volume}{96}
  (\bibinfo{year}{2001}) \bibinfo{pages}{1077--1087}.
\bibitem[{Airoldi et~al.(2008)Airoldi, Blei, Fienberg, and
  Xing}]{airoldi2008mixed}
\bibinfo{author}{E.~M. Airoldi}, \bibinfo{author}{D.~M. Blei},
  \bibinfo{author}{S.~E. Fienberg}, \bibinfo{author}{E.~P. Xing},
\newblock \bibinfo{title}{Mixed membership stochastic blockmodels},
\newblock \bibinfo{journal}{J. Mach. Learn. Res.} \bibinfo{volume}{9}
  (\bibinfo{year}{2008}) \bibinfo{pages}{1981--2014}.
\bibitem[{Fan et~al.(2016)Fan, Xu, and Cao}]{fan2016copula}
\bibinfo{author}{X.~Fan}, \bibinfo{author}{R.~Y.~D. Xu},
  \bibinfo{author}{L.~Cao},
\newblock \bibinfo{title}{Copula mixed-membership stochastic block model},
\newblock in: \bibinfo{editor}{G.~Brewka} (Ed.),
  \bibinfo{booktitle}{Proceedings of the Twenty-Fifth International Joint
  Conference on Artificial Intelligence}, \bibinfo{publisher}{Association for
  Computing Machinery}, \bibinfo{year}{2016}, pp. \bibinfo{pages}{1462--1468}.
\bibitem[{Pal and Coates(2019)}]{pal2019scalable}
\bibinfo{author}{S.~Pal}, \bibinfo{author}{M.~Coates},
\newblock \bibinfo{title}{Scalable {MCMC} in degree corrected stochastic block
  model},
\newblock in: \bibinfo{editor}{S.~Sanei}, \bibinfo{editor}{L.~Hanzo} (Eds.),
  \bibinfo{booktitle}{2019 IEEE International Conference on Acoustics, Speech
  and Signal Processing}, \bibinfo{publisher}{IEEE}, \bibinfo{year}{2019}, pp.
  \bibinfo{pages}{5461--5465}.
\bibitem[{Aicher et~al.(2015)Aicher, Jacobs, and Clauset}]{aicher2015learning}
\bibinfo{author}{C.~Aicher}, \bibinfo{author}{A.~Z. Jacobs},
  \bibinfo{author}{A.~Clauset},
\newblock \bibinfo{title}{Learning latent block structure in weighted
  networks},
\newblock \bibinfo{journal}{J. Complex Netw.} \bibinfo{volume}{3}
  (\bibinfo{year}{2015}) \bibinfo{pages}{221--248}.
\bibitem[{Peixoto(2018)}]{peixoto2018nonparametric}
\bibinfo{author}{T.~P. Peixoto},
\newblock \bibinfo{title}{Nonparametric weighted stochastic block models},
\newblock \bibinfo{journal}{Phys. Rev. E} \bibinfo{volume}{97}
  (\bibinfo{year}{2018}) \bibinfo{pages}{012306}.
\bibitem[{Peixoto(2015)}]{peixoto2015inferring}
\bibinfo{author}{T.~P. Peixoto},
\newblock \bibinfo{title}{Inferring the mesoscale structure of layered,
  edge-valued, and time-varying networks},
\newblock \bibinfo{journal}{Phys. Rev. E} \bibinfo{volume}{92}
  (\bibinfo{year}{2015}) \bibinfo{pages}{042807}.
\bibitem[{Stanley et~al.(2016)Stanley, Shai, Taylor, and
  Mucha}]{stanley2016clustering}
\bibinfo{author}{N.~Stanley}, \bibinfo{author}{S.~Shai},
  \bibinfo{author}{D.~Taylor}, \bibinfo{author}{P.~J. Mucha},
\newblock \bibinfo{title}{Clustering network layers with the strata multilayer
  stochastic block model},
\newblock \bibinfo{journal}{IEEE Trans. Netw. Sci. Eng.} \bibinfo{volume}{3}
  (\bibinfo{year}{2016}) \bibinfo{pages}{95--105}.
\bibitem[{Paul and Chen(2016)}]{paul2016consistent}
\bibinfo{author}{S.~Paul}, \bibinfo{author}{Y.~Chen},
\newblock \bibinfo{title}{Consistent community detection in multi-relational
  data through restricted multi-layer stochastic blockmodel},
\newblock \bibinfo{journal}{Electron. J. Stat.} \bibinfo{volume}{10}
  (\bibinfo{year}{2016}) \bibinfo{pages}{3807--3870}.
\bibitem[{Valles-Catala et~al.(2016)Valles-Catala, Massucci, Guimera, and
  Sales-Pardo}]{valles2016multilayer}
\bibinfo{author}{T.~Valles-Catala}, \bibinfo{author}{F.~A. Massucci},
  \bibinfo{author}{R.~Guimera}, \bibinfo{author}{M.~Sales-Pardo},
\newblock \bibinfo{title}{Multilayer stochastic block models reveal the
  multilayer structure of complex networks},
\newblock \bibinfo{journal}{Phys. Rev. X} \bibinfo{volume}{6}
  (\bibinfo{year}{2016}) \bibinfo{pages}{011036}.
\bibitem[{De~Bacco et~al.(2017)De~Bacco, Power, Larremore, and
  Moore}]{debacco2017community}
\bibinfo{author}{C.~De~Bacco}, \bibinfo{author}{E.~A. Power},
  \bibinfo{author}{D.~B. Larremore}, \bibinfo{author}{C.~Moore},
\newblock \bibinfo{title}{Community detection, link prediction, and layer
  interdependence in multilayer networks},
\newblock \bibinfo{journal}{Phys. Rev. E} \bibinfo{volume}{95}
  (\bibinfo{year}{2017}) \bibinfo{pages}{042317}.
\bibitem[{Fu et~al.(2009)Fu, Song, and Xing}]{fu2009dynamic}
\bibinfo{author}{W.~Fu}, \bibinfo{author}{L.~Song}, \bibinfo{author}{E.~P.
  Xing},
\newblock \bibinfo{title}{Dynamic mixed membership blockmodel for evolving
  networks},
\newblock in: \bibinfo{editor}{A.~Danyluk}, \bibinfo{editor}{L.~Bottou},
  \bibinfo{editor}{M.~Littman} (Eds.), \bibinfo{booktitle}{Proceedings of the
  26th Annual International Conference on Machine Learning},
  \bibinfo{publisher}{Association for Computing Machinery},
  \bibinfo{year}{2009}, pp. \bibinfo{pages}{329--336}.
\bibitem[{Xu and Hero(2014)}]{xu2014dynamic}
\bibinfo{author}{K.~S. Xu}, \bibinfo{author}{A.~O. Hero},
\newblock \bibinfo{title}{Dynamic stochastic blockmodels for time-evolving
  social networks},
\newblock \bibinfo{journal}{IEEE J. Sel. Top. Signal Process.}
  \bibinfo{volume}{8} (\bibinfo{year}{2014}) \bibinfo{pages}{552--562}.
\bibitem[{Peel and Clauset(2015)}]{peel2015detecting}
\bibinfo{author}{L.~Peel}, \bibinfo{author}{A.~Clauset},
\newblock \bibinfo{title}{Detecting change points in the large-scale structure
  of evolving networks},
\newblock in: \bibinfo{editor}{B.~Bonet}, \bibinfo{editor}{S.~Koenig} (Eds.),
  \bibinfo{booktitle}{Proceedings of the Twenty-Ninth AAAI Conference on
  Artificial Intelligence}, \bibinfo{publisher}{Association for the Advancement
  of Artificial Intelligence}, \bibinfo{year}{2015}, pp.
  \bibinfo{pages}{2914--2920}.
\bibitem[{Corneli et~al.(2016)Corneli, Latouche, and Rossi}]{corneli2016exact}
\bibinfo{author}{M.~Corneli}, \bibinfo{author}{P.~Latouche},
  \bibinfo{author}{F.~Rossi},
\newblock \bibinfo{title}{Exact {ICL} maximization in a non-stationary temporal
  extension of the stochastic block model for dynamic networks},
\newblock \bibinfo{journal}{Neurocomputing} \bibinfo{volume}{192}
  (\bibinfo{year}{2016}) \bibinfo{pages}{81--91}.
\bibitem[{Ghasemian et~al.(2016)Ghasemian, Zhang, Clauset, Moore, and
  Peel}]{ghasemian2016detectability}
\bibinfo{author}{A.~Ghasemian}, \bibinfo{author}{P.~Zhang},
  \bibinfo{author}{A.~Clauset}, \bibinfo{author}{C.~Moore},
  \bibinfo{author}{L.~Peel},
\newblock \bibinfo{title}{Detectability thresholds and optimal algorithms for
  community structure in dynamic networks},
\newblock \bibinfo{journal}{Phys. Rev. X} \bibinfo{volume}{6}
  (\bibinfo{year}{2016}) \bibinfo{pages}{031005}.
\bibitem[{Matias and Miele(2017)}]{matias2017statistical}
\bibinfo{author}{C.~Matias}, \bibinfo{author}{V.~Miele},
\newblock \bibinfo{title}{Statistical clustering of temporal networks through a
  dynamic stochastic block model},
\newblock \bibinfo{journal}{J. R. Stat. Soc. Series B Stat. Methodol.)}
  \bibinfo{volume}{79} (\bibinfo{year}{2017}) \bibinfo{pages}{1119--1141}.
\bibitem[{Peixoto and Rosvall(2017)}]{peixoto2017modelling}
\bibinfo{author}{T.~P. Peixoto}, \bibinfo{author}{M.~Rosvall},
\newblock \bibinfo{title}{Modelling sequences and temporal networks with
  dynamic community structures},
\newblock \bibinfo{journal}{Nat. Commun.} \bibinfo{volume}{8}
  (\bibinfo{year}{2017}) \bibinfo{pages}{582}.
\bibitem[{Zhang et~al.(2017)Zhang, Moore, and Newman}]{zhang2017random}
\bibinfo{author}{X.~Zhang}, \bibinfo{author}{C.~Moore},
  \bibinfo{author}{M.~E.~J. Newman},
\newblock \bibinfo{title}{Random graph models for dynamic networks},
\newblock \bibinfo{journal}{Eur. Phys. J. B} \bibinfo{volume}{90}
  (\bibinfo{year}{2017}) \bibinfo{pages}{1--14}.
\bibitem[{Peel(2015)}]{peel2015active}
\bibinfo{author}{L.~Peel},
\newblock \bibinfo{title}{Active discovery of network roles for predicting the
  classes of network nodes},
\newblock \bibinfo{journal}{J. Complex Netw.} \bibinfo{volume}{3}
  (\bibinfo{year}{2015}) \bibinfo{pages}{431--449}.
\bibitem[{Hric et~al.(2016)Hric, Peixoto, and Fortunato}]{hric2016network}
\bibinfo{author}{D.~Hric}, \bibinfo{author}{T.~P. Peixoto},
  \bibinfo{author}{S.~Fortunato},
\newblock \bibinfo{title}{Network structure, metadata, and the prediction of
  missing nodes and annotations},
\newblock \bibinfo{journal}{Phys. Rev. X} \bibinfo{volume}{6}
  (\bibinfo{year}{2016}) \bibinfo{pages}{031038}.
\bibitem[{Newman and Clauset(2016)}]{newman2016structure}
\bibinfo{author}{M.~E.~J. Newman}, \bibinfo{author}{A.~Clauset},
\newblock \bibinfo{title}{Structure and inference in annotated networks},
\newblock \bibinfo{journal}{Nat. Commun.} \bibinfo{volume}{7}
  (\bibinfo{year}{2016}) \bibinfo{pages}{11863}.
\bibitem[{Peixoto(2019)}]{peixoto2019bayesian}
\bibinfo{author}{T.~P. Peixoto},
\newblock \bibinfo{title}{Bayesian stochastic blockmodeling},
\newblock in: \bibinfo{editor}{P.~Doreian}, \bibinfo{editor}{V.~Batagelj},
  \bibinfo{editor}{A.~Ferligoj} (Eds.), \bibinfo{booktitle}{Advances in network
  clustering and blockmodeling}, \bibinfo{publisher}{Wiley Online Library},
  \bibinfo{year}{2019}, pp. \bibinfo{pages}{289--332}.
\bibitem[{Funke and Becker(2019)}]{funke2019stochastic}
\bibinfo{author}{T.~Funke}, \bibinfo{author}{T.~Becker},
\newblock \bibinfo{title}{Stochastic block models: {A} comparison of variants
  and inference methods},
\newblock \bibinfo{journal}{PLOS ONE} \bibinfo{volume}{14}
  (\bibinfo{year}{2019}) \bibinfo{pages}{e0215296}.
\bibitem[{White et~al.(1976)White, Boorman, and Breiger}]{white1976social}
\bibinfo{author}{H.~C. White}, \bibinfo{author}{S.~A. Boorman},
  \bibinfo{author}{R.~L. Breiger},
\newblock \bibinfo{title}{Social structure from multiple networks. {I}.
  {B}lockmodels of roles and positions},
\newblock \bibinfo{journal}{Am. J. Sociol.} \bibinfo{volume}{81}
  (\bibinfo{year}{1976}) \bibinfo{pages}{730--780}.
\bibitem[{Arabie et~al.(1978)Arabie, Boorman, and
  Levitt}]{arabie1978constructing}
\bibinfo{author}{P.~Arabie}, \bibinfo{author}{S.~A. Boorman},
  \bibinfo{author}{P.~R. Levitt},
\newblock \bibinfo{title}{Constructing blockmodels: {H}ow and why},
\newblock \bibinfo{journal}{J. Math. Psychol.} \bibinfo{volume}{17}
  (\bibinfo{year}{1978}) \bibinfo{pages}{21--63}.
\bibitem[{Wasserman and Anderson(1987)}]{wasserman1987stochastic}
\bibinfo{author}{S.~Wasserman}, \bibinfo{author}{C.~Anderson},
\newblock \bibinfo{title}{Stochastic a posteriori blockmodels: {C}onstruction
  and assessment},
\newblock \bibinfo{journal}{Soc. Netw.} \bibinfo{volume}{9}
  (\bibinfo{year}{1987}) \bibinfo{pages}{1--36}.
\bibitem[{Peixoto(2017)}]{peixoto2017nonparametric}
\bibinfo{author}{T.~P. Peixoto},
\newblock \bibinfo{title}{Nonparametric {B}ayesian inference of the
  microcanonical stochastic block model},
\newblock \bibinfo{journal}{Phys. Rev. E} \bibinfo{volume}{95}
  (\bibinfo{year}{2017}) \bibinfo{pages}{012317}.
\bibitem[{Karrer and Newman(2011)}]{karrer2011stochastic}
\bibinfo{author}{B.~Karrer}, \bibinfo{author}{M.~E. Newman},
\newblock \bibinfo{title}{Stochastic blockmodels and community structure in
  networks},
\newblock \bibinfo{journal}{Phys. Rev. E} \bibinfo{volume}{83}
  (\bibinfo{year}{2011}) \bibinfo{pages}{016107}.
\bibitem[{Gulikers et~al.(2017)Gulikers, Lelarge, and
  Massouli{\'e}}]{gulikers2017spectral}
\bibinfo{author}{L.~Gulikers}, \bibinfo{author}{M.~Lelarge},
  \bibinfo{author}{L.~Massouli{\'e}},
\newblock \bibinfo{title}{A spectral method for community detection in
  moderately sparse degree-corrected stochastic block models},
\newblock \bibinfo{journal}{Adv. Appl. Probab.} \bibinfo{volume}{49}
  (\bibinfo{year}{2017}) \bibinfo{pages}{686--721}.
\bibitem[{Chen et~al.(2018)Chen, Li, and Xu}]{chen2018convexified}
\bibinfo{author}{Y.~Chen}, \bibinfo{author}{X.~Li}, \bibinfo{author}{J.~Xu},
\newblock \bibinfo{title}{Convexified modularity maximization for
  degree-corrected stochastic block models},
\newblock \bibinfo{journal}{Ann. Stat.} \bibinfo{volume}{46}
  (\bibinfo{year}{2018}) \bibinfo{pages}{1573--1602}.
\bibitem[{Newman and Girvan(2004)}]{newman2004finding}
\bibinfo{author}{M.~E. Newman}, \bibinfo{author}{M.~Girvan},
\newblock \bibinfo{title}{Finding and evaluating community structure in
  networks},
\newblock \bibinfo{journal}{Phys. Rev. E} \bibinfo{volume}{69}
  (\bibinfo{year}{2004}) \bibinfo{pages}{026113}.
\bibitem[{Newman(2016)}]{newman2016equivalence}
\bibinfo{author}{M.~E. Newman},
\newblock \bibinfo{title}{Equivalence between modularity optimization and
  maximum likelihood methods for community detection},
\newblock \bibinfo{journal}{Phys. Rev. E} \bibinfo{volume}{94}
  (\bibinfo{year}{2016}) \bibinfo{pages}{052315}.
\bibitem[{Zhang and Peixoto(2020)}]{zhang2020statistical}
\bibinfo{author}{L.~Zhang}, \bibinfo{author}{T.~P. Peixoto},
\newblock \bibinfo{title}{Statistical inference of assortative community
  structures},
\newblock \bibinfo{journal}{Phys. Rev. Res.} \bibinfo{volume}{2}
  (\bibinfo{year}{2020}) \bibinfo{pages}{043271}.
\bibitem[{Peixoto(2012)}]{peixoto2012entropy}
\bibinfo{author}{T.~P. Peixoto},
\newblock \bibinfo{title}{Entropy of stochastic blockmodel ensembles},
\newblock \bibinfo{journal}{Phys. Rev. E} \bibinfo{volume}{85}
  (\bibinfo{year}{2012}) \bibinfo{pages}{056122}.
\bibitem[{Rissanen(1978)}]{rissanen1978modeling}
\bibinfo{author}{J.~Rissanen},
\newblock \bibinfo{title}{Modeling by shortest data description},
\newblock \bibinfo{journal}{Automatica} \bibinfo{volume}{14}
  (\bibinfo{year}{1978}) \bibinfo{pages}{465--471}.
\bibitem[{Rosvall and Bergstrom(2007)}]{rosvall2007information}
\bibinfo{author}{M.~Rosvall}, \bibinfo{author}{C.~T. Bergstrom},
\newblock \bibinfo{title}{An information-theoretic framework for resolving
  community structure in complex networks},
\newblock \bibinfo{journal}{Proc. Natl. Acad. Sci.} \bibinfo{volume}{104}
  (\bibinfo{year}{2007}) \bibinfo{pages}{7327--7331}.
\bibitem[{Peixoto(2013)}]{peixoto2013parsimonious}
\bibinfo{author}{T.~P. Peixoto},
\newblock \bibinfo{title}{Parsimonious module inference in large networks},
\newblock \bibinfo{journal}{Phys. Rev. Lett.} \bibinfo{volume}{110}
  (\bibinfo{year}{2013}) \bibinfo{pages}{148701}.
\bibitem[{Peixoto(2014{\natexlab{a}})}]{peixoto2014efficient}
\bibinfo{author}{T.~P. Peixoto},
\newblock \bibinfo{title}{Efficient {M}onte {C}arlo and greedy heuristic for
  the inference of stochastic block models},
\newblock \bibinfo{journal}{Phys. Rev. E} \bibinfo{volume}{89}
  (\bibinfo{year}{2014}{\natexlab{a}}) \bibinfo{pages}{012804}.
\bibitem[{Peixoto(2014{\natexlab{b}})}]{peixoto2014hierarchical}
\bibinfo{author}{T.~P. Peixoto},
\newblock \bibinfo{title}{Hierarchical block structures and high-resolution
  model selection in large networks},
\newblock \bibinfo{journal}{Phys. Rev. X} \bibinfo{volume}{4}
  (\bibinfo{year}{2014}{\natexlab{b}}) \bibinfo{pages}{011047}.
\bibitem[{Peixoto(2015)}]{peixoto2015model}
\bibinfo{author}{T.~P. Peixoto},
\newblock \bibinfo{title}{Model selection and hypothesis testing for
  large-scale network models with overlapping groups},
\newblock \bibinfo{journal}{Phys. Rev. X} \bibinfo{volume}{5}
  (\bibinfo{year}{2015}) \bibinfo{pages}{011033}.
\bibitem[{Lee and Wilkinson(2019)}]{lee2019review}
\bibinfo{author}{C.~Lee}, \bibinfo{author}{D.~J. Wilkinson},
\newblock \bibinfo{title}{A review of stochastic block models and extensions
  for graph clustering},
\newblock \bibinfo{journal}{Appl. Netw. Sci.} \bibinfo{volume}{4}
  (\bibinfo{year}{2019}) \bibinfo{pages}{122}.
\bibitem[{Jin et~al.(2021)Jin, Yu, Jiao, Pan, He, Wu, Yu, and
  Zhang}]{jin2021survey}
\bibinfo{author}{D.~Jin}, \bibinfo{author}{Z.~Yu}, \bibinfo{author}{P.~Jiao},
  \bibinfo{author}{S.~Pan}, \bibinfo{author}{D.~He}, \bibinfo{author}{J.~Wu},
  \bibinfo{author}{P.~S. Yu}, \bibinfo{author}{W.~Zhang}, \bibinfo{title}{A
  survey of community detection approaches: {F}rom statistical modeling to deep
  learning}, \bibinfo{year}{2021}. \bibinfo{note}{{e}-print arXiv:2101.01669}.
\bibitem[{Abbe(2017)}]{abbe2017community}
\bibinfo{author}{E.~Abbe},
\newblock \bibinfo{title}{Community detection and stochastic block models:
  recent developments},
\newblock \bibinfo{journal}{J. Mach. Learn. Res.} \bibinfo{volume}{18}
  (\bibinfo{year}{2017}) \bibinfo{pages}{6446--6531}.
\bibitem[{Newman(2010)}]{newman2010networks}
\bibinfo{author}{M.~E.~J. Newman}, \bibinfo{title}{Networks: {A}n
  introduction}, \bibinfo{publisher}{Oxford University Press},
  \bibinfo{year}{2010}.
\bibitem[{Fosdick et~al.(2018)Fosdick, Larremore, Nishimura, and
  Ugander}]{fosdick2018configuring}
\bibinfo{author}{B.~K. Fosdick}, \bibinfo{author}{D.~B. Larremore},
  \bibinfo{author}{J.~Nishimura}, \bibinfo{author}{J.~Ugander},
\newblock \bibinfo{title}{Configuring random graph models with fixed degree
  sequences},
\newblock \bibinfo{journal}{SIAM Review} \bibinfo{volume}{60}
  (\bibinfo{year}{2018}) \bibinfo{pages}{315--355}.
\bibitem[{Bianconi(2009)}]{bianconi2009entropy}
\bibinfo{author}{G.~Bianconi},
\newblock \bibinfo{title}{Entropy of network ensembles},
\newblock \bibinfo{journal}{Phys. Rev. E} \bibinfo{volume}{79}
  (\bibinfo{year}{2009}) \bibinfo{pages}{036114}.
\bibitem[{Squartini et~al.(2015)Squartini, de~Mol, den Hollander, and
  Garlaschelli}]{squartini2015breaking}
\bibinfo{author}{T.~Squartini}, \bibinfo{author}{J.~de~Mol},
  \bibinfo{author}{F.~den Hollander}, \bibinfo{author}{D.~Garlaschelli},
\newblock \bibinfo{title}{Breaking of ensemble equivalence in networks},
\newblock \bibinfo{journal}{Phys. Rev. Lett.} \bibinfo{volume}{115}
  (\bibinfo{year}{2015}) \bibinfo{pages}{268701}.
\bibitem[{Xing et~al.(2010)Xing, Fu, and Song}]{xing2010state}
\bibinfo{author}{E.~P. Xing}, \bibinfo{author}{W.~Fu},
  \bibinfo{author}{L.~Song},
\newblock \bibinfo{title}{A state-space mixed membership blockmodel for dynamic
  network tomography},
\newblock \bibinfo{journal}{Ann. Appl. Stat.} \bibinfo{volume}{4}
  (\bibinfo{year}{2010}) \bibinfo{pages}{535--566}.
\bibitem[{Yan(2016)}]{yan2016bayesian}
\bibinfo{author}{X.~Yan},
\newblock \bibinfo{title}{Bayesian model selection of stochastic block models},
\newblock in: \bibinfo{editor}{V.~S.~S. Subrahmanian},
  \bibinfo{editor}{J.~Rokne}, \bibinfo{editor}{R.~Kumar},
  \bibinfo{editor}{J.~Caverlee}, \bibinfo{editor}{H.~Tong} (Eds.),
  \bibinfo{booktitle}{Proceedings of the 2016 IEEE/ACM International Conference
  on Advances in Social Networks Analysis and Mining},
  \bibinfo{publisher}{IEEE}, \bibinfo{year}{2016}, pp.
  \bibinfo{pages}{323--328}.
\bibitem[{Wang and Bickel(2017)}]{wang2017likelihood}
\bibinfo{author}{Y.~X.~R. Wang}, \bibinfo{author}{P.~J. Bickel},
\newblock \bibinfo{title}{Likelihood-based model selection for stochastic block
  models},
\newblock \bibinfo{journal}{Ann. Stat.} \bibinfo{volume}{45}
  (\bibinfo{year}{2017}) \bibinfo{pages}{500--528}.
\bibitem[{Hu et~al.(2020)Hu, Qin, Yan, and Zhao}]{hu2020corrected}
\bibinfo{author}{J.~Hu}, \bibinfo{author}{H.~Qin}, \bibinfo{author}{T.~Yan},
  \bibinfo{author}{Y.~Zhao},
\newblock \bibinfo{title}{Corrected {B}ayesian information criterion for
  stochastic block models},
\newblock \bibinfo{journal}{J. Am. Stat. Assoc.} \bibinfo{volume}{115}
  (\bibinfo{year}{2020}) \bibinfo{pages}{1771--1783}.
\bibitem[{Newman and Reinert(2016)}]{newman2016estimating}
\bibinfo{author}{M.~E.~J. Newman}, \bibinfo{author}{G.~Reinert},
\newblock \bibinfo{title}{Estimating the number of communities in a network},
\newblock \bibinfo{journal}{Phys. Rev. Lett.} \bibinfo{volume}{117}
  (\bibinfo{year}{2016}) \bibinfo{pages}{078301}.
\bibitem[{C{\^o}me and Latouche(2015)}]{come2015model}
\bibinfo{author}{E.~C{\^o}me}, \bibinfo{author}{P.~Latouche},
\newblock \bibinfo{title}{Model selection and clustering in stochastic block
  models based on the exact integrated complete data likelihood},
\newblock \bibinfo{journal}{Stat. Model.} \bibinfo{volume}{15}
  (\bibinfo{year}{2015}) \bibinfo{pages}{564--589}.
\bibitem[{Hastings(1970)}]{hastings1970monte}
\bibinfo{author}{W.~K. Hastings},
\newblock \bibinfo{title}{Monte carlo sampling methods using {M}arkov chains
  and their applications},
\newblock \bibinfo{journal}{Biometrika} \bibinfo{volume}{57}
  (\bibinfo{year}{1970}) \bibinfo{pages}{97--109}.
\bibitem[{Chib and Greenberg(1995)}]{chib1995understanding}
\bibinfo{author}{S.~Chib}, \bibinfo{author}{E.~Greenberg},
\newblock \bibinfo{title}{Understanding the {M}etropolis-{H}astings algorithm},
\newblock \bibinfo{journal}{Am. Stat.} \bibinfo{volume}{49}
  (\bibinfo{year}{1995}) \bibinfo{pages}{327--335}.
\bibitem[{Gopalan et~al.(2012)Gopalan, Gerrish, Freedman, Blei, and
  Mimno}]{gopalan2012scalable}
\bibinfo{author}{P.~K. Gopalan}, \bibinfo{author}{S.~Gerrish},
  \bibinfo{author}{M.~Freedman}, \bibinfo{author}{D.~Blei},
  \bibinfo{author}{D.~Mimno},
\newblock \bibinfo{title}{Scalable inference of overlapping communities},
\newblock in: \bibinfo{editor}{F.~Pereira}, \bibinfo{editor}{C.~J.~C. Burges},
  \bibinfo{editor}{L.~Bottou}, \bibinfo{editor}{K.~Q. Weinberger} (Eds.),
  \bibinfo{booktitle}{Advances in Neural Information Processing Systems},
  volume~\bibinfo{volume}{25}, \bibinfo{publisher}{Neural Information
  Processing Systems Foundation, Inc.}, \bibinfo{year}{2012}, pp.
  \bibinfo{pages}{2249--2257}.
\bibitem[{Kim et~al.(2013)Kim, Gopalan, Blei, and Sudderth}]{kim2013efficient}
\bibinfo{author}{D.~I. Kim}, \bibinfo{author}{P.~K. Gopalan},
  \bibinfo{author}{D.~Blei}, \bibinfo{author}{E.~Sudderth},
\newblock \bibinfo{title}{Efficient online inference for {B}ayesian
  nonparametric relational models},
\newblock in: \bibinfo{booktitle}{Advances in Neural Information Processing
  Systems}, volume~\bibinfo{volume}{26}, \bibinfo{publisher}{Neural Information
  Processing Systems Foundation, Inc.}, \bibinfo{year}{2013}, pp.
  \bibinfo{pages}{962--970}.
\bibitem[{Hayashi et~al.(2016)Hayashi, Konishi, and
  Kawamoto}]{hayashi2016tractable}
\bibinfo{author}{K.~Hayashi}, \bibinfo{author}{T.~Konishi},
  \bibinfo{author}{T.~Kawamoto}, \bibinfo{title}{A tractable fully {B}ayesian
  method for the stochastic block model}, \bibinfo{year}{2016}.
  \bibinfo{note}{{e}-print arXiv:1602.02256}.
\bibitem[{Matias et~al.(2018)Matias, Rebafka, and
  Villers}]{matias2018semiparametric}
\bibinfo{author}{C.~Matias}, \bibinfo{author}{T.~Rebafka},
  \bibinfo{author}{F.~Villers},
\newblock \bibinfo{title}{A semiparametric extension of the stochastic block
  model for longitudinal networks},
\newblock \bibinfo{journal}{Biometrika} \bibinfo{volume}{105}
  (\bibinfo{year}{2018}) \bibinfo{pages}{665--680}.
\bibitem[{Perozzi et~al.(2014)Perozzi, Al-Rfou, and
  Skiena}]{perozzi2014deepwalk}
\bibinfo{author}{B.~Perozzi}, \bibinfo{author}{R.~Al-Rfou},
  \bibinfo{author}{S.~Skiena},
\newblock \bibinfo{title}{Deep{W}alk: {O}nline learning of social
  representations},
\newblock in: \bibinfo{editor}{S.~Macskassy}, \bibinfo{editor}{C.~Perlich},
  \bibinfo{editor}{J.~Leskovec}, \bibinfo{editor}{W.~Wang},
  \bibinfo{editor}{R.~Ghani} (Eds.), \bibinfo{booktitle}{Proceedings of the
  20th ACM SIGKDD international conference on Knowledge discovery and data
  mining}, \bibinfo{publisher}{Association for Computing Machinery},
  \bibinfo{year}{2014}, p. \bibinfo{pages}{701–710}.
\bibitem[{Coifman et~al.(2005)Coifman, Lafon, Lee, Maggioni, Nadler, Warner,
  and Zucker}]{coifman2005geometric}
\bibinfo{author}{R.~R. Coifman}, \bibinfo{author}{S.~Lafon},
  \bibinfo{author}{A.~B. Lee}, \bibinfo{author}{M.~Maggioni},
  \bibinfo{author}{B.~Nadler}, \bibinfo{author}{F.~Warner},
  \bibinfo{author}{S.~W. Zucker},
\newblock \bibinfo{title}{Geometric diffusions as a tool for harmonic analysis
  and structure definition of data: {D}iffusion maps},
\newblock \bibinfo{journal}{Proc. Natl. Acad. Sci. USA} \bibinfo{volume}{102}
  (\bibinfo{year}{2005}) \bibinfo{pages}{7426--7431}.
\bibitem[{Grover and Leskovec(2016)}]{grover2016node2vec}
\bibinfo{author}{A.~Grover}, \bibinfo{author}{J.~Leskovec},
\newblock \bibinfo{title}{node2vec: Scalable feature learning for networks},
\newblock in: \bibinfo{editor}{B.~Krishnapuram}, \bibinfo{editor}{M.~Shah},
  \bibinfo{editor}{A.~Smola}, \bibinfo{editor}{C.~Aggarwal},
  \bibinfo{editor}{D.~Shen}, \bibinfo{editor}{R.~Rastogi} (Eds.),
  \bibinfo{booktitle}{Proceedings of the 22nd {ACM SIGKDD} international
  conference on {K}nowledge discovery and data mining},
  \bibinfo{publisher}{Association for Computing Machinery},
  \bibinfo{year}{2016}, pp. \bibinfo{pages}{855--864}.
\bibitem[{Singhal(2012)}]{singhal2012introducing}
\bibinfo{author}{A.~Singhal}, \bibinfo{title}{Introducing the {K}nowledge
  {G}raph: things, not strings}, \bibinfo{year}{2012}. \bibinfo{note}{Google
  Blog. Available at:
  \url{https://blog.google/products/search/introducing-knowledge-graph-things-not/}}.
\bibitem[{Ugander et~al.(2011)Ugander, Karrer, Backstrom, and
  Marlow}]{ugander2011anatomy}
\bibinfo{author}{J.~Ugander}, \bibinfo{author}{B.~Karrer},
  \bibinfo{author}{L.~Backstrom}, \bibinfo{author}{C.~Marlow},
  \bibinfo{title}{The anatomy of the {F}acebook social graph},
  \bibinfo{year}{2011}. \bibinfo{note}{{e}-print arXiv:1111.4503}.
\bibitem[{Madey(2020)}]{madey2014linkedin}
\bibinfo{author}{D.~Madey}, \bibinfo{title}{{LinkedIn’s Vision for an
  Economic Graph: A Conversation With Jeff Weiner and Thomas Friedman}},
  \bibinfo{year}{2020}. \bibinfo{note}{The Economic Graph blog. Available at:
  \url{http://blog.linkedin.com/2014/06/30/linkedins-vision-for-an-economic-graph-a-conversation-with-jeff-weiner-and-thomas-friedman/}}.
\bibitem[{Walle(2020)}]{walle2020introducing}
\bibinfo{author}{T.~Walle}, \bibinfo{title}{{Introducing: The Real World
  Graph}}, \bibinfo{year}{2020}. \bibinfo{note}{Unacast blog. Available at:
  \url{https://www.unacast.com/post/introducing-the-real-world-graph}}.
\bibitem[{Milinovich(2017)}]{milinovich2017introducing}
\bibinfo{author}{J.~Milinovich}, \bibinfo{title}{{Introducing the Pinterest
  Taste Graph and enhanced targeting}}, \bibinfo{year}{2017}.
  \bibinfo{note}{Pinterest blog. Available at:
  \url{https://business.pinterest.com/en/blog/introducing-the-pinterest-taste-graph-and-enhanced-targeting}}.
\bibitem[{Sharma(2008)}]{sharma2009dikw}
\bibinfo{author}{N.~Sharma}, \bibinfo{title}{{The Origin of Data Information
  Knowledge Wisdom (DIKW) Hierarchy}}, \bibinfo{year}{2008}.
  \bibinfo{note}{Google Inc. Archived at:
  \url{https://doi.org/10.17605/OSF.IO/N7G9X}}.
\bibitem[{Ballandies et~al.(2018)Ballandies, Dapp, and
  Pournaras}]{ballandies2018decrypting}
\bibinfo{author}{M.~C. Ballandies}, \bibinfo{author}{M.~M. Dapp},
  \bibinfo{author}{E.~Pournaras}, \bibinfo{title}{Decrypting distributed ledger
  design-taxonomy, classification and blockchain community evaluation},
  \bibinfo{year}{2018}. \bibinfo{note}{{e}-print arXiv:1811.03419}.
\bibitem[{Wang et~al.(2019)Wang, Zha, Ni, Liu, Guo, Niu, and
  Zheng}]{wang2019survey}
\bibinfo{author}{X.~Wang}, \bibinfo{author}{X.~Zha}, \bibinfo{author}{W.~Ni},
  \bibinfo{author}{R.~P. Liu}, \bibinfo{author}{Y.~J. Guo},
  \bibinfo{author}{X.~Niu}, \bibinfo{author}{K.~Zheng},
\newblock \bibinfo{title}{Survey on blockchain for {I}nternet of {T}hings},
\newblock \bibinfo{journal}{Comput. Commun.} \bibinfo{volume}{136}
  (\bibinfo{year}{2019}) \bibinfo{pages}{10--29}.
\bibitem[{Zhou et~al.(2019)Zhou, Yu, Chen, and Kuo}]{zhou2019cyber}
\bibinfo{author}{Y.~Zhou}, \bibinfo{author}{F.~R. Yu},
  \bibinfo{author}{J.~Chen}, \bibinfo{author}{Y.~Kuo},
\newblock \bibinfo{title}{{Cyber-Physical-Social Systems: A State-of-the-Art
  Survey, Challenges and Opportunities}},
\newblock \bibinfo{journal}{IEEE Commun. Surv. Tutor.} \bibinfo{volume}{22}
  (\bibinfo{year}{2019}) \bibinfo{pages}{389--425}.
\bibitem[{Pticek et~al.(2016)Pticek, Podobnik, and Jezic}]{pticek2016beyond}
\bibinfo{author}{M.~Pticek}, \bibinfo{author}{V.~Podobnik},
  \bibinfo{author}{G.~Jezic},
\newblock \bibinfo{title}{Beyond the internet of things: the social networking
  of machines},
\newblock \bibinfo{journal}{Int. J. Distrib. Sens. Netw.} \bibinfo{volume}{12}
  (\bibinfo{year}{2016}) \bibinfo{pages}{8178417}.
\bibitem[{Skala et~al.(2015)Skala, Davidovic, Afgan, Sovic, and
  Sojat}]{skala2015scalable}
\bibinfo{author}{K.~Skala}, \bibinfo{author}{D.~Davidovic},
  \bibinfo{author}{E.~Afgan}, \bibinfo{author}{I.~Sovic},
  \bibinfo{author}{Z.~Sojat},
\newblock \bibinfo{title}{Scalable distributed computing hierarchy: {C}loud,
  fog and dew computing},
\newblock \bibinfo{journal}{Open J. Cloud Comput. (OJCC)} \bibinfo{volume}{2}
  (\bibinfo{year}{2015}) \bibinfo{pages}{16--24}.
\bibitem[{Guberovi{\'c} et~al.(2021)Guberovi{\'c}, Lipi{\'c}, and
  {\v{C}}avrak}]{guberovic2021dew}
\bibinfo{author}{E.~Guberovi{\'c}}, \bibinfo{author}{T.~Lipi{\'c}},
  \bibinfo{author}{I.~{\v{C}}avrak},
\newblock \bibinfo{title}{Dew {I}ntelligence: {F}ederated learning
  perspective},
\newblock in: \bibinfo{editor}{H.~Shahriar} (Ed.), \bibinfo{booktitle}{2021
  IEEE 45th Annual Computers, Software, and Applications Conference (COMPSAC)},
  \bibinfo{publisher}{IEEE}, \bibinfo{year}{2021}, pp.
  \bibinfo{pages}{1819--1824}.
\bibitem[{Wang et~al.(2020)Wang, Han, Leung, Niyato, Yan, and
  Chen}]{wang2020convergence}
\bibinfo{author}{X.~Wang}, \bibinfo{author}{Y.~Han}, \bibinfo{author}{V.~C.
  Leung}, \bibinfo{author}{D.~Niyato}, \bibinfo{author}{X.~Yan},
  \bibinfo{author}{X.~Chen},
\newblock \bibinfo{title}{Convergence of edge computing and deep learning: {A}
  comprehensive survey},
\newblock \bibinfo{journal}{IEEE Commun. Surv. Tutor.} \bibinfo{volume}{22}
  (\bibinfo{year}{2020}) \bibinfo{pages}{869--904}.
\bibitem[{Deng et~al.(2020)Deng, Zhao, Fang, Yin, Dustdar, and
  Zomaya}]{deng2020edge}
\bibinfo{author}{S.~Deng}, \bibinfo{author}{H.~Zhao},
  \bibinfo{author}{W.~Fang}, \bibinfo{author}{J.~Yin},
  \bibinfo{author}{S.~Dustdar}, \bibinfo{author}{A.~Y. Zomaya},
\newblock \bibinfo{title}{Edge intelligence: the confluence of edge computing
  and artificial intelligence},
\newblock \bibinfo{journal}{IEEE Internet Things J.} \bibinfo{volume}{7}
  (\bibinfo{year}{2020}) \bibinfo{pages}{7457--7469}.
\bibitem[{Lim et~al.(2020)Lim, Luong, Hoang, Jiao, Liang, Yang, Niyato, and
  Miao}]{lim2020federated}
\bibinfo{author}{W.~Y.~B. Lim}, \bibinfo{author}{N.~C. Luong},
  \bibinfo{author}{D.~T. Hoang}, \bibinfo{author}{Y.~Jiao},
  \bibinfo{author}{Y.-C. Liang}, \bibinfo{author}{Q.~Yang},
  \bibinfo{author}{D.~Niyato}, \bibinfo{author}{C.~Miao},
\newblock \bibinfo{title}{Federated learning in mobile edge networks: {A}
  comprehensive survey},
\newblock \bibinfo{journal}{IEEE Commun. Surv. Tutor.} \bibinfo{volume}{22}
  (\bibinfo{year}{2020}) \bibinfo{pages}{2031--2063}.
\bibitem[{Wang et~al.(2020)Wang, Zhang, Wang, Ma, and Liu}]{wang2020deep}
\bibinfo{author}{F.~Wang}, \bibinfo{author}{M.~Zhang},
  \bibinfo{author}{X.~Wang}, \bibinfo{author}{X.~Ma}, \bibinfo{author}{J.~Liu},
\newblock \bibinfo{title}{{Deep Learning for Edge Computing Applications: A
  State-of-the-Art Survey}},
\newblock \bibinfo{journal}{IEEE Access} \bibinfo{volume}{8}
  (\bibinfo{year}{2020}) \bibinfo{pages}{58322--58336}.
\bibitem[{Tsvetkova et~al.(2017)Tsvetkova, Yasseri, Meyer, Pickering, Engen,
  Walland, L{\"u}ders, F{\o}lstad, and Bravos}]{tsvetkova2017understanding}
\bibinfo{author}{M.~Tsvetkova}, \bibinfo{author}{T.~Yasseri},
  \bibinfo{author}{E.~T. Meyer}, \bibinfo{author}{J.~B. Pickering},
  \bibinfo{author}{V.~Engen}, \bibinfo{author}{P.~Walland},
  \bibinfo{author}{M.~L{\"u}ders}, \bibinfo{author}{A.~F{\o}lstad},
  \bibinfo{author}{G.~Bravos},
\newblock \bibinfo{title}{Understanding human-machine networks: a
  cross-disciplinary survey},
\newblock \bibinfo{journal}{ACM Comput. Surv.} \bibinfo{volume}{50}
  (\bibinfo{year}{2017}) \bibinfo{pages}{1--35}.
\bibitem[{Rahwan et~al.(2019)Rahwan, Cebrian, Obradovich, Bongard, Bonnefon,
  Breazeal, Crandall, Christakis, Couzin, Jackson et~al.}]{rahwan2019machine}
\bibinfo{author}{I.~Rahwan}, \bibinfo{author}{M.~Cebrian},
  \bibinfo{author}{N.~Obradovich}, \bibinfo{author}{J.~Bongard},
  \bibinfo{author}{J.-F. Bonnefon}, \bibinfo{author}{C.~Breazeal},
  \bibinfo{author}{J.~W. Crandall}, \bibinfo{author}{N.~A. Christakis},
  \bibinfo{author}{I.~D. Couzin}, \bibinfo{author}{M.~O. Jackson}, et~al.,
\newblock \bibinfo{title}{Machine behaviour},
\newblock \bibinfo{journal}{Nature} \bibinfo{volume}{568}
  (\bibinfo{year}{2019}) \bibinfo{pages}{477--486}.
\bibitem[{Lazer et~al.(2009)Lazer, Pentland, Adamic, Aral, Barabasi, Brewer,
  Christakis, Contractor, Fowler, Gutmann et~al.}]{lazer2009social}
\bibinfo{author}{D.~Lazer}, \bibinfo{author}{A.~Pentland},
  \bibinfo{author}{L.~Adamic}, \bibinfo{author}{S.~Aral},
  \bibinfo{author}{A.-L. Barabasi}, \bibinfo{author}{D.~Brewer},
  \bibinfo{author}{N.~Christakis}, \bibinfo{author}{N.~Contractor},
  \bibinfo{author}{J.~Fowler}, \bibinfo{author}{M.~Gutmann}, et~al.,
\newblock \bibinfo{title}{Computational social science.},
\newblock \bibinfo{journal}{Science} \bibinfo{volume}{323}
  (\bibinfo{year}{2009}) \bibinfo{pages}{721--723}.
\bibitem[{Holme and Liljeros(2015)}]{holme2015mechanistic}
\bibinfo{author}{P.~Holme}, \bibinfo{author}{F.~Liljeros},
\newblock \bibinfo{title}{Mechanistic models in computational social science},
\newblock \bibinfo{journal}{Front. Phys.} \bibinfo{volume}{3}
  (\bibinfo{year}{2015}) \bibinfo{pages}{78}.
\bibitem[{Schweitzer(2018)}]{schweitzer2018sociophysics}
\bibinfo{author}{F.~Schweitzer},
\newblock \bibinfo{title}{Sociophysics},
\newblock \bibinfo{journal}{Phys. Today} \bibinfo{volume}{71}
  (\bibinfo{year}{2018}) \bibinfo{pages}{40--46}.
\bibitem[{Bhattacharya and Kaski(2019)}]{bhattacharya2019social}
\bibinfo{author}{K.~Bhattacharya}, \bibinfo{author}{K.~Kaski},
\newblock \bibinfo{title}{Social physics: uncovering human behaviour from
  communication},
\newblock \bibinfo{journal}{Adv. Phys. X} \bibinfo{volume}{4}
  (\bibinfo{year}{2019}) \bibinfo{pages}{1527723}.
\bibitem[{Blumenstock et~al.(2015)Blumenstock, Cadamuro, and
  On}]{blumenstock2015predicting}
\bibinfo{author}{J.~Blumenstock}, \bibinfo{author}{G.~Cadamuro},
  \bibinfo{author}{R.~On},
\newblock \bibinfo{title}{Predicting poverty and wealth from mobile phone
  metadata},
\newblock \bibinfo{journal}{Science} \bibinfo{volume}{350}
  (\bibinfo{year}{2015}) \bibinfo{pages}{1073--1076}.
\bibitem[{Zhang et~al.(2016)Zhang, Brackbill, Yang, Becker, Herbert, and
  Centola}]{zhang2016support}
\bibinfo{author}{J.~Zhang}, \bibinfo{author}{D.~Brackbill},
  \bibinfo{author}{S.~Yang}, \bibinfo{author}{J.~Becker},
  \bibinfo{author}{N.~Herbert}, \bibinfo{author}{D.~Centola},
\newblock \bibinfo{title}{Support or competition? {H}ow online social networks
  increase physical activity: a randomized controlled trial},
\newblock \bibinfo{journal}{Prev. Med. Rep.} \bibinfo{volume}{4}
  (\bibinfo{year}{2016}) \bibinfo{pages}{453--458}.
\bibitem[{Jebb et~al.(2018)Jebb, Tay, Diener, and Oishi}]{jebb2018happiness}
\bibinfo{author}{A.~T. Jebb}, \bibinfo{author}{L.~Tay},
  \bibinfo{author}{E.~Diener}, \bibinfo{author}{S.~Oishi},
\newblock \bibinfo{title}{Happiness, income satiation and turning points around
  the world},
\newblock \bibinfo{journal}{Nat. Hum. Behav.} \bibinfo{volume}{2}
  (\bibinfo{year}{2018}) \bibinfo{pages}{33--38}.
\bibitem[{Bogomolov et~al.(2015)Bogomolov, Lepri, Staiano, Letouz{\'e}, Oliver,
  Pianesi, and Pentland}]{bogomolov2015moves}
\bibinfo{author}{A.~Bogomolov}, \bibinfo{author}{B.~Lepri},
  \bibinfo{author}{J.~Staiano}, \bibinfo{author}{E.~Letouz{\'e}},
  \bibinfo{author}{N.~Oliver}, \bibinfo{author}{F.~Pianesi},
  \bibinfo{author}{A.~Pentland},
\newblock \bibinfo{title}{Moves on the street: {C}lassifying crime hotspots
  using aggregated anonymized data on people dynamics},
\newblock \bibinfo{journal}{Big Data} \bibinfo{volume}{3}
  (\bibinfo{year}{2015}) \bibinfo{pages}{148--158}.
\bibitem[{Ginsberg et~al.(2009)Ginsberg, Mohebbi, Patel, Brammer, Smolinski,
  and Brilliant}]{ginsberg2009detecting}
\bibinfo{author}{J.~Ginsberg}, \bibinfo{author}{M.~H. Mohebbi},
  \bibinfo{author}{R.~S. Patel}, \bibinfo{author}{L.~Brammer},
  \bibinfo{author}{M.~S. Smolinski}, \bibinfo{author}{L.~Brilliant},
\newblock \bibinfo{title}{Detecting influenza epidemics using search engine
  query data},
\newblock \bibinfo{journal}{Nature} \bibinfo{volume}{457}
  (\bibinfo{year}{2009}) \bibinfo{pages}{1012--1014}.
\bibitem[{Wesolowski et~al.(2012)Wesolowski, Eagle, Tatem, Smith, Noor, Snow,
  and Buckee}]{wesolowski2012quantifying}
\bibinfo{author}{A.~Wesolowski}, \bibinfo{author}{N.~Eagle},
  \bibinfo{author}{A.~J. Tatem}, \bibinfo{author}{D.~L. Smith},
  \bibinfo{author}{A.~M. Noor}, \bibinfo{author}{R.~W. Snow},
  \bibinfo{author}{C.~O. Buckee},
\newblock \bibinfo{title}{Quantifying the impact of human mobility on malaria},
\newblock \bibinfo{journal}{Science} \bibinfo{volume}{338}
  (\bibinfo{year}{2012}) \bibinfo{pages}{267--270}.
\bibitem[{Wilson et~al.(2016)Wilson, zu~Erbach-Schoenberg, Albert, Power,
  Tudge, Gonzalez, Guthrie, Chamberlain, Brooks, Hughes
  et~al.}]{wilson2016rapid}
\bibinfo{author}{R.~Wilson}, \bibinfo{author}{E.~zu~Erbach-Schoenberg},
  \bibinfo{author}{M.~Albert}, \bibinfo{author}{D.~Power},
  \bibinfo{author}{S.~Tudge}, \bibinfo{author}{M.~Gonzalez},
  \bibinfo{author}{S.~Guthrie}, \bibinfo{author}{H.~Chamberlain},
  \bibinfo{author}{C.~Brooks}, \bibinfo{author}{C.~Hughes}, et~al.,
\newblock \bibinfo{title}{Rapid and near real-time assessments of population
  displacement using mobile phone data following disasters: the 2015 {N}epal
  {E}arthquake},
\newblock \bibinfo{journal}{PLOS Curr.} \bibinfo{volume}{8}
  (\bibinfo{year}{2016})
  \bibinfo{pages}{ecurrents.dis.d073fbece328e4c39087bc086d694b5c}.
\bibitem[{Ofli et~al.(2016)Ofli, Meier, Imran, Castillo, Tuia, Rey, Briant,
  Millet, Reinhard, Parkan et~al.}]{ofli2016combining}
\bibinfo{author}{F.~Ofli}, \bibinfo{author}{P.~Meier},
  \bibinfo{author}{M.~Imran}, \bibinfo{author}{C.~Castillo},
  \bibinfo{author}{D.~Tuia}, \bibinfo{author}{N.~Rey},
  \bibinfo{author}{J.~Briant}, \bibinfo{author}{P.~Millet},
  \bibinfo{author}{F.~Reinhard}, \bibinfo{author}{M.~Parkan}, et~al.,
\newblock \bibinfo{title}{Combining human computing and machine learning to
  make sense of big (aerial) data for disaster response},
\newblock \bibinfo{journal}{Big Data} \bibinfo{volume}{4}
  (\bibinfo{year}{2016}) \bibinfo{pages}{47--59}.
\bibitem[{Hobbs et~al.(2016)Hobbs, Burke, Christakis, and
  Fowler}]{hobbs2016online}
\bibinfo{author}{W.~R. Hobbs}, \bibinfo{author}{M.~Burke},
  \bibinfo{author}{N.~A. Christakis}, \bibinfo{author}{J.~H. Fowler},
\newblock \bibinfo{title}{Online social integration is associated with reduced
  mortality risk},
\newblock \bibinfo{journal}{Proc. Natl. Acad. Sci. USA} \bibinfo{volume}{113}
  (\bibinfo{year}{2016}) \bibinfo{pages}{12980--12984}.
\bibitem[{Podobnik et~al.(2017)Podobnik, Jusup, Kovac, and
  Stanley}]{podobnik2017predicting}
\bibinfo{author}{B.~Podobnik}, \bibinfo{author}{M.~Jusup},
  \bibinfo{author}{D.~Kovac}, \bibinfo{author}{H.~E. Stanley},
\newblock \bibinfo{title}{Predicting the rise of {EU} right-wing populism in
  response to unbalanced immigration},
\newblock \bibinfo{journal}{Complexity} \bibinfo{volume}{2017}
  (\bibinfo{year}{2017}).
\bibitem[{Lepri et~al.(2017)Lepri, Staiano, Sangokoya, Letouz{\'e}, and
  Oliver}]{lepri2017tyranny}
\bibinfo{author}{B.~Lepri}, \bibinfo{author}{J.~Staiano},
  \bibinfo{author}{D.~Sangokoya}, \bibinfo{author}{E.~Letouz{\'e}},
  \bibinfo{author}{N.~Oliver},
\newblock \bibinfo{title}{The tyranny of data? {T}he bright and dark sides of
  data-driven decision-making for social good},
\newblock in: \bibinfo{editor}{T.~Cerquitelli}, \bibinfo{editor}{D.~Quercia},
  \bibinfo{editor}{F.~Pasquale} (Eds.), \bibinfo{booktitle}{Transparent data
  mining for big and small data}, \bibinfo{publisher}{Springer},
  \bibinfo{year}{2017}, pp. \bibinfo{pages}{3--24}.
\bibitem[{Abebe and Goldner(2018)}]{abebe2018mechanism}
\bibinfo{author}{R.~Abebe}, \bibinfo{author}{K.~Goldner},
\newblock \bibinfo{title}{Mechanism design for social good},
\newblock \bibinfo{journal}{AI Matters} \bibinfo{volume}{4}
  (\bibinfo{year}{2018}) \bibinfo{pages}{27--34}.
\bibitem[{Shi et~al.(2020)Shi, Wang, and Fang}]{shi2020artificial}
\bibinfo{author}{Z.~R. Shi}, \bibinfo{author}{C.~Wang},
  \bibinfo{author}{F.~Fang}, \bibinfo{title}{Artificial intelligence for social
  good: {A} survey}, \bibinfo{year}{2020}. \bibinfo{note}{{e}-print
  arXiv:2001.01818}.
\bibitem[{Vinuesa et~al.(2020)Vinuesa, Azizpour, Leite, Balaam, Dignum,
  Domisch, Fell{\"a}nder, Langhans, Tegmark, and Nerini}]{vinuesa2020role}
\bibinfo{author}{R.~Vinuesa}, \bibinfo{author}{H.~Azizpour},
  \bibinfo{author}{I.~Leite}, \bibinfo{author}{M.~Balaam},
  \bibinfo{author}{V.~Dignum}, \bibinfo{author}{S.~Domisch},
  \bibinfo{author}{A.~Fell{\"a}nder}, \bibinfo{author}{S.~D. Langhans},
  \bibinfo{author}{M.~Tegmark}, \bibinfo{author}{F.~F. Nerini},
\newblock \bibinfo{title}{The role of artificial intelligence in achieving the
  {S}ustainable {D}evelopment {G}oals},
\newblock \bibinfo{journal}{Nat. Commun.} \bibinfo{volume}{11}
  (\bibinfo{year}{2020}) \bibinfo{pages}{1--10}.
\bibitem[{De~Montjoye et~al.(2013)De~Montjoye, Hidalgo, Verleysen, and
  Blondel}]{demontjoye2013unique}
\bibinfo{author}{Y.-A. De~Montjoye}, \bibinfo{author}{C.~A. Hidalgo},
  \bibinfo{author}{M.~Verleysen}, \bibinfo{author}{V.~D. Blondel},
\newblock \bibinfo{title}{Unique in the crowd: {T}he privacy bounds of human
  mobility},
\newblock \bibinfo{journal}{Sci. Rep.} \bibinfo{volume}{3}
  (\bibinfo{year}{2013}) \bibinfo{pages}{1376}.
\bibitem[{De~Montjoye et~al.(2015)De~Montjoye, Radaelli, Singh, and
  Pentland}]{demontjoye2015unique}
\bibinfo{author}{Y.-A. De~Montjoye}, \bibinfo{author}{L.~Radaelli},
  \bibinfo{author}{V.~K. Singh}, \bibinfo{author}{A.~S. Pentland},
\newblock \bibinfo{title}{Unique in the shopping mall: {O}n the
  reidentifiability of credit card metadata},
\newblock \bibinfo{journal}{Science} \bibinfo{volume}{347}
  (\bibinfo{year}{2015}) \bibinfo{pages}{536--539}.
\bibitem[{Pasquale(2015)}]{pasquale2015black}
\bibinfo{author}{F.~Pasquale}, \bibinfo{title}{The black box society},
  \bibinfo{publisher}{Harvard University Press}, \bibinfo{year}{2015}.
\bibitem[{Ruths and Pfeffer(2014)}]{ruths2014social}
\bibinfo{author}{D.~Ruths}, \bibinfo{author}{J.~Pfeffer},
\newblock \bibinfo{title}{Social media for large studies of behavior},
\newblock \bibinfo{journal}{Science} \bibinfo{volume}{346}
  (\bibinfo{year}{2014}) \bibinfo{pages}{1063--1064}.
\bibitem[{Xie et~al.(2020)Xie, Ras, van Gerven, and Doran}]{xie2020explainable}
\bibinfo{author}{N.~Xie}, \bibinfo{author}{G.~Ras}, \bibinfo{author}{M.~van
  Gerven}, \bibinfo{author}{D.~Doran}, \bibinfo{title}{Explainable deep
  learning: {A} field guide for the uninitiated}, \bibinfo{year}{2020}.
  \bibinfo{note}{{e}-print arXiv:2004.14545}.
\bibitem[{Floridi(2019)}]{floridi2019establishing}
\bibinfo{author}{L.~Floridi},
\newblock \bibinfo{title}{Establishing the rules for building trustworthy
  {AI}},
\newblock \bibinfo{journal}{Nat. Mach. Intell.} \bibinfo{volume}{1}
  (\bibinfo{year}{2019}) \bibinfo{pages}{261--262}.
\bibitem[{Theodorou and Dignum(2020)}]{theodorou2020towards}
\bibinfo{author}{A.~Theodorou}, \bibinfo{author}{V.~Dignum},
\newblock \bibinfo{title}{Towards ethical and socio-legal governance in {AI}},
\newblock \bibinfo{journal}{Nat. Mach. Intell.} \bibinfo{volume}{2}
  (\bibinfo{year}{2020}) \bibinfo{pages}{10--12}.
\bibitem[{Ala-Pietil\"{a} et~al.(2019)}]{alapietila2019ethics}
\bibinfo{author}{P.~Ala-Pietil\"{a}}, et~al., \bibinfo{title}{Ethics guidelines
  for trustworthy ai}, \bibinfo{year}{2019}. \bibinfo{note}{Report by
  High-Level Expert Group on Artificial Intelligence. Available at:
  \url{https://ec.europa.eu/newsroom/dae/document.cfm?doc_id=60419}. Archived
  at: \url{https://doi.org/10.17605/OSF.IO/N7G9X}}.
\bibitem[{Zdeborov{\'a}(2017)}]{zdeborova2017machine}
\bibinfo{author}{L.~Zdeborov{\'a}},
\newblock \bibinfo{title}{Machine learning: {N}ew tool in the box},
\newblock \bibinfo{journal}{Nat. Phys.} \bibinfo{volume}{13}
  (\bibinfo{year}{2017}) \bibinfo{pages}{420--421}.
\bibitem[{Shirado and Christakis(2017)}]{shirado2017locally}
\bibinfo{author}{H.~Shirado}, \bibinfo{author}{N.~A. Christakis},
\newblock \bibinfo{title}{Locally noisy autonomous agents improve global human
  coordination in network experiments},
\newblock \bibinfo{journal}{Nature} \bibinfo{volume}{545}
  (\bibinfo{year}{2017}) \bibinfo{pages}{370--374}.
\bibitem[{Crandall et~al.(2018)Crandall, Oudah, Ishowo-Oloko, Abdallah,
  Bonnefon, Cebrian, Shariff, Goodrich, Rahwan
  et~al.}]{crandall2018cooperating}
\bibinfo{author}{J.~W. Crandall}, \bibinfo{author}{M.~Oudah},
  \bibinfo{author}{F.~Ishowo-Oloko}, \bibinfo{author}{S.~Abdallah},
  \bibinfo{author}{J.-F. Bonnefon}, \bibinfo{author}{M.~Cebrian},
  \bibinfo{author}{A.~Shariff}, \bibinfo{author}{M.~A. Goodrich},
  \bibinfo{author}{I.~Rahwan}, et~al.,
\newblock \bibinfo{title}{Cooperating with machines},
\newblock \bibinfo{journal}{Nat. Commun.} \bibinfo{volume}{9}
  (\bibinfo{year}{2018}) \bibinfo{pages}{233}.
\bibitem[{Mohammadi et~al.(2018)Mohammadi, Al-Fuqaha, Sorour, and
  Guizani}]{mohammadi2018deep}
\bibinfo{author}{M.~Mohammadi}, \bibinfo{author}{A.~Al-Fuqaha},
  \bibinfo{author}{S.~Sorour}, \bibinfo{author}{M.~Guizani},
\newblock \bibinfo{title}{Deep learning for {IoT} big data and streaming
  analytics: {A} survey},
\newblock \bibinfo{journal}{IEEE Commun. Surv. Tutor.} \bibinfo{volume}{20}
  (\bibinfo{year}{2018}) \bibinfo{pages}{2923--2960}.
\bibitem[{Zhang et~al.(2019)Zhang, Yao, Sun, and Tay}]{zhang2019deep}
\bibinfo{author}{S.~Zhang}, \bibinfo{author}{L.~Yao}, \bibinfo{author}{A.~Sun},
  \bibinfo{author}{Y.~Tay},
\newblock \bibinfo{title}{Deep learning based recommender system: {A} survey
  and new perspectives},
\newblock \bibinfo{journal}{ACM Comput. Surv.} \bibinfo{volume}{52}
  (\bibinfo{year}{2019}) \bibinfo{pages}{1--38}.
\bibitem[{Raghu and Schmidt(2020)}]{raghu2020survey}
\bibinfo{author}{M.~Raghu}, \bibinfo{author}{E.~Schmidt}, \bibinfo{title}{A
  survey of deep learning for scientific discovery}, \bibinfo{year}{2020}.
  \bibinfo{note}{{e}-print arXiv:2003.11755}.
\bibitem[{Mehta et~al.(2019)Mehta, Bukov, Wang, Day, Richardson, Fisher, and
  Schwab}]{mehta2019high}
\bibinfo{author}{P.~Mehta}, \bibinfo{author}{M.~Bukov}, \bibinfo{author}{C.-H.
  Wang}, \bibinfo{author}{A.~G. Day}, \bibinfo{author}{C.~Richardson},
  \bibinfo{author}{C.~K. Fisher}, \bibinfo{author}{D.~J. Schwab},
\newblock \bibinfo{title}{A high-bias, low-variance introduction to machine
  learning for physicists},
\newblock \bibinfo{journal}{Phys. Rep.} \bibinfo{volume}{810}
  (\bibinfo{year}{2019}) \bibinfo{pages}{1--124}.
\bibitem[{Bengio(2009)}]{bengio2009learning}
\bibinfo{author}{Y.~Bengio}, \bibinfo{title}{Learning deep architectures for
  AI}, \bibinfo{publisher}{Now Publishers}, \bibinfo{year}{2009}.
\bibitem[{Bengio et~al.(2013)Bengio, Courville, and
  Vincent}]{bengio2013representation}
\bibinfo{author}{Y.~Bengio}, \bibinfo{author}{A.~Courville},
  \bibinfo{author}{P.~Vincent},
\newblock \bibinfo{title}{Representation learning: {A} review and new
  perspectives},
\newblock \bibinfo{journal}{IEEE Trans. Pattern Anal. Mach. Intell.}
  \bibinfo{volume}{35} (\bibinfo{year}{2013}) \bibinfo{pages}{1798--1828}.
\bibitem[{Deng and Yu(2014)}]{deng2014deep}
\bibinfo{author}{L.~Deng}, \bibinfo{author}{D.~Yu},
\newblock \bibinfo{title}{Deep learning: methods and applications},
\newblock \bibinfo{journal}{Found. Trends Signal Process.} \bibinfo{volume}{7}
  (\bibinfo{year}{2014}) \bibinfo{pages}{197--387}.
\bibitem[{LeCun et~al.(2015)LeCun, Bengio, and Hinton}]{lecun2015deep}
\bibinfo{author}{Y.~LeCun}, \bibinfo{author}{Y.~Bengio},
  \bibinfo{author}{G.~Hinton},
\newblock \bibinfo{title}{Deep learning},
\newblock \bibinfo{journal}{Nature} \bibinfo{volume}{521}
  (\bibinfo{year}{2015}) \bibinfo{pages}{436--444}.
\bibitem[{Schmidhuber(2015)}]{schmidhuber2015deep}
\bibinfo{author}{J.~Schmidhuber},
\newblock \bibinfo{title}{Deep learning in neural networks: {A}n overview},
\newblock \bibinfo{journal}{Neural Netw.} \bibinfo{volume}{61}
  (\bibinfo{year}{2015}) \bibinfo{pages}{85--117}.
\bibitem[{Goodfellow et~al.(2016)Goodfellow, Bengio, Courville, and
  Bengio}]{goodfellow2016deep}
\bibinfo{author}{I.~Goodfellow}, \bibinfo{author}{Y.~Bengio},
  \bibinfo{author}{A.~Courville}, \bibinfo{author}{Y.~Bengio},
  \bibinfo{title}{Deep learning}, \bibinfo{publisher}{The MIT Press},
  \bibinfo{year}{2016}.
\bibitem[{Liu et~al.(2017)Liu, Wang, Liu, Zeng, Liu, and
  Alsaadi}]{liu2017survey}
\bibinfo{author}{W.~Liu}, \bibinfo{author}{Z.~Wang}, \bibinfo{author}{X.~Liu},
  \bibinfo{author}{N.~Zeng}, \bibinfo{author}{Y.~Liu}, \bibinfo{author}{F.~E.
  Alsaadi},
\newblock \bibinfo{title}{A survey of deep neural network architectures and
  their applications},
\newblock \bibinfo{journal}{Neurocomputing} \bibinfo{volume}{234}
  (\bibinfo{year}{2017}) \bibinfo{pages}{11--26}.
\bibitem[{Fan et~al.(2019)Fan, Ma, and Zhong}]{fan2019selective}
\bibinfo{author}{J.~Fan}, \bibinfo{author}{C.~Ma}, \bibinfo{author}{Y.~Zhong},
  \bibinfo{title}{A selective overview of deep learning}, \bibinfo{year}{2019}.
  \bibinfo{note}{{e}-print arXiv:1904.05526}.
\bibitem[{Bullock et~al.(2020)Bullock, Luccioni, Pham, Lam, and
  Luengo-Oroz}]{bullock2020mapping}
\bibinfo{author}{J.~Bullock}, \bibinfo{author}{A.~Luccioni},
  \bibinfo{author}{K.~H. Pham}, \bibinfo{author}{C.~S.~N. Lam},
  \bibinfo{author}{M.~Luengo-Oroz}, \bibinfo{title}{Mapping the landscape of
  artificial intelligence applications against {COVID}-19},
  \bibinfo{year}{2020}. \bibinfo{note}{{e}-print arXiv:2003.11336}.
\bibitem[{Barua(2020)}]{barua2020understanding}
\bibinfo{author}{S.~Barua}, \bibinfo{title}{Understanding {C}oronanomics: {T}he
  economic implications of the coronavirus ({COVID}-19) pandemic},
  \bibinfo{year}{2020}. \bibinfo{note}{Social Science Research Network (SSRN)
  Paper No. 3566477. Available at: \url{https://ssrn.com/abstract=3566477}}.
\bibitem[{Baker et~al.(2020)Baker, Bloom, Davis, and Terry}]{baker2020covid}
\bibinfo{author}{S.~R. Baker}, \bibinfo{author}{N.~Bloom},
  \bibinfo{author}{S.~J. Davis}, \bibinfo{author}{S.~J. Terry},
  \bibinfo{title}{Covid-induced economic uncertainty}, \bibinfo{year}{2020}.
  \bibinfo{note}{National Bureau of Economic Research (NBER) Working Paper No.
  26983. Available at: \url{https://www.nber.org/papers/w26983}}.
\bibitem[{Nicola et~al.(2020)Nicola, Alsafi, Sohrabi, Kerwan, Al-Jabir,
  Iosifidis, Agha, and Agha}]{nicola2020socio}
\bibinfo{author}{M.~Nicola}, \bibinfo{author}{Z.~Alsafi},
  \bibinfo{author}{C.~Sohrabi}, \bibinfo{author}{A.~Kerwan},
  \bibinfo{author}{A.~Al-Jabir}, \bibinfo{author}{C.~Iosifidis},
  \bibinfo{author}{M.~Agha}, \bibinfo{author}{R.~Agha},
\newblock \bibinfo{title}{The socio-economic implications of the coronavirus
  and {COVID}-19 pandemic: a review},
\newblock \bibinfo{journal}{Int. J. Surg.} \bibinfo{volume}{78}
  (\bibinfo{year}{2020}) \bibinfo{pages}{185--193}.
\bibitem[{Zhang et~al.(2020)Zhang, Lyu, Liu, Zhang, Wang, and
  Luo}]{zhang2020monitoring}
\bibinfo{author}{Y.~Zhang}, \bibinfo{author}{H.~Lyu}, \bibinfo{author}{Y.~Liu},
  \bibinfo{author}{X.~Zhang}, \bibinfo{author}{Y.~Wang},
  \bibinfo{author}{J.~Luo}, \bibinfo{title}{Monitoring depression trend on
  {T}witter during the {COVID}-19 pandemic}, \bibinfo{year}{2020}.
  \bibinfo{note}{{e}-print arXiv:2007.00228}.
\bibitem[{Oliver et~al.(2018)Oliver, Odena, Raffel, Cubuk, and
  Goodfellow}]{oliver2018realistic}
\bibinfo{author}{A.~Oliver}, \bibinfo{author}{A.~Odena}, \bibinfo{author}{C.~A.
  Raffel}, \bibinfo{author}{E.~D. Cubuk}, \bibinfo{author}{I.~Goodfellow},
\newblock \bibinfo{title}{Realistic evaluation of deep semi-supervised learning
  algorithms},
\newblock in: \bibinfo{editor}{S.~Bengio}, \bibinfo{editor}{H.~Wallach},
  \bibinfo{editor}{H.~Larochelle}, \bibinfo{editor}{K.~Grauman},
  \bibinfo{editor}{N.~Cesa-Bianchi}, \bibinfo{editor}{R.~Garnett} (Eds.),
  \bibinfo{booktitle}{Advances in neural information processing systems},
  \bibinfo{year}{2018}, pp. \bibinfo{pages}{3235--3246}.
\bibitem[{Ouali et~al.(2020)Ouali, Hudelot, and Tami}]{ouali2020overview}
\bibinfo{author}{Y.~Ouali}, \bibinfo{author}{C.~Hudelot},
  \bibinfo{author}{M.~Tami}, \bibinfo{title}{An overview of deep
  semi-supervised learning}, \bibinfo{year}{2020}. \bibinfo{note}{{e}-print
  arXiv:2006.05278}.
\bibitem[{Kingma and Welling(2019)}]{kingma2019introduction}
\bibinfo{author}{D.~P. Kingma}, \bibinfo{author}{M.~Welling},
  \bibinfo{title}{An introduction to variational autoencoders},
  \bibinfo{year}{2019}. \bibinfo{note}{{e}-print arXiv:1906.02691}.
\bibitem[{Gui et~al.(2020)Gui, Sun, Wen, Tao, and Ye}]{gui2020review}
\bibinfo{author}{J.~Gui}, \bibinfo{author}{Z.~Sun}, \bibinfo{author}{Y.~Wen},
  \bibinfo{author}{D.~Tao}, \bibinfo{author}{J.~Ye}, \bibinfo{title}{A review
  on generative adversarial networks: {A}lgorithms, theory, and applications},
  \bibinfo{year}{2020}. \bibinfo{note}{{e}-print arXiv:2001.06937}.
\bibitem[{Fran{\c{c}}ois-Lavet et~al.(2018)Fran{\c{c}}ois-Lavet, Henderson,
  Islam, Bellemare, and Pineau}]{franccois2018introduction}
\bibinfo{author}{V.~Fran{\c{c}}ois-Lavet}, \bibinfo{author}{P.~Henderson},
  \bibinfo{author}{R.~Islam}, \bibinfo{author}{M.~G. Bellemare},
  \bibinfo{author}{J.~Pineau}, \bibinfo{title}{An introduction to deep
  reinforcement learning}, \bibinfo{year}{2018}. \bibinfo{note}{{e}-print
  arXiv:1811.12560}.
\bibitem[{Li(2017)}]{li2017deep}
\bibinfo{author}{Y.~Li}, \bibinfo{title}{Deep reinforcement learning: An
  overview}, \bibinfo{year}{2017}. \bibinfo{note}{{e}-print arXiv:1701.07274}.
\bibitem[{Arulkumaran et~al.(2017)Arulkumaran, Deisenroth, Brundage, and
  Bharath}]{arulkumaran2017deep}
\bibinfo{author}{K.~Arulkumaran}, \bibinfo{author}{M.~P. Deisenroth},
  \bibinfo{author}{M.~Brundage}, \bibinfo{author}{A.~A. Bharath},
\newblock \bibinfo{title}{Deep reinforcement learning: {A} brief survey},
\newblock \bibinfo{journal}{IEEE Signal Process. Mag.} \bibinfo{volume}{34}
  (\bibinfo{year}{2017}) \bibinfo{pages}{26--38}.
\bibitem[{Achiam(2018)}]{achiam2018spinning}
\bibinfo{author}{J.~Achiam}, \bibinfo{title}{Spinning up in deep {RL}},
  \bibinfo{year}{2018}. \bibinfo{note}{Available at:
  \url{https://spinningup.openai.com/en/latest/#}}.
\bibitem[{Brockman et~al.(2016)Brockman, Cheung, Pettersson, Schneider,
  Schulman, Tang, and Zaremba}]{brockman2016openai}
\bibinfo{author}{G.~Brockman}, \bibinfo{author}{V.~Cheung},
  \bibinfo{author}{L.~Pettersson}, \bibinfo{author}{J.~Schneider},
  \bibinfo{author}{J.~Schulman}, \bibinfo{author}{J.~Tang},
  \bibinfo{author}{W.~Zaremba}, \bibinfo{title}{Open{AI} {G}ym},
  \bibinfo{year}{2016}. \bibinfo{note}{{e}-print arXiv:1606.01540}.
\bibitem[{Dhariwal et~al.(2017)Dhariwal, Hesse, Klimov, Nichol, Plappert,
  Radford, Schulman, Sidor, Wu, and Zhokhov}]{dhariwal2017openai}
\bibinfo{author}{P.~Dhariwal}, \bibinfo{author}{C.~Hesse},
  \bibinfo{author}{O.~Klimov}, \bibinfo{author}{A.~Nichol},
  \bibinfo{author}{M.~Plappert}, \bibinfo{author}{A.~Radford},
  \bibinfo{author}{J.~Schulman}, \bibinfo{author}{S.~Sidor},
  \bibinfo{author}{Y.~Wu}, \bibinfo{author}{P.~Zhokhov},
  \bibinfo{title}{Open{AI} {B}aselines}, \bibinfo{year}{2017}.
  \bibinfo{note}{Available at: \url{https://github.com/openai/baselines}}.
\bibitem[{Stooke and Abbeel(2019)}]{stooke2019rlpyt}
\bibinfo{author}{A.~Stooke}, \bibinfo{author}{P.~Abbeel},
  \bibinfo{title}{{r}lpyt: {A} research code base for deep reinforcement
  learning in pytorch}, \bibinfo{year}{2019}. \bibinfo{note}{{e}-print
  arXiv:1909.01500}.
\bibitem[{Hernandez-Leal et~al.(2019)Hernandez-Leal, Kartal, and
  Taylor}]{hernandez2019survey}
\bibinfo{author}{P.~Hernandez-Leal}, \bibinfo{author}{B.~Kartal},
  \bibinfo{author}{M.~E. Taylor},
\newblock \bibinfo{title}{A survey and critique of multiagent deep
  reinforcement learning},
\newblock \bibinfo{journal}{Auton. Agent Multi Agent Syst.}
  \bibinfo{volume}{33} (\bibinfo{year}{2019}) \bibinfo{pages}{750--797}.
\bibitem[{Hernandez-Leal et~al.(2020)Hernandez-Leal, Kartal, and
  Taylor}]{hernandez2020very}
\bibinfo{author}{P.~Hernandez-Leal}, \bibinfo{author}{B.~Kartal},
  \bibinfo{author}{M.~E. Taylor},
\newblock \bibinfo{title}{A very condensed survey and critique of multiagent
  deep reinforcement learning},
\newblock in: \bibinfo{editor}{A.~El~Fallah-Seghrouchni},
  \bibinfo{editor}{G.~Sukthankar}, \bibinfo{editor}{B.~An},
  \bibinfo{editor}{N.~Yorke-Smith} (Eds.), \bibinfo{booktitle}{Proceedings of
  the 19th International Conference on Autonomous Agents and MultiAgent
  Systems}, \bibinfo{year}{2020}, pp. \bibinfo{pages}{2146--2148}.
\bibitem[{Roscher et~al.(2020)Roscher, Bohn, Duarte, and
  Garcke}]{roscher2020explainable}
\bibinfo{author}{R.~Roscher}, \bibinfo{author}{B.~Bohn}, \bibinfo{author}{M.~F.
  Duarte}, \bibinfo{author}{J.~Garcke},
\newblock \bibinfo{title}{Explainable machine learning for scientific insights
  and discoveries},
\newblock \bibinfo{journal}{IEEE Access} \bibinfo{volume}{8}
  (\bibinfo{year}{2020}) \bibinfo{pages}{42200--42216}.
\bibitem[{Iten et~al.(2020)Iten, Metger, Wilming, Del~Rio, and
  Renner}]{iten2020discovering}
\bibinfo{author}{R.~Iten}, \bibinfo{author}{T.~Metger},
  \bibinfo{author}{H.~Wilming}, \bibinfo{author}{L.~Del~Rio},
  \bibinfo{author}{R.~Renner},
\newblock \bibinfo{title}{Discovering physical concepts with neural networks},
\newblock \bibinfo{journal}{Phys. Rev. Lett.} \bibinfo{volume}{124}
  (\bibinfo{year}{2020}) \bibinfo{pages}{010508}.
\bibitem[{Zdeborov{\'a}(2020)}]{zdeborova2020understanding}
\bibinfo{author}{L.~Zdeborov{\'a}},
\newblock \bibinfo{title}{Understanding deep learning is also a job for
  physicists},
\newblock \bibinfo{journal}{Nat. Phys.} \bibinfo{volume}{16}
  (\bibinfo{year}{2020}) \bibinfo{pages}{602--604}.
\bibitem[{Udrescu and Tegmark(2020)}]{udrescu2020ai}
\bibinfo{author}{S.-M. Udrescu}, \bibinfo{author}{M.~Tegmark},
\newblock \bibinfo{title}{{AI Feynman}: {A} physics-inspired method for
  symbolic regression},
\newblock \bibinfo{journal}{Sci. Adv.} \bibinfo{volume}{6}
  (\bibinfo{year}{2020}) \bibinfo{pages}{eaay2631}.
\bibitem[{Wu and Tegmark(2019)}]{wu2019toward}
\bibinfo{author}{T.~Wu}, \bibinfo{author}{M.~Tegmark},
\newblock \bibinfo{title}{Toward an artificial intelligence physicist for
  unsupervised learning},
\newblock \bibinfo{journal}{Phys. Rev. E} \bibinfo{volume}{100}
  (\bibinfo{year}{2019}) \bibinfo{pages}{033311}.
\bibitem[{Choudhary et~al.(2019)Choudhary, Lindner, Holliday, Miller, Sinha,
  and Ditto}]{choudhary2019physics}
\bibinfo{author}{A.~Choudhary}, \bibinfo{author}{J.~F. Lindner},
  \bibinfo{author}{E.~G. Holliday}, \bibinfo{author}{S.~T. Miller},
  \bibinfo{author}{S.~Sinha}, \bibinfo{author}{W.~L. Ditto},
  \bibinfo{title}{Physics enhanced neural networks predict order and chaos},
  \bibinfo{year}{2019}. \bibinfo{note}{{e}-print arXiv:1912.01958}.
\bibitem[{Koch-Janusz and Ringel(2018)}]{koch2018mutual}
\bibinfo{author}{M.~Koch-Janusz}, \bibinfo{author}{Z.~Ringel},
\newblock \bibinfo{title}{Mutual information, neural networks and the
  renormalization group},
\newblock \bibinfo{journal}{Nat. Phys.} \bibinfo{volume}{14}
  (\bibinfo{year}{2018}) \bibinfo{pages}{578--582}.
\bibitem[{Carleo et~al.(2019)Carleo, Cirac, Cranmer, Daudet, Schuld, Tishby,
  Vogt-Maranto, and Zdeborov{\'a}}]{carleo2019machine}
\bibinfo{author}{G.~Carleo}, \bibinfo{author}{I.~Cirac},
  \bibinfo{author}{K.~Cranmer}, \bibinfo{author}{L.~Daudet},
  \bibinfo{author}{M.~Schuld}, \bibinfo{author}{N.~Tishby},
  \bibinfo{author}{L.~Vogt-Maranto}, \bibinfo{author}{L.~Zdeborov{\'a}},
\newblock \bibinfo{title}{Machine learning and the physical sciences},
\newblock \bibinfo{journal}{Rev. Mod. Phys.} \bibinfo{volume}{91}
  (\bibinfo{year}{2019}) \bibinfo{pages}{045002}.
\bibitem[{Bahri et~al.(2020)Bahri, Kadmon, Pennington, Schoenholz,
  Sohl-Dickstein, and Ganguli}]{bahri2020statistical}
\bibinfo{author}{Y.~Bahri}, \bibinfo{author}{J.~Kadmon},
  \bibinfo{author}{J.~Pennington}, \bibinfo{author}{S.~S. Schoenholz},
  \bibinfo{author}{J.~Sohl-Dickstein}, \bibinfo{author}{S.~Ganguli},
\newblock \bibinfo{title}{Statistical mechanics of deep learning},
\newblock \bibinfo{journal}{Annu. Rev. Condens. Matter Phys.}
  \bibinfo{volume}{11} (\bibinfo{year}{2020}) \bibinfo{pages}{501--528}.
\bibitem[{Mhaskar et~al.(2016)Mhaskar, Liao, and Poggio}]{mhaskar2016learning}
\bibinfo{author}{H.~Mhaskar}, \bibinfo{author}{Q.~Liao},
  \bibinfo{author}{T.~Poggio}, \bibinfo{title}{Learning functions: when is deep
  better than shallow}, \bibinfo{year}{2016}. \bibinfo{note}{{e}-print
  arXiv:1603.00988}.
\bibitem[{Lin et~al.(2017)Lin, Tegmark, and Rolnick}]{lin2017does}
\bibinfo{author}{H.~W. Lin}, \bibinfo{author}{M.~Tegmark},
  \bibinfo{author}{D.~Rolnick},
\newblock \bibinfo{title}{Why does deep and cheap learning work so well?},
\newblock \bibinfo{journal}{J. Stat. Phys.} \bibinfo{volume}{168}
  (\bibinfo{year}{2017}) \bibinfo{pages}{1223--1247}.
\bibitem[{Lin(2018)}]{lin2018generalization}
\bibinfo{author}{S.-B. Lin},
\newblock \bibinfo{title}{Generalization and expressivity for deep nets},
\newblock \bibinfo{journal}{IEEE Trans. Neural Netw. Learn. Syst.}
  \bibinfo{volume}{30} (\bibinfo{year}{2018}) \bibinfo{pages}{1392--1406}.
\bibitem[{Decelle et~al.(2019)Decelle, Martin-Mayor, and
  Seoane}]{decelle2019learning}
\bibinfo{author}{A.~Decelle}, \bibinfo{author}{V.~Martin-Mayor},
  \bibinfo{author}{B.~Seoane},
\newblock \bibinfo{title}{Learning a local symmetry with neural networks},
\newblock \bibinfo{journal}{Phys. Rev. E} \bibinfo{volume}{100}
  (\bibinfo{year}{2019}) \bibinfo{pages}{050102}.
\bibitem[{Pennington and Bahri(2017)}]{pennington2017geometry}
\bibinfo{author}{J.~Pennington}, \bibinfo{author}{Y.~Bahri},
\newblock \bibinfo{title}{Geometry of neural network loss surfaces via random
  matrix theory},
\newblock in: \bibinfo{editor}{D.~Precup}, \bibinfo{editor}{Y.~W. Teh} (Eds.),
  \bibinfo{booktitle}{Proceedings of the 34th International Conference on
  Machine Learning}, \bibinfo{year}{2017}, pp. \bibinfo{pages}{2798--2806}.
\bibitem[{Saxe et~al.(2019)Saxe, Bansal, Dapello, Advani, Kolchinsky, Tracey,
  and Cox}]{saxe2019information}
\bibinfo{author}{A.~M. Saxe}, \bibinfo{author}{Y.~Bansal},
  \bibinfo{author}{J.~Dapello}, \bibinfo{author}{M.~Advani},
  \bibinfo{author}{A.~Kolchinsky}, \bibinfo{author}{B.~D. Tracey},
  \bibinfo{author}{D.~D. Cox},
\newblock \bibinfo{title}{On the information bottleneck theory of deep
  learning},
\newblock \bibinfo{journal}{“J. Stat. Mech. Theory Exp.}
  \bibinfo{volume}{2019} (\bibinfo{year}{2019}) \bibinfo{pages}{124020}.
\bibitem[{Goldfeld and Polyanskiy(2020)}]{goldfeld2020information}
\bibinfo{author}{Z.~Goldfeld}, \bibinfo{author}{Y.~Polyanskiy},
\newblock \bibinfo{title}{The information bottleneck problem and its
  applications in machine learning},
\newblock \bibinfo{journal}{IEEE J. Sel. Areas Inform. Theory}
  \bibinfo{volume}{1} (\bibinfo{year}{2020}) \bibinfo{pages}{19--38}.
\bibitem[{Piran et~al.(2020)Piran, Shwartz-Ziv, and Tishby}]{piran2020dual}
\bibinfo{author}{Z.~Piran}, \bibinfo{author}{R.~Shwartz-Ziv},
  \bibinfo{author}{N.~Tishby}, \bibinfo{title}{The dual information
  bottleneck}, \bibinfo{year}{2020}. \bibinfo{note}{{e}-print
  arXiv:2006.04641}.
\bibitem[{Achille et~al.(2019)Achille, Paolini, and
  Soatto}]{achille2019information}
\bibinfo{author}{A.~Achille}, \bibinfo{author}{G.~Paolini},
  \bibinfo{author}{S.~Soatto}, \bibinfo{title}{Where is the information in a
  deep neural network?}, \bibinfo{year}{2019}. \bibinfo{note}{{e}-print
  arXiv:1905.12213}.
\bibitem[{Hafez-Kolahi et~al.(2019)Hafez-Kolahi, Kasaei, and
  Soleymani-Baghshah}]{hafez2019compressed}
\bibinfo{author}{H.~Hafez-Kolahi}, \bibinfo{author}{S.~Kasaei},
  \bibinfo{author}{M.~Soleymani-Baghshah}, \bibinfo{title}{Do compressed
  representations generalize better?}, \bibinfo{year}{2019}.
  \bibinfo{note}{{e}-print arXiv:1909.09706}.
\bibitem[{D'Angelo and B{\"o}ttcher(2020)}]{dangelo2020learning}
\bibinfo{author}{F.~D'Angelo}, \bibinfo{author}{L.~B{\"o}ttcher},
\newblock \bibinfo{title}{Learning the {I}sing model with generative neural
  networks},
\newblock \bibinfo{journal}{Phys. Rev. Res.} \bibinfo{volume}{2}
  (\bibinfo{year}{2020}) \bibinfo{pages}{023266}.
\bibitem[{Boucher(2020)}]{boucher2020montrealai}
\bibinfo{author}{V.~Boucher}, \bibinfo{title}{{MONTR\`{E}AL.AI} academy:
  artificial intelligence 101 first world-class overview of {AI} for all},
  \bibinfo{year}{2020}. \bibinfo{note}{VIP AI 101 cheat sheet. Available at:
  \url{https://www.montreal.ai/ai4all.pdf}. Archived at:
  \url{https://doi.org/10.17605/OSF.IO/N7G9X}}.
\bibitem[{Russell and Norvig(2002)}]{russell2002artificial}
\bibinfo{author}{S.~Russell}, \bibinfo{author}{P.~Norvig},
  \bibinfo{title}{Artificial intelligence: a modern approach},
  \bibinfo{publisher}{Prentice Hall}, \bibinfo{year}{2002}.
\bibitem[{Jordan(2019)}]{jordan2019artificial}
\bibinfo{author}{M.~I. Jordan},
\newblock \bibinfo{title}{Artificial intelligence---the revolution hasn't
  happened yet},
\newblock \bibinfo{journal}{Harvard Data Sci. Rev.} \bibinfo{volume}{1}
  (\bibinfo{year}{2019}).
\bibitem[{Searle(1980)}]{searle1980minds}
\bibinfo{author}{J.~R. Searle},
\newblock \bibinfo{title}{Minds, brains, and programs},
\newblock in: \bibinfo{editor}{D.~J. Levitin} (Ed.),
  \bibinfo{booktitle}{Foundations of cognitive psychology: core readings},
  \bibinfo{publisher}{The MIT Press}, \bibinfo{year}{1980}, pp.
  \bibinfo{pages}{417--457}.
\bibitem[{Fjelland(2020)}]{fjelland2020general}
\bibinfo{author}{R.~Fjelland},
\newblock \bibinfo{title}{Why general artificial intelligence will not be
  realized},
\newblock \bibinfo{journal}{Humanit. Soc. Sci. Commun.} \bibinfo{volume}{7}
  (\bibinfo{year}{2020}) \bibinfo{pages}{10}.
\bibitem[{Shalev-Shwartz and Ben-David(2014)}]{shalev2014understanding}
\bibinfo{author}{S.~Shalev-Shwartz}, \bibinfo{author}{S.~Ben-David},
  \bibinfo{title}{Understanding machine learning: From theory to algorithms},
  \bibinfo{publisher}{Cambridge University Press}, \bibinfo{year}{2014}.
\bibitem[{Abu-Mostafa et~al.(2012)Abu-Mostafa, Magdon-Ismail, and
  Lin}]{abumostafa2012learning}
\bibinfo{author}{Y.~S. Abu-Mostafa}, \bibinfo{author}{M.~Magdon-Ismail},
  \bibinfo{author}{H.-T. Lin}, \bibinfo{title}{Learning from data: a short
  course}, \bibinfo{publisher}{AMLBook}, \bibinfo{year}{2012}.
\bibitem[{Domingos(2012)}]{domingos2012few}
\bibinfo{author}{P.~Domingos},
\newblock \bibinfo{title}{A few useful things to know about machine learning},
\newblock \bibinfo{journal}{Commun. ACM} \bibinfo{volume}{55}
  (\bibinfo{year}{2012}) \bibinfo{pages}{78--87}.
\bibitem[{Bennett and Parrado-Hern{\'a}ndez(2006)}]{bennett2006interplay}
\bibinfo{author}{K.~P. Bennett}, \bibinfo{author}{E.~Parrado-Hern{\'a}ndez},
\newblock \bibinfo{title}{The interplay of optimization and machine learning
  research},
\newblock \bibinfo{journal}{J. Mach. Learn. Res.} \bibinfo{volume}{7}
  (\bibinfo{year}{2006}) \bibinfo{pages}{1265--1281}.
\bibitem[{Esteva et~al.(2019)Esteva, Robicquet, Ramsundar, Kuleshov, DePristo,
  Chou, Cui, Corrado, Thrun, and Dean}]{esteva2019guide}
\bibinfo{author}{A.~Esteva}, \bibinfo{author}{A.~Robicquet},
  \bibinfo{author}{B.~Ramsundar}, \bibinfo{author}{V.~Kuleshov},
  \bibinfo{author}{M.~DePristo}, \bibinfo{author}{K.~Chou},
  \bibinfo{author}{C.~Cui}, \bibinfo{author}{G.~Corrado},
  \bibinfo{author}{S.~Thrun}, \bibinfo{author}{J.~Dean},
\newblock \bibinfo{title}{A guide to deep learning in healthcare},
\newblock \bibinfo{journal}{Nat. Med.} \bibinfo{volume}{25}
  (\bibinfo{year}{2019}) \bibinfo{pages}{24--29}.
\bibitem[{Hochreiter and Schmidhuber(1997)}]{hochreiter1997long}
\bibinfo{author}{S.~Hochreiter}, \bibinfo{author}{J.~Schmidhuber},
\newblock \bibinfo{title}{Long short-term memory},
\newblock \bibinfo{journal}{Neural Comput.} \bibinfo{volume}{9}
  (\bibinfo{year}{1997}) \bibinfo{pages}{1735--1780}.
\bibitem[{Chung et~al.(2014)Chung, Gulcehre, Cho, and
  Bengio}]{chung2014empirical}
\bibinfo{author}{J.~Chung}, \bibinfo{author}{C.~Gulcehre},
  \bibinfo{author}{K.~Cho}, \bibinfo{author}{Y.~Bengio},
  \bibinfo{title}{Empirical evaluation of gated recurrent neural networks on
  sequence modeling}, \bibinfo{year}{2014}. \bibinfo{note}{{e}-print
  arXiv:1412.3555}.
\bibitem[{Vaswani et~al.(2017)Vaswani, Shazeer, Parmar, Uszkoreit, Jones,
  Gomez, Kaiser, and Polosukhin}]{vaswani2017attention}
\bibinfo{author}{A.~Vaswani}, \bibinfo{author}{N.~Shazeer},
  \bibinfo{author}{N.~Parmar}, \bibinfo{author}{J.~Uszkoreit},
  \bibinfo{author}{L.~Jones}, \bibinfo{author}{A.~N. Gomez},
  \bibinfo{author}{L.~Kaiser}, \bibinfo{author}{I.~Polosukhin},
  \bibinfo{title}{Attention is all you need}, \bibinfo{year}{2017}.
  \bibinfo{note}{{e}-print arXiv:1706.03762}.
\bibitem[{Qin et~al.(2017)Qin, Song, Chen, Cheng, Jiang, and
  Cottrell}]{qin2017dual}
\bibinfo{author}{Y.~Qin}, \bibinfo{author}{D.~Song}, \bibinfo{author}{H.~Chen},
  \bibinfo{author}{W.~Cheng}, \bibinfo{author}{G.~Jiang},
  \bibinfo{author}{G.~Cottrell}, \bibinfo{title}{A dual-stage attention-based
  recurrent neural network for time series prediction}, \bibinfo{year}{2017}.
  \bibinfo{note}{{e}-print arXiv:1704.02971}.
\bibitem[{Ward et~al.(2020)Ward, Joyner, Lickfold, Rowe, Guo, and
  Bennamoun}]{ward2020practical}
\bibinfo{author}{I.~R. Ward}, \bibinfo{author}{J.~Joyner},
  \bibinfo{author}{C.~Lickfold}, \bibinfo{author}{S.~Rowe},
  \bibinfo{author}{Y.~Guo}, \bibinfo{author}{M.~Bennamoun}, \bibinfo{title}{A
  practical guide to graph neural networks}, \bibinfo{year}{2020}.
  \bibinfo{note}{{e}-print arXiv:2010.05234}.
\bibitem[{Stone et~al.(2016)Stone, Brooks, Brynjolfsson, Calo, Etzioni, Hager,
  Hirschberg, Kalyanakrishnan, Kamar, Kraus et~al.}]{stone2016artificial}
\bibinfo{author}{P.~Stone}, \bibinfo{author}{R.~Brooks},
  \bibinfo{author}{E.~Brynjolfsson}, \bibinfo{author}{R.~Calo},
  \bibinfo{author}{O.~Etzioni}, \bibinfo{author}{G.~Hager},
  \bibinfo{author}{J.~Hirschberg}, \bibinfo{author}{S.~Kalyanakrishnan},
  \bibinfo{author}{E.~Kamar}, \bibinfo{author}{S.~Kraus}, et~al.,
  \bibinfo{title}{Artificial intelligence and life in 2030: the one hundred
  year study on artificial intelligence}, \bibinfo{year}{2016}.
  \bibinfo{note}{Available at:
  \url{https://ai100.stanford.edu/sites/g/files/sbiybj9861/f/ai100report10032016fnl_singles.pdf}}.
\bibitem[{Turkina(2018)}]{turkina2018importance}
\bibinfo{author}{E.~Turkina},
\newblock \bibinfo{title}{The importance of networking to entrepreneurship:
  {M}ontreal's artificial intelligence cluster and its born-global firm element
  ai},
\newblock \bibinfo{journal}{J. Small Bus. Entrepreneurship}
  \bibinfo{volume}{30} (\bibinfo{year}{2018}) \bibinfo{pages}{1--8}.
\bibitem[{Cetinic et~al.(2018)Cetinic, Lipic, and Grgic}]{cetinic2018fine}
\bibinfo{author}{E.~Cetinic}, \bibinfo{author}{T.~Lipic},
  \bibinfo{author}{S.~Grgic},
\newblock \bibinfo{title}{Fine-tuning convolutional neural networks for fine
  art classification},
\newblock \bibinfo{journal}{Expert Syst. Appl.} \bibinfo{volume}{114}
  (\bibinfo{year}{2018}) \bibinfo{pages}{107--118}.
\bibitem[{Petangoda et~al.(2020)Petangoda, Monk, and
  Deisenroth}]{petangoda2020foliated}
\bibinfo{author}{J.~Petangoda}, \bibinfo{author}{N.~A. Monk},
  \bibinfo{author}{M.~P. Deisenroth}, \bibinfo{title}{A foliated view of
  transfer learning}, \bibinfo{year}{2020}. \bibinfo{note}{{e}-print
  arXiv:2008.00546}.
\bibitem[{Battaglia et~al.(2016)Battaglia, Pascanu, Lai, Rezende, and
  Kavukcuoglu}]{battaglia2016interaction}
\bibinfo{author}{P.~W. Battaglia}, \bibinfo{author}{R.~Pascanu},
  \bibinfo{author}{M.~Lai}, \bibinfo{author}{D.~Rezende},
  \bibinfo{author}{K.~Kavukcuoglu}, \bibinfo{title}{Interaction networks for
  learning about objects, relations and physics}, \bibinfo{year}{2016}.
  \bibinfo{note}{{e}-print arXiv:1612.00222}.
\bibitem[{Lake et~al.(2017)Lake, Ullman, Tenenbaum, and
  Gershman}]{lake2017building}
\bibinfo{author}{B.~M. Lake}, \bibinfo{author}{T.~D. Ullman},
  \bibinfo{author}{J.~B. Tenenbaum}, \bibinfo{author}{S.~J. Gershman},
\newblock \bibinfo{title}{Building machines that learn and think like people},
\newblock \bibinfo{journal}{Behav. Brain Sci.} \bibinfo{volume}{40}
  (\bibinfo{year}{2017}) \bibinfo{pages}{e253}.
\bibitem[{Lake et~al.(2019)Lake, Salakhutdinov, and
  Tenenbaum}]{lake2019omniglot}
\bibinfo{author}{B.~M. Lake}, \bibinfo{author}{R.~Salakhutdinov},
  \bibinfo{author}{J.~B. Tenenbaum},
\newblock \bibinfo{title}{The {O}mniglot challenge: a 3-year progress report},
\newblock \bibinfo{journal}{Curr. Opin. Behav. Sci.} \bibinfo{volume}{29}
  (\bibinfo{year}{2019}) \bibinfo{pages}{97--104}.
\bibitem[{Hospedales et~al.(2020)Hospedales, Antoniou, Micaelli, and
  Storkey}]{hospedales2020meta}
\bibinfo{author}{T.~Hospedales}, \bibinfo{author}{A.~Antoniou},
  \bibinfo{author}{P.~Micaelli}, \bibinfo{author}{A.~Storkey},
  \bibinfo{title}{Meta-learning in neural networks: {A} survey},
  \bibinfo{year}{2020}. \bibinfo{note}{{e}-print arXiv:2004.05439}.
\bibitem[{Wang et~al.(2020)Wang, Yao, Kwok, and Ni}]{wang2020generalizing}
\bibinfo{author}{Y.~Wang}, \bibinfo{author}{Q.~Yao}, \bibinfo{author}{J.~T.
  Kwok}, \bibinfo{author}{L.~M. Ni},
\newblock \bibinfo{title}{Generalizing from a few examples: {A} survey on
  few-shot learning},
\newblock \bibinfo{journal}{ACM Comput. Surv.} \bibinfo{volume}{53}
  (\bibinfo{year}{2020}) \bibinfo{pages}{1--34}.
\bibitem[{Koch(2015)}]{koch2015siamese}
\bibinfo{author}{G.~Koch}, \bibinfo{title}{Siamese neural networks for one-shot
  image recognition}, \bibinfo{year}{2015}. \bibinfo{note}{Master thesis,
  University of Toronto. Available at:
  \url{http://www.cs.toronto.edu/~gkoch/files/msc-thesis.pdf}}.
\bibitem[{Vinyals et~al.(2016)Vinyals, Blundell, Lillicrap, Kavukcuoglu, and
  Wierstra}]{vinyals2016matching}
\bibinfo{author}{O.~Vinyals}, \bibinfo{author}{C.~Blundell},
  \bibinfo{author}{T.~Lillicrap}, \bibinfo{author}{K.~Kavukcuoglu},
  \bibinfo{author}{D.~Wierstra},
\newblock \bibinfo{title}{Matching networks for one shot learning},
\newblock \bibinfo{journal}{{e}-print arXiv:1606.04080}
  (\bibinfo{year}{2016}).
\bibitem[{Snell et~al.(2017)Snell, Swersky, and Zemel}]{snell2017prototypical}
\bibinfo{author}{J.~Snell}, \bibinfo{author}{K.~Swersky},
  \bibinfo{author}{R.~S. Zemel},
\newblock \bibinfo{title}{Prototypical networks for few-shot learning},
\newblock \bibinfo{journal}{{e}-print arXiv:1703.05175}
  (\bibinfo{year}{2017}).
\bibitem[{Sung et~al.(2018)Sung, Yang, Zhang, Xiang, Torr, and
  Hospedales}]{sung2018learning}
\bibinfo{author}{F.~Sung}, \bibinfo{author}{Y.~Yang},
  \bibinfo{author}{L.~Zhang}, \bibinfo{author}{T.~Xiang},
  \bibinfo{author}{P.~H. Torr}, \bibinfo{author}{T.~M. Hospedales},
\newblock \bibinfo{title}{Learning to compare: {R}elation network for few-shot
  learning},
\newblock in: \bibinfo{editor}{D.~Forsyth}, \bibinfo{editor}{I.~Laptev},
  \bibinfo{editor}{A.~Oliva}, \bibinfo{editor}{D.~Ramanan},
  \bibinfo{editor}{M.~S. Brown}, \bibinfo{editor}{B.~Morse},
  \bibinfo{editor}{S.~Peleg} (Eds.), \bibinfo{booktitle}{2018 IEEE/CVF
  Conference on Computer Vision and Pattern Recognition (CVPR)},
  \bibinfo{publisher}{IEEE Computer Society}, \bibinfo{year}{2018}, pp.
  \bibinfo{pages}{1199--1208}.
\bibitem[{Santoro et~al.(2016)Santoro, Bartunov, Botvinick, Wierstra, and
  Lillicrap}]{santoro2016meta}
\bibinfo{author}{A.~Santoro}, \bibinfo{author}{S.~Bartunov},
  \bibinfo{author}{M.~Botvinick}, \bibinfo{author}{D.~Wierstra},
  \bibinfo{author}{T.~Lillicrap},
\newblock \bibinfo{title}{Meta-learning with memory-augmented neural networks},
\newblock in: \bibinfo{editor}{M.~F. Balcan}, \bibinfo{editor}{K.~Q.
  Weinberger} (Eds.), \bibinfo{booktitle}{Proceedings of the 33rd International
  Conference on Machine Learning}, volume~\bibinfo{volume}{48},
  \bibinfo{publisher}{Proceedings of Machine Learning Research},
  \bibinfo{year}{2016}, pp. \bibinfo{pages}{1842--1850}.
\bibitem[{Ravi and Larochelle(2016)}]{ravi2016optimization}
\bibinfo{author}{S.~Ravi}, \bibinfo{author}{H.~Larochelle},
  \bibinfo{title}{Optimization as a model for few-shot learning},
  \bibinfo{year}{2016}. \bibinfo{note}{International Conference on Learning
  Representations 2017. Available at:
  \url{https://openreview.net/pdf?id=rJY0-Kcll}. Archived at:
  \url{https://doi.org/10.17605/OSF.IO/N7G9X}}.
\bibitem[{Finn et~al.(2017)Finn, Abbeel, and Levine}]{finn2017model}
\bibinfo{author}{C.~Finn}, \bibinfo{author}{P.~Abbeel},
  \bibinfo{author}{S.~Levine}, \bibinfo{title}{Model-agnostic meta-learning for
  fast adaptation of deep networks}, \bibinfo{year}{2017}.
  \bibinfo{note}{{e}-print arXiv:1703.03400}.
\bibitem[{Raghu et~al.(2019)Raghu, Raghu, Bengio, and Vinyals}]{raghu2019rapid}
\bibinfo{author}{A.~Raghu}, \bibinfo{author}{M.~Raghu},
  \bibinfo{author}{S.~Bengio}, \bibinfo{author}{O.~Vinyals},
  \bibinfo{title}{Rapid learning or feature reuse? {T}owards understanding the
  effectiveness of {MAML}}, \bibinfo{year}{2019}. \bibinfo{note}{{e}-print
  arXiv:1909.09157}.
\bibitem[{Liang and Yan(2019)}]{liang2019survey}
\bibinfo{author}{X.~Liang}, \bibinfo{author}{Z.~Yan},
\newblock \bibinfo{title}{A survey on game theoretical methods in human-machine
  networks},
\newblock \bibinfo{journal}{Future Gener. Comput. Syst.} \bibinfo{volume}{92}
  (\bibinfo{year}{2019}) \bibinfo{pages}{674--693}.
\bibitem[{Foerster et~al.(2017)Foerster, Chen, Al-Shedivat, Whiteson, Abbeel,
  and Mordatch}]{foerster2017learning}
\bibinfo{author}{J.~N. Foerster}, \bibinfo{author}{R.~Y. Chen},
  \bibinfo{author}{M.~Al-Shedivat}, \bibinfo{author}{S.~Whiteson},
  \bibinfo{author}{P.~Abbeel}, \bibinfo{author}{I.~Mordatch},
  \bibinfo{title}{Learning with opponent-learning awareness},
  \bibinfo{year}{2017}. \bibinfo{note}{{e}-print arXiv:1709.04326}.
\bibitem[{Press and Dyson(2012)}]{press2012iterated}
\bibinfo{author}{W.~H. Press}, \bibinfo{author}{F.~J. Dyson},
\newblock \bibinfo{title}{Iterated {P}risoner's {D}ilemma contains strategies
  that dominate any evolutionary opponent},
\newblock \bibinfo{journal}{Proc. Natl. Acad. Sci. USA} \bibinfo{volume}{109}
  (\bibinfo{year}{2012}) \bibinfo{pages}{10409--10413}.
\bibitem[{Stewart and Plotkin(2012)}]{stewart2012extortion}
\bibinfo{author}{A.~J. Stewart}, \bibinfo{author}{J.~B. Plotkin},
\newblock \bibinfo{title}{Extortion and cooperation in the {P}risoner’s
  {D}ilemma},
\newblock \bibinfo{journal}{Proc. Natl. Acad. Sci. USA} \bibinfo{volume}{109}
  (\bibinfo{year}{2012}) \bibinfo{pages}{10134--10135}.
\bibitem[{Rabinowitz et~al.(2018)Rabinowitz, Perbet, Song, Zhang, Eslami, and
  Botvinick}]{rabinowitz2018machine}
\bibinfo{author}{N.~Rabinowitz}, \bibinfo{author}{F.~Perbet},
  \bibinfo{author}{F.~Song}, \bibinfo{author}{C.~Zhang},
  \bibinfo{author}{S.~M.~A. Eslami}, \bibinfo{author}{M.~Botvinick},
\newblock \bibinfo{title}{Machine theory of mind},
\newblock in: \bibinfo{editor}{J.~Dy}, \bibinfo{editor}{A.~Krause} (Eds.),
  \bibinfo{booktitle}{Proceedings of the 35th International Conference on
  Machine Learning}, volume~\bibinfo{volume}{80},
  \bibinfo{publisher}{Proceedings of Machine Learning Research},
  \bibinfo{year}{2018}, pp. \bibinfo{pages}{4218--4227}.
\bibitem[{Glynatsi and Knight(2020)}]{glynatsi2020using}
\bibinfo{author}{N.~E. Glynatsi}, \bibinfo{author}{V.~A. Knight},
\newblock \bibinfo{title}{Using a theory of mind to find best responses to
  memory-one strategies},
\newblock \bibinfo{journal}{Sci. Rep.} \bibinfo{volume}{10}
  (\bibinfo{year}{2020}) \bibinfo{pages}{17287}.
\bibitem[{Knight et~al.(2016)Knight, Campbell, Harper, Langner, Campbell,
  Campbell, Carney, Chorley, Davidson-Pilon, Glass et~al.}]{knight2016open}
\bibinfo{author}{V.~A. Knight}, \bibinfo{author}{O.~Campbell},
  \bibinfo{author}{M.~Harper}, \bibinfo{author}{K.~M. Langner},
  \bibinfo{author}{J.~Campbell}, \bibinfo{author}{T.~Campbell},
  \bibinfo{author}{A.~Carney}, \bibinfo{author}{M.~Chorley},
  \bibinfo{author}{C.~Davidson-Pilon}, \bibinfo{author}{K.~Glass}, et~al.,
\newblock \bibinfo{title}{An open framework for the reproducible study of the
  iterated prisoner's dilemma},
\newblock \bibinfo{journal}{J. Open Res. Softw.} \bibinfo{volume}{4}
  (\bibinfo{year}{2016}) \bibinfo{pages}{e35}.
\bibitem[{Harper et~al.(2017)Harper, Knight, Jones, Koutsovoulos, Glynatsi, and
  Campbell}]{harper2017reinforcement}
\bibinfo{author}{M.~Harper}, \bibinfo{author}{V.~Knight},
  \bibinfo{author}{M.~Jones}, \bibinfo{author}{G.~Koutsovoulos},
  \bibinfo{author}{N.~E. Glynatsi}, \bibinfo{author}{O.~Campbell},
\newblock \bibinfo{title}{Reinforcement learning produces dominant strategies
  for the {I}terated {P}risoner’s {D}ilemma},
\newblock \bibinfo{journal}{PLOS ONE} \bibinfo{volume}{12}
  (\bibinfo{year}{2017}) \bibinfo{pages}{e0188046}.
\bibitem[{Au and Nau(2006)}]{au2006accident}
\bibinfo{author}{T.-C. Au}, \bibinfo{author}{D.~Nau},
\newblock \bibinfo{title}{Accident or intention: that is the question (in the
  {N}oisy {I}terated {P}risoner's {D}ilemma)},
\newblock in: \bibinfo{editor}{H.~Nakashima}, \bibinfo{editor}{M.~Wellman},
  \bibinfo{editor}{G.~Weiss}, \bibinfo{editor}{P.~Stone} (Eds.),
  \bibinfo{booktitle}{Proceedings of the fifth international joint conference
  on Autonomous agents and multiagent systems}, \bibinfo{publisher}{Association
  for Computing Machinery}, \bibinfo{year}{2006}, pp.
  \bibinfo{pages}{561--568}.
\bibitem[{Leibo et~al.(2017)Leibo, Zambaldi, Lanctot, Marecki, and
  Graepel}]{leibo2017multi}
\bibinfo{author}{J.~Z. Leibo}, \bibinfo{author}{V.~Zambaldi},
  \bibinfo{author}{M.~Lanctot}, \bibinfo{author}{J.~Marecki},
  \bibinfo{author}{T.~Graepel},
\newblock \bibinfo{title}{Multi-agent reinforcement learning in sequential
  social dilemmas},
\newblock \bibinfo{journal}{{e}-print arXiv:1702.03037}
  (\bibinfo{year}{2017}).
\bibitem[{Peysakhovich and Lerer(2018)}]{peysakhovich2018towards}
\bibinfo{author}{A.~Peysakhovich}, \bibinfo{author}{A.~Lerer},
\newblock \bibinfo{title}{Towards {AI} that can solve social dilemmas},
\newblock in: \bibinfo{editor}{C.~Amato}, \bibinfo{editor}{T.~Graepel},
  \bibinfo{editor}{J.~Leibo}, \bibinfo{editor}{F.~Oliehoek},
  \bibinfo{editor}{K.~Tuyls} (Eds.), \bibinfo{booktitle}{The 2018 AAAI Spring
  Symposium Series}, volume \bibinfo{volume}{SS-18},
  \bibinfo{publisher}{Association for the Advancement of Artificial
  Intelligence}, \bibinfo{year}{2018}, pp. \bibinfo{pages}{649--654}.
\bibitem[{Lowe et~al.(2017)Lowe, Wu, Tamar, Harb, Abbeel, and
  Mordatch}]{lowe2017multi}
\bibinfo{author}{R.~Lowe}, \bibinfo{author}{Y.~Wu}, \bibinfo{author}{A.~Tamar},
  \bibinfo{author}{J.~Harb}, \bibinfo{author}{P.~Abbeel},
  \bibinfo{author}{I.~Mordatch},
\newblock \bibinfo{title}{Multi-agent actor-critic for mixed
  cooperative-competitive environments},
\newblock \bibinfo{journal}{{e}-print arXiv:1706.02275}
  (\bibinfo{year}{2017}).
\bibitem[{Park et~al.(2019)Park, Cho, and Kim}]{park2019multi}
\bibinfo{author}{Y.~J. Park}, \bibinfo{author}{Y.~S. Cho},
  \bibinfo{author}{S.~B. Kim},
\newblock \bibinfo{title}{Multi-agent reinforcement learning with approximate
  model learning for competitive games},
\newblock \bibinfo{journal}{PLOS ONE} \bibinfo{volume}{14}
  (\bibinfo{year}{2019}) \bibinfo{pages}{e0222215}.
\bibitem[{Bachrach et~al.(2020)Bachrach, Everett, Hughes, Lazaridou, Leibo,
  Lanctot, Johanson, Czarnecki, and Graepel}]{bachrach2020negotiating}
\bibinfo{author}{Y.~Bachrach}, \bibinfo{author}{R.~Everett},
  \bibinfo{author}{E.~Hughes}, \bibinfo{author}{A.~Lazaridou},
  \bibinfo{author}{J.~Z. Leibo}, \bibinfo{author}{M.~Lanctot},
  \bibinfo{author}{M.~Johanson}, \bibinfo{author}{W.~M. Czarnecki},
  \bibinfo{author}{T.~Graepel},
\newblock \bibinfo{title}{Negotiating team formation using deep reinforcement
  learning},
\newblock \bibinfo{journal}{Artif. Intell.} \bibinfo{volume}{288}
  (\bibinfo{year}{2020}) \bibinfo{pages}{103356}.
\bibitem[{Degris et~al.(2012)Degris, White, and Sutton}]{degris2012off}
\bibinfo{author}{T.~Degris}, \bibinfo{author}{M.~White}, \bibinfo{author}{R.~S.
  Sutton},
\newblock \bibinfo{title}{Off-policy actor-critic},
\newblock \bibinfo{journal}{{e}-print arXiv:1205.4839}  (\bibinfo{year}{2012}).
\bibitem[{Haarnoja et~al.(2018)Haarnoja, Zhou, Hartikainen, Tucker, Ha, Tan,
  Kumar, Zhu, Gupta, Abbeel et~al.}]{haarnoja2018soft}
\bibinfo{author}{T.~Haarnoja}, \bibinfo{author}{A.~Zhou},
  \bibinfo{author}{K.~Hartikainen}, \bibinfo{author}{G.~Tucker},
  \bibinfo{author}{S.~Ha}, \bibinfo{author}{J.~Tan},
  \bibinfo{author}{V.~Kumar}, \bibinfo{author}{H.~Zhu},
  \bibinfo{author}{A.~Gupta}, \bibinfo{author}{P.~Abbeel}, et~al.,
\newblock \bibinfo{title}{Soft actor-critic algorithms and applications},
\newblock \bibinfo{journal}{{e}-print arXiv:1812.05905}
  (\bibinfo{year}{2018}).
\bibitem[{Konda and Tsitsiklis(2000)}]{konda2000actor}
\bibinfo{author}{V.~R. Konda}, \bibinfo{author}{J.~N. Tsitsiklis},
\newblock \bibinfo{title}{Actor-critic algorithms},
\newblock in: \bibinfo{editor}{S.~A. Solla}, \bibinfo{editor}{T.~K. Leen},
  \bibinfo{editor}{K.-R. M{\"u}ller} (Eds.), \bibinfo{booktitle}{Advances in
  neural information processing systems}, volume~\bibinfo{volume}{12},
  \bibinfo{organization}{The MIT Press}, \bibinfo{year}{2000}, pp.
  \bibinfo{pages}{1008--1014}.
\bibitem[{Kraemer and Banerjee(2016)}]{kraemer2016multi}
\bibinfo{author}{L.~Kraemer}, \bibinfo{author}{B.~Banerjee},
\newblock \bibinfo{title}{Multi-agent reinforcement learning as a rehearsal for
  decentralized planning},
\newblock \bibinfo{journal}{Neurocomputing} \bibinfo{volume}{190}
  (\bibinfo{year}{2016}) \bibinfo{pages}{82--94}.
\bibitem[{OroojlooyJadid and Hajinezhad(2019)}]{oroojlooyjadid2019review}
\bibinfo{author}{A.~OroojlooyJadid}, \bibinfo{author}{D.~Hajinezhad},
\newblock \bibinfo{title}{A review of cooperative multi-agent deep
  reinforcement learning},
\newblock \bibinfo{journal}{{e}-print arXiv:1908.03963}
  (\bibinfo{year}{2019}).
\bibitem[{Zhang et~al.(2019)Zhang, Yang, and Ba{\c{s}}ar}]{zhang2019multi}
\bibinfo{author}{K.~Zhang}, \bibinfo{author}{Z.~Yang},
  \bibinfo{author}{T.~Ba{\c{s}}ar},
\newblock \bibinfo{title}{Multi-agent reinforcement learning: A selective
  overview of theories and algorithms},
\newblock \bibinfo{journal}{{e}-print arXiv:1911.10635}
  (\bibinfo{year}{2019}).
\bibitem[{Li et~al.(2020)Li, Jin, Wang, Yan, and Zha}]{li2020f2a2}
\bibinfo{author}{W.~Li}, \bibinfo{author}{B.~Jin}, \bibinfo{author}{X.~Wang},
  \bibinfo{author}{J.~Yan}, \bibinfo{author}{H.~Zha},
\newblock \bibinfo{title}{{F2A2}: {F}lexible {F}ully-decentralized
  {A}pproximate {A}ctor-critic for cooperative multi-agent reinforcement
  learning},
\newblock \bibinfo{journal}{{e}-print arXiv:2004.11145}
  (\bibinfo{year}{2020}).
\bibitem[{Hansen et~al.(2004)Hansen, Bernstein, and
  Zilberstein}]{hansen2004dynamic}
\bibinfo{author}{E.~A. Hansen}, \bibinfo{author}{D.~S. Bernstein},
  \bibinfo{author}{S.~Zilberstein},
\newblock \bibinfo{title}{Dynamic programming for partially observable
  stochastic games},
\newblock in: \bibinfo{editor}{A.~G. Cohn} (Ed.),
  \bibinfo{booktitle}{Proceedings of the 19th national conference on
  {A}rtifical intelligence}, volume~\bibinfo{volume}{4},
  \bibinfo{publisher}{AAAI Press}, \bibinfo{year}{2004}, pp.
  \bibinfo{pages}{709--715}.
\bibitem[{Rovatsos(2019)}]{rovatsos2019we}
\bibinfo{author}{M.~Rovatsos},
\newblock \bibinfo{title}{We may not cooperate with friendly machines},
\newblock \bibinfo{journal}{Nat. Mach. Intell.} \bibinfo{volume}{1}
  (\bibinfo{year}{2019}) \bibinfo{pages}{497--498}.
\bibitem[{Ishowo-Oloko et~al.(2019)Ishowo-Oloko, Bonnefon, Soroye, Crandall,
  Rahwan, and Rahwan}]{ishowo2019behavioural}
\bibinfo{author}{F.~Ishowo-Oloko}, \bibinfo{author}{J.-F. Bonnefon},
  \bibinfo{author}{Z.~Soroye}, \bibinfo{author}{J.~Crandall},
  \bibinfo{author}{I.~Rahwan}, \bibinfo{author}{T.~Rahwan},
\newblock \bibinfo{title}{Behavioural evidence for a transparency-efficiency
  tradeoff in human-machine cooperation},
\newblock \bibinfo{journal}{Nat. Mach. Intell.} \bibinfo{volume}{1}
  (\bibinfo{year}{2019}) \bibinfo{pages}{517--521}.
\bibitem[{Yoeli et~al.(2013)Yoeli, Hoffman, Rand, and
  Nowak}]{yoeli2013powering}
\bibinfo{author}{E.~Yoeli}, \bibinfo{author}{M.~Hoffman},
  \bibinfo{author}{D.~G. Rand}, \bibinfo{author}{M.~A. Nowak},
\newblock \bibinfo{title}{Powering up with indirect reciprocity in a
  large-scale field experiment},
\newblock \bibinfo{journal}{Proc. Natl. Acad. Sci. USA} \bibinfo{volume}{110}
  (\bibinfo{year}{2013}) \bibinfo{pages}{10424--10429}.
\bibitem[{Lerer and Peysakhovich(2017)}]{lerer2017maintaining}
\bibinfo{author}{A.~Lerer}, \bibinfo{author}{A.~Peysakhovich},
\newblock \bibinfo{title}{Maintaining cooperation in complex social dilemmas
  using deep reinforcement learning},
\newblock \bibinfo{journal}{{e}-print arXiv:1707.01068}
  (\bibinfo{year}{2017}).
\bibitem[{Bonnefon and Rahwan(2020)}]{bonnefon2020machine}
\bibinfo{author}{J.-F. Bonnefon}, \bibinfo{author}{I.~Rahwan},
\newblock \bibinfo{title}{Machine thinking, fast and slow},
\newblock \bibinfo{journal}{Trends Cogn. Sci.} \bibinfo{volume}{24}
  (\bibinfo{year}{2020}) \bibinfo{pages}{1019--1027}.
\bibitem[{Karandikar et~al.(1998)Karandikar, Mookherjee, Ray, and
  Vega-Redondo}]{karandikar1998evolving}
\bibinfo{author}{R.~Karandikar}, \bibinfo{author}{D.~Mookherjee},
  \bibinfo{author}{D.~Ray}, \bibinfo{author}{F.~Vega-Redondo},
\newblock \bibinfo{title}{Evolving aspirations and cooperation},
\newblock \bibinfo{journal}{J. Econ. Theory} \bibinfo{volume}{80}
  (\bibinfo{year}{1998}) \bibinfo{pages}{292--331}.
\bibitem[{Stoyanovich et~al.(2020)Stoyanovich, Van~Bavel, and
  West}]{stoyanovich2020imperative}
\bibinfo{author}{J.~Stoyanovich}, \bibinfo{author}{J.~J. Van~Bavel},
  \bibinfo{author}{T.~V. West},
\newblock \bibinfo{title}{The imperative of interpretable machines},
\newblock \bibinfo{journal}{Nat. Mach. Intell.} \bibinfo{volume}{2}
  (\bibinfo{year}{2020}) \bibinfo{pages}{197--199}.
\bibitem[{Kearns et~al.(2006)Kearns, Suri, and
  Montfort}]{kearns2006experimental}
\bibinfo{author}{M.~Kearns}, \bibinfo{author}{S.~Suri},
  \bibinfo{author}{N.~Montfort},
\newblock \bibinfo{title}{An experimental study of the coloring problem on
  human subject networks},
\newblock \bibinfo{journal}{Science} \bibinfo{volume}{313}
  (\bibinfo{year}{2006}) \bibinfo{pages}{824--827}.
\bibitem[{Ferrara(2017)}]{ferrara2017disinformation}
\bibinfo{author}{E.~Ferrara},
\newblock \bibinfo{title}{Disinformation and social bot operations in the run
  up to the 2017 {F}rench presidential election},
\newblock \bibinfo{journal}{{e}-print arXiv:1707.00086}
  (\bibinfo{year}{2017}).
\bibitem[{Lazer et~al.(2018)Lazer, Baum, Benkler, Berinsky, Greenhill, Menczer,
  Metzger, Nyhan, Pennycook, Rothschild et~al.}]{lazer2018science}
\bibinfo{author}{D.~M. Lazer}, \bibinfo{author}{M.~A. Baum},
  \bibinfo{author}{Y.~Benkler}, \bibinfo{author}{A.~J. Berinsky},
  \bibinfo{author}{K.~M. Greenhill}, \bibinfo{author}{F.~Menczer},
  \bibinfo{author}{M.~J. Metzger}, \bibinfo{author}{B.~Nyhan},
  \bibinfo{author}{G.~Pennycook}, \bibinfo{author}{D.~Rothschild}, et~al.,
\newblock \bibinfo{title}{The science of fake news},
\newblock \bibinfo{journal}{Science} \bibinfo{volume}{359}
  (\bibinfo{year}{2018}) \bibinfo{pages}{1094--1096}.
\bibitem[{Stella et~al.(2018)Stella, Ferrara, and De~Domenico}]{stella2018bots}
\bibinfo{author}{M.~Stella}, \bibinfo{author}{E.~Ferrara},
  \bibinfo{author}{M.~De~Domenico},
\newblock \bibinfo{title}{Bots increase exposure to negative and inflammatory
  content in online social systems},
\newblock \bibinfo{journal}{Proc. Natl. Acad. Sci. USA} \bibinfo{volume}{115}
  (\bibinfo{year}{2018}) \bibinfo{pages}{12435--12440}.
\bibitem[{Stewart et~al.(2019)Stewart, Mosleh, Diakonova, Arechar, Rand, and
  Plotkin}]{stewart2019information}
\bibinfo{author}{A.~J. Stewart}, \bibinfo{author}{M.~Mosleh},
  \bibinfo{author}{M.~Diakonova}, \bibinfo{author}{A.~A. Arechar},
  \bibinfo{author}{D.~G. Rand}, \bibinfo{author}{J.~B. Plotkin},
\newblock \bibinfo{title}{Information gerrymandering and undemocratic
  decisions},
\newblock \bibinfo{journal}{Nature} \bibinfo{volume}{573}
  (\bibinfo{year}{2019}) \bibinfo{pages}{117--121}.
\bibitem[{Crawford and Calo(2016)}]{crawford2016there}
\bibinfo{author}{K.~Crawford}, \bibinfo{author}{R.~Calo},
\newblock \bibinfo{title}{There is a blind spot in {AI} research},
\newblock \bibinfo{journal}{Nature} \bibinfo{volume}{538}
  (\bibinfo{year}{2016}) \bibinfo{pages}{311--313}.
\bibitem[{D'Orsogna and Perc(2015)}]{dorsogna2015statistical}
\bibinfo{author}{M.~R. D'Orsogna}, \bibinfo{author}{M.~Perc},
\newblock \bibinfo{title}{Statistical physics of crime: {A} review},
\newblock \bibinfo{journal}{Phys. Life Rev.} \bibinfo{volume}{12}
  (\bibinfo{year}{2015}) \bibinfo{pages}{1--21}.
\bibitem[{Ball(2012)}]{ball2012why}
\bibinfo{author}{P.~Ball}, \bibinfo{title}{Why society is a complex matter},
  \bibinfo{publisher}{Springer}, \bibinfo{year}{2012}.
\bibitem[{Helbing et~al.(2015)Helbing, Brockmann, Chadefaux, Donnay, Blanke,
  Woolley-Meza, Moussaid, Johansson, Krause, Schutte
  et~al.}]{helbing2015saving}
\bibinfo{author}{D.~Helbing}, \bibinfo{author}{D.~Brockmann},
  \bibinfo{author}{T.~Chadefaux}, \bibinfo{author}{K.~Donnay},
  \bibinfo{author}{U.~Blanke}, \bibinfo{author}{O.~Woolley-Meza},
  \bibinfo{author}{M.~Moussaid}, \bibinfo{author}{A.~Johansson},
  \bibinfo{author}{J.~Krause}, \bibinfo{author}{S.~Schutte}, et~al.,
\newblock \bibinfo{title}{Saving human lives: {W}hat complexity science and
  information systems can contribute},
\newblock \bibinfo{journal}{J. Stat. Phys.} \bibinfo{volume}{158}
  (\bibinfo{year}{2015}) \bibinfo{pages}{735--781}.
\bibitem[{Wilson and Kelling(1982)}]{wilson1982broken}
\bibinfo{author}{J.~Q. Wilson}, \bibinfo{author}{G.~L. Kelling},
\newblock \bibinfo{title}{Broken windows},
\newblock \bibinfo{journal}{Atlantic Monthly} \bibinfo{volume}{249}
  (\bibinfo{year}{1982}) \bibinfo{pages}{29--38}.
\bibitem[{Kuznetsov(2004)}]{kuznetsov2004elements}
\bibinfo{author}{Y.~A. Kuznetsov}, \bibinfo{title}{Elements of applied
  bifurcation theory}, \bibinfo{publisher}{Springer}, \bibinfo{year}{2004}.
\bibitem[{Stanley(1971)}]{stanley1971introduction}
\bibinfo{author}{H.~E. Stanley}, \bibinfo{title}{Introduction to Phase
  Transitions and Critical Phenomena}, \bibinfo{publisher}{Oxford University
  Press}, \bibinfo{year}{1971}.
\bibitem[{Castellano et~al.(2009)Castellano, Fortunato, and
  Loreto}]{castellano2009statistical}
\bibinfo{author}{C.~Castellano}, \bibinfo{author}{S.~Fortunato},
  \bibinfo{author}{V.~Loreto},
\newblock \bibinfo{title}{Statistical physics of social dynamics},
\newblock \bibinfo{journal}{Rev. Mod. Phys.} \bibinfo{volume}{81}
  (\bibinfo{year}{2009}) \bibinfo{pages}{591--646}.
\bibitem[{Cohen and Felson(1979)}]{cohen1979social}
\bibinfo{author}{L.~E. Cohen}, \bibinfo{author}{M.~Felson},
\newblock \bibinfo{title}{Social change and crime rate trends: {A} routine
  activity approach},
\newblock \bibinfo{journal}{Am. Sociol. Rev.}  (\bibinfo{year}{1979})
  \bibinfo{pages}{588--608}.
\bibitem[{Short et~al.(2008)Short, D'orsogna, Pasour, Tita, Brantingham,
  Bertozzi, and Chayes}]{short2008statistical}
\bibinfo{author}{M.~B. Short}, \bibinfo{author}{M.~R. D'orsogna},
  \bibinfo{author}{V.~B. Pasour}, \bibinfo{author}{G.~E. Tita},
  \bibinfo{author}{P.~J. Brantingham}, \bibinfo{author}{A.~L. Bertozzi},
  \bibinfo{author}{L.~B. Chayes},
\newblock \bibinfo{title}{A statistical model of criminal behavior},
\newblock \bibinfo{journal}{Math. Mod. Meth. Appl. Sci.} \bibinfo{volume}{18}
  (\bibinfo{year}{2008}) \bibinfo{pages}{1249--1267}.
\bibitem[{Johnson et~al.(1997)Johnson, Bowers, and
  Hirschfield}]{johnson1997new}
\bibinfo{author}{S.~D. Johnson}, \bibinfo{author}{K.~Bowers},
  \bibinfo{author}{A.~Hirschfield},
\newblock \bibinfo{title}{New insights into the spatial and temporal
  distribution of repeat victimization},
\newblock \bibinfo{journal}{Brit. J. Criminol.} \bibinfo{volume}{37}
  (\bibinfo{year}{1997}) \bibinfo{pages}{224--241}.
\bibitem[{Townsley et~al.(2000)Townsley, Homel, and
  Chaseling}]{townsley2000repeat}
\bibinfo{author}{M.~Townsley}, \bibinfo{author}{R.~Homel},
  \bibinfo{author}{J.~Chaseling},
\newblock \bibinfo{title}{Repeat burglary victimisation: {S}patial and temporal
  patterns},
\newblock \bibinfo{journal}{Austr. New. Zeal. J. Criminol.}
  \bibinfo{volume}{33} (\bibinfo{year}{2000}) \bibinfo{pages}{37--63}.
\bibitem[{Johnson and Bowers(2004)}]{johnson2004stability}
\bibinfo{author}{S.~D. Johnson}, \bibinfo{author}{K.~J. Bowers},
\newblock \bibinfo{title}{The stability of space-time clusters of burglary},
\newblock \bibinfo{journal}{Brit. J. Criminol.} \bibinfo{volume}{44}
  (\bibinfo{year}{2004}) \bibinfo{pages}{55--65}.
\bibitem[{Short et~al.(2009)Short, D’orsogna, Brantingham, and
  Tita}]{short2009measuring}
\bibinfo{author}{M.~B. Short}, \bibinfo{author}{M.~R. D’orsogna},
  \bibinfo{author}{P.~J. Brantingham}, \bibinfo{author}{G.~E. Tita},
\newblock \bibinfo{title}{Measuring and modeling repeat and near-repeat
  burglary effects},
\newblock \bibinfo{journal}{J. Quant. Criminol.} \bibinfo{volume}{25}
  (\bibinfo{year}{2009}) \bibinfo{pages}{325--339}.
\bibitem[{Green(1995)}]{green1995cleaning}
\bibinfo{author}{L.~Green},
\newblock \bibinfo{title}{Cleaning up drug hot spots in {O}akland,
  {C}alifornia: {T}he displacement and diffusion effects},
\newblock \bibinfo{journal}{Just. Quart.} \bibinfo{volume}{12}
  (\bibinfo{year}{1995}) \bibinfo{pages}{737--754}.
\bibitem[{Braga(2001)}]{braga2001effects}
\bibinfo{author}{A.~A. Braga},
\newblock \bibinfo{title}{The effects of hot spots policing on crime},
\newblock \bibinfo{journal}{Ann. Am. Acad. Pol. Soc. Sci.}
  \bibinfo{volume}{578} (\bibinfo{year}{2001}) \bibinfo{pages}{104--125}.
\bibitem[{Braga et~al.(2014)Braga, Papachristos, and Hureau}]{braga2014effects}
\bibinfo{author}{A.~A. Braga}, \bibinfo{author}{A.~V. Papachristos},
  \bibinfo{author}{D.~M. Hureau},
\newblock \bibinfo{title}{The effects of hot spots policing on crime: {A}n
  updated systematic review and meta-analysis},
\newblock \bibinfo{journal}{Just. Quart.} \bibinfo{volume}{31}
  (\bibinfo{year}{2014}) \bibinfo{pages}{633--663}.
\bibitem[{Weisburd et~al.(2006)Weisburd, Wyckoff, Ready, Eck, Hinkle, and
  Gajewski}]{weisburd2006does}
\bibinfo{author}{D.~Weisburd}, \bibinfo{author}{L.~A. Wyckoff},
  \bibinfo{author}{J.~Ready}, \bibinfo{author}{J.~E. Eck},
  \bibinfo{author}{J.~C. Hinkle}, \bibinfo{author}{F.~Gajewski},
\newblock \bibinfo{title}{Does crime just move around the corner? {A}
  controlled study of spatial displacement and diffusion of crime control
  benefits},
\newblock \bibinfo{journal}{Criminology} \bibinfo{volume}{44}
  (\bibinfo{year}{2006}) \bibinfo{pages}{549--592}.
\bibitem[{Taniguchi et~al.(2009)Taniguchi, Rengert, and
  McCord}]{taniguchi2009size}
\bibinfo{author}{T.~A. Taniguchi}, \bibinfo{author}{G.~F. Rengert},
  \bibinfo{author}{E.~S. McCord},
\newblock \bibinfo{title}{Where size matters: {A}gglomeration economies of
  illegal drug markets in {P}hiladelphia},
\newblock \bibinfo{journal}{Just. Quart.} \bibinfo{volume}{26}
  (\bibinfo{year}{2009}) \bibinfo{pages}{670--694}.
\bibitem[{Becker(1968)}]{becker1968crime}
\bibinfo{author}{G.~S. Becker},
\newblock \bibinfo{title}{Crime and punishment: {A}n economic approach},
\newblock \bibinfo{journal}{J. Political Econ.} \bibinfo{volume}{76}
  (\bibinfo{year}{1968}) \bibinfo{pages}{169--217}.
\bibitem[{Doob and Webster(2003)}]{doob2003sentence}
\bibinfo{author}{A.~N. Doob}, \bibinfo{author}{C.~M. Webster},
\newblock \bibinfo{title}{Sentence severity and crime: {A}ccepting the null
  hypothesis},
\newblock \bibinfo{journal}{Crime Just.} \bibinfo{volume}{30}
  (\bibinfo{year}{2003}) \bibinfo{pages}{143--195}.
\bibitem[{Johnson et~al.(2007)Johnson, Bernasco, Bowers, Elffers, Ratcliffe,
  Rengert, and Townsley}]{johnson2007space}
\bibinfo{author}{S.~D. Johnson}, \bibinfo{author}{W.~Bernasco},
  \bibinfo{author}{K.~J. Bowers}, \bibinfo{author}{H.~Elffers},
  \bibinfo{author}{J.~Ratcliffe}, \bibinfo{author}{G.~Rengert},
  \bibinfo{author}{M.~Townsley},
\newblock \bibinfo{title}{Space-time patterns of risk: a cross national
  assessment of residential burglary victimization},
\newblock \bibinfo{journal}{J. Quant. Criminol.} \bibinfo{volume}{23}
  (\bibinfo{year}{2007}) \bibinfo{pages}{201--219}.
\bibitem[{Rengert et~al.(1999)Rengert, Piquero, and
  Jones}]{rengert1999distance}
\bibinfo{author}{G.~F. Rengert}, \bibinfo{author}{A.~R. Piquero},
  \bibinfo{author}{P.~R. Jones},
\newblock \bibinfo{title}{Distance decay reexamined},
\newblock \bibinfo{journal}{Criminology} \bibinfo{volume}{37}
  (\bibinfo{year}{1999}) \bibinfo{pages}{427--446}.
\bibitem[{Bernasco and Luykx(2003)}]{bernasco2003effects}
\bibinfo{author}{W.~Bernasco}, \bibinfo{author}{F.~Luykx},
\newblock \bibinfo{title}{Effects of attractiveness, opportunity and
  accessibility to burglars on residential burglary rates of urban
  neighborhoods},
\newblock \bibinfo{journal}{Criminology} \bibinfo{volume}{41}
  (\bibinfo{year}{2003}) \bibinfo{pages}{981--1002}.
\bibitem[{Bernasco and Nieuwbeerta(2005)}]{bernasco2005residential}
\bibinfo{author}{W.~Bernasco}, \bibinfo{author}{P.~Nieuwbeerta},
\newblock \bibinfo{title}{How do residential burglars select target areas? a
  new approach to the analysis of criminal location choice},
\newblock \bibinfo{journal}{Brit. J. Criminol.} \bibinfo{volume}{45}
  (\bibinfo{year}{2005}) \bibinfo{pages}{296--315}.
\bibitem[{Short et~al.(2010{\natexlab{a}})Short, Brantingham, Bertozzi, and
  Tita}]{short2010dissipation}
\bibinfo{author}{M.~B. Short}, \bibinfo{author}{P.~J. Brantingham},
  \bibinfo{author}{A.~L. Bertozzi}, \bibinfo{author}{G.~E. Tita},
\newblock \bibinfo{title}{Dissipation and displacement of hotspots in
  reaction-diffusion models of crime},
\newblock \bibinfo{journal}{Proc. Natl. Acad. Sci. USA} \bibinfo{volume}{107}
  (\bibinfo{year}{2010}{\natexlab{a}}) \bibinfo{pages}{3961--3965}.
\bibitem[{Short et~al.(2010{\natexlab{b}})Short, Bertozzi, and
  Brantingham}]{short2010nonlinear}
\bibinfo{author}{M.~B. Short}, \bibinfo{author}{A.~L. Bertozzi},
  \bibinfo{author}{P.~J. Brantingham},
\newblock \bibinfo{title}{Nonlinear patterns in urban crime: {H}otspots,
  bifurcations, and suppression},
\newblock \bibinfo{journal}{SIAM J. Appl. Dyn. Syst.} \bibinfo{volume}{9}
  (\bibinfo{year}{2010}{\natexlab{b}}) \bibinfo{pages}{462--483}.
\bibitem[{Cross and Hohenberg(1993)}]{cross1993pattern}
\bibinfo{author}{M.~C. Cross}, \bibinfo{author}{P.~C. Hohenberg},
\newblock \bibinfo{title}{Pattern formation outside of equilibrium},
\newblock \bibinfo{journal}{Rev. Mod. Phys.} \bibinfo{volume}{65}
  (\bibinfo{year}{1993}) \bibinfo{pages}{851}.
\bibitem[{Rodriguez and Bertozzi(2010)}]{rodriguez2010local}
\bibinfo{author}{N.~Rodriguez}, \bibinfo{author}{A.~Bertozzi},
\newblock \bibinfo{title}{Local existence and uniqueness of solutions to a
  {PDE} model for criminal behavior},
\newblock \bibinfo{journal}{Math. Mod. Meth. Appl. Sci.} \bibinfo{volume}{20}
  (\bibinfo{year}{2010}) \bibinfo{pages}{1425--1457}.
\bibitem[{Jones et~al.(2010)Jones, Brantingham, and
  Chayes}]{jones2010statistical}
\bibinfo{author}{P.~A. Jones}, \bibinfo{author}{P.~J. Brantingham},
  \bibinfo{author}{L.~R. Chayes},
\newblock \bibinfo{title}{Statistical models of criminal behavior: the effects
  of law enforcement actions},
\newblock \bibinfo{journal}{Math. Mod. Meth. Appl. Sci.} \bibinfo{volume}{20}
  (\bibinfo{year}{2010}) \bibinfo{pages}{1397--1423}.
\bibitem[{Cantrell et~al.(2012)Cantrell, Cosner, and
  Man{\'a}sevich}]{cantrell2012global}
\bibinfo{author}{R.~S. Cantrell}, \bibinfo{author}{C.~Cosner},
  \bibinfo{author}{R.~Man{\'a}sevich},
\newblock \bibinfo{title}{Global bifurcation of solutions for crime modeling
  equations},
\newblock \bibinfo{journal}{SIAM J. Math. Anal.} \bibinfo{volume}{44}
  (\bibinfo{year}{2012}) \bibinfo{pages}{1340--1358}.
\bibitem[{Berestycki et~al.(2013)Berestycki, Rodriguez, and
  Ryzhik}]{berestycki2013traveling}
\bibinfo{author}{H.~Berestycki}, \bibinfo{author}{N.~Rodriguez},
  \bibinfo{author}{L.~Ryzhik},
\newblock \bibinfo{title}{Traveling wave solutions in a reaction-diffusion
  model for criminal activity},
\newblock \bibinfo{journal}{Multiscale Model. Simul.} \bibinfo{volume}{11}
  (\bibinfo{year}{2013}) \bibinfo{pages}{1097--1126}.
\bibitem[{Zipkin et~al.(2014)Zipkin, Short, and Bertozzi}]{zipkin2014cops}
\bibinfo{author}{J.~R. Zipkin}, \bibinfo{author}{M.~B. Short},
  \bibinfo{author}{A.~L. Bertozzi},
\newblock \bibinfo{title}{Cops on the dots in a mathematical model of urban
  crime and police response},
\newblock \bibinfo{journal}{Disc. Cont. Dyn. Syst. B} \bibinfo{volume}{19}
  (\bibinfo{year}{2014}) \bibinfo{pages}{1479--1506}.
\bibitem[{Nuno et~al.(2008)Nuno, Herrero, and Primicerio}]{nuno2008triangle}
\bibinfo{author}{J.~C. Nuno}, \bibinfo{author}{M.~A. Herrero},
  \bibinfo{author}{M.~Primicerio},
\newblock \bibinfo{title}{A triangle model of criminality},
\newblock \bibinfo{journal}{Physica A} \bibinfo{volume}{387}
  (\bibinfo{year}{2008}) \bibinfo{pages}{2926--2936}.
\bibitem[{Nuno et~al.(2011)Nuno, Herrero, and
  Primicerio}]{nuno2011mathematical}
\bibinfo{author}{J.~C. Nuno}, \bibinfo{author}{M.~A. Herrero},
  \bibinfo{author}{M.~Primicerio},
\newblock \bibinfo{title}{A mathematical model of a criminal-prone society},
\newblock \bibinfo{journal}{Disc. Cont. Dyn. Syst. S} \bibinfo{volume}{4}
  (\bibinfo{year}{2011}) \bibinfo{pages}{193--207}.
\bibitem[{McMillon et~al.(2014)McMillon, Simon, and
  Morenoff}]{mcmillon2014modeling}
\bibinfo{author}{D.~McMillon}, \bibinfo{author}{C.~P. Simon},
  \bibinfo{author}{J.~Morenoff},
\newblock \bibinfo{title}{Modeling the underlying dynamics of the spread of
  crime},
\newblock \bibinfo{journal}{PLOS ONE} \bibinfo{volume}{9}
  (\bibinfo{year}{2014}) \bibinfo{pages}{e88923}.
\bibitem[{Sooknanan et~al.(2012)Sooknanan, Bhatt, and
  Comissiong}]{sooknanan2012criminals}
\bibinfo{author}{J.~Sooknanan}, \bibinfo{author}{B.~Bhatt},
  \bibinfo{author}{D.~M.~G. Comissiong},
\newblock \bibinfo{title}{Criminals treated as predators to be harvested: a two
  prey one predator model with group defense, prey migration and switching},
\newblock \bibinfo{journal}{J. Math. Res.} \bibinfo{volume}{4}
  (\bibinfo{year}{2012}) \bibinfo{pages}{92--106}.
\bibitem[{Mohler et~al.(2011)Mohler, Short, Brantingham, Schoenberg, and
  Tita}]{mohler2011self}
\bibinfo{author}{G.~O. Mohler}, \bibinfo{author}{M.~B. Short},
  \bibinfo{author}{P.~J. Brantingham}, \bibinfo{author}{F.~P. Schoenberg},
  \bibinfo{author}{G.~E. Tita},
\newblock \bibinfo{title}{Self-exciting point process modeling of crime},
\newblock \bibinfo{journal}{J. Am. Stat. Assoc.} \bibinfo{volume}{106}
  (\bibinfo{year}{2011}) \bibinfo{pages}{100--108}.
\bibitem[{Daley and Vere-Jones(2003)}]{daley2003introduction}
\bibinfo{author}{D.~J. Daley}, \bibinfo{author}{D.~Vere-Jones},
  \bibinfo{title}{An introduction to the theory of point processes, volume 1:
  Elementary theory and methods}, \bibinfo{publisher}{Springer},
  \bibinfo{year}{2003}.
\bibitem[{Bowers et~al.(2004)Bowers, Johnson, and
  Pease}]{bowers2004prospective}
\bibinfo{author}{K.~J. Bowers}, \bibinfo{author}{S.~D. Johnson},
  \bibinfo{author}{K.~Pease},
\newblock \bibinfo{title}{Prospective hot-spotting: the future of crime
  mapping?},
\newblock \bibinfo{journal}{Brit. J. Criminol.} \bibinfo{volume}{44}
  (\bibinfo{year}{2004}) \bibinfo{pages}{641--658}.
\bibitem[{Lewis et~al.(2012)Lewis, Mohler, Brantingham, and
  Bertozzi}]{lewis2012self}
\bibinfo{author}{E.~Lewis}, \bibinfo{author}{G.~Mohler}, \bibinfo{author}{P.~J.
  Brantingham}, \bibinfo{author}{A.~L. Bertozzi},
\newblock \bibinfo{title}{Self-exciting point process models of civilian deaths
  in {I}raq},
\newblock \bibinfo{journal}{Secur. J.} \bibinfo{volume}{25}
  (\bibinfo{year}{2012}) \bibinfo{pages}{244--264}.
\bibitem[{Mohler and Short(2012)}]{mohler2012geographic}
\bibinfo{author}{G.~O. Mohler}, \bibinfo{author}{M.~B. Short},
\newblock \bibinfo{title}{Geographic profiling from kinetic models of criminal
  behavior},
\newblock \bibinfo{journal}{SIAM J. Appl. Math.} \bibinfo{volume}{72}
  (\bibinfo{year}{2012}) \bibinfo{pages}{163--180}.
\bibitem[{Axelrod(1984)}]{axelrod1984evolution}
\bibinfo{author}{R.~Axelrod}, \bibinfo{title}{The Evolution of Cooperation},
  \bibinfo{publisher}{Basic Books}, \bibinfo{year}{1984}.
\bibitem[{Hardin(1968)}]{hardin1968tragedy}
\bibinfo{author}{G.~Hardin},
\newblock \bibinfo{title}{The tragedy of the commons},
\newblock \bibinfo{journal}{Science} \bibinfo{volume}{162}
  (\bibinfo{year}{1968}) \bibinfo{pages}{1243--1248}.
\bibitem[{Short et~al.(2010)Short, Brantingham, and
  D’orsogna}]{short2010cooperation}
\bibinfo{author}{M.~B. Short}, \bibinfo{author}{P.~J. Brantingham},
  \bibinfo{author}{M.~R. D’orsogna},
\newblock \bibinfo{title}{Cooperation and punishment in an adversarial game:
  {H}ow defectors pave the way to a peaceful society},
\newblock \bibinfo{journal}{Phys. Rev. E} \bibinfo{volume}{82}
  (\bibinfo{year}{2010}) \bibinfo{pages}{066114}.
\bibitem[{Lynch and Addington(2006)}]{lynch2006understanding}
\bibinfo{author}{J.~P. Lynch}, \bibinfo{author}{L.~A. Addington},
  \bibinfo{title}{Understanding Crime Statistics: Revisiting the Divergence of
  the NCVS and the UCR}, \bibinfo{publisher}{Cambridge University Press},
  \bibinfo{year}{2006}.
\bibitem[{Fehr(2004)}]{fehr2004dont}
\bibinfo{author}{E.~Fehr},
\newblock \bibinfo{title}{Don't lose your reputation},
\newblock \bibinfo{journal}{Nature} \bibinfo{volume}{432}
  (\bibinfo{year}{2004}) \bibinfo{pages}{449--450}.
\bibitem[{Szolnoki and Perc(2017)}]{szolnoki2017second}
\bibinfo{author}{A.~Szolnoki}, \bibinfo{author}{M.~Perc},
\newblock \bibinfo{title}{Second-order free-riding on antisocial punishment
  restores the effectiveness of prosocial punishment},
\newblock \bibinfo{journal}{Phys. Rev. X} \bibinfo{volume}{7}
  (\bibinfo{year}{2017}) \bibinfo{pages}{041027}.
\bibitem[{Schlag(1998)}]{schlag1998imitate}
\bibinfo{author}{K.~H. Schlag},
\newblock \bibinfo{title}{Why imitate, and if so, how? {A} boundedly rational
  approach to multi-armed bandits},
\newblock \bibinfo{journal}{J. Econ. Theory} \bibinfo{volume}{78}
  (\bibinfo{year}{1998}) \bibinfo{pages}{130--156}.
\bibitem[{Short et~al.(2013)Short, Pitcher, and D'orsogna}]{short2013external}
\bibinfo{author}{M.~B. Short}, \bibinfo{author}{A.~B. Pitcher},
  \bibinfo{author}{M.~R. D'orsogna},
\newblock \bibinfo{title}{External conversions of player strategy in an
  evolutionary game: {A} cost-benefit analysis through optimal control},
\newblock \bibinfo{journal}{Eur. J. Appl. Math.} \bibinfo{volume}{24}
  (\bibinfo{year}{2013}) \bibinfo{pages}{131}.
\bibitem[{D'Orsogna et~al.(2013)D'Orsogna, Kendall, McBride, and
  Short}]{dorsogna2013criminal}
\bibinfo{author}{M.~R. D'Orsogna}, \bibinfo{author}{R.~Kendall},
  \bibinfo{author}{M.~McBride}, \bibinfo{author}{M.~B. Short},
\newblock \bibinfo{title}{Criminal defectors lead to the emergence of
  cooperation in an experimental, adversarial game},
\newblock \bibinfo{journal}{PLOS ONE} \bibinfo{volume}{8}
  (\bibinfo{year}{2013}) \bibinfo{pages}{e61458}.
\bibitem[{Tsebelis(1990)}]{tsebelis1990penalty}
\bibinfo{author}{G.~Tsebelis},
\newblock \bibinfo{title}{Penalty has no impact on crime: {A} game-theoretic
  analysis},
\newblock \bibinfo{journal}{Ration. Soc.} \bibinfo{volume}{2}
  (\bibinfo{year}{1990}) \bibinfo{pages}{255--286}.
\bibitem[{Perc et~al.(2013)Perc, Donnay, and Helbing}]{perc2013understanding}
\bibinfo{author}{M.~Perc}, \bibinfo{author}{K.~Donnay},
  \bibinfo{author}{D.~Helbing},
\newblock \bibinfo{title}{Understanding recurrent crime as system-immanent
  collective behavior},
\newblock \bibinfo{journal}{PLOS ONE} \bibinfo{volume}{8}
  (\bibinfo{year}{2013}) \bibinfo{pages}{e76063}.
\bibitem[{Perc and Szolnoki(2015)}]{perc2015double}
\bibinfo{author}{M.~Perc}, \bibinfo{author}{A.~Szolnoki},
\newblock \bibinfo{title}{A double-edged sword: {B}enefits and pitfalls of
  heterogeneous punishment in evolutionary inspection games},
\newblock \bibinfo{journal}{Sci. Rep.} \bibinfo{volume}{5}
  (\bibinfo{year}{2015}) \bibinfo{pages}{11027}.
\bibitem[{McFadden(1974)}]{mcfadden1974conditional}
\bibinfo{author}{D.~McFadden},
\newblock \bibinfo{title}{Conditional logit analysis of qualitative choice
  behavior},
\newblock in: \bibinfo{editor}{P.~Zarembka} (Ed.),
  \bibinfo{booktitle}{Frontiers in econometrics}, \bibinfo{publisher}{Academic
  Press}, \bibinfo{year}{1974}, pp. \bibinfo{pages}{105--142}.
\bibitem[{Szab{\'o} and Fath(2007)}]{szabo2007evolutionary}
\bibinfo{author}{G.~Szab{\'o}}, \bibinfo{author}{G.~Fath},
\newblock \bibinfo{title}{Evolutionary games on graphs},
\newblock \bibinfo{journal}{Phys. Rep.} \bibinfo{volume}{446}
  (\bibinfo{year}{2007}) \bibinfo{pages}{97--216}.
\bibitem[{Perc(2017)}]{perc2017stability}
\bibinfo{author}{M.~Perc},
\newblock \bibinfo{title}{Stability of subsystem solutions in agent-based
  models},
\newblock \bibinfo{journal}{Eur. J. Phys.} \bibinfo{volume}{39}
  (\bibinfo{year}{2017}) \bibinfo{pages}{014001}.
\bibitem[{Dorogovtsev and Mendes(2002)}]{dorogovtsev2002evolution}
\bibinfo{author}{S.~N. Dorogovtsev}, \bibinfo{author}{J.~F. Mendes},
\newblock \bibinfo{title}{Evolution of networks},
\newblock \bibinfo{journal}{Adv. Phys.} \bibinfo{volume}{51}
  (\bibinfo{year}{2002}) \bibinfo{pages}{1079--1187}.
\bibitem[{Estrada(2012)}]{estrada2012structure}
\bibinfo{author}{E.~Estrada}, \bibinfo{title}{The Structure of Complex
  Networks: Theory and Applications}, \bibinfo{publisher}{Oxford University
  Press}, \bibinfo{year}{2012}.
\bibitem[{Mallory(2011)}]{mallory2011understanding}
\bibinfo{author}{S.~L. Mallory}, \bibinfo{title}{Understanding organized
  crime}, \bibinfo{publisher}{Jones \& Bartlett Publishers},
  \bibinfo{year}{2011}.
\bibitem[{Ribeiro et~al.(2018)Ribeiro, Alves, Martins, Lenzi, and
  Perc}]{ribeiro2018dynamical}
\bibinfo{author}{H.~V. Ribeiro}, \bibinfo{author}{L.~G.~A. Alves},
  \bibinfo{author}{A.~F. Martins}, \bibinfo{author}{E.~K. Lenzi},
  \bibinfo{author}{M.~Perc},
\newblock \bibinfo{title}{The dynamical structure of political corruption
  networks},
\newblock \bibinfo{journal}{J. Complex Netw.} \bibinfo{volume}{6}
  (\bibinfo{year}{2018}) \bibinfo{pages}{989--1003}.
\bibitem[{Gambetta(1988)}]{gambetta1998trust}
\bibinfo{author}{D.~Gambetta}, \bibinfo{title}{Trust: Making and breaking
  cooperative relations}, \bibinfo{publisher}{B. Blackwell},
  \bibinfo{year}{1988}.
\bibitem[{Beittel(2009)}]{beittel2009mexico}
\bibinfo{author}{J.~S. Beittel}, \bibinfo{title}{Mexico's drug-related
  violence}, \bibinfo{year}{2009}. \bibinfo{note}{Report by Congressional
  Research Service. Archived at: \url{https://doi.org/10.17605/OSF.IO/N7G9X}}.
\bibitem[{Raphael and Winter-Ebmer(2001)}]{raphael2001identifying}
\bibinfo{author}{S.~Raphael}, \bibinfo{author}{R.~Winter-Ebmer},
\newblock \bibinfo{title}{Identifying the effect of unemployment on crime},
\newblock \bibinfo{journal}{J. Law Econ.} \bibinfo{volume}{44}
  (\bibinfo{year}{2001}) \bibinfo{pages}{259--283}.
\bibitem[{Lin(2008)}]{lin2008does}
\bibinfo{author}{M.-J. Lin},
\newblock \bibinfo{title}{Does unemployment increase crime? {E}vidence from
  {US} data 1974--2000},
\newblock \bibinfo{journal}{J. Hum. Resour.} \bibinfo{volume}{43}
  (\bibinfo{year}{2008}) \bibinfo{pages}{413--436}.
\bibitem[{LaFree(1999)}]{lafree1999declining}
\bibinfo{author}{G.~LaFree},
\newblock \bibinfo{title}{Declining violent crime rates in the 1990s:
  {P}redicting crime booms and busts},
\newblock \bibinfo{journal}{Ann. Rev. Sociol.} \bibinfo{volume}{25}
  (\bibinfo{year}{1999}) \bibinfo{pages}{145--168}.
\bibitem[{Curtis(1997)}]{curtis1997improbable}
\bibinfo{author}{R.~Curtis},
\newblock \bibinfo{title}{The improbable transformation of inner-city
  neighborhoods: {C}rime, violence, drugs, and youth in the 1990s},
\newblock \bibinfo{journal}{J. Crim. Law Criminol.} \bibinfo{volume}{88}
  (\bibinfo{year}{1997}) \bibinfo{pages}{1233}.
\bibitem[{LaFree(1998)}]{lafree1998losing}
\bibinfo{author}{G.~LaFree}, \bibinfo{title}{Losing legitimacy: Street crime
  and the decline of social institutions in America},
  \bibinfo{publisher}{Westview Press}, \bibinfo{year}{1998}.
\bibitem[{Corsaro and McGarrell(2009)}]{corsaro2009testing}
\bibinfo{author}{N.~Corsaro}, \bibinfo{author}{E.~F. McGarrell},
\newblock \bibinfo{title}{Testing a promising homicide reduction strategy:
  {R}e-assessing the impact of the {I}ndianapolis ``pulling levers''
  intervention},
\newblock \bibinfo{journal}{J. Exp. Criminol.} \bibinfo{volume}{5}
  (\bibinfo{year}{2009}) \bibinfo{pages}{63}.
\bibitem[{McGarrell et~al.(2010)McGarrell, Corsaro, Hipple, and
  Bynum}]{mcgarrell2010project}
\bibinfo{author}{E.~F. McGarrell}, \bibinfo{author}{N.~Corsaro},
  \bibinfo{author}{N.~K. Hipple}, \bibinfo{author}{T.~S. Bynum},
\newblock \bibinfo{title}{Project safe neighborhoods and violent crime trends
  in {US} cities: {A}ssessing violent crime impact},
\newblock \bibinfo{journal}{J. Quant. Criminol.} \bibinfo{volume}{26}
  (\bibinfo{year}{2010}) \bibinfo{pages}{165--190}.
\bibitem[{Calderoni et~al.(2021)Calderoni, Campedelli, Szekely, Paolucci, and
  Andrighetto}]{calderoni2021recruitment}
\bibinfo{author}{F.~Calderoni}, \bibinfo{author}{G.~M. Campedelli},
  \bibinfo{author}{A.~Szekely}, \bibinfo{author}{M.~Paolucci},
  \bibinfo{author}{G.~Andrighetto},
\newblock \bibinfo{title}{Recruitment into organized crime: {A}n agent-based
  approach testing the impact of different policies},
\newblock \bibinfo{journal}{J. Quant. Criminol.}  (\bibinfo{year}{2021}).
\bibitem[{G{\'o}mez-Sorzano(2007)}]{gomez2007decomposing}
\bibinfo{author}{G.~A. G{\'o}mez-Sorzano},
\newblock \bibinfo{title}{Decomposing violence: crime cycles in the twentieth
  century in the {U}nited {S}tates},
\newblock \bibinfo{journal}{Appl. Econ. Int. Dev.} \bibinfo{volume}{7}
  (\bibinfo{year}{2007}).
\bibitem[{Zimring(2006)}]{zimring2006thegreat}
\bibinfo{author}{F.~E. Zimring}, \bibinfo{title}{The great American crime
  decline}, \bibinfo{publisher}{Oxford University Press}, \bibinfo{year}{2006}.
\bibitem[{Papachristos et~al.(2012)Papachristos, Braga, and
  Hureau}]{papachristos2012social}
\bibinfo{author}{A.~V. Papachristos}, \bibinfo{author}{A.~A. Braga},
  \bibinfo{author}{D.~M. Hureau},
\newblock \bibinfo{title}{Social networks and the risk of gunshot injury},
\newblock \bibinfo{journal}{J. Urban Health} \bibinfo{volume}{89}
  (\bibinfo{year}{2012}) \bibinfo{pages}{992--1003}.
\bibitem[{Papachristos et~al.(2013)Papachristos, Hureau, and
  Braga}]{papachristos2013corner}
\bibinfo{author}{A.~V. Papachristos}, \bibinfo{author}{D.~M. Hureau},
  \bibinfo{author}{A.~A. Braga},
\newblock \bibinfo{title}{The corner and the crew: {T}he influence of geography
  and social networks on gang violence},
\newblock \bibinfo{journal}{Am. Sociol. Rev.} \bibinfo{volume}{78}
  (\bibinfo{year}{2013}) \bibinfo{pages}{417--447}.
\bibitem[{Albert et~al.(2000)Albert, Jeong, and Barab{\'a}si}]{albert2000error}
\bibinfo{author}{R.~Albert}, \bibinfo{author}{H.~Jeong}, \bibinfo{author}{A.-L.
  Barab{\'a}si},
\newblock \bibinfo{title}{Error and attack tolerance of complex networks},
\newblock \bibinfo{journal}{Nature} \bibinfo{volume}{406}
  (\bibinfo{year}{2000}) \bibinfo{pages}{378--382}.
\bibitem[{Duijn et~al.(2014)Duijn, Kashirin, and Sloot}]{duijn2014relative}
\bibinfo{author}{P.~A. Duijn}, \bibinfo{author}{V.~Kashirin},
  \bibinfo{author}{P.~M. Sloot},
\newblock \bibinfo{title}{The relative ineffectiveness of criminal network
  disruption},
\newblock \bibinfo{journal}{Sci. Rep.} \bibinfo{volume}{4}
  (\bibinfo{year}{2014}) \bibinfo{pages}{4238}.
\bibitem[{Hegemann et~al.(2011)Hegemann, Smith, Barbaro, Bertozzi, Reid, and
  Tita}]{hegemann2011geographical}
\bibinfo{author}{R.~A. Hegemann}, \bibinfo{author}{L.~M. Smith},
  \bibinfo{author}{A.~B. Barbaro}, \bibinfo{author}{A.~L. Bertozzi},
  \bibinfo{author}{S.~E. Reid}, \bibinfo{author}{G.~E. Tita},
\newblock \bibinfo{title}{Geographical influences of an emerging network of
  gang rivalries},
\newblock \bibinfo{journal}{Physica A} \bibinfo{volume}{390}
  (\bibinfo{year}{2011}) \bibinfo{pages}{3894--3914}.
\bibitem[{van Gennip et~al.(2013)van Gennip, Hunter, Ahn, Elliott, Luh,
  Halvorson, Reid, Valasik, Wo, Tita et~al.}]{vangennip2013community}
\bibinfo{author}{Y.~van Gennip}, \bibinfo{author}{B.~Hunter},
  \bibinfo{author}{R.~Ahn}, \bibinfo{author}{P.~Elliott},
  \bibinfo{author}{K.~Luh}, \bibinfo{author}{M.~Halvorson},
  \bibinfo{author}{S.~Reid}, \bibinfo{author}{M.~Valasik},
  \bibinfo{author}{J.~Wo}, \bibinfo{author}{G.~E. Tita}, et~al.,
\newblock \bibinfo{title}{Community detection using spectral clustering on
  sparse geosocial data},
\newblock \bibinfo{journal}{SIAM J. Appl. Math.} \bibinfo{volume}{73}
  (\bibinfo{year}{2013}) \bibinfo{pages}{67--83}.
\bibitem[{Barbaro et~al.(2013)Barbaro, Chayes, and
  D’Orsogna}]{barbaro2013territorial}
\bibinfo{author}{A.~B. Barbaro}, \bibinfo{author}{L.~Chayes},
  \bibinfo{author}{M.~R. D’Orsogna},
\newblock \bibinfo{title}{Territorial developments based on graffiti: {A}
  statistical mechanics approach},
\newblock \bibinfo{journal}{Physica A} \bibinfo{volume}{392}
  (\bibinfo{year}{2013}) \bibinfo{pages}{252--270}.
\bibitem[{Thompson et~al.(2012)Thompson, Offler, Hirsch, Every, Thomas, and
  Dawson}]{thompson2012broken}
\bibinfo{author}{K.~Thompson}, \bibinfo{author}{N.~Offler},
  \bibinfo{author}{L.~Hirsch}, \bibinfo{author}{D.~Every},
  \bibinfo{author}{M.~J. Thomas}, \bibinfo{author}{D.~Dawson},
\newblock \bibinfo{title}{From broken windows to a renovated research agenda:
  {A} review of the literature on vandalism and graffiti in the rail industry},
\newblock \bibinfo{journal}{Transp. Res. Part A Policy Pract.}
  \bibinfo{volume}{46} (\bibinfo{year}{2012}) \bibinfo{pages}{1280--1290}.
\bibitem[{Rose-Ackerman(1975)}]{rose1975economics}
\bibinfo{author}{S.~Rose-Ackerman},
\newblock \bibinfo{title}{The economics of corruption},
\newblock \bibinfo{journal}{J. Public Econ.} \bibinfo{volume}{4}
  (\bibinfo{year}{1975}) \bibinfo{pages}{187--203}.
\bibitem[{Shleifer and Vishny(1993)}]{shleifer1993corruption}
\bibinfo{author}{A.~Shleifer}, \bibinfo{author}{R.~W. Vishny},
\newblock \bibinfo{title}{Corruption},
\newblock \bibinfo{journal}{Q. J. Econ.} \bibinfo{volume}{108}
  (\bibinfo{year}{1993}) \bibinfo{pages}{599--617}.
\bibitem[{Mauro(1995)}]{mauro1995corruption}
\bibinfo{author}{P.~Mauro},
\newblock \bibinfo{title}{Corruption and growth},
\newblock \bibinfo{journal}{Q. J. Econ.} \bibinfo{volume}{110}
  (\bibinfo{year}{1995}) \bibinfo{pages}{681--712}.
\bibitem[{Bardhan(1997)}]{bardhan1997corruption}
\bibinfo{author}{P.~Bardhan},
\newblock \bibinfo{title}{Corruption and development: a review of issues},
\newblock \bibinfo{journal}{J. Econ. Lit.} \bibinfo{volume}{35}
  (\bibinfo{year}{1997}) \bibinfo{pages}{1320--1346}.
\bibitem[{Shao et~al.(2007)Shao, Ivanov, Podobnik, and
  Stanley}]{shao2007quantitative}
\bibinfo{author}{J.~Shao}, \bibinfo{author}{P.~C. Ivanov},
  \bibinfo{author}{B.~Podobnik}, \bibinfo{author}{H.~E. Stanley},
\newblock \bibinfo{title}{Quantitative relations between corruption and
  economic factors},
\newblock \bibinfo{journal}{Eur. Phys. J. B} \bibinfo{volume}{56}
  (\bibinfo{year}{2007}) \bibinfo{pages}{157--166}.
\bibitem[{Haque et~al.(2008)Haque, Kneller et~al.}]{haque2008public}
\bibinfo{author}{M.~E. Haque}, \bibinfo{author}{R.~Kneller}, et~al.,
  \bibinfo{title}{Public investment and growth: The role of corruption},
  \bibinfo{year}{2008}. \bibinfo{note}{Centre for Growth and Business Cycle
  Research discussion paper series 098. Archived at:
  \url{https://doi.org/10.17605/OSF.IO/N7G9X}}.
\bibitem[{Gupta et~al.(2002)Gupta, Davoodi, and Alonso-Terme}]{gupta2002does}
\bibinfo{author}{S.~Gupta}, \bibinfo{author}{H.~Davoodi},
  \bibinfo{author}{R.~Alonso-Terme},
\newblock \bibinfo{title}{Does corruption affect income inequality and
  poverty?},
\newblock \bibinfo{journal}{Econ. Gov.} \bibinfo{volume}{3}
  (\bibinfo{year}{2002}) \bibinfo{pages}{23--45}.
\bibitem[{Xu and Chen(2005)}]{xu2005criminal}
\bibinfo{author}{J.~Xu}, \bibinfo{author}{H.~Chen},
\newblock \bibinfo{title}{Criminal network analysis and visualization},
\newblock \bibinfo{journal}{Commun. ACM} \bibinfo{volume}{48}
  (\bibinfo{year}{2005}) \bibinfo{pages}{100--107}.
\bibitem[{Del~Vicario et~al.(2016)Del~Vicario, Bessi, Zollo, Petroni, Scala,
  Caldarelli, Stanley, and Quattrociocchi}]{delvicario2016spreading}
\bibinfo{author}{M.~Del~Vicario}, \bibinfo{author}{A.~Bessi},
  \bibinfo{author}{F.~Zollo}, \bibinfo{author}{F.~Petroni},
  \bibinfo{author}{A.~Scala}, \bibinfo{author}{G.~Caldarelli},
  \bibinfo{author}{H.~E. Stanley}, \bibinfo{author}{W.~Quattrociocchi},
\newblock \bibinfo{title}{The spreading of misinformation online},
\newblock \bibinfo{journal}{Proc. Natl. Acad. Sci. USA} \bibinfo{volume}{113}
  (\bibinfo{year}{2016}) \bibinfo{pages}{554--559}.
\bibitem[{Luna-Pla and Nicol{\'a}s-Carlock(2020)}]{luna2020corruption}
\bibinfo{author}{I.~Luna-Pla}, \bibinfo{author}{J.~R. Nicol{\'a}s-Carlock},
\newblock \bibinfo{title}{Corruption and complexity: a scientific framework for
  the analysis of corruption networks},
\newblock \bibinfo{journal}{Appl. Netw. Sci} \bibinfo{volume}{5}
  (\bibinfo{year}{2020}) \bibinfo{pages}{13}.
\bibitem[{Nicol{\'a}s-Carlock and Luna-Pla(2021)}]{nicolas2021conspiracy}
\bibinfo{author}{J.~R. Nicol{\'a}s-Carlock}, \bibinfo{author}{I.~Luna-Pla},
\newblock \bibinfo{title}{Conspiracy of corporate networks in corruption
  scandals},
\newblock \bibinfo{journal}{Front. Phys.} \bibinfo{volume}{9}
  (\bibinfo{year}{2021}) \bibinfo{pages}{301}.
\bibitem[{Guimera and Amaral(2005)}]{guimera2005cartography}
\bibinfo{author}{R.~Guimera}, \bibinfo{author}{L.~A.~N. Amaral},
\newblock \bibinfo{title}{Cartography of complex networks: modules and
  universal roles},
\newblock \bibinfo{journal}{J. Stat. Mech. Theory Exp.} \bibinfo{volume}{2005}
  (\bibinfo{year}{2005}) \bibinfo{pages}{P02001}.
\bibitem[{Ferrara et~al.(2014)Ferrara, De~Meo, Catanese, and
  Fiumara}]{ferrara2014detecting}
\bibinfo{author}{E.~Ferrara}, \bibinfo{author}{P.~De~Meo},
  \bibinfo{author}{S.~Catanese}, \bibinfo{author}{G.~Fiumara},
\newblock \bibinfo{title}{Detecting criminal organizations in mobile phone
  networks},
\newblock \bibinfo{journal}{Expert Syst. Appl.} \bibinfo{volume}{41}
  (\bibinfo{year}{2014}) \bibinfo{pages}{5733--5750}.
\bibitem[{Bogomolov et~al.(2014)Bogomolov, Lepri, Staiano, Oliver, Pianesi, and
  Pentland}]{bogomolov2014once}
\bibinfo{author}{A.~Bogomolov}, \bibinfo{author}{B.~Lepri},
  \bibinfo{author}{J.~Staiano}, \bibinfo{author}{N.~Oliver},
  \bibinfo{author}{F.~Pianesi}, \bibinfo{author}{A.~Pentland},
\newblock \bibinfo{title}{Once upon a crime: towards crime prediction from
  demographics and mobile data},
\newblock in: \bibinfo{editor}{A.~A. Salah}, \bibinfo{editor}{J.~Cohn},
  \bibinfo{editor}{B.~Schuller}, \bibinfo{editor}{O.~Aran},
  \bibinfo{editor}{L.-P. Morency}, \bibinfo{editor}{P.~R. Cohen} (Eds.),
  \bibinfo{booktitle}{Proceedings of the 16th international conference on
  multimodal interaction}, \bibinfo{publisher}{Association for Computing
  Machinery}, \bibinfo{year}{2014}, pp. \bibinfo{pages}{427--434}.
\bibitem[{Herrmann et~al.(2008)Herrmann, Th{\"o}ni, and
  G{\"a}chter}]{herrmann2008antisocial}
\bibinfo{author}{B.~Herrmann}, \bibinfo{author}{C.~Th{\"o}ni},
  \bibinfo{author}{S.~G{\"a}chter},
\newblock \bibinfo{title}{Antisocial punishment across societies},
\newblock \bibinfo{journal}{Science} \bibinfo{volume}{319}
  (\bibinfo{year}{2008}) \bibinfo{pages}{1362--1367}.
\bibitem[{Rand et~al.(2009)Rand, Dreber, Ellingsen, Fudenberg, and
  Nowak}]{rand2009positive}
\bibinfo{author}{D.~G. Rand}, \bibinfo{author}{A.~Dreber},
  \bibinfo{author}{T.~Ellingsen}, \bibinfo{author}{D.~Fudenberg},
  \bibinfo{author}{M.~A. Nowak},
\newblock \bibinfo{title}{Positive interactions promote public cooperation},
\newblock \bibinfo{journal}{Science} \bibinfo{volume}{325}
  (\bibinfo{year}{2009}) \bibinfo{pages}{1272--1275}.
\bibitem[{Milinski et~al.(2002)Milinski, Semmann, and
  Krambeck}]{milinski2002reputation}
\bibinfo{author}{M.~Milinski}, \bibinfo{author}{D.~Semmann},
  \bibinfo{author}{H.-J. Krambeck},
\newblock \bibinfo{title}{Reputation helps solve the `tragedy of the commons'},
\newblock \bibinfo{journal}{Nature} \bibinfo{volume}{415}
  (\bibinfo{year}{2002}) \bibinfo{pages}{424--426}.
\bibitem[{Rand and Nowak(2011)}]{rand2011evolution}
\bibinfo{author}{D.~G. Rand}, \bibinfo{author}{M.~A. Nowak},
\newblock \bibinfo{title}{The evolution of antisocial punishment in optional
  public goods games},
\newblock \bibinfo{journal}{Nat. Commun.} \bibinfo{volume}{2}
  (\bibinfo{year}{2011}) \bibinfo{pages}{434}.
\bibitem[{G{\"a}chter et~al.(2008)G{\"a}chter, Renner, and
  Sefton}]{gachter2008long}
\bibinfo{author}{S.~G{\"a}chter}, \bibinfo{author}{E.~Renner},
  \bibinfo{author}{M.~Sefton},
\newblock \bibinfo{title}{The long-run benefits of punishment},
\newblock \bibinfo{journal}{Science} \bibinfo{volume}{322}
  (\bibinfo{year}{2008}) \bibinfo{pages}{1510--1510}.
\bibitem[{Boyd et~al.(2010)Boyd, Gintis, and Bowles}]{boyd2010coordinated}
\bibinfo{author}{R.~Boyd}, \bibinfo{author}{H.~Gintis},
  \bibinfo{author}{S.~Bowles},
\newblock \bibinfo{title}{Coordinated punishment of defectors sustains
  cooperation and can proliferate when rare},
\newblock \bibinfo{journal}{Science} \bibinfo{volume}{328}
  (\bibinfo{year}{2010}) \bibinfo{pages}{617--620}.
\bibitem[{Perc and Szolnoki(2012)}]{perc2012self}
\bibinfo{author}{M.~Perc}, \bibinfo{author}{A.~Szolnoki},
\newblock \bibinfo{title}{Self-organization of punishment in structured
  populations},
\newblock \bibinfo{journal}{New J. Phys.} \bibinfo{volume}{14}
  (\bibinfo{year}{2012}) \bibinfo{pages}{043013}.
\bibitem[{Hilbe and Sigmund(2010)}]{hilbe2010incentives}
\bibinfo{author}{C.~Hilbe}, \bibinfo{author}{K.~Sigmund},
\newblock \bibinfo{title}{Incentives and opportunism: from the carrot to the
  stick},
\newblock \bibinfo{journal}{Proc. R. Soc. B} \bibinfo{volume}{277}
  (\bibinfo{year}{2010}) \bibinfo{pages}{2427--2433}.
\bibitem[{Szolnoki and Perc(2010)}]{szolnoki2010reward}
\bibinfo{author}{A.~Szolnoki}, \bibinfo{author}{M.~Perc},
\newblock \bibinfo{title}{Reward and cooperation in the spatial public goods
  game},
\newblock \bibinfo{journal}{EPL (Europhys. Lett.)} \bibinfo{volume}{92}
  (\bibinfo{year}{2010}) \bibinfo{pages}{38003}.
\bibitem[{Szolnoki and Perc(2012)}]{szolnoki2012evolutionary}
\bibinfo{author}{A.~Szolnoki}, \bibinfo{author}{M.~Perc},
\newblock \bibinfo{title}{Evolutionary advantages of adaptive rewarding},
\newblock \bibinfo{journal}{New J. Physics} \bibinfo{volume}{14}
  (\bibinfo{year}{2012}) \bibinfo{pages}{093016}.
\bibitem[{Szolnoki and Perc(2015)}]{szolnoki2015antisocial}
\bibinfo{author}{A.~Szolnoki}, \bibinfo{author}{M.~Perc},
\newblock \bibinfo{title}{Antisocial pool rewarding does not deter public
  cooperation},
\newblock \bibinfo{journal}{Proc. R. Soc. B} \bibinfo{volume}{282}
  (\bibinfo{year}{2015}) \bibinfo{pages}{20151975}.
\bibitem[{Fang et~al.(2019)Fang, Benko, Perc, Xu, and
  Tan}]{fang2019synergistic}
\bibinfo{author}{Y.~Fang}, \bibinfo{author}{T.~P. Benko},
  \bibinfo{author}{M.~Perc}, \bibinfo{author}{H.~Xu}, \bibinfo{author}{Q.~Tan},
\newblock \bibinfo{title}{Synergistic third-party rewarding and punishment in
  the public goods game},
\newblock \bibinfo{journal}{Proc. R. Soc. A} \bibinfo{volume}{475}
  (\bibinfo{year}{2019}) \bibinfo{pages}{20190349}.
\bibitem[{Hu et~al.(2020)Hu, He, Weng, Chen, and Perc}]{hu2020rewarding}
\bibinfo{author}{L.~Hu}, \bibinfo{author}{N.~He}, \bibinfo{author}{Q.~Weng},
  \bibinfo{author}{X.~Chen}, \bibinfo{author}{M.~Perc},
\newblock \bibinfo{title}{Rewarding endowments lead to a win-win in the
  evolution of public cooperation and the accumulation of common resources},
\newblock \bibinfo{journal}{Chaos Solitons Fractals} \bibinfo{volume}{134}
  (\bibinfo{year}{2020}) \bibinfo{pages}{109694}.
\bibitem[{Szolnoki and Perc(2013)}]{szolnoki2013correlation}
\bibinfo{author}{A.~Szolnoki}, \bibinfo{author}{M.~Perc},
\newblock \bibinfo{title}{Correlation of positive and negative reciprocity
  fails to confer an evolutionary advantage: {P}hase transitions to elementary
  strategies},
\newblock \bibinfo{journal}{Phys. Rev. X} \bibinfo{volume}{3}
  (\bibinfo{year}{2013}) \bibinfo{pages}{041021}.
\bibitem[{Sigmund et~al.(2010)Sigmund, De~Silva, Traulsen, and
  Hauert}]{sigmund2010social}
\bibinfo{author}{K.~Sigmund}, \bibinfo{author}{H.~De~Silva},
  \bibinfo{author}{A.~Traulsen}, \bibinfo{author}{C.~Hauert},
\newblock \bibinfo{title}{Social learning promotes institutions for governing
  the commons},
\newblock \bibinfo{journal}{Nature} \bibinfo{volume}{466}
  (\bibinfo{year}{2010}) \bibinfo{pages}{861--863}.
\bibitem[{Szolnoki et~al.(2011)Szolnoki, Szab{\'o}, and
  Perc}]{szolnoki2011phase}
\bibinfo{author}{A.~Szolnoki}, \bibinfo{author}{G.~Szab{\'o}},
  \bibinfo{author}{M.~Perc},
\newblock \bibinfo{title}{Phase diagrams for the spatial public goods game with
  pool punishment},
\newblock \bibinfo{journal}{Phys. Rev. E} \bibinfo{volume}{83}
  (\bibinfo{year}{2011}) \bibinfo{pages}{036101}.
\bibitem[{Perc(2012)}]{perc2012sustainable}
\bibinfo{author}{M.~Perc},
\newblock \bibinfo{title}{Sustainable institutionalized punishment requires
  elimination of second-order free-riders},
\newblock \bibinfo{journal}{Sci. Rep.} \bibinfo{volume}{2}
  (\bibinfo{year}{2012}) \bibinfo{pages}{1--6}.
\bibitem[{Chen and Perc(2014)}]{chen2014optimal}
\bibinfo{author}{X.~Chen}, \bibinfo{author}{M.~Perc},
\newblock \bibinfo{title}{Optimal distribution of incentives for public
  cooperation in heterogeneous interaction environments},
\newblock \bibinfo{journal}{Front. Behav. Neurosci.} \bibinfo{volume}{8}
  (\bibinfo{year}{2014}) \bibinfo{pages}{248}.
\bibitem[{Berenji et~al.(2014)Berenji, Chou, and
  D'Orsogna}]{berenji2014recidivism}
\bibinfo{author}{B.~Berenji}, \bibinfo{author}{T.~Chou}, \bibinfo{author}{M.~R.
  D'Orsogna},
\newblock \bibinfo{title}{Recidivism and rehabilitation of criminal offenders:
  {A} carrot and stick evolutionary game},
\newblock \bibinfo{journal}{PLOS ONE} \bibinfo{volume}{9}
  (\bibinfo{year}{2014}) \bibinfo{pages}{e85531}.
\bibitem[{Dorogovtsev et~al.(2008)Dorogovtsev, Goltsev, and
  Mendes}]{dorogovtsev2008critical}
\bibinfo{author}{S.~N. Dorogovtsev}, \bibinfo{author}{A.~V. Goltsev},
  \bibinfo{author}{J.~F. Mendes},
\newblock \bibinfo{title}{Critical phenomena in complex networks},
\newblock \bibinfo{journal}{Rev. Mod. Phys.} \bibinfo{volume}{80}
  (\bibinfo{year}{2008}) \bibinfo{pages}{1275}.
\bibitem[{Scheffer et~al.(2012)Scheffer, Carpenter, Lenton, Bascompte, Brock,
  Dakos, Van~de Koppel, Van~de Leemput, Levin, Van~Nes
  et~al.}]{scheffer2012anticipating}
\bibinfo{author}{M.~Scheffer}, \bibinfo{author}{S.~R. Carpenter},
  \bibinfo{author}{T.~M. Lenton}, \bibinfo{author}{J.~Bascompte},
  \bibinfo{author}{W.~Brock}, \bibinfo{author}{V.~Dakos},
  \bibinfo{author}{J.~Van~de Koppel}, \bibinfo{author}{I.~A. Van~de Leemput},
  \bibinfo{author}{S.~A. Levin}, \bibinfo{author}{E.~H. Van~Nes}, et~al.,
\newblock \bibinfo{title}{Anticipating critical transitions},
\newblock \bibinfo{journal}{Science} \bibinfo{volume}{338}
  (\bibinfo{year}{2012}) \bibinfo{pages}{344--348}.
\bibitem[{Yu et~al.(2016)Yu, Xiao, Zhou, Wang, Wang, Kurths, and
  Schellnhuber}]{yu2016system}
\bibinfo{author}{Y.~Yu}, \bibinfo{author}{G.~Xiao}, \bibinfo{author}{J.~Zhou},
  \bibinfo{author}{Y.~Wang}, \bibinfo{author}{Z.~Wang},
  \bibinfo{author}{J.~Kurths}, \bibinfo{author}{H.~J. Schellnhuber},
\newblock \bibinfo{title}{System crash as dynamics of complex networks},
\newblock \bibinfo{journal}{Proc. Natl. Acad. Sci. USA} \bibinfo{volume}{113}
  (\bibinfo{year}{2016}) \bibinfo{pages}{11726--11731}.
\bibitem[{Cohen et~al.(2000)Cohen, Erez, Ben-Avraham, and
  Havlin}]{cohen2000resilience}
\bibinfo{author}{R.~Cohen}, \bibinfo{author}{K.~Erez},
  \bibinfo{author}{D.~Ben-Avraham}, \bibinfo{author}{S.~Havlin},
\newblock \bibinfo{title}{Resilience of the internet to random breakdowns},
\newblock \bibinfo{journal}{Phys. Rev. Lett.} \bibinfo{volume}{85}
  (\bibinfo{year}{2000}) \bibinfo{pages}{4626}.
\bibitem[{Majdandzic et~al.(2016)Majdandzic, Braunstein, Curme, Vodenska,
  Levy-Carciente, Stanley, and Havlin}]{majdandzic2016multiple}
\bibinfo{author}{A.~Majdandzic}, \bibinfo{author}{L.~A. Braunstein},
  \bibinfo{author}{C.~Curme}, \bibinfo{author}{I.~Vodenska},
  \bibinfo{author}{S.~Levy-Carciente}, \bibinfo{author}{H.~E. Stanley},
  \bibinfo{author}{S.~Havlin},
\newblock \bibinfo{title}{Multiple tipping points and optimal repairing in
  interacting networks},
\newblock \bibinfo{journal}{Nat. Commun.} \bibinfo{volume}{7}
  (\bibinfo{year}{2016}) \bibinfo{pages}{10850}.
\bibitem[{Majdandzic et~al.(2014)Majdandzic, Podobnik, Buldyrev, Kenett,
  Havlin, and Stanley}]{majdandzic2014spontaneous}
\bibinfo{author}{A.~Majdandzic}, \bibinfo{author}{B.~Podobnik},
  \bibinfo{author}{S.~V. Buldyrev}, \bibinfo{author}{D.~Y. Kenett},
  \bibinfo{author}{S.~Havlin}, \bibinfo{author}{H.~E. Stanley},
\newblock \bibinfo{title}{Spontaneous recovery in dynamical networks},
\newblock \bibinfo{journal}{Nat. Phys.} \bibinfo{volume}{10}
  (\bibinfo{year}{2014}) \bibinfo{pages}{34--38}.
\bibitem[{Greenhill(2016)}]{greenhill2016open}
\bibinfo{author}{K.~M. Greenhill},
\newblock \bibinfo{title}{Open arms behind barred doors: fear, hypocrisy and
  policy schizophrenia in the {E}uropean migration crisis},
\newblock \bibinfo{journal}{Eur. Law J.} \bibinfo{volume}{22}
  (\bibinfo{year}{2016}) \bibinfo{pages}{317--332}.
\bibitem[{{\v{Z}}i{\v{z}}ek(2016)}]{vzivzek2016against}
\bibinfo{author}{S.~{\v{Z}}i{\v{z}}ek}, \bibinfo{title}{Against the double
  blackmail: {R}efugees, terror and other troubles with the neighbours},
  \bibinfo{publisher}{Penguin UK}, \bibinfo{year}{2016}.
\bibitem[{Popper(2013)}]{popper2013open}
\bibinfo{author}{K.~Popper}, \bibinfo{title}{The Open Society and Its Enemies.
  New One-Volume Edition}, \bibinfo{publisher}{Princeton University Press},
  \bibinfo{year}{2013}.
\bibitem[{Podobnik et~al.(2019)Podobnik, Kirbis, Koprcina, and
  Stanley}]{podobnik2019emergence}
\bibinfo{author}{B.~Podobnik}, \bibinfo{author}{I.~S. Kirbis},
  \bibinfo{author}{M.~Koprcina}, \bibinfo{author}{H.~Stanley},
\newblock \bibinfo{title}{Emergence of the unified right- and left-wing
  populism---{W}hen radical societal changes become more important than
  ideology},
\newblock \bibinfo{journal}{Physica A} \bibinfo{volume}{517}
  (\bibinfo{year}{2019}) \bibinfo{pages}{459--474}.
\bibitem[{Durlauf(1999)}]{durlauf1999howcan}
\bibinfo{author}{S.~N. Durlauf},
\newblock \bibinfo{title}{How can statistical mechanics contribute to social
  science?},
\newblock \bibinfo{journal}{Proc. Natl. Acad. Sci. USA} \bibinfo{volume}{96}
  (\bibinfo{year}{1999}) \bibinfo{pages}{10582--10584}.
\bibitem[{Weidlich(1991)}]{weidlich1991physics}
\bibinfo{author}{W.~Weidlich},
\newblock \bibinfo{title}{Physics and social science---the approach of
  synergetics},
\newblock \bibinfo{journal}{Phys. Rep.} \bibinfo{volume}{204}
  (\bibinfo{year}{1991}) \bibinfo{pages}{1--163}.
\bibitem[{Weidlich(2006)}]{weidlich2006sociodynamics}
\bibinfo{author}{W.~Weidlich}, \bibinfo{title}{Sociodynamics: {A} systematic
  approach to mathematical modelling in the social sciences},
  \bibinfo{publisher}{Courier Corporation}, \bibinfo{year}{2006}.
\bibitem[{Galam(2008)}]{galam2008sociophysics}
\bibinfo{author}{S.~Galam},
\newblock \bibinfo{title}{Sociophysics: {A} review of {G}alam models},
\newblock \bibinfo{journal}{Int. J. Mod. Phys. C} \bibinfo{volume}{19}
  (\bibinfo{year}{2008}) \bibinfo{pages}{409--440}.
\bibitem[{Bansak et~al.(2018)Bansak, Ferwerda, Hainmueller, Dillon, Hangartner,
  Lawrence, and Weinstein}]{bansak2018improving}
\bibinfo{author}{K.~Bansak}, \bibinfo{author}{J.~Ferwerda},
  \bibinfo{author}{J.~Hainmueller}, \bibinfo{author}{A.~Dillon},
  \bibinfo{author}{D.~Hangartner}, \bibinfo{author}{D.~Lawrence},
  \bibinfo{author}{J.~Weinstein},
\newblock \bibinfo{title}{Improving refugee integration through data-driven
  algorithmic assignment},
\newblock \bibinfo{journal}{Science} \bibinfo{volume}{359}
  (\bibinfo{year}{2018}) \bibinfo{pages}{325--329}.
\bibitem[{Van~Lange et~al.(2013)Van~Lange, Joireman, Parks, and
  Van~Dijk}]{vanlange2013psychology}
\bibinfo{author}{P.~A. Van~Lange}, \bibinfo{author}{J.~Joireman},
  \bibinfo{author}{C.~D. Parks}, \bibinfo{author}{E.~Van~Dijk},
\newblock \bibinfo{title}{The psychology of social dilemmas: {A} review},
\newblock \bibinfo{journal}{Organ. Behav. Hum. Decis. Process.}
  \bibinfo{volume}{120} (\bibinfo{year}{2013}) \bibinfo{pages}{125--141}.
\bibitem[{Hauert and Szab{\'o}(2005)}]{hauert2005game}
\bibinfo{author}{C.~Hauert}, \bibinfo{author}{G.~Szab{\'o}},
\newblock \bibinfo{title}{Game theory and physics},
\newblock \bibinfo{journal}{Am. J. Phys.} \bibinfo{volume}{73}
  (\bibinfo{year}{2005}) \bibinfo{pages}{405--414}.
\bibitem[{Newman and Watts(1999)}]{newman1999renormalization}
\bibinfo{author}{M.~E.~J. Newman}, \bibinfo{author}{D.~J. Watts},
\newblock \bibinfo{title}{Renormalization group analysis of the small-world
  network model},
\newblock \bibinfo{journal}{Phys. Lett. A} \bibinfo{volume}{263}
  (\bibinfo{year}{1999}) \bibinfo{pages}{341--346}.
\bibitem[{Hinrichsen(2000)}]{hinrichsen2000nonequi}
\bibinfo{author}{H.~Hinrichsen},
\newblock \bibinfo{title}{Non-equilibrium critical phenomena and phase
  transitions into absorbing states},
\newblock \bibinfo{journal}{Adv. Phys.} \bibinfo{volume}{49}
  (\bibinfo{year}{2000}) \bibinfo{pages}{815--958}.
\bibitem[{Szab{\'o} and Hauert(2002)}]{szabo2002phase}
\bibinfo{author}{G.~Szab{\'o}}, \bibinfo{author}{C.~Hauert},
\newblock \bibinfo{title}{Phase transitions and volunteering in spatial public
  goods games},
\newblock \bibinfo{journal}{Phys. Rev. Lett.} \bibinfo{volume}{89}
  (\bibinfo{year}{2002}) \bibinfo{pages}{118101}.
\bibitem[{Perc(2016)}]{perc2016phase}
\bibinfo{author}{M.~Perc},
\newblock \bibinfo{title}{Phase transitions in models of human cooperation},
\newblock \bibinfo{journal}{Phys. Lett. A} \bibinfo{volume}{380}
  (\bibinfo{year}{2016}) \bibinfo{pages}{2803--2808}.
\bibitem[{Scheffer et~al.(2009)Scheffer, Bascompte, Brock, Brovkin, Carpenter,
  Dakos, Held, Van~Nes, Rietkerk, and Sugihara}]{scheffer2009early}
\bibinfo{author}{M.~Scheffer}, \bibinfo{author}{J.~Bascompte},
  \bibinfo{author}{W.~A. Brock}, \bibinfo{author}{V.~Brovkin},
  \bibinfo{author}{S.~R. Carpenter}, \bibinfo{author}{V.~Dakos},
  \bibinfo{author}{H.~Held}, \bibinfo{author}{E.~H. Van~Nes},
  \bibinfo{author}{M.~Rietkerk}, \bibinfo{author}{G.~Sugihara},
\newblock \bibinfo{title}{Early-warning signals for critical transitions},
\newblock \bibinfo{journal}{Nature} \bibinfo{volume}{461}
  (\bibinfo{year}{2009}) \bibinfo{pages}{53--59}.
\bibitem[{Scheffer and Carpenter(2003)}]{scheffer2003catastrophic}
\bibinfo{author}{M.~Scheffer}, \bibinfo{author}{S.~R. Carpenter},
\newblock \bibinfo{title}{Catastrophic regime shifts in ecosystems: linking
  theory to observation},
\newblock \bibinfo{journal}{Trends Ecol. Evol.} \bibinfo{volume}{18}
  (\bibinfo{year}{2003}) \bibinfo{pages}{648--656}.
\bibitem[{Levy(2005)}]{levy2005social}
\bibinfo{author}{M.~Levy},
\newblock \bibinfo{title}{Social phase transitions},
\newblock \bibinfo{journal}{J. Econ. Behav. Organ.} \bibinfo{volume}{57}
  (\bibinfo{year}{2005}) \bibinfo{pages}{71--87}.
\bibitem[{Levy(2008)}]{levy2008stock}
\bibinfo{author}{M.~Levy},
\newblock \bibinfo{title}{Stock market crashes as social phase transitions},
\newblock \bibinfo{journal}{J. Econ. Dyn. Control} \bibinfo{volume}{32}
  (\bibinfo{year}{2008}) \bibinfo{pages}{137--155}.
\bibitem[{(2016)}]{unknown2016standard}
\bibinfo{title}{Standard {E}urobarometer 85---{S}pring 2016},
  \bibinfo{year}{2016}. \bibinfo{note}{Survey conducted by TNS opinion \&
  social at the request of the European Commission, Directorate-General for
  Communication. Archived at: \url{https://doi.org/10.17605/OSF.IO/N7G9X}}.
\bibitem[{White and Geballe(1979)}]{white1979long}
\bibinfo{author}{R.~M. White}, \bibinfo{author}{T.~H. Geballe},
  \bibinfo{title}{Long range order in solids}, \bibinfo{publisher}{Academic
  Press}, \bibinfo{year}{1979}.
\bibitem[{Bertotti(1998)}]{bertotti1998hysteresis}
\bibinfo{author}{G.~Bertotti}, \bibinfo{title}{Hysteresis in magnetism: for
  physicists, materials scientists, and engineers},
  \bibinfo{publisher}{Academic Press}, \bibinfo{year}{1998}.
\bibitem[{Podobnik et~al.(2020)Podobnik, Koro{\v{s}}ak, Klemen, Sto{\v{z}}er,
  Dolen{\v{s}}ek, Rupnik, Ivanov, Holme, and Jusup}]{podobnik2020beta}
\bibinfo{author}{B.~Podobnik}, \bibinfo{author}{D.~Koro{\v{s}}ak},
  \bibinfo{author}{M.~S. Klemen}, \bibinfo{author}{A.~Sto{\v{z}}er},
  \bibinfo{author}{J.~Dolen{\v{s}}ek}, \bibinfo{author}{M.~S. Rupnik},
  \bibinfo{author}{P.~C. Ivanov}, \bibinfo{author}{P.~Holme},
  \bibinfo{author}{M.~Jusup},
\newblock \bibinfo{title}{$\beta$-cells operate collectively to help maintain
  glucose homeostasis},
\newblock \bibinfo{journal}{Biophys. J.} \bibinfo{volume}{118}
  (\bibinfo{year}{2020}) \bibinfo{pages}{2588--2595}.
\bibitem[{Witt(1997)}]{witt1997self}
\bibinfo{author}{U.~Witt},
\newblock \bibinfo{title}{Self-organization and economics---what is new?},
\newblock \bibinfo{journal}{Struct. Chang. Econ. Dyn.} \bibinfo{volume}{8}
  (\bibinfo{year}{1997}) \bibinfo{pages}{489--507}.
\bibitem[{Pfeifer et~al.(2007)Pfeifer, Lungarella, and Iida}]{pfeifer2007self}
\bibinfo{author}{R.~Pfeifer}, \bibinfo{author}{M.~Lungarella},
  \bibinfo{author}{F.~Iida},
\newblock \bibinfo{title}{Self-organization, embodiment, and biologically
  inspired robotics},
\newblock \bibinfo{journal}{Science} \bibinfo{volume}{318}
  (\bibinfo{year}{2007}) \bibinfo{pages}{1088--1093}.
\bibitem[{Helbing(2012)}]{helbing2012social}
\bibinfo{author}{D.~Helbing}, \bibinfo{title}{Social self-organization:
  {A}gent-based simulations and experiments to study emergent social behavior},
  \bibinfo{publisher}{Springer}, \bibinfo{year}{2012}.
\bibitem[{Park(2015)}]{park2015europe}
\bibinfo{author}{J.~Park}, \bibinfo{title}{Europe's migrant crisis},
  \bibinfo{year}{2015}. \bibinfo{note}{Council on Foreign Relations. Available
  at: \url{https://www.cfr.org/backgrounder/europes-migration-crisis}. Archived
  at: \url{https://doi.org/10.17605/OSF.IO/N7G9X}}.
\bibitem[{Mueller et~al.(2014)Mueller, Gray, and Kosec}]{mueller2014heat}
\bibinfo{author}{V.~Mueller}, \bibinfo{author}{C.~Gray},
  \bibinfo{author}{K.~Kosec},
\newblock \bibinfo{title}{Heat stress increases long-term human migration in
  rural {P}akistan},
\newblock \bibinfo{journal}{Nat. Clim. Change} \bibinfo{volume}{4}
  (\bibinfo{year}{2014}) \bibinfo{pages}{182--185}.
\bibitem[{Forst(2003)}]{forst2003toleration}
\bibinfo{author}{R.~Forst},
\newblock \bibinfo{title}{Toleration, justice and reason},
\newblock in: \bibinfo{editor}{D.~Castiglione}, \bibinfo{editor}{C.~Mckinnon}
  (Eds.), \bibinfo{booktitle}{The culture of toleration in diverse societies},
  \bibinfo{publisher}{Manchester University Press}, \bibinfo{year}{2003}, pp.
  \bibinfo{pages}{71--85}.
\bibitem[{Sullivan et~al.(1993)Sullivan, Piereson, and
  Marcus}]{sullivan1993political}
\bibinfo{author}{J.~L. Sullivan}, \bibinfo{author}{J.~Piereson},
  \bibinfo{author}{G.~E. Marcus}, \bibinfo{title}{Political tolerance and
  {A}merican democracy}, \bibinfo{publisher}{University of Chicago Press},
  \bibinfo{year}{1993}.
\bibitem[{Kunovich(2013)}]{kunovich2013labor}
\bibinfo{author}{R.~M. Kunovich},
\newblock \bibinfo{title}{Labor market competition and anti-immigrant
  sentiment: {O}ccupations as contexts},
\newblock \bibinfo{journal}{Int. Migr. Rev.} \bibinfo{volume}{47}
  (\bibinfo{year}{2013}) \bibinfo{pages}{643--685}.
\bibitem[{Paas and Halapuu(2012)}]{paas2012attitudes}
\bibinfo{author}{T.~Paas}, \bibinfo{author}{V.~Halapuu},
\newblock \bibinfo{title}{Attitudes towards immigrants and the integration of
  ethnically diverse societies.},
\newblock \bibinfo{journal}{East. J. Eur. Stud.} \bibinfo{volume}{3}
  (\bibinfo{year}{2012}) \bibinfo{pages}{161--176}.
\bibitem[{Esipova and Ray(2017)}]{esipova2017syrian}
\bibinfo{author}{N.~Esipova}, \bibinfo{author}{J.~Ray}, \bibinfo{title}{Syrian
  refugees not welcome in {E}astern {E}urope}, \bibinfo{year}{2017}.
  \bibinfo{note}{Gallup World Poll 2016. Available at:
  \url{https://news.gallup.com/poll/209828/syrian-refugees-not-welcome-eastern-europe.aspx}.
  Archived at: \url{https://doi.org/10.17605/OSF.IO/N7G9X}}.
\bibitem[{Nowak and Sigmund(1998)}]{nowak1998evolution}
\bibinfo{author}{M.~A. Nowak}, \bibinfo{author}{K.~Sigmund},
\newblock \bibinfo{title}{Evolution of indirect reciprocity by image scoring},
\newblock \bibinfo{journal}{Nature} \bibinfo{volume}{393}
  (\bibinfo{year}{1998}) \bibinfo{pages}{573--577}.
\bibitem[{Riolo et~al.(2001)Riolo, Cohen, and Axelrod}]{riolo2001evolution}
\bibinfo{author}{R.~L. Riolo}, \bibinfo{author}{M.~D. Cohen},
  \bibinfo{author}{R.~Axelrod},
\newblock \bibinfo{title}{Evolution of cooperation without reciprocity},
\newblock \bibinfo{journal}{Nature} \bibinfo{volume}{414}
  (\bibinfo{year}{2001}) \bibinfo{pages}{441--443}.
\bibitem[{Semyonov et~al.(2006)Semyonov, Raijman, and
  Gorodzeisky}]{semyonov2006rise}
\bibinfo{author}{M.~Semyonov}, \bibinfo{author}{R.~Raijman},
  \bibinfo{author}{A.~Gorodzeisky},
\newblock \bibinfo{title}{The rise of anti-foreigner sentiment in {E}uropean
  societies, 1988-2000},
\newblock \bibinfo{journal}{Am. Sociol. Rev.} \bibinfo{volume}{71}
  (\bibinfo{year}{2006}) \bibinfo{pages}{426--449}.
\bibitem[{Gorodzeisky and Semyonov(2016)}]{gorodzeisky2016not}
\bibinfo{author}{A.~Gorodzeisky}, \bibinfo{author}{M.~Semyonov},
\newblock \bibinfo{title}{Not only competitive threat but also racial
  prejudice: {S}ources of anti-immigrant attitudes in {E}uropean societies},
\newblock \bibinfo{journal}{Int. J. Public Opin. Res.} \bibinfo{volume}{28}
  (\bibinfo{year}{2016}) \bibinfo{pages}{331--354}.
\bibitem[{Hooghe and Dassonneville(2018)}]{hooghe2018explaining}
\bibinfo{author}{M.~Hooghe}, \bibinfo{author}{R.~Dassonneville},
\newblock \bibinfo{title}{Explaining the {T}rump vote: {T}he effect of racist
  resentment and anti-immigrant sentiments},
\newblock \bibinfo{journal}{PS Political Sci. Politics} \bibinfo{volume}{51}
  (\bibinfo{year}{2018}) \bibinfo{pages}{528--534}.
\bibitem[{Podobnik et~al.(2017)Podobnik, Jusup, Wang, and
  Stanley}]{podobnik2017fear}
\bibinfo{author}{B.~Podobnik}, \bibinfo{author}{M.~Jusup},
  \bibinfo{author}{Z.~Wang}, \bibinfo{author}{H.~E. Stanley},
\newblock \bibinfo{title}{How fear of future outcomes affects social dynamics},
\newblock \bibinfo{journal}{J. Stat. Phys.} \bibinfo{volume}{167}
  (\bibinfo{year}{2017}) \bibinfo{pages}{1007--1019}.
\bibitem[{Chiswick(1978)}]{chiswick1978effect}
\bibinfo{author}{B.~R. Chiswick},
\newblock \bibinfo{title}{The effect of americanization on the earnings of
  foreign-born men},
\newblock \bibinfo{journal}{J. Political Econ.} \bibinfo{volume}{86}
  (\bibinfo{year}{1978}) \bibinfo{pages}{897--921}.
\bibitem[{Borjas(1994)}]{borjas1994economics}
\bibinfo{author}{G.~J. Borjas},
\newblock \bibinfo{title}{The economics of immigration},
\newblock \bibinfo{journal}{J. Econ. Lit.} \bibinfo{volume}{32}
  (\bibinfo{year}{1994}) \bibinfo{pages}{1667--1717}.
\bibitem[{B{\"u}chel and Frick(2005)}]{buchel2005immigrants}
\bibinfo{author}{F.~B{\"u}chel}, \bibinfo{author}{J.~R. Frick},
\newblock \bibinfo{title}{Immigrants' economic performance across
  {E}urope---does immigration policy matter?},
\newblock \bibinfo{journal}{Popul. Res. Policy Rev.} \bibinfo{volume}{24}
  (\bibinfo{year}{2005}) \bibinfo{pages}{175--212}.
\bibitem[{Axelrod(1997)}]{axelrod1997dissemination}
\bibinfo{author}{R.~Axelrod},
\newblock \bibinfo{title}{The dissemination of culture: {A} model with local
  convergence and global polarization},
\newblock \bibinfo{journal}{J. Confl. Resolut.} \bibinfo{volume}{41}
  (\bibinfo{year}{1997}) \bibinfo{pages}{203--226}.
\bibitem[{Onnela and Reed-Tsochas(2010)}]{onnela2010spontaneous}
\bibinfo{author}{J.-P. Onnela}, \bibinfo{author}{F.~Reed-Tsochas},
\newblock \bibinfo{title}{Spontaneous emergence of social influence in online
  systems},
\newblock \bibinfo{journal}{Proc. Natl. Acad. Sci. USA} \bibinfo{volume}{107}
  (\bibinfo{year}{2010}) \bibinfo{pages}{18375--18380}.
\bibitem[{Castellano et~al.(2000)Castellano, Marsili, and
  Vespignani}]{castellano2000nonequilibrium}
\bibinfo{author}{C.~Castellano}, \bibinfo{author}{M.~Marsili},
  \bibinfo{author}{A.~Vespignani},
\newblock \bibinfo{title}{Nonequilibrium phase transition in a model for social
  influence},
\newblock \bibinfo{journal}{Phys. Rev. Lett.} \bibinfo{volume}{85}
  (\bibinfo{year}{2000}) \bibinfo{pages}{3536}.
\bibitem[{Weisbuch et~al.(2002)Weisbuch, Deffuant, Amblard, and
  Nadal}]{weisbuch2002meet}
\bibinfo{author}{G.~Weisbuch}, \bibinfo{author}{G.~Deffuant},
  \bibinfo{author}{F.~Amblard}, \bibinfo{author}{J.-P. Nadal},
\newblock \bibinfo{title}{Meet, discuss, and segregate!},
\newblock \bibinfo{journal}{Complexity} \bibinfo{volume}{7}
  (\bibinfo{year}{2002}) \bibinfo{pages}{55--63}.
\bibitem[{Klemm et~al.(2003{\natexlab{a}})Klemm, Egu{\'\i}luz, Toral, and
  San~Miguel}]{klemm2003nonequilibrium}
\bibinfo{author}{K.~Klemm}, \bibinfo{author}{V.~M. Egu{\'\i}luz},
  \bibinfo{author}{R.~Toral}, \bibinfo{author}{M.~San~Miguel},
\newblock \bibinfo{title}{Nonequilibrium transitions in complex networks: {A}
  model of social interaction},
\newblock \bibinfo{journal}{Phys. Rev. E} \bibinfo{volume}{67}
  (\bibinfo{year}{2003}{\natexlab{a}}) \bibinfo{pages}{026120}.
\bibitem[{Klemm et~al.(2003{\natexlab{b}})Klemm, Egu{\'\i}luz, Toral, and
  San~Miguel}]{klemm2003global}
\bibinfo{author}{K.~Klemm}, \bibinfo{author}{V.~M. Egu{\'\i}luz},
  \bibinfo{author}{R.~Toral}, \bibinfo{author}{M.~San~Miguel},
\newblock \bibinfo{title}{Global culture: {A} noise-induced transition in
  finite systems},
\newblock \bibinfo{journal}{Phys. Rev. E} \bibinfo{volume}{67}
  (\bibinfo{year}{2003}{\natexlab{b}}) \bibinfo{pages}{045101}.
\bibitem[{Weisbuch(2004)}]{weisbuch2004bounded}
\bibinfo{author}{G.~Weisbuch},
\newblock \bibinfo{title}{Bounded confidence and social networks},
\newblock \bibinfo{journal}{Eur. Phys. J. B} \bibinfo{volume}{38}
  (\bibinfo{year}{2004}) \bibinfo{pages}{339--343}.
\bibitem[{Lee et~al.(2017)Lee, Holme, and Lee}]{lee2017modeling}
\bibinfo{author}{E.~Lee}, \bibinfo{author}{P.~Holme}, \bibinfo{author}{S.~H.
  Lee},
\newblock \bibinfo{title}{Modeling the dynamics of dissent},
\newblock \bibinfo{journal}{Physica A} \bibinfo{volume}{486}
  (\bibinfo{year}{2017}) \bibinfo{pages}{262--272}.
\bibitem[{Juul and Porter(2019)}]{juul2019hipsters}
\bibinfo{author}{J.~S. Juul}, \bibinfo{author}{M.~A. Porter},
\newblock \bibinfo{title}{Hipsters on networks: {H}ow a minority group of
  individuals can lead to an antiestablishment majority},
\newblock \bibinfo{journal}{Phys. Rev. E} \bibinfo{volume}{99}
  (\bibinfo{year}{2019}) \bibinfo{pages}{022313}.
\bibitem[{Touboul(2019)}]{touboul2019hipster}
\bibinfo{author}{J.~D. Touboul},
\newblock \bibinfo{title}{The hipster effect: {W}hen anti-conformists all look
  the same},
\newblock \bibinfo{journal}{Discrete Contin. Dyn. Syst. Ser. B}
  \bibinfo{volume}{24} (\bibinfo{year}{2019}) \bibinfo{pages}{4379}.
\bibitem[{Dittrich et~al.(2000)Dittrich, Liljeros, Soulier, and
  Banzhaf}]{dittrich2000spontaneous}
\bibinfo{author}{P.~Dittrich}, \bibinfo{author}{F.~Liljeros},
  \bibinfo{author}{A.~Soulier}, \bibinfo{author}{W.~Banzhaf},
\newblock \bibinfo{title}{Spontaneous group formation in the seceder model},
\newblock \bibinfo{journal}{Phys. Rev. Lett.} \bibinfo{volume}{84}
  (\bibinfo{year}{2000}) \bibinfo{pages}{3205}.
\bibitem[{Dittrich and Banzhaf(2001)}]{dittrich2001survival}
\bibinfo{author}{P.~Dittrich}, \bibinfo{author}{W.~Banzhaf},
\newblock \bibinfo{title}{Survival of the unfittest?---the seceder model and
  its fitness landscape},
\newblock in: \bibinfo{editor}{J.~Kelemen}, \bibinfo{editor}{P.~Sos{\'i}k}
  (Eds.), \bibinfo{booktitle}{Advances in Artificial Life},
  \bibinfo{publisher}{Springer}, \bibinfo{year}{2001}, pp.
  \bibinfo{pages}{100--109}.
\bibitem[{Gr{\"o}nlund and Holme(2004)}]{gronlund2004networking}
\bibinfo{author}{A.~Gr{\"o}nlund}, \bibinfo{author}{P.~Holme},
\newblock \bibinfo{title}{Networking the seceder model: {G}roup formation in
  social and economic systems},
\newblock \bibinfo{journal}{Phys. Rev. E} \bibinfo{volume}{70}
  (\bibinfo{year}{2004}) \bibinfo{pages}{036108}.
\bibitem[{Holme and Gr\"onlund(2005)}]{youth2005modeling}
\bibinfo{author}{P.~Holme}, \bibinfo{author}{A.~Gr\"onlund},
\newblock \bibinfo{title}{Modeling the dynamics of youth subcultures},
\newblock \bibinfo{journal}{J. Artif. Soc. Soc. Simul.} \bibinfo{volume}{8}
  (\bibinfo{year}{2005}) \bibinfo{pages}{3}.
\bibitem[{Mudde(2004)}]{mudde2004populist}
\bibinfo{author}{C.~Mudde},
\newblock \bibinfo{title}{The populist zeitgeist},
\newblock \bibinfo{journal}{Gov. Oppos.} \bibinfo{volume}{39}
  (\bibinfo{year}{2004}) \bibinfo{pages}{541--563}.
\bibitem[{Betz(1993)}]{betz1993two}
\bibinfo{author}{H.-G. Betz},
\newblock \bibinfo{title}{The two faces of radical right-wing populism in
  {W}estern {E}urope},
\newblock \bibinfo{journal}{Rev. Politics} \bibinfo{volume}{55}
  (\bibinfo{year}{1993}) \bibinfo{pages}{663--685}.
\bibitem[{Oesch(2008)}]{oesch2008explaining}
\bibinfo{author}{D.~Oesch},
\newblock \bibinfo{title}{Explaining workers' support for right-wing populist
  parties in {W}estern {E}urope: {E}vidence from {A}ustria, {B}elgium,
  {F}rance, {N}orway, and {S}witzerland},
\newblock \bibinfo{journal}{Int. Political Sci. Rev.} \bibinfo{volume}{29}
  (\bibinfo{year}{2008}) \bibinfo{pages}{349--373}.
\bibitem[{Arzheimer and Carter(2006)}]{arzheimer2006political}
\bibinfo{author}{K.~Arzheimer}, \bibinfo{author}{E.~Carter},
\newblock \bibinfo{title}{Political opportunity structures and right-wing
  extremist party success},
\newblock \bibinfo{journal}{Eur. J. Political Res.} \bibinfo{volume}{45}
  (\bibinfo{year}{2006}) \bibinfo{pages}{419--443}.
\bibitem[{Bj{\o}rklund(2007)}]{bjorklund2007unemployment}
\bibinfo{author}{T.~Bj{\o}rklund},
\newblock \bibinfo{title}{Unemployment and the radical right in {S}candinavia:
  beneficial or non-beneficial for electoral support?},
\newblock \bibinfo{journal}{Comp. Eur. Politics} \bibinfo{volume}{5}
  (\bibinfo{year}{2007}) \bibinfo{pages}{245--263}.
\bibitem[{Smith(2010)}]{smith2010does}
\bibinfo{author}{J.~M. Smith},
\newblock \bibinfo{title}{Does crime pay? {I}ssue ownership, political
  opportunity, and the populist right in {W}estern {E}urope},
\newblock \bibinfo{journal}{Comp. Political Stud.} \bibinfo{volume}{43}
  (\bibinfo{year}{2010}) \bibinfo{pages}{1471--1498}.
\bibitem[{Lawton and Ackrill(2016)}]{lawton2016hard}
\bibinfo{author}{C.~Lawton}, \bibinfo{author}{R.~Ackrill}, \bibinfo{title}{Hard
  evidence: how areas with low immigration voted mainly for {B}rexit},
  \bibinfo{year}{2016}. \bibinfo{note}{Available at:
  \url{https://theconversation.com/hard-evidence-how-areas-with-low-immigration-voted-mainly-for-brexit-62138}.
  Archived at: \url{https://doi.org/10.17605/OSF.IO/N7G9X}}.
\bibitem[{Deffuant et~al.(2000)Deffuant, Neau, Amblard, and
  Weisbuch}]{deffuant2000mixing}
\bibinfo{author}{G.~Deffuant}, \bibinfo{author}{D.~Neau},
  \bibinfo{author}{F.~Amblard}, \bibinfo{author}{G.~Weisbuch},
\newblock \bibinfo{title}{Mixing beliefs among interacting agents},
\newblock \bibinfo{journal}{Adv. Complex Syst.} \bibinfo{volume}{3}
  (\bibinfo{year}{2000}) \bibinfo{pages}{87--98}.
\bibitem[{Abramowitz and Saunders(2008)}]{abramowitz2008polarization}
\bibinfo{author}{A.~I. Abramowitz}, \bibinfo{author}{K.~L. Saunders},
\newblock \bibinfo{title}{Is polarization a myth?},
\newblock \bibinfo{journal}{J. Politics} \bibinfo{volume}{70}
  (\bibinfo{year}{2008}) \bibinfo{pages}{542--555}.
\bibitem[{Saavedra et~al.(2007)Saavedra, Efstathiou, and
  Reed-Tsochas}]{saavedra2007identifying}
\bibinfo{author}{S.~Saavedra}, \bibinfo{author}{J.~Efstathiou},
  \bibinfo{author}{F.~Reed-Tsochas},
\newblock \bibinfo{title}{Identifying the underlying structure and dynamic
  interactions in a voting network},
\newblock \bibinfo{journal}{Physica A} \bibinfo{volume}{377}
  (\bibinfo{year}{2007}) \bibinfo{pages}{672--688}.
\bibitem[{Jasny et~al.(2015)Jasny, Waggle, and Fisher}]{jasny2015empirical}
\bibinfo{author}{L.~Jasny}, \bibinfo{author}{J.~Waggle}, \bibinfo{author}{D.~R.
  Fisher},
\newblock \bibinfo{title}{An empirical examination of echo chambers in us
  climate policy networks},
\newblock \bibinfo{journal}{Nat. Clim. Change} \bibinfo{volume}{5}
  (\bibinfo{year}{2015}) \bibinfo{pages}{782--786}.
\bibitem[{Williams et~al.(2015)Williams, McMurray, Kurz, and
  Lambert}]{williams2015network}
\bibinfo{author}{H.~T. Williams}, \bibinfo{author}{J.~R. McMurray},
  \bibinfo{author}{T.~Kurz}, \bibinfo{author}{F.~H. Lambert},
\newblock \bibinfo{title}{Network analysis reveals open forums and echo
  chambers in social media discussions of climate change},
\newblock \bibinfo{journal}{Glob. Environ. Change} \bibinfo{volume}{32}
  (\bibinfo{year}{2015}) \bibinfo{pages}{126--138}.
\bibitem[{Boutyline and Willer(2017)}]{boutyline2017social}
\bibinfo{author}{A.~Boutyline}, \bibinfo{author}{R.~Willer},
\newblock \bibinfo{title}{The social structure of political echo chambers:
  {V}ariation in ideological homophily in online networks},
\newblock \bibinfo{journal}{Political Psychol.} \bibinfo{volume}{38}
  (\bibinfo{year}{2017}) \bibinfo{pages}{551--569}.
\bibitem[{Usher et~al.(2018)Usher, Holcomb, and Littman}]{usher2018twitter}
\bibinfo{author}{N.~Usher}, \bibinfo{author}{J.~Holcomb},
  \bibinfo{author}{J.~Littman},
\newblock \bibinfo{title}{Twitter makes it worse: {P}olitical journalists,
  gendered echo chambers, and the amplification of gender bias},
\newblock \bibinfo{journal}{Int. J. Press/Politics} \bibinfo{volume}{23}
  (\bibinfo{year}{2018}) \bibinfo{pages}{324--344}.
\bibitem[{Cinelli et~al.(2021)Cinelli, Morales, Galeazzi, Quattrociocchi, and
  Starnini}]{cinelli2021echo}
\bibinfo{author}{M.~Cinelli}, \bibinfo{author}{G.~D.~F. Morales},
  \bibinfo{author}{A.~Galeazzi}, \bibinfo{author}{W.~Quattrociocchi},
  \bibinfo{author}{M.~Starnini},
\newblock \bibinfo{title}{The echo chamber effect on social media},
\newblock \bibinfo{journal}{Proc. Natl. Acad. Sci. USA} \bibinfo{volume}{118}
  (\bibinfo{year}{2021}) \bibinfo{pages}{e2023301118}.
\bibitem[{Flaxman et~al.(2016)Flaxman, Goel, and Rao}]{flaxman2016filter}
\bibinfo{author}{S.~Flaxman}, \bibinfo{author}{S.~Goel}, \bibinfo{author}{J.~M.
  Rao},
\newblock \bibinfo{title}{Filter bubbles, echo chambers, and online news
  consumption},
\newblock \bibinfo{journal}{Public Opin. Q.} \bibinfo{volume}{80}
  (\bibinfo{year}{2016}) \bibinfo{pages}{298--320}.
\bibitem[{Ben-Naim et~al.(2003)Ben-Naim, Krapivsky, and
  Redner}]{ben2003bifurcations}
\bibinfo{author}{E.~Ben-Naim}, \bibinfo{author}{P.~L. Krapivsky},
  \bibinfo{author}{S.~Redner},
\newblock \bibinfo{title}{Bifurcations and patterns in compromise processes},
\newblock \bibinfo{journal}{Physica D} \bibinfo{volume}{183}
  (\bibinfo{year}{2003}) \bibinfo{pages}{190--204}.
\bibitem[{Amblard and Deffuant(2004)}]{amblard2004role}
\bibinfo{author}{F.~Amblard}, \bibinfo{author}{G.~Deffuant},
\newblock \bibinfo{title}{The role of network topology on extremism propagation
  with the relative agreement opinion dynamics},
\newblock \bibinfo{journal}{Physica A} \bibinfo{volume}{343}
  (\bibinfo{year}{2004}) \bibinfo{pages}{725--738}.
\bibitem[{Stauffer and Meyer-Ortmanns(2004)}]{stauffer2004simulation}
\bibinfo{author}{D.~Stauffer}, \bibinfo{author}{H.~Meyer-Ortmanns},
\newblock \bibinfo{title}{Simulation of consensus model of {D}effuant et al. on
  a {B}arabasi-{A}lbert network},
\newblock \bibinfo{journal}{Int. J. Mod. Phys. C} \bibinfo{volume}{15}
  (\bibinfo{year}{2004}) \bibinfo{pages}{241--246}.
\bibitem[{Ben-Naim(2005)}]{ben2005opinion}
\bibinfo{author}{E.~Ben-Naim},
\newblock \bibinfo{title}{Opinion dynamics: rise and fall of political
  parties},
\newblock \bibinfo{journal}{EPL (Europhys. Lett.)} \bibinfo{volume}{69}
  (\bibinfo{year}{2005}) \bibinfo{pages}{671}.
\bibitem[{Holme and Newman(2006)}]{holme2006nonequilibrium}
\bibinfo{author}{P.~Holme}, \bibinfo{author}{M.~E.~J. Newman},
\newblock \bibinfo{title}{Nonequilibrium phase transition in the coevolution of
  networks and opinions},
\newblock \bibinfo{journal}{Phys. Rev. E} \bibinfo{volume}{74}
  (\bibinfo{year}{2006}) \bibinfo{pages}{056108}.
\bibitem[{Gross and Blasius(2008)}]{gross2008adaptive}
\bibinfo{author}{T.~Gross}, \bibinfo{author}{B.~Blasius},
\newblock \bibinfo{title}{Adaptive coevolutionary networks: a review},
\newblock \bibinfo{journal}{J. R. Soc. Interface} \bibinfo{volume}{5}
  (\bibinfo{year}{2008}) \bibinfo{pages}{259--271}.
\bibitem[{Kozma and Barrat(2008)}]{kozma2008consensus}
\bibinfo{author}{B.~Kozma}, \bibinfo{author}{A.~Barrat},
\newblock \bibinfo{title}{Consensus formation on adaptive networks},
\newblock \bibinfo{journal}{Phys. Rev. E} \bibinfo{volume}{77}
  (\bibinfo{year}{2008}) \bibinfo{pages}{016102}.
\bibitem[{Baumann et~al.(2020)Baumann, Lorenz-Spreen, Sokolov, and
  Starnini}]{baumann2020modeling}
\bibinfo{author}{F.~Baumann}, \bibinfo{author}{P.~Lorenz-Spreen},
  \bibinfo{author}{I.~M. Sokolov}, \bibinfo{author}{M.~Starnini},
\newblock \bibinfo{title}{Modeling echo chambers and polarization dynamics in
  social networks},
\newblock \bibinfo{journal}{Phys. Rev. Lett.} \bibinfo{volume}{124}
  (\bibinfo{year}{2020}) \bibinfo{pages}{048301}.
\bibitem[{Suchecki et~al.(2005)Suchecki, Egu{\'\i}luz, and
  San~Miguel}]{suchecki2005voter}
\bibinfo{author}{K.~Suchecki}, \bibinfo{author}{V.~M. Egu{\'\i}luz},
  \bibinfo{author}{M.~San~Miguel},
\newblock \bibinfo{title}{Voter model dynamics in complex networks: {R}ole of
  dimensionality, disorder, and degree distribution},
\newblock \bibinfo{journal}{Phys. Rev. E} \bibinfo{volume}{72}
  (\bibinfo{year}{2005}) \bibinfo{pages}{036132}.
\bibitem[{Mobilia et~al.(2007)Mobilia, Petersen, and Redner}]{mobilia2007role}
\bibinfo{author}{M.~Mobilia}, \bibinfo{author}{A.~Petersen},
  \bibinfo{author}{S.~Redner},
\newblock \bibinfo{title}{On the role of zealotry in the voter model},
\newblock \bibinfo{journal}{J. Stat. Mech. Theory Exp.} \bibinfo{volume}{2007}
  (\bibinfo{year}{2007}) \bibinfo{pages}{P08029}.
\bibitem[{Masuda et~al.(2010)Masuda, Gibert, and
  Redner}]{masuda2010heterogeneous}
\bibinfo{author}{N.~Masuda}, \bibinfo{author}{N.~Gibert},
  \bibinfo{author}{S.~Redner},
\newblock \bibinfo{title}{Heterogeneous voter models},
\newblock \bibinfo{journal}{Phys. Rev. E} \bibinfo{volume}{82}
  (\bibinfo{year}{2010}) \bibinfo{pages}{010103}.
\bibitem[{Fern{\'a}ndez-Gracia et~al.(2014)Fern{\'a}ndez-Gracia, Suchecki,
  Ramasco, San~Miguel, and Egu{\'\i}luz}]{fernandez2014voter}
\bibinfo{author}{J.~Fern{\'a}ndez-Gracia}, \bibinfo{author}{K.~Suchecki},
  \bibinfo{author}{J.~J. Ramasco}, \bibinfo{author}{M.~San~Miguel},
  \bibinfo{author}{V.~M. Egu{\'\i}luz},
\newblock \bibinfo{title}{Is the voter model a model for voters?},
\newblock \bibinfo{journal}{Phys. Rev. Lett.} \bibinfo{volume}{112}
  (\bibinfo{year}{2014}) \bibinfo{pages}{158701}.
\bibitem[{Redner(2019)}]{redner2019reality}
\bibinfo{author}{S.~Redner},
\newblock \bibinfo{title}{Reality-inspired voter models: {A} mini-review},
\newblock \bibinfo{journal}{C. R. Phys.} \bibinfo{volume}{20}
  (\bibinfo{year}{2019}) \bibinfo{pages}{275--292}.
\bibitem[{Steels(1995)}]{steels1995self}
\bibinfo{author}{L.~Steels},
\newblock \bibinfo{title}{A self-organizing spatial vocabulary},
\newblock \bibinfo{journal}{Artif. Life} \bibinfo{volume}{2}
  (\bibinfo{year}{1995}) \bibinfo{pages}{319--332}.
\bibitem[{Baronchelli et~al.(2006)Baronchelli, Felici, Loreto, Caglioti, and
  Steels}]{baronchelli2006sharp}
\bibinfo{author}{A.~Baronchelli}, \bibinfo{author}{M.~Felici},
  \bibinfo{author}{V.~Loreto}, \bibinfo{author}{E.~Caglioti},
  \bibinfo{author}{L.~Steels},
\newblock \bibinfo{title}{Sharp transition towards shared vocabularies in
  multi-agent systems},
\newblock \bibinfo{journal}{J. Stat. Mech. Theory Exp.} \bibinfo{volume}{2006}
  (\bibinfo{year}{2006}) \bibinfo{pages}{P06014}.
\bibitem[{Loreto and Steels(2007)}]{loreto2007emergence}
\bibinfo{author}{V.~Loreto}, \bibinfo{author}{L.~Steels},
\newblock \bibinfo{title}{Emergence of language},
\newblock \bibinfo{journal}{Nat. Phys.} \bibinfo{volume}{3}
  (\bibinfo{year}{2007}) \bibinfo{pages}{758--760}.
\bibitem[{Baronchelli et~al.(2008)Baronchelli, Loreto, and
  Steels}]{baronchelli2008depth}
\bibinfo{author}{A.~Baronchelli}, \bibinfo{author}{V.~Loreto},
  \bibinfo{author}{L.~Steels},
\newblock \bibinfo{title}{In-depth analysis of the {N}aming {G}ame dynamics:
  the homogeneous mixing case},
\newblock \bibinfo{journal}{Int. J. Mod. Phys. C} \bibinfo{volume}{19}
  (\bibinfo{year}{2008}) \bibinfo{pages}{785--812}.
\bibitem[{Lu et~al.(2009)Lu, Korniss, and Szymanski}]{lu2009naming}
\bibinfo{author}{Q.~Lu}, \bibinfo{author}{G.~Korniss}, \bibinfo{author}{B.~K.
  Szymanski},
\newblock \bibinfo{title}{The naming game in social networks: community
  formation and consensus engineering},
\newblock \bibinfo{journal}{J. Econ. Interact. Coord.} \bibinfo{volume}{4}
  (\bibinfo{year}{2009}) \bibinfo{pages}{221--235}.
\bibitem[{DeGiuli(2019)}]{degiuli}
\bibinfo{author}{E.~DeGiuli},
\newblock \bibinfo{title}{Random language model},
\newblock \bibinfo{journal}{Phys. Rev. Lett.} \bibinfo{volume}{122}
  (\bibinfo{year}{2019}) \bibinfo{pages}{128301}.
\bibitem[{Vespignani(2012)}]{vespignani2012modelling}
\bibinfo{author}{A.~Vespignani},
\newblock \bibinfo{title}{Modelling dynamical processes in complex
  socio-technical systems},
\newblock \bibinfo{journal}{Nat. Phys.} \bibinfo{volume}{8}
  (\bibinfo{year}{2012}) \bibinfo{pages}{32--39}.
\bibitem[{Neumann et~al.(2009)Neumann, Noda, and
  Kawaoka}]{neumann2009emergence}
\bibinfo{author}{G.~Neumann}, \bibinfo{author}{T.~Noda},
  \bibinfo{author}{Y.~Kawaoka},
\newblock \bibinfo{title}{Emergence and pandemic potential of swine-origin h1n1
  influenza virus},
\newblock \bibinfo{journal}{Nature} \bibinfo{volume}{459}
  (\bibinfo{year}{2009}) \bibinfo{pages}{931--939}.
\bibitem[{Liu et~al.(2013)Liu, Shi, Shi, Wang, Xiao, Li, Bi, Wu, Li, Yan
  et~al.}]{liu2013origin}
\bibinfo{author}{D.~Liu}, \bibinfo{author}{W.~Shi}, \bibinfo{author}{Y.~Shi},
  \bibinfo{author}{D.~Wang}, \bibinfo{author}{H.~Xiao},
  \bibinfo{author}{W.~Li}, \bibinfo{author}{Y.~Bi}, \bibinfo{author}{Y.~Wu},
  \bibinfo{author}{X.~Li}, \bibinfo{author}{J.~Yan}, et~al.,
\newblock \bibinfo{title}{Origin and diversity of novel avian influenza {A
  H7N9} viruses causing human infection: phylogenetic, structural, and
  coalescent analyses},
\newblock \bibinfo{journal}{Lancet} \bibinfo{volume}{381}
  (\bibinfo{year}{2013}) \bibinfo{pages}{1926--1932}.
\bibitem[{Kucharski and Edmunds(2014)}]{kucharski2014case}
\bibinfo{author}{A.~J. Kucharski}, \bibinfo{author}{W.~J. Edmunds},
\newblock \bibinfo{title}{Case fatality rate for {E}bola virus disease in west
  {A}frica},
\newblock \bibinfo{journal}{Lancet} \bibinfo{volume}{384}
  (\bibinfo{year}{2014}) \bibinfo{pages}{1260}.
\bibitem[{Corti et~al.(2015)Corti, Zhao, Pedotti, Simonelli, Agnihothram, Fett,
  Fernandez-Rodriguez, Foglierini, Agatic, Vanzetta
  et~al.}]{corti2015prophylactic}
\bibinfo{author}{D.~Corti}, \bibinfo{author}{J.~Zhao},
  \bibinfo{author}{M.~Pedotti}, \bibinfo{author}{L.~Simonelli},
  \bibinfo{author}{S.~Agnihothram}, \bibinfo{author}{C.~Fett},
  \bibinfo{author}{B.~Fernandez-Rodriguez}, \bibinfo{author}{M.~Foglierini},
  \bibinfo{author}{G.~Agatic}, \bibinfo{author}{F.~Vanzetta}, et~al.,
\newblock \bibinfo{title}{Prophylactic and postexposure efficacy of a potent
  human monoclonal antibody against {MERS} coronavirus},
\newblock \bibinfo{journal}{Proc. Natl. Acad. Sci. USA} \bibinfo{volume}{112}
  (\bibinfo{year}{2015}) \bibinfo{pages}{10473--10478}.
\bibitem[{Kermack and McKendrick(1927)}]{kermack1927contribution}
\bibinfo{author}{W.~O. Kermack}, \bibinfo{author}{A.~G. McKendrick},
\newblock \bibinfo{title}{A contribution to the mathematical theory of
  epidemics},
\newblock \bibinfo{journal}{Proc. R. Soc. Lond. A} \bibinfo{volume}{115}
  (\bibinfo{year}{1927}) \bibinfo{pages}{700--721}.
\bibitem[{Abbey(1952)}]{abbey1952examination}
\bibinfo{author}{H.~Abbey},
\newblock \bibinfo{title}{An examination of the {Reed-Frost} theory of
  epidemics},
\newblock \bibinfo{journal}{Hum. Biol.} \bibinfo{volume}{24}
  (\bibinfo{year}{1952}) \bibinfo{pages}{201--233}.
\bibitem[{Li and Shuai(2009)}]{li2009global}
\bibinfo{author}{M.~Y. Li}, \bibinfo{author}{Z.~Shuai},
\newblock \bibinfo{title}{Global stability of an epidemic model in a patchy
  environment},
\newblock \bibinfo{journal}{Can. Appl. Math. Q.} \bibinfo{volume}{17}
  (\bibinfo{year}{2009}) \bibinfo{pages}{175--187}.
\bibitem[{Guimera et~al.(2005)Guimera, Mossa, Turtschi, and
  Amaral}]{guimera2005worldwide}
\bibinfo{author}{R.~Guimera}, \bibinfo{author}{S.~Mossa},
  \bibinfo{author}{A.~Turtschi}, \bibinfo{author}{L.~N. Amaral},
\newblock \bibinfo{title}{The worldwide air transportation network: {A}nomalous
  centrality, community structure, and cities' global roles},
\newblock \bibinfo{journal}{Proc. Natl. Acad. Sci. USA} \bibinfo{volume}{102}
  (\bibinfo{year}{2005}) \bibinfo{pages}{7794--7799}.
\bibitem[{Hufnagel et~al.(2004)Hufnagel, Brockmann, and
  Geisel}]{hufnagel2004PNAS}
\bibinfo{author}{L.~Hufnagel}, \bibinfo{author}{D.~Brockmann},
  \bibinfo{author}{T.~Geisel},
\newblock \bibinfo{title}{Forecast and control of epidemics in a globalized
  world},
\newblock \bibinfo{journal}{Proc. Natl. Acad. Sci. USA} \bibinfo{volume}{101}
  (\bibinfo{year}{2004}) \bibinfo{pages}{15124--15129}.
\bibitem[{Colizza et~al.(2006)Colizza, Barrat, Barth{\'e}lemy, and
  Vespignani}]{colizza2006PNAS}
\bibinfo{author}{V.~Colizza}, \bibinfo{author}{A.~Barrat},
  \bibinfo{author}{M.~Barth{\'e}lemy}, \bibinfo{author}{A.~Vespignani},
\newblock \bibinfo{title}{The role of the airline transportation network in the
  prediction and predictability of global epidemics},
\newblock \bibinfo{journal}{Proc. Natl. Acad. Sci. USA} \bibinfo{volume}{103}
  (\bibinfo{year}{2006}) \bibinfo{pages}{2015--2020}.
\bibitem[{Bajardi et~al.(2011)Bajardi, Poletto, Ramasco, Tizzoni, Colizza, and
  Vespignani}]{bajardi2011human}
\bibinfo{author}{P.~Bajardi}, \bibinfo{author}{C.~Poletto},
  \bibinfo{author}{J.~J. Ramasco}, \bibinfo{author}{M.~Tizzoni},
  \bibinfo{author}{V.~Colizza}, \bibinfo{author}{A.~Vespignani},
\newblock \bibinfo{title}{Human mobility networks, travel restrictions, and the
  global spread of 2009 {H1N1} pandemic},
\newblock \bibinfo{journal}{PLOS ONE} \bibinfo{volume}{6}
  (\bibinfo{year}{2011}) \bibinfo{pages}{e16591}.
\bibitem[{Wang and Wu(2018)}]{lin2018NC}
\bibinfo{author}{L.~Wang}, \bibinfo{author}{J.~T. Wu},
\newblock \bibinfo{title}{Characterizing the dynamics underlying global spread
  of epidemics},
\newblock \bibinfo{journal}{Nat. Commun.} \bibinfo{volume}{9}
  (\bibinfo{year}{2018}) \bibinfo{pages}{218}.
\bibitem[{Balcan et~al.(2009)Balcan, Colizza, Gon{\c{c}}alves, Hu, Ramasco, and
  Vespignani}]{balcan2009PNAS}
\bibinfo{author}{D.~Balcan}, \bibinfo{author}{V.~Colizza},
  \bibinfo{author}{B.~Gon{\c{c}}alves}, \bibinfo{author}{H.~Hu},
  \bibinfo{author}{J.~J. Ramasco}, \bibinfo{author}{A.~Vespignani},
\newblock \bibinfo{title}{Multiscale mobility networks and the spatial
  spreading of infectious diseases},
\newblock \bibinfo{journal}{Proc. Natl. Acad. Sci. USA} \bibinfo{volume}{106}
  (\bibinfo{year}{2009}) \bibinfo{pages}{21484--21489}.
\bibitem[{Poletto et~al.(2012)Poletto, Tizzoni, and Colizza}]{poletto2012SR}
\bibinfo{author}{C.~Poletto}, \bibinfo{author}{M.~Tizzoni},
  \bibinfo{author}{V.~Colizza},
\newblock \bibinfo{title}{Heterogeneous length of stay of hosts' movements and
  spatial epidemic spread},
\newblock \bibinfo{journal}{Sci. Rep.} \bibinfo{volume}{2}
  (\bibinfo{year}{2012}) \bibinfo{pages}{476}.
\bibitem[{Chinazzi et~al.(2020)Chinazzi, Davis, Ajelli, Gioannini, Litvinova,
  Merler, y~Piontti, Mu, Rossi, Sun et~al.}]{vespignani2020science1}
\bibinfo{author}{M.~Chinazzi}, \bibinfo{author}{J.~T. Davis},
  \bibinfo{author}{M.~Ajelli}, \bibinfo{author}{C.~Gioannini},
  \bibinfo{author}{M.~Litvinova}, \bibinfo{author}{S.~Merler},
  \bibinfo{author}{A.~P. y~Piontti}, \bibinfo{author}{K.~Mu},
  \bibinfo{author}{L.~Rossi}, \bibinfo{author}{K.~Sun}, et~al.,
\newblock \bibinfo{title}{The effect of travel restrictions on the spread of
  the 2019 novel coronavirus ({COVID-19}) outbreak},
\newblock \bibinfo{journal}{Science} \bibinfo{volume}{368}
  (\bibinfo{year}{2020}) \bibinfo{pages}{395--400}.
\bibitem[{Soriano-Pa{\~n}os et~al.(2018)Soriano-Pa{\~n}os, Lotero, Arenas, and
  G{\'o}mez-Garde{\~n}es}]{soriano2018PRX}
\bibinfo{author}{D.~Soriano-Pa{\~n}os}, \bibinfo{author}{L.~Lotero},
  \bibinfo{author}{A.~Arenas}, \bibinfo{author}{J.~G{\'o}mez-Garde{\~n}es},
\newblock \bibinfo{title}{Spreading processes in multiplex metapopulations
  containing different mobility networks},
\newblock \bibinfo{journal}{Phys. Rev. X} \bibinfo{volume}{8}
  (\bibinfo{year}{2018}) \bibinfo{pages}{031039}.
\bibitem[{Bedford et~al.(2015)Bedford, Riley, Barr, Broor, Chadha, Cox,
  Daniels, Gunasekaran, Hurt, Kelso et~al.}]{bedford2015nature}
\bibinfo{author}{T.~Bedford}, \bibinfo{author}{S.~Riley},
  \bibinfo{author}{I.~G. Barr}, \bibinfo{author}{S.~Broor},
  \bibinfo{author}{M.~Chadha}, \bibinfo{author}{N.~J. Cox},
  \bibinfo{author}{R.~S. Daniels}, \bibinfo{author}{C.~P. Gunasekaran},
  \bibinfo{author}{A.~C. Hurt}, \bibinfo{author}{A.~Kelso}, et~al.,
\newblock \bibinfo{title}{Global circulation patterns of seasonal influenza
  viruses vary with antigenic drift},
\newblock \bibinfo{journal}{Nature} \bibinfo{volume}{523}
  (\bibinfo{year}{2015}) \bibinfo{pages}{217--220}.
\bibitem[{Gonzalez et~al.(2008)Gonzalez, Hidalgo, and
  Barabasi}]{gonzalez2008nature}
\bibinfo{author}{M.~C. Gonzalez}, \bibinfo{author}{C.~A. Hidalgo},
  \bibinfo{author}{A.-L. Barabasi},
\newblock \bibinfo{title}{Understanding individual human mobility patterns},
\newblock \bibinfo{journal}{nature} \bibinfo{volume}{453}
  (\bibinfo{year}{2008}) \bibinfo{pages}{779--782}.
\bibitem[{Wesolowski et~al.(2012)Wesolowski, Eagle, Tatem, Smith, Noor, Snow,
  and Buckee}]{wesolowski2012science}
\bibinfo{author}{A.~Wesolowski}, \bibinfo{author}{N.~Eagle},
  \bibinfo{author}{A.~J. Tatem}, \bibinfo{author}{D.~L. Smith},
  \bibinfo{author}{A.~M. Noor}, \bibinfo{author}{R.~W. Snow},
  \bibinfo{author}{C.~O. Buckee},
\newblock \bibinfo{title}{Quantifying the impact of human mobility on malaria},
\newblock \bibinfo{journal}{Science} \bibinfo{volume}{338}
  (\bibinfo{year}{2012}) \bibinfo{pages}{267--270}.
\bibitem[{Lai et~al.(2020)Lai, Ruktanonchai, Zhou, Prosper, Luo, Floyd,
  Wesolowski, Santillana, Zhang, Du et~al.}]{lai2020effect}
\bibinfo{author}{S.~Lai}, \bibinfo{author}{N.~W. Ruktanonchai},
  \bibinfo{author}{L.~Zhou}, \bibinfo{author}{O.~Prosper},
  \bibinfo{author}{W.~Luo}, \bibinfo{author}{J.~R. Floyd},
  \bibinfo{author}{A.~Wesolowski}, \bibinfo{author}{M.~Santillana},
  \bibinfo{author}{C.~Zhang}, \bibinfo{author}{X.~Du}, et~al.,
\newblock \bibinfo{title}{Effect of non-pharmaceutical interventions to contain
  {COVID-19} in {C}hina},
\newblock \bibinfo{journal}{Nature} \bibinfo{volume}{585}
  (\bibinfo{year}{2020}) \bibinfo{pages}{410--413}.
\bibitem[{Castro et~al.(2021)Castro, Generous, Luo, Pastore~y Piontti,
  Martinez, Gomes, Osthus, Fairchild, Ziemann, Vespignani
  et~al.}]{castro2021using}
\bibinfo{author}{L.~A. Castro}, \bibinfo{author}{N.~Generous},
  \bibinfo{author}{W.~Luo}, \bibinfo{author}{A.~Pastore~y Piontti},
  \bibinfo{author}{K.~Martinez}, \bibinfo{author}{M.~F. Gomes},
  \bibinfo{author}{D.~Osthus}, \bibinfo{author}{G.~Fairchild},
  \bibinfo{author}{A.~Ziemann}, \bibinfo{author}{A.~Vespignani}, et~al.,
\newblock \bibinfo{title}{Using heterogeneous data to identify signatures of
  dengue outbreaks at fine spatio-temporal scales across {B}razil},
\newblock \bibinfo{journal}{PLOS Negl. Trop. Dis.} \bibinfo{volume}{15}
  (\bibinfo{year}{2021}) \bibinfo{pages}{e0009392}.
\bibitem[{Nouvellet et~al.(2021)Nouvellet, Bhatia, Cori, Ainslie, Baguelin,
  Bhatt, Boonyasiri, Brazeau, Cattarino, Cooper et~al.}]{nouvellet2021NC}
\bibinfo{author}{P.~Nouvellet}, \bibinfo{author}{S.~Bhatia},
  \bibinfo{author}{A.~Cori}, \bibinfo{author}{K.~E. Ainslie},
  \bibinfo{author}{M.~Baguelin}, \bibinfo{author}{S.~Bhatt},
  \bibinfo{author}{A.~Boonyasiri}, \bibinfo{author}{N.~F. Brazeau},
  \bibinfo{author}{L.~Cattarino}, \bibinfo{author}{L.~V. Cooper}, et~al.,
\newblock \bibinfo{title}{Reduction in mobility and {COVID-19} transmission},
\newblock \bibinfo{journal}{Nat. Commun.} \bibinfo{volume}{12}
  (\bibinfo{year}{2021}) \bibinfo{pages}{1090}.
\bibitem[{Xu et~al.(2021)Xu, Wang, and Pei}]{xu2021JTM}
\bibinfo{author}{X.-K. Xu}, \bibinfo{author}{L.~Wang},
  \bibinfo{author}{S.~Pei},
\newblock \bibinfo{title}{Multiscale mobility explains differential
  associations between the gross domestic product and {COVID-19} transmission
  in {C}hinese cities},
\newblock \bibinfo{journal}{J. Travel Med.} \bibinfo{volume}{28}
  (\bibinfo{year}{2021}) \bibinfo{pages}{taaa236}.
\bibitem[{Salath{\'e}(2018)}]{salathe2018digital}
\bibinfo{author}{M.~Salath{\'e}},
\newblock \bibinfo{title}{Digital epidemiology: what is it, and where is it
  going?},
\newblock \bibinfo{journal}{Life Sci. Soc. Policy} \bibinfo{volume}{14}
  (\bibinfo{year}{2018}) \bibinfo{pages}{1--5}.
\bibitem[{Cervellin et~al.(2017)Cervellin, Comelli, and
  Lippi}]{cervellin2017google}
\bibinfo{author}{G.~Cervellin}, \bibinfo{author}{I.~Comelli},
  \bibinfo{author}{G.~Lippi},
\newblock \bibinfo{title}{Is {Google Trends} a reliable tool for digital
  epidemiology? {I}nsights from different clinical settings},
\newblock \bibinfo{journal}{J. Epidemiol. Glob. Health} \bibinfo{volume}{7}
  (\bibinfo{year}{2017}) \bibinfo{pages}{185--189}.
\bibitem[{Bengtsson et~al.(2011)Bengtsson, Lu, Thorson, Garfield, and
  Von~Schreeb}]{bengtsson2011improved}
\bibinfo{author}{L.~Bengtsson}, \bibinfo{author}{X.~Lu},
  \bibinfo{author}{A.~Thorson}, \bibinfo{author}{R.~Garfield},
  \bibinfo{author}{J.~Von~Schreeb},
\newblock \bibinfo{title}{Improved response to disasters and outbreaks by
  tracking population movements with mobile phone network data: a
  post-earthquake geospatial study in {H}aiti},
\newblock \bibinfo{journal}{PLOS Med.} \bibinfo{volume}{8}
  (\bibinfo{year}{2011}) \bibinfo{pages}{e1001083}.
\bibitem[{Polgreen et~al.(2008)Polgreen, Chen, Pennock, Nelson, and
  Weinstein}]{polgreen2008using}
\bibinfo{author}{P.~M. Polgreen}, \bibinfo{author}{Y.~Chen},
  \bibinfo{author}{D.~M. Pennock}, \bibinfo{author}{F.~D. Nelson},
  \bibinfo{author}{R.~A. Weinstein},
\newblock \bibinfo{title}{Using internet searches for influenza surveillance},
\newblock \bibinfo{journal}{Clin. Infect. Dis.} \bibinfo{volume}{47}
  (\bibinfo{year}{2008}) \bibinfo{pages}{1443--1448}.
\bibitem[{Lazer et~al.(2014)Lazer, Kennedy, King, and
  Vespignani}]{lazer2014parable}
\bibinfo{author}{D.~Lazer}, \bibinfo{author}{R.~Kennedy},
  \bibinfo{author}{G.~King}, \bibinfo{author}{A.~Vespignani},
\newblock \bibinfo{title}{The parable of {Google Flu}: traps in big data
  analysis},
\newblock \bibinfo{journal}{Science} \bibinfo{volume}{343}
  (\bibinfo{year}{2014}) \bibinfo{pages}{1203--1205}.
\bibitem[{Santillana et~al.(2014)Santillana, Zhang, Althouse, and
  Ayers}]{santillana2014can}
\bibinfo{author}{M.~Santillana}, \bibinfo{author}{D.~W. Zhang},
  \bibinfo{author}{B.~M. Althouse}, \bibinfo{author}{J.~W. Ayers},
\newblock \bibinfo{title}{What can digital disease detection learn from (an
  external revision to) {Google Flu Trends}?},
\newblock \bibinfo{journal}{Am. J. Prev. Med.} \bibinfo{volume}{47}
  (\bibinfo{year}{2014}) \bibinfo{pages}{341--347}.
\bibitem[{Yang et~al.(2015)Yang, Santillana, and Kou}]{yang2015accurate}
\bibinfo{author}{S.~Yang}, \bibinfo{author}{M.~Santillana},
  \bibinfo{author}{S.~C. Kou},
\newblock \bibinfo{title}{Accurate estimation of influenza epidemics using
  {G}oogle search data via {ARGO}},
\newblock \bibinfo{journal}{Proc. Natl. Acad. Sci. USA} \bibinfo{volume}{112}
  (\bibinfo{year}{2015}) \bibinfo{pages}{14473--14478}.
\bibitem[{Lu et~al.(2019)Lu, Hattab, Clemente, Biggerstaff, and
  Santillana}]{lu2019improved}
\bibinfo{author}{F.~S. Lu}, \bibinfo{author}{M.~W. Hattab},
  \bibinfo{author}{C.~L. Clemente}, \bibinfo{author}{M.~Biggerstaff},
  \bibinfo{author}{M.~Santillana},
\newblock \bibinfo{title}{Improved state-level influenza nowcasting in the
  {United States} leveraging {I}nternet-based data and network approaches},
\newblock \bibinfo{journal}{Nat. Commun.} \bibinfo{volume}{10}
  (\bibinfo{year}{2019}) \bibinfo{pages}{147}.
\bibitem[{Clemente et~al.(2019)Clemente, Lu, and
  Santillana}]{clemente2019improved}
\bibinfo{author}{L.~Clemente}, \bibinfo{author}{F.~Lu},
  \bibinfo{author}{M.~Santillana},
\newblock \bibinfo{title}{Improved real-time influenza surveillance: using
  internet search data in eight {Latin American} countries},
\newblock \bibinfo{journal}{JMIR Public Health Surveill.} \bibinfo{volume}{5}
  (\bibinfo{year}{2019}) \bibinfo{pages}{e12214}.
\bibitem[{Althouse et~al.(2011)Althouse, Ng, and
  Cummings}]{althouse2011prediction}
\bibinfo{author}{B.~M. Althouse}, \bibinfo{author}{Y.~Y. Ng},
  \bibinfo{author}{D.~A.~T. Cummings},
\newblock \bibinfo{title}{Prediction of dengue incidence using search query
  surveillance},
\newblock \bibinfo{journal}{PLOS Negl. Trop. Dis.} \bibinfo{volume}{5}
  (\bibinfo{year}{2011}) \bibinfo{pages}{e1258}.
\bibitem[{Chan et~al.(2011)Chan, Sahai, Conrad, and Brownstein}]{chan2011using}
\bibinfo{author}{E.~H. Chan}, \bibinfo{author}{V.~Sahai},
  \bibinfo{author}{C.~Conrad}, \bibinfo{author}{J.~S. Brownstein},
\newblock \bibinfo{title}{Using web search query data to monitor dengue
  epidemics: a new model for neglected tropical disease surveillance},
\newblock \bibinfo{journal}{PLOS Negl. Trop. Dis.} \bibinfo{volume}{5}
  (\bibinfo{year}{2011}) \bibinfo{pages}{e1206}.
\bibitem[{Alasaad(2013)}]{alasaad2013war}
\bibinfo{author}{S.~Alasaad},
\newblock \bibinfo{title}{War diseases revealed by the social media: massive
  leishmaniasis outbreak in the {S}yrian {S}pring},
\newblock \bibinfo{journal}{Parasites Vectors} \bibinfo{volume}{6}
  (\bibinfo{year}{2013}) \bibinfo{pages}{94}.
\bibitem[{Ocampo et~al.(2013)Ocampo, Chunara, and Brownstein}]{ocampo2013using}
\bibinfo{author}{A.~J. Ocampo}, \bibinfo{author}{R.~Chunara},
  \bibinfo{author}{J.~S. Brownstein},
\newblock \bibinfo{title}{Using search queries for malaria surveillance,
  {T}hailand},
\newblock \bibinfo{journal}{Malar. J.} \bibinfo{volume}{12}
  (\bibinfo{year}{2013}) \bibinfo{pages}{390}.
\bibitem[{Milinovich et~al.(2014)Milinovich, Avril, Clements, Brownstein, Tong,
  and Hu}]{milinovich2014using}
\bibinfo{author}{G.~J. Milinovich}, \bibinfo{author}{S.~M. Avril},
  \bibinfo{author}{A.~C. Clements}, \bibinfo{author}{J.~S. Brownstein},
  \bibinfo{author}{S.~Tong}, \bibinfo{author}{W.~Hu},
\newblock \bibinfo{title}{Using internet search queries for infectious disease
  surveillance: screening diseases for suitability},
\newblock \bibinfo{journal}{BMC Infect. Dis.} \bibinfo{volume}{14}
  (\bibinfo{year}{2014}) \bibinfo{pages}{690}.
\bibitem[{Messina et~al.(2014)Messina, Brady, Pigott, Brownstein, Hoen, and
  Hay}]{messina2014global}
\bibinfo{author}{J.~P. Messina}, \bibinfo{author}{O.~J. Brady},
  \bibinfo{author}{D.~M. Pigott}, \bibinfo{author}{J.~S. Brownstein},
  \bibinfo{author}{A.~G. Hoen}, \bibinfo{author}{S.~I. Hay},
\newblock \bibinfo{title}{A global compendium of human dengue virus
  occurrence},
\newblock \bibinfo{journal}{Sci. Data} \bibinfo{volume}{1}
  (\bibinfo{year}{2014}) \bibinfo{pages}{140004}.
\bibitem[{Gluskin et~al.(2014)Gluskin, Johansson, Santillana, and
  Brownstein}]{gluskin2014evaluation}
\bibinfo{author}{R.~T. Gluskin}, \bibinfo{author}{M.~A. Johansson},
  \bibinfo{author}{M.~Santillana}, \bibinfo{author}{J.~S. Brownstein},
\newblock \bibinfo{title}{Evaluation of {I}nternet-based dengue query data:
  {Google Dengue Trends}},
\newblock \bibinfo{journal}{PLOS Negl. Trop. Dis.} \bibinfo{volume}{8}
  (\bibinfo{year}{2014}) \bibinfo{pages}{e2713}.
\bibitem[{Yang et~al.(2017)Yang, Kou, Lu, Brownstein, Brooke, and
  Santillana}]{yang2017advances}
\bibinfo{author}{S.~Yang}, \bibinfo{author}{S.~C. Kou},
  \bibinfo{author}{F.~Lu}, \bibinfo{author}{J.~S. Brownstein},
  \bibinfo{author}{N.~Brooke}, \bibinfo{author}{M.~Santillana},
\newblock \bibinfo{title}{Advances in using {I}nternet searches to track
  dengue},
\newblock \bibinfo{journal}{PLOS Comput. Biol.} \bibinfo{volume}{13}
  (\bibinfo{year}{2017}) \bibinfo{pages}{e1005607}.
\bibitem[{Strauss et~al.(2017)Strauss, Castro, Reintjes, and
  Torres}]{strauss2017google}
\bibinfo{author}{R.~A. Strauss}, \bibinfo{author}{J.~S. Castro},
  \bibinfo{author}{R.~Reintjes}, \bibinfo{author}{J.~R. Torres},
\newblock \bibinfo{title}{Google dengue trends: {A}n indicator of epidemic
  behavior. {T}he {V}enezuelan {C}ase},
\newblock \bibinfo{journal}{Int. J. Med. Inform.} \bibinfo{volume}{104}
  (\bibinfo{year}{2017}) \bibinfo{pages}{26--30}.
\bibitem[{Ho et~al.(2018)Ho, Carvajal, Bautista, Capistrano, Viacrusis,
  Hernandez, and Watanabe}]{ho2018using}
\bibinfo{author}{H.~T. Ho}, \bibinfo{author}{T.~M. Carvajal},
  \bibinfo{author}{J.~R. Bautista}, \bibinfo{author}{J.~D.~R. Capistrano},
  \bibinfo{author}{K.~M. Viacrusis}, \bibinfo{author}{L.~F.~T. Hernandez},
  \bibinfo{author}{K.~Watanabe},
\newblock \bibinfo{title}{Using {G}oogle trends to examine the spatio-temporal
  incidence and behavioral patterns of dengue disease: a case study in
  metropolitan {M}anila, {P}hilippines},
\newblock \bibinfo{journal}{Trop. Med. Infect. Dis.} \bibinfo{volume}{3}
  (\bibinfo{year}{2018}) \bibinfo{pages}{118}.
\bibitem[{Gesualdo et~al.(2013)Gesualdo, Stilo, Gonfiantini, Pandolfi, Velardi,
  and Tozzi}]{gesualdo2013influenza}
\bibinfo{author}{F.~Gesualdo}, \bibinfo{author}{G.~Stilo},
  \bibinfo{author}{M.~V. Gonfiantini}, \bibinfo{author}{E.~Pandolfi},
  \bibinfo{author}{P.~Velardi}, \bibinfo{author}{A.~E. Tozzi},
\newblock \bibinfo{title}{Influenza-like illness surveillance on {T}witter
  through automated learning of na{\"\i}ve language},
\newblock \bibinfo{journal}{PLOS ONE} \bibinfo{volume}{8}
  (\bibinfo{year}{2013}) \bibinfo{pages}{e82489}.
\bibitem[{Broniatowski et~al.(2013)Broniatowski, Paul, and
  Dredze}]{broniatowski2013national}
\bibinfo{author}{D.~A. Broniatowski}, \bibinfo{author}{M.~J. Paul},
  \bibinfo{author}{M.~Dredze},
\newblock \bibinfo{title}{National and local influenza surveillance through
  {T}witter: an analysis of the 2012-2013 influenza epidemic},
\newblock \bibinfo{journal}{PLOS ONE} \bibinfo{volume}{8}
  (\bibinfo{year}{2013}) \bibinfo{pages}{e83672}.
\bibitem[{de~Almeida Marques-Toledo et~al.(2017)de~Almeida Marques-Toledo,
  Degener, Vinhal, Coelho, Meira, Code{\c{c}}o, and
  Teixeira}]{dealmeida2017dengue}
\bibinfo{author}{C.~de~Almeida Marques-Toledo}, \bibinfo{author}{C.~M.
  Degener}, \bibinfo{author}{L.~Vinhal}, \bibinfo{author}{G.~Coelho},
  \bibinfo{author}{W.~Meira}, \bibinfo{author}{C.~T. Code{\c{c}}o},
  \bibinfo{author}{M.~M. Teixeira},
\newblock \bibinfo{title}{Dengue prediction by the web: {T}weets are a useful
  tool for estimating and forecasting {D}engue at country and city level},
\newblock \bibinfo{journal}{PLOS Negl. Trop. Dis.} \bibinfo{volume}{11}
  (\bibinfo{year}{2017}) \bibinfo{pages}{e0005729}.
\bibitem[{Deiner et~al.(2019)Deiner, Fathy, Kim, Niemeyer, Ramirez, Ackley,
  Liu, Lietman, and Porco}]{deiner2019facebook}
\bibinfo{author}{M.~S. Deiner}, \bibinfo{author}{C.~Fathy},
  \bibinfo{author}{J.~Kim}, \bibinfo{author}{K.~Niemeyer},
  \bibinfo{author}{D.~Ramirez}, \bibinfo{author}{S.~F. Ackley},
  \bibinfo{author}{F.~Liu}, \bibinfo{author}{T.~M. Lietman},
  \bibinfo{author}{T.~C. Porco},
\newblock \bibinfo{title}{Facebook and {T}witter vaccine sentiment in response
  to measles outbreaks},
\newblock \bibinfo{journal}{Health Inform. J.} \bibinfo{volume}{25}
  (\bibinfo{year}{2019}) \bibinfo{pages}{1116--1132}.
\bibitem[{Hu et~al.(2021)Hu, Wang, Luo, Yan, Zhang, Huang, Liu, Ly, Kacker, and
  Li}]{hu2021revealing}
\bibinfo{author}{T.~Hu}, \bibinfo{author}{S.~Wang}, \bibinfo{author}{W.~Luo},
  \bibinfo{author}{Y.~Yan}, \bibinfo{author}{M.~Zhang},
  \bibinfo{author}{X.~Huang}, \bibinfo{author}{R.~Liu},
  \bibinfo{author}{K.~Ly}, \bibinfo{author}{V.~Kacker},
  \bibinfo{author}{Z.~Li},
\newblock \bibinfo{title}{Revealing public opinion towards {COVID-19} vaccines
  using {T}witter data in the {U}nited {S}tates: a spatiotemporal perspective},
\newblock \bibinfo{journal}{J. Med. Internet Res.} \bibinfo{volume}{23}
  (\bibinfo{year}{2021}) \bibinfo{pages}{e30854}.
\bibitem[{Salath{\'e} and Khandelwal(2011)}]{salathe2011assessing}
\bibinfo{author}{M.~Salath{\'e}}, \bibinfo{author}{S.~Khandelwal},
\newblock \bibinfo{title}{Assessing vaccination sentiments with online social
  media: implications for infectious disease dynamics and control},
\newblock \bibinfo{journal}{PLOS Comput. Biol.} \bibinfo{volume}{7}
  (\bibinfo{year}{2011}) \bibinfo{pages}{e1002199}.
\bibitem[{Salath{\'e} et~al.(2013)Salath{\'e}, Vu, Khandelwal, and
  Hunter}]{salathe2013dynamics}
\bibinfo{author}{M.~Salath{\'e}}, \bibinfo{author}{D.~Q. Vu},
  \bibinfo{author}{S.~Khandelwal}, \bibinfo{author}{D.~R. Hunter},
\newblock \bibinfo{title}{The dynamics of health behavior sentiments on a large
  online social network},
\newblock \bibinfo{journal}{EPJ Data Sci.} \bibinfo{volume}{2}
  (\bibinfo{year}{2013}) \bibinfo{pages}{4}.
\bibitem[{Bian et~al.(2012)Bian, Topaloglu, and Yu}]{bian2012towards}
\bibinfo{author}{J.~Bian}, \bibinfo{author}{U.~Topaloglu},
  \bibinfo{author}{F.~Yu},
\newblock \bibinfo{title}{Towards large-scale twitter mining for drug-related
  adverse events},
\newblock in: \bibinfo{editor}{C.~C. Yang}, \bibinfo{editor}{H.~Chen},
  \bibinfo{editor}{H.~Wactlar}, \bibinfo{editor}{C.~Combi},
  \bibinfo{editor}{X.~Tang} (Eds.), \bibinfo{booktitle}{Proceedings of the 2012
  international workshop on Smart health and wellbeing},
  \bibinfo{publisher}{Association for Computing Machinery},
  \bibinfo{year}{2012}, pp. \bibinfo{pages}{25--32}.
\bibitem[{Jurdak et~al.(2015)Jurdak, Zhao, Liu, AbouJaoude, Cameron, and
  Newth}]{jurdak2015understanding}
\bibinfo{author}{R.~Jurdak}, \bibinfo{author}{K.~Zhao},
  \bibinfo{author}{J.~Liu}, \bibinfo{author}{M.~AbouJaoude},
  \bibinfo{author}{M.~Cameron}, \bibinfo{author}{D.~Newth},
\newblock \bibinfo{title}{Understanding human mobility from {T}witter},
\newblock \bibinfo{journal}{PLOS ONE} \bibinfo{volume}{10}
  (\bibinfo{year}{2015}) \bibinfo{pages}{e0131469}.
\bibitem[{Blanford et~al.(2015)Blanford, Huang, Savelyev, and
  MacEachren}]{blanford2015geo}
\bibinfo{author}{J.~I. Blanford}, \bibinfo{author}{Z.~Huang},
  \bibinfo{author}{A.~Savelyev}, \bibinfo{author}{A.~M. MacEachren},
\newblock \bibinfo{title}{Geo-located tweets. enhancing mobility maps and
  capturing cross-border movement},
\newblock \bibinfo{journal}{PLOS ONE} \bibinfo{volume}{10}
  (\bibinfo{year}{2015}) \bibinfo{pages}{e0129202}.
\bibitem[{(2020)}]{unknown2020facebook}
\bibinfo{title}{Facebook {D}ata for {G}ood {A}nnual {R}eport 2020},
  \bibinfo{year}{2020}. \bibinfo{note}{Available at:
  \url{https://dataforgood.fb.com/wp-content/uploads/2021/01/Facebook-Data-for-Good-2020-Annual-Report-1.pdf}.
  Archived at: \url{https://doi.org/10.17605/OSF.IO/N7G9X}}.
\bibitem[{Kuchler et~al.(2021)Kuchler, Russel, and Stroebel}]{kuchler2021jue}
\bibinfo{author}{T.~Kuchler}, \bibinfo{author}{D.~Russel},
  \bibinfo{author}{J.~Stroebel},
\newblock \bibinfo{title}{{JUE Insight}: {T}he geographic spread of {COVID-19}
  correlates with the structure of social networks as measured by {F}acebook},
\newblock \bibinfo{journal}{J. Urban Econ.}  (\bibinfo{year}{2021})
  \bibinfo{pages}{103314}.
\bibitem[{Spelta et~al.(2020)Spelta, Flori, Pierri, Bonaccorsi, and
  Pammolli}]{spelta2020after}
\bibinfo{author}{A.~Spelta}, \bibinfo{author}{A.~Flori},
  \bibinfo{author}{F.~Pierri}, \bibinfo{author}{G.~Bonaccorsi},
  \bibinfo{author}{F.~Pammolli},
\newblock \bibinfo{title}{After the lockdown: simulating mobility, public
  health and economic recovery scenarios},
\newblock \bibinfo{journal}{Sci. Rep.} \bibinfo{volume}{10}
  (\bibinfo{year}{2020}) \bibinfo{pages}{16950}.
\bibitem[{Maas et~al.(2019)Maas, Iyer, Gros, Park, McGorman, Nayak, and
  Dow}]{maas2019facebook}
\bibinfo{author}{P.~Maas}, \bibinfo{author}{S.~Iyer},
  \bibinfo{author}{A.~Gros}, \bibinfo{author}{W.~Park},
  \bibinfo{author}{L.~McGorman}, \bibinfo{author}{C.~Nayak},
  \bibinfo{author}{P.~A. Dow},
\newblock \bibinfo{title}{Facebook {D}isaster {M}aps: Aggregate insights for
  crisis response \& recovery.},
\newblock in: \bibinfo{editor}{Z.~Franco}, \bibinfo{editor}{J.~J.
  Gonz\'{a}lez}, \bibinfo{editor}{J.~H. Can\'{o}s} (Eds.),
  \bibinfo{booktitle}{Proceedings of the 16th International Conference on
  Information Systems for Crisis Response And Management},
  \bibinfo{publisher}{International Association for Information Systems for
  Crisis Response and Management}, \bibinfo{year}{2019}, pp.
  \bibinfo{pages}{836--847}.
\bibitem[{Cook et~al.(2011)Cook, Conrad, Fowlkes, and
  Mohebbi}]{cook2011assessing}
\bibinfo{author}{S.~Cook}, \bibinfo{author}{C.~Conrad}, \bibinfo{author}{A.~L.
  Fowlkes}, \bibinfo{author}{M.~H. Mohebbi},
\newblock \bibinfo{title}{Assessing {G}oogle flu trends performance in the
  {U}nited {S}tates during the 2009 influenza virus {A(H1N1)} pandemic},
\newblock \bibinfo{journal}{PLOS ONE} \bibinfo{volume}{6}
  (\bibinfo{year}{2011}) \bibinfo{pages}{e23610}.
\bibitem[{Butler(2013)}]{butler2013google}
\bibinfo{author}{D.~Butler},
\newblock \bibinfo{title}{When {G}oogle got flu wrong: {US} outbreak foxes a
  leading web-based method for tracking seasonal flu},
\newblock \bibinfo{journal}{Nature} \bibinfo{volume}{494}
  (\bibinfo{year}{2013}) \bibinfo{pages}{155--157}.
\bibitem[{Olson et~al.(2013)Olson, Konty, Paladini, Viboud, and
  Simonsen}]{olson2013reassessing}
\bibinfo{author}{D.~R. Olson}, \bibinfo{author}{K.~J. Konty},
  \bibinfo{author}{M.~Paladini}, \bibinfo{author}{C.~Viboud},
  \bibinfo{author}{L.~Simonsen},
\newblock \bibinfo{title}{Reassessing {G}oogle {F}lu {T}rends data for
  detection of seasonal and pandemic influenza: a comparative epidemiological
  study at three geographic scales},
\newblock \bibinfo{journal}{PLOS Comput. Biol.} \bibinfo{volume}{9}
  (\bibinfo{year}{2013}) \bibinfo{pages}{e1003256}.
\bibitem[{Santillana et~al.(2015)Santillana, Nguyen, Dredze, Paul, Nsoesie, and
  Brownstein}]{santillana2015combining}
\bibinfo{author}{M.~Santillana}, \bibinfo{author}{A.~T. Nguyen},
  \bibinfo{author}{M.~Dredze}, \bibinfo{author}{M.~J. Paul},
  \bibinfo{author}{E.~O. Nsoesie}, \bibinfo{author}{J.~S. Brownstein},
\newblock \bibinfo{title}{Combining search, social media, and traditional data
  sources to improve influenza surveillance},
\newblock \bibinfo{journal}{PLOS Comput. Biol.} \bibinfo{volume}{11}
  (\bibinfo{year}{2015}) \bibinfo{pages}{e1004513}.
\bibitem[{Smolinski et~al.(2015)Smolinski, Crawley, Baltrusaitis, Chunara,
  Olsen, W{\'o}jcik, Santillana, Nguyen, and Brownstein}]{smolinski2015flu}
\bibinfo{author}{M.~S. Smolinski}, \bibinfo{author}{A.~W. Crawley},
  \bibinfo{author}{K.~Baltrusaitis}, \bibinfo{author}{R.~Chunara},
  \bibinfo{author}{J.~M. Olsen}, \bibinfo{author}{O.~W{\'o}jcik},
  \bibinfo{author}{M.~Santillana}, \bibinfo{author}{A.~Nguyen},
  \bibinfo{author}{J.~S. Brownstein},
\newblock \bibinfo{title}{Flu {N}ear {Y}ou: {C}rowdsourced symptom reporting
  spanning 2 influenza seasons},
\newblock \bibinfo{journal}{Am. J. Public Health} \bibinfo{volume}{105}
  (\bibinfo{year}{2015}) \bibinfo{pages}{2124--2130}.
\bibitem[{Santillana(2016)}]{santillana2016perspectives}
\bibinfo{author}{M.~Santillana},
\newblock \bibinfo{title}{Editorial commentary: {P}erspectives on the future of
  internet search engines and biosurveillance systems},
\newblock \bibinfo{journal}{Clin. Infect. Dis.} \bibinfo{volume}{64}
  (\bibinfo{year}{2016}) \bibinfo{pages}{42--43}.
\bibitem[{McGough et~al.(2017)McGough, Brownstein, Hawkins, and
  Santillana}]{mcgough2017forecasting}
\bibinfo{author}{S.~F. McGough}, \bibinfo{author}{J.~S. Brownstein},
  \bibinfo{author}{J.~B. Hawkins}, \bibinfo{author}{M.~Santillana},
\newblock \bibinfo{title}{Forecasting {Z}ika incidence in the 2016 {L}atin
  {A}merica outbreak combining traditional disease surveillance with search,
  social media, and news report data},
\newblock \bibinfo{journal}{PLOS Negl. Trop. Dis.} \bibinfo{volume}{11}
  (\bibinfo{year}{2017}) \bibinfo{pages}{e0005295}.
\bibitem[{Brownstein et~al.(2008)Brownstein, Freifeld, Reis, and
  Mandl}]{brownstein2008surveillance}
\bibinfo{author}{J.~S. Brownstein}, \bibinfo{author}{C.~C. Freifeld},
  \bibinfo{author}{B.~Y. Reis}, \bibinfo{author}{K.~D. Mandl},
\newblock \bibinfo{title}{Surveillance {S}ans {F}rontieres: {I}nternet-based
  emerging infectious disease intelligence and the {HealthMap} project},
\newblock \bibinfo{journal}{PLOS Med.} \bibinfo{volume}{5}
  (\bibinfo{year}{2008}) \bibinfo{pages}{e151}.
\bibitem[{Baltrusaitis et~al.(2018)Baltrusaitis, Brownstein, Scarpino, Bakota,
  Crawley, Conidi, Gunn, Gray, Zink, and
  Santillana}]{baltrusaitis2018comparison}
\bibinfo{author}{K.~Baltrusaitis}, \bibinfo{author}{J.~S. Brownstein},
  \bibinfo{author}{S.~V. Scarpino}, \bibinfo{author}{E.~Bakota},
  \bibinfo{author}{A.~W. Crawley}, \bibinfo{author}{G.~Conidi},
  \bibinfo{author}{J.~Gunn}, \bibinfo{author}{J.~Gray},
  \bibinfo{author}{A.~Zink}, \bibinfo{author}{M.~Santillana},
\newblock \bibinfo{title}{Comparison of crowd-sourced, electronic health
  records based, and traditional health-care based influenza-tracking systems
  at multiple spatial resolutions in the {U}nited {S}tates of {A}merica},
\newblock \bibinfo{journal}{BMC Infect. Dis.} \bibinfo{volume}{18}
  (\bibinfo{year}{2018}) \bibinfo{pages}{403}.
\bibitem[{Liu et~al.(2020)Liu, Clemente, Poirier, Ding, Chinazzi, Davis,
  Vespignani, and Santillana}]{liu2020real}
\bibinfo{author}{D.~Liu}, \bibinfo{author}{L.~Clemente},
  \bibinfo{author}{C.~Poirier}, \bibinfo{author}{X.~Ding},
  \bibinfo{author}{M.~Chinazzi}, \bibinfo{author}{J.~Davis},
  \bibinfo{author}{A.~Vespignani}, \bibinfo{author}{M.~Santillana},
\newblock \bibinfo{title}{Real-time forecasting of the {COVID-19} outbreak in
  {C}hinese provinces: machine learning approach using novel digital data and
  estimates from mechanistic models},
\newblock \bibinfo{journal}{J. Med. Internet Res.} \bibinfo{volume}{22}
  (\bibinfo{year}{2020}) \bibinfo{pages}{e20285}.
\bibitem[{Lu et~al.(2018)Lu, Hou, Baltrusaitis, Shah, Leskovec, Hawkins,
  Brownstein, Conidi, Gunn, Gray et~al.}]{lu2018accurate}
\bibinfo{author}{F.~S. Lu}, \bibinfo{author}{S.~Hou},
  \bibinfo{author}{K.~Baltrusaitis}, \bibinfo{author}{M.~Shah},
  \bibinfo{author}{J.~Leskovec}, \bibinfo{author}{J.~Hawkins},
  \bibinfo{author}{J.~Brownstein}, \bibinfo{author}{G.~Conidi},
  \bibinfo{author}{J.~Gunn}, \bibinfo{author}{J.~Gray}, et~al.,
\newblock \bibinfo{title}{Accurate influenza monitoring and forecasting using
  novel internet data streams: a case study in the {B}oston {M}etropolis},
\newblock \bibinfo{journal}{JMIR Public Health Surveill.} \bibinfo{volume}{4}
  (\bibinfo{year}{2018}) \bibinfo{pages}{e4}.
\bibitem[{Pastor-Satorras et~al.(2015)Pastor-Satorras, Castellano, Van~Mieghem,
  and Vespignani}]{pastor2015RMP}
\bibinfo{author}{R.~Pastor-Satorras}, \bibinfo{author}{C.~Castellano},
  \bibinfo{author}{P.~Van~Mieghem}, \bibinfo{author}{A.~Vespignani},
\newblock \bibinfo{title}{Epidemic processes in complex networks},
\newblock \bibinfo{journal}{Rev. Mod. Phys.} \bibinfo{volume}{87}
  (\bibinfo{year}{2015}) \bibinfo{pages}{925}.
\bibitem[{Pastor-Satorras and Vespignani(2001)}]{pastor2001PRL}
\bibinfo{author}{R.~Pastor-Satorras}, \bibinfo{author}{A.~Vespignani},
\newblock \bibinfo{title}{Epidemic spreading in scale-free networks},
\newblock \bibinfo{journal}{Phys. Rev. Lett.} \bibinfo{volume}{86}
  (\bibinfo{year}{2001}) \bibinfo{pages}{3200}.
\bibitem[{Castellano and Pastor-Satorras(2010)}]{castellano2010PRL}
\bibinfo{author}{C.~Castellano}, \bibinfo{author}{R.~Pastor-Satorras},
\newblock \bibinfo{title}{Thresholds for epidemic spreading in networks},
\newblock \bibinfo{journal}{Phys. Rev. Lett.} \bibinfo{volume}{105}
  (\bibinfo{year}{2010}) \bibinfo{pages}{218701}.
\bibitem[{Cohen et~al.(2003)Cohen, Havlin, and Ben-Avraham}]{cohen2003PRL}
\bibinfo{author}{R.~Cohen}, \bibinfo{author}{S.~Havlin},
  \bibinfo{author}{D.~Ben-Avraham},
\newblock \bibinfo{title}{Efficient immunization strategies for computer
  networks and populations},
\newblock \bibinfo{journal}{Phys. Rev. Lett.} \bibinfo{volume}{91}
  (\bibinfo{year}{2003}) \bibinfo{pages}{247901}.
\bibitem[{Colizza and Vespignani(2007)}]{colizza2007PRL}
\bibinfo{author}{V.~Colizza}, \bibinfo{author}{A.~Vespignani},
\newblock \bibinfo{title}{Invasion threshold in heterogeneous metapopulation
  networks},
\newblock \bibinfo{journal}{Phys. Rev. Lett.} \bibinfo{volume}{99}
  (\bibinfo{year}{2007}) \bibinfo{pages}{148701}.
\bibitem[{Colizza and Vespignani(2008)}]{colizza2008JTB}
\bibinfo{author}{V.~Colizza}, \bibinfo{author}{A.~Vespignani},
\newblock \bibinfo{title}{Epidemic modeling in metapopulation systems with
  heterogeneous coupling pattern: {T}heory and simulations},
\newblock \bibinfo{journal}{J. Theor. Biol.} \bibinfo{volume}{251}
  (\bibinfo{year}{2008}) \bibinfo{pages}{450--467}.
\bibitem[{Balcan and Vespignani(2011)}]{balcan2011NP}
\bibinfo{author}{D.~Balcan}, \bibinfo{author}{A.~Vespignani},
\newblock \bibinfo{title}{Phase transitions in contagion processes mediated by
  recurrent mobility patterns},
\newblock \bibinfo{journal}{Nat. Phys.} \bibinfo{volume}{7}
  (\bibinfo{year}{2011}) \bibinfo{pages}{581--586}.
\bibitem[{Balcan and Vespignani(2012)}]{balcan2012JTB}
\bibinfo{author}{D.~Balcan}, \bibinfo{author}{A.~Vespignani},
\newblock \bibinfo{title}{Invasion threshold in structured populations with
  recurrent mobility patterns},
\newblock \bibinfo{journal}{J. Theor. Biol.} \bibinfo{volume}{293}
  (\bibinfo{year}{2012}) \bibinfo{pages}{87--100}.
\bibitem[{Fraser et~al.(2009)Fraser, Donnelly, Cauchemez, Hanage, Van~Kerkhove,
  Hollingsworth, Griffin, Baggaley, Jenkins, Lyons et~al.}]{fraser2009science}
\bibinfo{author}{C.~Fraser}, \bibinfo{author}{C.~A. Donnelly},
  \bibinfo{author}{S.~Cauchemez}, \bibinfo{author}{W.~P. Hanage},
  \bibinfo{author}{M.~D. Van~Kerkhove}, \bibinfo{author}{T.~D. Hollingsworth},
  \bibinfo{author}{J.~Griffin}, \bibinfo{author}{R.~F. Baggaley},
  \bibinfo{author}{H.~E. Jenkins}, \bibinfo{author}{E.~J. Lyons}, et~al.,
\newblock \bibinfo{title}{Pandemic potential of a strain of influenza
  {A(H1N1)}: early findings},
\newblock \bibinfo{journal}{Science} \bibinfo{volume}{324}
  (\bibinfo{year}{2009}) \bibinfo{pages}{1557--1561}.
\bibitem[{Du et~al.(2020)Du, Wang, Cauchemez, Xu, Wang, Cowling, and
  Meyers}]{du2020EIDrisk}
\bibinfo{author}{Z.~Du}, \bibinfo{author}{L.~Wang},
  \bibinfo{author}{S.~Cauchemez}, \bibinfo{author}{X.~Xu},
  \bibinfo{author}{X.~Wang}, \bibinfo{author}{B.~J. Cowling},
  \bibinfo{author}{L.~A. Meyers},
\newblock \bibinfo{title}{Risk for transportation of coronavirus disease from
  {W}uhan to other cities in {C}hina},
\newblock \bibinfo{journal}{Emerg. Infect. Dis.} \bibinfo{volume}{26}
  (\bibinfo{year}{2020}) \bibinfo{pages}{1049--1052}.
\bibitem[{Wu et~al.(2020)Wu, Leung, and Leung}]{wu2020lancet}
\bibinfo{author}{J.~T. Wu}, \bibinfo{author}{K.~Leung}, \bibinfo{author}{G.~M.
  Leung},
\newblock \bibinfo{title}{Nowcasting and forecasting the potential domestic and
  international spread of the {2019-nCoV} outbreak originating in {W}uhan,
  {C}hina: a modelling study},
\newblock \bibinfo{journal}{Lancet} \bibinfo{volume}{395}
  (\bibinfo{year}{2020}) \bibinfo{pages}{689--697}.
\bibitem[{Gautreau et~al.(2008)Gautreau, Barrat, and
  Barthelemy}]{gautreau2008global}
\bibinfo{author}{A.~Gautreau}, \bibinfo{author}{A.~Barrat},
  \bibinfo{author}{M.~Barthelemy},
\newblock \bibinfo{title}{Global disease spread: statistics and estimation of
  arrival times},
\newblock \bibinfo{journal}{J. Theor. Biol.} \bibinfo{volume}{251}
  (\bibinfo{year}{2008}) \bibinfo{pages}{509--522}.
\bibitem[{Tomba and Wallinga(2008)}]{tomba2008simple}
\bibinfo{author}{G.~S. Tomba}, \bibinfo{author}{J.~Wallinga},
\newblock \bibinfo{title}{A simple explanation for the low impact of border
  control as a countermeasure to the spread of an infectious disease},
\newblock \bibinfo{journal}{Math. Biosci.} \bibinfo{volume}{214}
  (\bibinfo{year}{2008}) \bibinfo{pages}{70--72}.
\bibitem[{Brockmann and Helbing(2013)}]{brockmann2013science}
\bibinfo{author}{D.~Brockmann}, \bibinfo{author}{D.~Helbing},
\newblock \bibinfo{title}{The hidden geometry of complex, network-driven
  contagion phenomena},
\newblock \bibinfo{journal}{Science} \bibinfo{volume}{342}
  (\bibinfo{year}{2013}) \bibinfo{pages}{1337--1342}.
\bibitem[{Ross(1996)}]{ross1996stochastic}
\bibinfo{author}{S.~M. Ross}, \bibinfo{title}{Stochastic processes},
  \bibinfo{publisher}{John Wiley \& Sons}, \bibinfo{year}{1996}.
\bibitem[{Grubaugh et~al.(2017)Grubaugh, Ladner, Kraemer, Dudas, Tan,
  Gangavarapu, Wiley, White, Th{\'e}z{\'e}, Magnani
  et~al.}]{grubaugh2017nature}
\bibinfo{author}{N.~D. Grubaugh}, \bibinfo{author}{J.~T. Ladner},
  \bibinfo{author}{M.~U. Kraemer}, \bibinfo{author}{G.~Dudas},
  \bibinfo{author}{A.~L. Tan}, \bibinfo{author}{K.~Gangavarapu},
  \bibinfo{author}{M.~R. Wiley}, \bibinfo{author}{S.~White},
  \bibinfo{author}{J.~Th{\'e}z{\'e}}, \bibinfo{author}{D.~M. Magnani}, et~al.,
\newblock \bibinfo{title}{Genomic epidemiology reveals multiple introductions
  of {Z}ika virus into the {U}nited {S}tates},
\newblock \bibinfo{journal}{Nature} \bibinfo{volume}{546}
  (\bibinfo{year}{2017}) \bibinfo{pages}{401--405}.
\bibitem[{da~Silva~Filipe et~al.(2020)da~Silva~Filipe, Shepherd, Williams,
  Hughes, Aranday-Cortes, Asamaphan, Ashraf, Balcazar, Brunker, Campbell
  et~al.}]{filipe2020natmicrob}
\bibinfo{author}{A.~da~Silva~Filipe}, \bibinfo{author}{J.~G. Shepherd},
  \bibinfo{author}{T.~Williams}, \bibinfo{author}{J.~Hughes},
  \bibinfo{author}{E.~Aranday-Cortes}, \bibinfo{author}{P.~Asamaphan},
  \bibinfo{author}{S.~Ashraf}, \bibinfo{author}{C.~Balcazar},
  \bibinfo{author}{K.~Brunker}, \bibinfo{author}{A.~Campbell}, et~al.,
\newblock \bibinfo{title}{Genomic epidemiology reveals multiple introductions
  of {SARS-CoV-2} from mainland {E}urope into {S}cotland},
\newblock \bibinfo{journal}{Nat. Microbiol.} \bibinfo{volume}{6}
  (\bibinfo{year}{2020}) \bibinfo{pages}{112--122}.
\bibitem[{Cover and Thomas(2006)}]{cover2006elements}
\bibinfo{author}{T.~M. Cover}, \bibinfo{author}{J.~A. Thomas},
  \bibinfo{title}{Elements of information theory, 2nd edition},
  \bibinfo{publisher}{John Wiley \& Sons}, \bibinfo{year}{2006}.
\bibitem[{van~de Schoot et~al.(2021)van~de Schoot, Depaoli, King, Kramer,
  M{\"a}rtens, Tadesse, Vannucci, Gelman, Veen, Willemsen
  et~al.}]{vandeschoot2021bayesian}
\bibinfo{author}{R.~van~de Schoot}, \bibinfo{author}{S.~Depaoli},
  \bibinfo{author}{R.~King}, \bibinfo{author}{B.~Kramer},
  \bibinfo{author}{K.~M{\"a}rtens}, \bibinfo{author}{M.~G. Tadesse},
  \bibinfo{author}{M.~Vannucci}, \bibinfo{author}{A.~Gelman},
  \bibinfo{author}{D.~Veen}, \bibinfo{author}{J.~Willemsen}, et~al.,
\newblock \bibinfo{title}{Bayesian statistics and modelling},
\newblock \bibinfo{journal}{Nat. Rev. Methods Primers} \bibinfo{volume}{1}
  (\bibinfo{year}{2021}) \bibinfo{pages}{1}.
\bibitem[{Lambert(2018)}]{lambert2018student}
\bibinfo{author}{B.~Lambert}, \bibinfo{title}{A student’s guide to {B}ayesian
  statistics}, \bibinfo{publisher}{Sage}, \bibinfo{year}{2018}.
\bibitem[{Gelman et~al.(2017)Gelman, Simpson, and Betancourt}]{gelman2017prior}
\bibinfo{author}{A.~Gelman}, \bibinfo{author}{D.~Simpson},
  \bibinfo{author}{M.~Betancourt},
\newblock \bibinfo{title}{The prior can often only be understood in the context
  of the likelihood},
\newblock \bibinfo{journal}{Entropy} \bibinfo{volume}{19}
  (\bibinfo{year}{2017}) \bibinfo{pages}{555}.
\bibitem[{Lemoine(2019)}]{lemoine2019moving}
\bibinfo{author}{N.~P. Lemoine},
\newblock \bibinfo{title}{Moving beyond noninformative priors: why and how to
  choose weakly informative priors in {B}ayesian analyses},
\newblock \bibinfo{journal}{Oikos} \bibinfo{volume}{128} (\bibinfo{year}{2019})
  \bibinfo{pages}{912--928}.
\bibitem[{Balcan et~al.(2009)Balcan, Hu, Goncalves, Bajardi, Poletto, Ramasco,
  Paolotti, Perra, Tizzoni, Van~den Broeck et~al.}]{balcan2009seasonal}
\bibinfo{author}{D.~Balcan}, \bibinfo{author}{H.~Hu},
  \bibinfo{author}{B.~Goncalves}, \bibinfo{author}{P.~Bajardi},
  \bibinfo{author}{C.~Poletto}, \bibinfo{author}{J.~J. Ramasco},
  \bibinfo{author}{D.~Paolotti}, \bibinfo{author}{N.~Perra},
  \bibinfo{author}{M.~Tizzoni}, \bibinfo{author}{W.~Van~den Broeck}, et~al.,
\newblock \bibinfo{title}{Seasonal transmission potential and activity peaks of
  the new influenza {A(H1N1)}: a {M}onte {C}arlo likelihood analysis based on
  human mobility},
\newblock \bibinfo{journal}{BMC Med.} \bibinfo{volume}{7}
  (\bibinfo{year}{2009}) \bibinfo{pages}{45}.
\bibitem[{Li et~al.(2020)Li, Guan, Wu, Wang, Zhou, Tong, Ren, Leung, Lau, Wong
  et~al.}]{cowling2020NEJM}
\bibinfo{author}{Q.~Li}, \bibinfo{author}{X.~Guan}, \bibinfo{author}{P.~Wu},
  \bibinfo{author}{X.~Wang}, \bibinfo{author}{L.~Zhou},
  \bibinfo{author}{Y.~Tong}, \bibinfo{author}{R.~Ren}, \bibinfo{author}{K.~S.
  Leung}, \bibinfo{author}{E.~H. Lau}, \bibinfo{author}{J.~Y. Wong}, et~al.,
\newblock \bibinfo{title}{Early transmission dynamics in {W}uhan, {C}hina, of
  novel coronavirus---infected pneumonia},
\newblock \bibinfo{journal}{N. Engl. J. Med.} \bibinfo{volume}{382}
  (\bibinfo{year}{2020}) \bibinfo{pages}{1199--1207}.
\bibitem[{Imai et~al.(2020)Imai, Dorigatti, Cori, Donnelly, Riley, and
  Ferguson}]{imai2020Report2}
\bibinfo{author}{N.~Imai}, \bibinfo{author}{I.~Dorigatti},
  \bibinfo{author}{A.~Cori}, \bibinfo{author}{C.~Donnelly},
  \bibinfo{author}{S.~Riley}, \bibinfo{author}{N.~M. Ferguson},
  \bibinfo{title}{Estimating the potential total number of novel {C}oronavirus
  cases in {W}uhan {C}ity, {C}hina}, \bibinfo{year}{2020}.
  \bibinfo{note}{Report by Imperial College London. Available at:
  \url{https://spiral.imperial.ac.uk/handle/10044/1/77150}. Archived at:
  \url{https://doi.org/10.17605/OSF.IO/N7G9X}}.
\bibitem[{Li et~al.(2020)Li, Pei, Chen, Song, Zhang, Yang, and
  Shaman}]{li2020science}
\bibinfo{author}{R.~Li}, \bibinfo{author}{S.~Pei}, \bibinfo{author}{B.~Chen},
  \bibinfo{author}{Y.~Song}, \bibinfo{author}{T.~Zhang},
  \bibinfo{author}{W.~Yang}, \bibinfo{author}{J.~Shaman},
\newblock \bibinfo{title}{Substantial undocumented infection facilitates the
  rapid dissemination of novel coronavirus ({SARS-CoV-2})},
\newblock \bibinfo{journal}{Science} \bibinfo{volume}{368}
  (\bibinfo{year}{2020}) \bibinfo{pages}{489--493}.
\bibitem[{Sanche et~al.(2020)Sanche, Lin, Xu, Romero-Severson, Hengartner, and
  Ke}]{sanche2020high}
\bibinfo{author}{S.~Sanche}, \bibinfo{author}{Y.~T. Lin},
  \bibinfo{author}{C.~Xu}, \bibinfo{author}{E.~Romero-Severson},
  \bibinfo{author}{N.~Hengartner}, \bibinfo{author}{R.~Ke},
\newblock \bibinfo{title}{High contagiousness and rapid spread of severe acute
  respiratory syndrome coronavirus 2},
\newblock \bibinfo{journal}{Emerg. Infect. Dis.} \bibinfo{volume}{26}
  (\bibinfo{year}{2020}) \bibinfo{pages}{1470--1477}.
\bibitem[{Luo et~al.(2018)Luo, Gao, and Cassels}]{luo2018large}
\bibinfo{author}{W.~Luo}, \bibinfo{author}{P.~Gao},
  \bibinfo{author}{S.~Cassels},
\newblock \bibinfo{title}{A large-scale location-based social network to
  understanding the impact of human geo-social interaction patterns on
  vaccination strategies in an urbanized area},
\newblock \bibinfo{journal}{Comput. Environ. Urban Syst.} \bibinfo{volume}{72}
  (\bibinfo{year}{2018}) \bibinfo{pages}{78--87}.
\bibitem[{Yin et~al.(2021)Yin, Zhang, Li, Liu, Chen, Luo, Lai, Li, Tang, Ning
  et~al.}]{yin2021data}
\bibinfo{author}{L.~Yin}, \bibinfo{author}{H.~Zhang}, \bibinfo{author}{Y.~Li},
  \bibinfo{author}{K.~Liu}, \bibinfo{author}{T.~Chen},
  \bibinfo{author}{W.~Luo}, \bibinfo{author}{S.~Lai}, \bibinfo{author}{Y.~Li},
  \bibinfo{author}{X.~Tang}, \bibinfo{author}{L.~Ning}, et~al.,
\newblock \bibinfo{title}{A data driven agent-based model that recommends
  non-pharmaceutical interventions to suppress {C}oronavirus disease 2019
  resurgence in megacities},
\newblock \bibinfo{journal}{J. R. Soc. Interface} \bibinfo{volume}{18}
  (\bibinfo{year}{2021}) \bibinfo{pages}{20210112}.
\bibitem[{Chang et~al.(2021)Chang, Pierson, Koh, Gerardin, Redbird, Grusky, and
  Leskovec}]{chang2021nature}
\bibinfo{author}{S.~Chang}, \bibinfo{author}{E.~Pierson},
  \bibinfo{author}{P.~W. Koh}, \bibinfo{author}{J.~Gerardin},
  \bibinfo{author}{B.~Redbird}, \bibinfo{author}{D.~Grusky},
  \bibinfo{author}{J.~Leskovec},
\newblock \bibinfo{title}{Mobility network models of {COVID-19} explain
  inequities and inform reopening},
\newblock \bibinfo{journal}{Nature} \bibinfo{volume}{589}
  (\bibinfo{year}{2021}) \bibinfo{pages}{82--87}.
\bibitem[{Mossong et~al.(2008)Mossong, Hens, Jit, Beutels, Auranen,
  Mikolajczyk, Massari, Salmaso, Tomba, Wallinga et~al.}]{mossong2008social}
\bibinfo{author}{J.~Mossong}, \bibinfo{author}{N.~Hens},
  \bibinfo{author}{M.~Jit}, \bibinfo{author}{P.~Beutels},
  \bibinfo{author}{K.~Auranen}, \bibinfo{author}{R.~Mikolajczyk},
  \bibinfo{author}{M.~Massari}, \bibinfo{author}{S.~Salmaso},
  \bibinfo{author}{G.~S. Tomba}, \bibinfo{author}{J.~Wallinga}, et~al.,
\newblock \bibinfo{title}{Social contacts and mixing patterns relevant to the
  spread of infectious diseases},
\newblock \bibinfo{journal}{PLOS Med.} \bibinfo{volume}{5}
  (\bibinfo{year}{2008}) \bibinfo{pages}{e74}.
\bibitem[{Prem et~al.(2017)Prem, Cook, and Jit}]{prem2017projecting}
\bibinfo{author}{K.~Prem}, \bibinfo{author}{A.~R. Cook},
  \bibinfo{author}{M.~Jit},
\newblock \bibinfo{title}{Projecting social contact matrices in 152 countries
  using contact surveys and demographic data},
\newblock \bibinfo{journal}{PLOS Comput. Biol.} \bibinfo{volume}{13}
  (\bibinfo{year}{2017}) \bibinfo{pages}{e1005697}.
\bibitem[{Mistry et~al.(2021)Mistry, Litvinova, y~Piontti, Chinazzi, Fumanelli,
  Gomes, Haque, Liu, Mu, Xiong et~al.}]{mistry2021inferring}
\bibinfo{author}{D.~Mistry}, \bibinfo{author}{M.~Litvinova},
  \bibinfo{author}{A.~P. y~Piontti}, \bibinfo{author}{M.~Chinazzi},
  \bibinfo{author}{L.~Fumanelli}, \bibinfo{author}{M.~F. Gomes},
  \bibinfo{author}{S.~A. Haque}, \bibinfo{author}{Q.-H. Liu},
  \bibinfo{author}{K.~Mu}, \bibinfo{author}{X.~Xiong}, et~al.,
\newblock \bibinfo{title}{Inferring high-resolution human mixing patterns for
  disease modeling},
\newblock \bibinfo{journal}{Nat. Commun.} \bibinfo{volume}{12}
  (\bibinfo{year}{2021}) \bibinfo{pages}{323}.
\bibitem[{O'Driscoll et~al.(2021)O'Driscoll, Dos~Santos, Wang, Cummings, Azman,
  Paireau, Fontanet, Cauchemez, and Salje}]{odriscoll2021age}
\bibinfo{author}{M.~O'Driscoll}, \bibinfo{author}{G.~R. Dos~Santos},
  \bibinfo{author}{L.~Wang}, \bibinfo{author}{D.~A. Cummings},
  \bibinfo{author}{A.~S. Azman}, \bibinfo{author}{J.~Paireau},
  \bibinfo{author}{A.~Fontanet}, \bibinfo{author}{S.~Cauchemez},
  \bibinfo{author}{H.~Salje},
\newblock \bibinfo{title}{Age-specific mortality and immunity patterns of
  {SARS-CoV-2}},
\newblock \bibinfo{journal}{Nature} \bibinfo{volume}{590}
  (\bibinfo{year}{2021}) \bibinfo{pages}{140--145}.
\bibitem[{Adam et~al.(2020)Adam, Wu, Wong, Lau, Tsang, Cauchemez, Leung, and
  Cowling}]{adam2020clustering}
\bibinfo{author}{D.~C. Adam}, \bibinfo{author}{P.~Wu}, \bibinfo{author}{J.~Y.
  Wong}, \bibinfo{author}{E.~H. Lau}, \bibinfo{author}{T.~K. Tsang},
  \bibinfo{author}{S.~Cauchemez}, \bibinfo{author}{G.~M. Leung},
  \bibinfo{author}{B.~J. Cowling},
\newblock \bibinfo{title}{Clustering and superspreading potential of
  {SARS-CoV-2} infections in {H}ong {K}ong},
\newblock \bibinfo{journal}{Nat. Med.} \bibinfo{volume}{26}
  (\bibinfo{year}{2020}) \bibinfo{pages}{1714--1719}.
\bibitem[{Wong and Collins(2020)}]{wong2020evidence}
\bibinfo{author}{F.~Wong}, \bibinfo{author}{J.~J. Collins},
\newblock \bibinfo{title}{Evidence that coronavirus superspreading is
  fat-tailed},
\newblock \bibinfo{journal}{Proc. Natl. Acad. Sci. USA} \bibinfo{volume}{117}
  (\bibinfo{year}{2020}) \bibinfo{pages}{29416--29418}.
\bibitem[{Lloyd-Smith et~al.(2005)Lloyd-Smith, Schreiber, Kopp, and
  Getz}]{lloyd2005superspreading}
\bibinfo{author}{J.~O. Lloyd-Smith}, \bibinfo{author}{S.~J. Schreiber},
  \bibinfo{author}{P.~E. Kopp}, \bibinfo{author}{W.~M. Getz},
\newblock \bibinfo{title}{Superspreading and the effect of individual variation
  on disease emergence},
\newblock \bibinfo{journal}{Nature} \bibinfo{volume}{438}
  (\bibinfo{year}{2005}) \bibinfo{pages}{355--359}.
\bibitem[{Perra(2021)}]{perra2021non}
\bibinfo{author}{N.~Perra},
\newblock \bibinfo{title}{Non-pharmaceutical interventions during the
  {COVID-19} pandemic: {A} review},
\newblock \bibinfo{journal}{Phys. Rep.} \bibinfo{volume}{913}
  (\bibinfo{year}{2021}) \bibinfo{pages}{1--52}.
\bibitem[{Zhang et~al.(2020)Zhang, Litvinova, Liang, Wang, Wang, Zhao, Wu,
  Merler, Viboud, Vespignani et~al.}]{zhang2020changes}
\bibinfo{author}{J.~Zhang}, \bibinfo{author}{M.~Litvinova},
  \bibinfo{author}{Y.~Liang}, \bibinfo{author}{Y.~Wang},
  \bibinfo{author}{W.~Wang}, \bibinfo{author}{S.~Zhao},
  \bibinfo{author}{Q.~Wu}, \bibinfo{author}{S.~Merler},
  \bibinfo{author}{C.~Viboud}, \bibinfo{author}{A.~Vespignani}, et~al.,
\newblock \bibinfo{title}{Changes in contact patterns shape the dynamics of the
  {COVID-19} outbreak in {C}hina},
\newblock \bibinfo{journal}{Science} \bibinfo{volume}{368}
  (\bibinfo{year}{2020}) \bibinfo{pages}{1481--1486}.
\bibitem[{Ali et~al.(2020)Ali, Wang, Lau, Xu, Du, Wu, Leung, and
  Cowling}]{ali2020serial}
\bibinfo{author}{S.~T. Ali}, \bibinfo{author}{L.~Wang}, \bibinfo{author}{E.~H.
  Lau}, \bibinfo{author}{X.-K. Xu}, \bibinfo{author}{Z.~Du},
  \bibinfo{author}{Y.~Wu}, \bibinfo{author}{G.~M. Leung},
  \bibinfo{author}{B.~J. Cowling},
\newblock \bibinfo{title}{Serial interval of {SARS-CoV-2} was shortened over
  time by nonpharmaceutical interventions},
\newblock \bibinfo{journal}{Science} \bibinfo{volume}{369}
  (\bibinfo{year}{2020}) \bibinfo{pages}{1106--1109}.
\bibitem[{Paolotti et~al.(2014)Paolotti, Carnahan, Colizza, Eames, Edmunds,
  Gomes, Koppeschaar, Rehn, Smallenburg, Turbelin et~al.}]{paolotti2014web}
\bibinfo{author}{D.~Paolotti}, \bibinfo{author}{A.~Carnahan},
  \bibinfo{author}{V.~Colizza}, \bibinfo{author}{K.~Eames},
  \bibinfo{author}{J.~Edmunds}, \bibinfo{author}{G.~Gomes},
  \bibinfo{author}{C.~Koppeschaar}, \bibinfo{author}{M.~Rehn},
  \bibinfo{author}{R.~Smallenburg}, \bibinfo{author}{C.~Turbelin}, et~al.,
\newblock \bibinfo{title}{Web-based participatory surveillance of infectious
  diseases: the {I}nfluenzanet participatory surveillance experience},
\newblock \bibinfo{journal}{Clin. Microbiol. Infect.} \bibinfo{volume}{20}
  (\bibinfo{year}{2014}) \bibinfo{pages}{17--21}.
\bibitem[{Thomas and Cook(2006)}]{thomas2006visual}
\bibinfo{author}{J.~J. Thomas}, \bibinfo{author}{K.~A. Cook},
\newblock \bibinfo{title}{A visual analytics agenda},
\newblock \bibinfo{journal}{IEEE Comput. Graph. Appl.} \bibinfo{volume}{26}
  (\bibinfo{year}{2006}) \bibinfo{pages}{10--13}.
\bibitem[{Luo(2016)}]{luo2016visual}
\bibinfo{author}{W.~Luo},
\newblock \bibinfo{title}{Visual analytics of geo-social interaction patterns
  for epidemic control},
\newblock \bibinfo{journal}{Int. J. Health Geogr.} \bibinfo{volume}{15}
  (\bibinfo{year}{2016}) \bibinfo{pages}{28}.
\bibitem[{Waters et~al.(2016)Waters, Zalasiewicz, Summerhayes, Barnosky,
  Poirier, Ga{\l}uszka, Cearreta, Edgeworth, Ellis, Ellis
  et~al.}]{waters2016anthropocene}
\bibinfo{author}{C.~N. Waters}, \bibinfo{author}{J.~Zalasiewicz},
  \bibinfo{author}{C.~Summerhayes}, \bibinfo{author}{A.~D. Barnosky},
  \bibinfo{author}{C.~Poirier}, \bibinfo{author}{A.~Ga{\l}uszka},
  \bibinfo{author}{A.~Cearreta}, \bibinfo{author}{M.~Edgeworth},
  \bibinfo{author}{E.~C. Ellis}, \bibinfo{author}{M.~Ellis}, et~al.,
\newblock \bibinfo{title}{The {A}nthropocene is functionally and
  stratigraphically distinct from the {H}olocene},
\newblock \bibinfo{journal}{Science} \bibinfo{volume}{351}
  (\bibinfo{year}{2016}) \bibinfo{pages}{aad2622}.
\bibitem[{Pellow(1999)}]{pellow1999framing}
\bibinfo{author}{D.~N. Pellow},
\newblock \bibinfo{title}{Framing emerging environmental movement tactics:
  mobilizing consensus, demobilizing conflict},
\newblock \bibinfo{journal}{Sociol. Forum} \bibinfo{volume}{14}
  (\bibinfo{year}{1999}) \bibinfo{pages}{659--683}.
\bibitem[{Molina and Rowland(1974)}]{molina1974stratospheric}
\bibinfo{author}{M.~J. Molina}, \bibinfo{author}{F.~S. Rowland},
\newblock \bibinfo{title}{Stratospheric sink for chlorofluoromethanes: chlorine
  atom-catalysed destruction of ozone},
\newblock \bibinfo{journal}{Nature} \bibinfo{volume}{249}
  (\bibinfo{year}{1974}) \bibinfo{pages}{810--812}.
\bibitem[{Farman et~al.(1985)Farman, Gardiner, and Shanklin}]{farman1985large}
\bibinfo{author}{J.~C. Farman}, \bibinfo{author}{B.~G. Gardiner},
  \bibinfo{author}{J.~D. Shanklin},
\newblock \bibinfo{title}{Large losses of total ozone in {A}ntarctica reveal
  seasonal {ClO}$_x$/{NO}$_x$ interaction},
\newblock \bibinfo{journal}{Nature} \bibinfo{volume}{315}
  (\bibinfo{year}{1985}) \bibinfo{pages}{207--210}.
\bibitem[{{Ozone Secretariat UNEP}(2006)}]{united2006handbook}
\bibinfo{author}{{Ozone Secretariat UNEP}}, \bibinfo{title}{Handbook for the
  {M}ontreal protocol on substances that deplete the ozone layer, 7th edition},
  \bibinfo{publisher}{United Nations Environment Programme},
  \bibinfo{year}{2006}.
\bibitem[{{World Meteorological Organization (WMO)}(2018)}]{wmo2018scientific}
\bibinfo{author}{{World Meteorological Organization (WMO)}},
  \bibinfo{title}{Scientific {A}ssessment of {O}zone {D}epletion: 2018},
  \bibinfo{year}{2018}. \bibinfo{note}{Global Ozone Research and Monitoring
  Project---Report No.~58. Available at:
  \url{https://ozone.unep.org/science/assessment/sap}. Archived at:
  \url{https://doi.org/10.17605/OSF.IO/N7G9X}}.
\bibitem[{Hoffman(1927)}]{hoffman1927deaths}
\bibinfo{author}{F.~L. Hoffman},
\newblock \bibinfo{title}{Deaths from lead poisoning},
\newblock \bibinfo{journal}{Bulletin of the United States Bureau of Labor
  Statistics} \bibinfo{volume}{426} (\bibinfo{year}{1927})
  \bibinfo{pages}{1--45}.
\bibitem[{Patterson(1965)}]{patterson1965contaminated}
\bibinfo{author}{C.~C. Patterson},
\newblock \bibinfo{title}{Contaminated and natural lead environments of man},
\newblock \bibinfo{journal}{Arch. Environ. Health} \bibinfo{volume}{11}
  (\bibinfo{year}{1965}) \bibinfo{pages}{344--360}.
\bibitem[{Yoshinaga(2012)}]{yoshinaga2012lead}
\bibinfo{author}{J.~Yoshinaga},
\newblock \bibinfo{title}{Lead in the {J}apanese living environment},
\newblock \bibinfo{journal}{Environ. Health Prev. Med.} \bibinfo{volume}{17}
  (\bibinfo{year}{2012}) \bibinfo{pages}{433--443}.
\bibitem[{Deryabina et~al.(2015)Deryabina, Kuchmel, Nagorskaya, Hinton,
  Beasley, Lerebours, and Smith}]{deryabina2015long}
\bibinfo{author}{T.~G. Deryabina}, \bibinfo{author}{S.~V. Kuchmel},
  \bibinfo{author}{L.~L. Nagorskaya}, \bibinfo{author}{T.~G. Hinton},
  \bibinfo{author}{J.~C. Beasley}, \bibinfo{author}{A.~Lerebours},
  \bibinfo{author}{J.~T. Smith},
\newblock \bibinfo{title}{Long-term census data reveal abundant wildlife
  populations at {C}hernobyl},
\newblock \bibinfo{journal}{Curr. Biol.} \bibinfo{volume}{25}
  (\bibinfo{year}{2015}) \bibinfo{pages}{R824--R826}.
\bibitem[{De~Bruijn et~al.(2003)De~Bruijn, Hansen, Johansson, Luotamo, Munn,
  Musset, Olsen, Olsson, Paya-Perez, Pedersen, Rasmussen, and
  Sokull-Kluttgen}]{debruijn2003technical}
\bibinfo{author}{J.~De~Bruijn}, \bibinfo{author}{B.~G. Hansen},
  \bibinfo{author}{S.~Johansson}, \bibinfo{author}{M.~Luotamo},
  \bibinfo{author}{S.~J. Munn}, \bibinfo{author}{C.~Musset},
  \bibinfo{author}{S.~I. Olsen}, \bibinfo{author}{H.~Olsson},
  \bibinfo{author}{A.~B. Paya-Perez}, \bibinfo{author}{F.~Pedersen},
  \bibinfo{author}{K.~Rasmussen}, \bibinfo{author}{B.~Sokull-Kluttgen},
  \bibinfo{title}{Technical guidance document on risk assessment},
  \bibinfo{howpublished}{European Communities}, \bibinfo{year}{2003}.
  \bibinfo{note}{Available at:
  \url{https://echa.europa.eu/documents/10162/16960216/tgdpart2\_2ed\_en.pdf}.
  Archived at: \url{https://doi.org/10.17605/OSF.IO/N7G9X}}.
\bibitem[{Parasuraman(2011)}]{parasuraman2011toxicological}
\bibinfo{author}{S.~Parasuraman},
\newblock \bibinfo{title}{Toxicological screening},
\newblock \bibinfo{journal}{J. Pharmacol. Pharmacother.} \bibinfo{volume}{2}
  (\bibinfo{year}{2011}) \bibinfo{pages}{74--79}.
\bibitem[{Witeska and Jezierska(2003)}]{witeska2003effects}
\bibinfo{author}{M.~Witeska}, \bibinfo{author}{B.~Jezierska},
\newblock \bibinfo{title}{The effects of environmental factors on metal
  toxicity to fish (review)},
\newblock \bibinfo{journal}{Fresenius Environ. Bull.} \bibinfo{volume}{12}
  (\bibinfo{year}{2003}) \bibinfo{pages}{824--829}.
\bibitem[{Wang(1987)}]{wang1987factors}
\bibinfo{author}{W.~Wang},
\newblock \bibinfo{title}{Factors affecting metal toxicity to (and accumulation
  by) aquatic organisms---{O}verview},
\newblock \bibinfo{journal}{Environ. Int.} \bibinfo{volume}{13}
  (\bibinfo{year}{1987}) \bibinfo{pages}{437--457}.
\bibitem[{Norwood et~al.(2003)Norwood, Borgmann, Dixon, and
  Wallace}]{norwood2003effects}
\bibinfo{author}{W.~P. Norwood}, \bibinfo{author}{U.~Borgmann},
  \bibinfo{author}{D.~G. Dixon}, \bibinfo{author}{A.~Wallace},
\newblock \bibinfo{title}{Effects of metal mixtures on aquatic biota: a review
  of observations and methods},
\newblock \bibinfo{journal}{Hum. Ecol. Risk Assess.} \bibinfo{volume}{9}
  (\bibinfo{year}{2003}) \bibinfo{pages}{795--811}.
\bibitem[{Llanos et~al.(2019)Llanos, Leal, Luu, Jost, Stadler, and
  Restrepo}]{llanos2019exploration}
\bibinfo{author}{E.~J. Llanos}, \bibinfo{author}{W.~Leal},
  \bibinfo{author}{D.~H. Luu}, \bibinfo{author}{J.~Jost},
  \bibinfo{author}{P.~F. Stadler}, \bibinfo{author}{G.~Restrepo},
\newblock \bibinfo{title}{Exploration of the chemical space and its three
  historical regimes},
\newblock \bibinfo{journal}{Proc. Natl. Acad. Sci. USA} \bibinfo{volume}{116}
  (\bibinfo{year}{2019}) \bibinfo{pages}{12660--12665}.
\bibitem[{Galloway and Lewis(2016)}]{galloway2016marine}
\bibinfo{author}{T.~S. Galloway}, \bibinfo{author}{C.~N. Lewis},
\newblock \bibinfo{title}{Marine microplastics spell big problems for future
  generations},
\newblock \bibinfo{journal}{Proc. Natl. Acad. Sci. USA} \bibinfo{volume}{113}
  (\bibinfo{year}{2016}) \bibinfo{pages}{2331--2333}.
\bibitem[{Vanbergen and {the Insect Pollinators
  Initiative}(2013)}]{vanbergen2013threats}
\bibinfo{author}{A.~J. Vanbergen}, \bibinfo{author}{{the Insect Pollinators
  Initiative}},
\newblock \bibinfo{title}{Threats to an ecosystem service: pressures on
  pollinators},
\newblock \bibinfo{journal}{Front. Ecol. Environ.} \bibinfo{volume}{11}
  (\bibinfo{year}{2013}) \bibinfo{pages}{251--259}.
\bibitem[{Barrick et~al.(2017)Barrick, Ch{\^a}tel, Bruneau, and
  Mouneyrac}]{barrick2017role}
\bibinfo{author}{A.~Barrick}, \bibinfo{author}{A.~Ch{\^a}tel},
  \bibinfo{author}{M.~Bruneau}, \bibinfo{author}{C.~Mouneyrac},
\newblock \bibinfo{title}{The role of high-throughput screening in
  ecotoxicology and engineered nanomaterials},
\newblock \bibinfo{journal}{Environ. Toxicol. Chem.} \bibinfo{volume}{36}
  (\bibinfo{year}{2017}) \bibinfo{pages}{1704--1714}.
\bibitem[{Murphy et~al.(2018)Murphy, Nisbet, Antczak, Garcia-Reyero, Gergs,
  Lika, Mathews, Muller, Nacci, Peace et~al.}]{murphy2018incorporating}
\bibinfo{author}{C.~A. Murphy}, \bibinfo{author}{R.~M. Nisbet},
  \bibinfo{author}{P.~Antczak}, \bibinfo{author}{N.~Garcia-Reyero},
  \bibinfo{author}{A.~Gergs}, \bibinfo{author}{K.~Lika},
  \bibinfo{author}{T.~Mathews}, \bibinfo{author}{E.~B. Muller},
  \bibinfo{author}{D.~Nacci}, \bibinfo{author}{A.~Peace}, et~al.,
\newblock \bibinfo{title}{Incorporating suborganismal processes into dynamic
  energy budget models for ecological risk assessment},
\newblock \bibinfo{journal}{Integr. Environ. Assess. Manag.}
  \bibinfo{volume}{14} (\bibinfo{year}{2018}) \bibinfo{pages}{615--624}.
\bibitem[{Ankley et~al.(2010)Ankley, Bennett, Erickson, Hoff, Hornung, Johnson,
  Mount, Nichols, Russom, Schmieder et~al.}]{ankley2010adverse}
\bibinfo{author}{G.~T. Ankley}, \bibinfo{author}{R.~S. Bennett},
  \bibinfo{author}{R.~J. Erickson}, \bibinfo{author}{D.~J. Hoff},
  \bibinfo{author}{M.~W. Hornung}, \bibinfo{author}{R.~D. Johnson},
  \bibinfo{author}{D.~R. Mount}, \bibinfo{author}{J.~W. Nichols},
  \bibinfo{author}{C.~L. Russom}, \bibinfo{author}{P.~K. Schmieder}, et~al.,
\newblock \bibinfo{title}{Adverse outcome pathways: a conceptual framework to
  support ecotoxicology research and risk assessment},
\newblock \bibinfo{journal}{Environ. Toxicol. Chem.} \bibinfo{volume}{29}
  (\bibinfo{year}{2010}) \bibinfo{pages}{730--741}.
\bibitem[{Marques et~al.(2018)Marques, Augustine, Lika, Pecquerie, Domingos,
  and Kooijman}]{marques2018amp}
\bibinfo{author}{G.~M. Marques}, \bibinfo{author}{S.~Augustine},
  \bibinfo{author}{K.~Lika}, \bibinfo{author}{L.~Pecquerie},
  \bibinfo{author}{T.~Domingos}, \bibinfo{author}{S.~A. L.~M. Kooijman},
\newblock \bibinfo{title}{The {AmP} project: comparing species on the basis of
  dynamic energy budget parameters},
\newblock \bibinfo{journal}{PLOS Comput. Biol.} \bibinfo{volume}{14}
  (\bibinfo{year}{2018}) \bibinfo{pages}{e1006100}.
\bibitem[{Von~Bertalanffy(1957)}]{vonbertalanffy1957quantitative}
\bibinfo{author}{L.~Von~Bertalanffy},
\newblock \bibinfo{title}{Quantitative laws in metabolism and growth},
\newblock \bibinfo{journal}{Q. Rev. Biol.} \bibinfo{volume}{32}
  (\bibinfo{year}{1957}) \bibinfo{pages}{217--231}.
\bibitem[{Jager(2011)}]{jager2011some}
\bibinfo{author}{T.~Jager},
\newblock \bibinfo{title}{Some good reasons to ban {EC\textit{x}} and related
  concepts in ecotoxicology},
\newblock \bibinfo{journal}{Environ. Sci. Technol.} \bibinfo{volume}{45}
  (\bibinfo{year}{2011}) \bibinfo{pages}{8180--8181}.
\bibitem[{Klanjscek et~al.(2012)Klanjscek, Nisbet, Priester, and
  Holden}]{klanjscek2012modeling}
\bibinfo{author}{T.~Klanjscek}, \bibinfo{author}{R.~M. Nisbet},
  \bibinfo{author}{J.~H. Priester}, \bibinfo{author}{P.~A. Holden},
\newblock \bibinfo{title}{Modeling physiological processes that relate toxicant
  exposure and bacterial population dynamics},
\newblock \bibinfo{journal}{PLOS ONE} \bibinfo{volume}{7}
  (\bibinfo{year}{2012}) \bibinfo{pages}{e26955}.
\bibitem[{Klanjscek et~al.(2013)Klanjscek, Nisbet, Priester, and
  Holden}]{klanjscek2013dynamic}
\bibinfo{author}{T.~Klanjscek}, \bibinfo{author}{R.~M. Nisbet},
  \bibinfo{author}{J.~H. Priester}, \bibinfo{author}{P.~A. Holden},
\newblock \bibinfo{title}{Dynamic energy budget approach to modeling mechanisms
  of {CdSe} quantum dot toxicity},
\newblock \bibinfo{journal}{Ecotoxicology} \bibinfo{volume}{22}
  (\bibinfo{year}{2013}) \bibinfo{pages}{319--330}.
\bibitem[{Fablet et~al.(2011)Fablet, Pecquerie, De~Pontual, H{\o}ie, Millner,
  Mosegaard, and Kooijman}]{fablet2011shedding}
\bibinfo{author}{R.~Fablet}, \bibinfo{author}{L.~Pecquerie},
  \bibinfo{author}{H.~De~Pontual}, \bibinfo{author}{H.~H{\o}ie},
  \bibinfo{author}{R.~Millner}, \bibinfo{author}{H.~Mosegaard},
  \bibinfo{author}{S.~A. L.~M. Kooijman},
\newblock \bibinfo{title}{Shedding light on fish otolith biomineralization
  using a bioenergetic approach},
\newblock \bibinfo{journal}{PLOS ONE} \bibinfo{volume}{6}
  (\bibinfo{year}{2011}) \bibinfo{pages}{e27055}.
\bibitem[{Pecquerie et~al.(2012)Pecquerie, Fablet, De~Pontual, Bonhommeau,
  Alunno-Bruscia, Petitgas, and Kooijman}]{pecquerie2012reconstructing}
\bibinfo{author}{L.~Pecquerie}, \bibinfo{author}{R.~Fablet},
  \bibinfo{author}{H.~De~Pontual}, \bibinfo{author}{S.~Bonhommeau},
  \bibinfo{author}{M.~Alunno-Bruscia}, \bibinfo{author}{P.~Petitgas},
  \bibinfo{author}{S.~A. L.~M. Kooijman},
\newblock \bibinfo{title}{Reconstructing individual food and growth histories
  from biogenic carbonates},
\newblock \bibinfo{journal}{Mar. Ecol. Prog. Ser.} \bibinfo{volume}{447}
  (\bibinfo{year}{2012}) \bibinfo{pages}{151--164}.
\bibitem[{Klanjscek et~al.(2017)Klanjscek, Muller, Holden, and
  Nisbet}]{klanjscek2017host}
\bibinfo{author}{T.~Klanjscek}, \bibinfo{author}{E.~B. Muller},
  \bibinfo{author}{P.~A. Holden}, \bibinfo{author}{R.~M. Nisbet},
\newblock \bibinfo{title}{Host--symbiont interaction model explains
  non-monotonic response of soybean growth and seed production to nano-{CeO2}
  exposure},
\newblock \bibinfo{journal}{Environ. Sci. Technol.} \bibinfo{volume}{51}
  (\bibinfo{year}{2017}) \bibinfo{pages}{4944--4950}.
\bibitem[{Jager et~al.(2011)Jager, Albert, Preuss, and
  Ashauer}]{jager2011general}
\bibinfo{author}{T.~Jager}, \bibinfo{author}{C.~Albert}, \bibinfo{author}{T.~G.
  Preuss}, \bibinfo{author}{R.~Ashauer},
\newblock \bibinfo{title}{General unified threshold model of survival---a
  toxicokinetic-toxicodynamic framework for ecotoxicology},
\newblock \bibinfo{journal}{Environ. Sci. Technol.} \bibinfo{volume}{45}
  (\bibinfo{year}{2011}) \bibinfo{pages}{2529--2540}.
\bibitem[{{OECD}(2003)}]{oecd2003current}
\bibinfo{author}{{OECD}}, \bibinfo{title}{Current approaches in the statistical
  analysis of ecotoxicity data: {A} guidance to application},
  \bibinfo{howpublished}{OECD Publishing}, \bibinfo{year}{2003}.
  \bibinfo{note}{OECD Series on Testing and Assessment, No.~54. Available at:
  \url{https://doi.org/10.1787/9789264085275-en}}.
\bibitem[{Jager et~al.(2010)Jager, Vandenbrouck, Baas, De~Coen, and
  Kooijman}]{jager2010biology}
\bibinfo{author}{T.~Jager}, \bibinfo{author}{T.~Vandenbrouck},
  \bibinfo{author}{J.~Baas}, \bibinfo{author}{W.~M. De~Coen},
  \bibinfo{author}{S.~A. L.~M. Kooijman},
\newblock \bibinfo{title}{A biology-based approach for mixture toxicity of
  multiple endpoints over the life cycle},
\newblock \bibinfo{journal}{Ecotoxicology} \bibinfo{volume}{19}
  (\bibinfo{year}{2010}) \bibinfo{pages}{351--361}.
\bibitem[{Jager et~al.(2014)Jager, Gudmundsd\'{o}ttir, and
  Cedergreen}]{jager2014dynamic}
\bibinfo{author}{T.~Jager}, \bibinfo{author}{E.~M. Gudmundsd\'{o}ttir},
  \bibinfo{author}{N.~Cedergreen},
\newblock \bibinfo{title}{Dynamic modeling of sublethal mixture toxicity in the
  nematode \textit{{C}aenorhabditis elegans}},
\newblock \bibinfo{journal}{Environ. Sci. Technol.} \bibinfo{volume}{48}
  (\bibinfo{year}{2014}) \bibinfo{pages}{7026--7033}.
\bibitem[{Margerit et~al.(2016)Margerit, Gomez, and
  Gilbin}]{margerit2016dynamic}
\bibinfo{author}{A.~Margerit}, \bibinfo{author}{E.~Gomez},
  \bibinfo{author}{R.~Gilbin},
\newblock \bibinfo{title}{Dynamic energy-based modeling of uranium and cadmium
  joint toxicity to \textit{{C}aenorhabditis elegans}},
\newblock \bibinfo{journal}{Chemosphere} \bibinfo{volume}{146}
  (\bibinfo{year}{2016}) \bibinfo{pages}{405--412}.
\bibitem[{Jager et~al.(2013)Jager, Martin, and Zimmer}]{jager2013debkiss}
\bibinfo{author}{T.~Jager}, \bibinfo{author}{B.~T. Martin},
  \bibinfo{author}{E.~I. Zimmer},
\newblock \bibinfo{title}{{DEBkiss} or the quest for the simplest generic model
  of animal life history},
\newblock \bibinfo{journal}{J. Theor. Biol.} \bibinfo{volume}{328}
  (\bibinfo{year}{2013}) \bibinfo{pages}{9--18}.
\bibitem[{Ledder et~al.(2004)Ledder, Logan, and Joern}]{ledder2004dynamic}
\bibinfo{author}{G.~Ledder}, \bibinfo{author}{J.~D. Logan},
  \bibinfo{author}{A.~Joern},
\newblock \bibinfo{title}{Dynamic energy budget models with size-dependent
  hazard rates},
\newblock \bibinfo{journal}{J. Math. Biol.} \bibinfo{volume}{48}
  (\bibinfo{year}{2004}) \bibinfo{pages}{605--622}.
\bibitem[{Martin et~al.(2012)Martin, Zimmer, Grimm, and
  Jager}]{martin2012dynamic}
\bibinfo{author}{B.~T. Martin}, \bibinfo{author}{E.~I. Zimmer},
  \bibinfo{author}{V.~Grimm}, \bibinfo{author}{T.~Jager},
\newblock \bibinfo{title}{Dynamic {E}nergy {B}udget theory meets
  individual-based modelling: a generic and accessible implementation},
\newblock \bibinfo{journal}{Methods Ecol. Evol.} \bibinfo{volume}{3}
  (\bibinfo{year}{2012}) \bibinfo{pages}{445--449}.
\bibitem[{De~Roos(2008)}]{deroos2008demographic}
\bibinfo{author}{A.~M. De~Roos},
\newblock \bibinfo{title}{Demographic analysis of continuous-time life-history
  models},
\newblock \bibinfo{journal}{Ecol. Lett.} \bibinfo{volume}{11}
  (\bibinfo{year}{2008}) \bibinfo{pages}{1--15}.
\bibitem[{Beekman et~al.(2019)Beekman, Thompson, and
  Jusup}]{beekman2019thermodynamic}
\bibinfo{author}{M.~Beekman}, \bibinfo{author}{M.~Thompson},
  \bibinfo{author}{M.~Jusup},
\newblock \bibinfo{title}{Thermodynamic constraints and the evolution of
  parental provisioning in vertebrates},
\newblock \bibinfo{journal}{Behav. Ecol.} \bibinfo{volume}{30}
  (\bibinfo{year}{2019}) \bibinfo{pages}{583--591}.
\bibitem[{Klanjscek et~al.(2006)Klanjscek, Caswell, Neubert, and
  Nisbet}]{klanjscek2006integrating}
\bibinfo{author}{T.~Klanjscek}, \bibinfo{author}{H.~Caswell},
  \bibinfo{author}{M.~G. Neubert}, \bibinfo{author}{R.~M. Nisbet},
\newblock \bibinfo{title}{Integrating dynamic energy budgets into matrix
  population models},
\newblock \bibinfo{journal}{Ecol. Model.} \bibinfo{volume}{196}
  (\bibinfo{year}{2006}) \bibinfo{pages}{407--420}.
\bibitem[{Ijima et~al.(2019)Ijima, Jusup, Takada, Akita, Matsuda, and
  Klanjscek}]{ijima2019effects}
\bibinfo{author}{H.~Ijima}, \bibinfo{author}{M.~Jusup},
  \bibinfo{author}{T.~Takada}, \bibinfo{author}{T.~Akita},
  \bibinfo{author}{H.~Matsuda}, \bibinfo{author}{T.~Klanjscek},
\newblock \bibinfo{title}{Effects of environmental change and early-life
  stochasticity on {P}acific bluefin tuna population growth},
\newblock \bibinfo{journal}{Mar. Environ. Res.} \bibinfo{volume}{149}
  (\bibinfo{year}{2019}) \bibinfo{pages}{18--26}.
\bibitem[{De~Roos et~al.(1992)De~Roos, Diekmann, and Metz}]{deroos1992studying}
\bibinfo{author}{A.~M. De~Roos}, \bibinfo{author}{O.~Diekmann},
  \bibinfo{author}{J.~A.~J. Metz},
\newblock \bibinfo{title}{Studying the dynamics of structured population
  models: a versatile technique and its application to {D}aphnia},
\newblock \bibinfo{journal}{Am. Nat} \bibinfo{volume}{139}
  (\bibinfo{year}{1992}) \bibinfo{pages}{123--147}.
\bibitem[{De~Roos and Persson(2001)}]{deroos2001physiologically}
\bibinfo{author}{A.~M. De~Roos}, \bibinfo{author}{L.~Persson},
\newblock \bibinfo{title}{Physiologically structured models---from versatile
  technique to ecological theory},
\newblock \bibinfo{journal}{Oikos} \bibinfo{volume}{94} (\bibinfo{year}{2001})
  \bibinfo{pages}{51--71}.
\bibitem[{Smallegange et~al.(2017)Smallegange, Caswell, Toorians, and
  De~Roos}]{smallegange2017mechanistic}
\bibinfo{author}{I.~M. Smallegange}, \bibinfo{author}{H.~Caswell},
  \bibinfo{author}{M.~E.~M. Toorians}, \bibinfo{author}{A.~M. De~Roos},
\newblock \bibinfo{title}{Mechanistic description of population dynamics using
  dynamic energy budget theory incorporated into integral projection models},
\newblock \bibinfo{journal}{Methods Ecol. Evol.} \bibinfo{volume}{8}
  (\bibinfo{year}{2017}) \bibinfo{pages}{146--154}.
\bibitem[{Hunter and Caswell(2005)}]{hunter2005use}
\bibinfo{author}{C.~M. Hunter}, \bibinfo{author}{H.~Caswell},
\newblock \bibinfo{title}{The use of the vec-permutation matrix in spatial
  matrix population models},
\newblock \bibinfo{journal}{Ecol. Model.} \bibinfo{volume}{188}
  (\bibinfo{year}{2005}) \bibinfo{pages}{15--21}.
\bibitem[{Phillips et~al.(2006)Phillips, Anderson, and
  Schapire}]{phillips2006maximum}
\bibinfo{author}{S.~J. Phillips}, \bibinfo{author}{R.~P. Anderson},
  \bibinfo{author}{R.~E. Schapire},
\newblock \bibinfo{title}{Maximum entropy modeling of species geographic
  distributions},
\newblock \bibinfo{journal}{Ecol. Model.} \bibinfo{volume}{190}
  (\bibinfo{year}{2006}) \bibinfo{pages}{231--259}.
\bibitem[{Phillips et~al.(2017)Phillips, Anderson, Dud{\'\i}k, Schapire, and
  Blair}]{phillips2017opening}
\bibinfo{author}{S.~J. Phillips}, \bibinfo{author}{R.~P. Anderson},
  \bibinfo{author}{M.~Dud{\'\i}k}, \bibinfo{author}{R.~E. Schapire},
  \bibinfo{author}{M.~E. Blair},
\newblock \bibinfo{title}{Opening the black box: {A}n open-source release of
  {M}axent},
\newblock \bibinfo{journal}{Ecography} \bibinfo{volume}{40}
  (\bibinfo{year}{2017}) \bibinfo{pages}{887--893}.
\bibitem[{Faurby et~al.(2020)Faurby, Silvestro, Werdelin, and
  Antonelli}]{faurby2020brain}
\bibinfo{author}{S.~Faurby}, \bibinfo{author}{D.~Silvestro},
  \bibinfo{author}{L.~Werdelin}, \bibinfo{author}{A.~Antonelli},
\newblock \bibinfo{title}{Brain expansion in early hominins predicts carnivore
  extinctions in {E}ast {A}frica},
\newblock \bibinfo{journal}{Ecol. Lett.} \bibinfo{volume}{23}
  (\bibinfo{year}{2020}) \bibinfo{pages}{537--544}.
\bibitem[{Kennedy et~al.(2019)Kennedy, Oakleaf, Theobald, Baruch-Mordo, and
  Kiesecker}]{kennedy2019managing}
\bibinfo{author}{C.~M. Kennedy}, \bibinfo{author}{J.~R. Oakleaf},
  \bibinfo{author}{D.~M. Theobald}, \bibinfo{author}{S.~Baruch-Mordo},
  \bibinfo{author}{J.~Kiesecker},
\newblock \bibinfo{title}{Managing the middle: {A} shift in conservation
  priorities based on the global human modification gradient},
\newblock \bibinfo{journal}{Glob. Change Biol.} \bibinfo{volume}{25}
  (\bibinfo{year}{2019}) \bibinfo{pages}{811--826}.
\bibitem[{Belle et~al.(2018)Belle, Kingston, Burgess, Sandwith, Ali, and
  MacKinnon}]{belle2018protected}
\bibinfo{author}{E.~Belle}, \bibinfo{author}{N.~Kingston},
  \bibinfo{author}{N.~Burgess}, \bibinfo{author}{T.~Sandwith},
  \bibinfo{author}{N.~Ali}, \bibinfo{author}{K.~MacKinnon},
  \bibinfo{title}{Protected {P}lanet {R}eport 2018},
  \bibinfo{howpublished}{United Nations Environment Programme},
  \bibinfo{year}{2018}. \bibinfo{note}{Available at:
  \url{https://livereport.protectedplanet.net/pdf/Protected_Planet_Report_2018.pdf}}.
\bibitem[{Holloway and Humphreys(2019)}]{holloway2019business}
\bibinfo{author}{J.~C. Holloway}, \bibinfo{author}{C.~Humphreys},
  \bibinfo{title}{The business of tourism}, \bibinfo{publisher}{Sage},
  \bibinfo{year}{2019}.
\bibitem[{Scheffer et~al.(2000)Scheffer, Brock, and
  Westley}]{scheffer2000socioeconomic}
\bibinfo{author}{M.~Scheffer}, \bibinfo{author}{W.~Brock},
  \bibinfo{author}{F.~Westley},
\newblock \bibinfo{title}{Socioeconomic mechanisms preventing optimum use of
  ecosystem services: an interdisciplinary theoretical analysis},
\newblock \bibinfo{journal}{Ecosystems} \bibinfo{volume}{3}
  (\bibinfo{year}{2000}) \bibinfo{pages}{451--471}.
\bibitem[{Scheffer et~al.(2001)Scheffer, Carpenter, Foley, Folke, and
  Walker}]{scheffer2001catastrophic}
\bibinfo{author}{M.~Scheffer}, \bibinfo{author}{S.~Carpenter},
  \bibinfo{author}{J.~A. Foley}, \bibinfo{author}{C.~Folke},
  \bibinfo{author}{B.~Walker},
\newblock \bibinfo{title}{Catastrophic shifts in ecosystems},
\newblock \bibinfo{journal}{Nature} \bibinfo{volume}{413}
  (\bibinfo{year}{2001}) \bibinfo{pages}{591--596}.
\bibitem[{Klanjscek et~al.(2016)Klanjscek, Muller, and
  Nisbet}]{klanjscek2016feedbacks}
\bibinfo{author}{T.~Klanjscek}, \bibinfo{author}{E.~B. Muller},
  \bibinfo{author}{R.~M. Nisbet},
\newblock \bibinfo{title}{Feedbacks and tipping points in organismal response
  to oxidative stress},
\newblock \bibinfo{journal}{J. Theor. Biol.} \bibinfo{volume}{404}
  (\bibinfo{year}{2016}) \bibinfo{pages}{361--374}.
\bibitem[{Marn et~al.(2020)Marn, Jusup, Kooijman, and
  Klanjscek}]{marn2020quantifying}
\bibinfo{author}{N.~Marn}, \bibinfo{author}{M.~Jusup}, \bibinfo{author}{S.~A.
  L.~M. Kooijman}, \bibinfo{author}{T.~Klanjscek},
\newblock \bibinfo{title}{Quantifying impacts of plastic debris on marine
  wildlife identifies ecological breakpoints},
\newblock \bibinfo{journal}{Ecol. Lett.} \bibinfo{volume}{23}
  (\bibinfo{year}{2020}) \bibinfo{pages}{1479--1487}.
\bibitem[{Hamilton and Butler(2001)}]{hamilton2001outport}
\bibinfo{author}{L.~C. Hamilton}, \bibinfo{author}{M.~J. Butler},
\newblock \bibinfo{title}{Outport adaptations: {S}ocial indicators through
  {N}ewfoundland's cod crisis},
\newblock \bibinfo{journal}{Hum. Ecol. Rev.} \bibinfo{volume}{8}
  (\bibinfo{year}{2001}) \bibinfo{pages}{1--11}.
\bibitem[{Hamilton et~al.(2004)Hamilton, Haedrich, and
  Duncan}]{hamilton2004above}
\bibinfo{author}{L.~C. Hamilton}, \bibinfo{author}{R.~L. Haedrich},
  \bibinfo{author}{C.~M. Duncan},
\newblock \bibinfo{title}{Above and below the water: social/ecological
  transformation in northwest {N}ewfoundland},
\newblock \bibinfo{journal}{Popul. Environ.} \bibinfo{volume}{25}
  (\bibinfo{year}{2004}) \bibinfo{pages}{195--215}.
\bibitem[{Myers et~al.(1997)Myers, Hutchings, and Barrowman}]{myers1997fish}
\bibinfo{author}{R.~A. Myers}, \bibinfo{author}{J.~A. Hutchings},
  \bibinfo{author}{N.~J. Barrowman},
\newblock \bibinfo{title}{Why do fish stocks collapse? the example of cod in
  {A}tlantic {C}anada},
\newblock \bibinfo{journal}{Ecol. Appl.} \bibinfo{volume}{7}
  (\bibinfo{year}{1997}) \bibinfo{pages}{91--106}.
\bibitem[{Frank et~al.(2011)Frank, Petrie, Fisher, and
  Leggett}]{frank2011transient}
\bibinfo{author}{K.~T. Frank}, \bibinfo{author}{B.~Petrie},
  \bibinfo{author}{J.~A.~D. Fisher}, \bibinfo{author}{W.~C. Leggett},
\newblock \bibinfo{title}{Transient dynamics of an altered large marine
  ecosystem},
\newblock \bibinfo{journal}{Nature} \bibinfo{volume}{477}
  (\bibinfo{year}{2011}) \bibinfo{pages}{86--89}.
\bibitem[{Rose and Rowe(2015)}]{rose2015northern}
\bibinfo{author}{G.~A. Rose}, \bibinfo{author}{S.~Rowe},
\newblock \bibinfo{title}{Northern cod comeback},
\newblock \bibinfo{journal}{Can. J. Fish. Aquat. Sci.} \bibinfo{volume}{72}
  (\bibinfo{year}{2015}) \bibinfo{pages}{1789--1798}.
\bibitem[{De~Roos and Persson(2002)}]{deroos2002size}
\bibinfo{author}{A.~M. De~Roos}, \bibinfo{author}{L.~Persson},
\newblock \bibinfo{title}{Size-dependent life-history traits promote
  catastrophic collapses of top predators},
\newblock \bibinfo{journal}{Proc. Natl. Acad. Sci. USA} \bibinfo{volume}{99}
  (\bibinfo{year}{2002}) \bibinfo{pages}{12907--12912}.
\bibitem[{King et~al.(2015)King, Cavender-Bares, Balvanera, Mwampamba, and
  Polasky}]{king2015trade}
\bibinfo{author}{E.~King}, \bibinfo{author}{J.~Cavender-Bares},
  \bibinfo{author}{P.~Balvanera}, \bibinfo{author}{T.~H. Mwampamba},
  \bibinfo{author}{S.~Polasky},
\newblock \bibinfo{title}{Trade-offs in ecosystem services and varying
  stakeholder preferences: evaluating conflicts, obstacles, and opportunities},
\newblock \bibinfo{journal}{Ecol. Soc.} \bibinfo{volume}{20}
  (\bibinfo{year}{2015}) \bibinfo{pages}{25}.
\bibitem[{Okayama(2016)}]{okayama2016general}
\bibinfo{author}{H.~Okayama},
\newblock \bibinfo{title}{General strategy for understanding intracellular
  molecular interaction cascades that elicit stimulus-invoked biological
  processes},
\newblock \bibinfo{journal}{Proc. Jpn. Acad. Ser. B Phys. Biol. Sci.}
  \bibinfo{volume}{92} (\bibinfo{year}{2016}) \bibinfo{pages}{372--385}.
\bibitem[{Baross and Hoffman(1985)}]{baross1985submarine}
\bibinfo{author}{J.~A. Baross}, \bibinfo{author}{S.~E. Hoffman},
\newblock \bibinfo{title}{Submarine hydrothermal vents and associated gradient
  environments as sites for the origin and evolution of life},
\newblock \bibinfo{journal}{Orig. Life Evol. Biosph.} \bibinfo{volume}{15}
  (\bibinfo{year}{1985}) \bibinfo{pages}{327--345}.
\bibitem[{Lindeman(1942)}]{lindeman1942trophic}
\bibinfo{author}{R.~L. Lindeman},
\newblock \bibinfo{title}{The trophic-dynamic aspect of ecology},
\newblock \bibinfo{journal}{Ecology} \bibinfo{volume}{23}
  (\bibinfo{year}{1942}) \bibinfo{pages}{399--417}.
\bibitem[{Pachauri et~al.(2014)Pachauri, Allen, Barros, Broome, Cramer, Christ,
  Church, Clarke, Dahe, Dasgupta et~al.}]{pachauri2014climate}
\bibinfo{author}{R.~K. Pachauri}, \bibinfo{author}{M.~R. Allen},
  \bibinfo{author}{V.~R. Barros}, \bibinfo{author}{J.~Broome},
  \bibinfo{author}{W.~Cramer}, \bibinfo{author}{R.~Christ},
  \bibinfo{author}{J.~A. Church}, \bibinfo{author}{L.~Clarke},
  \bibinfo{author}{Q.~Dahe}, \bibinfo{author}{P.~Dasgupta}, et~al.,
  \bibinfo{title}{Climate change 2014: synthesis report. Contribution of
  Working Groups I, II and III to the fifth assessment report of the
  Intergovernmental Panel on Climate Change}, \bibinfo{publisher}{IPCC},
  \bibinfo{year}{2014}.
\bibitem[{Holland(1986)}]{holland1986statistics}
\bibinfo{author}{P.~W. Holland},
\newblock \bibinfo{title}{Statistics and causal inference},
\newblock \bibinfo{journal}{J. Am. Stat. Assoc.} \bibinfo{volume}{81}
  (\bibinfo{year}{1986}) \bibinfo{pages}{945--960}.
\bibitem[{Carleton and Hsiang(2016)}]{carleton2016social}
\bibinfo{author}{T.~A. Carleton}, \bibinfo{author}{S.~M. Hsiang},
\newblock \bibinfo{title}{Social and economic impacts of climate},
\newblock \bibinfo{journal}{Science} \bibinfo{volume}{353}
  (\bibinfo{year}{2016}) \bibinfo{pages}{aad9837}.
\bibitem[{Guo et~al.(2014)Guo, Gasparrini, Armstrong, Li, Tawatsupa, Tobias,
  Lavigne, Coelho, Leone, Pan et~al.}]{guo2014global}
\bibinfo{author}{Y.~Guo}, \bibinfo{author}{A.~Gasparrini},
  \bibinfo{author}{B.~Armstrong}, \bibinfo{author}{S.~Li},
  \bibinfo{author}{B.~Tawatsupa}, \bibinfo{author}{A.~Tobias},
  \bibinfo{author}{E.~Lavigne}, \bibinfo{author}{M.~d. S. Z.~S. Coelho},
  \bibinfo{author}{M.~Leone}, \bibinfo{author}{X.~Pan}, et~al.,
\newblock \bibinfo{title}{Global variation in the effects of ambient
  temperature on mortality: a systematic evaluation},
\newblock \bibinfo{journal}{Epidemiology} \bibinfo{volume}{25}
  (\bibinfo{year}{2014}) \bibinfo{pages}{781--789}.
\bibitem[{Burgess et~al.(2013)Burgess, Deschenes, Donaldson, and
  Greenstone}]{burgess2013unequal}
\bibinfo{author}{R.~Burgess}, \bibinfo{author}{O.~Deschenes},
  \bibinfo{author}{D.~Donaldson}, \bibinfo{author}{M.~Greenstone},
  \bibinfo{title}{The unequal effects of weather and climate change: {E}vidence
  from mortality in {I}ndia}, \bibinfo{year}{2013}. \bibinfo{note}{Available
  at:
  \url{https://cla.umn.edu/sites/cla.umn.edu/files/burgess_deschenes_donaldson_greenstone_unequal_effects.pdf}.
  Archived at: \url{https://doi.org/10.17605/OSF.IO/N7G9X}}.
\bibitem[{Hsiang and Narita(2012)}]{hsiang2012adaptation}
\bibinfo{author}{S.~M. Hsiang}, \bibinfo{author}{D.~Narita},
\newblock \bibinfo{title}{Adaptation to cyclone risk: {E}vidence from the
  global cross-section},
\newblock \bibinfo{journal}{Clim. Chang. Econ.} \bibinfo{volume}{3}
  (\bibinfo{year}{2012}) \bibinfo{pages}{1250011}.
\bibitem[{Anttila-Hughes and Hsiang(2013)}]{anttila2013destruction}
\bibinfo{author}{J.~Anttila-Hughes}, \bibinfo{author}{S.~Hsiang},
  \bibinfo{title}{Destruction, disinvestment, and death: {E}conomic and human
  losses following environmental disaster}, \bibinfo{year}{2013}.
  \bibinfo{note}{Social Science Research Network (SSRN) Paper No. 2220501.
  Available at: \url{https://ssrn.com/abstract=2220501}}.
\bibitem[{Schlenker and Roberts(2009)}]{schlenker2009nonlinear}
\bibinfo{author}{W.~Schlenker}, \bibinfo{author}{M.~J. Roberts},
\newblock \bibinfo{title}{Nonlinear temperature effects indicate severe damages
  to {US} crop yields under climate change},
\newblock \bibinfo{journal}{Proc. Natl. Acad. Sci. USA} \bibinfo{volume}{106}
  (\bibinfo{year}{2009}) \bibinfo{pages}{15594--15598}.
\bibitem[{Hidalgo et~al.(2010)Hidalgo, Naidu, Nichter, and
  Richardson}]{hidalgo2010economic}
\bibinfo{author}{F.~D. Hidalgo}, \bibinfo{author}{S.~Naidu},
  \bibinfo{author}{S.~Nichter}, \bibinfo{author}{N.~Richardson},
\newblock \bibinfo{title}{Economic determinants of land invasions},
\newblock \bibinfo{journal}{Rev. Econ. Stat.} \bibinfo{volume}{92}
  (\bibinfo{year}{2010}) \bibinfo{pages}{505--523}.
\bibitem[{Graff~Zivin and Neidell(2014)}]{graff2014temperature}
\bibinfo{author}{J.~Graff~Zivin}, \bibinfo{author}{M.~Neidell},
\newblock \bibinfo{title}{Temperature and the allocation of time:
  {I}mplications for climate change},
\newblock \bibinfo{journal}{J. Labor Econ.} \bibinfo{volume}{32}
  (\bibinfo{year}{2014}) \bibinfo{pages}{1--26}.
\bibitem[{Graff~Zivin et~al.(2018)Graff~Zivin, Hsiang, and
  Neidell}]{graff2018temperature}
\bibinfo{author}{J.~Graff~Zivin}, \bibinfo{author}{S.~M. Hsiang},
  \bibinfo{author}{M.~Neidell},
\newblock \bibinfo{title}{Temperature and human capital in the short and long
  run},
\newblock \bibinfo{journal}{J. Assoc. Environ. Resour. Econ.}
  \bibinfo{volume}{5} (\bibinfo{year}{2018}) \bibinfo{pages}{77--105}.
\bibitem[{Auffhammer(2018)}]{auffhammer2018climate}
\bibinfo{author}{M.~Auffhammer}, \bibinfo{title}{Climate adaptive response
  estimation: {S}hort and long run impacts of climate change on residential
  electricity and natural gas consumption using big data},
  \bibinfo{year}{2018}. \bibinfo{note}{National Bureau of Economic Research
  (NBER) Working Paper No. 24397. Available at:
  \url{https://www.nber.org/papers/w24397}}.
\bibitem[{Burke et~al.(2015)Burke, Hsiang, and Miguel}]{burke2015global}
\bibinfo{author}{M.~Burke}, \bibinfo{author}{S.~M. Hsiang},
  \bibinfo{author}{E.~Miguel},
\newblock \bibinfo{title}{Global non-linear effect of temperature on economic
  production},
\newblock \bibinfo{journal}{Nature} \bibinfo{volume}{527}
  (\bibinfo{year}{2015}) \bibinfo{pages}{235--239}.
\bibitem[{Deryugina and Hsiang(2014)}]{deryugina2014does}
\bibinfo{author}{T.~Deryugina}, \bibinfo{author}{S.~M. Hsiang},
  \bibinfo{title}{Does the environment still matter? {D}aily temperature and
  income in the {U}nited {S}tates}, \bibinfo{year}{2014}.
  \bibinfo{note}{National Bureau of Economic Research (NBER) Working Paper No.
  20750. Available at: \url{https://www.nber.org/papers/w20750}}.
\bibitem[{Zhang et~al.(2018)Zhang, Deschenes, Meng, and
  Zhang}]{zhang2018temperature}
\bibinfo{author}{P.~Zhang}, \bibinfo{author}{O.~Deschenes},
  \bibinfo{author}{K.~Meng}, \bibinfo{author}{J.~Zhang},
\newblock \bibinfo{title}{Temperature effects on productivity and factor
  reallocation: {E}vidence from a half million {C}hinese manufacturing plants},
\newblock \bibinfo{journal}{J. Environ. Econ. Manage.} \bibinfo{volume}{88}
  (\bibinfo{year}{2018}) \bibinfo{pages}{1--17}.
\bibitem[{Baylis(2020)}]{baylis2020temperature}
\bibinfo{author}{P.~Baylis},
\newblock \bibinfo{title}{Temperature and temperament: {E}vidence from
  {T}witter},
\newblock \bibinfo{journal}{J. Public Econ.} \bibinfo{volume}{184}
  (\bibinfo{year}{2020}) \bibinfo{pages}{104161}.
\bibitem[{Ranson(2014)}]{ranson2014crime}
\bibinfo{author}{M.~Ranson},
\newblock \bibinfo{title}{Crime, weather, and climate change},
\newblock \bibinfo{journal}{J. Environ. Econ. Manage.} \bibinfo{volume}{67}
  (\bibinfo{year}{2014}) \bibinfo{pages}{274--302}.
\bibitem[{Hsiang et~al.(2011)Hsiang, Meng, and Cane}]{hsiang2011civil}
\bibinfo{author}{S.~M. Hsiang}, \bibinfo{author}{K.~C. Meng},
  \bibinfo{author}{M.~A. Cane},
\newblock \bibinfo{title}{Civil conflicts are associated with the global
  climate},
\newblock \bibinfo{journal}{Nature} \bibinfo{volume}{476}
  (\bibinfo{year}{2011}) \bibinfo{pages}{438--441}.
\bibitem[{Bohra-Mishra et~al.(2014)Bohra-Mishra, Oppenheimer, and
  Hsiang}]{bohra2014nonlinear}
\bibinfo{author}{P.~Bohra-Mishra}, \bibinfo{author}{M.~Oppenheimer},
  \bibinfo{author}{S.~M. Hsiang},
\newblock \bibinfo{title}{Nonlinear permanent migration response to climatic
  variations but minimal response to disasters},
\newblock \bibinfo{journal}{Proc. Natl. Acad. Sci. USA} \bibinfo{volume}{111}
  (\bibinfo{year}{2014}) \bibinfo{pages}{9780--9785}.
\bibitem[{Steffen et~al.(2020)Steffen, Richardson, Rockstr{\"o}m, Schellnhuber,
  Dube, Dutreuil, Lenton, and Lubchenco}]{steffen2020emergence}
\bibinfo{author}{W.~Steffen}, \bibinfo{author}{K.~Richardson},
  \bibinfo{author}{J.~Rockstr{\"o}m}, \bibinfo{author}{H.~J. Schellnhuber},
  \bibinfo{author}{O.~P. Dube}, \bibinfo{author}{S.~Dutreuil},
  \bibinfo{author}{T.~M. Lenton}, \bibinfo{author}{J.~Lubchenco},
\newblock \bibinfo{title}{The emergence and evolution of {E}arth {S}ystem
  {S}cience},
\newblock \bibinfo{journal}{Nat. Rev. Earth Environ.} \bibinfo{volume}{1}
  (\bibinfo{year}{2020}) \bibinfo{pages}{54--63}.
\bibitem[{Seneviratne et~al.(2012)Seneviratne, Nicholls, Easterling, Goodess,
  Kanae, Kossin, Luo, Marengo, McInnes, Rahimi et~al.}]{seneviratne2012changes}
\bibinfo{author}{S.~Seneviratne}, \bibinfo{author}{N.~Nicholls},
  \bibinfo{author}{D.~Easterling}, \bibinfo{author}{C.~Goodess},
  \bibinfo{author}{S.~Kanae}, \bibinfo{author}{J.~Kossin},
  \bibinfo{author}{Y.~Luo}, \bibinfo{author}{J.~Marengo},
  \bibinfo{author}{K.~McInnes}, \bibinfo{author}{M.~Rahimi}, et~al.,
\newblock \bibinfo{title}{Changes in climate extremes and their impacts on the
  natural physical environment},
\newblock in: \bibinfo{editor}{C.~B. Field}, \bibinfo{editor}{V.~Barros},
  \bibinfo{editor}{T.~F. Stocker}, \bibinfo{editor}{D.~Qin},
  \bibinfo{editor}{D.~J. Dokken}, \bibinfo{editor}{K.~L. Ebi},
  \bibinfo{editor}{M.~D. Mastrandrea}, \bibinfo{editor}{K.~J. Mach},
  \bibinfo{editor}{G.-K. Plattner}, \bibinfo{editor}{S.~K. Allen},
  \bibinfo{editor}{M.~Tignor}, \bibinfo{editor}{P.~M. Midgley} (Eds.),
  \bibinfo{booktitle}{Managing the Risks of Extreme Events and Disasters to
  Advance Climate Change Adaptation. A Special Report of Working Groups I and
  II of the Intergovernmental Panel on Climate Change (IPCC)},
  \bibinfo{publisher}{Cambridge University Press}, \bibinfo{year}{2012}, pp.
  \bibinfo{pages}{109--230}.
\bibitem[{Lenton et~al.(2008)Lenton, Held, Kriegler, Hall, Lucht, Rahmstorf,
  and Schellnhuber}]{lenton2008tipping}
\bibinfo{author}{T.~M. Lenton}, \bibinfo{author}{H.~Held},
  \bibinfo{author}{E.~Kriegler}, \bibinfo{author}{J.~W. Hall},
  \bibinfo{author}{W.~Lucht}, \bibinfo{author}{S.~Rahmstorf},
  \bibinfo{author}{H.~J. Schellnhuber},
\newblock \bibinfo{title}{Tipping elements in the {E}arth's climate system},
\newblock \bibinfo{journal}{Proc. Natl. Acad. Sci. USA} \bibinfo{volume}{105}
  (\bibinfo{year}{2008}) \bibinfo{pages}{1786--1793}.
\bibitem[{Gladwell(2006)}]{gladwell2006tipping}
\bibinfo{author}{M.~Gladwell}, \bibinfo{title}{The tipping point: How little
  things can make a big difference}, \bibinfo{publisher}{Little, Brown},
  \bibinfo{year}{2006}.
\bibitem[{Rahmstorf(2002)}]{rahmstorf2002ocean}
\bibinfo{author}{S.~Rahmstorf},
\newblock \bibinfo{title}{Ocean circulation and climate during the past 120,000
  years},
\newblock \bibinfo{journal}{Nature} \bibinfo{volume}{419}
  (\bibinfo{year}{2002}) \bibinfo{pages}{207--214}.
\bibitem[{Dakos et~al.(2008)Dakos, Scheffer, van Nes, Brovkin, Petoukhov, and
  Held}]{dakos2008slowing}
\bibinfo{author}{V.~Dakos}, \bibinfo{author}{M.~Scheffer},
  \bibinfo{author}{E.~H. van Nes}, \bibinfo{author}{V.~Brovkin},
  \bibinfo{author}{V.~Petoukhov}, \bibinfo{author}{H.~Held},
\newblock \bibinfo{title}{Slowing down as an early warning signal for abrupt
  climate change},
\newblock \bibinfo{journal}{Proc. Natl. Acad. Sci. USA} \bibinfo{volume}{105}
  (\bibinfo{year}{2008}) \bibinfo{pages}{14308--14312}.
\bibitem[{Caesar et~al.(2018)Caesar, Rahmstorf, Robinson, Feulner, and
  Saba}]{caesar2018observed}
\bibinfo{author}{L.~Caesar}, \bibinfo{author}{S.~Rahmstorf},
  \bibinfo{author}{A.~Robinson}, \bibinfo{author}{G.~Feulner},
  \bibinfo{author}{V.~Saba},
\newblock \bibinfo{title}{Observed fingerprint of a weakening {A}tlantic
  {O}cean overturning circulation},
\newblock \bibinfo{journal}{Nature} \bibinfo{volume}{556}
  (\bibinfo{year}{2018}) \bibinfo{pages}{191--196}.
\bibitem[{Doyle et~al.(2016)Doyle, Sreenivasan, Szymanski, and
  Korniss}]{doyle2016social}
\bibinfo{author}{C.~Doyle}, \bibinfo{author}{S.~Sreenivasan},
  \bibinfo{author}{B.~K. Szymanski}, \bibinfo{author}{G.~Korniss},
\newblock \bibinfo{title}{Social consensus and tipping points with opinion
  inertia},
\newblock \bibinfo{journal}{Physica A} \bibinfo{volume}{443}
  (\bibinfo{year}{2016}) \bibinfo{pages}{316--323}.
\bibitem[{Centola(2010)}]{centola2010spread}
\bibinfo{author}{D.~Centola},
\newblock \bibinfo{title}{The spread of behavior in an online social network
  experiment},
\newblock \bibinfo{journal}{Science} \bibinfo{volume}{329}
  (\bibinfo{year}{2010}) \bibinfo{pages}{1194--1197}.
\bibitem[{Lenton et~al.(2019)Lenton, Rockstr{\"o}m, Gaffney, Rahmstorf,
  Richardson, Steffen, and Schellnhuber}]{lenton2019climate}
\bibinfo{author}{T.~M. Lenton}, \bibinfo{author}{J.~Rockstr{\"o}m},
  \bibinfo{author}{O.~Gaffney}, \bibinfo{author}{S.~Rahmstorf},
  \bibinfo{author}{K.~Richardson}, \bibinfo{author}{W.~Steffen},
  \bibinfo{author}{H.~J. Schellnhuber},
\newblock \bibinfo{title}{Climate tipping points—too risky to bet against},
\newblock \bibinfo{journal}{Nature} \bibinfo{volume}{575}
  (\bibinfo{year}{2019}) \bibinfo{pages}{592--595}.
\bibitem[{Otto et~al.(2020)Otto, Donges, Cremades, Bhowmik, Hewitt, Lucht,
  Rockstr{\"o}m, Allerberger, McCaffrey, Doe et~al.}]{otto2020social}
\bibinfo{author}{I.~M. Otto}, \bibinfo{author}{J.~F. Donges},
  \bibinfo{author}{R.~Cremades}, \bibinfo{author}{A.~Bhowmik},
  \bibinfo{author}{R.~J. Hewitt}, \bibinfo{author}{W.~Lucht},
  \bibinfo{author}{J.~Rockstr{\"o}m}, \bibinfo{author}{F.~Allerberger},
  \bibinfo{author}{M.~McCaffrey}, \bibinfo{author}{S.~S. Doe}, et~al.,
\newblock \bibinfo{title}{Social tipping dynamics for stabilizing {E}arth’s
  climate by 2050},
\newblock \bibinfo{journal}{Proc. Natl. Acad. Sci. USA} \bibinfo{volume}{117}
  (\bibinfo{year}{2020}) \bibinfo{pages}{2354--2365}.
\bibitem[{Carpenter and Brock(2006)}]{carpenter2006rising}
\bibinfo{author}{S.~R. Carpenter}, \bibinfo{author}{W.~A. Brock},
\newblock \bibinfo{title}{Rising variance: a leading indicator of ecological
  transition},
\newblock \bibinfo{journal}{Ecology Lett.} \bibinfo{volume}{9}
  (\bibinfo{year}{2006}) \bibinfo{pages}{311--318}.
\bibitem[{Livina and Lenton(2007)}]{livina2007modified}
\bibinfo{author}{V.~N. Livina}, \bibinfo{author}{T.~M. Lenton},
\newblock \bibinfo{title}{A modified method for detecting incipient
  bifurcations in a dynamical system},
\newblock \bibinfo{journal}{Geophys. Res. Lett.} \bibinfo{volume}{34}
  (\bibinfo{year}{2007}) \bibinfo{pages}{L03712}.
\bibitem[{Prettyman et~al.(2018)Prettyman, Kuna, and
  Livina}]{prettyman2018novel}
\bibinfo{author}{J.~Prettyman}, \bibinfo{author}{T.~Kuna},
  \bibinfo{author}{V.~Livina},
\newblock \bibinfo{title}{A novel scaling indicator of early warning signals
  helps anticipate tropical cyclones},
\newblock \bibinfo{journal}{EPL (Europhy. Lett.)} \bibinfo{volume}{121}
  (\bibinfo{year}{2018}) \bibinfo{pages}{10002}.
\bibitem[{Wang et~al.(2012)Wang, Dearing, Langdon, Zhang, Yang, Dakos, and
  Scheffer}]{wang2012flickering}
\bibinfo{author}{R.~Wang}, \bibinfo{author}{J.~A. Dearing},
  \bibinfo{author}{P.~G. Langdon}, \bibinfo{author}{E.~Zhang},
  \bibinfo{author}{X.~Yang}, \bibinfo{author}{V.~Dakos},
  \bibinfo{author}{M.~Scheffer},
\newblock \bibinfo{title}{Flickering gives early warning signals of a critical
  transition to a eutrophic lake state},
\newblock \bibinfo{journal}{Nature} \bibinfo{volume}{492}
  (\bibinfo{year}{2012}) \bibinfo{pages}{419--422}.
\bibitem[{Guttal and Jayaprakash(2008)}]{guttal2008changing}
\bibinfo{author}{V.~Guttal}, \bibinfo{author}{C.~Jayaprakash},
\newblock \bibinfo{title}{Changing skewness: an early warning signal of regime
  shifts in ecosystems},
\newblock \bibinfo{journal}{Ecol. Lett.} \bibinfo{volume}{11}
  (\bibinfo{year}{2008}) \bibinfo{pages}{450--460}.
\bibitem[{Biggs et~al.(2009)Biggs, Carpenter, and Brock}]{biggs2009turning}
\bibinfo{author}{R.~Biggs}, \bibinfo{author}{S.~R. Carpenter},
  \bibinfo{author}{W.~A. Brock},
\newblock \bibinfo{title}{Turning back from the brink: detecting an impending
  regime shift in time to avert it},
\newblock \bibinfo{journal}{Proc. Natl. Acad. Sci. USA} \bibinfo{volume}{106}
  (\bibinfo{year}{2009}) \bibinfo{pages}{826--831}.
\bibitem[{Dakos et~al.(2012)Dakos, Carpenter, Brock, Ellison, Guttal, Ives,
  Kefi, Livina, Seekell, van Nes et~al.}]{dakos2012methods}
\bibinfo{author}{V.~Dakos}, \bibinfo{author}{S.~R. Carpenter},
  \bibinfo{author}{W.~A. Brock}, \bibinfo{author}{A.~M. Ellison},
  \bibinfo{author}{V.~Guttal}, \bibinfo{author}{A.~R. Ives},
  \bibinfo{author}{S.~Kefi}, \bibinfo{author}{V.~Livina},
  \bibinfo{author}{D.~A. Seekell}, \bibinfo{author}{E.~H. van Nes}, et~al.,
\newblock \bibinfo{title}{Methods for detecting early warnings of critical
  transitions in time series illustrated using simulated ecological data},
\newblock \bibinfo{journal}{PLOS ONE} \bibinfo{volume}{7}
  (\bibinfo{year}{2012}) \bibinfo{pages}{e41010}.
\bibitem[{Richardson(1922)}]{richardson1922weather}
\bibinfo{author}{L.~F. Richardson}, \bibinfo{title}{Weather prediction by
  numerical process}, \bibinfo{publisher}{Cambridge university press},
  \bibinfo{year}{1922}.
\bibitem[{Callendar(1938)}]{callendar1938artificial}
\bibinfo{author}{G.~S. Callendar},
\newblock \bibinfo{title}{The artificial production of carbon dioxide and its
  influence on temperature},
\newblock \bibinfo{journal}{Q. J. R. Meteorol. Soc.} \bibinfo{volume}{64}
  (\bibinfo{year}{1938}) \bibinfo{pages}{223--240}.
\bibitem[{Manabe and Wetherald(1967)}]{manabe1967thermal}
\bibinfo{author}{S.~Manabe}, \bibinfo{author}{R.~T. Wetherald},
\newblock \bibinfo{title}{Thermal equilibrium of the atmosphere with a given
  distribution of relative humidity},
\newblock \bibinfo{journal}{J. Atmos. Sci.} \bibinfo{volume}{24}
  (\bibinfo{year}{1967}) \bibinfo{pages}{241--259}.
\bibitem[{{Working Group 1 of the IPCC}(1990)}]{ipcc1990policymakers}
\bibinfo{author}{{Working Group 1 of the IPCC}},
\newblock \bibinfo{title}{Policymakers summary},
\newblock in: \bibinfo{editor}{J.~T. Houghton}, \bibinfo{editor}{G.~J.
  Jenkins}, \bibinfo{editor}{J.~J. Ephraums} (Eds.),
  \bibinfo{booktitle}{Climate Change: The IPCC Scientific Assessment},
  \bibinfo{publisher}{Cambridge University Press}, \bibinfo{year}{1990}, pp.
  \bibinfo{pages}{vii--xxxiv}.
\bibitem[{Rockstr{\"o}m et~al.(2009)Rockstr{\"o}m, Steffen, Noone, Persson,
  Chapin, Lambin, Lenton, Scheffer, Folke, Schellnhuber
  et~al.}]{rockstrom2009safe}
\bibinfo{author}{J.~Rockstr{\"o}m}, \bibinfo{author}{W.~Steffen},
  \bibinfo{author}{K.~Noone}, \bibinfo{author}{{\AA}.~Persson},
  \bibinfo{author}{F.~S. Chapin}, \bibinfo{author}{E.~F. Lambin},
  \bibinfo{author}{T.~M. Lenton}, \bibinfo{author}{M.~Scheffer},
  \bibinfo{author}{C.~Folke}, \bibinfo{author}{H.~J. Schellnhuber}, et~al.,
\newblock \bibinfo{title}{A safe operating space for humanity},
\newblock \bibinfo{journal}{Nature} \bibinfo{volume}{461}
  (\bibinfo{year}{2009}) \bibinfo{pages}{472--475}.
\bibitem[{Medhaug et~al.(2017)Medhaug, Stolpe, Fischer, and
  Knutti}]{medhaug2017reconciling}
\bibinfo{author}{I.~Medhaug}, \bibinfo{author}{M.~B. Stolpe},
  \bibinfo{author}{E.~M. Fischer}, \bibinfo{author}{R.~Knutti},
\newblock \bibinfo{title}{Reconciling controversies about the `global warming
  hiatus'},
\newblock \bibinfo{journal}{Nature} \bibinfo{volume}{545}
  (\bibinfo{year}{2017}) \bibinfo{pages}{41--47}.
\bibitem[{Neelin(2010)}]{neelin2010climate}
\bibinfo{author}{J.~D. Neelin}, \bibinfo{title}{Climate change and climate
  modeling}, \bibinfo{publisher}{Cambridge University Press},
  \bibinfo{year}{2010}.
\bibitem[{Zebiak and Cane(1987)}]{zebiak1987model}
\bibinfo{author}{S.~E. Zebiak}, \bibinfo{author}{M.~A. Cane},
\newblock \bibinfo{title}{A model of {E}l {N}i{\~n}o--southern oscillation},
\newblock \bibinfo{journal}{Mon. Weather Rev.} \bibinfo{volume}{115}
  (\bibinfo{year}{1987}) \bibinfo{pages}{2262--2278}.
\bibitem[{Parson and Fisher-Vanden(1997)}]{parson1997integrated}
\bibinfo{author}{E.~A. Parson}, \bibinfo{author}{K.~Fisher-Vanden},
\newblock \bibinfo{title}{Integrated assessment models of global climate
  change},
\newblock \bibinfo{journal}{Annu. Rev. Energ. Environ.} \bibinfo{volume}{22}
  (\bibinfo{year}{1997}) \bibinfo{pages}{589--628}.
\bibitem[{van Vuuren et~al.(2011)van Vuuren, Lowe, Stehfest, Gohar, Hof, Hope,
  Warren, Meinshausen, and Plattner}]{vanvuuren2011well}
\bibinfo{author}{D.~P. van Vuuren}, \bibinfo{author}{J.~Lowe},
  \bibinfo{author}{E.~Stehfest}, \bibinfo{author}{L.~Gohar},
  \bibinfo{author}{A.~F. Hof}, \bibinfo{author}{C.~Hope},
  \bibinfo{author}{R.~Warren}, \bibinfo{author}{M.~Meinshausen},
  \bibinfo{author}{G.-K. Plattner},
\newblock \bibinfo{title}{How well do integrated assessment models simulate
  climate change?},
\newblock \bibinfo{journal}{Clim. Change} \bibinfo{volume}{104}
  (\bibinfo{year}{2011}) \bibinfo{pages}{255--285}.
\bibitem[{Hope et~al.(1993)Hope, Anderson, and Wenman}]{hope1993policy}
\bibinfo{author}{C.~Hope}, \bibinfo{author}{J.~Anderson},
  \bibinfo{author}{P.~Wenman},
\newblock \bibinfo{title}{Policy analysis of the greenhouse effect: an
  application of the {PAGE} model},
\newblock \bibinfo{journal}{Energy Policy} \bibinfo{volume}{21}
  (\bibinfo{year}{1993}) \bibinfo{pages}{327--338}.
\bibitem[{Holme and Rocha(2021)}]{holme2021ncc}
\bibinfo{author}{P.~Holme}, \bibinfo{author}{J.~C. Rocha},
  \bibinfo{title}{Networks of climate change: Connecting causes and
  consequences}, \bibinfo{year}{2021}. \bibinfo{note}{{e}-print
  arXiv:2105.12537}.
\bibitem[{Tsonis and Roebber(2004)}]{tsonis2004architecture}
\bibinfo{author}{A.~A. Tsonis}, \bibinfo{author}{P.~J. Roebber},
\newblock \bibinfo{title}{The architecture of the climate network},
\newblock \bibinfo{journal}{Physica A} \bibinfo{volume}{333}
  (\bibinfo{year}{2004}) \bibinfo{pages}{497--504}.
\bibitem[{Tsonis et~al.(2006)Tsonis, Swanson, and Roebber}]{tsonis2006networks}
\bibinfo{author}{A.~A. Tsonis}, \bibinfo{author}{K.~L. Swanson},
  \bibinfo{author}{P.~J. Roebber},
\newblock \bibinfo{title}{What do networks have to do with climate?},
\newblock \bibinfo{journal}{Bull. Am. Meteorol. Soc.} \bibinfo{volume}{87}
  (\bibinfo{year}{2006}) \bibinfo{pages}{585--596}.
\bibitem[{Tsonis and Swanson(2008)}]{tsonis2008topology}
\bibinfo{author}{A.~A. Tsonis}, \bibinfo{author}{K.~L. Swanson},
\newblock \bibinfo{title}{Topology and predictability of {E}l {N}i{\~n}o and
  {L}a {N}i{\~n}a networks},
\newblock \bibinfo{journal}{Phys. Rev. Lett.} \bibinfo{volume}{100}
  (\bibinfo{year}{2008}) \bibinfo{pages}{228502}.
\bibitem[{Yamasaki et~al.(2008)Yamasaki, Gozolchiani, and
  Havlin}]{yamasaki2008climate}
\bibinfo{author}{K.~Yamasaki}, \bibinfo{author}{A.~Gozolchiani},
  \bibinfo{author}{S.~Havlin},
\newblock \bibinfo{title}{Climate networks around the globe are significantly
  affected by {E}l {N}i{\~n}o},
\newblock \bibinfo{journal}{Phys. Rev. Lett.} \bibinfo{volume}{100}
  (\bibinfo{year}{2008}) \bibinfo{pages}{228501}.
\bibitem[{Gozolchiani et~al.(2008)Gozolchiani, Yamasaki, Gazit, and
  Havlin}]{gozolchiani2008pattern}
\bibinfo{author}{A.~Gozolchiani}, \bibinfo{author}{K.~Yamasaki},
  \bibinfo{author}{O.~Gazit}, \bibinfo{author}{S.~Havlin},
\newblock \bibinfo{title}{Pattern of climate network blinking links follows
  {E}l {N}i{\~n}o events},
\newblock \bibinfo{journal}{EPL (Europhys. Lett.)} \bibinfo{volume}{83}
  (\bibinfo{year}{2008}) \bibinfo{pages}{28005}.
\bibitem[{Gozolchiani et~al.(2011)Gozolchiani, Havlin, and
  Yamasaki}]{gozolchiani2011emergence}
\bibinfo{author}{A.~Gozolchiani}, \bibinfo{author}{S.~Havlin},
  \bibinfo{author}{K.~Yamasaki},
\newblock \bibinfo{title}{Emergence of {E}l {N}i{\~n}o as an autonomous
  component in the climate network},
\newblock \bibinfo{journal}{Phys. Rev. Lett.} \bibinfo{volume}{107}
  (\bibinfo{year}{2011}) \bibinfo{pages}{148501}.
\bibitem[{Ludescher et~al.(2013)Ludescher, Gozolchiani, Bogachev, Bunde,
  Havlin, and Schellnhuber}]{ludescher2013improved}
\bibinfo{author}{J.~Ludescher}, \bibinfo{author}{A.~Gozolchiani},
  \bibinfo{author}{M.~I. Bogachev}, \bibinfo{author}{A.~Bunde},
  \bibinfo{author}{S.~Havlin}, \bibinfo{author}{H.~J. Schellnhuber},
\newblock \bibinfo{title}{Improved {E}l {N}i{\~n}o forecasting by cooperativity
  detection},
\newblock \bibinfo{journal}{Proc. Natl. Acad. Sci. USA} \bibinfo{volume}{110}
  (\bibinfo{year}{2013}) \bibinfo{pages}{11742--11745}.
\bibitem[{Meng et~al.(2017)Meng, Fan, Ashkenazy, and
  Havlin}]{meng2017percolation}
\bibinfo{author}{J.~Meng}, \bibinfo{author}{J.~Fan},
  \bibinfo{author}{Y.~Ashkenazy}, \bibinfo{author}{S.~Havlin},
\newblock \bibinfo{title}{Percolation framework to describe {E}l {N}i{\~n}o
  conditions},
\newblock \bibinfo{journal}{Chaos} \bibinfo{volume}{27} (\bibinfo{year}{2017})
  \bibinfo{pages}{035807}.
\bibitem[{Fan et~al.(2017)Fan, Meng, Ashkenazy, Havlin, and
  Schellnhuber}]{fan2017network}
\bibinfo{author}{J.~Fan}, \bibinfo{author}{J.~Meng},
  \bibinfo{author}{Y.~Ashkenazy}, \bibinfo{author}{S.~Havlin},
  \bibinfo{author}{H.~J. Schellnhuber},
\newblock \bibinfo{title}{Network analysis reveals strongly localized impacts
  of {E}l {N}i{\~n}o},
\newblock \bibinfo{journal}{Proc. Natl. Acad. Sci. USA} \bibinfo{volume}{114}
  (\bibinfo{year}{2017}) \bibinfo{pages}{7543--7548}.
\bibitem[{Meng et~al.(2018)Meng, Fan, Ashkenazy, Bunde, and
  Havlin}]{meng2018forecasting}
\bibinfo{author}{J.~Meng}, \bibinfo{author}{J.~Fan},
  \bibinfo{author}{Y.~Ashkenazy}, \bibinfo{author}{A.~Bunde},
  \bibinfo{author}{S.~Havlin},
\newblock \bibinfo{title}{Forecasting the magnitude and onset of {E}l
  {N}i{\~n}o based on climate network},
\newblock \bibinfo{journal}{New J. Phys.} \bibinfo{volume}{20}
  (\bibinfo{year}{2018}) \bibinfo{pages}{043036}.
\bibitem[{Boers et~al.(2014)Boers, Bookhagen, Barbosa, Marwan, Kurths, and
  Marengo}]{boers2014prediction}
\bibinfo{author}{N.~Boers}, \bibinfo{author}{B.~Bookhagen},
  \bibinfo{author}{H.~M. Barbosa}, \bibinfo{author}{N.~Marwan},
  \bibinfo{author}{J.~Kurths}, \bibinfo{author}{J.~Marengo},
\newblock \bibinfo{title}{Prediction of extreme floods in the eastern {C}entral
  {A}ndes based on a complex networks approach},
\newblock \bibinfo{journal}{Nat. Commun.} \bibinfo{volume}{5}
  (\bibinfo{year}{2014}) \bibinfo{pages}{5199}.
\bibitem[{Boers et~al.(2019)Boers, Goswami, Rheinwalt, Bookhagen, Hoskins, and
  Kurths}]{boers2019complex}
\bibinfo{author}{N.~Boers}, \bibinfo{author}{B.~Goswami},
  \bibinfo{author}{A.~Rheinwalt}, \bibinfo{author}{B.~Bookhagen},
  \bibinfo{author}{B.~Hoskins}, \bibinfo{author}{J.~Kurths},
\newblock \bibinfo{title}{Complex networks reveal global pattern of
  extreme-rainfall teleconnections},
\newblock \bibinfo{journal}{Nature} \bibinfo{volume}{566}
  (\bibinfo{year}{2019}) \bibinfo{pages}{373--377}.
\bibitem[{Stolbova et~al.(2016)Stolbova, Surovyatkina, Bookhagen, and
  Kurths}]{stolbova2016tipping}
\bibinfo{author}{V.~Stolbova}, \bibinfo{author}{E.~Surovyatkina},
  \bibinfo{author}{B.~Bookhagen}, \bibinfo{author}{J.~Kurths},
\newblock \bibinfo{title}{Tipping elements of the {I}ndian monsoon:
  {P}rediction of onset and withdrawal},
\newblock \bibinfo{journal}{Geophys. Res. Lett.} \bibinfo{volume}{43}
  (\bibinfo{year}{2016}) \bibinfo{pages}{3982--3990}.
\bibitem[{Van Der~Mheen et~al.(2013)Van Der~Mheen, Dijkstra, Gozolchiani,
  Den~Toom, Feng, Kurths, and Hernandez-Garcia}]{vandermheen2013interaction}
\bibinfo{author}{M.~Van Der~Mheen}, \bibinfo{author}{H.~A. Dijkstra},
  \bibinfo{author}{A.~Gozolchiani}, \bibinfo{author}{M.~Den~Toom},
  \bibinfo{author}{Q.~Feng}, \bibinfo{author}{J.~Kurths},
  \bibinfo{author}{E.~Hernandez-Garcia},
\newblock \bibinfo{title}{Interaction network based early warning indicators
  for the {A}tlantic {MOC} collapse},
\newblock \bibinfo{journal}{Geophys. Res. Lett.} \bibinfo{volume}{40}
  (\bibinfo{year}{2013}) \bibinfo{pages}{2714--2719}.
\bibitem[{Feng et~al.(2014)Feng, Viebahn, and Dijkstra}]{feng2014deep}
\bibinfo{author}{Q.~Y. Feng}, \bibinfo{author}{J.~P. Viebahn},
  \bibinfo{author}{H.~A. Dijkstra},
\newblock \bibinfo{title}{Deep ocean early warning signals of an {A}tlantic
  {MOC} collapse},
\newblock \bibinfo{journal}{Geophys. Res. Lett.} \bibinfo{volume}{41}
  (\bibinfo{year}{2014}) \bibinfo{pages}{6009--6015}.
\bibitem[{Feng and Dijkstra(2014)}]{feng2014north}
\bibinfo{author}{Q.~Y. Feng}, \bibinfo{author}{H.~Dijkstra},
\newblock \bibinfo{title}{Are {N}orth {A}tlantic multidecadal {SST} anomalies
  westward propagating?},
\newblock \bibinfo{journal}{Geophys. Res. Lett.} \bibinfo{volume}{41}
  (\bibinfo{year}{2014}) \bibinfo{pages}{541--546}.
\bibitem[{Zhou et~al.(2015)Zhou, Gozolchiani, Ashkenazy, and
  Havlin}]{zhou2015teleconnection}
\bibinfo{author}{D.~Zhou}, \bibinfo{author}{A.~Gozolchiani},
  \bibinfo{author}{Y.~Ashkenazy}, \bibinfo{author}{S.~Havlin},
\newblock \bibinfo{title}{Teleconnection paths via climate network direct link
  detection},
\newblock \bibinfo{journal}{Phys. Rev. Lett.} \bibinfo{volume}{115}
  (\bibinfo{year}{2015}) \bibinfo{pages}{268501}.
\bibitem[{Ying et~al.(2020)Ying, Zhou, Han, Chen, Ye, and Xue}]{ying2020rossby}
\bibinfo{author}{N.~Ying}, \bibinfo{author}{D.~Zhou}, \bibinfo{author}{Z.~Han},
  \bibinfo{author}{Q.~Chen}, \bibinfo{author}{Q.~Ye}, \bibinfo{author}{Z.~Xue},
\newblock \bibinfo{title}{Rossby waves detection in the {CO2} and temperature
  multilayer climate network},
\newblock \bibinfo{journal}{Geophys. Res. Lett.} \bibinfo{volume}{47}
  (\bibinfo{year}{2020}) \bibinfo{pages}{e2019GL086507}.
\bibitem[{Donges et~al.(2015)Donges, Heitzig, Beronov, Wiedermann, Runge, Feng,
  Tupikina, Stolbova, Donner, Marwan et~al.}]{donges2015unified}
\bibinfo{author}{J.~F. Donges}, \bibinfo{author}{J.~Heitzig},
  \bibinfo{author}{B.~Beronov}, \bibinfo{author}{M.~Wiedermann},
  \bibinfo{author}{J.~Runge}, \bibinfo{author}{Q.~Y. Feng},
  \bibinfo{author}{L.~Tupikina}, \bibinfo{author}{V.~Stolbova},
  \bibinfo{author}{R.~V. Donner}, \bibinfo{author}{N.~Marwan}, et~al.,
\newblock \bibinfo{title}{Unified functional network and nonlinear time series
  analysis for complex systems science: {T}he pyunicorn package},
\newblock \bibinfo{journal}{Chaos} \bibinfo{volume}{25} (\bibinfo{year}{2015})
  \bibinfo{pages}{113101}.
\bibitem[{Fan et~al.(2018)Fan, Meng, Ashkenazy, Havlin, and
  Schellnhuber}]{fan2018climate}
\bibinfo{author}{J.~Fan}, \bibinfo{author}{J.~Meng},
  \bibinfo{author}{Y.~Ashkenazy}, \bibinfo{author}{S.~Havlin},
  \bibinfo{author}{H.~J. Schellnhuber},
\newblock \bibinfo{title}{Climate network percolation reveals the expansion and
  weakening of the tropical component under global warming},
\newblock \bibinfo{journal}{Proc. Natl. Acad. Sci. USA} \bibinfo{volume}{115}
  (\bibinfo{year}{2018}) \bibinfo{pages}{E12128--E12134}.
\bibitem[{Quiroga et~al.(2002)Quiroga, Kreuz, and
  Grassberger}]{quiroga2002event}
\bibinfo{author}{R.~Q. Quiroga}, \bibinfo{author}{T.~Kreuz},
  \bibinfo{author}{P.~Grassberger},
\newblock \bibinfo{title}{Event synchronization: a simple and fast method to
  measure synchronicity and time delay patterns},
\newblock \bibinfo{journal}{Phys. Rev. E} \bibinfo{volume}{66}
  (\bibinfo{year}{2002}) \bibinfo{pages}{041904}.
\bibitem[{Donner et~al.(2010)Donner, Zou, Donges, Marwan, and
  Kurths}]{donner2010recurrence}
\bibinfo{author}{R.~V. Donner}, \bibinfo{author}{Y.~Zou},
  \bibinfo{author}{J.~F. Donges}, \bibinfo{author}{N.~Marwan},
  \bibinfo{author}{J.~Kurths},
\newblock \bibinfo{title}{Recurrence networks---a novel paradigm for nonlinear
  time series analysis},
\newblock \bibinfo{journal}{New J. Phys.} \bibinfo{volume}{12}
  (\bibinfo{year}{2010}) \bibinfo{pages}{033025}.
\bibitem[{Dijkstra(2005)}]{dijkstra2005nonlinear}
\bibinfo{author}{H.~A. Dijkstra}, \bibinfo{title}{Nonlinear physical
  oceanography: a dynamical systems approach to the large scale ocean
  circulation and {E}l {N}i\~{n}o}, \bibinfo{publisher}{Springer Science \&
  Business Media}, \bibinfo{year}{2005}.
\bibitem[{Clarke(2008)}]{clarke2008introduction}
\bibinfo{author}{A.~J. Clarke}, \bibinfo{title}{An introduction to the dynamics
  of {E}l {N}i{\~n}o and the {S}outhern {O}scillation},
  \bibinfo{publisher}{Elsevier}, \bibinfo{year}{2008}.
\bibitem[{Duan and Wei(2013)}]{duan2013spring}
\bibinfo{author}{W.~Duan}, \bibinfo{author}{C.~Wei},
\newblock \bibinfo{title}{The `spring predictability barrier' for {ENSO}
  predictions and its possible mechanism: results from a fully coupled model},
\newblock \bibinfo{journal}{Int. J. Climatol.} \bibinfo{volume}{33}
  (\bibinfo{year}{2013}) \bibinfo{pages}{1280--1292}.
\bibitem[{Ludescher et~al.(2014)Ludescher, Gozolchiani, Bogachev, Bunde,
  Havlin, and Schellnhuber}]{ludescher2014very}
\bibinfo{author}{J.~Ludescher}, \bibinfo{author}{A.~Gozolchiani},
  \bibinfo{author}{M.~I. Bogachev}, \bibinfo{author}{A.~Bunde},
  \bibinfo{author}{S.~Havlin}, \bibinfo{author}{H.~J. Schellnhuber},
\newblock \bibinfo{title}{Very early warning of next {E}l {N}i{\~n}o},
\newblock \bibinfo{journal}{Proc. Natl. Acad. Sci. USA} \bibinfo{volume}{111}
  (\bibinfo{year}{2014}) \bibinfo{pages}{2064--2066}.
\bibitem[{Kirtman(2003)}]{kirtman2003cola}
\bibinfo{author}{B.~P. Kirtman},
\newblock \bibinfo{title}{The {COLA} anomaly coupled model: {E}nsemble {ENSO}
  prediction},
\newblock \bibinfo{journal}{Mon. Weather Rev.} \bibinfo{volume}{131}
  (\bibinfo{year}{2003}) \bibinfo{pages}{2324--2341}.
\bibitem[{Chen et~al.(2004)Chen, Cane, Kaplan, Zebiak, and
  Huang}]{chen2004predictability}
\bibinfo{author}{D.~Chen}, \bibinfo{author}{M.~A. Cane},
  \bibinfo{author}{A.~Kaplan}, \bibinfo{author}{S.~E. Zebiak},
  \bibinfo{author}{D.~Huang},
\newblock \bibinfo{title}{Predictability of {E}l {N}i{\~n}o over the past 148
  years},
\newblock \bibinfo{journal}{Nature} \bibinfo{volume}{428}
  (\bibinfo{year}{2004}) \bibinfo{pages}{733--736}.
\bibitem[{Larkin and Harrison(2005)}]{larkin2005definition}
\bibinfo{author}{N.~K. Larkin}, \bibinfo{author}{D.~Harrison},
\newblock \bibinfo{title}{On the definition of {E}l {N}i{\~n}o and associated
  seasonal average {US} weather anomalies},
\newblock \bibinfo{journal}{Geophys. Res. Lett.} \bibinfo{volume}{32}
  (\bibinfo{year}{2005}) \bibinfo{pages}{L13705}.
\bibitem[{Kao and Yu(2009)}]{kao2009contrasting}
\bibinfo{author}{H.-Y. Kao}, \bibinfo{author}{J.-Y. Yu},
\newblock \bibinfo{title}{Contrasting eastern-{P}acific and central-{P}acific
  types of {ENSO}},
\newblock \bibinfo{journal}{J. Clim.} \bibinfo{volume}{22}
  (\bibinfo{year}{2009}) \bibinfo{pages}{615--632}.
\bibitem[{Ashok et~al.(2007)Ashok, Behera, Rao, Weng, and
  Yamagata}]{ashok2007nino}
\bibinfo{author}{K.~Ashok}, \bibinfo{author}{S.~K. Behera},
  \bibinfo{author}{S.~A. Rao}, \bibinfo{author}{H.~Weng},
  \bibinfo{author}{T.~Yamagata},
\newblock \bibinfo{title}{El {N}i{\~n}o {M}odoki and its possible
  teleconnection},
\newblock \bibinfo{journal}{J. Geophys. Res. Oceans} \bibinfo{volume}{112}
  (\bibinfo{year}{2007}) \bibinfo{pages}{C11007}.
\bibitem[{Lu et~al.(2020)Lu, Yuan, Chen, and Gong}]{lu2020impacts}
\bibinfo{author}{Z.~Lu}, \bibinfo{author}{N.~Yuan}, \bibinfo{author}{L.~Chen},
  \bibinfo{author}{Z.~Gong},
\newblock \bibinfo{title}{On the impacts of {E}l {N}i{\~n}o events: {A} new
  monitoring approach using complex network analysis},
\newblock \bibinfo{journal}{Geophys. Res. Lett.} \bibinfo{volume}{47}
  (\bibinfo{year}{2020}) \bibinfo{pages}{e2019GL086533}.
\bibitem[{Roxy et~al.(2017)Roxy, Ghosh, Pathak, Athulya, Mujumdar, Murtugudde,
  Terray, and Rajeevan}]{roxy2017threefold}
\bibinfo{author}{M.~K. Roxy}, \bibinfo{author}{S.~Ghosh},
  \bibinfo{author}{A.~Pathak}, \bibinfo{author}{R.~Athulya},
  \bibinfo{author}{M.~Mujumdar}, \bibinfo{author}{R.~Murtugudde},
  \bibinfo{author}{P.~Terray}, \bibinfo{author}{M.~Rajeevan},
\newblock \bibinfo{title}{A threefold rise in widespread extreme rain events
  over central {I}ndia},
\newblock \bibinfo{journal}{Nat. Commun.} \bibinfo{volume}{8}
  (\bibinfo{year}{2017}) \bibinfo{pages}{1--11}.
\bibitem[{Son et~al.(2020)Son, Wang, Tseng, Schuler, Becker, and
  Yoon}]{son2020climate}
\bibinfo{author}{R.~Son}, \bibinfo{author}{S.-Y.~S. Wang},
  \bibinfo{author}{W.-L. Tseng}, \bibinfo{author}{C.~W.~B. Schuler},
  \bibinfo{author}{E.~Becker}, \bibinfo{author}{J.-H. Yoon},
\newblock \bibinfo{title}{Climate diagnostics of the extreme floods in {P}eru
  during early 2017},
\newblock \bibinfo{journal}{Clim. Dyn.} \bibinfo{volume}{54}
  (\bibinfo{year}{2020}) \bibinfo{pages}{935--945}.
\bibitem[{Fischer and Knutti(2016)}]{fischer2016observed}
\bibinfo{author}{E.~M. Fischer}, \bibinfo{author}{R.~Knutti},
\newblock \bibinfo{title}{Observed heavy precipitation increase confirms theory
  and early models},
\newblock \bibinfo{journal}{Nat. Clim. Change} \bibinfo{volume}{6}
  (\bibinfo{year}{2016}) \bibinfo{pages}{986--991}.
\bibitem[{Borodina et~al.(2017)Borodina, Fischer, and
  Knutti}]{borodina2017models}
\bibinfo{author}{A.~Borodina}, \bibinfo{author}{E.~M. Fischer},
  \bibinfo{author}{R.~Knutti},
\newblock \bibinfo{title}{Models are likely to underestimate increase in heavy
  rainfall in the extratropical regions with high rainfall intensity},
\newblock \bibinfo{journal}{Geophys. Res. Lett.} \bibinfo{volume}{44}
  (\bibinfo{year}{2017}) \bibinfo{pages}{7401--7409}.
\bibitem[{Guterres(2019)}]{guterres2019stressing}
\bibinfo{author}{A.~Guterres}, \bibinfo{title}{Stressing air pollution kills 7
  million people annually, {S}ecretary-{G}eneral urges governments to build
  green economy, in message for {W}orld {E}nvironment {D}ay},
  \bibinfo{year}{2019}. \bibinfo{note}{Press release
  SG/SM/19607-ENV/DEV/1957-OBV/1887. Available at:
  \url{https://www.un.org/press/en/2019/sgsm19607.doc.htm}}.
\bibitem[{Zhang et~al.(2019)Zhang, Fan, Chen, Ashkenazy, and
  Havlin}]{zhang2019significant}
\bibinfo{author}{Y.~Zhang}, \bibinfo{author}{J.~Fan},
  \bibinfo{author}{X.~Chen}, \bibinfo{author}{Y.~Ashkenazy},
  \bibinfo{author}{S.~Havlin},
\newblock \bibinfo{title}{Significant impact of {R}ossby waves on air pollution
  detected by network analysis},
\newblock \bibinfo{journal}{Geophys. Res. Lett.} \bibinfo{volume}{46}
  (\bibinfo{year}{2019}) \bibinfo{pages}{12476--12485}.
\bibitem[{Scheffer(2009)}]{scheffer2009critical}
\bibinfo{author}{M.~Scheffer}, \bibinfo{title}{Critical transitions in nature
  and society}, \bibinfo{publisher}{Princeton University Press},
  \bibinfo{year}{2009}.
\bibitem[{Stanley(1971)}]{stanley1971phase}
\bibinfo{author}{H.~E. Stanley}, \bibinfo{title}{Introduction to phase
  transitions and critical phenomena}, \bibinfo{publisher}{Oxford University
  Press}, \bibinfo{year}{1971}.
\bibitem[{Peters and Neelin(2006)}]{peters2006critical}
\bibinfo{author}{O.~Peters}, \bibinfo{author}{J.~D. Neelin},
\newblock \bibinfo{title}{Critical phenomena in atmospheric precipitation},
\newblock \bibinfo{journal}{Nat. Phys.} \bibinfo{volume}{2}
  (\bibinfo{year}{2006}) \bibinfo{pages}{393--396}.
\bibitem[{Privman(1990)}]{privman1990finite}
\bibinfo{author}{V.~Privman}, \bibinfo{title}{Finite size scaling and numerical
  simulation of statistical systems}, \bibinfo{publisher}{World Scientific},
  \bibinfo{year}{1990}.
\bibitem[{Held and Soden(2006)}]{held2006robust}
\bibinfo{author}{I.~M. Held}, \bibinfo{author}{B.~J. Soden},
\newblock \bibinfo{title}{Robust responses of the hydrological cycle to global
  warming},
\newblock \bibinfo{journal}{J. Clim.} \bibinfo{volume}{19}
  (\bibinfo{year}{2006}) \bibinfo{pages}{5686--5699}.
\bibitem[{Fu et~al.(2006)Fu, Johanson, Wallace, and Reichler}]{fu2006enhanced}
\bibinfo{author}{Q.~Fu}, \bibinfo{author}{C.~M. Johanson},
  \bibinfo{author}{J.~M. Wallace}, \bibinfo{author}{T.~Reichler},
\newblock \bibinfo{title}{Enhanced mid-latitude tropospheric warming in
  satellite measurements},
\newblock \bibinfo{journal}{Science} \bibinfo{volume}{312}
  (\bibinfo{year}{2006}) \bibinfo{pages}{1179--1179}.
\bibitem[{Lu et~al.(2007)Lu, Vecchi, and Reichler}]{lu2007expansion}
\bibinfo{author}{J.~Lu}, \bibinfo{author}{G.~A. Vecchi},
  \bibinfo{author}{T.~Reichler},
\newblock \bibinfo{title}{Expansion of the {H}adley cell under global warming},
\newblock \bibinfo{journal}{Geophys. Res. Lett.} \bibinfo{volume}{34}
  (\bibinfo{year}{2007}) \bibinfo{pages}{L06805}.
\bibitem[{Seo et~al.(2014)Seo, Frierson, and Son}]{seo2014mechanism}
\bibinfo{author}{K.-H. Seo}, \bibinfo{author}{D.~M. Frierson},
  \bibinfo{author}{J.-H. Son},
\newblock \bibinfo{title}{A mechanism for future changes in {H}adley
  circulation strength in {CMIP5} climate change simulations},
\newblock \bibinfo{journal}{Geophys. Res. Lett.} \bibinfo{volume}{41}
  (\bibinfo{year}{2014}) \bibinfo{pages}{5251--5258}.
\bibitem[{Mitas and Clement(2005)}]{mitas2005has}
\bibinfo{author}{C.~M. Mitas}, \bibinfo{author}{A.~Clement},
\newblock \bibinfo{title}{Has the {H}adley cell been strengthening in recent
  decades?},
\newblock \bibinfo{journal}{Geophys. Res. Lett.} \bibinfo{volume}{32}
  (\bibinfo{year}{2005}) \bibinfo{pages}{L03809}.
\bibitem[{Stauffer and Aharony(2003)}]{aharony2003introduction}
\bibinfo{author}{D.~Stauffer}, \bibinfo{author}{A.~Aharony},
  \bibinfo{title}{Introduction to percolation theory},
  \bibinfo{publisher}{Taylor \& Francis}, \bibinfo{year}{2003}.
\bibitem[{Fan et~al.(2020)Fan, Meng, Liu, Saberi, Kurths, and
  Nagler}]{fan2020universal}
\bibinfo{author}{J.~Fan}, \bibinfo{author}{J.~Meng}, \bibinfo{author}{Y.~Liu},
  \bibinfo{author}{A.~A. Saberi}, \bibinfo{author}{J.~Kurths},
  \bibinfo{author}{J.~Nagler},
\newblock \bibinfo{title}{Universal gap scaling in percolation},
\newblock \bibinfo{journal}{Nat. Phys.} \bibinfo{volume}{16}
  (\bibinfo{year}{2020}) \bibinfo{pages}{455--461}.
\bibitem[{Kossin et~al.(2014)Kossin, Emanuel, and Vecchi}]{kossin2014poleward}
\bibinfo{author}{J.~P. Kossin}, \bibinfo{author}{K.~A. Emanuel},
  \bibinfo{author}{G.~A. Vecchi},
\newblock \bibinfo{title}{The poleward migration of the location of tropical
  cyclone maximum intensity},
\newblock \bibinfo{journal}{Nature} \bibinfo{volume}{509}
  (\bibinfo{year}{2014}) \bibinfo{pages}{349--352}.
\bibitem[{Dijkstra et~al.(2019)Dijkstra, Petersik, Hernandez-Garcia, and
  Lopez}]{dijkstra2019application}
\bibinfo{author}{H.~Dijkstra}, \bibinfo{author}{P.~Petersik},
  \bibinfo{author}{E.~Hernandez-Garcia}, \bibinfo{author}{C.~Lopez},
\newblock \bibinfo{title}{The application of machine learning techniques to
  improve {E}l {N}i{\~n}o prediction skill},
\newblock \bibinfo{journal}{Front. Phys.} \bibinfo{volume}{7}
  (\bibinfo{year}{2019}) \bibinfo{pages}{153}.
\bibitem[{Luo et~al.(2017)Luo, Steptoe, Chang, Link, Clarke, and
  Maciejewski}]{luo2017impact}
\bibinfo{author}{W.~Luo}, \bibinfo{author}{M.~Steptoe},
  \bibinfo{author}{Z.~Chang}, \bibinfo{author}{R.~Link},
  \bibinfo{author}{L.~Clarke}, \bibinfo{author}{R.~Maciejewski},
\newblock \bibinfo{title}{Impact of spatial scales on the intercomparison of
  climate scenarios},
\newblock \bibinfo{journal}{IEEE Comput. Graph. Appl.} \bibinfo{volume}{37}
  (\bibinfo{year}{2017}) \bibinfo{pages}{40--49}.
\bibitem[{Berezovsky(2019)}]{berezovsky2019structure}
\bibinfo{author}{J.~Berezovsky},
\newblock \bibinfo{title}{The structure of musical harmony as an ordered phase
  of sound: {A} statistical mechanics approach to music theory},
\newblock \bibinfo{journal}{Sci. Adv.} \bibinfo{volume}{5}
  (\bibinfo{year}{2019}) \bibinfo{pages}{eaav8490}.
\bibitem[{Kim et~al.(2014)Kim, Son, and Jeong}]{kim2014large}
\bibinfo{author}{D.~Kim}, \bibinfo{author}{S.-W. Son},
  \bibinfo{author}{H.~Jeong},
\newblock \bibinfo{title}{Large-scale quantitative analysis of painting arts},
\newblock \bibinfo{journal}{Sci. Rep.} \bibinfo{volume}{4}
  (\bibinfo{year}{2014}) \bibinfo{pages}{7370}.
\bibitem[{Laws(2002)}]{laws2002physics}
\bibinfo{author}{K.~Laws}, \bibinfo{title}{Physics and the {A}rt of {D}ance:
  {U}nderstanding {M}ovement}, \bibinfo{publisher}{Oxford University Press},
  \bibinfo{year}{2002}.
\bibitem[{Gandica et~al.(2021)Gandica, Lansing, Chung, Thurner, and
  Chew}]{gandica2021bali}
\bibinfo{author}{Y.~Gandica}, \bibinfo{author}{J.~S. Lansing},
  \bibinfo{author}{N.~N. Chung}, \bibinfo{author}{S.~Thurner},
  \bibinfo{author}{L.~Y. Chew}, \bibinfo{title}{Bali's ancient rice terraces:
  {A} {H}amiltonian approach}, \bibinfo{year}{2021}. \bibinfo{note}{{e}-print
  arXiv:2103.04466}.
\bibitem[{Ahn et~al.(2011)Ahn, Ahnert, Bagrow, and
  Barab{\'a}si}]{ahn2011flavor}
\bibinfo{author}{Y.-Y. Ahn}, \bibinfo{author}{S.~E. Ahnert},
  \bibinfo{author}{J.~P. Bagrow}, \bibinfo{author}{A.-L. Barab{\'a}si},
\newblock \bibinfo{title}{Flavor network and the principles of food pairing},
\newblock \bibinfo{journal}{Sci. Rep.} \bibinfo{volume}{1}
  (\bibinfo{year}{2011}) \bibinfo{pages}{196}.
\bibitem[{Lee et~al.(2010)Lee, Ramenzoni, and Holme}]{lee2010emergence}
\bibinfo{author}{S.~Lee}, \bibinfo{author}{V.~C. Ramenzoni},
  \bibinfo{author}{P.~Holme},
\newblock \bibinfo{title}{Emergence of collective memories},
\newblock \bibinfo{journal}{PLOS ONE} \bibinfo{volume}{5}
  (\bibinfo{year}{2010}) \bibinfo{pages}{e12522}.
\bibitem[{Braha(2012)}]{braha2012global}
\bibinfo{author}{D.~Braha},
\newblock \bibinfo{title}{Global civil unrest: contagion, self-organization,
  and prediction},
\newblock \bibinfo{journal}{PLOS ONE} \bibinfo{volume}{7}
  (\bibinfo{year}{2012}) \bibinfo{pages}{e48596}.

\end{thebibliography}



\end{document}